\documentclass[12pt]{article}
\usepackage{graphicx}
\usepackage{aas_macros}
\usepackage[english]{babel}
\usepackage{latexsym}

\usepackage{amsmath}
\usepackage{amsfonts}
\usepackage{amssymb}

\newcommand{\sbf}[1]{\mbox{{\scriptsize$\bf{#1}$}}}
\def\lfrac#1#2{{#1/#2}}
\def\E{{\mathcal E}}
\def\Ds{D_s}

\begin{document}

\title{Electron-positron pairs in physics and astrophysics: from heavy nuclei to black holes}

\author{Remo Ruffini$^{1,2,3}$, Gregory Vereshchagin$^1$ and She-Sheng Xue$^1$ \\ \textit{{\small $^1$ ICRANet and ICRA, p.le della Repubblica 10, 65100 Pescara, Italy,}} \\\textit{{\small $^2$ Dip. di Fisica, Universit\`a di Roma ``La Sapienza'', Piazzale Aldo Moro 5, I-00185 Roma, Italy,}}
\\\textit{{\small $^3$ ICRANet, Universit\'e de Nice Sophia Antipolis, Grand Ch\^ateau, BP 2135, 28, avenue de Valrose,}}
\\\textit{{\small 06103 NICE CEDEX 2, France.}}}

\date{}
\thispagestyle{empty}
\maketitle

\begin{abstract}
Due to the interaction of physics and astrophysics we are witnessing in these
years a splendid synthesis of theoretical, experimental and observational
results originating from three fundamental physical processes. They were
originally proposed by Dirac, by Breit and Wheeler and by Sauter, Heisenberg,
Euler and Schwinger. For almost seventy years they have all three been followed
by a continued effort of experimental verification on Earth-based experiments.
The Dirac process, $e^+e^-\rightarrow2\gamma$, has been by far the most
successful. It has obtained extremely accurate experimental verification and
has led as well to an enormous number of new physics in possibly one of the
most fruitful experimental avenues by introduction of storage rings in Frascati
and followed by the largest accelerators worldwide: DESY, SLAC etc. The
Breit--Wheeler process, $2\gamma\rightarrow e^+e^-$, although conceptually
simple, being the inverse process of the Dirac one, has been by far one of the
most difficult to be verified experimentally. Only recently, through the
technology based on free electron X-ray laser and its numerous applications in
Earth-based experiments, some first indications of its possible verification
have been reached. The vacuum polarization process in strong electromagnetic
field, pioneered by Sauter, Heisenberg, Euler and Schwinger, introduced the
concept of critical electric field $E_c=m_e^2c^3/(e\hbar)$. It has been
searched without success for more than forty years by heavy-ion collisions in
many of the leading particle accelerators worldwide.

The novel situation today is that these same processes can be
studied on a much more grandiose scale during the gravitational
collapse leading to the formation of a black hole being observed in
Gamma Ray Bursts (GRBs). This report is dedicated  to the scientific
race. The theoretical and experimental work developed in Earth-based
laboratories is confronted with the theoretical interpretation of
space-based observations of phenomena originating on cosmological
scales. What has become clear in the last ten years is that all the
three above mentioned processes, duly extended in the general
relativistic framework, are necessary for the understanding of the
physics of the gravitational collapse to a black hole. Vice versa,
the natural arena where these processes can be observed in mutual
interaction and on an unprecedented scale, is indeed the realm of
relativistic astrophysics.

We systematically analyze the conceptual developments which have followed the
basic work of Dirac and Breit--Wheeler. We also recall how the seminal work of
Born and Infeld inspired the work by Sauter, Heisenberg and Euler on effective
Lagrangian leading to the estimate of the rate for the process of
electron--positron production in a constant electric field. In addition of
reviewing the intuitive semi-classical treatment of quantum mechanical
tunneling for describing the process of electron--positron production, we
recall the calculations in \emph{Quantum Electro-Dynamics} of the Schwinger
rate and effective Lagrangian for constant electromagnetic fields. We also
review the electron--positron production in both time-alternating
electromagnetic fields, studied by Brezin, Itzykson, Popov, Nikishov and
Narozhny, and the corresponding processes relevant for pair production at the
focus of coherent laser beams as well as electron beam--laser collision. We
finally report some current developments based on the general JWKB approach
which allows to compute the Schwinger rate in spatially varying and time
varying electromagnetic fields.

We also recall the pioneering work of Landau and Lifshitz, and Racah on the
collision of charged particles as well as experimental success of AdA and ADONE
in the production of electron--positron pairs.

We then turn to the possible experimental verification of these phenomena. We
review: (A) the experimental verification of the $e^+e^-\rightarrow 2\gamma$
process studied by Dirac. We also briefly recall the very successful
experiments of $e^+e^-$ annihilation to hadronic channels, in addition to the
Dirac electromagnetic channel; (B) ongoing Earth based experiments to detect
electron--positron production in strong fields by focusing coherent laser beams
and by electron beam--laser collisions; and (C) the multiyear attempts to
detect electron--positron production in Coulomb fields for a large atomic
number $Z>137$ in heavy ion collisions. These attempts follow the classical
theoretical work of Popov and Zeldovich, and Greiner and their schools.

We then turn to astrophysics. We first review the basic work on the energetics
and electrodynamical properties of an electromagnetic black hole and the
application of the Schwinger formula around Kerr--Newman black holes as
pioneered by Damour and Ruffini. We only focus on black hole masses larger than
the critical mass of neutron stars, for convenience assumed to coincide with
the Rhoades and Ruffini upper limit of 3.2 $M_\odot$. In this case the electron
Compton wavelength is much smaller than the spacetime curvature and all
previous results invariantly expressed can be applied following well
established rules of the equivalence principle. We derive the corresponding
rate of electron--positron pair production and introduce the concept of
dyadosphere. We review recent progress in describing the evolution of optically
thick electron--positron plasma in presence of supercritical electric field,
which is relevant both in astrophysics as well as ongoing laser beam
experiments. In particular we review recent progress based on the
Vlasov-Boltzmann-Maxwell equations to study the feedback of the created
electron--positron pairs on the original constant electric field. We evidence
the existence of plasma oscillations and its interaction with photons leading
to energy and number equipartition of photons, electrons and positrons. We
finally review the recent progress obtained by using the Boltzmann equations to
study the evolution of an electron--positron-photon plasma towards thermal
equilibrium and determination of its characteristic timescales. The crucial
difference introduced by the correct evaluation of the role of two and three
body collisions, direct and inverse, is especially evidenced. We then present
some general conclusions.

The results reviewed in this report are going to be submitted to decisive tests
in the forthcoming years both in physics and astrophysics. To mention only a
few of the fundamental steps in testing in physics we recall the starting of
experimental facilities at the National Ignition Facility at the Lawrence
Livermore National Laboratory as well as corresponding French Laser the Mega
Joule project. In astrophysics these results will be tested in galactic and
extragalactic black holes observed in binary X-ray sources, active galactic
nuclei, microquasars and in the process of gravitational collapse to a neutron
star and also of two neutron stars to a black hole giving origin to GRBs. The
astrophysical description of the stellar precursors and the initial physical
conditions leading to a gravitational collapse process will be the subject of a
forthcoming report. As of today no theoretical description has yet been found
to explain either the emission of the remnant for supernova or the formation of
a charged black hole for GRBs. Important current progress toward the
understanding of such phenomena as well as of the electrodynamical structure of
neutron stars, the supernova explosion and the theories of GRBs will be
discussed in the above mentioned forthcoming report. What is important to
recall at this stage is only that both the supernovae and GRBs processes are
among the most energetic and transient phenomena ever observed in the Universe:
a supernova can reach energy of $\sim 10^{54}$ ergs on a time scale of a few
months and GRBs can have emission of up to $\sim 10^{54}$ ergs in a time scale
as short as of a few seconds. The central role of neutron stars in the
description of supernovae, as well as of black holes and the electron--positron
plasma, in the description of GRBs, pioneered by one of us (RR) in 1975, are
widely recognized. Only the theoretical basis to address these topics are
discussed in the present report.
\end{abstract}
%\begin{keyword}
%Critical field \sep pair production \sep plasma oscillations \sep
%black holes
%\PACS05.60.Gg \sep25.75.Dw \sep52.27.Ep \sep95.30.Cq
%\end{keyword}

\tableofcontents

\section{Introduction}

The annihilation of electron--positron pair into two photons, and its inverse
process -- the production of electron--positron pair by the collision of two
photons were first studied in the framework of quantum mechanics by P.A.M.
Dirac and by G. Breit and J.A. Wheeler in the 1930s
\cite{1930PCPS...26..361D,1934PhRv...46.1087B}.

A third fundamental  process was pioneered by the work of Fritz Sauter and
Oscar Klein, pointing to the possibility of creating an electron--positron pair
from the vacuum in a constant electromagnetic field. This became known as the
`Klein paradox' and such a process named as {\it vacuum polarization}. It would
occur for an electric field stronger than the critical value
\begin{equation}
E_c\equiv{\frac{m_e^2c^3}{ e\hbar}}\simeq 1.3\cdot 10^{16}\, {\rm V/cm}.
\label{critical1}
\end{equation}
where $m_e$, $e$, $c$ and $\hbar$ are respectively the electron mass and charge, the speed of light and the Planck's constant.

The experimental difficulties to verify the existence of such three processes
became immediately clear. While the process studied by Dirac was almost
immediately observed \cite{1934PCPS...30..347K} and the electron--positron
collisions became possibly the best tested and prolific phenomenon ever
observed in physics. The Breit--Wheeler process, on the contrary, is still
today waiting a direct observational verification. Similarly the vacuum
polarization process defied dedicated attempts for almost fifty years in
experiments in nuclear physics laboratories and accelerators all over the
world, see Section \ref{chap-pair-application}.

From the theoretical point of view the conceptual changes implied by these
processes became immediately clear. They were by vastness and depth only
comparable to the modifications of the linear gravitational theory of Newton
introduced by the nonlinear general relativistic equations of Einstein. In the
work of Euler, Oppenheimer and Debye, Born and his school it became clear that
the existence of the Breit--Wheeler process was conceptually modifying the
linearity of the Maxwell theory. In fact the creation of the electron--positron
pair out of the two photons modifies the concept of superposition of the linear
electromagnetic Maxwell equations and impose the necessity to transit to a
nonlinear theory of electrodynamics. In a certain sense the Breit--Wheeler
process was having for electrodynamics the same fundamental role of
Gedankenexperiment that the equivalence principle had for gravitation. Two
different attempts to study these nonlinearities in the electrodynamics were
made: one by Born and Infeld
\cite{1933Natur.132..282B,1934RSPSA.143..410B,1934RSPSA.144..425B} and one by
Euler and Heisenberg \cite{1936ZPhy...98..714H}. These works prepared the even
greater revolution of Quantum Electro-Dynamics by Tomonaga
\cite{1946PThPh...1...27T}, Feynman
\cite{1948RvMP...20..367F,1949PhRv...76..749F,1949PhRv...76..769F}, Schwinger
\cite{1948PhRv...74.1439S,1949PhRv...75..651S,1949PhRv...76..790S} and Dyson
\cite{1949PhRv...75..486D,1949PhRv...75.1736D}.

In Section \ref{pairproduction} we review the fundamental contributions to the
electron--positron pair creation and annihilation and to the concept of the
critical electric field. In Section \ref{Diracep} we review the Dirac
derivation \cite{1930PCPS...26..361D} of the electron--positron annihilation
process obtained within the perturbation theory in the framework of
relativistic quantum mechanics and his derivation of the classical formula for
the cross-section $\sigma_{e^+e^-}^{\rm{lab}}$ in the rest frame of the
electron
\begin{equation}
\sigma_{e^+e^-}^{\rm{lab}}=\pi\left(\frac{\alpha\hbar}{m_e\,c}\right)^{2}(\hat\gamma-1)^{-1}\left\{\frac{\hat\gamma^2+4\hat\gamma+1}{\hat\gamma^2-1}\ln [\hat\gamma+(\hat\gamma^2-1)^{1/2}]-\frac{\hat\gamma+3}{(\hat\gamma^{2}-1)^{1/2}}\right\},\nonumber
\end{equation}
where $\hat\gamma\equiv{\mathcal E_{+}}/m_e\,c^{2}\geq1$ is the energy of the
positron and $\alpha=e^2/(\hbar c)$ is as usual the fine structure constant,
and we recall the corresponding formula for the center of mass reference frame.
In Section \ref{BW} we recall the main steps in the classical Breit--Wheeler
work \cite{1934PhRv...46.1087B} on the production of a real electron--positron
pair in the collision of two photons, following the same method used by Dirac
and leading to the evaluation of the total cross-section
$\sigma_{\gamma\gamma}$ in the center of mass of the system
\begin{equation}
\sigma_{\gamma\gamma} = \frac{\pi}{2}\left(\frac{\alpha\hbar}{m_e\,c}\right)^2
(1-\hat\beta^2)\Big[ 2\hat\beta(\hat\beta^2-2)+(3-\hat\beta^4)\ln \big(\frac{1+\hat\beta}{1-\hat\beta}\big)\Big],
\quad\text{with}\quad\hat\beta=\frac{c|{\bf p}|}{{\mathcal E}},\nonumber
\end{equation}
where $\hat\beta$ is the reduced velocity of the electron or the positron. In
Section \ref{higher} we recall the basic higher order processes, compared to
the Dirac and Breit--Wheeler ones, leading to pair creation. In Section
\ref{sauter} we recall the famous Klein paradox
\cite{1929ZPhy...53..157K,1931ZPhy...73..547S} and the possible tunneling
between the positive and negative energy states leading to the concept of level
crossing and pair creation by analogy to the Gamow tunneling \cite{Gamow1931}
in the nuclear potential barrier. We then turn to the celebrated Sauter work
\cite{1931ZPhy...69..742S} showing the possibility of creating a pair in a
uniform electric field $E$. We recover in Section \ref{includingE} a JWKB
approximation in order to reproduce and improve on the Sauter result by
obtaining the classical Sauter exponential term as well as the prefactor
\begin{equation}              \!\!\!\!\!\!
 \frac{\Gamma _{\rm JWKB}}{V}\simeq D_s\frac{\alpha E^2}{ 2\pi^2\hbar}
e^{-\lfrac{\pi E_c}{E}},\nonumber
\end{equation}
where $D_s=2$ for a spin-$1/2$ particle and $D_s=1$ for spin-$0$,
$V$ is the volume. Finally, in Section \ref{includingB} the case of
a simultaneous presence of an electric and a magnetic field $B$ is
presented leading to the estimate of pair production rate
\begin{equation}
%\!\!\!\!\!\!\!\!\!\!\!\!\!
\frac{\Gamma _{\rm JWKB}}{V}\simeq
\frac{\alpha   \beta \varepsilon  }{\pi\hbar }\coth\left(\frac{\pi  \beta }{ \varepsilon }\right)
\exp\left(-\frac{\pi E_c}{ \varepsilon }\right),\quad {\rm  spin-1/2 \hskip0.2cm particle} \nonumber
\end{equation}
and
\begin{equation}
%\!\!\!\!\!\!\!\!\!\!\!\!\!
\frac{\Gamma _{\rm JWKB}}{V}\simeq  \frac{\alpha  \beta  \varepsilon }{2\pi\hbar }
\sinh^{-1}\left(\frac{\pi  \beta }{ \varepsilon }\right)
\exp\left(-\frac{\pi E_c}{ \varepsilon }\right),\quad {\rm  spin-0 \hskip0.2cm particle},\nonumber
\end{equation}
where
\begin{eqnarray}
 \!\!\!\!\!\!\!\!\!\!\!\!
\varepsilon & \equiv\!&
\sqrt{(S^2+P^2)^{1/2}+ S},\nonumber \\
\beta & \equiv\!&
\sqrt{(S^2+P^2)^{1/2}- S},\nonumber
\end{eqnarray}
where the scalar $S$ and the pseudoscalar $P$ are
\begin{equation}    \!\!\!\!\!\!
 S \equiv
\frac{1}{4}F_{\mu \nu }F^{\mu \nu }=
\frac{1}{2}({\bf E}^2-{\bf B}^2);
\quad
  P\equiv
\frac{1}{4}F_{\mu \nu }\tilde F^{\mu \nu }=
 {\bf E}\cdot {\bf B},\nonumber
\end{equation}
where $\tilde F^{\mu \nu }\equiv  \epsilon ^{\mu \nu  \lambda  \kappa }F_{ \lambda  \kappa }$ is the dual field tensor.

In Section \ref{nonlinearEM} we first recall the seminal work of
Hans Euler \cite{1936AnP...418..398E} pointing out for the first
time the necessity of nonlinear character of electromagnetism
introducing the classical Euler Lagrangian
\begin{equation}
{\mathcal L}=\frac{{\bf E}^2-{\bf B}^2}{8\pi}+\frac{1}{\alpha}\frac{1}{E_0^2}\left[a_E\left({\bf E}^2-{\bf B}^2\right)^2+b_E\left({\bf E}\cdot {\bf B}\right)^2\right],\nonumber
\end{equation}
where
\begin{equation}
a_E=-1/(360\pi^2), \quad b_E=-7/(360\pi^2),\nonumber
\end{equation}
a first order perturbation to the Maxwell Lagrangian. In Section
\ref{Bornwork} we review the alternative theoretical approach of
nonlinear electrodynamics by Max Born \cite{1934RSPSA.143..410B} and
his collaborators, to the more ambitious attempt to obtain the
correct nonlinear Lagrangian of Electro-Dynamics. The motivation of
Born was to attempt a theory free of divergences in the observable
properties of an elementary particle, what has become known as
`unitarian' standpoint versus the `dualistic' standpoint in
description of elementary particles and fields. We recall how the
Born Lagrangian was formulated
\begin{equation}
\mathcal{L}=\sqrt{1+2S-P^{2}}-1,\nonumber
\end{equation}
and one of the first solutions derived by Born and Infeld
\cite{1934RSPSA.144..425B}. We also recall one of the interesting aspects of
the courageous approach of Born had been to formulate this Lagrangian within a
unified theory of gravitation and electromagnetism following Einstein program.
Indeed, we also recall the very interesting solution within the Born theory
obtained by Hoffmann \cite{1935PhRv...47..877H,1937PhRv...51..765H}. Still in
the work of Born \cite{1934RSPSA.143..410B} the seminal idea of describing the
nonlinear vacuum properties of this novel electrodynamics by an effective
dielectric constant and magnetic permeability functions of the field arisen. We
then review in Section \ref{hew-effecitve} the work of Heisenberg and Euler
\cite{1936ZPhy...98..714H} adopting the general approach of Born and
generalizing to the presence of a real and imaginary part of the electric
permittivity and magnetic permeability. They obtain an integral expression of
the effective Lagrangian given by
\begin{eqnarray}
 \Delta {\mathcal L}_{\rm eff}&=&\frac{e^2}{16\pi^2\hbar c}\int^\infty_0
e^{-s}\frac{ds}{s^3}\Big[is^2\,\bar E \bar B
\frac{\cos(s[\bar E^2-\bar B^2+2i(\bar E\bar B)]^{1/2})
+{\rm c.c.}}{\cos(s[\bar E^2-\bar B^2+2i(\bar E\bar B)]^{1/2})-{\rm c.c.}}\nonumber\\
&&+ \left(\frac{m_e^2c^3}{e\hbar}\right)^2
+\frac{s^2}{3}(|\bar B|^2-|\bar E|^2)\Big],\nonumber
\end{eqnarray}
where
$\bar E,\bar B$ are the dimensionless reduced fields in the unit of the critical field $E_c$,
\begin{equation}
\bar E=\frac{|{\bf E}|}{E_c},~~~~\bar B=\frac{|{\bf B}|}{E_c}.\nonumber
\end{equation}
obtaining the real part and the crucial imaginary term which relates to the
pair production in a given electric field. It is shown how these results give
as a special case the previous result obtained by Euler (\ref{euler lagra}). In
Section \ref{weiss-effecitve} the work by Weisskopf \cite{Weisskopf1936}
working on a spin-0 field fulfilling the Klein--Gordon equation, in contrast to
the spin 1/2 field studied by Heisenberg and Euler, confirms the
Euler-Heisenberg result. Weisskopf obtains explicit expression of pair creation
in an arbitrary strong magnetic field and in an electric field described by
$\bar E$ and $\bar B$ expansion.

For the first time Heisenberg and Euler provided a description of
the vacuum properties by the characteristic scale of strong field
$E_c$ and the effective Lagrangian of nonlinear electromagnetic
fields. In 1951, Schwinger
\cite{1951PhRv...82..664S,1954PhRv...93..615S,1954PhRv...94.1362S}
made an elegant quantum field theoretic reformulation of this
discovery in the QED framework. This played an important role in
understanding the properties of the QED theory in strong
electromagnetic fields. The QED theory in strong coupling regime,
i.e., in the regime of strong electromagnetic fields, is still a
vast arena awaiting for experimental verification as well as of
further theoretical understanding.

In Section \ref{chap-pair-theory} after recalling some general properties of
QED in Section \ref{qedintro} and some basic processes in Section
\ref{qedprocesse} we proceed to the consideration of the Dirac and the
Breit--Wheeler processes in QED in Secton \ref{DiracBWqed}. Then we discuss
some higher order processes, namely double pair production in Section
\ref{doub}, electron-nucleus bremsstrahlung and pair production by a photon in
the field of a nucleus in Section \ref{bremss}, and finally pair production by
two ions in Section \ref{ionion}. In Section \ref{qedpair} the classical result
for the vacuum to vacuum decay via pair creation in uniform electric field by
Schwinger is recalled
\begin{equation} \!\!\!\!\!\!\!\!
\frac{ \Gamma }{V}
=\frac{\alpha E^2}{ \pi^2}\sum_{n=1}^\infty \frac{1}{ n^2}\exp
\left(-\frac{n\pi E_c}{ E}\right).\nonumber
\end{equation}
This formula generalizes and encompasses the previous results
reviewed in our report: the JWKB results, discussed in Section
\ref{semi}, and the Sauter exponential factor
(\ref{wkbprobability}), and the Heisenberg-Euler imaginary part of
the effective Lagrangian. We then recall the generalization of this
formula to the case of a constant electromagnetic fields. Such
results were further generalized to spatially nonuniform and
time-dependent electromagnetic fields by Nikishov
\cite{Nikishov1969}, Vanyashin and Terent'ev
\cite{1965JETP...21..375V}, Popov
\cite{1971ZhPmR..13..261P,1972JETP...34..709P,2001JETPL..74..133P},
Narozhny and Nikishov \cite{Narozhnyi:1970uv} and Batalin and
Fradkin \cite{1970TMP.....5.1080B}. We then conclude this argument
by giving the real and imaginary parts for the effective Lagrangian
for arbitrary constant electromagnetic field recently published by
Ruffini and Xue \cite{2006JKPSP}. This result generalizes the
previous result obtained by Weisskopf in strong fields. In weak
field it gives the Euler-Heisenberg effective Lagrangian. As we will
see in the Section \ref{Xray} much attention has been given
experimentally to the creation of pairs in the rapidly changing
electric fields. A fundamental contribution in this field studying
pair production rates in an oscillating electric field was given by
Brezin and Itzykson \cite{1970PhRvD...2.1191B} and we recover in
Section \ref{alternating} their main results which apply both to the
case of bosons and fermions. We recall how similar results were
independently obtained two years later by Popov
\cite{1972JETP...35..659P}. In Section \ref{lighttheoryquantum} we
recall an alternative physical process considering the quantum
theory of the interaction of free electron with the field of a
strong electromagnetic waves: an ultrarelativistic electron absorbs
multiple photons and emits only a single photon in the reaction
\cite{1996PhRvL..76.3116B}:
\begin{equation}
e+n\omega \rightarrow e' + \gamma .
\nonumber
\end{equation}
This process appears to be of the great relevance as we will see in the next
Section for the nonlinear effects originating from laser beam experiments.
Particularly important appears to be the possibility outlined by Burke et al.
\cite{1997PhRvL..79.1626B} that the high-energy photon $\gamma$ created in the
first process propagates through the laser field, it interacts with laser
photons $n\omega$ to produce an electron--positron pair
\begin{equation}
\gamma+n\omega \rightarrow e^+ +e^- .
\nonumber
\end{equation}
We also refer to the papers by Narozhny and Popov
\cite{Nikishov1964,Nikishov1964a,Nikishov1965,1967JETP...25.1135N,Nikishov1979,Nikishov1965a}
studying the dependence of this process on the status of the
polarization of the photons.

We point out the great relevance of departing from the case of the uniform
electromagnetic field originally considered by Sauter, Heisenberg and Euler,
and Schwinger. We also recall some of the classical works of Brezin and
Itzykson and Popov on time varying fields. The space variation of the field was
also considered in the classical papers of Nikishov and Narozhny as well as in
the work of Wang and Wong. Finally, we recall the work of Khriplovich
\cite{2000NCimB.115..761K} studying the vacuum polarization around a
Reissner--Nordstr\"om black hole. A more recent approach using the worldline
formalism, sometimes called the string-inspired formalism, was advanced by
Dunne and Schubert \cite{2001PhR...355...73S,2005PhRvD..72j5004D}.

In Section \ref{inhomogeneousfield}, after recalling studies of pair production
in inhomogeneous electromagnetic fields in the literature by
\cite{2005PhRvD..72j5004D, 2006PhRvD..73f5028D, 2006PhRvD..74f5015D,
2002PhRvD..65j5002K, 2006PhRvD..73f5020K, 2007PhRvD..75d5013K}, we present a
brief review of our recent work \cite{kleinert:025011} where the general
formulas for pair production rate as functions of either crossing energy level
or classical turning point, and total production rate are obtained in external
electromagnetic fields which vary either in one space direction $E(z)$ or in
time $E(t)$. In Sections \ref{semivar} and \ref{time}, these formulas are
explicitly derived in the JWKB approximation and generalized to the case of
three-dimensional electromagnetic configurations. We apply these formulas to
several cases of such inhomogeneous electric field configurations, which are
classified into two categories. In the first category, we study two cases: a
semi-confined field $E(z)\not=0$ for $z\lesssim \ell $ and the Sauter field
\begin{eqnarray}
E(z)=E_0/{\rm cosh}^2\left({z}/{\ell }\right),~~~~~\nonumber
V(z)&=&- \sigma_s\, m_ec^2\tanh\left({z}/{\ell }\right),
\nonumber
\end{eqnarray}
where $\ell$ is width in the $z$-direction, and
\begin{equation}
\sigma_s\equiv
%|
eE_0
%|
\ell /m_ec^2=(\ell /\lambda_C)(E_0/E_c).
\nonumber
\end{equation}
In these two cases the pairs produced are not confined by the
electric potential and can reach an infinite distance. The resultant
pair production rate varies as a function of space coordinate. The
result we obtained is drastically different from the Schwinger rate
in homogeneous electric fields without any boundary. We clearly show
that the approximate application of the Schwinger rate to electric
fields limited within finite size of space overestimates the total
number of pairs produced, particularly when the finite size is
comparable with the Compton wavelength $\lambda_C$, see Figs.
\ref{TUNNf} and \ref{rratef} where it is clearly shown how the rate
of pair creation far from being constant goes to zero at both
boundaries. The same situation is also found for the case of the
semi-confined field $z(z)\not=0$ for $|z|\lesssim \ell $, see
Eq.~(\ref{hfc}). In the second category, we study a linearly rising
electric field $E(z)\sim z$, corresponding to a harmonic potential
$V(z)\sim z^2$, see Figs.~\ref{sauterf}. In this case the energy
spectra of bound states are discrete and thus energy crossing levels
for tunneling are discrete. To obtain the total number of pairs
created, using the general formulas for pair production rate, we
need to sum over all discrete energy crossing levels, see
Eq.~(\ref{gwkbpb1}), provided these energy levels are not occupied.
Otherwise, the pair production would stop due to the Pauli
principle.

In Section \ref{chap-pair-application} we focus on the phenomenology of
electron--positron pair creation and annihilation experiments. There are three
different aspects which are examined: the verification of the process
(\ref{ee2gamma}) initially studied by Dirac, the process (\ref{2gammaee})
studied by Breit and Wheeler, and then the classical work of vacuum
polarization process around a supercritical nucleus, following the Sauter,
Euler, Heisenberg and Schwinger work. We first recall in Section \ref{e+e-phys}
how the process (\ref{ee2gamma}) predicted by Dirac was almost immediately
discovered by Klemperer \cite{1934PCPS...30..347K}. Following this discovery
the electron--positron collisions have become possibly the most prolific field
of research in the domain of particle physics. The crucial step experimentally
was the creation of the first electron--positron collider the ``Anello
d'Accumulazione'' (AdA) was built by the theoretical proposal of Bruno Touschek
in Frascati (Rome) in 1960 \cite{2004PhP.....6..156B}. Following the success of
AdA (luminosity $\sim 10^{25}$/(cm$^2$ sec), beam energy $\sim$0.25GeV), it was
decided to build in the Frascati National Laboratory a storage ring of the same
kind, Adone. Electron-positron colliders have been built and proposed for this
purpose all over the world (CERN, SLAC, INP, DESY, KEK and IHEP). The aim here
is just to recall the existence of this enormous field of research which
appeared following the original Dirac idea. The main cross-sections
(\ref{muon}) and (\ref{hadron}) are recalled and the diagram (Fig.
\ref{resonances}) summarizing this very great success of particle physics is
presented. While the Dirac process (\ref{ee2gamma}) has been by far one of the
most prolific in physics, the Breit--Wheeler process (\ref{2gammaee}) has been
one of the most elusive for direct observations. In Earth-bound experiments the
major effort today is directed to evidence this phenomenon in very strong and
coherent electromagnetic field in lasers. In this process collision of many
photons may lead in the future to pair creation. This topic is discussed in
Section \ref{Xray}. Alternative evidence for the Breit--Wheeler process can
come from optically thick electron--positron plasma which may be created either
in the future in Earth-bound experiments, or currently observed in
astrophysics, see Section \ref{aksenov}. One additional way to probe the
existence of the Breit--Wheeler process is by establishing in astrophysics an
upper limits to observable high-energy photons, as a function of distance,
propagating in the Universe as pioneered by Nikishov \cite{Nikishov1961}, see
Section \ref{BWastro}. We then recall in Section \ref{light} how the crucial
experimental breakthrough came from the idea of John Madey
\cite{1977PhRvL..38..892D} of self-amplified spontaneous emission in an
undulator, which results when charges interact with the synchrotron radiation
they emit \cite{2002PhRvL..88t4801T}. Such X-ray free electron lasers have been
constructed among others at DESY and SLAC and  focus energy onto a small spot
hopefully with the size of the X-ray laser wavelength $\lambda \simeq O(0.1)$nm
\cite{Nuhn2000}, and obtain a very large electric field $ E\sim 1/\lambda$,
much larger than those obtainable with any optical laser of the same power.
This technique can be used to achieve a very strong electric field near to its
critical value for observable electron--positron pair production in vacuum. No
pair can be created by a single laser beam. It is then assumed that each X-ray
laser pulse is split into two equal parts and recombined to form a standing
wave with a frequency $\omega$. We then recall how for a laser pulse with
wavelength  $\lambda$ about $1\mu m$ and the theoretical diffraction limit
$\sigma_{\rm laser}\simeq \lambda$ being reached, the critical intensity laser
beam would be
\begin{equation}
I^c_{\rm laser}=\frac{c}{ 4\pi} E_c^2 \simeq 4.6\cdot 10^{29}W/{\rm cm}^2. \nonumber
\end{equation}
In Section \ref{alternatingfield} we recall the theoretical formula for the
probability of pair production in time-alternating electric field in two
limiting cases of large frequency and small frequency. It is interesting that
in the limit of large field and small frequency the production rate approach
the one of the Sauter, Heisenberg, Euler and Schwinger, discussed in Section
\ref{chap-pair-theory}. In the following Section \ref{Xrayfree} we recall the
actually reached experimental limits quoted by Ringwald
\cite{2001PhLB..510..107R} for a X-ray laser and give a reference to the
relevant literature. In Section \ref{circpolar} we summarize some of the most
recent theoretical estimates for pair production by a circularly polarized
laser beam by Narozhny, Popov and their collaborators. In this case the field
invariants (\ref{lightlike}) are not vanishing and pair creation can be
achieved by a single laser beam. They computed the total number of
electron--positron pairs produced as a function of intensity and focusing
parameter of the laser. Particularly interesting is their analysis of the case
of two counter-propagating focused laser pulses with circular polarizations,
pair production becomes experimentally observable when the laser intensity
$I_{\rm laser}\sim 10^{26}W/{\rm cm}^2$ for each beam, which is about $1\sim 2$
orders of magnitude lower than for a single focused laser pulse, and more than
$3$ orders of magnitude lower than the critical intensity
(\ref{criticaldensity}). Equally interesting are the considerations which first
appear in treating this problem that the back reaction of the pairs created on
the field has to be taken into due account. We give the essential references
and we will see in Section \ref{time-independent} how indeed this feature
becomes of paramount importance in the field of astrophysics. We finally review
in Section \ref{lasertech} the technological situation attempting to increase
both the frequency and the intensity of laser beams.

The difficulty of evidencing the Breit--Wheeler process even when the
high-energy photon beams have a center of mass energy larger than the
energy-threshold $2m_ec^2= 1.02$ MeV was clearly recognized since the early
days. We discuss the crucial role of the effective nonlinear terms originating
in strong electromagnetic laser fields: the interaction needs not to be limited
to initial states of two photons
\cite{1962JMP.....3...59R,1971PhRvL..26.1072R}. A collective state of many
interacting laser photons occurs. We turn then in Section \ref{light} to an
even more complex and interesting procedure: the interaction of an
ultrarelativistic electron beam with a terawatt laser pulse, performed at SLAC
\cite{1996NIMPA.383..309K}, when strong electromagnetic fields are involved. A
first nonlinear Compton scattering process occurs in which the
ultrarelativistic electrons absorb multiple photons from the laser field and
emit a single photon via the process (\ref{process1}). The theory of this
process has been given in Section \ref{lighttheoryquantum}. The second is a
drastically improved Breit--Wheeler process (\ref{process2}) by which the
high-energy photon $\gamma$, created in the first process, propagates through
the laser field and interacts with laser photons $n\omega$ to produce an
electron--positron pair \cite{1997PhRvL..79.1626B}. In Section
\ref{electronlaserexp} we describe the status of this very exciting experiments
which give the first evidence for the observation in the laboratory of the
Breit--Wheeler process although in a somewhat indirect form. Having determined
the theoretical basis as well as attempts to verify experimentally the
Breit--Wheeler formula we turn in Section \ref{BWastro} to a most important
application of the Breit--Wheeler process in the framework of cosmology. As
pointed out by Nikishov \cite{Nikishov1961} the existence of background photons
in cosmology puts a stringent cutoff on the maximum trajectory of the
high-energy photons in cosmology.

Having reviewed both the theoretical and observational evidence of the Dirac
and Breit--Wheeler processes of creation and annihilation of electron--positron
pairs we turn then to one of the most conspicuous field of theoretical and
experimental physics dealing with the process of electron--positron pair
creation by vacuum polarization in the field of a heavy nuclei. This topic has
originated one of the vastest experimental and theoretical physics activities
in the last forty years, especially by the process of collisions of heavy ions.
We first review in Section \ref{Z137} the $Z=137$ catastrophe, a collapse to
the center, in semi-classical approach, following the Pomeranchuk work
\cite{Pomeranchuk1945} based on the imposing the quantum conditions on the
classical treatment of the motion of two relativistic particles in circular
orbits. We then proceed showing in Section \ref{finitez173} how the
introduction of the finite size of the nucleus, following the classical work of
Popov and Zeldovich \cite{1971SvPhU..14..673Z}, leads to the critical charge of
a nucleus of $Z_{cr}=173$ above which a bare nucleus would lead to the level
crossing between the bound state and negative energy states of electrons in the
field of a bare nucleus. We then review in Section \ref{ad} the recent
theoretical progress in analyzing the pair creation process in a Coulomb field,
taking into account radial dependence and time variability of electric field.
We finally recall in Section \ref{heavy} the attempt to use heavy-ion
collisions to form transient superheavy ``quasimolecules'': a long-lived
metastable nuclear complex with $Z>Z_{cr}$. It was expected that the two heavy
ions of charges respectively $Z_1$ and $Z_2$ with $Z_1+Z_2>Z_{cr}$ would reach
small inter-nuclear distances well within the electron's orbiting radii. The
electrons would not distinguish between the two nuclear centers and they would
evolve as if they were bounded by nuclear ``quasimolecules'' with nuclear
charge $Z_1+Z_2$. Therefore, it was expected that electrons would evolve
quasi-statically through a series of well defined nuclear ``quasimolecules''
states in the two-center field of the nuclei as the inter-nuclear separation
decreases and then increases again. When heavy-ion collision occurs the two
nuclei come into contact and some deep inelastic reaction occurs determining
the duration $\Delta t_s$ of this contact. Such ``sticking time'' is expected
to depend on the nuclei involved in the reaction and on the beam energy.
Theoretical attempts have been proposed to study the nuclear aspects of
heavy-ion collisions at energies very close to the Coulomb barrier and search
for conditions, which would serve as a trigger for prolonged nuclear reaction
times, to enhance the amplitude of pair production. The sticking time $\Delta
t_s$ should be larger than $1\sim 2\cdot 10^{-21}$ sec \cite{Greiner1999} in
order to have significant pair production. Up to now no success has been
achieved in justifying theoretically such a long sticking time. In reality the
characteristic sticking time has been found of the order of $\Delta t\sim
10^{-23}$ sec, hundred times shorter than the needed to activate the pair
creation process. We finally recall in Section \ref{heavyexp} the
Darmstadt-Brookhaven dialogue between the Orange and the Epos groups and the
Apex group at Argonne in which the claim for discovery of electron--positron
pair creation by vacuum polarization in heavy-ion collisions was finally
retracted. Out of the three fundamental processes addressed in this report, the
Dirac electron--positron annihilation and the Breit--Wheeler electron--positron
creation from two photons have found complete theoretical descriptions within
Quantum Electro-Dynamics. The first one is very likely the best tested process
in physical science, while the second has finally obtained the first indirect
experimental evidence. The third process, the one of the vacuum polarization
studied by Sauter, Euler, Heisenberg and Schwinger, presents in Earth-bound
experiments presents a situation ``terra incognita''.

We turn then to astrophysics, where, in the process of gravitational collapse
to a black hole and in its outcomes these three processes will be for the first
time verified on a much larger scale, involving particle numbers of the order
of $10^{60}$, seeing both the Dirac process and the Breit--Wheeler process at
work in symbiotic form and electron--positron plasma created from the
``blackholic energy'' during the process of gravitational collapse. It is
becoming more and more clear that the gravitational collapse process to a
Kerr--Newman black hole is possibly the most complex problem ever addressed in
physics and astrophysics. What is most important for this report is that it
gives for the first time the opportunity to see the above three processes
simultaneously at work under ultrarelativistic special and general relativistic
regimes. The process of gravitational collapse is characterized by the
timescale $\Delta t_{g}=GM/c^3\simeq 5\cdot 10^{-6}M/M_\odot$ sec and the
energy involved are of the order of $\Delta E=10^{54}M/M_\odot$ ergs. It is
clear that this is one of the most energetic and most transient phenomena in
physics and astrophysics and needs for its correct description such a highly
time varying treatment. Our approach in Section \ref{blackhole} is to gain
understanding of this process by separating the different components and
describing 1) the basic energetic process of an already formed black hole, 2)
the vacuum polarization process of an already formed black hole, 3) the basic
formula of the gravitational collapse recovering the Tolman-Oppenheimer-Snyder
solutions and evolving to the gravitational collapse of charged and uncharged
shells. This will allow among others to obtain a better understanding of the
role of irreducible mass of the black hole and the maximum blackholic energy
extractable from the gravitational collapse. We will as well address some
conceptual issues between general relativity and thermodynamics which have been
of interest to theoretical physicists in the last forty years. Of course in
these brief chapter we will be only recalling some of these essential themes
and refer to the literature where in-depth analysis can be found. In Section
\ref{testKN} we recall the Kerr--Newman metric and the associated
electromagnetic field. We then recall the classical work of Carter
\cite{Carter1968} integrating the Hamilton-Jacobi equations for charged
particle motions in the above given metric and electromagnetic field. We then
recall in Section \ref{extractbh} the introduction of the effective potential
techniques in order to obtain explicit expression for the trajectory of a
particle in a Kerr--Newman geometry, and especially the introduction of the
reversible--irreversible transformations which lead then to the
Christodoulou-Ruffini mass formula of the black hole
\begin{equation}
M^{2}c^{4}=\left(  M_{\mathrm{ir}}c^{2}+{\frac{c^{2}Q^{2}}{4GM_{\mathrm{ir}}}%
}\right)  ^{2}+{\frac{L^{2}c^{8}}{4G^{2}M_{\mathrm{ir}}^{2}}},\nonumber%
\end{equation}
where $M_{\mathrm{ir}}$ is the irreducible mass of a black hole, $Q$ and $L$
are its charge and angular momentum. We then recall in Section \ref{druffini}
the positive and negative root states of the Hamilton--Jacobi equations as well
as their quantum limit. We finally introduce in Section \ref{ruffini} the
vacuum polarization process in the Kerr--Newman geometry as derived by Damour
and Ruffini \cite{1975PhRvL..35..463D} by using a spatially orthonormal tetrad
which made the application of the Schwinger formalism in this general
relativistic treatment almost straightforward. We then recall in Section
\ref{dyadosphere} the definition of a dyadosphere in a Reissner--Nordstr\"om
geometry, a region extending from the horizon radius
\begin{eqnarray}
r_{+}&=&1.47 \cdot 10^5\mu (1+\sqrt{1-\xi^2})\hskip0.1cm {\rm cm}
\nonumber
\end{eqnarray}
out to an outer radius
\begin{align}
r^\star=\left(\frac{\hbar}{m_ec}\right)^{1/2}\left(\frac{GM}{
c^2}\right)^{1/2} \left(\frac{m_{\rm p}}{m_e}\right)^{1/2}\left(\frac{e}
{q_{\rm p}}\right)^{1/2}\left(\frac{Q}{\sqrt{G}M}\right)^{1/2}= \nonumber \\
=1.12\cdot 10^8\sqrt{\mu\xi} \hskip0.1cm {\rm cm},
\nonumber
\end{align}
where the dimensionless mass and charge parameters $\mu=\frac{M} {M_{\odot}}$,
$\xi=\frac{Q}{(M\sqrt{G})}\le 1$. In Section \ref{dyadotorus} the definition of
a dyadotorus in a Kerr--Newman metric is recalled. We have focused on the
theoretically well defined problem of pair creation in the electric field of an
already formed black hole. Having set the background for the blackholic energy
we recall some fundamental features of the dynamical process of the
gravitational collapse. In Section \ref{gravcoll} we address some specific
issues on the dynamical formation of the black hole, recalling first the
Oppenheimer-Snyder solution \cite{1939PhRv...56..455O} and then considering its
generalization to the charged nonrotating case using the classical work of W.
Israel and V. de la Cruz \cite{1966NCimB..44....1I,1967NCimA..51..744C}. In
Section \ref{TOSsolution} we recover the classical Tolman-Oppenheimer-Snyder
solution in a more transparent way than it is usually done in the literature.
In the Section \ref{collapse} we are studying using the Israel-de la Cruz
formalism the collapse of a charged shell to a black hole for selected cases of
a charged shell collapsing on itself or collapsing in an already formed
Reissner--Nordstr\"om black hole. Such elegant and powerful formalism has
allowed to obtain for the first time all the analytic equations for such large
variety of possibilities of the process of the gravitational collapse. The
theoretical analysis of the collapsing shell considered in the previous section
allows to reach a deeper understanding of the mass formula of black holes at
least in the case of a Reissner--Nordstr\"om black hole. This allows as well to
give in Section \ref{energyextr} an expression of the irreducible mass of the
black hole only in terms of its kinetic energy of the initial rest mass
undergoing gravitational collapse and its gravitational energy and kinetic
energy $T_{+}$ at the crossing of the black hole horizon $r_{+}$
\begin{equation}
M_{\mathrm{ir}}=M_{0}-\tfrac{M_{0}^{2}}{2r_{+}}+T_{+}. \nonumber%
\end{equation}
Similarly strong, in view of their generality, are the considerations in
Section \ref{subandoverbh} which indicate a sharp difference between the vacuum
polarization process in an overcritical $E\gg E_c$ and undercritical $E\ll E_c$
black hole. For $E\gg E_c$ the electron--positron plasma created will be
optically thick with average particle energy 10 MeV. For $E\ll E_c$ the process
of the radiation will be optically thin and the characteristic energy will be
of the order of $10^{21}$ eV. This argument will be further developed in a
forthcoming report. In Section \ref{bhthermodyn} we show how the expression of
the irreducible mass obtained in the previous Section leads to a theorem
establishing an upper limit to 50\% of the total mass energy initially at rest
at infinity which can be extracted from any process of gravitational collapse
independent of the details. These results also lead to some general
considerations which have been sometimes claimed in reconciling general
relativity and thermodynamics.

The conditions encountered in the vacuum polarization process around black
holes lead to a number of electron--positron pairs created of the order of
$10^{60}$ confined in the dyadosphere volume, of the order of a few hundred
times to the horizon of the black hole. Under these conditions the plasma is
expected to be optically thick and is very different from the nuclear
collisions and laser case where pairs are very few and therefore optically
thin. We turn then in Section \ref{time-independent}, to discuss a new
phenomenon: the plasma oscillations, following the dynamical evolution of pair
production in an external electric field close to the critical value. In
particular, we will examine: (i) the back reaction of pair production on the
external electric field; (ii) the screening effect of pairs on the electric
field; (iii) the motion of pairs and their interactions with the created photon
fields. In Secs. \ref{semi-classical} and \ref{kineticplasma}, we review
semi-classical and kinetic theories describing the plasma oscillations using
respectively the Dirac-Maxwell equations and the Boltzmann-Vlasov equations.
The electron--positron pairs, after they are created, coherently oscillate back
and forth giving origin to an oscillating electric field. The oscillations last
for at least a few hundred Compton times. We review the damping due to the
quantum decoherence. The energy from collective motion of the classical
electric field and pairs flows to the quantum fluctuations of these fields.
This process is quantitatively discussed by using the quantum Boltzmann-Vlasov
equation in Sections \ref{quanvlasov} and \ref {quantumdecoherence}. The
damping due to collision decoherence is quantitatively discussed in Sections
\ref{collisiondecoherence} and \ref{PO} by using Boltzmann-Vlasov equation with
particle collisions terms. This damping determines the energy flows from
collective motion of the classical electric field and pairs to the kinetic
energy of non-collective motion of particles of these fields due to collisions.
In Section \ref{PO}, we particularly address the study of the influence of the
collision processes $e^{+}e^{-}\rightleftarrows \gamma\gamma$ on the plasma
oscillations in supercritical electric field \cite{2003PhLB..559...12R}. It is
shown that the plasma oscillation is mildly affected by a small number of
photons creation in the early evolution during a few hundred Compton times (see
Fig.~\ref{Oscillation}). In the later evolution of $10^{3-4}$ Compton times,
the oscillating electric field is damped to its critical value with a large
number of photons created. An equipartition of number and energy between
electron--positron pairs and photons is reached (see Fig. \ref {Oscillation}).
In Section~\ref {electrofluidodynamics}, we introduce an approach based on the
following three equations: the number density continuity equation, the
energy-momentum conservation equation and the Maxwell equations. We describe
the plasma oscillation for both overcritical electric field $E>E_c$  and
undercritical electric field $E < E_c$ \cite{2007PhLA..371..399R}. In
additional of reviewing the result well known in the literature for $E>E_c$ we
review some novel result for the case $E<E_{c}$. It was traditionally assumed
that electron--positron pairs, created by the vacuum polarization process, move
as charged particles in external uniform electric field reaching arbitrary
large Lorentz factors. It is reviewed how recent computations show the
existence of plasma oscillations of the electron--positron pairs also for
$E\lesssim E_{c}$. For both cases we quote the maximum Lorentz factors $\gamma
_{\max }$ reached by the electrons and positrons as well as the length of
oscillations. Two specific cases are given. For $E_{0}=10E_{c}$ the length of
oscillations 10 $\hbar /(m_e c)$, and $E_{0}=0.15E_{c}$ the length of
oscillations 10$^{7}$ $\hbar /(m_e c)$. We also review the asymptotic behavior
in time, $t\rightarrow\infty$, of the plasma oscillations by the phase portrait
technique. Finally we review some recent results which differentiate the case
$E>E_{c}$ from the one $E<E_{c}$ with respect to the creation of the rest mass
of the pair versus their kinetic energy. For $E>E_{c}$ the vacuum polarization
process transforms the electromagnetic energy of the field mainly in the rest
mass of pairs, with moderate contribution to their kinetic energy.

We then turn in Section \ref{aksenov} to the last physical process needed in
ascertaining the reaching of equilibrium of an optically thick
electron--positron plasma. The average energy of electrons and positrons we
illustrate is $0.1<\epsilon<10$ MeV. These bounds are necessary from the one
hand to have significant amount of electron--positron pairs to make the plasma
optically thick, and from the other hand to avoid production of other particles
such as muons. As we will see in the next report these are indeed the relevant
parameters for the creation of ultrarelativistic regimes to be encountered in
pair creation process during the formation phase of a black hole. We then
review the problem of evolution of optically thick, nonequilibrium
electron--positron plasma, towards an equilibrium state, following
\cite{2007PhRvL..99l5003A,2008AIPC..966..191A}. These results have been mainly
obtained by two of us (RR and GV) in recent publications and all relevant
previous results are also reviewed in this Section \ref{aksenov}. We have
integrated directly relativistic Boltzmann equations with all binary and triple
interactions between electrons, positrons and photons two kinds of equilibrium
are found: kinetic and thermal ones. Kinetic equilibrium is obtained on a
timescale of few $(\sigma_T n_\pm c)^{-1}$, where $\sigma_T$ and $n_\pm$ are
Thomson's cross-section and electron--positron concentrations respectively,
when detailed balance is established between all binary interactions in plasma.
Thermal equilibrium is reached on a timescale of few $(\alpha\sigma_T n_\pm
c)^{-1}$, when all binary and triple, direct and inverse interactions are
balanced. In Section \ref{qualitplasma} basic plasma parameters are
illustrated. The computational scheme as well as the discretization procedure
are discussed in Section \ref{computsch}. Relevant conservation laws are given
in Section \ref{conslaws}. Details on binary interactions, consisting of
Compton, M{\o }ller and Bhabha scatterings, Dirac pair annihilation and
Breit--Wheeler pair creation processes, and triple interactions, consisting of
relativistic bremsstrahlung, double Compton process, radiative pair production
and three photon annihilation process, are presented in Section \ref{binaryint}
and \ref{tripleint}, respectively. In Section \ref{binaryint} collisional
integrals with binary interactions are computed from first principles, using
QED matrix elements. In Section \ref{coulcut} Coulomb scattering and the
corresponding cutoff in collisional integrals are discussed. Numerical results
are presented in Section \ref{numresp} where the time dependence of energy and
number densities as well as chemical potential and temperature of
electron--positron-photon plasma is shown, together with particle spectra. The
most interesting result of this analysis is to have differentiate the role of
binary and triple interactions. The detailed balance in binary interactions
following the classical work of Ehlers \cite{1973rela.conf....1E} leads to a
distribution function of the form of the Fermi-Dirac for electron--positron
pairs or of the Bose-Einstein for the photons. This is the reason we refer in
the text to such conditions as the Ehlers equilibrium conditions. The crucial
role of the direct and inverse three-body interactions is well summarized in
fig. \ref{ntot}, panel A from which it is clear that the inverse three-body
interactions are essential in reaching thermal equilibrium. If the latter are
neglected, the system deflates to the creation of electron--positron pairs all
the way down to the threshold of $0.5$MeV. This last result which is referred
as the Cavallo--Rees scenario \cite{1978MNRAS.183..359C} is simply due to
improper neglection of the inverse triple reaction terms.

In Section \ref{remarks} we present some general remarks.

Here and in the following we will use Latin indices running from 1
to 3, Greek indices running from 0 to 3, and we will adopt the
Einstein summation rule.

\section{The fundamental contributions to the electron--positron
pair creation and annihilation and the concept of critical
electric field}\label{pairproduction}
%Early quantum electrodynamics

In this Section we recall the annihilation process of an electron--positron
pair with the production of two photons
\begin{equation}\label{ee2gamma}
e^{+}+e^{-}\rightarrow\gamma_{1}+\gamma_{2},
\end{equation}
studied by Dirac in \cite{1930PCPS...26..361D}, the Breit--Wheeler process of
electron--positron pair production by light-light collisions
\cite{1934PhRv...46.1087B}
\begin{equation}\label{2gammaee}
\gamma_{1}+\gamma_{2}\rightarrow e^{+}+e^{-}.
\end{equation}
and the vacuum polarization in external electric field, introduced
by Sauter \cite{1931ZPhy...69..742S}. These three results, obtained
in the mid-30's of the last century
\cite{1987PhUsp..30...791M,1995eqe..book.....M}, played a crucial
role in the development of the \emph{Quantum Electro-Dynamics}
(QED).

\subsection{Dirac's electron--positron annihilation}
\label{Diracep}

Dirac had proposed his theory of the electron \cite{1928RSPSA.117..610D,1930RSPSA.126..360D} in the framework of relativistic quantum theory. Such a theory predicted the existence of positive and negative energy states. Only the positive energy states could correspond to the electrons. The negative energy states had to have a physical meaning since transitions were considered to be possible from positive to negative energy states. It was proposed by Dirac \cite{1930RSPSA.126..360D} that nearly all possible states of negative energy are occupied with just one electron in accordance with Pauli's exclusion principle and that the unoccupied states, `holes' in the negative energy states should be regarded as `positrons'\footnote{Actually initially \cite{1930PCPS...26..361D,1930RSPSA.126..360D} Dirac believed that these `holes' in negative energy spectrum describe protons, but later he realized that these holes represent particles with the same mass as of electron but with opposite charge, `anti-electrons' \cite{1931RSPSA.133...60D}. The discovery of these anti-electrons was made by Anderson in 1932 \cite{1932PhRv...41..405A} and named by him `positrons' \cite{1933PhRv...43..491A}.}. Historical review of this exciting discovery is given in \cite{1982hdqt.book.....M}.

Adopting his time-dependent perturbation theory \cite{1926RSPSA.112..661D} in
the framework of relativistic Quantum Mechanics Dirac pointed out in
\cite{1930PCPS...26..361D} the necessity of the annihilation process of
electron--positron pair into two photons (\ref{ee2gamma}). He considered an
electron under the simultaneous influence of two incident beams of radiation,
which induce transition of the electron to states of negative energy, then he
calculated the transition probability per unit time, using the well established
validity of the Einstein emission and absorption coefficients, which connect
spontaneous and stimulated emission probabilities. He obtained the explicit
expression of the cross-section of the annihilation process.

Such process is \emph{spontaneous}, i.e. it occurs necessarily for any pair of electron and positron independently of their energy. The process does not need any previously existing radiation. The derivation of the cross-section, considering the stimulated emission process, was simplified by the fact that the electromagnetic field could be treated as an external classical perturbation and did not need to be quantized \cite{1954qtr..book.....H}.

Dirac started from his wave equation \cite{1928RSPSA.117..610D} for the spinor field $\Psi$:
\begin{equation}\label{Pauli's equation}
\left\{\frac{E}{c}+\frac{e}{c}\,A_{0}+\mbox{\boldmath$\alpha$}\cdot\left({\bf p}+\frac{e}{c}\,{\bf A}\right)+\beta_D m_e c\right\}\Psi=0,
\end{equation}
where $m_e$ and $e$ are electron's mass and charge, ${\bf A}$ is electromagnetic vector potential, and the matrices $\mbox{\boldmath$\alpha$}$ and $\beta_D$ are:
\begin{equation}
\mbox{\boldmath$\alpha$} = \left(\begin{array}{cc}
  0 & \mbox{\boldmath$\sigma$}\\
  \mbox{\boldmath$\sigma$} & 0
\end{array}\right)\qquad\text{and}
\qquad
\beta_D = \left(\begin{array}{cc}
  I & 0\\
  0 & -I
\end{array}\right),
\end{equation}
where $\mbox{\boldmath$\sigma$}$ and $I$ are respectively the Pauli's and unit matrices. By choosing a gauge in which $A_{0}$ vanishes he obtained:
\begin{equation}\label{2p2epotential}
{\bf A}=
{\bf a}_{1}\,e^{i\omega_{1}\left[t-{\bf l}_{1}\cdot{\bf x}/c\right]}+
{\bf a}^{*}_{1}\,e^{-i\omega_{1}\left[t-{\bf l}_{1}\cdot{\bf x}/c\right]}+
{\bf a}_{2}\,e^{i\omega_{2}\left[t-{\bf l}_{2}\cdot{\bf x}/c\right]}+
{\bf a}^{*}_{2}\,e^{-i\omega_{2}\left[t-{\bf l}_{2}\cdot{\bf x}/c\right]},
\end{equation}
where $\omega_{1}$ and $\omega_{2}$ are respectively the frequencies of the two beams, ${\bf l}_{1}$ and ${\bf l}_{2}$ are the unit vectors in their direction of motion and ${\bf a}_{1}$ and ${\bf a}_{2}$ are the polarization vectors, the modulus of which are the amplitudes of the two beams.

Dirac solved Eq.~(\ref{Pauli's equation}) by a perturbation method,
finding a solution of the form
$\psi=\psi_{0}+\psi_{1}+\psi_{2}+\dots$, where $\psi_{0}$ is the
solution in the free case, and  $\psi_{1}$ is the first order
perturbation containing the field ${\bf A}$, or, explicitly
$-\frac{e}{c}\,\mbox{\boldmath$\alpha$}\cdot{\bf A}$. He then
computed the explicit expression of the second order expansion term
$\psi_{2}$, which represents electrons that have made the double
photon emission process and decay into negative energy states. He
evaluated the transition amplitude for the stimulated transition
process, which reads
\begin{equation}\label{DBW density}
w_{e^{+}+e^{-}\rightarrow \gamma_{1}+\gamma_{2}}=\frac{16e^{2}|{\bf a}_1|^2 |{\bf a}_2|^2}
{|{\mathcal E}'|m_e\,c^{2}} K_{12}\frac{1-\cos(\delta {\mathcal E}'t/h)}{(\delta {\mathcal E}')^{2}},
\end{equation}
where ${\mathcal E}'=m_e\,c^{2}-\nu_{1}-\nu_{2}$, $\nu_{1}$ and $\nu_{2}$ are the photons' frequencies and
\begin{equation}\label{Dirac K}
K_{12}=-({\bf m}_{1}\cdot{\bf m}_{2})^{2}+\frac{1}{4}\left[1-({\bf m}_{1}\cdot{\bf m}_{2})({\bf n}_{1}\cdot{\bf n}_{2})+({\bf m}_{1}\cdot{\bf n}_{2})({\bf m}_{1}\cdot{\bf n}_2)\right]\frac{\nu_{1}+\nu_{2}}{m_ec^{2}},
\end{equation}
is a dimensionless number depending on the unit vectors in the
directions of the two photon's polarization vectors ${\bf m}_{1}$
and ${\bf m}_{2}$. The quantities ${\bf n}_{1}$ and ${\bf n}_{2}$
are respectively given by ${\bf n}_{1,2}={\bf l}_{1,2}\times {\bf
m}_{1,2}$. Introducing the intensity of the two incident beams
\begin{equation}
I_{1}=\frac{\nu_{1}^{2}}{2\pi\,c}|k_{1}|^{2},\qquad I_{2}=\frac{\nu_{2}^{2}}{2\pi\,c}|k_{2}|^{2},
\end{equation}
where ${\bf k}_{1,2}=\omega_{1,2}{\bf l}_{1,2}$.
Dirac obtained from the above transition amplitude the transition probability
\begin{equation}
P_{e^{+}+e^{-}\rightarrow \gamma_{1}+\gamma_{2}}=\frac{8\pi^{2}c^{2}e^{4}}
{|{\mathcal E}'|m_e\,c^{2}\nu_{1}^{2}\nu_{2}^{2}}K_{12}\frac{1-\cos(\delta {\mathcal E}'t/h)}{(\delta {\mathcal E}')^{2}}.
\end{equation}
In order to evaluate the spontaneous emission probability Dirac uses the relation between the Einstein coefficients $A_E$ and $B_E$ which is of the form
\begin{equation}\label{EinCoeff}
A_E/B_E={2\pi h}/{c^{2}}(\nu_{1,2}/2\pi)^{3}.
\end{equation}
Integrating on all possible directions of emission he obtains the total
probability per unit time in the rest frame of the electron
\begin{equation}\label{Dirac cross-section}
\sigma_{e^+e^-}^{\rm{lab}}=\pi\left(\frac{\alpha\hbar}{m_e\,c}\right)^{2}(\hat\gamma-1)^{-1}\left\{\frac{\hat\gamma^2+4\hat\gamma+1}{\hat\gamma^2-1}\ln [\hat\gamma+(\hat\gamma^2-1)^{1/2}]-\frac{\hat\gamma+3}{(\hat\gamma^{2}-1)^{1/2}}\right\},
\end{equation}
where $\hat\gamma\equiv{\mathcal E_{+}}/m_e\,c^{2}\geq1$ is the energy of the
positron and $\alpha=e^2/(\hbar c)$ is the fine structure constant. Again,
historically Dirac was initially confused about the negative energy states
interpretation as we recalled. Although he derived the correct formula, he was
doubtful about the presence in it of the mass of the electron or of the mass of
the proton. Of course today this has been clarified and this derivation is
fully correct if one uses the mass of the electron and applied this formula to
description of electron--positron annihilation.
%Klein Nishina
The limit for high-energy pairs ($\hat\gamma\gg 1$) is
\begin{equation}
\sigma_{e^+e^-}^{\rm{lab}}\simeq\frac{\pi}{\hat\gamma}\left(\frac{\alpha\hbar}{m_e\,c}\right)^2\left[\ln\left(2\hat\gamma\right)-1\right];
\label{Dirac section gg1}
\end{equation}

The corresponding center of mass formula is
\begin{equation}
\sigma_{e^+ e^-} = \frac{\pi}{4\hat\beta^2}\left(\frac{\alpha\hbar}{m_e\,c}\right)^2
(1-\hat\beta^2)\Big[ 2\hat\beta(\hat\beta^2-2)+(3-\hat\beta^4)\ln \big(\frac{1+\hat\beta}{1-\hat\beta}\big)\Big],
\label{Dirac section0}
\end{equation}
where $\hat\beta$ is the reduced velocity of the electron or the
positron.

\subsection{Breit--Wheeler pair production}\label{BW}

We now turn to the equally important derivation on the production of an
electron--positron pair in the collision of two real photons given by Breit and
Wheeler \cite{1934PhRv...46.1087B}. According to Dirac's theory of the
electron, this process is caused by a transition of an electron from a negative
energy state to a positive energy under the influence of two light quanta on
the vacuum. This process differently from the one considered by Dirac, which
occurs spontaneously, has a threshold due to the fact that electron and
positron mass is not zero. In other words in the center of mass of the system
there must be  sufficient available energy to create an electron--positron
pair. This energy must be larger than twice of electron rest mass energy.

Breit and Wheeler, following the discovery of the positron by
Anderson \cite{1933PhRv...43..491A}, studied the effect of two light
waves upon an electron in a negative energy state, represented by a
normalized Dirac wave function $\psi^{(0)}$. Like in the previous
case studied by Dirac \cite{1930PCPS...26..361D} the light waves
have frequencies $\omega_i$, wave vectors ${\bf k}_i$ and vector
potentials (\ref{2p2epotential}).
%\begin{equation}\label{2p2epotential}
%{\bf A}={\bf a}^{*}e^{-i(\omega_it-{\bf k}_i\cdot {\bf x})}+{\bf a}e^{i(\omega_it-{\bf k}_i\cdot {\bf %x})},\qquad i=1,2
%\end{equation}
%The coefficients ${\bf a}^{*}$ and ${\bf a}$ are quantum mechanical amplitudes of light waves.
Under the influence of the light waves, the initial electron wave
function $\psi^{(0)}$ is changed after some time $t$ into a final
wave function $\psi^{(t)}$. The method adopted is the time-dependent
perturbation \cite{1926RSPSA.112..661D} (for details see
\cite{Landau1981a}) to solve the Dirac equation with the
time-dependent potential $e{\bf A}(t)$ (\ref{2p2epotential}). The
transition amplitude was calculated by an expansion in powers of
${\bf a}_{1,2}$ up to $O(\alpha^2)$. The wave function $\psi^{(t)}$
contains a term representing an electron in a positive energy state.
The associated density is found to be
\begin{align}\label{BW density}
w_{\gamma_{1}+\gamma_{2}\rightarrow e^{+}+e^{-}} & =\left(\frac{\alpha\hbar}{m_e\,c}\right)^2|{\bf a}_1|^2 |{\bf a}_2|^2 K_{12}\,\frac{|1-\exp (-it\delta {\mathcal E}/\hbar)|^2}{\left(\delta{\mathcal E}\right)^2}
\\
& =\left(\frac{\alpha\hbar}{m_e\,c}\right)^22|{\bf a}_1|^2 |{\bf a}_2|^2 K_{12}\,\frac{|1-\cos (\delta{\mathcal E}t/\hbar)|^2}{\left(\delta{\mathcal E}\right)^2},
\end{align}
where $K_{12}$ is the dimensionless number already obtained by Dirac,
Eq.~(\ref{Dirac K}), depending on initial momenta and spin of the wave function
$\psi^{(0)}$ and the polarizations of the quanta. This quantity is actually the
squared transition matrix in the momenta and spin of initial and final states
of light and electron--positron. The squared amplitudes $|{\bf a}_{1,2}|^2$ in
Eq.~(\ref{BW density}) are determined by the intensities $I_{1,2}$ of the two
light beams as
\begin{equation}\label{bwaa}
|{\bf a}_{1,2}|^2=\frac{2\pi c}{\omega_{1,2}^2}\,I_{1,2}.
\end{equation}
The quantity $\delta{\mathcal E}$ in Eq.~(\ref{BW density}) is the difference
in energies between initial light states and final electron--positron states.
Indicating by ${\mathcal E}^{(-)}=-c(p_1^2+m_e^2c^2)^{1/2}$, where $p_1$ is the
four-momentum of the positron, the negative energy of the electron in its
initial state and the corresponding quantity for the electron ${\mathcal
E}_{2}=-c(p_2^2+m_e^2c^2)^{1/2}$, where $p_2$ is the 4-momentum of the
electron, $\delta{\mathcal E}$ is given by
\begin{equation}\label{BW energy}
\delta{\mathcal E}=c(p_2^2+m_e^2c^2)^{1/2}+{\mathcal E}_1-\hbar\omega_1-\hbar\omega_2,\qquad\text{where}\qquad
{\mathcal E}_1=-{\mathcal E}^{(-)},
\end{equation}
and ${\bf p}_2=-{\bf p}_1+{\bf k}_1+{\bf k}_2$ is the final momentum of the electron. From this energy and momentum conservation it follows
\begin{equation}\label{BW relation}
d(\delta{\mathcal E})=c^2\Big[\frac{|{\bf p}_1|}{{\mathcal E}_1}-\frac{{\bf p}_1\cdot{\bf p}_2}{(|{\bf p}_1|{\mathcal E}_2)}\Big]dp_1.
\end{equation}
It is then possible to sum the probability densities (\ref{BW density}) over all possible initial electron states of negative energy in the volume $V$. An integral over the phase space
$\int 2|{\bf p}_1|^2d|{\bf p}_1|d\Omega_1 V/(2\pi\hbar)^3$ must be performed.
The effective collision area for the head-on collision of two light quanta was shown by Breit and Wheeler to be
\begin{equation}\label{BW section}
\sigma_{\gamma\gamma} = 2\left(\frac{\alpha\hbar}{m\,c}\right)^2\int\frac{c|{\bf p}_1|^2}{\hbar\omega_1
\hbar\omega_2}K_{12}\Big[\frac{|{\bf p}_1|}{{\mathcal E}_1}-\frac{{\bf p}_1\cdot{\bf p}_2}{|{\bf p}_1|{\mathcal E}_2}\Big]^{-1}
d\Omega_1,
\end{equation}
where $\Omega_1$ is the solid angler, which fulfills the total
energy conservation $\delta{\mathcal E}=0$.

In the center of mass of the system, the momenta of the electron and the positron are equal and opposite ${\bf p}_1=-{\bf p}_2$. In that frame the momenta of the photons in the initial state are ${\bf k}_1=-{\bf k}_2$. As a consequence, the energies of the electron and the positron are equal: ${\mathcal E}_1={\mathcal E}_2={\mathcal E}$, and so are the energies of the photons:
$\hbar\omega_1=\hbar\omega_2={\mathcal E}_\gamma={\mathcal E}$.
The total cross-section of the process is then
\begin{equation}\label{BW section1}
\sigma_{\gamma\gamma} = 2\left(\frac{\alpha\hbar}{m_e\,c}\right)^2\frac{c|{\bf p}|}{{\mathcal E}}\int K_{12} d\Omega_1,
\end{equation}
where $|{\bf p}|=|{\bf p}_1|=|{\bf p}_2|$, and ${\mathcal E}=(c^2|{\bf p}|^2+m_e^2c^4)^{1/2}$.
Therefore, the necessary kinematic condition in order for the process (\ref{2gammaee}) taking place is that the energy of the two colliding photons be larger than the threshold $2m_ec^2$, i.e.,
\begin{equation}\label{BW threshold}
{\mathcal E}_\gamma > m_e c^{2}.
\end{equation}

From Eq.~(\ref{BW section1}) the total cross-section in the center
of mass of the system is
\begin{equation}\label{BW section0}
\sigma_{\gamma\gamma} =
\frac{\pi}{2}\left(\frac{\alpha\hbar}{m_e\,c}\right)^2
(1-\hat\beta^2)\Big[ 2\hat\beta(\hat\beta^2-2)+(3-\hat\beta^4)\ln
\big(\frac{1+\hat\beta}{1-\hat\beta}\big)\Big],
\quad\text{with}\quad\hat\beta=\frac{c|{\bf p}|}{{\mathcal E}}.
\end{equation}
%This result agreed with a calculation performed by Karplus and Neumann \cite{KarNeu51}, in the framework of the scattering theory. They calculated the imaginary part of the light-light scattering amplitude up to order $O(\alpha^2)$, this order in perturbation theory gives exactly the result above.
In modern QED cross-sections (\ref{Dirac section gg1}) and (\ref{BW section0})
emerge form two tree-level Feynman diagrams (see, for example, the textbook
\cite{Itzykson2006} and Section \ref{chap-pair-theory}).

For ${\mathcal E} \gg m_ec^2$, the total effective cross-section is
approximately proportional to
\begin{equation}
\label{bwsection3}
\sigma_{\gamma\gamma} \simeq \pi\left(\frac{\alpha\hbar}{m_ec}\right)^2
\left(\frac{m_ec^2}{{\mathcal E}}\right)^2.
\end{equation}

The cross-section in line (\ref{BW section0}) can be easily
generalized to an arbitrary reference frame, in which the two
photons $k_1$ and $k_2$ cross with arbitrary relative directions.
The Lorentz invariance of the scalar product of their 4-momenta
$(k_1 k_2)$ gives $\omega_1\omega_2={\mathcal E}_\gamma^2$. Since
${\mathcal E}_\gamma={\mathcal E}=m_ec^2/\sqrt{1-\hat\beta^2}$, to
obtain the total cross-section in the arbitrary frame ${\mathcal
K}$, we must therefore make the following substitution
\cite{1982els..book.....B}
\begin{equation}\label{substitute}
\hat\beta\rightarrow \sqrt{1-m_e^2c^4/(\omega_1\omega_2)},
\end{equation}
in Eq.~(\ref{BW section0}).

\subsection{Collisional $e^+e^-$ pair creation near nuclei: Bethe and Heitler, Landau and Lifshitz, Sauter, and Racah}
\label{higher}

After having recalled in the previous sections the classical works of Dirac on
the reaction (\ref{ee2gamma}) and Breit--Wheeler on the reaction
(\ref{2gammaee}) it is appropriate to return for a moment on the discovery of
electron--positron pairs from observations of cosmic rays. The history of this
discovery sees as major actors on one side Carl Anderson
\cite{1933PhRv...43..491A} at Caltech and on the other side Patrick Maynard
Stuart Blackett and Giuseppe Occhialini \cite{1933RSPSA.139..699B} at the
Cavendish laboratory. A fascinating reconstruction of their work can be found
e.g. in \cite{1982hdqt.book.....M}. The scene was however profoundly influenced
by a fierce conceptual battle between Robert A. Millikan at Caltech and Arthur
Compton at Chicago on the mechanism of production of these cosmic rays. For a
refreshing memory of these heated discussions and a role also of Sir James
Hopwood Jeans see e.g. \cite{timeurl}. The contention by Millikan was that the
electron--positron pairs had to come from photons originating between the
stars, while Jeans located their source on the stars. Compton on the contrary
insisted on their origin from the collision of charged particles in the Earth
atmosphere. Moreover, at the same time there were indications that similar
process of charged particles would occur by the scattering of the radiation
from polonium-beryllium, see e.g. Joliot and Curie \cite{Curie1932}.

It was therefore a natural outcome that out of this scenario two major
theoretical developments occurred. One development inquired electron--positron
pair creation by the interaction of photons with nuclei following the reaction:
\begin{equation}
\gamma+Z\longrightarrow Z+e^++e^-,
\label{gi2p}
\end{equation}
major contributors were Oppenheimer and Plesset
\cite{1933PhRv...44...53O}, Heitler \cite{1933ZPhy...84..145H},
Bethe-Heitler \cite{1934RSPSA.146...83B}, Sauter
\cite{1934AnP...412..404S} and Racah \cite{Racah1934}. Heitler
\cite{1933ZPhy...84..145H} obtained an order of magnitude estimate
of the total cross-section of this process
\begin{equation}
\sigma_{Z\gamma\rightarrow Ze^+e^-}\simeq\alpha Z^2\left(\frac{e^2}{m_ec^2}\right)^2.
\label{ome}
\end{equation}
In the ultrarelativistic case $\epsilon_\pm\gg m_e$ the total cross-section for pair production by a photon with a given energy $\omega$ is \cite{1934RSPSA.146...83B}
\begin{equation}
\sigma=\frac{28}{9}Z^2\alpha r_e^2\left(\log\frac{2\omega}{m_e}-\frac{109}{42}\right).
\label{sigmaZ1Z2}
\end{equation}

\begin{figure}[!ht]
    \centering
        \includegraphics[width=5cm]{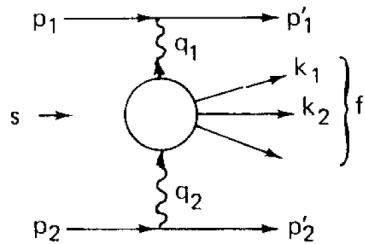}
    \caption{The sketch of two photon particle production. Reproduced from \cite{1975PhR....15..181B}.}
    \label{twophoton}
\end{figure}

The second development was the study of the reaction
\begin{equation}
Z_1+Z_2\longrightarrow Z_1+Z_2+e^++e^-.
\label{2pe+e-}
\end{equation}
with the fundamental contribution of Landau and Lifshitz
\cite{Landau1934} and Racah \cite{Racah1934,Racah1937}. This process
is an example of two photon pair production, see. Fig.
\ref{twophoton}. The 4-momenta of particles $Z_1$ and $Z_2$ are
respectively $p_1$ and $p_2$.
The total pair production cross-section is \cite{Landau1934}
\begin{equation}
\sigma_\mathrm{Landau}=\frac{28}{27\pi}r_e^2(Z_1Z_2\alpha)^2L_\gamma^3,
\label{sigmaL}
\end{equation}
where $L_\gamma=\log\gamma$. Racah \cite{Racah1937} gives next to leading terms
\begin{equation}
\sigma_\mathrm{Racah}=\frac{28}{27\pi}r_e^2(Z_1Z_2\alpha)^2(L_\gamma^3-2.2L_\gamma^2+3.84L_\gamma-1.636).
\label{sigmaR}
\end{equation}
The differential cross-section is given in Section \ref{bremss}. The
differential distributions of electrons and positrons in a wide
energy range was computed by Bhabha in \cite{1935RSPSA.152..559B}.

In parallel progress on the reaction
\begin{equation}
e^-+Z\longrightarrow e^-+Z+\gamma,
\label{eibr}
\end{equation}
was made by Sommerfeld \cite{1931AnP...403..257S}, Heitler \cite{1933ZPhy...84..145H} and later by Bethe and Heitler \cite{1934RSPSA.146...83B}.

Once the exact cross-section of the process (\ref{gi2p}) was known,
the corresponding cross-section for the process (\ref{eibr}) was
found by an elegant method, called the equivalent photons method
\cite{1975PhR....15..181B,2005ARNPS..55..271B}. The idea to treat
the field of a fast charged particle in a way similar to
electromagnetic radiation with particular frequency spectrum goes
back to Fermi \cite{Fermi1924}. In such a way electromagnetic
interaction of this particle e.g. with a nucleus is reduced to the
interaction of this radiation with the nucleus. This idea was
successfully applied to the calculation of the cross-section of
interaction of relativistic charged particles by Weizs{\"a}cker
\cite{1934ZPhy...88..612W} and Williams \cite{1934PhRv...45..729W}.
In fact, this method establishes the relation between the
high-energy photon induced cross-section $d\sigma_{\gamma
X\rightarrow Y}$ to the corresponding cross-section induced by a
charged particle $d\sigma_{e X\rightarrow Y}$ by the relation which
is expressed by
\begin{equation}
    d\sigma_{eX\rightarrow Y}=\int\frac{n(\omega)}{\omega}d\sigma_{\gamma X\rightarrow Y}d\omega,
    \label{equivphoton}
\end{equation}
where $n(\omega)$ is the spectrum of equivalent photons. Its simple generalization,
\begin{equation}
\sigma_{ee\rightarrow Y}=\int\frac{d\omega_1}{\omega_1}\frac{d\omega_2}{\omega_2}n(\omega_1)n(\omega_2)d\sigma_{\gamma_1\gamma_2\rightarrow Y}.
\label{sigmaeegg}
\end{equation}
Generally speaking, the equivalent photon approximation consists in
ignoring that in such a case intermediate (virtual) photons are a)
off mass shell and b) no longer transversely polarized. In the early
years this spectrum was estimated on the ground of semi-classical
approximations \cite{1934ZPhy...88..612W,1979PhRvD..19..100O} as
\begin{equation}
    n(\omega)=\frac{2\alpha}{\pi}\ln\left(\frac{E}{\omega}\right),
\end{equation}
where $E$ is relativistic charged particle energy. This logarithmic
dependence of the equivalent photon spectrum on the particle energy
is characteristic of the Coulomb field. Racah \cite{Racah1934}
applied this method to compute the bremsstrahlung cross-section in
the process (\ref{eibr}), which is given in Section \ref{bremss}.
Bethe and Heitler \cite{1934RSPSA.146...83B}, obtained the same
formula and computed the effect of the screening of the electrons of
the nucleus. They found the screening is significant when the energy
of relativistic particle is not too high ($E\simeq mc^2$), where $m$
is the mass of the particle. Finally, Bethe and Heitler discussed
the energy loss of charged particles in a medium.

Racah \cite{Racah1937} used the equivalent photons method to compute
from (\ref{sigmaeegg}) the cross-section of pair creation at
collision of two charged particles (\ref{2pe+e-}). Unlike Landau and
Lifshitz result \cite{Landau1934} $\sigma\sim \log^3(2E)$ which is
valid only for $\log2E\gg1$ the cross-section of Racah contains more
terms of different powers of the logarithm, see Section
\ref{ionion}.

\subsection{Klein paradox and Sauter work}

\label{sauter}

\begin{figure}[!ptb]
%\begin{center}
\begin{picture}(10,350)
\put(60,50){\includegraphics[width=9cm]{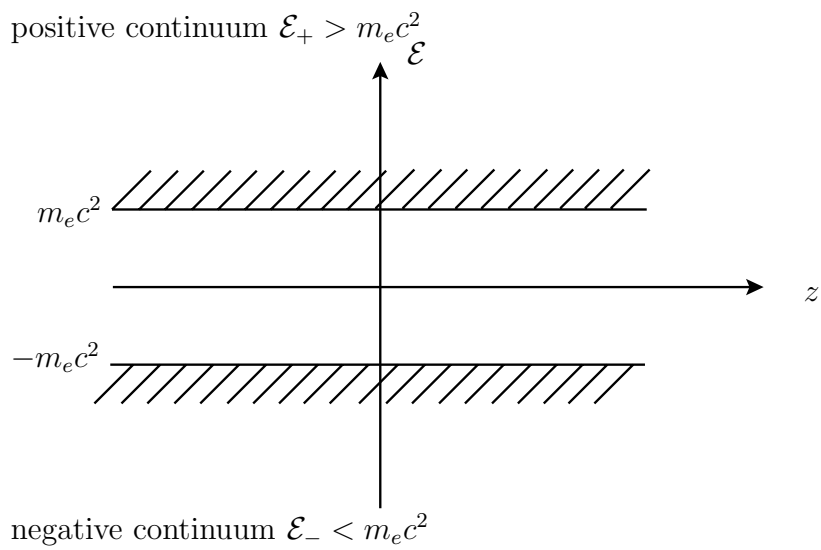}}
\put(330,130){$z$}
\put(180,220){${\mathcal E}$}
\put(30,230){positive continuum ${\mathcal E}_+>m_ec^2$}
\put(30,40){negative continuum ${\mathcal E}_-<m_ec^2$}
\put(40,160){$m_ec^2$}
\put(30,105){$-m_ec^2$}
\end{picture}
%\end{center}
\caption{The mass-gap $2m_ec^2$  that separates the positive continuum spectrum ${\mathcal E}_+$
from the negative continuum spectrum ${\mathcal E}_-$.}%
\label{gap}%
\end{figure}

Every relativistic wave equation of a free particle of mass $m_e$,
momentum ${\bf p}$ and energy ${\mathcal E}$, admits ``positive
energy'' and ``negative energy'' solutions. In Minkowski space such
a solution is symmetric with respect to the zero energy and the wave
function given by
\begin{equation}
\psi^{\pm}({\bf x},t)\sim e^{\frac{i}{\hbar}({\bf k}\cdot{\bf x}-{\mathcal E}_{\pm}t)}
\label{freeparticle}
\end{equation}
describes a relativistic particle, whose energy, mass and momentum satisfy,
\begin{equation}
{\mathcal E}_{\pm}^2=m_e^2c^4 +c^2|{\bf p}|^2;\quad {\mathcal E}_\pm=\pm\sqrt{m_e^2c^4 +c^2|{\bf p}|^2}.
\label{klein0}
\end{equation}
This gives rise to  the familiar positive and negative energy spectrum
(${\mathcal E}_\pm$) of positive and negative energy states $\psi^{\pm}({\bf
x},t)$ of the relativistic particle, as represented in Fig.~\ref{gap}. In such
a situation, in absence of external field and at zero temperature, all the
quantum states are stable; that is, there is no possibility of ``positive''
(``negative'') energy states decaying into ``negative'' (``positive'') energy
states since there is an energy gap $2m_ec^2$ separating the negative energy
spectrum from the positive energy spectrum. This stability condition was
implemented by Dirac by considering all negative energy states as fully filled.

A scalar field described by the wave function $\phi(x)$ satisfies the
Klein--Gordon equation
\cite{1926AnP...386..109S,1926ZPhy...39..226F,1926ZPhy...37..895K,1926ZPhy...40..117G}
\begin{eqnarray}
\left\{ \left[ i\hbar \partial _\mu+\frac{e}{c}A_\mu(z)\right] ^2
-m_e^2c^2
\right\}
\phi(x)=0.
\label{@KG0}
\end{eqnarray}
If there is only an electric field $E(z)$ in the $z$-direction
and varying only as a function of $z$, we can
choose a vector potential with the only nonzero component
$A_0(z)$ and potential energy
\begin{equation}
V(z)=-eA_0(z)
=e\int^z dz'E(z').
\label{@VEQ}
\end{equation}
For an electron of charge $-e$ by assuming
\[
\phi(x)=e^{-i{\mathcal E}t/\hbar }e^{i{\sbf p}_\perp {\sbf x}_\perp/\hbar } \phi(z),
\]
with a fixed transverse momentum ${\bf p}_\perp$ in the
$x,y$-direction and an energy eigenvalue ${\mathcal E}$, and
Eq.~(\ref{@KG0}) becomes simply
\begin{eqnarray}
\left[-\hbar ^2\frac{d^2}{dz^2}+p_\perp^2
+m_e^2c^2-\frac{1}{c^2} \left[\E-V(z) \right]^2\right]  \phi(z)=0.
\label{KGeq}\end{eqnarray}

Klein studied a relativistic particle moving in an external {\it
step function} potential $V(z)=V_0\Theta(z)$ and in this case
Eq.~(\ref{KGeq}) is modified as
\begin{equation}
[{\mathcal E}-V_0]^2=m_e^2c^4 +c^2|{\bf p}|^2; \quad {\mathcal E}_\pm=V_0\pm\sqrt{m_e^2c^4 +c^2|{\bf p}|^2},
\label{klein0bis}
\end{equation}
where $|{\bf p}|^2=|{\bf p}_z|^2+{\bf p}_\perp^2$.
He solved his relativistic wave equation \cite{1926AnP...386..109S,1926ZPhy...39..226F,1926ZPhy...37..895K,1926ZPhy...40..117G} by considering an incident free relativistic wave of positive energy states scattered by the constant potential $V_0$, leading to reflected and transmitted waves. He found a paradox that in the case $V_0\ge {\mathcal E}+m_ec^2$, the reflected flux is larger than the incident flux $j_{\rm ref} >j_{\rm inc}$, although the total flux is conserved, i.e. $j_{\rm inc}=j_{\rm ref}+j_{\rm tran}$. This is known as the Klein paradox \cite{1929ZPhy...53..157K,1931ZPhy...73..547S}. This implies that negative energy states have contributions to both the
transmitted flux $j_{\rm tran}$ and reflected flux $j_{\rm ref}$.

%\begin{figure}[!ptb]
%\begin{center}
%\includegraphics[height=8.8cm,width=12.8cm]{dmassgap}
%\end{center}
%\caption{The spectrum of allowed states for a free wave in flat space describing a quantum particle of mass
%$\mu$ is here represented in function of the frequency $\omega={\mathcal E}/\hbar$.
%In such a free field situation all the states
%are stable; that is, there is  no possibility of ``positive'' (``negative'') wave decaying into a ``negative''
%(``positive'') one. This figure
%is reproduced from Fig. I in Ref. \cite{1975mgm..conf..459D}}%
%\label{demourgap}%
%\end{figure}

\begin{figure}[!ptb]
\begin{center}
\includegraphics[width=12cm]{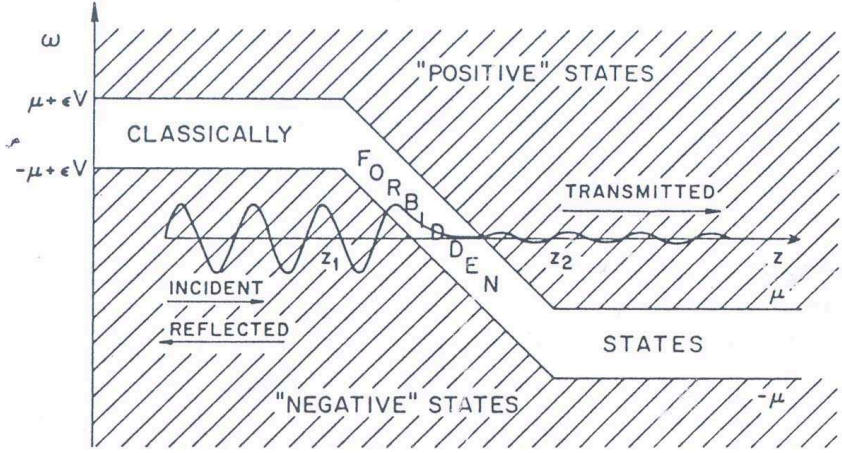}
\end{center}
\caption{In presence of a strong enough electric field the boundaries of the classically allowed states (``positive'' or ``negative'') can be so tilted that a ``negative'' is at the same level as a ``positive'' (level crossing). Therefore a ``negative'' wave-packet from the left will be partially transmitted, after an exponential damping due to the tunneling through the classically forbidden states, as s ``positive'' wave-packet outgoing to the right. This figure
is reproduced from Fig.~II in Ref.~\cite{1975mgm..conf..459D}, and $\mu=m_ec^2, \epsilon V=V(z), \omega={\mathcal E}$.}%
\label{demourgape}%
\end{figure}

Sauter studied this problem by considering a potential varying in
the $z$-direction corresponding to a constant electric field $E$ in
the $\hat {\bf z}={\bf z}/|{\bf z}|$-direction and considering spin
1/2 particles fulfilling the Dirac equation. In this case the energy
${\mathcal E}$ is shifted by the amount $V(z)=-e Ez$. He further
assumed an electric field $E$ uniform between $z_1$ and $z_2$ and
null outside. Fig.~\ref{demourgape} represents the corresponding
sketch of allowed states. The key point now, which is the essence of
the Klein paradox \cite{1929ZPhy...53..157K,1931ZPhy...73..547S}, is
that a level crossing between the positive and negative energy
levels occurs. Under this condition the above mentioned stability of
the ``positive energy'' states is lost for sufficiently strong
electric fields. The same is true for ``negative energy'' states.
Some ``positive energy'' and ``negative energy'' states have the
same energy levels. Thus, these ``negative energy'' waves incident
from the left will be both reflected back by the electric field and
partly transmitted to the right as a ```positive energy'' wave, as
shown in Fig.~\ref{demourgape} \cite{1975mgm..conf..459D}. This
transmission represents a quantum tunneling of the wave function
through the electric potential barrier, where classical states are
forbidden. This quantum tunneling phenomenon was pioneered by George
Gamow by the analysis of alpha particle emission or capture in the
nuclear potential barrier (Gamow wall) \cite{Gamow1931}. In the
latter case however the tunneling occurred between two states of
positive energy while in the Klein paradox and Sauter computation
the tunneling occurs for the first time between the positive and
negative energy states giving rise to the totally new concept of the
creation of particle-antiparticle pairs in the positive energy state
as we are going to show.

Sauter first solved the relativistic Dirac equation in the presence of the
constant electric field by the ansatz,
\begin{equation}
\psi_s({\bf x},t)= e^{\frac{i}{\hbar}(k_xx+k_yy-{\mathcal E}_\pm t)}\chi_{s_3}(z)
\label{Eparticle}
\end{equation}
where spinor function $\chi_{s_3}(z)$ obeys the following equation
($\gamma_0, \gamma_i $ are Dirac matrices)
\begin{equation}
\left[\hbar c\gamma_3\frac{d}{dz}+\gamma_0(V(z)-{\mathcal E}_\pm)+(m_ec^2+ic\gamma_2p_y
+ic\gamma_1 p_x)\right]\chi_{s_3}(z)=0,
\label{sauterfunction}
\end{equation}
and the solution $\chi_{s_3}(z)$ can be expressed in terms of
hypergeometric functions \cite{1931ZPhy...69..742S}. Using this wave
function $\psi_s({\bf x},t)$ (\ref{Eparticle}) and the flux
$ic\psi_s^\dagger\gamma_3\psi_s$, Sauter computed the transmitted
flux of positive energy states, the incident and reflected fluxes of
negative energy states, as well as exponential decaying flux of
classically forbidden states, as indicated in Fig.~\ref{demourgape}.
Using the regular matching conditions of the wave functions and
fluxes at boundaries of the potential, Sauter found that the
transmission coefficient $|T|^2$ of the wave through the electric
potential barrier from the negative energy state to positive energy
states:
\begin{equation}
|T|^2=\frac{|{\rm transmission\hskip0.2cm flux}|}{|{\rm incident\hskip0.2cm flux}|}
\sim e^{-\pi\frac{m_e^2c^3}{\hbar e E}}.
\label{transmission}
\end{equation}
This is the probability of negative energy states decaying to positive energy states, caused by an external electric field. The method that Sauter adopted to calculate the transmission coefficient $|T|^2$ is indeed the same as the one Gamow used to calculate quantum tunneling of the wave function through nuclear potential barrier, leading to the $\alpha$-particle emission \cite{Gamow1931}.

The simplest way to calculate the transmission coefficient $|T|^2$
(\ref{transmission}) is the JWKB (Jeffreys--Wentzel--Kramers--Brillouin)
approximation. The electric potential $V(z)$ is not a constant. The
corresponding solution of the Dirac equation is not straightforward, however it
can be found using the quasi-classical, JWKB approximation. Particle's energy
${\mathcal E}$, momentum ${\bf p}$ and mass $m_e$ satisfy,
\begin{equation}
[{\mathcal E}_\pm-V(z)]^2=m_e^2c^4 +c^2|{\bf p}|^2;\quad {\mathcal E}_\pm=V(z)\pm\sqrt{m_e^2c^4 +c^2|{\bf p}|^2},
\label{klein1}
\end{equation}
where the momentum $p_z(z)$ is spatially dependent. The momentum
$p_z>0$ for both negative and positive energy states and the wave
functions exhibit usual oscillatory behavior of propagating wave in
the $\hat {\bf z}$-direction, i.e. $\exp \frac{i}{\hbar}p_z z$.
Inside the electric potential barrier where are the classically
forbidden states, the momentum $p_z^2$ given by Eq.~(\ref{klein1})
becomes negative, and $p_z$ becomes imaginary, which means that the
wave function will have an exponential behavior, i.e. $\exp
-\frac{1}{\hbar}\int |p_z| dz $, instead of the oscillatory behavior
which characterizes the positive and negative energy states.
Therefore the transmission coefficient $|T|^2$ of the wave through
the one-dimensional potential barrier is given by
\begin{equation}
|T|^2 \propto \exp -\frac{2}{\hbar}\int_{z_-}^{z_+} |p_z| dz ,
\label{transmission1}
\end{equation}
where $z_-$ and $z_+$ are roots of the equation $p_z(z)=0$ defining the turning points of the classical trajectory, separating positive and negative energy states.

\subsection{A semi-classical description of pair production in quantum mechanics}\label{semi}

\subsubsection{An external constant electric field}\label{includingE}

\begin{figure}[!ht]
%\begin{center}
\begin{picture}(10,350)
\put(0,0){\includegraphics[width=11cm]{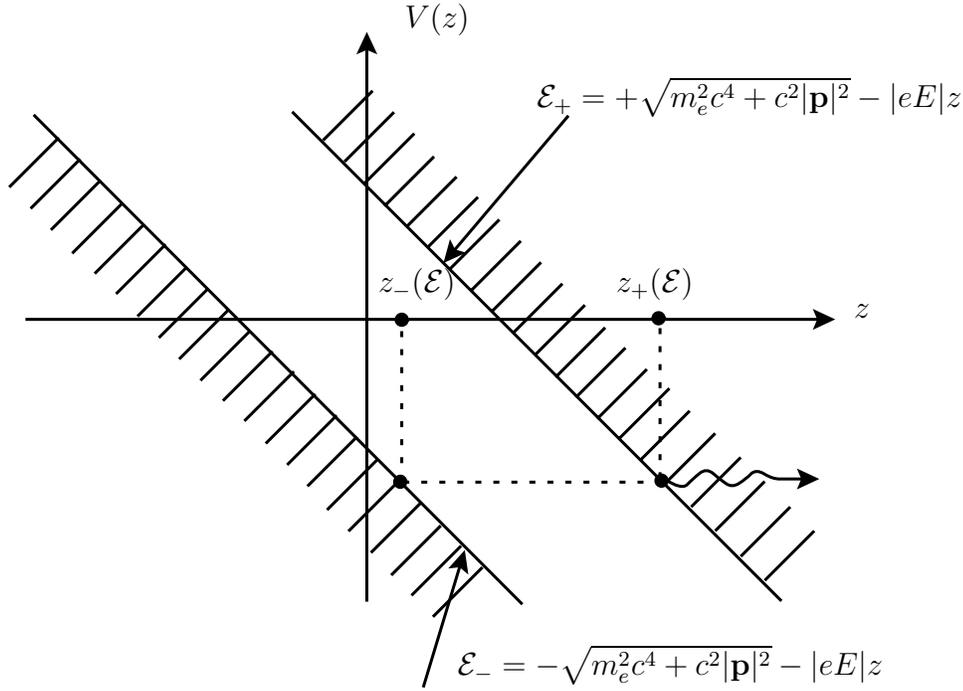}}
\put(320,140){$z$}
\put(150,250){$V(z)$}
\put(200,220){${\mathcal E}_+ =+\sqrt{m_e^2c^4 +c^2|{\bf p}|^2}-|eE|z$}
\put(170,5){${\mathcal E}_- =-\sqrt{m_e^2c^4 +c^2|{\bf p}|^2}-|eE|z$}
%\put(60,260){positive continuum}
%\put(60,50){negative continuum}
%\put(0,200){$m_ec^2$}
%\put(0,110){$-m_ec^2$}
\put(140,150){$z_-({\mathcal E})$}
\put(230,150){$z_+({\mathcal E})$}
\end{picture}
%\end{center}
\caption{
Energy-spectra ${\mathcal E}_\pm$ %once without and once
 with
an external electric field ${\bf E}$
along $\hat {\bf z}$-direction (for
$-\ell <z<\ell$ and $\ell\gg 1$).
Crossing energy-levels
appear, indicated by a dashed line
between two continuum energy-spectra ${\mathcal E}_-$ and
${\mathcal E}_+$. The
turning points $z_\pm({\mathcal E})$
for the crossing energy-levels ${\mathcal E}$
of Eq.~(\ref{crossing})
are  marked. This implies that virtual electrons at
these crossing energy-levels in the negative energy-spectrum
can quantum-mechanically tunnel toward infinity [$z\gg z_+({\mathcal E})$] as real electrons;
empty states left over in the negative energy-spectrum represent real positrons.
This is how quantum tunneling produces pairs of electrons and positrons.}%
\label{egap}%
\end{figure}

The phenomenon of pair production can be understood as a quantum
mechanical tunneling process of relativistic particles. The external
electric field modifies the positive and negative energy spectrum of
the free Hamiltonian. Let the field vector ${\bf E}$ point in the
$\hat {\bf z}$-direction. The electric potential is $A_0=-|E|z$
where $-\ell<z<+\ell$ and the length $\ell \gg \hbar/(m_e c)$, then
the positive and negative continuum energy spectra are
\begin{equation}
{\mathcal E}_\pm=|eE|z\pm\sqrt{(cp_z)^2+c^2{\bf p}_\perp^2+(m_ec^2)^2},
%({\mathcal E}_\pm-V(z))^2=(\epsilon_p\pm m_ec^2-|eE|z)^2=(cp_z)^2+c^2{\bf p}_\perp^2+(m_ec^2)^2,
\label{energyl+-const}
\end{equation}
where $p_z$ is the momentum in $\hat {\bf z}$-direction, ${\bf
p}_\perp$ transverse momenta. The energy spectra ${\mathcal E}_\pm$
(\ref{energyl+-const}) are sketched in Fig.~\ref{egap}. One finds
that crossing energy levels ${\mathcal E}$ between two energy
spectra ${\mathcal E}_-$ and ${\mathcal E}_+$ (\ref{energyl+-const})
appear, then quantum tunneling process occurs. The probability
amplitude for this process can be estimated by a semi-classical
calculation using JWKB method (see e.g.
\cite{Landau1981a,kleinert:025011}):
\begin{eqnarray}
  {\mathcal P}_{\rm JWKB}(|{\bf p}_\perp |) &\equiv  & \exp\left\{-{\frac{2}{\hbar}}\int_{z_-({\mathcal E}_-)}^{z_+({\mathcal E}_+)} p_zdz\right\},
\label{tprobability1}
\end{eqnarray}
where
\begin{eqnarray}
p_z &=& \sqrt{{\bf p}_\perp ^2+m_e^2c^2-({\mathcal E}-|eE|z)^2/c^2}
\label{px}
\end{eqnarray}
is the classical momentum. %[cf. Eq.~(\ref{transmission1})].
The limits of integration $z_\pm({\mathcal E}_\pm)$ are the turning
points of the classical orbit in imaginary time. They are determined
by setting $p_z=0$ in Eq.~(\ref{energyl+-const}). The solutions are
\begin{equation}
z_\pm({\mathcal E}_\pm)=\frac{c\big[{\bf p}_\perp ^2+m_e^2c^2\big]^{1/2}+{\mathcal E}_\pm}{|eE|},
%\label{crosspoint+-}
\end{equation} \!\!\!\!
At the turning points of the classical orbit, the crossing energy
level
\begin{equation}\label{crossing}
{\mathcal E}={\mathcal E}_+={\mathcal E}_-,
\end{equation}
as shown by dashed line in Fig.~\ref{egap}.
The {\em tunneling length\/} is
\begin{equation}
z_+({\mathcal E}_+) - z_-({\mathcal E}_-) =\frac{2m_ec^2}{|eE|}=2\frac{\hbar}{m_ec}\left(\frac{E_c}{E}\right),
\label{tunnelinglength}
\end{equation}
which is independent of crossing energy levels ${\mathcal E}$. The
critical
 electric field $E_c$ in Eq.~(\ref{critical1})
is the field at which
the tunneling length (\ref{tunnelinglength}) is
twice the Compton length $\lambda_C\equiv\hbar/m_ec$.

Changing the variable of integration from  $z$ to $y(z)$,
\begin{equation}
y(z)= \frac{{\mathcal E}-|eE|z}{c{\sqrt{{\bf p}_\perp ^2+m_e^2c^2}}},
%\label{y(x)}
\end{equation}
we obtain
\begin{equation}
y_-(z_-) =-1,~~~~
y_+(z_+) =+ 1
%\label{y+-}
\end{equation}
and
the JWKB probability amplitude (\ref{tprobability1}) becomes
\begin{eqnarray}
{\mathcal P}_{\rm JWKB}(|{\bf p}_\perp |) &= & \exp\left[ -\frac{2E_c}{E}  \left(1+\frac{{\bf p}_\perp ^2}{m_e^2c^2}\right)
\int^{+ 1}_{-1} dy\sqrt{1-y^2}\right]\nonumber\\
&=&
\exp\Big[-\frac{\pi E_c}{E}  \left(1+\frac{{\bf p}_\perp ^2}{m_e^2c^2}\right)\Big].
%\label{wwkbp}
\end{eqnarray}
Summing over initial and final spin states and integrating over the
transverse  phase space $\int d{\bf z}_\perp d{\bf
p}_\perp/(2\pi\hbar)^2$ yields the final result
\begin{eqnarray} \!\!\!\!\!\!
\!\!\!\!{\mathcal P}_{\rm JWKB} &\approx D_sV_\perp
e^{-{\pi c } \lfrac{m_e^2c^2}{|eE|\hbar }}\!
\int\frac{d^2{\bf p}_\perp}{(2\pi\hbar)^2}
e^{-{\pi c } \lfrac{{\bf p}_\perp ^2}{|eE|\hbar }}
= \nonumber\\
&=D_sV_\perp\frac{|eE|}{4\pi^2c\hbar}
e^{-\lfrac{\pi E_c}{E}},
\label{tprobability2}
\end{eqnarray}
where the transverse surface $V_\perp=\int d{\bf z}_\perp$. For the
constant electric field $E$ in $-\ell<z<+\ell$, crossing energy
levels ${\mathcal E}$ vary from the maximal energy potential
$V(-\ell)=+eE\ell$ to the minimal energy potential
$V(+\ell)=-eE\ell$. This probability Eq.~(\ref{tprobability2}) is
independent of crossing energy levels ${\mathcal E}$. We integrate
Eq.~(\ref{tprobability2}) over crossing energy levels $\int
d{\mathcal E}/m_ec^2$ and divide it by the time interval $\Delta
t\simeq h/m_ec^2$ during which quantum tunneling occurs, and find
the transition rate per unit time and volume
\begin{equation}              \!\!\!\!\!\!
 \frac{\Gamma _{\rm JWKB}}{V}\simeq D_s\frac{\alpha E^2}{ 2\pi^2\hbar}
e^{-\lfrac{\pi E_c}{E}},
\label{wkbprobability}
\end{equation}
where $D_s=2$ for a spin-$1/2$ particle and $D_s=1$ for spin-$0$,
$V$ is the volume. The JWKB result contains the Sauter exponential
$e^{-\pi E_c/E}$ \cite{1931ZPhy...69..742S} and reproduces as well
the prefactor of Heisenberg and Euler \cite{1936ZPhy...98..714H}.

Let us specify a quantitative condition for the validity of the
above ``semi-classical'' JWKB approximation, which is in fact
leading term of the expansion of wave function in powers of $\hbar$.
In order to have the next-leading term be much smaller than the
leading term, the de Broglie wavelength $\lambda (z)\equiv
\lfrac{2\pi\hbar}{p_z(z)}$ of wave function of the tunneling
particle must have only small spatial variations \cite{Landau1981a}:
\begin{equation}
\frac{1}{2\pi}\left|\frac{d\lambda(z)}{dz}\right|=
\frac{\hbar}{p_z^2(z)}\left|\frac{dp_z(z)}{dz}\right|\ll 1.
\label{wkbcondition}
\end{equation}
with $p_z(z)$ of Eq.~(\ref{px}). The electric potential $A_0=-e|z|Z$
must satisfy
\begin{equation}
\frac{\hbar}{2p^3_z}
\left|\frac{dA_0}{dz}\right|\simeq \frac{E}{E_c}\ll 1.
\label{wkbc1}
\end{equation}
so that the result (\ref{wkbprobability}) is valid only for $E\ll E_c$.

%Otherwise the potential barrier is too small to allow for a JWKB approximation of the tunneling rate.

\subsubsection{An additional constant magnetic field}
\label{includingB}

The result (\ref{wkbprobability})
can be generalized to
include a uniform magnetic field
 ${\bf B}$.
The calculation is simplest
by going into a Lorentz frame in which
${\bf B}$ and ${\bf E}$ are parallel to each other,
which  is always possible for uniform and static electromagnetic field.
This frame will be referred to a {\em center-of-fields frame\/},
and the associated fields will be denoted by
${\bf B}_{\rm CF}$ and ${\bf E}_{\rm CF}$.
Suppose the initial
${\bf B}$ and ${\bf E}$ are not parallel,
then we perform a Lorentz transformation
 with a velocity
determined by \cite{1975ctf..book.....L}
\begin{equation}
\frac{{\bf v}/c}{1+(|{\bf v}|/c)^2}=\frac{{\bf E}
\times{\bf B}}{|{\bf E}|^2+|{\bf B}|^2},
\label{eblandau}
\end{equation}
in the direction $\hat {\bf v}\equiv {\bf v}/|{\bf v}|$
as follows
\begin{eqnarray}
{\bf E}_{\rm CF}&=&({\bf E}\cdot{\hat{\bf v}}){\hat{\bf v}}
+\frac{{\hat{\bf v}}\times ({\bf E}\times{\hat{\bf v}})
+({\bf v}/c)\times{\bf B}}{[1-(|{\bf v}|/c)^2]^{1/2}},
\label{elorentz}\\
{\bf B}_{\rm CF}&=&({\bf B}\cdot{\hat{\bf v}}){\hat{\bf v}}
+\frac{{\hat{\bf v}}\times ({\bf B}\times{\hat{\bf v}})
-({\bf v}/c)\times{\bf E}}{[1-(|{\bf v}|/c)^2]^{1/2}}.
\label{blorentz}
\end{eqnarray}
The
fields
${\bf B}_{\rm CF}$ and ${\bf E}_{\rm CF}$
are now parallel.
As a consequence,
the wave
function factorizes into a Landau state and into a
spinor function, this last one first calculated by
Sauter (see Eqs.~(\ref{Eparticle}),(\ref{sauterfunction})).
The energy spectrum
in the JWKB approximation is still given by Eq.~(\ref{energyl+-const}),
but the squared transverse momenta ${\bf p}_\perp^2$ is
quantized due to the presence of the
magnetic field: they
are replaced
by the Landau energy levels whose transverse energies
have the discrete spectrum
\begin{equation}
c^2{\bf p}_\perp^2 =
2m_ec^2 \times \frac{{\bf p}_\perp^2}{2m_e}
\rightarrow
2m_ec^2 \times  g \frac{\hbar \omega _L}2
\left(n\!+\!\frac{1}{2}\!+\!\hat\sigma\right)
,~n=0,1,2,\cdot\cdot\cdot~ ,
\label{landaulevel}\end{equation}
where
$g= 2+ \alpha /\pi+\dots $ is the anomalous magnetic moment of the electron \cite{1948PhRv...73..416S,1949PhRv...76..769F,1972PhR.....3..193L,1974PhRvD..10.4007C,1977PhLB...68..191B},
$ \omega _L=e|{\bf B}_{\rm CF}|/m_e c$ the Landau frequency,
$\hat\sigma=\pm 1/2$ for a spin-$1/2$ particle ($\hat\sigma=0$ for a
 spin-$0$ particle) are eigenvalues of spinor operator
$\sigma_z$ in the $(\bf \hat z)$-direction, i.e., in the  common
direction of ${\bf E}_{\rm CF}$ and ${\bf B}_{\rm CF}$ in the
selected frame. The quantum number $n$ characterizing the Landau
levels is associated with harmonic oscillations in the plane
orthogonal to ${\bf E}_{\rm CF}$ and ${\bf B}_{\rm CF}$. Apart from
the replacement (\ref{landaulevel}), the JWKB calculation remains
the same as in the case of constant electric field
(\ref{wkbprobability}). We must only replace the integration over
the transverse phase space $\int dxdy d{\bf p}_\perp/(2\pi\hbar)^2$
in Eq.~(\ref{tprobability2}) by the sum over all Landau levels with
the degeneracy $V_\perp e|{\bf B}_{\rm CF}|/(2\pi\hbar c)$
\cite{Landau1981a}:
\begin{equation}
  \frac{V_\perp e|{\bf B}_{\rm CF}|}{2\pi\hbar c}\sum_{n\hat\sigma}
\exp \left[-\pi\frac{2c\hbar |e||{\bf B}_{\rm CF}|(n+1/2+\hat\sigma)+(m_ec^2)^2}{e|{\bf E}_{\rm CF}|c\hbar}\right].
%\label{tprobability2h}  \!\!\!\!\!\!\!
\end{equation}
The results are
\begin{equation}
  \frac{V_\perp e|{\bf B}_{\rm CF}|}{2\pi\hbar c}\coth\left(\frac{\pi |{\bf B}_{\rm CF}|}{|{\bf E}_{\rm CF}|}\right)
\exp\left(-\frac{\pi E_c}{|{\bf E}_{\rm CF}|}\right),\quad {\rm ~~~ spin}-1/2 \hskip0.2cm {\rm particle}
\label{wkbehfermion}
\end{equation}
and
\begin{equation}
%\!\!\!\!\!\!\!\!\!\!\!\!\!
  \frac{V_\perp e|{\bf B}_{\rm CF}|}{4\pi\hbar c}\sinh^{-1}\left(\frac{\pi |{\bf B}_{\rm CF}|}{|{\bf E}_{\rm CF}|}\right)
\exp\left(-\frac{\pi E_c}{|{\bf E}_{\rm CF}|}\right),\quad {\rm  spin}-0 \hskip0.2cm {\rm particle}.
\label{wkbehboson}
\end{equation}
We find the pair production rate per unit time and volume
\begin{equation}
%\!\!\!\!\!\!\!\!\!\!\!\!\!
\frac{\Gamma _{\rm JWKB}}{V}\simeq
\frac{\alpha |{\bf B}_{\rm CF}||{\bf E}_{\rm CF}|}{\pi\hbar }\coth\left(\frac{\pi |{\bf B}_{\rm CF}|}{ |{\bf E}_{\rm CF}|}\right)
\exp\left(-\frac{\pi E_c}{|{\bf E}_{\rm CF}|}\right),\quad {\rm  spin-1/2 \hskip0.2cm particle}
\label{wkbehfermion1}
\end{equation}
and
\begin{equation}
%\!\!\!\!\!\!\!\!\!\!\!\!\!
\frac{\Gamma _{\rm JWKB}}{V}\simeq  \frac{\alpha |{\bf B}_{\rm CF}||{\bf E}_{\rm CF}|}{2\pi\hbar }
\sinh^{-1}\left(\frac{\pi |{\bf B}_{\rm CF}|}{|{\bf E}_{\rm CF}|}\right)
\exp\left(-\frac{\pi E_c}{|{\bf E}_{\rm CF}|}\right),\quad {\rm  spin-0 \hskip0.2cm particle}.
\label{wkbehboson1}
\end{equation}

We can now go back to an arbitrary
Lorentz frame by expressing the result in terms of the two
 Lorentz invariants
that can be formed
from the
${\bf B}$ and ${\bf E}$ fields: the scalar $S$ and the pseudoscalar $P$
\begin{equation}    \!\!\!\!\!\!
 S \equiv
\frac{1}{4}F_{\mu \nu }F^{\mu \nu }=
\frac{1}{2}({\bf E}^2-{\bf B}^2);
\quad
  P\equiv
\frac{1}{4}F_{\mu \nu }\tilde F^{\mu \nu }=
 {\bf E}\cdot {\bf B},
\label{lightlike}
\end{equation}
where $\tilde F^{\mu \nu }\equiv  \epsilon ^{\mu \nu  \lambda  \kappa }F_{ \lambda  \kappa }$ is the dual field tensor. We define the invariants
$ \varepsilon $ and $ \beta $
as the solutions of the invariant equations
\begin{equation}
 \!\!\!\!\!\varepsilon ^2- \beta ^2\equiv {\bf E}^2-{\bf B}^2\equiv 2 S,~~~~
 \varepsilon \beta \equiv {\bf E}\,{\bf B}\equiv P,
\label{ab}
\end{equation}
and obtain
\begin{eqnarray}\label{fieldinvariant}
 \!\!\!\!\!\!\!\!\!\!\!\!
\varepsilon & \equiv &
\sqrt{(S^2+P^2)^{1/2}+ S}, \\
\beta & \equiv &
\sqrt{(S^2+P^2)^{1/2}- S}.
\end{eqnarray}
In the special frame
with parallel ${\bf B}_{\rm CF}$ and ${\bf E}_{\rm CF}$,
we see that
$ \beta  =|{\bf B}_{\rm CF}|$ and
$ \varepsilon =|{\bf E}_{\rm CF}|$, so that we can replace
(\ref{wkbehfermion1})
and
(\ref{wkbehboson1}) directly by the invariant
expressions\begin{equation}
%\!\!\!\!\!\!\!\!\!\!\!\!\!
\frac{\Gamma _{\rm JWKB}}{V}\simeq
\frac{\alpha   \beta \varepsilon  }{\pi\hbar }\coth\left(\frac{\pi  \beta }{ \varepsilon }\right)
\exp\left(-\frac{\pi E_c}{ \varepsilon }\right),\quad {\rm  spin-1/2 \hskip0.2cm particle}
\label{wkbehfermion2}
\end{equation}
and
\begin{equation}
%\!\!\!\!\!\!\!\!\!\!\!\!\!
\frac{\Gamma _{\rm JWKB}}{V}\simeq  \frac{\alpha  \beta  \varepsilon }{2\pi\hbar }
\sinh^{-1}\left(\frac{\pi  \beta }{ \varepsilon }\right)
\exp\left(-\frac{\pi E_c}{ \varepsilon }\right),\quad {\rm  spin-0 \hskip0.2cm particle},
\label{wkbehboson2}
\end{equation}
which are pair production rates in arbitrary constant
electromagnetic fields. We would like to point out that $S$ and $P$
in (\ref{lightlike}) are identically zero for any field
configuration in which
\begin{equation}
|{\bf E}|=|{\bf B}|;\quad {\bf E}\perp{\bf B}=0.
\label{epb}
\end{equation}
As example, for a plane wave of electromagnetic field, $ \varepsilon=\beta =0$ and no pairs are produced.

\section{Nonlinear electrodynamics and rate of pair creation}\label{nonlinearEM}

\subsection{Hans Euler and light-light scattering}\label{Eulerwork}

Hans Euler in his celebrated diplom thesis
\cite{1936AnP...418..398E} discussed at the University of Leipzig
called attention on the reaction
\[
\gamma_1\gamma_2\longrightarrow e^+e^-\longrightarrow \gamma_1'\gamma_2'
\]
He recalled that Halpern \cite{1933PhRv...44..855H} and Debye \cite{Debye1934}
first recognized that Dirac theory of electrons and the Dirac process
(\ref{ee2gamma}) and the Breit--Wheeler one (\ref{2gammaee}) had fundamental
implication for the light on light scattering and consequently implied a
modifications of the Maxwell equations.

If the energy of the photons is high enough then a real electron--positron pair
is created, following Breit and Wheeler \cite{1934PhRv...46.1087B}. Again, if
electron--positron pair does exist, two photons are created following
\cite{1930PCPS...26..361D}. In the case that the sum of energies of the two
photons are smaller than the threshold $2m_e c^2$ then the reaction (above)
still occurs through a virtual pair of electron and positron.

Under this condition the light-light scattering implies deviation from
superposition principle, and therefore the linear theory of electromagnetism
has to be substituted by a nonlinear one. Maxwell equations acquire nonlinear
corrections due to the Dirac theory of the electron.

Euler first attempted to describe this nonlinearity by an effective Lagrangian
representing the interaction term. He showed that the interaction term had to
contain the forth power of the field strengths and its derivatives
\begin{equation}
{\mathcal E}_{int}={\rm const} \int \left[FFFF+{\rm
const'}\frac{\partial F}{\partial x}\frac{\partial F}{\partial
x}FF+...\right], \label{llU}
\end{equation}
$F$ being symbolically the electromagnetic field strength. He also
estimated that the constants may be determined from dimensional
considerations. Since the interaction $U_{int}$ has the dimension of
energy density and contains electric charge in the forth power, the
constants up to numerical factors are
\begin{equation}
{\rm const}=\frac{\hbar c}{e^2}\frac{1}{E_e^2}, \quad\quad {\rm
const'}=\left(\frac{\hbar}{m_e c}\right)^2,
\end{equation}
where $E_e=e\left(\frac{e^2}{m_e c^2}\right)^{-2}=\alpha^{-1}E_c$,
namely ``the field strength at the edge of the electron''.

From these general qualitative considerations Euler made an important further
step taking into account that the Lagrangian (\ref{llU}) describing such a
process had necessarily be built from invariants constructed from the field
strengths, such as ${\bf E}^2-{\bf B}^2$ and ${\bf E}\cdot {\bf B}$ following a
precise procedure indicated by Max Born, see e.g. Pauli's book
\cite{Pauli1981}. Contrary to the usual Maxwell Lagrangian which is only a
function of $F_{\mu\nu}^2$ Euler first recognized that virtual
electron--positron loops are represented by higher powers in the field strength
corrections to the linear action of electromagnetism and written down the
Lagrangian with second order corrections
\begin{equation}\label{euler lagra}
{\mathcal L}=\frac{{\bf E}^2-{\bf B}^2}{8\pi}+\frac{1}{\alpha}\frac{1}{E_0^2}\left[a_E\left({\bf E}^2-{\bf B}^2\right)^2+b_E\left({\bf E}\cdot {\bf B}\right)^2\right],
\end{equation}
where
\begin{equation}\label{abconst}
a_E=-1/(360\pi^2), \quad b_E=-7/(360\pi^2).
\end{equation}
The crucial result of Euler has been to determine the values of the
coefficients (\ref{abconst}) using time-dependent perturbation
technique, e.g. \cite{Landau1981a} in Dirac theory.

Euler computed only the lowest order corrections in $\alpha$ to Maxwell
equations, namely ``the 1/137 fraction of the field strength at the edge of the
electron''. This perturbation method did not allow calculation of the tunneling
rate for electron--positron pair creation in strong electromagnetic field which
became the topic of the further work with Heisenberg
\cite{1936ZPhy...98..714H}.

\subsection{Born's nonlinear electromagnetism}\label{Bornwork}

A nonlinear theory of electrodynamics was independently proposed and developed
by Max Born \cite{1933Natur.132..282B,1934RSPSA.143..410B} and later by Born
and Infeld \cite{1934RSPSA.144..425B}. The main motivation in Born's approach
was the avoidance of infinities in an elementary particle description. Among
the classical discussions on the fundamental interactions this topic had
attracted attention of a large number of scientists. It was clear in fact from
the considerations of J.J. Thomson, Abraham Lorentz that a point-like electron
needed to have necessarily an infinite mass. The existence of a finite radius
was attempted by Poincare by introduction of non-electromagnetic stresses. Also
among the attempts we have to recall the theory of Mie
\cite{1912AnP...344....1M, 1912AnP...342..511M, 1913AnP...345....1M,
1914GNM.........23B} modifying the Maxwell theory by nonlinear terms. This
theory however had serious difficulty because solutions of Mie field equations
depend on the absolute value of the potentials.

Max Born developed his theory in collaboration with Infeld. This alternative to
the Maxwell theory is today called the Born-Infeld theory which still finds
interest in the framework of subnuclear physics. The coauthorship of Infeld is
felt by the general premise of the article in distinguishing the
\emph{unitarian} standpoint versus the \emph{dualistic} standpoint in the
description of particles and fields. ``In the dualistic standpoint the
particles are the sources of the field, are acted own by the field but are not
a part of the field. Their characteristic properties are inertia, measured by
specific constant, the mass'' \cite{1934RSPSA.144..425B}. The unitarian theory
developed by Thomson, Lorentz and Mie tends to describe the particle as a
point-like singularity but with finite mass-energy density fulfilling uniquely
an appropriate nonlinear field equations. It is interesting that this approach
was later developed in the classical book by Einstein and Infeld
\cite{Einstein1967} as well as in the classical paper by Einstein, Infeld and
Hoffmann \cite{1938AM....39..65E} on equations of motion in General Relativity.

In the Born-Infeld approach the emphasis is directed to a formalism
encompassing General Relativity. But for simplicity the field equations are
solved within the realm only of the electromagnetic field. A basic tensor
$a_{\alpha\beta} = g_{\alpha\beta} + f_{\alpha\beta}$ is introduced. Its
symmetric part $g_{\alpha\beta}$ is identified with a metric component and the
antisymmetric part $f_{\alpha\beta}$ with the electromagnetic field. Formally
therefore both the electromagnetic and gravitational fields are present
although the authors explicitly avoided to insert the part of the Lagrangian
describing the gravitational interaction and focused uniquely on the following
nonlinear Lagrangian
\begin{equation}\label{lagra born}
\mathcal{L}=\sqrt{1+2S-P^{2}}-1.
\end{equation}
The necessity to have the quadratic form of the $P$ term is due to obtain a
Lagrangian invariant under reflections as pointed out by W.~Pauli in his
classical book \cite{Pauli1981}. For small field strengths Lagrangian
(\ref{lagra born}) has the same form as (\ref{euler lagra}) obtained by Euler.

From the nonlinear Lagrangian (\ref{lagra born}) Born and Infeld calculated the
fields $\textbf{D}$ and $\textbf{H}$ through a tensor, $P^i_{0}=D^i$ and
$P^{ij}=-\epsilon^{ijk}H^k$, where
\begin{equation}\label{born1}
P_{\mu\nu}\equiv  \frac{\delta {\mathcal L}_{\rm Born}}{\delta F^{\mu\nu}}
=\frac{F_{\mu\nu}-P\tilde F_{\mu\nu}}{\sqrt{1+2S-P^{2}}},
\end{equation}
and introduced therefore an effective electric permittivity and magnetic permeability which are functions of $S$ and $P$. It is very interesting that Born and Infeld managed to obtain a solution for electrostatic field of a point particle ($P=0$) in which the radial component $D_r=e/r^2$ becomes infinite as $r\rightarrow 0$ but the radial component of $\textbf{E}$ field is perfectly finite and is given by the expression
\begin{equation}
E_r=\frac{e}{r_0^2\sqrt{1+(r/r_0)^4}},
\end{equation}
where $r_0$ is the ``radius'' of the electron.

Most important the integral of the electromagnetic energy is finite and given by
\begin{equation}
\int{\mathcal H}_{\rm Born}dV=\int (P_{\mu\nu}F^{\mu\nu}-{\mathcal L}_{\rm Born})dV=1.2361\frac{e}{r_0}^2,
\end{equation}
Equating this energy to $m_e c^2$ they obtain $r_0=1.2361 e^2/(m_e c^2)$.

The attempt therefore is to have a theoretical framework explaining
the mass of the electron solely by a modified nonlinear
electromagnetic field theory. This approach has not been followed by
the current theories in particle physics where the dualistic
approach is today adopted in which the charged particles are
described by half-integer spin fields and electro-magnetic
interactions by integer-spin fields.

The initial goal to develop a fully covariant theory of electrodynamics within
General Relativity although not developed by Born himself was not abandoned.
Hoffmann found an analytic solution \cite{1935PhRv...47..877H} to the coupled
system of the Einstein-Born-Infeld equations.

\subsection{The Euler-Heisenberg Lagrangian}\label{EulerHwork}

The two different approaches of Born and Infeld and of Euler present
strong analogies and substantial differences. The attempt of Born
and Infeld was to obtain at once a new nonlinear Lagrangian for
electromagnetic field replacing the Maxwell Lagrangian in order to
avoid the appearance of infinite self-energy for a classical
point-like electron.

The attempt of Euler \cite{1935NW.....23..246E} was more
conservative, to obtain the first order nonlinear perturbation
corrections to the Maxwell Lagrangian on the ground of the Dirac
theory of the electron.

Born and Infeld in addition introduced an effective dielectric
constant and an effective magnetic permeability of the vacuum out of
their nonlinear Lagrangian (\ref{lagra born}). This approach was
adopted as well in the classical work of Heisenberg and Euler
\cite{1936ZPhy...98..714H}. They introduced an effective Lagrangian
on the ground of the Dirac theory of the electron and expressed the
result in integral form duly taking away infinities, see Section
\ref{hew-effecitve}. This integral was explicitly performed in the
weak field limit and the special attention was given to the real
part, see Section \ref{weiss-effecitve} and the imaginary part, see
Section \ref{imm-effecitve}.

A successive work of Weisskopf \cite{Weisskopf1936} derived the same
equations of Heisenberg and Euler for the real part of the
dielectric constant and magnetic permeability by using instead of
the spin 1/2 particle of the Dirac equation the scalar relativistic
wave equation of Klein and Gordon. The results differ from the one
of spin 1/2 particle only by a factor 2 due to the Bose statistics,
see Section \ref{hew-effecitve}. The technique used by Weisskopf
refers to the case of magnetic field of arbitrary strengths and
describes the electric field perturbatively, see Section
\ref{weiss-effecitve}. As we will see in the following, the
Heisenberg and Euler integral can be found in the case of arbitrary
large both electric and magnetic fields, see Section
\ref{nonlinear}.%aga

\subsubsection{Real part of the effective Lagrangian}

\label{hew-effecitve}

We now recall how Heisenberg and Euler adopted the crucial idea of Max Born to
describe the nonlinear Lagrangian by the introduction of an effective
dielectric constant and magnetic permeability \cite{1936ZPhy...98..714H}. They
further extended this idea by adopting the most general case of a dielectric
constant containing real and imaginary part. Such an approach is generally
followed in the description of dissipative media. The crucial point was to
relate electron--positron pair creation process to imaginary part of the
Lagrangian.

Let ${\mathcal L}$ to be the Lagrangian density of electromagnetic
fields ${\bf E},{\bf B}$, a Legendre transformation produces the
Hamiltonian density:
\begin{equation}
{\mathcal H}=E_i{\frac{\delta {\mathcal L}}{ \delta E_i}} - {\mathcal L}.
\label{hlrelation}
\end{equation}
In  Maxwell's theory, the two densities are given by
\begin{equation}
{\mathcal L}_M={\frac{1}{8\pi}}({\bf  E}^2-{\bf B}^2),\hskip0.5cm {\mathcal H}_M
={\frac{1}{8\pi}}({\bf E}^2+{\bf B}^2).
\label{maxwell}
\end{equation}
To quantitatively describe nonlinear electromagnetic properties of
the vacuum based on the Dirac theory, Heisenberg and Euler
introduced an effective Lagrangian ${\mathcal L}_{\rm eff}$ of the
vacuum state and an associated Hamiltonian density
\begin{equation}
{\mathcal L}_{\rm eff}={\mathcal L}_M+\Delta {\mathcal L},\hskip0.5cm
{\mathcal H}_{\rm eff}={\mathcal H}_M + \Delta {\mathcal H}.
\label{effectivelh}
\end{equation}
Here ${\mathcal H}_{\rm eff}$ and ${\mathcal L}_{\rm eff}$ are complex
functions of ${\bf E}$ and ${\bf B}$. In Maxwell's theory, $\Delta {\mathcal
L}\equiv 0$ in the vacuum, so that $\bf  D={\bf E}$ and $\bf  H={\bf B}$.

Heisenberg and Euler derived the induced fields $\bf  D,\bf  H$ as the derivatives
\begin{equation}
D_i\equiv{\frac{\delta {\mathcal L}_{\rm eff}}{\delta E_i}},\hskip0.5cm H_i\equiv-{\frac{\delta {\mathcal L}_{\rm eff}}{\delta B_i}}.
\label{dh}
\end{equation}
Consequently, the vacuum behaves as a dielectric and permeable medium
\cite{1936ZPhy...98..714H,Weisskopf1936} in which,
\begin{equation}%aga
D_i=\epsilon_{ik}E_k,\hskip0.5cm H_i=\mu_{ik}B_k,
\label{dh1}
\end{equation}
where $\epsilon_{ik}$ and $\mu_{ik}$ are complex and field-dependent dielectric
and permeability tensors of the vacuum.

The discussions on complex dielectric and permeability tensors ($\epsilon_{ik}$
and $\mu_{ik}$) can be found for example in Ref.~\cite{1960ecm..book.....L}.
The effective Lagrangian and Hamiltonian densities in such a medium are given
by
\begin{equation}
{\mathcal L}_{\rm eff}={\frac{1}{8\pi}}({\bf E}\cdot {\bf  D}-{\bf B}\cdot {\bf  H}),\hskip0.5cm {\mathcal H}_{\rm eff}
={\frac{1}{8\pi}}({\bf E}\cdot {\bf  D}+{\bf B}\cdot {\bf  H}).
\label{effmaxwell}
\end{equation}
In this medium, the conservation of electromagnetic energy
has the form
\begin{equation}
-{\rm div}{\bf  S}={\bf E}\cdot{\frac{\partial {\bf  D}}{\partial t}}+{\bf B}\cdot {\frac{\partial {\bf  H}}{\partial t}},\hskip0.5cm {\bf  S}= c{\bf E}\times {\bf B},
\label{effcons}
\end{equation}
where $\bf  S$ is the Poynting vector describing the density of electromagnetic
energy flux. Consider complex and monochromatic electromagnetic field
\begin{equation}
{\bf E}={\bf E}(\omega) \exp-i(\omega t);\quad
{\bf B}={\bf B}(\omega) \exp-i(\omega t),
\label{monochromaticfields}
\end{equation}
of frequency $\omega$, and dielectric and permeability tensors are
frequency-dependent, i.e., $\epsilon_{ik}(\omega)$ and $\mu_{ik}(\omega)$.
Substituting these fields and tensors into the right-hand side of
Eq.~(\ref{effcons}), one obtains the dissipation of electromagnetic energy per
unit time into the medium
\begin{equation}
Q_{\rm dis}={\frac{\omega}{2}}\left\{{\rm Im}\left[\epsilon_{ik}(\omega)\right]E_i E^*_k
+{\rm Im}\left[\mu_{ik}(\omega)\right]B_iB^*_k\right\}.
\label{dissipation}
\end{equation}
This is nonzero if $\epsilon_{ik}(\omega)$ and $\mu_{ik}(\omega)$
contain an imaginary part. The dissipation of electromagnetic energy
in a medium is accompanied by heat production. In the light of the
third thermodynamical law of entropy increase, the energy lost
$Q_{\rm dis}$ of electromagnetic fields in the medium is always
positive, i.e., $Q_{\rm dis}>0$. As a consequence, ${\rm
Im}[\epsilon_{ik}(\omega)]>0$ and ${\rm Im}[\mu_{ik}(\omega)]>0$.
The real parts of $\epsilon_{ik}(\omega)$ and $\mu_{ik}(\omega)$
represent an electric and magnetic polarizability of the vacuum and
leads, for example, to the refraction of light in an electromagnetic
field, or to the elastic scattering of light from light. The
$n_{ij}(\omega)=\sqrt{\epsilon_{ik}(\omega)\mu_{kj}(\omega)}$ is the
reflection index of the medium. The field dependence of
$\epsilon_{ik}$ and $\mu_{ik}$ implies nonlinear electromagnetic
properties of the vacuum as a dielectric and permeable medium.

The effective Lagrangian density (\ref{effectivelh}) is a relativistically
invariant function of the
field strengths ${\bf E}$ and ${\bf B}$.
Since $({\bf E}^2-{\bf B}^2)$ and ${\bf E}\cdot {\bf B}$ are relativistic invariants, one can formally expand
$\Delta {\mathcal L}$ in powers of weak field strengths:
\begin{equation}
\Delta {\mathcal L} = \kappa_{2,0}
 ({\bf E}^2-{\bf B}^2)^2+\kappa_{0,2}
 ({\bf E}\cdot {\bf B})^2
+ \kappa _{3,0} ({\bf E}^2 -{\bf B}^2)^3 + \kappa _{1,2} ({\bf E}^2-
{\bf B}^2)({\bf E}\cdot {\bf B})^2+\dots~ , \label{leffective}
\end{equation}
where $\kappa_{i,j}$ are field-independent constants whose
subscripts indicate the powers of $({\bf E}^2-{\bf B}^2)$ and ${\bf
E}\cdot{\bf B}$, respectively. Note that the invariant ${\bf
E}\cdot{\bf B}$ appears only in even powers since it is odd under
parity and electromagnetism is parity invariant. The Lagrangian
density (\ref{leffective}) corresponds, via relation
(\ref{hlrelation}), to
\begin{eqnarray}
\Delta {\mathcal H}&=&\kappa_{2,0}
 ({\bf E}^2-{\bf B}^2)(3{\bf E}^2+{\bf B}^2)+ \kappa _{0,2}
 ({\bf E}\cdot {\bf B})^2 \nonumber\\
&&+ \kappa _{3,0} ({\bf E}^2 -{\bf B}^2)^2(5{\bf E}^2+{\bf B}^2)+
 \kappa _{1,2}(3{\bf E}^2- {\bf B}^2)({\bf E}\cdot {\bf B})^2+\dots~ .
\label{heffective}
\end{eqnarray}
To obtain ${\mathcal H}_{\rm eff}$ in Dirac's theory, one has to
calculate
\begin{equation} \Delta {\mathcal H}=\sum_k\left\{\psi^*_k,\Big[
\mbox{\boldmath$\alpha$}\cdot(-i h c{\bf \nabla} + e{\bf A}\
)) +\beta_D m_ec^2\Big]\psi_k\right\},
\label{wcal}
\end{equation}
where $\{\psi_k(x)\}$ are the wave functions of the occupied
negative energy states.
%check\mn{check this, you wrote different}
When performing the sum, one encounters infinities which were
removed by Dirac, Heisenberg, and Weisskopf
\cite{Weisskopf1936,1934PCPS...30..150D,1934ZPhy...90..209H,1934ZPhy...89...27W}
by a suitable subtraction.

Heisenberg
 \cite{1934ZPhy...90..209H}
expressed the Hamiltonian density in terms of the density matrix $
\rho (x,x')=\sum_k\psi^*_k(x)\psi_k(x')$ \cite{1934PCPS...30..150D}.
Heisenberg and Euler \cite{1936ZPhy...98..714H} calculated the
coefficients $ \kappa _{i,j}$. They did so by solving the Dirac
equation in the presence of parallel electric and magnetic fields
${\bf E}$ and ${\bf B}$ in a specific direction,
\begin{equation}
\psi_k(x)\rightarrow
\psi_{p_z,n,s_3}\equiv
e^{{\frac{i}{\hbar}}(zp_z-{\mathcal E}t)}u_{n}(y)\chi_{s_3}(x),
\hskip0.5cm  n=0,1,2,\dots~
\label{diracsolution}
\end{equation}
where $\{u_n(y)\}$ are the Landau states\footnote{Landau determined
the quantum states of a particle in an external magnetic field in
1930 \cite{Landau1981a,1982els..book.....B}.} depending on the
magnetic field and $\chi_{s_3}(x)$ are the spinor functions
calculated by Sauter \cite{1931ZPhy...69..742S}. Heisenberg and
Euler used the Euler-Maclaurin formula to perform the sum over $ n$,
and obtained for the additional Lagrangian in (\ref{effectivelh})
the integral representation
\begin{eqnarray}
 \Delta {\mathcal L}_{\rm eff}&=&\frac{e^2}{16\pi^2\hbar c}\int^\infty_0
e^{-s}\frac{ds}{s^3}\Big[is^2\,\bar E \bar B
\frac{\cos(s[\bar E^2-\bar B^2+2i(\bar E\bar B)]^{1/2})
+{\rm c.c.}}{\cos(s[\bar E^2-\bar B^2+2i(\bar E\bar B)]^{1/2})-{\rm c.c.}}\nonumber\\
&&+ \left(\frac{m_e^2c^3}{e\hbar}\right)^2
+\frac{s^2}{3}(|\bar B|^2-|\bar E|^2)\Big],
\label{effectiveint}
\end{eqnarray}
where
$\bar E,\bar B$ are the dimensionless reduced fields in the unit of the critical field $E_c$,
\begin{equation}
\bar E=\frac{|{\bf E}|}{E_c},~~~~\bar B=\frac{|{\bf B}|}{E_c}.
\label{dimenlessEB}
\end{equation}
Expanding this expression in powers of $ \alpha $ up to $ \alpha ^3$
yields
the following values for the four constants:
\begin{equation}
\kappa_{2,0}=\frac{\alpha}{360\pi^2}E_c^{-2},\hskip0.3cm \kappa _{0,2} = 7\kappa_{2,0} ,\hskip0.3cm
\kappa _{3,0}= \frac{2 \alpha}{315\pi^2}E_c^{-4},
\hskip0.3cm \kappa_{1,2} = {\frac{13}{2}} \kappa _{3,0} .
\label{lconstant}
\end{equation}
The above results will receive higher corrections in QED
and are correct only up to order $ \alpha ^2$.
Up to this order,
the field-dependent dielectric and permeability tensors $\epsilon_{ik}$
and $\mu_{ik}$ (\ref{dh1}) have the following real parts
for weak fields
\begin{eqnarray}
{\rm Re}(\epsilon_{ik})&=&\delta_{ik} +
\frac{\alpha}{180\pi^2}\big[2(\bar E^2-\bar B^2)\delta_{ik}+7\bar B_i\bar B_k\big]
+{\mathcal O}( \alpha ^2),
\nonumber\\
{\rm Re}(\mu_{ik})&=&\delta_{ik} +
\frac{\alpha}{180\pi^2}\big[2(\bar E^2-\bar B^2)\delta_{ik}+7\bar E_i\bar E_k\big]
+{\mathcal O}( \alpha ^2).
\label{dielectricity}
\end{eqnarray}

\subsubsection{Weisskopf effective Lagrangian}

\label{weiss-effecitve}

Weisskopf \cite{Weisskopf1936} adopted a simpler method. He
considered first the special case in which ${\bf E}=0, {\bf
B}\not=0$ and used the Landau states to find $\Delta {\mathcal H}$
of Eq.~(\ref{heffective}), extracting from this $\kappa_{2,0}$ and $
\kappa _{3,0}$. Then he  added a weak electric field ${\bf E}\not=0$
to calculate perturbatively its contributions to $\Delta {\mathcal
H}$ in the Born approximation (see for example \cite{Landau1981a}).
This led again to the coefficients (\ref{lconstant}),
(\ref{dielectricity}). In addition to the weak field expansion of
real part of effective Lagrangian, Weisskopf also obtained the
leading order term considering very large field strengths $\bar E\gg
1$ or $\bar B\gg 1$,
\begin{eqnarray}
\Delta {\mathcal L}_{\rm eff}\sim -\frac{e^2}{12\pi^2 \hbar c}E^2\ln \bar E;\quad
\Delta {\mathcal L}_{\rm eff}\sim \frac{e^2}{12\pi^2 \hbar c}B^2\ln \bar B,
\label{wstong}
\end{eqnarray}
We shall address this same problem in Section~\ref{nonlinear} in the
framework of QED \cite{2006JKPSP} and we will compare and contrast
our exact expressions with the one given by Weisskopf. The crucial
point stressed by Weisskopf is that if one limits to the analysis of
the real part of the dielectric constant and magnetic permeability
then the nonlinearity of effective electromagnetic Lagrangian
represent only small corrections even for field strengths which are
much higher than the critical field strength $E_c$. As we will show
however, the contribution of the imaginary part of the effective
Lagrangian diverges as pointed out by Heisenberg and Euler
\cite{1936ZPhy...98..714H}.

\subsubsection{Imaginary part of the effective Lagrangian}

\label{imm-effecitve}

Heisenberg and Euler \cite{1936ZPhy...98..714H} were the first
to  realize that for ${\bf E}\not=0$ the powers
series expansion (\ref{leffective}) is not convergent,
due to singularities of the integrand in (\ref{effectiveint}) at
$s=\pi/\bar E, 2\pi/\bar E,\dots~$. They concluded
that the powers series
expansion (\ref{leffective}) does not yield
all corrections to the Maxwell
Lagrangian, calling for a more careful evaluation
of the integral representation (\ref{effectiveint}).
Selecting an integration path that avoids
these singularities, they found an imaginary term.
Motivated by Sauter's work \cite{1931ZPhy...69..742S}
on Klein paradox \cite{1929ZPhy...53..157K,1931ZPhy...73..547S}, Heisenberg and Euler
estimated the size of the imaginary term  in the effective Lagrangian as
\begin{equation}
{\rm Im}{\mathcal L}_{\rm eff}=-{\frac{8}{\pi}}{\bar E}^2 m_ec^2\left(\frac{m_ec}{h}\right)^3 e^{-\pi/\bar E},
\label{herate}
\end{equation}
and pointed out that it is associated with pair production by the electric field. The exponential in this expression is exactly reproducing the Sauter result (\ref{transmission}). However, for the first time the pre-exponential factor is determined. This imaginary term
in the effective Lagrangian is related to the imaginary parts of field-dependent dielectric $\epsilon$
and permeability $\mu$ of the vacuum.

In 1950's, Schwinger
\cite{1951PhRv...82..664S,1954PhRv...93..615S,1954PhRv...94.1362S}
derived the same formula (\ref{effectiveint}) within the {\it
Quantum Electro-Dymanics} (QED). In the following sections, our
discussions and computations will focus on the Schwinger formula,
the real and imaginary parts of effective Lagrangian for arbitrary
values of electromagnetic field strength.

The consideration of Heisenberg and Euler were applied to a uniform
electric field. The exponential factor $e^{-\pi/{\bar E}}$ in
Eqs.~(\ref{transmission}) and (\ref{herate}) characterizes the
transmission coefficient of quantum tunneling, Heisenberg and Euler
\cite{1936ZPhy...98..714H} introduced the critical field strength
(\ref{dimenlessEB}). They compared it with the field strength $E_e$
of an electron at its classical radius, $E_e=e/r_e^2$ where
$r_e=\alpha \hbar/(m_ec)$. They found the field strength $E_e$ is
137 time larger than the critical field strength $E_c$, i.e.
$E_e=\alpha ^{-1}E_c$. At a critical radius
$r_c=\alpha^{1/2}\hbar/(m_ec)< r_e$, the field strength of the
electron would be equal to the critical field strength $E_c$. There
have been various attempts to reach the critical field: in
Secs.~\ref{Z137} and \ref{heavy} we will examine the possibility of
reaching such value around the bare nucleus. In
Section~\ref{dyadosphere} we will discuss the possibility of
reaching such a field in an astrophysical setting around a black
hole.

In conclusion, if an electric field attempts to tear an electron out of the
filled state the gap energy must be gained over the distance of two electron
radii. The virtual particles give an electron a radius of the order of the
Compton wavelength $\lambda_C$. Thus we expect  a significant creation of
electron--positron pairs if the work done by the electric field $E$ over twice
the Compton wave length $\hbar/m_ec$ is larger than $2m_ec^2$
\begin{equation}
eE\left({\frac{2\hbar}{ m_ec}}\right)>2m_ec^2 .
\nonumber
\end{equation}
This condition defines a critical electric field (\ref{critical1}) above which
pair creation becomes abundant.
To have an idea how large this critical electric field is, we
compare it with the value of the electric field required to ionize a hydrogen atom.
There the above inequality
holds for twice of the Bohr radius and the Rydberg energy
\begin{equation}
eE_{\rm ion}\left({\frac{2\hbar}{  {\alpha  m_ec}}}\right)> \alpha ^2{m_ec^2},
\nonumber
\end{equation}
where $E_{\rm ion}=m_e^2e^5/\hbar^4=5.14\times 10^9$ V/cm, so that $E_c= E^{\rm ion}/ \alpha^3 $ is about $10^6$
times as large,
a value that has so far
not been reached in a laboratory on Earth.

\section{Pair production and annihilation in QED}\label{chap-pair-theory}

\subsection{Quantum Electro-Dynamics}\label{qedintro}

\label{qedvac}

\emph{Quantum Electro-Dynamics} (QED), the quantum theory of
electrons, positrons, and photons, was established by by Tomonaga
\cite{1946PThPh...1...27T}, Feynman
\cite{1948RvMP...20..367F,1949PhRv...76..749F,1949PhRv...76..769F},
Schwinger
\cite{1948PhRv...74.1439S,1949PhRv...75..651S,1949PhRv...76..790S}
and Dyson \cite{1949PhRv...75..486D,1949PhRv...75.1736D} and others
in the 1940's and 1950's \cite{Schwinger1958}. For decades, both
theoretical computations and experimental tests have been developed
to great perfection. It is now one of the fundamental pillars of the
theory of the microscopic world. Many excellent monographs have been
written
\cite{1959ittq.book.....B,Bjorken1998,Bjorken1965,Feynman1998,Feynman1965,1982els..book.....B,Itzykson2006,Lee1990,1951PhRv...82..664S,1954PhRv...93..615S,1954PhRv...94.1362S,Schwinger1970,Schwinger1998,1995qtf..book.....W,Kleinert1990}%
, so the concepts of the theory and the techniques of calculation are well
explained. On the basis of this material, we review some aspects and
properties of the QED that are relevant to the subject of the present review.

QED combines a relativistic extension of quantum mechanics with a
quantized electromagnetic field. The nonrelativistic system has a
unique ground state, which is the state with no particle, the
\emph{vacuum state\/}. The excited states contain a fixed number of
electrons and an arbitrary number of photons. As electrons are
allowed to become relativistic, their number becomes also arbitrary,
and it is possible to create pairs of electrons and positrons.

In the modern functional integral description, the nonrelativistic
system is described by a given set of fluctuating particle orbits
running forward in time. If the theory is continued to an imaginary
time, in which case one speaks of a \emph{Euclidean formulation\/},
the nonrelativistic system corresponds to a canonical statistical
ensemble of trajectories.

In the relativistic system, the orbits form worldlines in
four-dimensional space-time which may run in any time direction, in
particular they may run backwards in time, in which case the
backward parts of a line correspond to positrons. The number of
lines is arbitrary and the Euclidean formulation corresponds to a
grand-canonical ensemble. The most efficient way of describing such
an ensemble is by a single fluctuating field \cite{Kleinert1990}.

The vacuum state contains no physical particles. It does, however,
harbor zero-point oscillations of the electron and photon fields. In
the worldline description, the vacuum is represented by a
grand-canonical ensemble of interacting closed world lines. These
are called \emph{virtual particles\/}. Thus the vacuum contains the
full complexity of a many-body problem so that one may rightfully
say that \emph{the vacuum is the world\/} \cite{Streater2000}. In
the Fourier decomposition of the fluctuating fields, virtual
particles correspond to Fourier components, or \emph{modes\/}, in
which the 4-vectors
of energy and momentum $k^{\mu}\equiv(k^{0},\mathbf{k})\equiv({\mathcal{E}%
},\mathbf{k})$ do not satisfy the mass-shell relation
\begin{equation}
k^{2}\equiv(k^{0})^{2}-c^{2}|\mathbf{k}|^{2}={\mathcal{E}}^{2}-c^{2}%
|\mathbf{k}|^{2}=m_{e}^{2}c^{4}, \label{massshellrelation}%
\end{equation}
valid for real particles.

The only way to evaluate physical consequences from QED is based on
the smallness of the electromagnetic interaction. It is
characterized by the dimensionless fine structure constant $\alpha$.
All theoretical results derived from QED are found in the form of
series expansions in powers of $\alpha$, which are expansions around
the non-interacting system. Unfortunately, all these expansions are
badly divergent (see e.g. Section 4.62 in \cite{Kleinert2008}). The
number of terms contributing to the same order of $\alpha$ grows
factorially fast, i.e., faster than any exponential, leading to a
zero radius of convergence. Fortunately, however, the coupling
$\alpha$ is so small that the series possess an apparent convergence
up to order $1/\alpha\approx137$, which is much higher than will be
calculable for a long time to come (see e.g. Section 4.62 in
\cite{Kleinert2008}). With this rather academic limitation,
perturbation expansions are well defined.

In perturbation expansions, all physical processes are expressible
in terms of \emph{Feynman diagrams\/}. These are graphic
representations of the interacting world lines of all particles.
Among these lines, there are some which run to infinity. They
satisfy the mass shell relation (\ref{massshellrelation}) and
describe real particles observable in the laboratory. Those which
remain inside a finite space-time region are virtual.

The presence of virtual particles in the perturbation expansions leads to
observable effects. Some of these have been measured and calculated with great
accuracy. The most famous examples are

\begin{enumerate}
\item the electrostatic polarizability of quantum fluctuations of the QED
vacuum has been measured in the Lamb shift \cite{1947PhRv...72..241L,1953PhRv...89...98T}.

\item the anomalous magnetic moment of the electron \cite{1948PhRv...73..416S,1949PhRv...76..769F,1972PhR.....3..193L,1974PhRvD..10.4007C,1977PhLB...68..191B}.

\item the dependence of the electric charge on the distance. It is observed by
measuring cross-sections of electron--positron collisions, most recently in the
L3-experiments at the \emph{Large Electron-Positron Collider\/} (LEP) at CERN
\cite{L3CERN}.

\item the \emph{Casimir effect\/} caused by virtual photons, i.e., by the
fluctuations of the electromagnetic field in the QED vacuum \cite{Casimir1948,Fierz1960}. It
causes an attractive force \cite{1958Phy....24..751S,1997PhRvL..78....5L,1998PhRvL..81.4549M} between two uncharged conducting
plates in the vacuum (see also
\cite{2001PhRvE..63e1101H,2000PhRvA..62e2109H,2004PhRvA..69b2117C,Xue:1987fa,Xue:1988xj,1986AcPSn..34.1084X,Zheng1993}).
\end{enumerate}

There are, of course, many other discussions of the effects of virtual
particles caused either by external boundary conditions or by external
classical fields \cite{1992PNAS...89.4091S,1992PNAS...8911118S,1993PNAS...90..958S,1993PNAS...90.2105S,1993PNAS...90.4505S,1993PNAS...90.7285S,1994PNAS...91.6473S,Bordag1999,Bordag1996,2004JPhA...37R.209M,Milton2001,2001PhLB..508..211X,2003PhRvD..68a3004X,2003MPLA...18.1325X,2005GReGr..37..857X}.

An interesting aspect of virtual particles both theoretically and
experimentally is the possibility that they can become real by the effect of
external fields. In this case, real particles are excited out of the vacuum. In
the previous Section~\ref{sauter} and \ref{hew-effecitve}, we have shown that
this possibility was first pointed out in the framework of quantum mechanics by
Klein, Sauter, Euler and Heisenberg
\cite{1929ZPhy...53..157K,1931ZPhy...73..547S,1931ZPhy...69..742S,1936ZPhy...98..714H}
who studied the behavior of the Dirac vacuum in a strong external electric
field. Afterward, Schwinger studied this process and derived the probability
(\textit{Schwinger formula}) in the field theory of Quantum Electro-Dynamics,
which will be described in this chapter. If the field is sufficiently strong,
the energy of the vacuum can be lowered by creating an electron--positron pair.
This makes the vacuum unstable. This is the
\emph{Sauter-Euler-Heisenberg-Schwinger process\/} for electron--positron pair
production. There are many reasons for the interest in the phenomenon of pair
production in a strong electric field. The most compelling one is that now both
laboratory conditions and astrophysical events provide possibilities for
observing this process.

In the following chapters, in addition to reviewing the \textit{Schwinger
formula} and QED-effective Lagrangian in constant electromagnetic fields, we
will also derive the probability of pair production in an alternating field,
and discuss theoretical studies of pair production in (i) electron-beam--laser
collisions and (ii) superstrong Coulomb potential. In addition, the plasma
oscillations of electron--positron pairs in electric fields will be reviewed in
Section \ref{time-independent}. The rest part of this chapter, we shall use
natural units $\hbar=c=1$.

%%%%%%%%%%%%%%%%%%%%%%%%%%%%%%%%%%%%%%%%%%%%%%%%%%%%%%%%%%%%%%%%%%%%%%%%%%%%%%%%%%%%%%

\subsection{Basic processes in Quantum Electro-Dynamics}\label{qedprocesse}

The total Lagrangian describing the interacting system of photons,
electrons, and positrons reads, see e.g. \cite{1982els..book.....B}
\begin{equation}
{\mathcal{L}}={\mathcal{L}}_{0}^{\gamma}+{\mathcal{L}}_{0}^{e^{+}e^{-}%
}+{\mathcal{L}}_{\mathrm{int}}, \label{qcdl}%
\end{equation}
where the free Lagrangians ${\mathcal{L}}_{0}^{e^{+}e^{-}}$ and ${\mathcal{L}%
}_{0}^{\gamma}$ for electrons and photons are expressed in terms of quantized
Dirac field $\psi(x)$ and quantized electromagnetic field $A_{\mu}(x)$ as
follows:
\begin{align}
{\mathcal{L}}_{0}^{e^{+}e^{-}}  &  =~\,\bar{\psi}(x)(i\gamma^{\mu}%
\partial_{\mu}-m_{e})\psi(x),\label{L0pair}\\
{\mathcal{L}}_{0}^{\gamma}~~  &  =-{\frac{1}{4}}F_{\mu\nu}(x)F^{\mu\nu
}(x)+\mathrm{gauge\!-\!fixing~term}. \label{L0gamma}%
\end{align}
Here $\gamma^{\mu}$ are the $4\times4$ Dirac matrices, $\bar{\psi}%
(x)\equiv\psi^{\dagger}(x)\gamma^{0}$, and $F_{\mu\nu}=\partial_{\mu}A_{\nu
}-\partial_{\nu}A_{\mu}$ denotes the electromagnetic field tensor. Minimal
coupling gives rise to the interaction Lagrangian
\begin{equation}
{\mathcal{L}}_{\mathrm{int}}=-ej^{\mu}(x)A_{\mu}(x),\quad\quad j^{\mu}%
(x)=\bar{\psi}(x)\gamma^{\mu}\psi(x). \label{L0int}%
\end{equation}

After quantization, the photon field is expanded into plane waves as
\begin{equation}
A_\mu(x)=\int \frac{d^3k}{2k_0(2\pi)^3}\sum^3_{\lambda=1}\left[a^{(\lambda)}({\bf k} )\epsilon_\mu^{(\lambda)}({\bf k} )e^{-ikx}
+a^{(\lambda)\dagger}({\bf k} )\epsilon_\mu^{(\lambda)*}({\bf k} )e^{ikx}\right],
\label{photonmodes}
\end{equation}
where $\epsilon_\mu^{(\lambda)}$ are polarization vectors, and
$a^{(\lambda)}$, $a^{(\lambda)\dagger}$
are annihilation and creation operators of photons.
The quantized fermion field $\psi(x)$ has the expansion
\begin{eqnarray}
\!\!\!\!\!\!\!\!\!\psi(x) &=& \int{\frac{d^3k}{ (2\pi)^3}} \frac{m}{ k^0}
\sum_{\alpha=1,2}\Big[b_\alpha({\bf k},s_3)
u^{(\alpha)}({\bf k},s_3)e^{-ikx}+d^\dagger_\alpha({\bf k},s_3)v^{(\alpha)}({\bf k},s_3)e^{ikx}\Big],\nonumber\\
\label{fermionmodes}
\end{eqnarray}
where the four-component spinors $u^{(\alpha)}({\bf k},s_3)$,
$v^{(\alpha)}( {\bf k},s_3)$ are positive and negative energy
solutions of the Dirac equation with momentum ${\bf k}$ and spin
component $s_3$. The operators $b({\bf k},s_3)$, $b^\dagger({\bf
k},s_3)$ annihilate and create electrons, the operators $d({\bf
k},s_3)$ and $d^\dagger({\bf k},s_3)$ do the same for positrons
\cite{1982els..book.....B}.

In the framework of QED the transition probability from an initial to a final
state for a given process is represented by the imaginary part of the unitary
S-matrix squared
\begin{equation}
{\mathcal P}_{f\leftarrow i}=\left\vert \left\langle \mathrm{{f,out}\left\vert
Im\,{\mathcal S}\right\vert {i,in}}\right\rangle \right\vert ^{2},
\end{equation}
where
\begin{equation}
\mathrm{Im\,{\mathcal S}}=(2\pi)^{4}\delta^{4}(P_{f}-P_{i})\left\vert M_{fi}\right\vert ,
\end{equation}
$M_{fi}$ is called matrix element and $\delta$-function stays for
energy-momentum conservation in the process.

When initial state contains two particles with energies $\epsilon_{1}$ and
$\epsilon_{2}$, and final state contain arbitrary number of particles having
3-momenta $\mathbf{p}_{i}^{\prime}$, the transition probability per unit time
and unit volume is given by%
\begin{equation}
\frac{d{\mathcal P}_{f\leftarrow i}}{dVdt}=(2\pi)^{4}\delta^{4}(P_{f}-P_{i})\left\vert
M_{fi}\right\vert ^{2}\frac{1}{4\epsilon_{1}\epsilon_{2}}%
%TCIMACRO{\dprod \limits_{i}}%
%BeginExpansion
{\displaystyle\prod\limits_{i}}
%EndExpansion
\frac{d^{3}p_{i}^{\prime}}{\left(  2\pi\right)  ^{3}2\epsilon_{i}}.
\label{dif_prob}%
\end{equation}
The Lorentz invariant differential cross-section for a given process
is then obtained from (\ref{dif_prob}) by dividing it on the flux
density of initial
particles%
\begin{equation}
d\sigma=(2\pi)^{4}\delta^{4}(P_{f}-P_{i})\left\vert M_{fi}\right\vert
^{2}\frac{1}{4I_{kin}}%
%TCIMACRO{\dprod \limits_{i}}%
%BeginExpansionv
{\displaystyle\prod\limits_{i}}
%EndExpansion
\frac{d^{3}p_{i}^{\prime}}{\left(  2\pi\right)  ^{3}2\epsilon_{i}},
\end{equation}
where $p_{1}$ and $p_{2}$\ are particles' 4-momenta, $m_{1}$ and $m_{2}$\ are
their masses respectively, $I_{kin}=\sqrt{\left(  p_{1}p_{2}\right)  ^{2}-m_{1}%
^{2}m_{2}^{2}}$.

It is useful to work with Mandelstam variables which are kinematic invariants
built from particles 4-momenta. Consider the process $A+B\longrightarrow C+D$.
Lorentz invariant variables can be constructed in the following way%
\begin{align}
s  &  =\left(  p_{A}+p_{B}\right)  ^{2}=\left(  p_{C}+p_{D}\right)
^{2},\nonumber\\
t  &  =\left(  p_{A}+p_{C}\right)  ^{2}=\left(  p_{B}+p_{D}\right)  ^{2},\\
u  &  =\left(  p_{B}+p_{C}\right)  ^{2}=\left(  p_{A}+p_{D}\right)
^{2}.\nonumber
\end{align}
Since any incoming particle can be regarded as outgoing antiparticle, it gives
rise to the crossing symmetry property of the scattering amplitude, which is
best reflected in the Mandelstam variables. In fact, reactions
$A+B\longrightarrow C+D$, $A+\bar{C}\longrightarrow\bar{B}+D$ or $A+\bar
{D}\longrightarrow C+\bar{B}$ where the bar denotes the antiparticle are just
different cross-channels of a single general reaction. The meaning of the
variables $s,t,u$ changes, but the amplitude is the same.

The S-matrix is computed through the interaction operator as%
\begin{equation}
{\mathcal S}={\mathcal T}\exp\left(  i\int{\mathcal{L}}_{\mathrm{int}}d^{4}x\right)  ,
\end{equation}
where ${\mathcal T}$ is the chronological operator. Perturbation theory is applied,
since the fine structure constant is small, while any additional interaction
in collision of particles contains the factor $\alpha$.

A\ simple and elegant way of computation of the S-matrix and consequently of
the matrix element $M_{fi}$\ is due to Feynman, who discovered a
graphical way to depict each QED process, in momentum representation.

In what follows we consider briefly the calculation for the case of
Compton scattering process \cite{1982els..book.....B}, which is
given by two Feynman diagrams. Conservation law for 4-momenta is
$p+k=p^{\prime}+k^{\prime}$, where $p$ and $k$ are 4-momenta of
electron and photon respectively, and invariant $I_{kin}=\frac{1}{4}
(s-m_e^{2})^{2}$. After the calculation of traces with
gamma-matrices, the final result, expressed in Mandelstam variables,
is
\begin{align}
|M_{fi}|^{2}  &  =2^{7}\pi^{2}e^{4}\left[  \frac{m_e^{2}}{s-m_e^{2}}+\frac{m_e^{2}%
}{u-m_e^{2}}+\left(
\frac{m_e^{2}}{s-m_e^{2}}+\frac{m_e^{2}}{u-m_e^{2}}\right)
^{2}\right. \nonumber\\
&  \left.  -\frac{1}{4}\left(  \frac{s-m_e^{2}}{u-m_e^{2}}+\frac{u-m_e^{2}}{s-m_e^{2}%
}\right)  \right]  , \label{M_fi_gamma1_1}%
\end{align}
$s=(p+k)^{2}$, $t=(p-p^{\prime})^{2}$ and$\ u=(p-k^{\prime})^{2}$. Since the
differential cross-section is independent of the azimuth of $\mathbf{p}%
_{1}^{\prime}$ relative to $\mathbf{p}_{1}$,\ it is obtained from
(\ref{M_fi_gamma1_1}) as%
\begin{equation}
d\sigma=\frac{1}{64\pi}|M_{fi}|^{2}\frac{dt}{I_{kin}^{2}}.
\label{dsigma}
\end{equation}

In the laboratory frame, where $s-m_e^{2}=2m_ew$, $u-m_e^{2}=-2m_ew^{\prime}$
and electron is at rest before the collision with photon, the differential
cross-section of Compton scattering is thus given by the Klein--Nishina formula
\cite{1929ZPhy...52..853K}
\begin{equation}
d\sigma=\frac{1}{2}\left(  \frac{e^{2}}{m}\right)  ^{2}\left(  \frac
{\omega^{\prime}}{\omega}\right)  ^{2}\left(  \frac{\omega}{\omega^{\prime}}+\frac{\omega^{\prime}}%
{\omega}-\sin^{2}\vartheta\right)  , \label{KN}%
\end{equation}
where $\omega$ and $\omega^{\prime}$\ are frequencies of photon
before and after the collision, $\vartheta$\ is the angle at which
the photon is scattered.

\subsection{The Dirac and the Breit--Wheeler processes in QED}\label{DiracBWqed}

We turn now to the formulas obtained within framework of quantum mechanics by
Dirac \cite{1930PCPS...26..361D} and Breit and Wheeler
\cite{1934PhRv...46.1087B} within QED. The crossing symmetry allows to readily
write the matrix element for the pair production (\ref{ee2gamma}) and pair
annihilation (\ref{2gammaee}) processes with the energy-momentum conservation
written as $p_++p_-=k_1+k_2$, where $p_+$ and $p_-$ are 4-momenta of the
positron and the electron, $k_1$ and $k_2$ are 4-momenta of two photons. It is
in fact given by the same formula (\ref{M_fi_gamma1_1}) with the substitution
$p\rightarrow p_{-},$ $p^{\prime }\rightarrow p_{+},$ $k\rightarrow k_{1},$
$k^{\prime}\rightarrow k_{2}$, but with different meaning of the kinematic
invariants $s=(p_{-}-k_{1})^{2}$, $t=(p_{-}+p_{+})^{2}$,$\
u=(p_{-}-k_{2})^{2}$. Matrix elements for Dirac and Breit--Wheeler processes
are the same. The differential cross-section of the Dirac process is obtained
from (\ref{M_fi_gamma1_1})\ with the exchange $s\leftrightarrow t$ and the
invariant $I_{kin}=\frac{1}{4}t(t-4m_e^{2})$, which leads to (\ref{Dirac
cross-section}). For the case of the Breit--Wheeler process with the invariant
$I_{kin}=\frac{1}{4}t^{2}$, the result is reduced to (\ref{BW section0}).

Since the Dirac pair annihilation process (\ref{ee2gamma}) is the inverse of
Breit--Wheeler pair production (\ref{2gammaee}), it is useful to compare the
cross-section of the two processes. We note that the squared transition
amplitude $|M_{fi}|^2$ must be the same for two processes, due to the CPT
invariance. The cross-sections could be different only by kinematics and
statistical factors. Let us consider the pair annihilation process in the
center of mass system where ${\mathcal E}={\mathcal E}_1+{\mathcal
E}_2={\mathcal E}'_1+{\mathcal E}'_2$ is the total energy, the initial and
final momenta are equal and opposite, ${\bf p}_1 =-{\bf p}_2\equiv {\bf p}$ and
${\bf p}'_1 =-{\bf p}'_2\equiv {\bf p}'$. The differential cross-section is
given by (\ref{dsigma}). For the Breit and Wheeler process (\ref{2gammaee}) of
two colliding photons with 4-momenta $k_1$ and $k_2$, the scalar
$I^2_{\gamma\gamma}=(k_1k_2)^2$. For the Dirac process (\ref{ee2gamma}) of
colliding electron and positron with 4-momenta $p_1$ and $p_2$, the scalar
$I^2_{e^+ e^-} = (p_1p_2)^2-m_e^4$. As results, one has
\begin{equation}\label{I}
\frac{d\sigma_{\gamma\gamma}}{d\sigma_{e^+ e^-}}
=\frac{I^2_{e^+ e^-}}{I^2_{\gamma\gamma}} = \frac{2(k_1k_2)-4m_e^2}{2(k_1k_2)}= \frac{{\mathcal E}^2-2m_e^2}{{\mathcal E}^2}
=\left(\frac{|{\bf p}|}{{\mathcal E}}\right)^2=\hat\beta^2,
\end{equation}
where momenta and energies are related by
\begin{equation}
(p_1+p_2)^2=(k_1+k_2)^2=2(k_1k_2)=2{\mathcal E}^2.
\nonumber
\end{equation}
Integrating Eq.~(\ref{I}) over all scattering angles yields the total cross-section. Whereas the previous $\sigma_{\gamma\gamma}$ required division by a Bose factor 2 for the two identical photons in the final state, the cross-section $\sigma_{e^+ e^-}$ has no such factor since the final electron and positron are not identical. Hence we obtain
\begin{equation}\label{sigmaee-sigmagg}
\sigma_{e^+ e^-} = \frac{1}{2\hat\beta^2}\sigma_{\gamma\gamma}.
\end{equation}
By re-expressing the kinematic quantities in the laboratory frame, one obtains the Dirac cross-section (\ref{Dirac cross-section}).

As shown in Eq.~(\ref{sigmaee-sigmagg}) in the center of mass of the
system, the two cross-sections $\sigma_{e^+ e^-}$ and
$\sigma_{\gamma\gamma}$ of the above described phenomena differ only
in the kinematics and statistical factor $1/(2\hat\beta^{2})$, which
is related to the fact that the resulting particles are massless or
massive. The process of electron and positron production by the
collision of two photons has a kinematic energy threshold, while the
process of electron and positron annihilation to two photons has not
such kinematic energy threshold. In the limit of high energy
neglecting the masses of the electron and positron,
$\hat\beta\rightarrow 1$, the difference between two cross-sections
$\sigma_{e^+ e^-}$ and $\sigma_{\gamma\gamma}$ is only the
statistical factor $1/2$.

The total cross-sections (\ref{Dirac section0}) of Breit--Wheeler's and Dirac's
process are of the same order of magnitude $\sim 10^{-25}{\rm cm}^2$ and have
the same energy dependence $1/{\mathcal E}^2$ above the energy threshold. The
energy threshold ($2m_ec^2$) have made until now technically impossible to
observe the pair production by the Breit--Wheeler process in laboratory
experiments at the intersection of two beams of X-rays. Another reason is of
course the smallness of the total cross-section (\ref{bwsection3})
($\sigma_{\gamma\gamma}\lesssim 10^{-25}$cm$^2$) and the experimental
limitations on the intensities $I_i$ (\ref{bwaa}) of the light beams. We shall
see however, that this Breit--Wheeler process occurs routinely in the
dyadosphere of a black hole. The observations of such phenomena in the
astrophysical setting are likely to give the first direct observational test of
the validity of the Breit--Wheeler process, see e.g. \cite{Ruffini2009}.

\subsection{Double-pair production}\label{doub}

Following the Breit--Wheeler pioneer work on the process (\ref{2gammaee}),
Cheng and Wu
\cite{1969PhRvL..22..666C,1969PhRvL..23.1311C,1969PhRv..182.1852C,1969PhRv..182.1868C,1969PhRv..182.1873C,1969PhRv..182.1899C}
considered the high-energy behavior of scattering amplitudes and cross-section
of two photon collision, up to higher order ${\mathcal O}(\alpha^4)$
\cite{1970PhRvD...1.3414C},
\begin{equation}\label{gamma4ee}
\gamma_{1}+\gamma_{2}\rightarrow e^{+}+e^{-}+ e^{+}+e^{-}.
\end{equation}
For this purpose, they calculated the two photon forward scattering
amplitude $M_{\gamma\gamma}$ (see Eq.~(\ref{dif_prob})) by taking
into account all relevant Feynman diagrams via two electron loops up
to the order ${\mathcal O}(\alpha^4)$. The total cross-section
$\sigma_{\gamma\gamma}$ for photon-photon scattering is related to
the photon-photon scattering amplitude $M_{\gamma\gamma}$ in the
forward direction by the optical theorem,
\begin{equation}\label{optict}
\sigma_{\gamma\gamma}(s)=\frac{1}{s}{\rm Im}M_{\gamma\gamma},
\end{equation}
where $s$ is the square of the total energy in the center of mass
system. They obtained the total cross-section of double pair
production (\ref{gamma4ee}) at high energy $s\gg 2m_e$,
\begin{equation}\label{cwresult}
\lim_{s\rightarrow \infty}\sigma_{\gamma\gamma}(s)=
\frac{\alpha^4}{36\pi m_e^2}[175\zeta(3)-38]\sim 6.5 \mu b,
\end{equation}
which is  independent of $s$ as well as helicities of the incoming photons. Up
to the $\alpha^4$ order, Eq.~(\ref{cwresult}) is the largest term in the total
cross-section for photon-photon scattering at very high energy. This can be
seen by comparing Eq.~(\ref{cwresult}) with the cross-section (\ref{BW
section0},\ref{bwsection3}) of the Breit--Wheeler process (\ref{2gammaee}),
which is the lowest-order process in a photon-photon collision and vanishes as
$s\rightarrow \infty$. Thus, although the Breit--Wheeler cross-section is of
lower order in $\alpha$, the Cheng-Wu cross-section (\ref{cwresult}) is larger
as the energy becomes sufficiently high. In particular, the Cheng-Wu
cross-section (\ref{cwresult}) exceeds the Breit--Wheeler one (\ref{BW
section0}) as the center of mass energy of the photon ${\mathcal E}\ge
0.24$GeV. Note that in the double pair production (\ref{gamma4ee}), the energy
threshold is ${\mathcal E}\ge 2m_e$ rather than ${\mathcal E}\ge m_e$ in the
one pair production of Breit--Wheeler.

In Ref.~\cite{1970PhRvD...1..467C,1970PhRvD...2.2103C}, using the same method
Cheng and Wu further calculated other cross-sections for high-energy
photon-photon scattering to double pion, muon and electron--positron pairs:

\begin{itemize}

\item the process of double muon pair production
$\gamma_{1}+\gamma_{2}\rightarrow \mu^{+}+\mu^{-}+ \mu^{+}+\mu^{-}$
and its cross-section can be obtained by replacing $m_e\rightarrow m_\mu$ in Eq.~(\ref{cwresult}), thus $\sigma_{\gamma\gamma}(s)\sim 1.5\cdot 10^{-4} \mu b$.

\item the process of double pion pair production
$\gamma_{1}+\gamma_{2}\rightarrow \pi^{+}+\pi^{-}+ \pi^{+}+\pi^{-}$
and its extremely small cross-section
\begin{equation}\label{cwpi}
\lim_{s\rightarrow \infty}\sigma_{\gamma\gamma}(s)=
\frac{\alpha^4}{144\pi m_\pi^2}[7\zeta(3)+10]\sim 0.23\cdot 10^{-5} \mu b,
\end{equation}

\item the process of one pion pair and one electron--positron pair production $\gamma_{1}+\gamma_{2}\rightarrow e^{+}+e^{-}+ \pi^{+}+\pi^{-}$
and its small cross-section
\begin{equation}\label{cwpiee}
\lim_{s\rightarrow \infty}\sigma_{\gamma\gamma}(s)=
\frac{2\alpha^4}{27\pi m_\pi^2}\left[\left(\ln\frac{m_\pi^2}{m_e^2}\right)^2
+\frac{16}{3}\ln\frac{m_\pi^2}{m_e^2}+\frac{163}{18}\right]\sim 0.26\cdot 10^{-3} \mu b,
\end{equation}
which is more than one hundred times larger than (\ref{cwpi}).

\item the process of one pion pair and one muon-antimuon pair production $\gamma_{1}+\gamma_{2}\rightarrow \mu^{+}+\mu^{-}+ \pi^{+}+\pi^{-}$
and its small cross-section can be obtained by replacing $m_e\rightarrow m_\mu$ in Eq.~(\ref{cwpiee}) everywhere.

\end{itemize}

\subsection{Electron-nucleus bremsstrahlung and pair production by a photon in the field of a nucleus}\label{bremss}

The other two important QED processes, related by the crossing symmetry are the
electron-nucleus bremsstrahlung (\ref{eibr}) and creation of electron--positron
pair by a photon in the field of a nucleus (\ref{gi2p}). These processes were
considered already in the early years of QED. They are of higher order,
comparing to the Compton scattering and the Breit--Wheeler processes,
respectively, and contain one more vertex connecting fermions with the photon.

The nonrelativistic cross-section for the process (\ref{eibr}) was derived by Sommerfeld \cite{1931AnP...403..257S}. Here we remind the basic results obtained in the relativistic case by Bethe and Heitler \cite{1934RSPSA.146...83B} and independently by Sauter \cite{1934AnP...412..404S}.
The Feynman diagram for bremsstrahlung can be imagined considering for the Compton scattering, but treating one of the photons as virtual one corresponding to an external field. We consider this process in Born approximation, and the momentum recoil of the nucleus is neglected. Integrating the differential cross-section over all directions of the photon and the outgoing electron one has, see e.g. \cite{1982els..book.....B}
\begin{equation}
\begin{array}{lr}
d\sigma=Z^2\alpha r_e^2\frac{d\omega}{\omega}\frac{p^\prime}{p}\left\{\frac{4}{3}-2\epsilon\epsilon^\prime\frac{p^2+p^{\prime2}}{p^2 p^{\prime2}}+m_e^2\left(l\frac{\epsilon^\prime}{p^3}+l^\prime\frac{\epsilon}{p^{\prime3}-\frac{ll^\prime}{pp^\prime}}\right)+\right.  \\ \left.+L\left[\frac{8\epsilon\epsilon^\prime}{3pp^\prime}+\frac{\omega^2}{p^3p^{\prime3}}\left(\epsilon^2\epsilon^{\prime2}+p^2p^{\prime2}+m_e^2\epsilon\epsilon^\prime\right)+\frac{m_e^2\omega}{2pp^\prime}\left(l\frac{\epsilon\epsilon^\prime+p^2}{p^3}-l^\prime\frac{\epsilon\epsilon^\prime+p^{\prime2}}{p^{\prime3}}\right)\right]\right\},
\end{array}
\end{equation}
where
\begin{equation}
L=\log\frac{\epsilon\epsilon^\prime+pp^\prime-m_e^2}{\epsilon\epsilon^\prime-pp^\prime-m_e^2}, \quad
l=\log\frac{\epsilon+p}{\epsilon-p}, \quad
l^\prime=\log\frac{\epsilon^\prime+p^\prime}{\epsilon^\prime-p^\prime},
\end{equation}
and $\omega$ is photon energy, $p$ and $p^\prime$ are electron momenta before and after the collision, respectively, $\epsilon$ and $\epsilon^\prime$ are its initial and final energies.

The averaged cross-section for the process of pair production by a photon in the field of a nucleus may be obtained by applying the transformation rules relating the processes (\ref{gi2p}) and (\ref{eibr}), see e.g. \cite{1982els..book.....B}. The result is
\begin{equation}
\begin{array}{lcr}
d\sigma = Z^2\alpha r_e^2\frac{p_+p_-}{\omega^3}d\epsilon_+\biggl\{-\frac{4}{3}-2\epsilon_+\epsilon_-\frac{p_+^2+p_-^2}{p_+^2p_-^2}+m_e^2\left(l_-\frac{\epsilon_+}{p_-^3}+l_+\frac{\epsilon_-}{p_+^3-\frac{l_+l_-}{p_+p_-}}\right)+  \\ +L\biggl[\frac{8\epsilon_+\epsilon_-}{3p_+p_-}+\frac{\omega^2}{p_+^3p_-^3}\left(\epsilon_+^2\epsilon_-^2+p_+^2p_-^2-m_e^2\epsilon_+\epsilon_-\right)- \\ -\frac{m_e^2\omega}{2p_+p_-}\left(l_+\frac{\epsilon_+\epsilon_- -p_+^2}{p_+^3}+l_-\frac{\epsilon_+\epsilon_- -p_-^2}{p_-^3}\right)\biggr]\biggr\},
\label{dsigmapm}
\end{array}
\end{equation}
where
\begin{equation}
L=\log\frac{\epsilon_+\epsilon_-+p_+p_-+m_e^2}{\epsilon_+\epsilon_- -p_+p_-+m_e^2}, \quad
l_\pm=\log\frac{\epsilon_\pm+p_\pm}{\epsilon_\pm-p_\pm},
\end{equation}
and $p_\pm$ and $\epsilon_\pm$ are momenta and energies of positron and electron respectively.

The total cross-section for this process is given in Section
\ref{higher}, see (\ref{sigmaZ1Z2}). In the ultrarelativistic
approximation relaxing the condition $Z\alpha\ll1$ the pair
production by the process (\ref{gi2p}) was treated by Bethe and
Maximon in \cite{1954PhRv...93..768B,1954PhRv...93..788D}. The total
cross-section (\ref{sigmaZ1Z2}) is becoming then
\begin{equation}
\sigma=\frac{28}{9}Z^2\alpha r_e^2\left(\log\frac{2\omega}{m_e}-\frac{109}{42}-f(Z\alpha)\right),
\label{sigmaZ1Z2r}
\end{equation}
where
\begin{equation}
f(Z\alpha)=\gamma_E+Re\Psi(1+iZ\alpha)=(Z\alpha)^2
\sum_{n=1}^\infty\frac{1}{n[n^2+(Z\alpha)^2]}.
\end{equation}

\subsection{Pair production in collision of two ions}\label{ionion}

The process of the pair production by two ions (\ref{2pe+e-}) in
ultrarelativistic approximation was considered by Landau and
Lifshitz \cite{Landau1934} and Racah \cite{Racah1937}. For the
modern review of these topics see \cite{2007PhR...453....1B}. The
corresponding differential cross-section with logarithmic accuracy
can be obtained from the differential cross-section (\ref{dsigmapm})
taking its ultrarelativistic approximation $\gamma\gg1$ for the
Lorentz factor of the relative motion of the two nuclei with charges
$Z_1$ and $Z_2$ respectively, and treating the real photon line in
the process (\ref{gi2p}) as a virtual photon corresponding to the
external field of the nucleus. One should then multiply the
cross-section by the spectrum of these equivalent photons, see
Section \ref{higher}, and the result is
\begin{equation}
d\sigma=\frac{8}{\pi}r_e^2(Z_1Z_2\alpha)^2\frac{d\epsilon_+d\epsilon_-}{(\epsilon_++\epsilon_-)^4}\left(\epsilon_+^2+\epsilon_-^2+\frac{2}{3}\epsilon_+\epsilon_-\right)\log\frac{\epsilon_+\epsilon_-}{m_e(\epsilon_++\epsilon_-)}\log\frac{m_e\gamma}{(\epsilon_++\epsilon_-)}.
\end{equation}

The total cross-section is given in Section \ref{higher}, see
(\ref{sigmaL}) and (\ref{sigmaR}). More recent results containing
higher order in $\alpha$ corrections are obtained in
\cite{2002PhLB..538...45B,2002PhRvA..66d2720B}, see also
\cite{2006PPNL....3..246V,2007PPNL....4...18V}.

\begin{figure}[!ht]
    \centering
        \includegraphics[width=5cm]{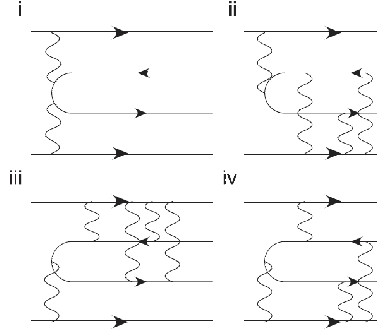}
    \caption{Classification of the $e^+e^-$ pair production by the number of photons attached to a nucleus. Reproduced from \cite{2007PhR...453....1B}.}
    \label{coulcor}
\end{figure}
Lepton pair production in relativistic ion collision to all orders
in $Z\alpha$ with logarithmic accuracy is studied in
\cite{2003JPhG...29.1227G} where the matrix elements are separated
in different classes, see Fig. \ref{coulcor}, according to numbers
of photon lines attached to a given nucleus
\begin{equation}
M=M_{(\mathrm{i})}+M_{(\mathrm{ii})}+M_{(\mathrm{iii})}+M_{(\mathrm{iv})}.
\end{equation}
The Born amplitude $\left|M_{(\mathrm{i})}\right|^2$ corresponding
to the lowest order in $Z\alpha$ with one photon line attached to
each nucleus was computed by Landau and Lifshitz \cite{Landau1934}
who obtained the famous $L_\gamma^3$ dependence of the cross-section
in (\ref{sigmaL}). It should be mentioned that the Racah formula
(\ref{sigmaR}) in contrast with the Landau and Lifshitz result
(\ref{sigmaL}) contains also terms proportional to $L_\gamma^2$ and
$L_\gamma$. These terms come from the absolute square of
$M_{(\mathrm{ii})}$ and $M_{(\mathrm{iii})}$ and their interference
with the Born amplitude $M_{(\mathrm{i})}$
\cite{2002PhRvA..66d2720B}. Their result for the Coulomb corrections
which is defined as the difference between the full cross-section
and the Born approximation, in order $L_\gamma^2$ is of the
``Bethe--Maximon'' type \cite{2002PhRvA..66d2720B}
\begin{equation}
\sigma_\mathrm{C}=\frac{28}{27\pi}r_e^2(Z_1Z_2\alpha)^2\left[f(Z_1\alpha)f(Z_2\alpha)\right](L_\gamma^2+{\mathcal O}(L_\gamma)).
\label{sigmaC}
\end{equation}
The Coulomb corrections here are up to $L_\gamma^2$-term in the
Racah formula (\ref{sigmaR}).

The calculations mentioned above were all made as early as in the
1930s. Clearly, at that time only $e^+e^-$ pair production was
discussed. However, these calculations can be considered for any
lepton pair production, for example $\mu^+\mu^-$ pair production, as
long as the total energy in the center of mass of the system is
large enough. However, simple substitution $m_e\longrightarrow m_l$,
where $l$ stands for any lepton is not sufficient since when the
energy reaches the inverse radius $\Lambda=1/R\sim30{\mathrm MeV}$
of the nucleus the electric field of the nucleus cannot be
approximated as a Coulomb field of a point-like particle
\cite{1998PhRvD..57.4025I}. The review of computation for lepton
pair production can be found in \cite{1973SvPhU..16..322G} and
\cite{1975PhR....15..181B}.

Another effect of large enough collision energy is multiple pair
production. Early work on this subject started with the observation
that the impact-parameter-dependent total pair production
probability computed in the lowest order perturbation theory is
larger than one. The analysis of Ref. \cite{2002PhLB..538...45B}
devoted to the study of the corresponding Feynman diagrams in the
high-energy limit leads to the probability of $N$ lepton pair
production obeying the Poisson distribution. For a review of this
topic see \cite{2007PhR...453....1B}.

Two photon particle pair production by collision of two electrons or
electron and positron (see Fig. \ref{twophoton}, where 4-momenta
$p_1$ and $p_2$ correspond to their momenta) were studied in storage
rings in Novosibirsk ($e^+e^-\rightarrow e^+e^-e^+e^-$
\cite{1971PhLB...34..663B}) and in Frascati at ADONE
($e^+e^-\rightarrow e^+e^-e^+e^-$ \cite{Bacci1972},
$e^+e^-\rightarrow e^+e^-\mu^+\mu^-$ \cite{1974PhRvL..32..385B},
$e^+e^-\rightarrow e^+e^-\pi^+\pi^-$ \cite{1974PhLB...48..380O}),
see also \cite{1980LNP...134....9B}. At high energy the total
cross-section of two photon production of lepton pairs is given by
\cite{Landau1934,1975PhR....15..181B}
\begin{equation}
\sigma_{e^\pm e^-\rightarrow e^\pm e^- l^+l^-}=\frac{28\alpha^4}{27\pi m_l^2}\left(\ln\frac{s}{m_e^2}\right)^2\ln\frac{s}{m_e^2};\quad l\equiv e \;\; {\mathrm or} \;\; \mu.
\end{equation}
see c.f. Eq. (\ref{sigmaL}).

\subsection{QED description of pair production}\label{qedpair}

We turn now to a Sauter-Heisenberg-Euler process in QED. An external
electromagnetic field is incorporated by adding to the quantum field
$A_\mu$ in (\ref{L0int}) an unquantized external vector potential
$A^{\rm e}_\mu$, so that the total interaction becomes
\begin{equation}
{\mathcal L}_{\rm int}+
{\mathcal L}^{\rm e}_{\rm int}
 = -e\bar\psi(x)\gamma^\mu
\psi(x)
\left[
A_\mu(x)+
A^{\rm e}_\mu(x)\right] .
\label{intc}
\end{equation}
Instead of an operator formulation, one can derive the quantum field
theory from a functional integral formulation, see e.g.
\cite{Kleinert2004}, in which the quantum mechanical partition
function is described by
\begin{equation}
Z[A^{\rm e}]=\int [{\mathcal D}\psi {\mathcal D}\bar\psi {\mathcal D}A_\mu]
\exp \left[ i\int d^4x ({\mathcal L} + {\mathcal L}^{\rm e}_{\rm int}
) \right],
\label{path}
\end{equation}
to be integrated over all fluctuating electromagnetic and
Grassmannian electron fields. The normalized quantity $Z[A^{\rm e}]$
gives the amplitude for the vacuum to vacuum transition in the
presence of the external classical electromagnetic field:
\begin{equation}
\langle {\rm out}, 0|0, {\rm in}\rangle = \frac{Z[A^{\rm e}]}{ Z[0]},
\label{vvamplitude}
\end{equation}
where $|0, {\rm in}\rangle$ is the initial vacuum state at the time
$t=t_-\rightarrow -\infty$, and $\langle {\rm out}, 0|$ is the final
vacuum state at the time $t=t_+\rightarrow +\infty$. By selecting
only the one-particle irreducible Feynman diagrams in the
perturbation expansion of $Z[A^{\rm e}]$ one obtains the effective
action as a functional of $A^{\rm e}$:
\begin{equation}
\Delta{\mathcal A}_{\rm eff}[A^{\rm e}]\equiv -i\ln \langle {\rm out}, 0|0, {\rm in}\rangle.
\label{eaction}
\end{equation}
In general, there exists no local effective Lagrangian density
$\Delta{\mathcal L}_{\rm eff}$ whose space-time integral is
$\Delta{\mathcal A}_{\rm eff}[A^{\rm e}]$. An infinite set of
derivatives would be needed, i.e., $\Delta{\mathcal L}_{\rm eff}$
would have the arguments $A^{\rm e}(x),\partial_\mu A^{\rm e}(x),
\partial_\mu\partial _\nu A^{\rm e}(x), \dots~$, containing
gradients of arbitrary high order. With presently available methods
it is possible to calculate a few terms in such a gradient
expansion, or a semi-classical approximation {\it \`a l\`a} JWKB for
an arbitrary but smooth space-time dependence (see Section~3.21ff in
Ref.~\cite{Kleinert2004}). Under the assumption that the external
field $A^{\rm e}(x)$ varies smoothly over a finite space-time
region, we may define an approximately local effective Lagrangian
$\Delta{\mathcal L}_{\rm eff}[A^{\rm e}(x)]$,
\begin{equation}
\Delta{\mathcal A}_{\rm eff}[A^{\rm e}]\simeq \int d^4x \Delta{\mathcal L}_{\rm eff}[A^{\rm e}(x)]
\approx V\Delta t \Delta{\mathcal L}_{\rm eff}[A^{\rm e}],
\label{effl}
\end{equation}
where $V$ is the spatial volume and time interval $\Delta t=t_+-t_-$.

For a large time interval $\Delta t=t_+-t_-\rightarrow \infty$, the
amplitude of the vacuum to vacuum transition (\ref{vvamplitude}) has
the form,
\begin{equation}
\langle {\rm out}, 0|0 ,{\rm in}\rangle = e^{-i(\Delta{\mathcal E}_0-i\Gamma/2)\Delta t},
\label{vvamplitude1}
\end{equation}
where $\Delta{\mathcal E}_0={\mathcal E}_0(A^{\rm e})-{\mathcal
E}_0(0)$ is the difference between the vacuum energies in the
presence and the absence of the external field, $\Gamma$ is the
vacuum decay rate, and $\Delta t$ the time over which the field is
nonzero. The probability that the vacuum remains as it is in the
presence of the external classical electromagnetic field is
\begin{equation}
|\langle {\rm out}, 0|0 ,{\rm in}\rangle|^2 =
    e^{-2{\rm Im}\Delta{\mathcal A}_{\rm eff}[A^{\rm e}]}.
\label{probability}
\end{equation}
This determines the decay rate of the vacuum in an external
electromagnetic field:
\begin{equation}
\frac{ \Gamma}{V}= \frac{2{\rm \,Im}\Delta{\mathcal A}_{\rm
eff}[A^{\rm e}]}{V\Delta t} \approx 2{\rm \,Im}\Delta{\mathcal
L}_{\rm eff}[A^{\rm e}]. \label{path21}
\end{equation}
The vacuum decay is caused by the production of electron and
positron pairs. The external field changes the energy density by
\begin{equation}
\frac{ \Delta{\mathcal E}_0}{V}= -\frac{{\rm \,Re}\Delta{\mathcal
A}_{\rm eff}[A^{\rm e}]}{V\Delta t} \approx -{\rm
\,Re}\Delta{\mathcal L}_{\rm eff}[A^{\rm e}]. \label{path21e}
\end{equation}

\subsubsection{Schwinger formula for pair production in
uniform electric field}\label{Schwingerformula}

The Dirac fields appears quadratically in the partition functional
(\ref{path}) and can be integrated out, leading to
\begin{equation}
Z[A^{\rm e}]=\int  {\mathcal D}A_\mu\,
{\rm Det}
\{i\!\!\not{\!\partial}-e[\not{\hspace{-5pt}A(x)}+
\not{\hspace{-5pt}A}^{\rm e}(x)]
-m_e+i\eta\};\quad
\not{\!\partial}\equiv\gamma^\mu\partial_\mu, \quad \not{\!A}\equiv\gamma^\mu A_\mu,
\end{equation}
where Det denotes the functional determinant of the Dirac operator.
Ignoring the fluctuations of the electromagnetic field, the result
is a functional of the external vector potential  $A^{\rm e}(x)$:
\begin{equation}
Z[A^{\rm e}]\approx {\rm const}\times
{\rm Det}
\{i\!\!\not{\!\partial}-e
\not{\hspace{-5pt}A}^{\rm e}(x)-m_e+i\eta\}.
\label{path1}
\end{equation}
The infinitesimal constant $i\eta$ with $ \eta >0$ specifies the
treatment of singularities in energy integrals. From
Eqs.~(\ref{vvamplitude})--(\ref{path1}), the effective action
(\ref{probability}) is given by
\begin{equation}
 \Delta{\mathcal A}_{\rm eff}[A^{\rm e}]=-i
{\rm Tr}\ln \left\{\left[ i\!\!\not{\!\partial}-
e\not{\hspace{-5pt}A}^{\rm e}(x)-m_e+i\eta\right]
\frac{1}{
 i\!\!\not{\!\partial}
-m_e+i\eta}\right\},
\label{eaction12}
\end{equation}
where Tr denotes the functional and Dirac trace. In physical units, this is of order $\hbar$.
The result may be expressed as a one-loop Feynman diagram, so that one speaks of one-loop approximation.
More convenient will be the equivalent
expression
\begin{equation}
\Delta{\mathcal A}_{\rm eff}[A^{\rm e}]=
-\frac{i}{2}\,{\rm Tr}\ln \left(\{[i\!\!\not{\!\partial}-
e\not{\hspace{-5pt}A}^{\rm e}(x)]^2-m_e^2+i\eta\}
\frac{1}{ -\partial ^2-m_e^2+i\eta}\right),
\label{eaction1}
\end{equation}
where
\begin{equation}
[i\!\!\not{\!\partial}- e\not{\hspace{-5pt}A}^{\rm e}(x)]^2=
[i\partial_\mu-eA^{\rm e}_\mu(x)]^2+\frac{e}{2}\sigma^{\mu\nu}F^{\rm
e}_{\mu\nu}, \label{sqrt}
\end{equation}
where $\sigma^{\mu\nu}\equiv \frac{i}{2}[\gamma^\mu,\gamma^\nu]$,\,
$F^{\rm e}_{\mu\nu}=\partial_\mu A^{\rm e}_\nu-\partial_\nu A^{\rm
e}_\mu$. Using the identity
\begin{equation}
\ln\frac{a_2}{ a_1}=\int_0^\infty \frac{ds}{ s}\big[ e^{is(a_1+i\eta)}-e^{is(a_2+i\eta)}\big],
\label{identityab}
\end{equation}
Eq.~(\ref{eaction1}) becomes
\begin{equation}
\Delta{\mathcal A}_{\rm eff}[A^{\rm e}]
=\frac{i}{2}\int_0^\infty \frac{ds}{ s}e^{-is(m_e^2-i\eta)}{\rm Tr}
\langle x| e^{ is\left\{[i\partial_\mu-eA^{\rm e}_\mu(x)]^2+\frac{e}{2}\sigma^{\mu\nu}F^{\rm e}_{\mu\nu}\right\}}
-e^{-is\partial ^2}|x\rangle ,
\label{probability01}
\end{equation}
where $\langle x|\{\cdot\cdot\cdot\}|x\rangle $ are the diagonal
matrix elements in the local basis $|x\rangle $. This is defined by
the matrix elements with the momentum eigenstates $|k\rangle$ being
plane waves: $\langle x|k\rangle= e^{-ikx}$. The symbol ${\rm Tr}$
denotes integral $\int d^4x$ in space-time and the trace in spinor
space. For constant electromagnetic fields, the integrand in
(\ref{probability01}) does not depend on $x$, and
$\sigma^{\mu\nu}F^{\rm e}_{\mu\nu}$ commutes with all other
operators. This will allows us to calculate the exponential in
Eq.~(\ref{probability01}) explicitly. The presence of $-i \eta $ in
the mass term ensures convergence of integral for
$s\rightarrow\infty$.

If only a constant electric field ${\bf E}$ is present, it may be
assumed to point along the ${\hat{\bf z}}$-axis, and one can choose
a gauge such that $A^{\rm e}_z=-Et$ is the only nonzero component of
$A^{\rm e}_\mu$. Then one finds
\begin{equation}
{\rm tr}\exp is\left[\frac{e}{2}\sigma^{\mu\nu}F^{\rm e}_{\mu\nu}\right]=4\cosh(seE),
\label{iz1}
\end{equation}
where the symbol ${\rm tr}$ denotes the trace in spinor space.
Using the commutation relation $[\partial _0,x^0]=1$, where $x^0=t$,
one computes the exponential term in the effective action
(\ref{probability01}) (c.e.g.~\cite{Itzykson2006})
\begin{equation}
\langle x| \exp is\left[(i\partial _\mu-eA^{\rm e}_\mu(x))^2+\frac{e}{2}
\sigma^{\mu\nu}F^{\rm e}_{\mu\nu}\right]
|x\rangle =\frac{eE}{ (2\pi)^2is}\coth(eEs).
\label{iz2}
\end{equation}
The second term in Eq.~(\ref{probability01}) is obtained by setting
$E= 0$ in Eq.~(\ref{iz2}), so that the effective action
(\ref{probability01}) yields,
\begin{equation}
\Delta{\mathcal A}_{\rm eff}=
\frac{1}{ 2(2\pi)^2}\int d^4x \int_0^\infty \frac{ds}{ s^3}
\left[eEs\coth(eEs)-1\right]
e^{-is(m^2_e-i\eta)}.
\label{oneloops1}
\end{equation}
Since the field is constant, the integral over $x$ gives a volume
factor, and the effective action (\ref{probability01}) can be
attributed to the space-time integral over an effective Lagrangian
(\ref{effl})
\begin{equation}
\Delta{\mathcal L}_{\rm eff}
=
\frac{1}{ 2(2\pi)^2}\int_0^\infty \frac{ds}{ s^3}
\left[eEs\coth(eEs)-1\right]
e^{-is(m^2_e-i\eta)}.
\label{oneloopl1us}
\end{equation}

By expanding the integrand in Eq.~(\ref{oneloopl1us}) in powers of
$e$, one obtains,
\begin{equation}
\frac{1}{ s^3}\left[eEs\coth(eEs)-1\right]
e^{-i s(m^2_e-i\eta)}= \left[\frac{e^2}{ 3s}E^2-\frac{e^4s}{ 45}E^4 +{\mathcal O}(e^6)\right]
e^{-i s(m^2_e-i\eta)}.
\label{oneloopl1s20}
\end{equation}
The small-$s$ divergence in the integrand,
\begin{equation}
\frac{e^2}{ 3}E^2
\frac{1}{ 2(2\pi)^2}\int_0^\infty \frac{ds}{ s}e^{-is(m^2_e-i\eta)},
\label{oneloopl1s2}
\end{equation}
is proportional to the electric term in the original Maxwell
Lagrangian. The divergent term (\ref{oneloopl1s2}) can therefore be
removed by a renormalization of the field $E$. Thus we subtract a
counterterm in Eq.~(\ref{oneloopl1us}) and form
\begin{equation}
\Delta{\mathcal L}_{\rm eff}
=
\frac{1}{ 2(2\pi)^2}\int_0^\infty \frac{ds}{ s^3}
\left[eEs\coth(eEs)-1-\frac{e^2}{ 3}E^2s^2\right]
e^{-is(m^2_e-i\eta)}.
\label{oneloopl1}
\end{equation}
Remembering Eq.~(\ref{path21}), we find
from (\ref{oneloopl1})  the decay rate of the vacuum per unit volume
\begin{equation}
\frac{ \Gamma }{V}
=\frac{1}{ (2\pi)^2}{\rm Im}\int_0^\infty \frac{ds}{ s^3}\left[eEs\coth(eEs)-1-\frac{e^2}{ 3}E^2s^2\right]
e^{-is(m_e^2-i\eta)}.
\label{gprobability}
\end{equation}

The integral (\ref{gprobability}) can be evaluated analytically by the
method of residues.
Since the integrand is even,
the integral can be extended
to the entire $s$-axis. After this, the
integration contour is deformed to enclose the
negative imaginary axis
and to pick up
the contributions of the poles of the $\coth$ function
at $s=n\pi/eE$.
The result is
\begin{equation} \!\!\!\!\!\!\!\!
\frac{ \Gamma }{V}
=\frac{\alpha E^2}{ \pi^2}\sum_{n=1}^\infty \frac{1}{ n^2}\exp
\left(-\frac{n\pi E_c}{ E}\right).
\label{probability1}
\end{equation}
This result, i.e. the {\it Schwinger formula}
\cite{1951PhRv...82..664S,1954PhRv...93..615S,1954PhRv...94.1362S}
is valid to lowest order in $\hbar$ for arbitrary constant electric
field strength.

An analogous calculation for a charged scalar field yields
\begin{equation}
\frac{ \Gamma }{V}
=\frac{\alpha E^2}{ 2\pi^2}\sum_{n=1}^\infty \frac{(-1)^{n+1}}{ n^2}\exp
\left(-\frac{n\pi E_c}{ E}\right),
\label{bosonrate}
\end{equation}
which generalizes the Weisskopf treatment being restricted to the
leading term $n=1$. These Schwinger results complete the derivation
of the probability for pair productions. The leading $n=1$ -terms of
(\ref{probability1}) and (\ref{bosonrate}) agrees with the JWKB
results we discuss in Section \ref{semi}, and thus the correct
Sauter exponential factor (\ref{transmission}) and Heisenberg-Euler
imaginary part of the effective Lagrangian (\ref{herate}).

Narozhny and Nikishov \cite{Narozhnyi:1970uv} have expressed
Eq.~(\ref{probability}) through the probability of one pair
production $P_1$, of $n$ pair production $P_n$ with
$n=1,2,3,\cdot\cdot\cdot$ as well as the average number of pair
productions
\begin{equation}
|\langle {\rm out}, 0|0 ,{\rm in}\rangle|^2 = 1- P_1-P_2-P_3-\cdot\cdot\cdot ,
\label{nprobability}
\end{equation}
where $P_n, (n=1,2,3,\cdot\cdot\cdot)$ is the probability of $n$
pair production, and the probability of one pair production is,
\begin{equation}
P_1 = V\Delta t\frac{\alpha E^2}{2\pi^2}\ln \left(1-e^{-\frac{\pi m_e^2}{eE}}\right)
e^{-2V\Delta t {\rm Im}\Delta{\mathcal L}_{\rm eff}[A^{\rm e}]} .
\label{nprobability1}
\end{equation}
The average number $\bar N$ of pair productions is then given by
\begin{equation}
\bar N=\sum_{n=1}^\infty nP_n =V\Delta t \frac{\alpha E^2}{ \pi^2}\exp
\left(-\frac{\pi E_c}{ E}\right),
\label{nprobabilitym1}
\end{equation}
which is the quantity directly related to the experimental measurements.

\subsubsection{Pair production in constant electromagnetic fields}\label{EandB}

Since the QED theory is gauge and Lorentz invariant, effective
action $\Delta{\mathcal A}_{\rm eff}$ and Lagrangian
$\Delta{\mathcal L}_{\rm eff}$ are expressed as functionals of the
scalar and pseudoscalar invariants $S,P$ (\ref{lightlike}). Thus
they must be invariant under the discrete duality transformation:
\begin{equation}
|{\bf B}|\rightarrow -i|{\bf E}|,\quad |{\bf E}|\rightarrow i|{\bf B}|,
\label{duality1}
\end{equation}
i.e.,
\begin{equation}
\beta\rightarrow -i\varepsilon ,\quad \varepsilon\rightarrow i\beta.
\label{duality2}
\end{equation}
This implies that in the case ${\bf E}=0$ and ${\bf B}\not= 0$, results
can be simply obtained by replacing $|{\bf E}|\rightarrow i|{\bf B}|$
in Eqs.~(\ref{iz2}), (\ref{oneloopl1}), (\ref{gprobability}):
\begin{equation}
\langle x| \exp is\left[(i\partial _\mu-eA^{\rm e}_\mu(x))^2+\frac{e}{2}\sigma^{\mu\nu}F^{\rm e}_{\mu\nu}\right]
|x\rangle =\frac{eB}{ (2\pi)^2is}\cot(eBs),
\label{iz2b}
\end{equation}
and
\begin{equation}
\Delta{\mathcal L}_{\rm eff}
=
\frac{1}{ 2(2\pi)^2}\int_0^\infty \frac{ds}{ s^3}
\left[eBs\cot(eBs)-1+\frac{e^2}{ 3}B^2s^2\right]
e^{-is(m^2_e-i\eta)}.
\label{oneloopl1b}
\end{equation}

In the presence of both constant electric and magnetic fields ${\bf E}$ and ${\bf B}$,
we adopt parallel ${\bf E}_{\rm CF}$ and ${\bf B}_{\rm CF}$ pointing
along the ${\hat{\bf z}}$-axis in the {\it center-of-fields frame}, as discussed after Eqs.~(\ref{eblandau}), (\ref{elorentz}), (\ref{blorentz}). We can choose a gauge such
that only $A^{\rm e}_z=-E_{\rm CF}t$, $A^{\rm e}_y=B_{\rm CF}x^1$ are nonzero.
Due to constant fields, the exponential in the effective action Eq.~(\ref{probability01})
can be factorized into a product of the magnetic part and the electric part.
Following the same method used to compute the electric part (\ref{iz1},\ref{iz2}), one
can compute the magnetic part by using the commutation relation
$[\partial _1,x^1]=1$, where $x^1=x$. Or one can make the substitution
(\ref{duality1}) to obtain the magnetic part, based on the discrete symmetry
of duality. As results, Eqs.~(\ref{iz1}), (\ref{iz2}) become
\begin{equation}
{\rm tr}\exp is\left[\frac{e}{2}\sigma^{\mu\nu}F^{\rm e}_{\mu\nu}\right]
=4\cosh(seE_{\rm CF})\cos(seB_{\rm CF}),
\label{iz1eb}
\end{equation}
and
\begin{align}
&\langle x| \exp is\left\{[i\partial _\mu-eA^{\rm e}_\mu(x)]^2+\frac{e}{2}\sigma^{\mu\nu}F^{\rm e}_{\mu\nu}\right\}
|x\rangle \nonumber\\
&=\frac{1}{(2\pi)^2}\frac{eE_{\rm CF}}{ is}\coth(seE_{\rm CF})
\frac{eB_{\rm CF}}{ s}\cot(seB_{\rm CF}).
\label{iz2eb}
\end{align}
In this special frame, the effective Lagrangian is then given by
\begin{align}
\Delta{\mathcal L}_{\rm eff}&=
\frac{1}{ 2(2\pi)^2}\int_0^\infty \frac{ds}{ s^3}
\Big[e^2E_{\rm CF}B_{\rm CF} s^2\coth(seE_{\rm CF} )
\cot(seB_{\rm CF})\nonumber\\
&-1-\frac{e^2}{ 3}(E^2_{\rm CF}-B^2_{\rm CF})s^2
\Big]
\cdot e^{-is(m^2_e-i\eta)}.
\label{onelooplcf}
\end{align}
Using definitions in Eqs.~(\ref{lightlike}), (\ref{ab}),
(\ref{fieldinvariant}), we obtain the effective Lagrangian
\begin{align}
\Delta{\mathcal L}_{\rm eff}
&\!=\!
\frac{1}{ 2(2\pi)^2}\int_0^\infty \frac{ds}{ s^3}
\Big[e^2\varepsilon\beta s^2\coth(e\varepsilon s )\cot(e\beta s )\nonumber\\
&-1-\frac{e^2}{ 3}(\varepsilon^2-\beta^2)s^2\Big]
e^{-is(m^2_e-i\eta)};
\label{oneloopl}
\end{align}
and the decay rate
\begin{align}
\frac{ \Gamma }{V}&=\frac{1}{ (2\pi)^2}{\rm Im}\int_0^\infty \frac{ds}{ s^3}
\Big[e^2\varepsilon\beta s^2\coth(e\varepsilon s )\cot(e\beta s )\nonumber\\
&-1-\frac{e^2}{ 3}(\varepsilon^2-\beta^2)s^2\Big]
e^{-is(m_e^2-i\eta)},
\label{gprobabilityeb}
\end{align}
in terms of the invariants $\varepsilon$ and $\beta$ (\ref {fieldinvariant}) for
arbitrary electromagnetic fields ${\bf E}$ and ${\bf B}$.

The integral (\ref{gprobabilityeb}) is evaluated as in Eq.~(\ref{probability1}) by the
method of residues, and yields \cite{1951PhRv...82..664S,1954PhRv...93..615S,1954PhRv...94.1362S}
\begin{equation}
\frac{ \Gamma }{V}=\frac{  \alpha   \varepsilon^2}{ \pi^2 }
\sum_{n=1}  \frac{1}{n^2}
\frac{ n\pi\beta / \varepsilon }
{\tanh {n\pi \beta/ \varepsilon}}\exp\left(-\frac{n\pi E_c}{ \varepsilon}\right),
\label{probabilityeh}
\end{equation}
which reduces for $\beta \rightarrow 0$ (${\bf B}=0$) correctly to (\ref{probability1}).
The $n=1$ -term is the JWKB approximation (\ref{wkbehfermion2}).

The analogous result for bosonic fields
is
\begin{equation}
\frac{ \Gamma }{V}=\frac{  \alpha   \varepsilon^2}{ 2\pi^2 }
\sum_{n=1}  \frac{(-1)^n}{n^2}
\frac{ n\pi\beta / \varepsilon }
{\sinh {n\pi \beta/ \varepsilon}}\exp\left(-\frac{n\pi E_c}{ \varepsilon}\right),
\label{probabilityehb}
\end{equation}
where the first term $n=1$ corresponds to the Euler-Heisenberg result (\ref{herate}).
Note that the magnetic field produces in the fermionic case an extra factor
$\lfrac{( n\pi\beta / \varepsilon )}
{\tanh ({n\pi \beta/ \varepsilon}})>1$ in each term
which enhances the decay rate. The bosonic series (\ref{probabilityehb}), on the other hand,
carries in each term  a suppression factor
$\lfrac{ (n\pi\beta / \varepsilon )}
{\sinh {n\pi \beta/ \varepsilon}}<1$.
The average number $\bar N$ (\ref{nprobabilitym1}) is given by
\begin{equation}
\bar N=\sum_{n=1}^\infty nP_n =V\Delta t  \frac{\alpha}{\pi}
\frac{ \alpha\beta \varepsilon }
{\tanh {\pi \beta/ \varepsilon}}\exp\left(-\frac{\pi E_c}{ \varepsilon}\right).
\label{nprobabilitym}
\end{equation}

The decay rate $ \Gamma /V$ gives the number of electron--positron pairs
produced per unit volume and time. The prefactor can be estimated on
dimensional grounds. It has the dimension of $E_c^2/\hbar $, i.e., $m^4
c^5/\hbar^4$. This arises from the energy of a pair $2m_ec^2$ divided by the
volume whose diameter is the Compton wavelength $\hbar/m_ec$, produced within a
Compton time
 $\hbar/m_ec^2$.
The exponential factor suppresses
pair production as long as the
electric field is much smaller
than the critical electric field $E_c$, in  which case
the JWKB results (\ref{wkbehfermion2}) and (\ref{wkbehboson2})
are good approximations.

The general results ({\ref{probabilityeh}),(\ref{probabilityehb})
was first obtained by Schwinger
\cite{1951PhRv...82..664S,1954PhRv...93..615S,1954PhRv...94.1362S}
for scalar and spinor electrodynamics (see also Nikishov
\cite{Nikishov1969}, Batalin and Fradkin
\cite{1965JETP...21..375V}). The method was extended to special
space-time-dependent fields in
Refs.~\cite{1971ZhPmR..13..261P,1972JETP...34..709P,2001JETPL..74..133P,Narozhnyi:1970uv,1970TMP.....5.1080B}.
The monographs
\cite{Itzykson2006,Kleinert2008,Greiner1985,1980ato..book.....G,Fradkin1991}
can be consulted about more detailed calculation, discussion and
bibliography.

\subsubsection{Effective nonlinear Lagrangian for arbitrary constant electromagnetic field}\label{nonlinear}

Starting from the integral form of Heisenberg and Euler Lagrangian (\ref{oneloopl}) we find explicitly real and imaginary parts of the effective Lagrangian $\Delta{\mathcal L}_{\rm eff}$
(\ref{oneloopl}) for arbitrary constant electromagnetic fields ${\bf E}$ and ${\bf B}$ \cite{2006JKPSP}.
The essential step is to reach a direct analytic form resulting from performing the integration. We use the expressions \cite{1994tisp.book.....G},
\begin{align}
e\varepsilon s\coth(e\varepsilon s )
&=\sum_{n=-\infty}^{\infty}\frac{s^2 }{(s^2+\tau^2_n)};\quad
\tau_n\equiv n\pi/e\varepsilon,\label{cosnm0}\\
e\beta s\cot(e\beta s )
&=\sum_{m=-\infty}^{\infty}\frac{s^2 }{(s^2-\tau^2_m)},\quad
\tau_m\equiv m\pi/e\beta,
\label{cosnm}
\end{align}
and obtain for the finite effective Lagrangian of Heisenberg and Euler integral representation,
\begin{align}
\Delta{\mathcal L}_{\rm eff}
&\!=\!
\frac{1}{ 2(2\pi)^2}\sum_{n,m=-\infty}^{\infty}{\!\!\!}'\int_0^\infty ds\frac{s}{ \tau^2_n+\tau^2_m}
\Big[\frac{\bar\delta_{m0} }{(s^2-\tau^2_m)}-\frac{\bar\delta_{n0} }{(s^2+\tau^2_n)}\Big]
e^{-is(m^2_e-i\eta)},
\label{onelooplfini}
\end{align}
where divergent terms $n\not=0,m=0$, $n=0,m\not=0$ and $n=m=0$ are excluded from the sum, as indicated by a prime.
The symbol
$\bar\delta_{ij}\equiv 1-\delta_{ij}$ denotes the complimentary Kronecker-$\delta$ which vanishes for $i=j$.
The divergent term with $n=m=0$ is eliminated by the zero-field subtraction in Eq.~(\ref{oneloopl}), while
the divergent terms $n\not=0,m=0$ and $n=0,m\not=0$
\begin{align}
\Delta{\mathcal L}^{\rm div}_{\rm eff}
=
\frac{1}{ 2(2\pi)^2}\int_0^\infty \frac{ds}{ s}e^{-is(m^2_e-i\eta)}2\left( \sum_{m=1}^{\infty}\frac{1}{\tau^2_m}-\sum_{n=1}^{\infty}
\frac{1}{\tau^2_n}\right),
\label{div}
\end{align}
are eliminated by the second subtraction in Eq.~(\ref{oneloopl}). This can be seen by performing the sums
\begin{equation}
\sum_{n=1}^\infty\frac{1}{\tau^k_n}=\left(\frac{e\varepsilon}{\pi}\right)^k\zeta(k);\quad
\sum_{n=1}^\infty\frac{1}{\tau^k_m}=\left(\frac{e\beta}{\pi}\right)^k\zeta(k),
\label{sumrule}
\end{equation}
where $\zeta(k)=\sum_n 1/n^k$ is the Riemann function.

The infinitesimal $i\eta$ accompanying the mass term in the $s$-integral (\ref{onelooplfini})
is equivalent to replacing
$e^{-is(m^2_e-i\eta)}$ by $e^{-is(1-i\eta)m^2_e}$. This implies that $s$
is to be integrated slightly below (above) the real axis for $s>0$ ($s<0$). Equivalently one may shift
the $\tau_m$ ($-\tau_m$) variables slightly upwards
(downwards) to $\tau_m+i\eta$ ($-\tau_m-i\eta$) in the complex plane.

In order to calculate the finite effective Lagrangian (\ref{onelooplfini}),
the factor $e^{-is(1-i\eta)m^2_e}$ is divided into its sin and cos parts:
\begin{align}
\Delta{\mathcal L}^{\sin}_{\rm eff}
&=
\frac{-i}{ 4(2\pi)^2}\sum_{n,m=-\infty}^{\infty}{\!\!\!}'\int_{-\infty}^\infty \frac{sds}{ \tau^2_n+\tau^2_m}
\Big[\frac{\bar\delta_{m0} }{(s^2-\tau^2_m)}-\frac{\bar\delta_{n0} }{(s^2+\tau^2_n)}\Big]\sin[s(1-i\eta)m^2_e];
\label{onelooplsin}\\
\Delta{\mathcal L}^{\cos}_{\rm eff}
&=
\frac{1}{ 2(2\pi)^2}\sum_{n,m=-\infty}^{\infty}{\!\!\!}'\int_0^\infty \frac{sds}{ \tau^2_n+\tau^2_m}
\Big[\frac{\bar\delta_{m0} }{(s^2-\tau^2_m)}-\frac{\bar\delta_{n0} }{(s^2+\tau^2_n)}\Big]
\cos[s(1-i\eta)m^2_e].
\label{onelooplcos0}
\end{align}
The sin part (\ref{onelooplsin}) has an even integrand allowing for
an extension of the $s$-integral over the entire $s$-axis. The
contours of integration can then be closed by infinite semicircles
in the half plane, the integration receives contributions from poles
$\pm \tau_m, \pm i\tau_n$, so that the residue theorem leads to,
\begin{align}
\Delta{\mathcal L}^{\sin}_{\rm eff} &= i\frac{\alpha\varepsilon\beta }{ 2\pi}
\sum^\infty_{n=1}\frac{1}{n}\coth\left(\frac{n\pi \beta}{ \varepsilon}\right) \exp(-n\pi E_c/\varepsilon)\label{sinE}\\
&-i\frac{\alpha\varepsilon\beta }{ 2\pi}\sum^\infty_{m=1}\frac{1}{m}\coth\left(\frac{m\pi \varepsilon}{ \beta}\right)
\exp (im\pi E_c/\beta)\label{sinB}
\end{align}
The first part (\ref{sinE}) leads to the exact non-perturbative Schwinger rate (\ref{probabilityeh})
for pair production.

The second term, as we see below, is canceled by the imaginary part of the cos term. In fact, shifting $s\rightarrow s-i\eta$, we rewrite the cos part of effective Lagrangian (\ref{onelooplcos0}) as
\begin{equation}
\Delta{\mathcal L}^{\cos}_{\rm eff}
=\frac{1}{2(2\pi)^2}
\sum_{n,m=-\infty}^{\infty}{\!\!\!}' \int_0^\infty ds  \frac{\cos(sm^2_e)}{\tau^2_n+\tau^2_m}
\left(\frac{s\bar\delta_{m0} }{ s^2-\tau^2_m-i\eta}-\frac{s\bar\delta_{n0} }{ s^2+\tau^2_n-i\eta}\right).
\label{onelooplcos}
\end{equation}
In the first term of magnetic part, singularities $s=\tau_m, (m>0)$ and $s=-\tau_m, (m<0)$ appear
in integrating $s$-axis. We decompose
\begin{align}
\frac{s}{ s^2-\tau^2_m-i\eta}=i\frac{\pi}{2}\delta(s-\tau_m)+i\frac{\pi}{2}\delta(s+\tau_m)
+{\mathcal P}\frac{s}{ s^2-\tau^2_m},\label{h+}
\end{align}
where ${\mathcal P}$ indicates the principle value under the integral. The integrals over the $\delta$-functions
give
\begin{equation}
\Delta_\delta{\mathcal L}^{\cos}_{\rm eff}=i\frac{\alpha\varepsilon\beta }{ 2\pi}\sum^\infty_{m=1}\frac{1}{m}\coth\left(\frac{m\pi \varepsilon}{ \beta}\right) \exp (im\pi E_c/\beta),
\label{hagensin}
\end{equation}
which exactly cancels the second part (\ref{sinB}) of the sin part $\Delta{\mathcal L}^{\sin}_{\rm eff}$.

It remains to find the principle-value integrals in Eq.~(\ref{onelooplcos}), which corresponds to the real part of the effective Lagrangian
\begin{equation}
(\Delta{\mathcal L}^{\cos}_{\rm eff})_{\mathcal P}
=\frac{1}{2(2\pi)^2}
\sum_{n,m=-\infty}^{\infty}{\!\!\!}' \frac{1}{\tau^2_n+\tau^2_m}{\mathcal P}\int_0^\infty ds
\cos(sm^2_e)
\left(\frac{s\bar\delta_{m0} }{ s^2-\tau^2_m}-\frac{s\bar\delta_{n0} }{ s^2+\tau^2_n}\right).
\label{onelooplcosp}
\end{equation}
We rewrite the cos function as $\cos(sm^2_e)=(e^{ism^2_e}+e^{-ism^2_e})/2$ and make the rotations of
integration contours by $\pm\pi/2$ respectively,
\begin{eqnarray}
(\Delta{\mathcal L}^{\cos}_{\rm eff})_{\mathcal P}
&=&\frac{1}{2(2\pi)^2}
\sum_{n,m=-\infty}^{\infty}{\!\!\!}' \frac{1}{\tau^2_n+\tau^2_m}\times\nonumber\\
&\times&{\mathcal P}\int_0^\infty d\tau
\left(\frac{\bar\delta_{m0}\tau e^{-\tau} }{ \tau^2-(i\tau_mm_e^2)^2}
-\frac{\bar\delta_{n0}\tau e^{-\tau} }{ \tau^2-(\tau_nm_e^2)^2}\right).
\label{onelooplcosp1}
\end{eqnarray}
Using the formulas (see Secs.~3.354, 8.211.1 and 8.211.2 in Ref.~\cite{1994tisp.book.....G})
\begin{equation}
J(z) \equiv {\mathcal P}\int^\infty_0 ds \frac{s e^{-s}}{ s^2-z^2}
= -\frac{1}{2}\Big[e^{-z}{\rm Ei}(z)
+ e^{z}{\rm Ei}(-z)\Big],
\label{J(z)1}
\end{equation}
where ${\rm Ei}(z)$ is the exponential-integral function,
\begin{equation}
{\rm Ei}(z) \equiv {\mathcal P}\int_{-\infty}^z dt \frac{e^t}{t}=\log(-z)+\sum_{k=1}^\infty\frac{z^k}{kk!},
\label{J(z)2}
\end{equation}
we obtain the principal-value integrals (\ref{onelooplcosp1}),
\begin{equation}
(\Delta{\mathcal L}^{\cos}_{\rm eff})_{\mathcal P}  =
\frac{1}{2(2\pi)^2}\sum_{n,m=-\infty}^{\infty}{\!\!\!}' \frac{1}{ \tau^2_m+\tau^2_n}
\Big[\bar\delta_{m0}J(i\tau_m m^2_e)-\bar\delta_{n0}J(\tau_n m^2_e)\Big].
\label{pertur}
\end{equation}

Having so obtained the real part of an effective Lagrangian for an arbitrary constant electromagnetic field we recover the usual approximate results by suitable expansion of the exact formula.
With the help of the series and asymptotic representation (see formula 8.215 in Ref.~\cite{1994tisp.book.....G})
of the exponential-integral function ${\rm Ei}(z)$ for
large $z$, corresponding to weak electromagnetic fields ($\varepsilon\ll 1, \beta\ll 1$),
\begin{equation}
J(z) =-\frac{1}{z^2}-\frac{6}{z^4}-\frac{120}{z^6}-\frac{5040}{z^8}-\frac{362880}{z^{10}}+\cdot\cdot\cdot ,
\label{J(z)3}
\end{equation}
and Eq.~(\ref{pertur}), we find,
\begin{align}
(\Delta{\mathcal L}^{\cos}_{\rm eff})_{\mathcal P}  &=
\frac{1}{2(2\pi)^2}\sum_{n,m=-\infty}^{\infty}{\!\!\!}' \frac{1}{ \tau^2_m+\tau^2_n}
\Big\{\bar\delta_{n0}\left[\frac{1}{\tau_n^2m_e^4}+\frac{6}{\tau_n^4m_e^8}+\frac{120}{\tau_n^6m_e^{12}}+\cdot\cdot\cdot\right]\nonumber\\
&+\bar\delta_{m0}\left[\frac{1}{\tau_m^2m_e^4}-\frac{6}{\tau_m^4m_e^8}+\frac{120}{\tau_m^6m_e^{12}}+\cdot\cdot\cdot\right]\Big\}.
\label{perexpansion}
\end{align}
Applying the summation formulas (\ref{sumrule}), the weak field
expansion (\ref{perexpansion}) is seen to agree with the Heisenberg
and Euler effective Lagrangian \cite{1936ZPhy...98..714H},
\begin{align}
(\Delta{\mathcal L}_{\rm eff})_{\mathcal P} &=
\frac{2 \alpha}{90 \pi E_c^2}\left\{
({\bf E}^2\!-\!{\bf B}^2)^2+7 ({\bf E}\cdot {\bf B})^2 \right\}\nonumber\\
&+\frac{2 \alpha}{315 \pi^2 E_c^4}\left\{
2({\bf E}^2\!-\!{\bf B}^2)^3+13({\bf E}^2\!-\!{\bf B}^2)
 ({\bf E}\cdot {\bf B})^2 \right\}+\cdot\cdot\cdot ,
\label{Kleinert1}
\end{align}
which is expressed in terms of a powers series of weak
electromagnetic fields up to $O(\alpha^3)$. The expansion coefficients
of the terms of order $n$
have the general form $m_e^4/(E_c)^n$. As long as the fields
are much smaller than $E_c$, the expansion
converges.

On the other hand, we can address the limiting form of the effective Lagrangian (\ref{pertur}) corresponding to electromagnetic fields ($\varepsilon\gg 1, \beta\gg 1$). We use the series and asymptotic representation (formulas 8.214.1 and 8.214.2 in Ref.~\cite{1994tisp.book.....G})
of the exponential-integral function ${\rm Ei}(z)$ for
small $z\ll 1$,
\begin{equation}
J(z) =-\frac{1}{2}\Big[e^z\ln(z)+e^{-z}\ln(-z)\Big]+\gamma_E +{\mathcal O}(z),
\label{smallz}
\end{equation}
with $\gamma_E=0.577216$ being the Euler-Mascheroni constant, we
obtain the leading terms in the strong field expansion of
Eq.~(\ref{pertur}),
\begin{equation}
(\Delta{\mathcal L}^{\rm cos}_{\rm eff})_{\mathcal P}
=\frac{1}{2(2\pi)^2}\sum_{n,m=-\infty}^{\infty '} \frac{1}{ \tau^2_m+\tau^2_n}\Big[\bar\delta_{n0}\ln(\tau_n m^2_e)-\bar\delta_{m0}\ln(\tau_m m^2_e)\Big]+\cdot\cdot\cdot .
\label{strongexp}
\end{equation}
In the case of vanishing magnetic field ${\bf B}=0$ and $m=0$ in Eq.~(\ref{strongexp}), we have,
\begin{equation}
(\Delta{\mathcal L}^{\rm cos}_{\rm eff})_{\mathcal P}  =\frac{1}{2(2\pi)^2}\sum_{n=1}^{\infty} \frac{1}{ \tau^2_n}
\ln(\tau_n m^2_e)+\cdot\cdot\cdot =\frac{\alpha E^2}{2\pi^2}\sum_{n=1}^{\infty} \frac{1}{ n^2}
\ln\left(\frac{n\pi E_c}{E}\right)+\cdot\cdot\cdot ,
\label{strongexpe}
\end{equation}
for a strong electric field $\bf E$. In the case of vanishing electric field ${\bf E}=0$ and $n=0$ in Eq.~(\ref{strongexp}),
we obtain for the strong magnetic field $\bf B$,
\begin{equation}
(\Delta{\mathcal L}^{\rm cos}_{\rm eff})_{\mathcal P}  =-\frac{1}{2(2\pi)^2}\sum_{m=1}^{\infty} \frac{1}{ \tau^2_m}
\ln(\tau_m m^2_e)+\cdot\cdot\cdot =-\frac{\alpha B^2}{2\pi^2}\sum_{m=1}^{\infty} \frac{1}{ m^2}
\ln\left(\frac{m\pi E_c}{B}\right)+\cdot\cdot\cdot .
\label{strongexpb}
\end{equation}
The ($m=1$) term is the one obtained by Weisskopf \cite{Weisskopf1936}.

We have presented in Eqs.~(\ref{sinE}), (\ref{sinB}), (\ref{hagensin}), (\ref{pertur})
closed form results for the one-loop effective Lagrangian $\Delta{\mathcal L}_{\rm eff}$ (\ref{oneloopl})
for arbitrary strength of constant electromagnetic fields.
The results will receive fluctuation corrections
from higher loop diagrams. These carry one or more factors $ \alpha $, $ \alpha ^2$, \dots~ and are thus
suppressed by factors $1/137$.
Thus results are valid for all field strengths
with an error no larger than roughly  1\%.
If we include, for example, the two-loop correction, the first term in the Heisenberg and Euler effective Lagrangian (\ref{Kleinert1}) becomes \cite{Kleinert2008}
\begin{equation}
(\Delta{\mathcal L}_{\rm eff})_{\mathcal P} =
\frac{2 \alpha}{90 \pi E_c^2}\left\{
\left(1+\frac{40 \alpha }{9\pi}\right)({\bf E}^2\!-\!{\bf B}^2)^2+7
\left(1+\frac{1315 \alpha }{252\pi} \right)({\bf E}\cdot {\bf B})^2 \right\}.
\label{Kleinert2}
\end{equation}
Readers can consult the recent review article \cite{Dunne:2004nc}, where
one finds discussions and computations of the effective Lagrangian at the two-loop level, and \cite{2000NuPhB.564..591D} for discussion of pair production rate.

\subsection{Theory of pair production in an alternating field}\label{alternating}

When the external electromagnetic field $F^{\rm e}_{\mu\nu}$ is
space-time-dependent, i.e., $F^{\rm e}_{\mu\nu}=F^{\rm
e}_{\mu\nu}({\bf x},t)$ the exponential in Eq.~(\ref{probability01})
can no longer be calculated exactly. In  this case, JWKB methods
have to be used to calculate pair production rates
\cite{1970PhRvD...2.1191B,1971ZhPmR..13..261P,1972JETP...34..709P,2001JETPL..74..133P,1972JETP...35..659P,1973JETPL..17..368M,Marinov1977,1971ZhPmR..13..261P,1972JETP...34..709P,2001JETPL..74..133P}.
The aim of this section is to show how one can use a semi-classical
JWKB approach to estimate the rate of pair production in an
oscillating electric field as first indicated by Brezin and Itzykson
in Ref.~\cite{1970PhRvD...2.1191B}. They evaluated the production
rate of charged boson pairs. The results they obtained can be
straightforwardly generalized to charged fermion case, since the
spins of charged particles contribute essentially with a counting
factor to the final results (see Secs.~\ref{semi} and
\ref{Schwingerformula}). Thus, let the electromagnetic potential be
$\hat {\bf z}$ directed, uniform in space and periodic in time with
frequency $\omega_0$:
\begin{equation}
A^{\rm e}_\mu(x)=(0,0,0,A(t)),\quad
A(t)=\frac{E}{\omega_0}\cos\omega_0 t.
\label{alterpotential}
\end{equation}
Then the electric field is $\hat {\bf z}$ directed, uniform in space and periodic in time as well. The electric field strength is given by $E(t)=-\dot A(t)=E\sin\omega_0 t$. It is
assumed that the electric field is adiabatically switched on and damped off in a time
$T^{\rm e}$, which is much larger than the period of oscillation $T_0=2\pi/\omega_0$. Suppose also
that $T_0$ is much larger than the Compton time $2\pi/\omega$ of the created particle , i.e.,
\begin{equation}
T^{\rm e}\gg T_0\gg \frac{2\pi}{\omega}\simeq \frac{2\pi}{m_e},
\label{thyrachy}
\end{equation}
where $\omega=\sqrt{|{\bf p}|^2+m_e^2}$, ${\bf p}$ being the 3-momentum of the created particle. Furthermore, $eE$ is
assumed to be much smaller than $m_e^2$, i.e., $E\ll E_c$ (see Eq.~(\ref{wkbc1})).

We have to study the time evolution of a scattered wave function
$\psi(t)$ representing the production of particle and antiparticle
pairs in the electromagnetic potential (\ref{alterpotential}). As
usual, an antiparticle can be thought of as a wave-packet moving
backward in time. Therefore, for large positive time (forward) only
positive energy modes ($\sim e^{-i\omega t}$) contribute to
$\psi(t)$. Similarly, for large negative times both positive energy
and negative energy modes ($\sim e^{i\omega t}$) contribute to
$\psi(t)$ which satisfies the differential equation
\cite{1970PhRvD...2.1191B}:
\begin{equation}
\left[\frac{d^2}{dt^2}+\omega^2(t)\right]\psi(t)=0,
\label{bscatt}
\end{equation}
where the ``variable frequency'' is defined as
\begin{equation}
\omega(t)\equiv \left\{m_e^2+{\bf p_\perp}^2+[p_z-eA(t)]^2\right\}^{1/2}.
\label{bscattw}
\end{equation}
The JWKB method suggests a general solution of the from
\begin{equation}
\psi(t)=\alpha(t)e^{-i\chi(t)}+\beta(t)e^{i\chi(t)},
\quad \chi(t)\equiv \int_0^tdt'\omega(t'),
\label{bscattwave}
\end{equation}
where the boundary conditions at large positive and negative times are:
\begin{equation}
\alpha(-\infty)=1,\quad \beta(+\infty)=0;
\quad \dot\chi(\pm\infty)=\omega.
\label{bscattcond}
\end{equation}
The backward scattering amplitude
$\beta(t)$ for large negative time ($t\rightarrow -\infty$) represents
the probability of antiparticle production.

The normalization condition $|\psi(t)|^2=1$ implies
\begin{equation}
\dot\alpha(t)e^{-i\chi(t)}+\dot\beta(t)e^{i\chi(t)}=0.
\end{equation}
Eq.~(\ref{bscatt}) can be written in terms of the scattering amplitudes as
\begin{equation}
\dot\alpha(t)e^{-i\chi(t)}-\dot\beta(t)e^{i\chi(t)}
=-\frac{\dot\omega(t)}{\omega(t)}\left[\alpha(t)e^{-i\chi(t)}-\beta(t)e^{i\chi(t)}\right],
\label{bscattwaven}
\end{equation}
or, which is the same,
\begin{eqnarray}
\dot\alpha(t)&=&-\frac{\dot\omega(t)}{2\omega(t)}\left[\alpha(t)-\beta(t)e^{i2\chi(t)}\right],
\label{bscattwaven1}\\
\dot\beta(t)&=&-\frac{\dot\omega(t)}{2\omega(t)}\left[\beta(t)-\alpha(t)e^{-i2\chi(t)}\right].
\label{bscattwaven2}
\end{eqnarray}
It follows from assumption (\ref{thyrachy}) that $\dot\omega(t)$ vanishes as
$|t|\rightarrow\infty$, i.e.,
\begin{equation}
\frac{\dot\omega(t)}{\omega^2(t)}=\frac{eE[p_z-eA(t)]}{\left\{m_e^2+{\bf p_\perp}^2+[p_z-eA(t)]^2\right\}^{3/2}} \ll 1.
\end{equation}
More precisely
\begin{equation}
\left|\frac{\dot\omega(t)}{\omega^2(t)}\right|
<\frac{eE}{m_e^2+{\bf p_\perp}^2}<\frac{eE}{m_e^2}\ll 1.
\label{bscattwr}
\end{equation}
Therefore, $\alpha(t)$ and $\beta(t)$ slowly vary in time and tend to constants
as $|t|\rightarrow\infty$. The phase $e^{i2\chi(t)}$ oscillates very rapidly as
compared to the variation of $\alpha(t)$ and $\beta(t)$, for
$\dot\chi(t)=\omega(t)\gg |\dot\omega(t)/\omega(t)|$.
In the zeroth order the oscillating terms in
Eqs.~(\ref{bscattwaven1}), (\ref{bscattwaven2}) are negligible and one finds
\begin{equation}
\alpha^{(0)}(t)=[\omega/\omega(t)]^{1/2}\simeq 1;\quad \beta^{(0)}(t)=0,
\label{1iteration}
\end{equation}
which duly satisfy the boundary conditions, and
\begin{equation}
\beta^{(1)}(t)=\int_t^\infty dt'\frac{\dot\omega(t')}{2\omega(t')}e^{-i2\chi(t')},
\label{bscattwavesolu}
\end{equation}
where (\ref{thyrachy}) and (\ref{bscattwr}) have been used.
$|\beta^{(1)}(-\infty)|^2$ gives information about the probability of particle-antiparticle pair
production. Namely, the probability of pair production per
unit volume and time is given by
\begin{eqnarray}
\tilde {\mathcal P} &=& \lim_{T^{\rm e}\rightarrow \infty}\frac{1}{T^{\rm e}}
\int \frac{d^3k}{(2\pi)^3}|\beta^{(1)}(-T^{\rm e})|^2\nonumber\\
&=&\int \frac{d^3k}{(2\pi)^3}\lim_{T^{\rm e}\rightarrow \infty}\frac{1}{T^{\rm e}}
\left|\int_{-T^{\rm e}/2}^{T^{\rm e}/2}
dt'\frac{\dot\omega(t')}{2\omega(t')}e^{-i2\chi(t')}\right|^2.
\label{bscattpro}
\end{eqnarray}
Since $\omega(t)$ is a periodic function with the same frequency $\omega_0$ as $A(t)$
one can make a Fourier series expansion:
\begin{equation}
\frac{\dot\omega(t)}{2\omega(t)}=\sum_{n=-\infty}^{+\infty}c_n e^{in\omega_0}.
\label{fouriou}
\end{equation}
Defined a renormalized frequency $\Omega$ via $\chi(t)=t\Omega$ one finds
\begin{equation}
\Omega\equiv \int^{2\pi}_0\frac{dx}{2\pi}\left[m_e^2+{\bf k_\perp}^2+
\left(k_3-\frac{eE}{\omega_0}\cos x\right)^2\right]^{1/2}.
\label{norw}
\end{equation}
so that
\begin{equation}
\lim_{T^{\rm e}\rightarrow \infty}\frac{1}{T^{\rm e}}
\left|\int_{-T^{\rm e}/2}^{T^{\rm e}/2}
dt'\frac{\dot\omega(t')}{2\omega(t')}e^{-i2\chi(t')}\right|^2
=2\pi\sum_n\delta(n\omega_0-2\Omega)|c_n|^2.
\label{bscattdel}
\end{equation}
Consequently, the probability of pair production (\ref{bscattpro}) is,
\begin{equation}
\tilde {\mathcal P}=\int \frac{d^3k}{(2\pi)^2}
\sum_n\delta(n\omega_0-2\Omega)|c_n|^2=\int \frac{d^3k}{(2\pi)^2\omega_0}
|c_{n^\circ}|^2,
\label{bscattpro1}
\end{equation}
where ${n^\circ}=2\Omega/\omega_0$ and $c_{n^\circ}$ are determined via
Eq.~(\ref{fouriou}) as
\begin{equation}
c_{n^\circ}=\int_{-\pi}^{\pi} \frac{dx}{2\pi}
\frac{\dot\omega(x)}{2\omega(x)}\exp\left\{\frac{2i}{\omega_0}\int^x_0dx'
\left[m_e^2+{\bf p_\perp}^2+\left(p_z-\frac{eE}{\omega_0}\cos(x')\right)^2\right]^{1/2}  \right\}.
\label{bscattpro2}
\end{equation}
The expression for $c_{n^\circ}$ contains a very rapidly oscillating
phase factor with frequency of the order of $m_e/\omega_0$, and it
decreases very rapidly in terms of imaginary time $\tau=-it$. Its
evaluation requires the application of the steepest-descent method
in the complex time $x=\omega_0 t$ plane. This is done by selecting
a proper contour turning in a neighborhood of the saddle point and
following the steepest-descent line, so as to find the main
contributions to the integral in Eq.~(\ref{bscattpro2}). The saddle
point originates from branch points and poles in
Eq.~(\ref{bscattpro2}), which are the zeros of $\omega(x)$.
Mathematical details can be found in
Ref.~\cite{1970PhRvD...2.1191B}. One finds
\begin{equation}
\tilde {\mathcal P} \simeq\frac{\omega_0}{9}\int \frac{d^3k}{(2\pi)^2}
e^{-2A}\cos^2B,
\label{bscattpro3}
\end{equation}
where
\begin{equation}
-A+iB=\frac{2i}{\omega_0}\int^{x_0}_0dx'
\left[m_e^2+{\bf p_\perp}^2+\left(p_z-\frac{eE}{\omega_0}\cos(x')\right)^2\right]^{1/2},
\label{complexab}
\end{equation}
and the saddle point is $x_0=1/\pi+i\sinh^{-1}[(\omega_0/eE)(m_e^2+{\bf p_\perp}^2)^{1/2}]$.

The exponential factor $e^{-2A}$ in Eq.~(\ref{bscattpro3}) indicates
that particle-antiparticle pairs tend to be emitted with small
momenta. This allows one to estimate the right-hand side of
Eq.~(\ref{bscattpro3}) as follow: (i) $p_z$ is set equal to zero,
moreover, the range of the $p_z$-integration is of the order of
$2eE/\omega_0$ as suggested by the classical equation of motion
(\ref{bscatt}); (ii) $\cos^2B$ is replaced by its average value
$1/2$. As a result, one obtains \cite{1970PhRvD...2.1191B},
\begin{equation}
\tilde {\mathcal P}\simeq\frac{(eE)^3}{18\pi\omega_0^2}\int_{\eta^{-1}}^\infty du u
\exp\left[-\frac{\pi eE}{\omega_0^2}u^2g(u)\right],
\label{bscattpro4}
\end{equation}
where $\eta^{-1}=m_e\omega_0/(eE)$, $u=(m_e^2+{\bf p_\perp}^2)\omega_0^2/(eE)^2$ and
\begin{equation}
g(z) = \frac{4}{\pi}\int_0^1dy \Big[\frac{1-y^2}{ 1 + z^{-2} y^2}\Big]^{1/2}=F\left(\frac{1}{2},\frac{1}{2};2;-z^{-2}\right),
\label{xlasereta}
\end{equation}
where $F(1/2,1/2;2;-z^{-2})={}_2F_1(1/2,1/2;2;-z^{-2})$ is the Gauss
hypergeometrical function. The function $u^2g(u)$ is monotonically
increasing:
\begin{equation}
\frac{eE}{\omega_0^2}u^2g(u)\ge \frac{eE}{\omega_0^2}\eta^{-2}g(\eta)
= \frac{m_e^2}{eE}g(\eta)\gg 1,
\label{intlimit}
\end{equation}
which indicates that the integral in (\ref{bscattpro4}) is strongly
dominated by values in a neighborhood of $u=\eta^{-1}$. This allows
one to approximately perform the integration and leads to the rate
of pair production of charged bosons \cite{1970PhRvD...2.1191B},
\begin{equation}
\tilde {\mathcal P}_{\rm boson}\simeq\frac{\alpha E^2}{2\pi}\frac{1}{g(\eta)+\frac{1}{2\eta} g'(\eta)}
\exp\left[-\frac{\pi m_e^2}{eE}g(\eta)\right].
\label{bscattpro5}
\end{equation}
Analogously, the rate of pair production of charged fermions can be
approximately obtained from Eq.~(\ref{bscattpro5}) by taking into
account two helicity states of fermions (see Secs.~\ref{semi} and
\ref{Schwingerformula}),
\begin{equation}
\tilde {\mathcal P}_{\rm fermion}\simeq\frac{\alpha E^2}{\pi}\frac{1}{g(\eta)+\frac{1}{2\eta} g'(\eta)}
\exp\left[-\frac{\pi m_e^2}{eE}g(\eta)\right].
\label{bscattpro5f}
\end{equation}
This formula has played an important role in recent studies of
electron and positron pair production by laser beams, which we will
discuss in some details in Section~\ref{Xray}. Momentum spectrum of
electrons and positrons, produced from the vacuum, was calculated in
\cite{1971ZhPmR..13..261P,1972JETP...34..709P,1972JETP...35..659P,2001JETPL..74..133P}.
For $\eta\gg1$ this distribution is concentrated along the direction
of electric field, while for $\eta\ll1$ it approaches isotropic one.

Unfortunately, it appears very difficult to produce a macroscopic electric
field with strength of the order of the critical value (\ref{critical1}) and
lifetime long enough ($\gg \hbar/(m_ec^2)$) in any ground laboratory to
directly observe the Sauter-Euler-Heisenberg-Schwinger process of
electron--positron pair production in vacuum. The same argument applies for the
production of any other pair of fermions or bosons. In the following Section,
we discuss some ideas to experimentally create a transient electric field
$E\lesssim E_c$ in Earth-bound laboratories, whose lifetime is expected to be
long enough (larger than $\hbar/m_ec^2$) for the pair production process to
take place.

\subsection{Nonlinear Compton scattering and Breit-Wheeler process}

In Section~\ref{BW}, we have discussed the Breit--Wheeler process
\cite{1934PhRv...46.1087B} in which an electron--positron pair is produced in
the collision of two real photons $\gamma_1 + \gamma_2 \rightarrow e^+ +e^-$
(\ref{2gammaee}). The cross-section they obtained is $O(r_e^2)$, where $r_e$ is
the classical electron radius, see Eq.~(\ref{bwsection3}). This lowest order
photon-photon pair production cross-section is so small that it is difficult to
observe creation of pairs in the collision of two high-energy photon beams,
even if their center of mass energy is larger than the energy-threshold
$2m_ec^2= 1.02$ MeV.

In the previous Sections we have seen that in strong electromagnetic fields in
lasers the effective nonlinear terms (\ref{Kleinert1}) become significant and
therefore, the interaction needs not to be limited to initial states of two
photons \cite{1962JMP.....3...59R, 1971PhRvL..26.1072R}. A collective state of
many interacting laser photons occurs.

We turn now to two important processes \cite{1996PhRvL..76.3116B,
1997PhRvL..79.1626B} emerging in the interaction of an ultrarelativistic
electron beam with a terawatt laser pulse, performed at SLAC
\cite{1996NIMPA.383..309K}, when strong electromagnetic fields are involved.
The first process is the nonlinear Compton scattering, in which an
ultrarelativistic electron absorbs multiple photons from the laser field, but
emits only a single photon via the process
\begin{equation}
e+n\omega \rightarrow e' + \gamma ,
\label{process1}
\end{equation}
where $\omega$ represents photons from the strong electromagnetic wave of the
laser beam (its frequency being $\omega$), $n$ indicates the number of absorbed
photons and $\gamma$ represents a high-energy emitted photon (see
Eq.~(\ref{process2}) for {\it cross symmetry}). The theory of this nonlinear
Compton effect (\ref{process1}) is given in Section \ref{lighttheoryquantum}.
The same process (\ref{process1}) has been expressed by Bamber et al.
\cite{1999PhRvD..60i2004B} in a  semi-classical framework. The second is the
nonlinear Breit--Wheeler process
\begin{equation}
\gamma+n\omega \rightarrow e^+ +e^- .
\label{process2}
\end{equation}
between this very high-energy photon $\gamma$ and multiple laser photons: the
high-energy photon $\gamma$, created in the first process, propagates through
the laser field and interacts with laser photons $n\omega$ to produce an
electron--positron pair \cite{1997PhRvL..79.1626B}.

In the electric field $E$ of an intense laser beam, an electron oscillates with
the frequency $\omega$ of the laser and its maximum velocity in unit of the
speed of light is given by
\begin{equation}
v_{\rm max}\gamma_{\rm max}=\frac{eE}{m\omega},\quad\quad \gamma_{\rm max}=1/\sqrt{1-v_{\rm max}^2}.
\label{maxv}
\end{equation}
In the case of weak electric field, $v_{\rm max}\ll 1$ and the nonrelativistic
electron emits the {\it dipole} radiation well described in linear and
perturbative QED. On the other hand, in the case of strong electric fields,
$v_{\rm max}\rightarrow 1$ and the ultrarelativistically oscillating electron
emits {\it multi-pole} radiation. The radiated power is then a nonlinear
function of the intensity of the incident laser beam. Using the maximum
velocity $v_{\rm max}$ of oscillating electrons in the electric field of laser
beam, one can characterize the effect of nonlinear Compton scattering by the
dimensionless parameter
\begin{equation}
\eta =v_{\rm max}\gamma_{\rm max}=\frac{eE_{\rm rms}}{ m\omega }=\frac{m_ec^2}{ \omega\hbar}\frac{E_{\rm rms}}{ E_c},
\label{eta}
\end{equation}
where the subscript `rms' means root-mean-square, with respect to the number of
interacting laser photons with scattered electron. The parameter $\eta$ can be
expressed as a Lorentz invariant,
\begin{equation}
\eta^2=\frac{e^2|\langle A_\mu A^\mu \rangle|}{ m_e^2},
\label{eta1}
\end{equation}
where $A_\mu$ is the gauge potential of laser wave, $\partial^\mu A_\mu=0$ and
the time-average is taken over one period of laser wave, $\langle A_\mu \rangle =0$ and
\begin{equation}
\langle A_\mu A^\mu \rangle=\langle (A_\mu  -\langle A_\mu \rangle)^2\rangle.
\label{aaverage}
\end{equation}
Eq.~(\ref{eta1}) shows that $\eta^2$ is the intensity parameter of laser
fields, and $\eta$ in (\ref{eta}) coincides with the parameter $\eta$
introduced in Eq.~(\ref{bscattpro4}) for the pair production in an alternating
electric field (see Section~\ref{alternating}).

%%%%%%%%%%%%%%%%%%%%%%%%%%%%%%%%%%%%%%%%%%%%%%%%%%%%%%%%%%%%%%%%%%%%%%%%%%%%%%%%%%%%%%%%%%%%%

\subsection{Quantum description of nonlinear Compton effect}\label{lighttheoryquantum}

In Refs.~\cite{1962JMP.....3...59R, 1971PhRvL..26.1072R,
Nikishov1964, Nikishov1964a, Nikishov1965, 1967JETP...25.1135N,
Nikishov1979, Nikishov1965a, Sengupta1952, 1964PhRv..133..705B,
Goldman1964, 1964PhL.....8..103G, Eberly1969, 1982els..book.....B},
the quantum theory of the interaction of free electrons with the
field of a strong electromagnetic wave has been studied. The
application of quantum perturbation theory to such interaction
requires not only that the interaction constant $\alpha$ should be
small but also that field should be sufficiently weak. The
characteristic quantity in this respect is the dimensionless
invariant ratio $\eta$, see (\ref{eta1}). The photon
emission processes occurring in the interaction of an electron with
the field of a strong electromagnetic wave have been discussed in
Ref.~\cite{1982els..book.....B} for any $\eta$ value. The method
used is based on an exact treatment of this interaction, while the
interaction of the electron with the newly emitted photons regarded
as a perturbation.

Laser beam is considered as a monochromatic plane wave, described by
the gauge potential $A_\mu(\phi)$ and $\phi=kx$, where wave vector
$k=(\omega, {\bf k})$ $(k^2=0)$ (see Eq.~(\ref{2p2epotential})). The
Dirac equation can be exactly solved \cite{Wolkow1935} for an
electron moving in this field of electromagnetic plane wave of an
arbitrary polarization and the normalized wave function of the
electron with momentum $p$ is given by (c.e.g.
\cite{1982els..book.....B}),
\begin{eqnarray}
\psi_p &=& \left[1+\frac{e}{2(kp)}\not\!{k}\not{\hspace{-5pt}A}\right]\frac{u(p)}{\sqrt{2q_0}}e^{i\Phi},\label{eplanes}\\
\Phi &=& -px-\int^{kx}_0\left[\frac{e}{(kp)}(pA)-\frac{e^2}{2(kp)}A^2\right]d\phi,
\label{ephase}
\end{eqnarray}
where $u(p)$ is the solution of free Dirac equation
$(\not\!{p}-m_e)u(p)=0$ and the time-average value of 4-vector,
\begin{equation}
q= p -\frac{e^2\langle A^2\rangle}{2(kp)}k ,
\label{tqmomen}
\end{equation}
is the kinetic momentum operator in the electron state $\psi_p$
(\ref{eplanes}) and the ``effective mass'' $m_*$ of the electron in
the field is
\begin{equation}
q^2=m_*^2, \quad m_*=m_e\sqrt{1+\eta^2 }, \label{effectivemass}
\end{equation}
where $\eta^2$ is given by (\ref{eta1}).
The electron becomes ``heavy'' in an oscillating electromagnetic field.

The $S$-matrix element for a transition of the electron from the
state $\psi_p$ to the state $\psi_{p'}$, with emission of a photon
having momentum $k'$ and polarization $\epsilon'$ is given by
(c.e.g. \cite{1982els..book.....B})
\begin{eqnarray}
S_{fi}&=&-ie\int \bar\psi_{p'}(\gamma \epsilon'{}^*)\psi_p\frac{e^{ik'x}}{\sqrt{2\omega'}}d^4x\label{eplaneS1}\\
&=&\frac{1}{2\omega'\cdot 2q_0\cdot 2q_0'}\sum_n M^{(n)}_{fi}(2\pi)^4i\delta^{(4)}(nk+q-q'-k'),
\label{eplaneS}
\end{eqnarray}
where the integrand in Eq.~(\ref{eplaneS1}) is expanded in Fourier
series and expansion coefficients are in terms of Bessel functions
$J_n$, the scattering amplitude $M^{(n)}_{fi}$ in
Eq.~(\ref{eplaneS}) is obtained\footnote{The explicit expression
$M^{(n)}_{fi}$ is not given here for its complexity, see for example
\cite{1982els..book.....B}.} by integrating over $x$.
Eq.~(\ref{eplaneS}) shows that $S_{fi}$ is an infinite sum of terms,
each corresponds to an energy-momentum conservation law
$nk+q=q'+k'$, indicating an electron ($q$) absorbs $n$-photons
($nk$) and emits another photon ($k'$) of frequency
\begin{equation}
\omega'=\frac{n\omega}{1+(n\omega/m_*)(1-\cos\theta)},
\label{gammaenergy}
\end{equation}
in the frame of reference where the electron is at rest (${\bf
q}=0,q_0=m_*$), and $\theta$ is the angle between $\bf k$ and $\bf
k'$. Given the $n$th term of the $S$-matrix $S_{fi}$
(\ref{eplaneS}), the differential probability per unit volume and
unit time yields,
\begin{equation}
d{\mathcal P}_{e\gamma}^{(n)}=
\frac{d^3{\bf k}'d^3{\bf q}'}{(2\pi)^6\cdot 2\omega'\cdot 2q_0\cdot
2q_0'}|M^{(n)}_{fi}|^2(2\pi)^4i\delta^{(4)}(nk+q-q'-k').
\label{eplaneP}
\end{equation}
Integrating over the phase space of final states $\int d^3{\bf
k}'d^3{\bf q}'$, one obtains the total probability of emission from
unit volume in unit time (circular polarization),
\begin{equation}
{\mathcal P}_{e\gamma}=\frac{e^2m_e^2}{4q_0}
\sum_{n=1}^\infty\int_0^{\kappa_n}\frac{d\kappa}{(1+\kappa)^2}
\left[-4J^2_n(z)+\eta^2\big(2+\frac{\kappa^2}{1+\kappa}\big)
(J^2_{n+1}+J^2_{n-1}-2J^2_n) \right],
\label{epprobability}
\end{equation}
where $\kappa=(kk')/(kp')$, $\kappa_n=2n(kp)/m_*^2$ and Bessel functions
$J_n(z)$,
\begin{equation}
z=2m_e^2\frac{\eta}{(1+\eta^2)^{1/2}}
\left[\frac{\kappa}{\kappa_n}
\left(1-\frac{\kappa}{\kappa_n}\right)\right]^{1/2},
\label{bargument}
\end{equation}
for any $\eta$ value. A systematic investigation of various quantum processes
in the field of a strong electromagnetic wave can be found in
\cite{Nikishov1964, Nikishov1964a, Nikishov1965, 1967JETP...25.1135N,
Nikishov1979, Nikishov1965a}, in particular photon emission and pair production
in the field of a plane wave with various polarizations are discussed.

We now turn to the Breit--Wheeler process for multi-photons (\ref{process2}). In this process, the pair production is attributed to the interaction of a high-energy photon with many laser photons in the electromagnetic laser wave. Actually, the Breit--Wheeler process for
multi-photons, see Eq. (\ref{process2}), is related to the nonlinear
Compton scattering process, see Eq. (\ref{process1}), by {\it crossing
symmetry}. By replacement $p\rightarrow -p$ and $k'\rightarrow -l$ and reverse
the common sign of the expression in Eq.~(\ref{eplaneS}), one obtains the
probability of pair production (\ref{process2}) by a photon $\gamma$ (momentum
$l$) colliding with $n$ laser photons (momentum $k$) per unit volume in unit
time (circular polarization) \cite{Nikishov1964, Nikishov1964a, Nikishov1965,
1967JETP...25.1135N, Nikishov1979, Nikishov1965a},
\begin{eqnarray}
{\mathcal P}_{\gamma\gamma}&=&\frac{e^2m_e^2}{16l_0}\sum_{n>n_0}^\infty\int_1^{\upsilon_n}
\frac{d\upsilon}{\upsilon^{3/2}(1+\upsilon)^{1/2}}\times\nonumber\\
&\times&\left[2J^2_n(z)+\eta^2(2\upsilon-1)(J^2_{n+1}+J^2_{n-1}-2J^2_n) \right],
\label{ggprobability}
\end{eqnarray}
where $\upsilon=(kl)^2/4(kq)(kq')$, $\upsilon_n=n/n_0$, $n_0=2m_*^2/(kl)$ and
Bessel functions $J_n(z)$,
\begin{equation}
z=4m_e^2\frac{\eta (1+\eta^2)^{1/2}}{(kl)}
\left[\frac{\upsilon}{\upsilon_n}\left(1-\frac{\upsilon}{\upsilon_n}\right)\right]^{1/2}.
\nonumber
\end{equation}
In Eq.~(\ref{ggprobability}), the number $n$ of laser photons must be larger
than $n_0$ ($n>n_0$), which is the energy threshold $n_0(kl)=2m_*^2$ for the
process (\ref{process2}) of pair production to occur.

\section{Semi-classical description of pair production in a general
electric field}\label{inhomogeneousfield}

As shown in previous sections, the rate of pair production may be split into an
exponential and a pre-exponential factor. The exponent is determined by the
classical trajectory of the tunneling particle in imaginary time which has the
smallest action. It plays the same role as the activation energy in a Boltzmann
factor with a ``temperature" $\hbar $. The pre-exponential factor is determined
by the quantum fluctuations of the path around that trajectory. At the
semi-classical level, the latter is obtained from the functional determinant of
the quadratic fluctuations. It can be calculated in closed form only for a few
classical paths \cite{Kleinert2004}. An efficient technique for doing this is
based on the JWKB wave functions, another on solving the Heisenberg equations
of motion for the position operator in the external field \cite{Kleinert2004}.

Given the difficulties in calculating the pre-exponential factor, only a few
nonuniform electric fields in space or in time have led to analytic results for
the pair production rate: (i) the electric field in the ${z}$-direction is
confined in the space $x<x_0$, i.e., ${\bf E}=E(x)\hat{\bf z}$ where
$E(x)=E_0\Theta(x_0-x)$ \cite{1988PhRvD..38.3593M,1989PhRvD..40.1667M}; (ii)
the electric field in the ${z}$-direction depends only on the light-cone
coordinate $z_+=(t+z)/\sqrt{2}$, i.e., ${\bf E}=E(z_+)\hat{\bf z}$
\cite{2000PhRvD..62l5005T,2003PhRvD..67a6003A}.
%\mn{list which field configs have been solved analytically. IT IS DONE!! READ IT}
If the nonuniform field has the form $E(z)=E_0/\cosh^2(z)$, which will be
referred as a Sauter field, the rate was calculated by solving the Dirac
equation \cite{Narozhnyi:1970uv} in the same way as Heisenberg and Euler did
for the constant electric field. For general space and time dependencies, only
the exponential factor can be written down easily --- the fluctuation factor is
usually hard to calculate \cite{1971SvPhU..14..673Z}. In the Coulomb field of
heavy nucleus whose size is finite and charge $Z$ is supercritical, the problem
becomes even more difficult for bound states being involved in pair production,
and a lot of effort has been spent on this issue \cite{1971SvPhU..14..673Z,
Greiner1985, 1978PhR....38..227R}.
%\mn{I add here refs of Greiner and Rafelski}

If the electric field has only a time dependence $E=E(t)$, both exponential and
pre-exponential factors were approximately computed by Brezin and Itzykson
using JWKB methods for the purely periodic field  $E(t)=E_0\cos\omega_0 t$
\cite{1970PhRvD...2.1191B}. The result was generalized by Popov in
Ref.~\cite{1973JETPL..17..368M,1974ZNatA..29.1267S} to more general
time-dependent fields $E(t)$. After this, several time-independent but
space-dependent fields were treated, for instance an electric field between
two conducting plates \cite{1988PhRvD..38..348W}, and an electric field around
a Reissner--Nordstr\"om black hole \cite{2000NCimB.115..761K}.

The semi-classical expansion was carried beyond the JWKB
approximation by calculating higher order corrections in powers of
$\hbar$ in Refs. \cite{2002PhRvD..65j5002K,2006PhRvD..73f5020K} and
\cite{2007PhRvD..75d5013K}. Unfortunately, these terms do not
comprise all corrections of the same orders $\hbar$ as explained in
\cite{kleinert:025011}.

An alternative approach to the same problems was recently proposed by using the
worldline formalism \cite{2001PhR...355...73S}, sometimes called the
``string-inspired formalism''. This formalism is closely related to Schwinger's
quantum field theoretic treatment of the tunneling problem, where the
evaluation of a fluctuation determinant is required involving the {\it
fields\/} of the particle pairs created from the vacuum. The worldline approach
is a special technique for calculating precisely this functional determinant.
Within the worldline formalism, Dunne and Schubert \cite{2005PhRvD..72j5004D}
calculated the exponential factor and Dunne et al. \cite{2006PhRvD..73f5028D}
gave the associated prefactor for various  field configurations: for instance a
spatially uniform, and single-pulse field with a temporal Sauter shape $\propto
1/\cosh^2  \omega  t$. For general $z$-dependencies, a numerical calculation
scheme was proposed in Ref.
\cite{2001NuPhB.613..353G,2002IJMPA..17..966G,2002NuPhB.646..158L,2003JHEP...06..018G}
and applied further in \cite{2005PhRvD..72f5001G}. For a multidimensional
extension of the techniques see Ref. \cite{2006PhRvD..74f5015D}.

%and other configurations
%e.g., periodic field $E(t)=E_0\cos\omega_0 t$, fields localized in space (Sauter field) and in time $E(t)=E_0/\cosh^2(t/T)$.
%Recently, classical solutions (instantons) and effective actions for a wide class of nonuniform electric field backgrounds
%are used to determine the exponential factor and pre-exponential factor is calculated by the Gaussian saddle
%approximation
%\cite{2005PhRvD..72j5004D,2001PhR...355...73S,2001NuPhB.613..353G,2002IJMPA..17..966G,2003JHEP...06..018G,2005JHEP...02..069B,2005PhRvD..72f5001G,2006PhRvD..74f5015D,2006PhRvD..73f5028D}. %\mn{ I ADD dont you want to say which configs?}
%The lowest semiclassical approximation
%to the pre-exponential factor for the Sauter field
%was improved only recently in
%an the integration of the quadratic part of effective action
%over the frequency and transverse momentum Ref.~\cite{2007PhRvD..75d5013K}.

In this Section, a general expression is derived for the pair
production rate in nonuniform electric fields $E(z)$ pointing in the
$z$-direction recently derived in
\cite{2008AIPC..966..213X,kleinert:025011}. A simple variable change
in all formulas leads to results for electric fields depending also
on time rather than space. As examples, three cases will be treated:
(i) a nonzero electric field confined to a region of size $\ell $,
i.e., $E(z)\not=0, |z|\lesssim \ell $ (Sauter field see
Eq.~(\ref{sauterv})); (ii) a nonzero electric field in a half space,
i.e., $E(z)\not=0, z\gtrsim 0 $ (see Eq.~(\ref{hv})); (iii) an
electric field increasing linearly like
 $E(z)\sim z$.
In addition, the process of negative energy electrons tunneling into
the bound states of an electric potential with the emission of
positrons will be studied, by considering the case: the electric
field $E(z)\sim z$ of harmonic potential $V(z)\sim z^2$.

\subsection{Semi-classical description of pair production}\label{semivar}

The phenomenon of pair production in an external electric field can
be understood as a quantum mechanical tunneling process of Dirac
electrons \cite{1928RSPSA.117..610D,1930PCPS...26..361D}. In the
original Dirac picture, the electric field bends the positive and
negative energy levels of the Hamiltonian, leading to a level
crossing and a tunneling of the electrons in the negative energy
band to the positive energy band. Let the field vector ${\bf E}(z)$
point in the ${z}$-direction. In the one-dimensional potential
energy (\ref{@VEQ}) the classical positive and negative energy
spectra are
\begin{equation}
{\mathcal E}_\pm(p_z,{p}_\perp;z)=\pm\sqrt{(cp_z)^2+c^2{ p}_\perp^2+(m_ec^2)^2}+V(z),
\label{energyl+-}
\end{equation}
where $p_z$ is the momentum in
the ${z}$-direction, ${\bf p}_\perp$ the momentum
orthogonal to it, and $p_\perp\equiv |{\bf p}_\perp|$.
For a given energy
${\mathcal E}$, the tunneling takes place
from $z_-$ to $z_+$ determined by $p_z=0$ in Eq.~(\ref{energyl+-})
\begin{equation}
{\mathcal E}={\mathcal E}_+(0,{p}_\perp;z_+)=
{\mathcal E}_-(0,{p}_\perp;z_-).
\end{equation}
The
points
$z_\pm$
are the {\em turning points\/}
of the classical trajectories
 crossing from the positive energy band
to the negative one at energy $\E$.
They satisfy
the equations
\begin{equation}
V(z_\pm)=\mp\sqrt{c^2p_\perp^2+m_e^2c^4}+{\mathcal E} .
\label{crosspoint+-}
\end{equation}
This energy level crossing $\E$ is shown in Fig.~\ref{sauterf} for
the Sauter potential $V(z)\propto \tanh (z/\ell)$.
% where an electric field creates positive-energy solutions running into the positive $z$-direction.
\begin{figure}[!th]
\begin{center}
\begin{picture}(105.64,184.645)
\def\fsz{\footnotesize}
\def\ssz{\scriptsize}
\def\tsz{\tiny}
\def\dst{\displaystyle}\unitlength1mm
\put(-40,0){\includegraphics[width=11cm,clip]{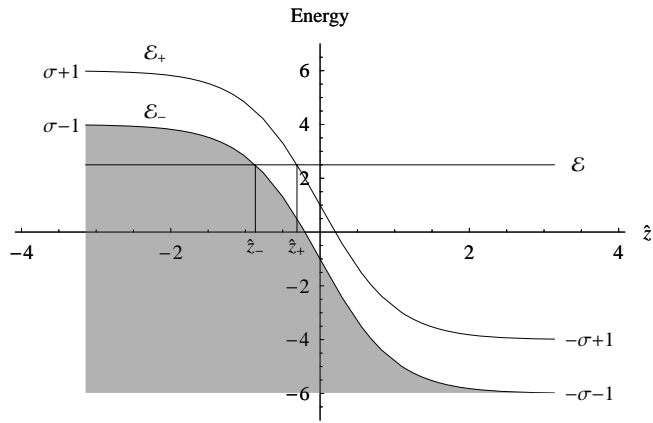}}
%\put(56.5,21.6){\footnotesize $/\ell$}
%\put(49.5,27.0){\footnotesize ${\E}$}
%\put(0,0){\input c:/emtex/texdraw/gratingf}
\end{picture}
\end{center}
\caption{
Positive- and negative-energy
spectra ${\mathcal E}_\pm(z)$
of Eq.~(\ref{energyl+-})
in units of $m_ec^2$,
with  $p_z=p_\perp =0$
as a function of $\hat z=z/\ell $
for the Sauter potential $V_\pm(z)$ (\ref{sauterv}) for $\sigma_s=5$.
%{\footnotesize\tt In figure you must indicate positions of $z_-$ and $z_+$!}
}%
\label{sauterf}%
\end{figure}

\subsubsection{JWKB transmission probability for Klein--Gordon Field}

The probability of quantum tunneling  in the $z$-direction is most easily
studied for a scalar field which satisfies the Klein--Gordon equation
(\ref{@KG0}). If there is only an electric field in the $z$-direction which
varies only along $z$, a vector potential with the only nonzero component
(\ref{@VEQ}) is chosen, and the ansatz $\phi(x)=e^{-i{\E}t/\hbar }e^{i{\sbf
p}_\perp {\sbf x}_\perp/\hbar } \phi_{{\sbf p}_\perp,\E}(z)$, is made with a
fixed momentum ${\bf p}_\perp$ in the $x,y$-direction and an energy ${\E}$, and
Eq.~(\ref{@KG0}) becomes simply
\begin{eqnarray}
\left[-\hbar ^2\frac{d^2}{dz^2}+p_\perp^2
+m_e^2c^2-\frac{1}{c^2} \left[\E-V(z) \right]^2\right]  \phi_{{\sbf p}_\perp,\E}(z)=0.
\label{KG2}
\end{eqnarray}
By expressing the wave function
$ \phi_{{\sbf p}_\perp,\E}(z)$ as an exponential
\begin{equation}
 \phi_{{\sbf p
}_\perp,\E}(z)
=   {\mathcal C}\,e^{iS_{{\sbf p}_\perp,\E}/\hbar },
%\label{@}
\end{equation}
where ${\mathcal C}$ is some normalization  constant, the wave
equation becomes a Riccati equation for $S_{{\sbf p}_\perp,\E}$:
\begin{equation} \label{n4.5}
-i\hbar \partial _z^2
S_{{\sbf p}_\perp,\E}(z)+
[\partial _z
S_{{\sbf p}_\perp,\E}(z)]^2-p_z^2(z)=0.
\end{equation}
where the function $p_z(z)$ is the solution of
the equation
\begin{eqnarray}
p_z^2(z)=\frac{1}{c^2}\left[\E-V(z) \right]^2-p_\perp^2-m_e^2c^2.
\label{WKBr2}
\end{eqnarray}
The solution
of Eq. (\ref{n4.5})  can be found iteratively as an expansion in powers of $\hbar$:
\begin{equation} \label{4.13}
S_{{\sbf p}_\perp,\E}(z)=
S^{(0)}_{{\sbf p}_\perp,\E}(z)-i\hbar
S^{(1)}_{{\sbf p}_\perp,\E}(z)+(-i\hbar )^2
S^{(2)}_{{\sbf p}_\perp,\E}(z)+\dots~.
\end{equation}
Neglecting the expansion terms after $ S^{(1)}_{{\sbf
p}_\perp,\E}(z)=-\log p^{1/2}_z(z) $ leads to the JWKB approximation
for the wave functions of positive and negative energies (see e.g.
\cite{Landau1981a,Kleinert2004})
\begin{eqnarray}
\phi^{\rm JWKB}_{{\sbf p}_\perp,\E}(z)=
\frac{\mathcal C}{p^{1/2}_z(z)}e^{iS^{(0)}_{{\sbf p}_\perp,\E}(z)/\hbar }.
\label{WKBs}
\end{eqnarray}
where  $S^{(0)}_{{\sbf p}_\perp,\E}(z)$ is the eikonal
\begin{eqnarray}
S^{(0)}_{{\sbf p}_\perp,\E}(z)=\int^zp_z(z')dz'.
\label{WKBs1}
\end{eqnarray}
Between the turning points $z_-<z<z_+$, whose positions are illustrated in
Fig.~\ref{sauterf}, the momentum $p_z(z)$ is imaginary and it is useful to
define the positive function
\begin{eqnarray}
 \kappa _z(z)\equiv  \sqrt{
p_\perp^2+m_e^2c^2-  \frac{1}{c^2}
\left[\E-V(z) \right]^2
}\geq 0.
\end{eqnarray}
The tunneling wave function in this regime is the linear combination
\begin{eqnarray}
\frac{{\mathcal C}}{2(\kappa_z)^{1/2}}
\exp\left[ -\frac{1}{\hbar}\int^{z}_{z_-}\kappa_zdz\right]+
\frac{\bar {\mathcal C}}{2(\kappa_z)^{1/2}}
\exp\left[ +\frac{1}{\hbar}\int^{z}_{z_-}\kappa_zdz\right].
\label{WKBt}
\end{eqnarray}
Outside the turning points, i.e., for $z<z_-$ and $z>z_+$, there
exist negative energy and positive energy solutions for $\E<\E_-$
and $\E>\E_+$ for positive $p_z$. On the left-hand side of $z_-$,
the general solution is a linear combination of an incoming wave
running to the right and outgoing wave running to the left:
\begin{eqnarray}
\frac{\mathcal C_+}{(p_z)^{1/2}}\exp
\left[
 \frac{i}{\hbar}\int^zp_zdz\right]  +
 \frac{\mathcal C_-}{(p_z)^{1/2}}\exp\left[  -\frac{i}{\hbar}\int^zp_zdz\right] .
\label{WKBin}
\end{eqnarray}
On the right-hand of $z_+$, there is only an outgoing wave
\begin{eqnarray}
\frac{\mathcal T}{(p_z)^{1/2}}\exp\left[
 \frac{i}{\hbar}\int^z_{z_+}p_zdz\right] ,
\label{WKBout}
\end{eqnarray}
The connection equations
can be solved by
\begin{equation}
\bar {\mathcal C}=0,~~
{\mathcal C_\pm}=e^{\pm i\pi/4}{\mathcal C}/2,~~ {\mathcal T}=
{\mathcal C}_+ \exp \left[ -\frac{1}{\hbar}\int_{z_-}^{z_+}\kappa_z dz\right].
\label{outa}
\end{equation}
The incident flux density is
\begin{eqnarray}
j_z\equiv \frac{\hbar}{2m_ei}\left[\phi^*\partial_z\phi - (\partial_z\phi^*)\phi\right]
=\frac{p_z}{m_e}\phi^*\phi=\frac{|{\mathcal C}_+|^2 }{m_e},
\label{influx}
\end{eqnarray}
which can be written as
\begin{eqnarray}
j_z(z)=v_z(z)n_-(z),
\end{eqnarray}
where $v_z(z)=p_z(z)/m_e$ is the velocity and $n_-(z)=\phi^*(z)\phi(z)$ the
density of the incoming particles. Note that the $z$-dependence of $v_z(z)$ and
$n_-(z)$ cancel each other. By analogy, the outgoing flux  density is
$|{\mathcal T}|^2 /m_e$.

\subsubsection{Rate of pair production}
From the considerations given above, the transmission probability
\begin{eqnarray}
{\mathcal P}_{\rm JWKB}\equiv \frac{\rm transmitted ~~ flux}{\rm incident ~~ flux }
\label{wdefine}
\end{eqnarray}
is found to be the simple exponential
\begin{eqnarray}
{\mathcal P}_{\rm JWKB}(p_\perp,{\mathcal E}) &= & \exp\left[-
{\frac{2}{\hbar}}\int_{z_-}^{z_+} \kappa _zdz\right].
\label{tprobability3}
\end{eqnarray}
% The  amplitude of inverse process, electron--positron annihilation,
%is also given by
%Eq.~(\ref{tprobability3}) for the theorem of ${\mathcal CPT}$-invariance.
In order to derive
 from
(\ref{wdefine})
the total rate of pair production in the electric field, it must be multiplied
with the incident particle flux density
at the entrance $z_-$ of the tunnel.
The particle velocity
at that point is
 $v_z=\partial \E /\partial p_z$, where
the relation between $\E $ and $z_-$ is given by
 Eq.~(\ref{crosspoint+-}):
\begin{equation}
-1= \frac{{\mathcal E}-V(z_-)}{{\sqrt{(c{p}_\perp) ^2+m_e^2c^4}}}.
\label{@Entr}\end{equation}
This must be multiplied
with the particle density
which is given by the phase space density
$d^3p/(2\pi \hbar )^3$.
The incident flux density at the tunnel entrance is therefore
\begin{eqnarray}
j_z(z_-)=
\Ds\int  \frac{\partial \E}{\partial p_z}
\frac{d^2{p}_\perp}{(2\pi\hbar)^2}
\frac{d{p}_z}{2\pi\hbar} =
\Ds\int  \frac{ d\E}{{2\pi\hbar}}
\frac{d^2{p}_\perp}{(2\pi\hbar)^2},
\label{nflux}
\end{eqnarray}
and the extra factor $\Ds$
is equal to $2$ for electrons with two spin orientations\footnote{By setting $\Ds$ equal to $1$ one can obtain the tunneling result also for
spin-$0$ particles although the Dirac picture is no longer applicable.}.

It is useful to change
 the variable of integration from  $z$ to $\zeta(z)$
defined by
\begin{equation}
\zeta({p}_\perp,\E;z)\equiv \frac{{\mathcal E}-V(z)}{{\sqrt{(c{p}_\perp) ^2+m_e^2c^4}}},
\label{y(x)}
\end{equation}
and to introduce the
notation for
the electric field
$E({p}_\perp,\E;\zeta)\equiv E[\bar z({p}_\perp,\E;\zeta)]$, where $\bar z(
{p}_\perp,\E;\zeta)$ is the inverse function
of
(\ref{y(x)}),
the equations
in (\ref{crosspoint+-})
reduce to
\begin{equation}
\zeta_-({p}_\perp,\E;z_-) =-1,~~~~
\zeta_+({p}_\perp,\E;z_+) = +1.
\label{y+-}
\end{equation}
In terms of the variable $\zeta$,
the JWKB transmission probability (\ref{tprobability3})
can be rewritten as
\begin{eqnarray}
{\mathcal P}_{\rm JWKB}(p_\perp,{\mathcal E}) = \exp\left\{ -\frac{2m^2_ec^3}{e \hbar E_0}
  \left[1+\frac{(c{p}_\perp) ^2}{m_e^2c^4}\right]
\int^{1}_{-1} d\zeta\frac{\sqrt{1-\zeta^2}}{ E({p}_\perp,\E;\zeta)/E_0}\right\}.
\label{wwkbp}
\end{eqnarray}
%where
%\begin{eqnarray}
%z(y)=\bar z[(1+(c{p}_\perp) ^2/(m_e^2c^4))^{1/2} y],
%\label{ey}
%\end{eqnarray}
%is the inverse function of Eq.~(\ref{y(x)}).
Here a standard field
strength $E_0$ has been introduced
to make the integral in the exponent
dimensionless, which is abbreviated
 by
\begin{eqnarray}
G(p_\perp,{\mathcal E})\equiv \frac{2}{\pi}\int^{1}_{-1} d\zeta\frac{\sqrt{1-\zeta^2}}{E({p}_\perp,\E;\zeta)/E_0}.
\label{gf}
 \end{eqnarray}
The first term in the exponent
of (\ref{wwkbp})
is equal to $2E_c/E_0$.

At the semi-classical level, tunneling takes place only if the
potential height is larger than $2m_ec^2$ and for energies $\E$ for
which there are two real turning points $z_\pm$. The total tunneling
rate is obtained by integrating over all incoming momenta and the
total area
 $V_\perp=\int dxdy$
of the incoming flux. The JWKB-rate per area is
\begin{eqnarray}
\frac{\Gamma _{\rm JWKB}}{V_\perp}&=&
\Ds \int\frac{ d\E}{2\pi \hbar }
\int\frac{d^2{p}_\perp}{(2\pi\hbar)^2}
{\mathcal P}_{\rm JWKB}(p_\perp,{\mathcal E}).
\label{gxy0}
\end{eqnarray}
 Using the
relation following from (\ref{@Entr})
\begin{equation}
d{\mathcal E}=eE(z_-)dz_-,
 \label{@REL}\end{equation}
the alternative expression is obtained
%Using Eqs.~(\ref{wwkbp})--(\ref{nflux}), the
%total rate of pair production per volume layer $dz_-\int dxdy$ at given crossing energy-level ${\mathcal E}$ is
\begin{eqnarray}
\frac{\Gamma _{\rm JWKB}}{V_\perp}&=&
\Ds \int \frac{dz_-}{2\pi \hbar }
\int\frac{d^2{p}_\perp}{(2\pi\hbar)^2}
%|
eE(z_-)
%|
{\mathcal P}_{\rm JWKB}(p_\perp,{\mathcal E}(z_-)),
\label{gxy}
\end{eqnarray}
 where
${\mathcal E}(z_-)$ is obtained by solving the differential equation
(\ref{@REL}).

The integral over $p_\perp$ cannot be done exactly.  At the
semi-classical level, this is fortunately not necessary. Since $E_c$
is proportional to $1/\hbar $, the exponential in (\ref{wwkbp})
restricts the transverse momentum ${p}_\perp$ to be small of the
order of $ \sqrt{\hbar }$, so that the integral in (\ref{gxy}) may
be calculated from an expansion of $G(p_\perp,{\mathcal E})$ up to
the order ${p}_\perp^2$:
%This makes integral over ${p}_\perp$ Gaussian
%If we denote $G(0,\E)$ by $G_0$, and $E({0},\E;\zeta)$ by $E_0(\E;\zeta)$, the expansion reads
\begin{eqnarray}
G(p_\perp,{\mathcal E})
& \simeq & \frac{2}{\pi}\int^{1}_{-1} d\zeta\frac{\sqrt{1-\zeta^2}}{ E(0,\E;\zeta)/E_0}
\left[1-\frac{1}{2}\frac{d E({ 0},\E,\zeta)/d\zeta}{ E({ 0},\E,\zeta)}\zeta \,\delta +\dots
\right]= \nonumber\\
& = & G(0,{\mathcal E})+
G_ \delta (0,{\mathcal E}) \delta+\dots ,
\label{gfhbar}
\end{eqnarray}
where
$\delta\equiv  \delta (p_\perp)\equiv (c{p}_\perp) ^2/(m_e^2c^4)$,
and
\begin{eqnarray}
G_ \delta (0,{\mathcal E}) &\equiv &-\frac{1}{\pi}\int^{1}_{-1}
d\zeta\frac{\zeta\sqrt{1-\zeta^2}}{ E^2(0,\E;\zeta)/E_0} E'(0,\E;\zeta)\nonumber\\
&=&-\frac{1}{2}
G(0,{\mathcal E})+\frac{1}{\pi}\int^{1}_{-1} d\zeta\frac{\zeta^2}{\sqrt{1-\zeta^2}}
\frac{d\zeta}{ E(0,\E,\zeta)/E_0}.
\label{ghbar}
\end{eqnarray}
The integral over ${\bf p}_\perp$
in (\ref{gxy}) is
approximately performed
as follows:
\begin{eqnarray}
&&\int \frac{d^2p_\perp}{(2\pi \hbar )^2}
e^{-\pi({E_c}/E_0)(1+ \delta )[
G (0,{\mathcal E})
+G_ \delta (0,{\mathcal E}) \delta ]}
= \label{tranp}\\
&&=
\frac{m_e^2c^2}{4\pi\hbar ^2}\int_0^\infty d \delta  \,
e^{-\pi({E_c}/E_0)[
G (0,{\mathcal E}) + \delta
\tilde G (0,{\mathcal E})
}
\approx\frac{eE_0}{4\pi^2\hbar c\tilde
G (0,{\mathcal E})
}
e^{-\pi({E_c}/E_0)
G (0,{\mathcal E})},
\nonumber
\end{eqnarray}
where
\begin{equation}
\tilde
G (0,{\mathcal E})
\equiv G (0,{\mathcal E})
+G_ \delta  (0,{\mathcal E})
.
%\label{@}
\end{equation}
%
% $ \alpha \equiv e^2/\hbar c$ is the fine structure constant.

The electric fields $E(p_\perp,\E;\zeta)$ at the tunnel entrance
$z_-$ in the prefactor of (\ref{gxy}) can be expanded similarly to
first order in $ \delta $. If $z_-^0$ denotes the solutions of
(\ref{@Entr}) at $p_\perp=0$, it is found that for small $\delta$:
\begin{equation}
 \Delta z_-\equiv z_--z_-^0
\approx
\frac{m_ec^2}{E(z^0_-)}
\frac{\delta}{2} .
\end{equation}
so that
\begin{eqnarray}
E(z_-)\simeq
E(z^0_-)-
{m_ec^2}\frac{E'(z^0_-)}{E(z^0_-)}\frac{\delta}2.
\label{ehbar}
\end{eqnarray}
Here the extra term proportional to $\delta$ can be neglected in the
semi-classical limit since it gives a contribution to the prefactor
of the order $\hbar $. Thus the JWKB-rate (\ref{gxy}) of pair
production per unit area is obtained
\begin{eqnarray}
\frac{\Gamma _{\rm JWKB}}{V_\perp}
\equiv \int dz \frac{\partial _z\Gamma _{\rm JWKB}(z)}{V_\perp}
\simeq \int dz\frac{\Ds e^2 E_0 E(z)}{8\pi^3\hbar ^2c
\,\tilde G(0,{\mathcal E}(z))
}
e^{-\pi ( E_c/{E_0})G(0,{\mathcal E}(z))},
\label{pgxy3}
\end{eqnarray}
where $z$ is short for  $z^0_-$. At this point it is useful to
return from the integral $\int dz_-
%|
eE(z_-)
%|
$ introduced in (\ref{gxy}) to the original energy integral $\int
d\E$ in (\ref{gxy0}), so that the final result is
\begin{eqnarray}
\frac{\Gamma _{\rm JWKB}}{V_\perp}\equiv
\int d\E \frac{\partial _{\E}\Gamma _{\rm JWKB}(z)}{V_\perp}
\simeq \Ds
\frac{%|
eE_0
%|
}{4\pi^2\hbar c}
\int \frac{d\E}{2\pi \hbar }
\frac{1}{
\tilde G(0,{\mathcal E})
}
%\left[1+\frac{\hbar}{2\pi^2 c[G(0,{\mathcal E}_-)-G_ \delta (0,{\mathcal E}_-)
%]}\frac{E_0 E'(z)}{E^2(z)}\right]
e^{-{\pi ( E_c/E_0)G(0,{\mathcal E})}},
\label{pgwk1}
\end{eqnarray}
where ${\mathcal E}$-integration is over all crossing energy levels.

These formula can be approximately applied to the three-dimensional
case of electric fields ${\bf E}(x,y,z)$ and potentials $V(x,y,z)$
at the points $(x,y,z)$ where the tunneling length
(\ref{tunnelinglength}) is much smaller than the variation lengths
$\delta x_\perp$ of electric potentials $V(x,y,z)$ in the
$xy$-plane,
\begin{eqnarray}
\frac{1}{z_+-z_-} \gg \frac{1}{V}\frac{\delta V}{\delta x_\perp}.
\label{3dcon}
\end{eqnarray}
At these points $(x,y,z)$, one can arrange the tunneling path $dz$
and momentum $p_z(x,z,z)$ in the direction of electric field,
corresponding perpendicular area $d^2V_\perp\equiv dxdy$ for
incident flux and perpendicular momentum ${\bf p}_\perp$. It is then
approximately reduced to a one-dimensional problem in the region of
size $~{\mathcal O}(a)$ around these points. The surfaces
$z_-(x_-,y_-,\E)$ and $z_+=(x_+,y_+,\E)$ associated with the
classical turning points are determined by Eqs.~(\ref{y(x)}) and
Eqs.~(\ref{y+-}) for a given energy $\E$. The JWKB-rate of pair
production (\ref{pgxy3}) can then be expressed as a volume integral
over the rate density per volume element
\begin{eqnarray}
\Gamma _{\rm JWKB}=
\int dxdydz
\frac{d^3\Gamma _{\rm JWKB}}{dx\,dy\,dz}=
\int dtdxdydz
\frac{d^4N_{\rm JWKB}}{dt\,dx\,dy\,dz}.
\label{3drate0}
\end{eqnarray}
On the right-hand side it is useful to rewrite the rate $ \Gamma
_{\rm JWKB}$ as the time derivative of the number of pair creation
events $dN_{\rm JWKB}/dt$, so that one obtains an event density in
four-space

\begin{eqnarray}
\frac{d^4N_{\rm JWKB}}{dt\,dx\,dy\,dz}
\approx \Ds \frac{e^2 E_0 E(z)}{8\pi^3\hbar
\,\tilde G(0,{\mathcal E}(z))
}
e^{-\pi ( E_c/{E_0})G(0,{\mathcal E}(z))},
\label{3drate}
\end{eqnarray}
Here $x,y$ and $z$ are related by the function $z=z_- (x,y,\E)$
which is obtained by solving (\ref{@REL}).

It is now useful to observe that the left-hand side of
(\ref{3drate}) is a Lorentz invariant quantity. In addition, it is
symmetric under the exchange of time and $z$, and this symmetry will
be exploited in the next section to  relate pair production
processes in a $z$-dependent electric field $E(z)$ to those in a
time-dependent field $E(t)$.

Attempts to go beyond the JWKB results (\ref{pgxy3}) or
(\ref{pgwk1}) require a great amount of work. Corrections will come
from three sources:
\begin{enumerate}
\item[I] from the
higher terms
of order in~$ (\hbar)^n$ with $n>1$
 in the expansion (\ref{4.13})
solving
the Riccati equation
(\ref{n4.5}).
\item[II]
from  the perturbative evaluation of the
integral over ${\bf p}_\perp$ in Eqs. (\ref{gxy0})
 or
(\ref{gxy})
when going beyond the Gaussian approximation.
\item[III]
from perturbative
 corrections
to the Gaussian
energy integral (\ref{pgwk1})
or the corresponding $z$-integral (\ref{pgxy3}).
\end{enumerate}
All these corrections contribute terms of higher order in $ \hbar$.

\label{@RATE}

\subsubsection{Including a smoothly varying ${\bf B}(z)$-field  parallel to ${\bf E}(z)$}

The results presented above can easily be extended for the presence of a
constant magnetic field ${\bf B}$ parallel to ${\bf E}(z)$. Then the wave
function factorizes into a Landau state and a spinor function first calculated
by Sauter \cite{1931ZPhy...69..742S}. In the JWKB approximation, the energy
spectrum is still given by Eq.~(\ref{energyl+-}), but the squared transverse
momenta $p_\perp^2$ is quantized and must be replaced by discrete values
corresponding to the Landau energy levels. From the known nonrelativistic
levels for the Hamiltonian $p_\perp^2/2m_e$ one extracts immediately the
replacements (\ref{landaulevel}). Apart from the replacement
(\ref{landaulevel}), the JWKB calculations remain the same. Thus one must only
replace the integration over the transverse momenta $\int d^2{
p}_\perp/(2\pi\hbar)^2$ in Eq.~(\ref{tranp}) by the sum over all Landau levels
with the degeneracy $ eB/(2\pi\hbar c)$. Thus, the right-hand side becomes
\begin{equation}
  \frac{eB}{2\pi\hbar c}e^{-\pi({E_c}/E_0)
G (0,{\mathcal E})}
\sum_{n,\hat\sigma}
e^{-\pi (B/E_0)(n+1/2+g\hat\sigma)
\tilde G (0,{\mathcal E})
},
\label{tprobability2h}  \!\!\!\!\!\!\!
\end{equation}
where $g$ and $\hat\sigma$ are as in (\ref{landaulevel}). The result is, for
spin-0 and spin-1/2:
\begin{eqnarray}
\frac{eE_0}{4\pi^2\hbar c\tilde G (0,{\mathcal E})}
e^{-\pi({E_c}/E_0)G (0,{\mathcal E})}
f_{0,1/2}(
 B
\tilde G (0,{\mathcal E})/E_0)
%\label{wkbehfermion2}
\end{eqnarray}
where
\begin{eqnarray}
f_{0}(x)\equiv
\frac{\pi x}{\sinh \pi x},\ \ \  \
f_{1/2}(x)\equiv
2\frac{ \pi x }{\sinh \pi x}
{ \cosh \frac{\pi gx}2}
%\label{wkbehboson}
\end{eqnarray}
In the limit $B\rightarrow 0$,
Eq.~(\ref{wkbehboson})
reduces
 to Eq.~(\ref{tranp}).

The result remains approximately valid if the magnetic field has a smooth
$z$-dependence varying little over a Compton wavelength $ \lambda _C$.

In the following only nonuniform electric fields without a magnetic
field are considered.

\subsection{Time-dependent electric fields}\label{time}

The semi-classical considerations given above can be applied with
little change to the different physical situation in which the
electric field along the $z$-direction depends only on time rather
than  $z$. Instead of the time $t$ itself it is better to work with
the zeroth length coordinate $x_0=ct$, as usual in relativistic
calculations. As an intermediate step consider for a vector
potential
\begin{equation}
A_\mu=(A_0(z),0,0,A_z(x_0)),
%\label{@}
\end{equation}
with the
electric field
\begin{equation}
E=-\partial _zA_0(z)-\partial_0A_z(x_0),~~~~x_0\equiv ct.
%\label{@}
\end{equation}
The associated Klein--Gordon equation  (\ref{@KG0}) reads
\begin{eqnarray}
\left\{ \left[ i\hbar \partial _0+\frac{e}{c}A_0(z)\right] ^2
+\hbar ^2\partial _{{\sbf x}_\perp}^2
-\left[ i\hbar \partial _z+\frac{e}{c}A_z(x_0)\right] ^2-m_e^2c^2\right\}
\phi(x)=0.
\label{@KG}\end{eqnarray}
The previous discussion
was valid under the assumption
$A_z(x_0)=0$, in which case the ansatz
\begin{align}
\phi(x)=e^{-i{\E}t/\hbar }e^{i{\sbf p}_\perp {\sbf x}_\perp}
\phi_{{\sbf p}_\perp,\E}(z),\nonumber
\end{align}
led to the field equation (\ref{KG2}). For the present discussion it
is useful to write the ansatz as $\phi(x)=e^{-ip_0x_0/\hbar
}e^{i{\sbf p}_\perp {\sbf x}_\perp/\hbar } \phi_{{\sbf
p}_\perp,p_0}(z)$ with $p_0=\E/c$, and Eq.~(\ref{KG2}) in the form
\begin{eqnarray}
\left\{\frac{1}{c^2}\left[\E-e\int^z dz'\,E(z')\right]^2-p_\perp^2
-m_e^2c^2
+\hbar ^2\frac{d^2}{dz^2}
\right\}  \phi_{{\sbf p}_\perp,p_0}(z)=0.
\label{KG2t}
\end{eqnarray}

Now assume that the electric field depends only on $x_0=ct$. Then
the ansatz $\phi(x)=e^{ip_z z/\hbar }e^{i{\sbf p}_\perp {\sbf
x}_\perp/\hbar } \phi_{{\sbf p}_\perp,p_z}(x_0)$ leads to the field
equation
\begin{eqnarray}
\left\{-\hbar ^2\frac{d^2}{d{x_0}^2}-p_\perp^2
-m_e^2c^2-\left[-p_z-\frac{e}{c}\int^{x_0} dx'_0 E(x'_0) \right]^2\right\}  \phi_{{\sbf p}_\perp,p_z}(x_0)=0.
\label{KG3}\end{eqnarray}

If Eq.~(\ref{KG3}) is compared with (\ref{KG2t}) it can be seen that
one arises from the other by interchanging
\begin{equation}
z\leftrightarrow x_0,~~~p_\perp\rightarrow ip_\perp, ~~~c\rightarrow ic,~~~E\rightarrow -iE.
\label{trantable}\end{equation}
With these exchanges, it may easy to calculate the decay rate of the
vacuum caused by a time-dependent electric field $E(x_0)$ using the
formulas derived above.

\subsection{Applications}\label{applicat}

Now formulas~(\ref{pgwk1}) or (\ref{pgxy3}) are applied to various external
field configurations capable of producing electron--positron pairs.

\subsubsection{Step-like electric field}

First one checks the result for the original case of a constant
electric field $E(z)\equiv eE_0$ where the potential energy is the
linear function  $V(z)=-eE_0z$. Here the function (\ref{gf}) becomes
trivial
\begin{eqnarray}
G(0,\E)=\frac{2}{\pi}\int^{1}_{-1} d\zeta\sqrt{1-\zeta^2}=1,\quad G_\delta(0,\E)=0,
\label{gfc}
\end{eqnarray}
which is independent of $\E$ (or $z_-$). The JWKB-rate for pair
production per unit time and volume is found from Eq.~(\ref{pgxy3})
to be
\begin{equation}
 \frac{\Gamma^{\rm EH} _{\rm JWKB}}{V}\simeq \Ds\frac{\alpha E_0^2}{ 2\pi^2\hbar}
e^{-\lfrac{\pi E_c}{E_0}}.
\label{xy2}
\end{equation}
where $V\equiv dz_- V_\perp $. This expression contains the exponential
$e^{-\pi E_c/E_0}$ found by Sauter \cite{1931ZPhy...69..742S}, and the correct
prefactor as calculated by Heisenberg and Euler \cite{1936ZPhy...98..714H}, and
by Schwinger
\cite{1951PhRv...82..664S,1954PhRv...93..615S,1954PhRv...94.1362S}.

In order to apply the transformation rules (\ref{trantable}) to
obtain the analogous result for the constant electric field in time,
one can rewrite Eq.~(\ref{xy2}) as
\begin{equation}
 \frac{dN_{\rm JWKB}}{dx_0V}\simeq \Ds\frac{\alpha E_0^2}{ 2\pi^2\hbar c}
e^{-\lfrac{\pi E_c}{E_0}},
\label{txy2}
\end{equation}
where $dN_{\rm JWKB}/dx_0=\Gamma^{\rm EH} _{\rm JWKB}/c$ and $N_{\rm
JWKB}$ is the number of pairs produced. Applying the transformation
rules (\ref{trantable}) to Eq.~(\ref{txy2}), one obtains the same
formula as Eq.~(\ref{xy2}).

\subsubsection{Sauter electric field}

Let us now consider the nontrivial Sauter electric field localized
within finite slab in the $xy$-plane with the width $\ell $ in the
$\hat {\bf z}$-direction. A field of this type can be produced,
e.g., between two opposite charged conducting plates. The electric
field $E(z)\hat{\bf z}$ in the $z$-direction and the associated
potential energy $V(z)$ are given by
\begin{eqnarray}
E(z)=E_0/{\rm cosh}^2\left({z}/{\ell }\right),~~~~~\label{sfield}
V(z)&=&- \sigma_s\, m_ec^2\tanh\left({z}/{\ell }\right),
\label{sauterv}
\end{eqnarray}
where
\begin{equation}
\sigma_s\equiv
%|
eE_0
%|
\ell /m_ec^2=(\ell /\lambda_C)(E_0/E_c).
\label{@gamm}\end{equation}
From now on natural units, in which energies are measured in units
of $m_ec^2$, are used. Figure~\ref{sauterf} shows the positive and
negative energy spectra ${\mathcal E}_\pm(z)$ of
Eq.~(\ref{energyl+-}) for $p_z=p_\perp=0$ in particular the energy
gap and energy level crossings. From Eq.~(\ref{crosspoint+-}) one
finds the classical turning points
\begin{equation}
z_\pm=   \ell   ~
{\rm arc tanh}\frac{\E\pm\sqrt{1+ \delta }}{ \sigma_s }=
\frac{\ell }2\ln\frac{ \sigma_s+  {\mathcal E}\pm \sqrt{1+ \delta }}{
 \sigma_s-  {\mathcal E} \mp
 \sqrt{1+ \delta }}.
\label{scrossing}
\end{equation}
Tunneling is possible for all energies satisfying
\begin{equation}
- \sqrt{1+ \delta }+ \sigma_s \ge  {\mathcal E} \ge  \sqrt{1+ \delta }-  \sigma_s ,
\label{crossingregime}
\end{equation}
for the strength parameter
$ \sigma_s >\sqrt{1+ \delta }$.

One may invert Eq.~(\ref{y(x)}) to find the relation between $\zeta$ and $z$:
\begin{equation}
z=z(p_\perp,\E;\zeta)= \ell   \,
{\rm arc tanh}\frac{\E+ \zeta \sqrt{1+ \delta }}
{ \sigma_s }=\frac{\ell }2\ln\frac{ \sigma_s+  {\mathcal E}+\zeta \sqrt{1+ \delta }
}{
 \sigma_s-  {\mathcal E} -
  \zeta \sqrt{1+ \delta }}.
\label{@zEqu}\end{equation}
In terms of the function $z(p_\perp,\E;\zeta)$, the Eq. (\ref{scrossing}) reads
simply $z_\pm=z(p_\perp,\E;\pm1)$.

Inserting (\ref{@zEqu}) into the equation for $E(z)$ in
Eq.~(\ref{sauterv}), one obtains
\begin{equation}
E(z)= E_0\left[1-\left(\frac{ \zeta\sqrt{1+\delta}- {\mathcal E}}{ \sigma_s}\right)^2\right]
\equiv E(p_\perp,\E;\zeta).
\label{sey}
\end{equation}
$G(0,{\mathcal E})$ and $G_ \delta (0,{\mathcal E})$ of
Eqs.~(\ref{gf}), (\ref{gfhbar}) and (\ref{ghbar}) are calculated:
\begin{eqnarray}
G(0,{\mathcal E})= 2 \sigma_s^2 - \sigma_s
\left[( \sigma_s -  {\mathcal E})^2 -1\right]^{1/2}- \sigma_s
\left[( \sigma_s +  {\mathcal E})^2 -1\right]^{1/2},
\label{sgf}
\end{eqnarray}
and
\begin{eqnarray}
G (0,{\mathcal E})+G_\delta (0,{\mathcal E})=\frac{\sigma_s}{2}\left\{
\left[( \sigma_s -  {\mathcal E})^2 -1\right]^{-1/2}+
\left[(\sigma_s + {\mathcal E})^2 -1\right]^{-1/2}\right\}.
\label{pgsgf}
\end{eqnarray}
Substituting the functions $G (0,{\mathcal E})$ and $G_\delta
(0,{\mathcal E})$ into Eqs.~(\ref{pgxy3}) and (\ref{pgwk1}), one
obtains the general expression for the pair production rate per
volume slice at a given tunnel entrance  point $z_-(\E)$ or the
associated energy $\E(z_-)$. The pair production rate per area is
obtained by integrating over all slices permitted by the energy
inequality (\ref{crossingregime}).

In Fig. \ref{TUNNf}, the slice dependence of the integrand in the
tunneling rate (\ref{pgxy3}) for the Sauter potential
(\ref{sauterv}) is shown and compared with the constant field
expression (\ref{xy2}) of Euler and Heisenberg, if it is evaluated
at the $z$-dependent electric field $E(z)$. This is done once as a
function of the tunnel entrance point $z$ and once as a function of
the associated energy $\E$. On each plot, the difference between the
two curves illustrates the nonlocality of the tunneling
process\footnote{Note that omitting the $z$-integral in the rate
formula (\ref{pgxy3}) does not justify calling the result a ``local
production rate'', as done in the abstract of Ref.
\cite{2005PhRvD..72f5001G}. The result is always nonlocal and
depends on {\em all\/} gradients of the electric field.}.
\begin{figure}[!tbhp]
\includegraphics[width=12cm]{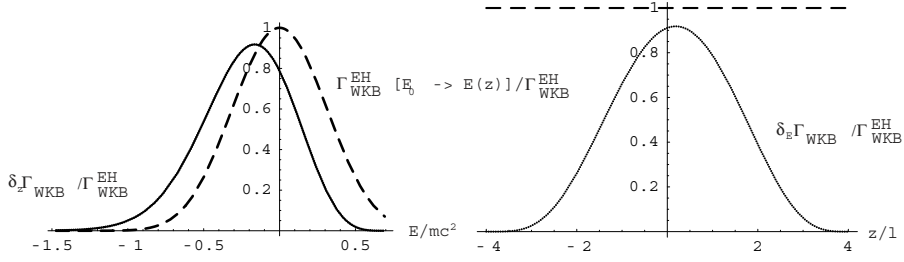}
\caption[]{The slice dependence of the integrand in the tunneling rate
(\ref{pgxy3}) for the Sauter potential  (\ref{sauterv}) is plotted: left, as a
function of the tunnel entrance $z$ (compare with numeric results plotted in
Fig. 1 of Ref. \cite{2005PhRvD..72f5001G}); right, as a function of the
associated energy $\E$, which is normalized by the Euler-Heisenberg rate
(\ref{xy2}). The dashed curve in left figure shows the Euler-Heisenberg
expression (\ref{xy2}) evaluated for the $z$-dependent field $E(z)$ to
illustrate the nonlocality of the production rate. The dashed curve in right
figure shows the Euler-Heisenberg  expression (\ref{xy2}) which is independent
of energy-level crossing $\E$. The dimensionless parameters are $ \sigma_s
=5,\,E_0/E_c=1 $. } \label{TUNNf}
\end{figure}
The integral is dominated by the region around $\E\sim 0$, where the tunneling
length is shortest [see Fig.~\ref{sauterf}] and tunneling probability is
largest. Both functions $G (0,{\mathcal E})$ and $G_\delta (0,{\mathcal E})$
have a symmetric peak at $\E=0$. Around the peak they can be expanded in powers
of $ {\mathcal E}$  as
\begin{eqnarray}
G (0,{\mathcal E}) & = & 2 [
\sigma_s^2 - \sigma_s(\sigma_s^2 -1)^{1/2}]+\frac{\sigma_s}{(\sigma_s^2 -1)^{3/2}}
{\E^2}
 + {\mathcal O}(\E^4)= \nonumber \\
& = &
G_0( \sigma_s )+ \frac{1}{2}G_2( \sigma_s )\,
{\E^2} +{\mathcal O}(\E^4)
,
\label{exppsgf}
\end{eqnarray}
and
\begin{eqnarray}
G (0,{\mathcal E})+
G_ \delta (0,{\mathcal E})
& = &
\frac{\sigma_s}{ ( \sigma_s ^2 -1)^{1/2}}+ \frac{1}{2}
\frac{(1 +2 \sigma_s^2)}{(\sigma_s ^2 -1)^{5/2}}
{\E^2} +{\mathcal O}(\E^4)
= \nonumber \\
& = &
\overline G_0( \sigma_s )+ \frac{1}{2}\overline G_2( \sigma_s )\,
{\E^2} +{\mathcal O}(\E^4) .
\label{exppgsgf}
\end{eqnarray}
The exponential $e^{-\lfrac{\pi G(0,{\mathcal E}) E_c}{E_0}}$ has a
Gaussian peak around $\E=0$ whose width is of the order of
$1/E_c\propto \hbar$. This implies that in the semi-classical limit,
one may perform only a Gaussian integral and neglect the
$\E$-dependence of the prefactor in (\ref{pgwk1}). Recalling that
$\E$ in this section is in natural units with $m_ec^2=1$, one should
replace $\int d\E$ by $ m_ec^2 \int d\E$ and can perform the
integral over $\E$ approximately as follows
\begin{eqnarray}\!\!\!\! \!\!
\frac{\Gamma _{\rm JWKB}}{V_\perp}
& \!\simeq \! &
 \Ds
\frac{
%|
eE_0
%|
m_ec^2 }{4\pi^2\hbar c}
\frac{1}{\overline G_0}
e^{-{\pi (E_c/E_0) G_0 }}
\!\!\int\! \frac{d\E}{2\pi \hbar }
e^{-{\pi (E_c/E_0)G''_0\, }\E^2/{2}}\!=\! \nonumber \\
& \!=\! & \Ds
\frac{
%|
eE_0
%|
}{4\pi^2\hbar c}
\frac{1}{\overline G_0} \frac{e^{-{\pi(E_c/E_0) G_0 }}}{
2\pi \hbar  \sqrt{G_0''E_c/2E_0} }
.
\label{pgwk4}
\end{eqnarray}
For convenience, the limits of integration over $E$ is extended from
the interval $(-1+ \sigma_s ,1- \sigma_s )$ to $(-\infty,\infty)$.
This introduces exponentially small errors and can be ignored.

Using the relation (\ref{@gamm}) one may replace $
%|
eE_0
%|
m_ec^2/\hbar c$ by $ e^2 E_0^22\ell / \sigma_s $,
and obtain
\begin{eqnarray}
\frac{\Gamma _{\rm JWKB}[\rm total]}{V_\perp \ell }
\simeq \Ds\frac{\alpha E^{2}_0 }{2\pi^2\hbar}
 \sqrt{\frac{E_0}{E_c}}\frac{(\sigma_s^2-1)^{5/4}}{\sigma_s^{5/2}}
e^{-\lfrac{\pi G_0( \sigma_s ) E_c}{E_0}}
.
\label{expgwkb1}
\end{eqnarray}
This approximate result agrees \footnote{See Eq.~(63) of Dunne and
Schubert \cite{2005PhRvD..72j5004D}, and replace there $\tilde
\gamma \rightarrow 1/ \sigma_s $. It agrees also with the later
paper by Dunne et al. \cite{2006PhRvD..73f5028D} apart from a factor
2.} with that obtained before with a different, somewhat more
complicated technique proposed by Dunne and Schubert
\cite{2005PhRvD..72j5004D} after the fluctuation determinant was
calculated exactly in \cite{2006PhRvD..73f5028D} with the help of
the Gelfand--Yaglom method, see Section 2.2 in
Ref.~\cite{Kleinert2004}. The advantage of knowing the exact
fluctuation determinant could not, however, be fully exploited since
the remaining integral was calculated only in the saddle point
approximation. The rate (\ref{expgwkb1}) agrees  with the leading
term of the expansion (42) of Kim and Page
\cite{2007PhRvD..75d5013K}. Note that the higher expansion terms
calculated by the latter authors do not yet lead to proper higher
order results since they are only of type II and III in the list
after Eq. (\ref{3drate}). The terms of equal order in $\hbar$ in the
expansion (\ref{4.13}) of the solution of the Riccati equation are
still missing.

Using the transformation rules (\ref{trantable}), it is
straightforward to obtain the pair production rate of the Sauter
type of  electric field depending on time rather than space.
According to the transformation rules (\ref{trantable}), one has to
replace ${\ell}\rightarrow c\delta T$, where $\delta T$ is the
characteristic time over which the electric field acts---the analog
of $\ell$ in (\ref{sauterv}). Thus the field (\ref{sauterv}) becomes
\begin{eqnarray}
E(t)=E_0/{\rm cosh}^2\left({t}/\delta T\right),~~~~~\label{tsfield}
V(t)&=&- \tilde\sigma_s\, m_ec^2\tanh\left({t}/\delta T\right).
\label{tsauterv}
\end{eqnarray}
According to the same rules, one must also replace
$\sigma_s\rightarrow i\tilde\sigma_s$, where
\begin{equation}
\tilde\sigma_s\equiv
%|
eE_0
%|
\delta T /m_ec.
\label{t@gamm}
\end{equation}
This brings
$G_0(\sigma_s)$ of Eq. (\ref{sgf})
to the form
\begin{eqnarray}
G_0(\sigma_s)\rightarrow G^t_0(\tilde\sigma_s)=
2 [\tilde\sigma_s(\tilde\sigma_s^2 - 1)^{1/2}-\tilde\sigma_s^2],
\label{gt0}
\end{eqnarray}
and yields the pair production rate
\begin{eqnarray}
\frac{\Gamma^z _{\rm JWKB}[\rm total]}{V_\perp \delta T }
\simeq \Ds\frac{\alpha E^{2}_0 }{2\pi^2\hbar}
 \sqrt{\frac{E_0}{E_c}}\left(\frac{\tilde\sigma_s^2+1}{\tilde\sigma_s^2}\right)^{5/4}
e^{-\lfrac{\pi G^t_0( \tilde\sigma_s ) E_c}{E_0}},
\label{texpgwkb1}
\end{eqnarray}
where $\Gamma^z _{\rm JWKB}[\rm total]=\partial N_{\rm
JWKB}/\partial z$ is the number of pairs produced per unit thickness
in a spatial shell parallel to the $xy$-plane. This is in agreement
with Ref. \cite{2006PhRvD..73f5028D}.

Note also that the constant field result (\ref{xy2}) of Euler and
Heisenberg cannot be deduced from (\ref{expgwkb1}) by simply taking
the limit $\ell \rightarrow \infty$ as one might have expected. The
reason is that the saddle point approximation (\ref{pgwk4}) to the
integral (\ref{pgwk1}) becomes invalid in this limit. Indeed, if $
\ell \propto \sigma_s$ is large in Eqs.~(\ref{sgf}) and
(\ref{pgsgf}), these become
\begin{eqnarray}
G(0,{\mathcal E})\rightarrow
G (0,{\mathcal E})+G_\delta (0,{\mathcal E})
\rightarrow  \frac{1}{1-\E^2/ \sigma_s ^2},
\end{eqnarray}
and the integral in (\ref{pgwk1}) becomes approximately
\begin{eqnarray}
e^{-{\pi ( E_c/E_0)}}
\int _{- \sigma_s }^ {+\sigma_s} \frac{d\E}{2\pi \hbar }
\left({1-\E^2/ \sigma_s ^2}\right)
e^{-{\pi ( E_c/E_0)(\E^2/ \sigma_s ^2)}}
\label{@XXX}\end{eqnarray}
For not too large $ \ell \propto \sigma_s$, the integral can be evaluated
in the leading Gaussian approximation
\begin{eqnarray}
\int _{-\infty }^ \infty \frac{d\E}{2\pi \hbar }
e^{-{\pi ( E_c/E_0)(\E^2/ \sigma_s ^2)}}=\frac{1}{2\pi \hbar } \sqrt{\frac{E_0}{E_c}}\sigma_s,
\end{eqnarray}
corresponding
to the previous result (\ref{expgwkb1})
for large-$ \sigma_s $.
For a constant field, however, where the integrands becomes flat,
the Gaussian approximation is no longer applicable. Instead one must
first
set
 $ \sigma_s \rightarrow \infty $ in
the integrand of (\ref{@XXX}), making it constant.
Then the integral
(\ref{@XXX})
 becomes\footnote{This treatment is analogous to that of the translational degree of freedom
in instanton calculations in Section 17.3.1
of \cite{Kleinert2004} [see in particular Eq.~(17.112)].}
\begin{eqnarray}
e^{-{\pi ( E_c/E_0)}}2\sigma_s/2\pi \hbar =
 e^{-{\pi ( E_c/E_0)}}2\ell
%|
eE_0
%|
/m_ec^2\,2\pi \hbar .
\end{eqnarray}
Inserting this into (\ref{pgwk1}) one recovers the constant field
result (\ref{xy2}). One must replace $2\ell  $ by $L$ to comply with
the relation (\ref{@REL}) from which one obtains
\begin{eqnarray}
\int d\E=\int dz
%|
eE(z)
%|
=
%|
eE_0
%|
\int dz/\cosh^2(z/\ell )=2\ell
%|
eE_0
%|
=L
%|
eE_0
%|
.
\nonumber
\end{eqnarray}

In order to see the boundary effect on the pair production rate,
this section is closed with a comparison
%\mn{Remo's interesting to see boundary effect}
between pair production rates in the constant field (\ref{xy2}) and
Sauter field (\ref{expgwkb1}) for the same field strength $E_0$ in
the volume $V_\perp \ell $. The ratio $R_{\rm rate}$ of pair
production rates (\ref{expgwkb1}) and (\ref{xy2}) in the volume
$V_\perp \ell $ is defined as
\begin{eqnarray}
R_{\rm rate}= \sqrt{\frac{E_0}{E_c}}e^{\lfrac{\pi E_c}{E_0}}\frac{(\sigma_s^2-1)^{5/4}}{\sigma_s^{5/2}}
e^{-\lfrac{\pi G_0( \sigma_s ) E_c}{E_0}}.
\label{compcs}
\end{eqnarray}
The soft boundary of the Sauter field (\ref{sauterv}) reduces its
pair production rate with respect to the pair production rate
(\ref{xy2}) computed in a constant field of width $L=2\ell$. The
reduction is shown quantitatively in Fig.~\ref{rratef}, where curves
are plotted for the rates (\ref{xy2}) and (\ref{expgwkb1}), and and
for their ratio (\ref{compcs}) at $E_0=E_c$ and
$\sigma_s=\ell/\lambda_C$ [recall (\ref{@gamm})]. One can see that
the reduction is significant if the width of the field slab shrinks
to the size of a
 Compton wavelength
$ \lambda _C$.
\begin{figure}[!th]
\begin{center}
\includegraphics[width=12cm]{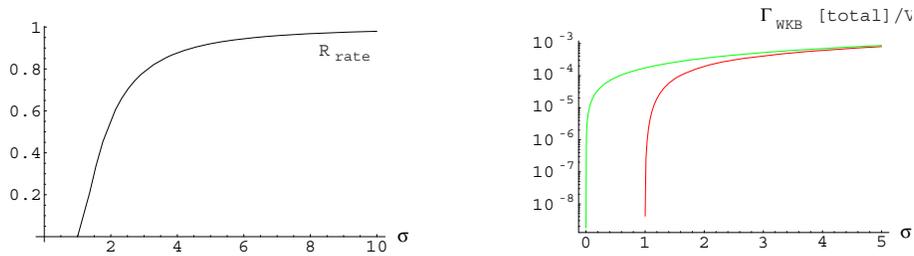}
\end{center}
\caption{Left: Ratio $R_{\rm rate}$
defined in  Eq.~(\ref{compcs}) is plotted as function of
$\sigma_s$ in the left figure.
Right:
Number of pairs created in slab
of Compton width per area and time as functions of $\sigma_s$.
Upper curve is
for the constant field (\ref{xy2}), lower
for the Sauter field (\ref{expgwkb1})).
Both plots
are for $E_0=E_c$ and $\sigma_s=\ell /\lambda_C$.
}
\label{rratef}%
\end{figure}

\subsubsection{Constant electric field in half space}

As a second application consider
an electric field
which is zero for $z<0$ and goes to $-E_0$
over a distance $\ell$
as follows:
\begin{eqnarray}
E(z)=-\frac{E_0}{2}\left[\tanh\left(\frac{z}{\ell}\right)+1\right],~~~\label{hfield}
V(z)= -\frac{\sigma_s}2 m_ec^2\left\{\ln\cosh\left(\frac{z}{\ell}\right)+\frac{z}{\ell}\right\},
\label{hv}
\end{eqnarray}
where $ \sigma_s \equiv eE_0\ell/m_ec^2$. In Fig.~\ref{halfspace},
the positive and negative energy spectra ${\mathcal E}_\pm(z)$
defined by Eq.~(\ref{energyl+-}) for $p_z=p_\perp=0$ are plotted to
show energy gap and level crossing.
\begin{figure}[!th]
\begin{center}
\includegraphics[width=12cm]{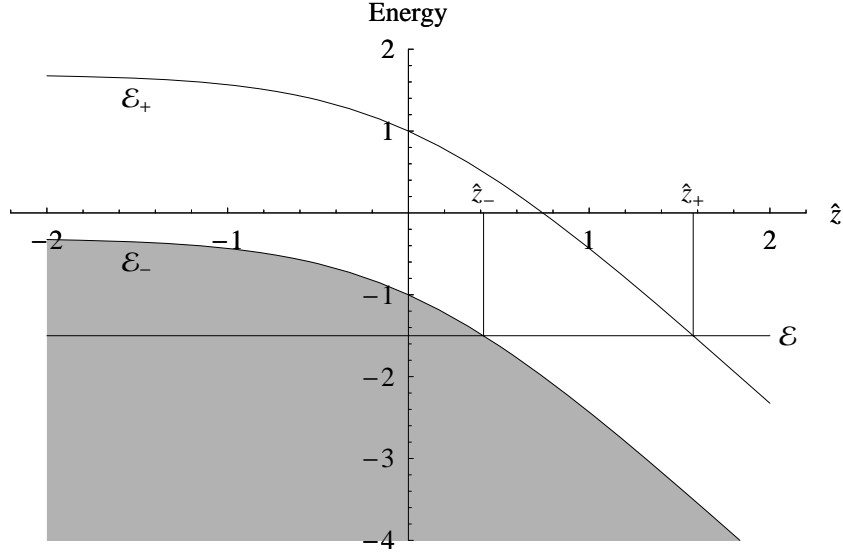}
\end{center}
\caption{Energies (\ref{energyl+-})
for a soft electric field step $E(z)$ of Eq.~(\ref{hfield}) and the potentials $V_\pm(z)$
(\ref{hv}) for $\sigma_s=5$.
Positive and negative-energies ${\mathcal E}_\pm(z)$
of Eq.~(\ref{energyl+-}) are plotted for $p_z= p_\perp =0$
as functions of $\hat z=z/\ell$. }%
\label{halfspace}%
\end{figure}
From Eq.~(\ref{crosspoint+-}) one finds now the
classical
turning points [instead of (\ref{scrossing})]
\begin{equation}
z_\pm = \frac{\ell}{2}\ln\left[2e^{({\mathcal E}\pm \sqrt{1+ \delta } )/\sigma_s}-1\right].
\label{hcrossing}
\end{equation}
For tunneling to take place,
the
 energy
${\mathcal E}$ has to satisfy
\begin{equation}
{\mathcal E} \le
 \sqrt{1+ \delta }
-\sigma_s\ln 2
,
\label{cregimeh}
\end{equation}
and $ \sigma_s $ must be larger than
$\sqrt{1+ \delta }\zeta $.
%Introducing the abbreviation
% $\theta(p_\perp,\E;\zeta)=(\hat\zeta -{\mathcal E})/\sigma_s$,
%this can be written as
%%
%\begin{equation}
%\theta\le \sigma_s \log 2.
%\end{equation}
%
Expressing $z/\ell$ in terms of $\zeta$
as
\begin{equation}
z=z(p_\perp,\E;\zeta)=
\frac{\ell}{2}\ln\left[2e^{({\mathcal E}+ \zeta \sqrt{1+ \delta } )/\sigma_s}
-1\right],
\label{hcrossing2}
\end{equation}
so that $z_\pm=z(p_\perp,\E;\pm1)$,
one finds
the electric field in the form
\begin{equation}
E(z)=E_0\left[1-{\frac{1}{2}}
e^{( \zeta\sqrt{1+ \delta }-\E)/ \sigma_s }\right]
\equiv
E(p_\perp,\E;\zeta)
%=E_0\left[1-e^{(\zeta \sqrt{1+ \delta }-1 )/ \sigma_s  }y\right],~~~~ y\equiv
%e^{(1-\E - \sigma_s \log 2)/ \sigma_s }
.
\label{hey}
\end{equation}
Inserting this into Eq.~(\ref{gf})
and expanding
$E_0/E(p_\perp,\E;\zeta)$ in
 powers
one obtains
\begin{eqnarray}
G (p_\perp,{\mathcal E})&=&
1+\sum_{n=1}^\infty \frac{e^{-n\E/ \sigma_s }}{2^n} \frac{2}{\pi}
\int^{1}_{-1}
 d\zeta \sqrt{1-\zeta^2}\,e^{n \hat\zeta/ \sigma_s }
=\nonumber\\
&=&1+\sum_{n=1}^\infty e^{-n\E  / \sigma_s }
I_1(n\sqrt{1+ \delta } / \sigma_s ),
\end{eqnarray}
where $I_1(x)$
is a modified Bessel function.
Expanding
$I_1(n\sqrt{1+ \delta } / \sigma_s )$ in powers of $ \delta $:
\begin{eqnarray}
I_1(n\sqrt{1+ \delta } / \sigma_s )=
I_1(n / \sigma_s )+
(n/ 4\sigma_s )
[
I_0(n / \sigma_s )+
I_2(n / \sigma_s )] \delta +\dots~,
\end{eqnarray}
one can identify
\begin{eqnarray}
G (0,{\mathcal E})&\!\!\!=\!\!\!& 1+
\sum_{n=1}^\infty e^{-n\E  / \sigma_s }
I_1(n/ \sigma_s ),  \\  ~
G (0,{\mathcal E})+
G _ \delta (0,{\mathcal E})
&\!\!\!=\!\!\!&1+      \frac{1}{2}
\sum_{n=1}^\infty e^{-n\E  / \sigma_s } [
(n/ \sigma_s )I_0(n/ \sigma_s )
-I_1(n/ \sigma_s )].
\end{eqnarray}
The integral over $\E$ in
Eq.~(\ref{pgwk1})
starts at $\E_<=1- \sigma_s \log 2$ where the integrand
rises from  0 to  $1$ as $\E$ exceeds a few units of $ \sigma_s $.
The derivative
of $e^{-\pi(E_c/E_0)G(0,\E)}$  drops from $1$  to
 $e^{-\pi(E_c/E_0)}$  over this  interval.
Hence the derivative
$\partial _\E
e^{-\pi(E_c/E_0)G(0,\E)}$ is  peaked
around some value $\bar \E$.
Thus
the integral
$\int d\E e^{-\pi(E_c/E_0)G(0,\E)}$
is performed by parts as
\begin{equation}
 \int d\E e^{-\pi(E_c/E_0)G(0,\E)}
 =
\E e^{-\pi(E_c/E_0)G(0,\E)}\Big|_{\E_<}^\infty-
 \int d\E \,\E\,\partial _\E e^{-\pi(E_c/E_0)G(0,\E)}.
 \label{hfc}
\end{equation}
The first term can be rewritten with the help of $d\E= eE_0dz$ as $
e^{-\pi(E_c/E_0)}|eE_0|\ell/2$, thus giving rise to the decay rate
(\ref{xy2}) in the volume $V_\perp \ell/2$ , and the second term
gives only a small correction to this. The second term in
Eq.~(\ref{hfc}) shows that the boundary effects reduce the pair
production rate compared with the pair production rate (\ref{xy2})
in the constant field without any boundary.
%\mn{Remo is interesting to see boundary effect}

\subsection{Tunneling into bound States}\label{tunnelingsec}

We turn now to the case in which
%It is interesting to see
%what happens if
instead of an outgoing wave as given by (\ref{WKBout}) there is a
bound state. A linearly rising electric field whose potential is
harmonic is considered:
\begin{eqnarray}
\label{hav}
E(z)=E_0\left(\frac{z}{\lambda_C}\right),~~~\label{hafield}
V(z)= \frac{e  E_0\lambda _C}{2}\left(\frac{z}{\lambda_C}\right)^2.
\end{eqnarray}
It will be convenient to parametrize the field strength $E_0$ in
terms of a dimensionless reduced electric field ${\epsilon}$ as
$E_0= \epsilon \hbar c/e\lambda_C^2=\epsilon E_c$. In
Fig.~\ref{hamo}, the positive and negative energy spectra ${\mathcal
E}_\pm(z)$ defined by Eq.~(\ref{energyl+-}) for $p_z=p_\perp=0$ are
plotted to show energy gap and level crossing for $\epsilon >0$
(left) and $\epsilon<0$ (right).
\begin{figure}[!th]
\begin{center}
\includegraphics[width=12cm]{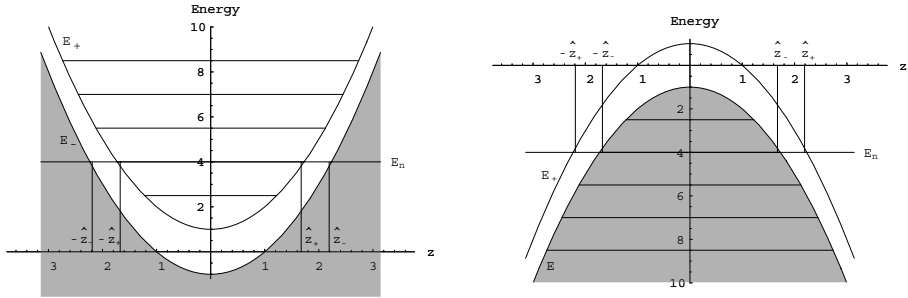}
\end{center}
\caption[]{
Positive- and negative-energy spectra ${\mathcal E}_\pm(z)$
of Eq.~(\ref{energyl+-}) for $p_z= p_\perp =0$
as a function of $\hat z\equiv z/\lambda_C$
for the linearly rising  electric field $E(z)$
with the harmonic potential (\ref{hav}). The
reduced field strengths are $ \epsilon =2$ (left figure) and
$ \epsilon =-2$ (right figure).
On the left, bound states
are filled and positrons escape to $z=\pm\infty$.
On the right,
bound electrons with negative energy
tunnel out of the well and escape with increasing energy to
$z=\pm\infty$.
}
\label{hamo}%
\end{figure}
If  $\epsilon$
is positive, Eq.~(\ref{crosspoint+-})
yields
for $z>0$,
\begin{equation}
z_\pm = \lambda_C \sqrt{ \frac{{2}}{\epsilon}}\left({\mathcal E}\mp \sqrt{1+ \delta } \right)^{1/2},
~~~~
z_+ < z_-,\
\label{@LHS}\end{equation}
 and mirror-reflected
turning points for $z<0$, obtained by exchanging $z_\pm \rightarrow
-z_\pm $ in (\ref{@LHS}). Negative energy electrons tunnel into the
potential well $-z_+ < z < + z_+$, where $ {\mathcal E} \ge \E_+ $,
forming bound states. The associated  positrons run off to infinity.
%\mn{I still do not understand the directions of electron and positron moving}

\subsubsection{JWKB transmission probability}

Due to the physical application to be discussed in the next section, here only
the tunneling process for $\epsilon > 0$ on the left-hand side of
Fig.~\ref{hamo} will be studied. One can consider the regime $z<0$ with the
turning pints $-z_-<-z_+$. The incident wave and flux for $z<-z_-$ pointing in
the positive $z$-direction are given by Eqs.~(\ref{WKBin}) and (\ref{influx}).
The wave function for $-z_-<z<-z_+$ has the form Eq.~(\ref{WKBt}) with the
replacement $z_-\rightarrow -z_-$. The transmitted wave is now no longer freely
propagating as in (\ref{WKBout}), but describes a bound state
of a positive energy electron: %\mn{check}
\begin{eqnarray}
\phi_{\E_n}(z)=\frac{\mathcal B}{(p_z)^{1/2}}\cos\left[ \frac{1}{\hbar}\int^z_{-z_+}p_zdz - \frac{\pi}{4}\right].
\label{WKBbo}
\end{eqnarray}
The Sommerfeld quantization condition
\begin{eqnarray}
\frac{1}{\hbar}\int_{-z_+}^{+z_+}p_z dz = \pi (n+\frac{1}{2}),~~~n=0,1,2,\dots~.
\label{quan}
\end{eqnarray}
fixes the
energies  $\E_n$. The connection
rules
for the wave functions (\ref{WKBt}) and (\ref{WKBbo})
at the turning point
 $-z_+$ determine
\begin{eqnarray}
{\mathcal B}=\sqrt{2}{\mathcal C}_+ e^{-i\pi n}\exp \left[ -\frac{1}{\hbar}\int_{-z_-}^{-z_+}\kappa_z dz\right].
\label{outbo}
\end{eqnarray}
Assuming the states $\phi_{\E_n}(z)$ to
 be initially unoccupied, the transmitted flux
to these states at the classical turning point $-z_+$ is
\begin{eqnarray}
\frac{\hbar}{m_e}\phi_{\E_n}(z)
\partial_z\phi^*_{\E_n}(z)\Big|_{z\rightarrow -z_+}=
\frac{|{\mathcal B}|^2}{(2m_e)}
=\frac{|{\mathcal C}_+|^2}{m_e}
\exp \left[ -\frac{2}{\hbar}\int_{-z_-}^{-z_+}\kappa_z dz\right].
\label{fluxbo}
\end{eqnarray}
From Eqs.~(\ref{influx}), (\ref{fluxbo}), and (\ref{wdefine}) one then finds
the JWKB transmission probability for positrons to fill these bound states
leaving a positron outside:
\begin{eqnarray}
{\mathcal P}_{\rm JWKB}(p_\perp,\E_n)=
\exp \left[ -\frac{2}{\hbar}\int_{-z_-}^{-z_+}\kappa_z dz\right]
%=\exp \left[ -\frac{2}{\hbar}\int^{z_-}_{z_+}\kappa_z dz\right]
,
\label{wdefine1}
\end{eqnarray}
which has the same form as Eq.~(\ref{tprobability2}). The same result is
obtained once more for $z>0$ with turning points $z_+<z_-$, which can be
obtained from (\ref{wdefine1}) by the mirror reflection $-z_\pm\leftrightarrow
z_\pm$.

\subsubsection{Energy spectrum of bound states}

From Eq.~(\ref{WKBr2}) for $p_z$ and Eq.~(\ref{hafield}) for the potential $V(z)$,
the eikonal (\ref{quan}) is calculated to determine the
energy spectrum $\E_n$ of bound states
\begin{eqnarray}
\frac{1}{\hbar}\int_{-z_+}^{+z_+}p_z dz &=& 2\frac{\epsilon}{\lambda_C^3}\int_{0}^{z_+}\left[(z^2-z_+^2)(z^2-z_-^2) \right]^{1/2} dz\nonumber\\
&=&\frac{2\epsilon z_+}{3\lambda_C^3}\left[(z_+^2+z_-^2)E(t)-(z_-^2-z_+^2)K(t)\right],
\label{quan1}
\end{eqnarray}
where $E(t)$, $K(t)$ are complete
elliptical integrals of the first and second kind, respectively, and $t\equiv z_+/z_-$.
The Sommerfeld quantization rule (\ref{quan}) becomes
\begin{eqnarray}
&&\frac{8}{3}\left[\frac{2({\mathcal E}_n- \sqrt{1+ \delta })}{\epsilon} \right]^{1/2}\left[\E_n E(t_n)-(\sqrt{1+\delta})K(t_n)\right]=\pi (n+\frac{1}{2}),\label{quan2} \\
&&t_n\equiv \left(\frac{{\mathcal E}_n- \sqrt{1+ \delta }}{{\mathcal E}_n + \sqrt{1+ \delta }} \right)^{1/2}.
\nonumber
\end{eqnarray}
For
any given
transverse momentum
$p_\perp= \sqrt{ \delta } $, this determines
the discrete energies
 $\E_n$.

\subsubsection{Rate of pair production}

By analogy
with Eqs.~(\ref{gxy0}) and (\ref{pgwk1}),
the transmission
probability (\ref{wdefine1}) must now be
integrated
over all
incident
particles with the flux (\ref{nflux})
to yield
the rate of pair production:
\begin{eqnarray}
\frac{\Gamma _{\rm JWKB}}{V_\perp}
&\!\!=\!\!&
2\Ds \sum_n\frac{ \omega_n}{2\pi}
\int\frac{d^2{p}_\perp}{(2\pi\hbar)^2}
{\mathcal P}_{\rm JWKB}(p_\perp,{\mathcal E}_n),\label{gwkbpb0}\\
&\!\!\approx\!\! &
2\Ds
\frac{|eE_0 |}{4\pi^2\hbar c}
\sum_n \frac{\omega_n}{2\pi}
\frac{1}{
G(0,{\mathcal E}_n)
+G_ \delta (0,{\mathcal E}_n)
}
%\left[1+\frac{\hbar}{2\pi^2 c[G(0,{\mathcal E}_-)-G_ \delta (0,{\mathcal E}_-)
%]}\frac{E_0 E'(z)}{E^2(z)}\right]
e^{-{\pi ( E_c/E_0)G(0,{\mathcal E}_n)}}.
\label{gwkbpb1}
\end{eqnarray}
In obtaining these expressions one has used the  energy conservation law to
perform the integral over $\E$. This receives only contributions for $\E=\E_n$
where  $\int d\E=\omega_n\hbar\equiv \E_n-\E_{n-1}$. The factor $2$ accounts
for the equal contributions from the two regimes $z>0$ and $z<0$. The previous
relation (\ref{@REL}) is now replaced by
\begin{equation}
\omega_n\hbar = |eE(z_-^n)|\Delta z_-^n.
%\label{@}
\end{equation}
Using Eq.~(\ref{y(x)}) and expressing $z/\lambda_C >0$ in terms of $\zeta$
as
\begin{equation}
z=z(p_\perp,\E_n;\zeta)=
\lambda_C \sqrt{\frac{{2}}{{\epsilon}}}\left({\mathcal E}_n-\zeta \sqrt{1+ \delta } \right)^{1/2},
\label{hacrossing2}
\end{equation}
one calculates $z_\pm=z(p_\perp,\E_n;\pm1)$,
and find
the electric field in the form
\begin{equation}
E(z)=E_0 \sqrt{\frac{{2}}{{\epsilon}}}\left({\mathcal E}_n-\zeta \sqrt{1+ \delta } \right)^{1/2}
\equiv
E(p_\perp,\E_n;\zeta)
%=E_0\left[1-e^{(\zeta \sqrt{1+ \delta }-1 )/ \sigma_s  }y\right],~~~~ y\equiv
%e^{(1-\E_n - \sigma_s \log 2)/ \sigma_s }
.
\label{hmey}
\end{equation}
Inserting this into Eq.~(\ref{gf}) one obtains
\begin{eqnarray}
G (p_\perp,{\mathcal E}_n)&=&\frac{2}{\pi}
 \sqrt{\frac{{\epsilon}}{{2}}}\int_{-1}^1d\zeta\frac{\sqrt{1-\zeta^2}}{[{\mathcal E}_n-\zeta \sqrt{1+ \delta }]^{1/2}},\nonumber\\
&=&
\frac{8}{3\pi}\sqrt{\frac{{\epsilon}}{{2}}}\frac{({\mathcal E}^\delta_n+1)^{1/2}}{(1+ \delta )^{1/4}}\left[(1-{\mathcal E}^\delta_n)
K (q^ \delta _n)+{\mathcal E}^\delta_nE(q^\delta_n)\right]
\label{hagf}
\end{eqnarray}
where ${\mathcal E}^\delta_n\equiv {\mathcal E}_n/(1+\delta)^{1/2}$
and $q^\delta_n=\sqrt{2/({\mathcal E}^\delta_n+1)}$.
Expanding
$G (p_\perp,{\mathcal E}_n)$ in powers of $\delta$ one finds
the zeroth order term
\begin{eqnarray}
G (0,{\mathcal E}_n)&=&
\frac{8}{3\pi}\sqrt{\frac{{\epsilon}}{{2}}}({\mathcal E}_n+1)^{1/2}
\left[(1-{\mathcal E}_n)K(q_n)+{\mathcal E}_nE(q_n)\right]
\label{hagf0}
\end{eqnarray}
and
the derivative
\begin{eqnarray} \!\!\!\!\!
G_\delta (0,{\mathcal E}_n)&=&
\frac{\sqrt{\epsilon}}{3\pi}\frac{q_n}{{\mathcal E}_n}{q_n-1}\left[(4-5q_n
+{\mathcal E}_n(7-6q_n))E(q_n)+\right.\nonumber\\
&+&\left.(1-{\mathcal E}_n-7{\mathcal E}_n^2)
(q_n-1)K(q_n)\right].
\label{hagf1}
\end{eqnarray}
where $q_n\equiv \sqrt{2/({\mathcal E}_n+1)}$.

%%%%%%%%%%%%%%%%%%%%%%%%%%%%%%%%%%%%%%%%%%%%%%%%%%%%%%%%%%%%%%%%%%%%%%%%%%%%%%%%%%%%%%%%%%%%

\section{Phenomenology of electron--positron pair creation and annihilation}
\label{chap-pair-application}

\subsection{$e^+ e^-$ annihilation experiments in particle physics}
\label{e+e-phys}

The $e^+ e^-\rightarrow \gamma+\gamma$ process predicted by Dirac
was almost immediately observed \cite{1934PCPS...30..347K}. The
$e^+e^-$ annihilation experiments have since became possibly the
most prolific field of research in the active domain of particle
physics. The Dirac pair annihilation process (\ref{ee2gamma}) has no
energy threshold and the energy release in the $e^+e^-$ collision is
larger than $2m_ec^2$. This process is the only one in the limit of
low energy. As the $e^+e^-$ energy increases the collision produces
not only photons through the Dirac process (\ref{process2}) but also
other particles. For early work in this direction, predicting
resonances for pions, K-mesons etc., see \cite{1961PhRv..124.1577C}.
Production of such particles are achievable and precisely conceived
in experimental particle physics, but hardly possible with the
vacuum polarization process {\bf}. In particular when the energy in
the center of mass is larger than twice muon mass $m_\mu$ about 100
times electron mass, the electron and positron electromagnetically
annihilate into two muons $e^+e^-\rightarrow \mu^+\mu^-$ via the
intermediation of a virtual photon. The cross-section in the center
of mass frame is given by \cite{1982PhR....81....1A}
\begin{equation}\label{muon}
\sigma_ {e^+e^-\rightarrow \mu^+\mu^-}=\frac{16\pi^2\alpha^2(\hbar c)^2}{q_{\rm cm}^2}
\,{\rm Im}\,\bar\omega_{\mu^+\mu^-}(q_{\rm cm}^2)
=\frac{4\pi\alpha^2(\hbar c)^2}{3q_{\rm cm}^2}
=\frac{86.8{\rm n b}}{q_{\rm cm}^2({\rm G e V})^2}
\end{equation}
where $\bar\omega_{\mu^+\mu^-}(q_{\rm cm}^2)$ is the muon part of
the vacuum polarization tensor and $q_{\rm cm}^2=c^2(p_++p_-)^2/4$
the square of energy of the center of mass frame, where $p_\pm$ are
4-momenta of positron and electron. At very high energy
$m_\mu^2/q_{\rm cm}^2\rightarrow 0$, ${\rm
Im}\,\bar\omega_{\mu^+\mu^-}(q_{\rm cm}^2)\rightarrow 1/(12\pi)$.

At very high energies of the order of several GeV, electron and
positron electromagnetically annihilate into hadrons, whose
cross-section has the same structure as the cross-section
(\ref{muon}) with $\bar\omega_{\mu^+\mu^-}(q_{\rm cm}^2)$ replaced
by the hadron part of the vacuum polarization tensor
$\bar\omega_{\rm hadrons}(q_{\rm cm}^2)$,
\begin{equation}\label{hadron}
\sigma_ {e^+e^-\rightarrow {\rm hadron}}=\frac{16\pi^2\alpha^2(\hbar c)^2}{q_{\rm cm}^2}
\,{\rm Im}\,\bar\omega_{\rm hardrons}(q_{\rm cm}^2).
\end{equation}
The two cross-sections (\ref{muon}) and (\ref{hadron}) are
comparable, of the order of a few tens of nanobarns
($10^{-33}$cm$^2$). It is traditional to call $R$ the ratio of
hadronic to electromagnetic annihilation cross-sections
\cite{1982PhR....81....1A},
\begin{equation}\label{Rhadron}
R(q_{\rm cm}^2)=\frac{\sigma_ {e^+e^-\rightarrow {\rm hadron}}}
{\sigma_{e^+e^-\rightarrow \mu^+\mu^-}}=12\pi{\rm Im}\,\bar\omega_{\rm hadrons}(q_{\rm cm}^2).
\end{equation}
As the energy $q_{\rm cm}^2$ of electron and positron collision increases and
reaches the mass--energy thresholds of constituents of hadrons, i.e.
``quarks'', narrow resonances occurs, see Fig. \ref{resonances} for the ratio
$R$ as a function of $\sqrt{q^2}$ measured at SLAC \cite{1976ARNPS..26...89S}.
These resonances correspond to production of particles such as $J/\psi$,
$\Upsilon$ etc.
\begin{figure}[!th]
\begin{center}
\includegraphics[width=10cm]{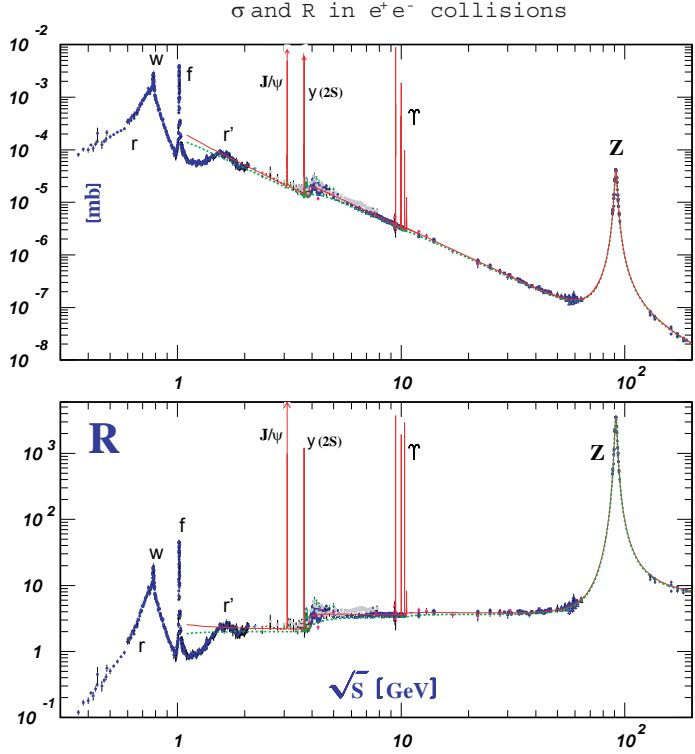}
\end{center}
\caption{The total cross section of $e^+e^-\rightarrow {\rm hadron}$
(\ref{hadron}) and the ratio $R=\sigma_ {e^+e^-\rightarrow {\rm
hadron}}/\sigma_{e^+e^-\rightarrow \mu^+\mu^-}$ (\ref{Rhadron}), where $s\equiv
q_{\rm cm}^2$. Reproduced from Ref. [C. Amsler et al. (Particle Data Group),
Physics Letters B667, 1 (2008)].} \label{resonances}
\end{figure}
This provides a fruitful investigation of hadron physics. For a review of this
topic see \cite{1975PhR....15..181B}.

As the center of mass energy $q_{\rm cm}^2$ reaches the electroweak scale
(several hundred GeVs), electron and positron annihilation process probes a
rich domain of investigating electroweak physics, see for instance
Refs.~\cite{1996ARNPS..46..533M,Barklow1997}. Recent experiments on $e^+e^-$
collisions at LEP, SLAC and the Tevatron allowed precision tests of the
electroweak Standard Model. In \cite{1999PhR...322..125A,2004PhR...403..189A}
the results of these precision tests together with implications on parameters,
in particular Higgs boson mass, as well as constraints for possible new physics
effects are discussed.

Electron and positron collisions are used to produce many particles in the
laboratory, which live too short to occur naturally. Several electron--positron
colliders have been built and proposed for this purpose all over the world
(CERN, SLAC, INP, DESY, KEK and IHEP), since the first electron--positron
collider the ``Anello d'Accumulazione'' (AdA) was built by the theoretical
proposal of Bruno Touschek in Frascati (Rome) in 1960
\cite{2004PhP.....6..156B}. Following the success of AdA (luminosity $\sim
10^{25}$/(cm$^2$ sec), beam energy $\sim$0.25GeV), it was decided to build in
the Frascati National Laboratory a storage ring of the same kind, Adone and
then Daphne (luminosity $\sim 10^{33}$/(cm$^2$sec), beam energy $\sim$0.51GeV),
with the aim of exploring the new energy range in subnuclear physics opened by
the possibility of observing particle-antiparticle interactions with center of
mass at rest. The biggest of all is CERN's Large Electron Positron (LEP)
collider \cite{L3CERN}, which began operation in the summer of 1989 and have
reached a maximal collision energy of 206.5 GeV. The detectors around the LEP
ring have been able to perform precise experiments, testing and extending our
knowledge of particles and their strong, electromagnetic and weak interactions,
as described by the Standard Model for elementary particle physics.

All these clearly show how the study of $e^+e^-$ reaction introduced by Dirac
have grown to be one of the most prolific field in particle physics and have
received remarkable verification in energies up to TeV in a succession of
machines increasing in energy.

\subsection{The Breit--Wheeler process in laser physics}\label{Xray}

While the Dirac process (\ref{ee2gamma}) has been by far one of the most
prolific in physics, the Breit--Wheeler process (\ref{2gammaee}) has been one
of the most elusive for direct observations. In Earth-bound experiments the
major effort today is directed to evidence this phenomenon in very strong and
coherent electromagnetic field in lasers. In this process collision of many
photons may lead in the future to pair creation. This topic is discussed in the
following Sections. Alternative evidence for the Breit--Wheeler process can
come from optically thick electron--positron plasma which may be created in the
future either in Earth-bound experiments, or currently observed in
astrophysics, see Section \ref{aksenov}. One additional way to probe the
existence of the Breit--Wheeler process is by establishing in astrophysics an
upper limits to observable high-energy photons, as a function of distance,
propagating in the Universe as pioneered by Nikishov \cite{Nikishov1961}, see
Section \ref{BWastro}.

We first briefly discuss the phenomenon of electron--positron pair production
at the focus of an X-ray free electron laser, as given in the review articles
\cite{2001PhLB..510..107R,2001hep.ph...12254R,2003hep.ph....4139R}. In the
early 1970's, the question was raised whether intense laser beams could be used
to produce a very strong electric field by focusing the laser beam onto a small
spot of size of the laser wavelength $\lambda$, so as to possibly study
electron--positron pair production in vacuum
\cite{1970SPhD...14..678B,1970PhRvD...2.1191B}. However, it was found that the
power density of all available and conceivable optical lasers
\cite{1994Sci...264..917P} is too small to have a sizable pair production rate
for observations \cite{1970SPhD...14..678B, 1970PhRvD...2.1191B,
1971ZhPmR..13..261P, 1972JETP...34..709P, 2001JETPL..74..133P,
1972PhRvD...6.2299T, 1972JETP...35..659P, 1973JETPL..17..368M, Marinov1977,
1974JETP...38..427N, 1973ZhPmR..18..435P, Mostepanenko:1974im, Popov1974}, since the wavelength of optical lasers and the size of focusing spot are too large to have a strong enough electric field.

Definite projects for the construction of X-ray free electron lasers (XFEL) have been
%%ar
set up at both DESY and SLAC. Both are based on the principle pioneered by John
Madey \cite{1977PhRvL..38..892D} of self-amplified spontaneous emission in an
undulator, which results when charges interact with the synchrotron radiation
they emit \cite{2002PhRvL..88t4801T}. At DESY the project is called XFEL and is
part of the design of the electron--positron collider TESLA
\cite{Materlik2008,Brinkmann:1997ws,Materlik1999,2004IJMPA..19.5097B}
%%ar
but is now being build as a stand-alone facility. At SLAC the
project so-called Linac Coherent Light Source (LCLS) has been
proposed \cite{Arthur1998,Lindau1999,lcls2008}. It has been pointed
out by several authors
\cite{Melissinos1999,Chen1998,1999PhRvL..83..256C,2003PlPhR..29..207T}
that having at hand an X-ray free electron laser, the experimental
study and application of strong field physics turn out to be
possible. One will use not only the strong energy and transverse
coherence of the X-ray laser beam, but also focus it onto a small
spot hopefully with the size of the X-ray laser wavelength $\lambda
\simeq O(0.1)$nm \cite{Nuhn2000}, and obtain a very large electric
field $ E\sim 1/\lambda$, much larger than those obtainable with any
optical laser of the same power.

Using the X-ray laser, we can hopefully achieve a very strong electric field
near to its critical value for observable electron--positron pair production in
vacuum. Electron-positron pair production at the focus of an X-ray laser has
been discussed in Ref.~\cite{Melissinos1999}, and an estimate of the
corresponding production rate has been presented in Ref.~\cite{Chen1998}. In
Ref.~\cite{2001PhLB..510..107R,2001hep.ph...12254R,2003hep.ph....4139R}, the
critical laser parameters, such as the laser power and focus spot size, are
determined in order for achieving an observable effect of pair production in
vacuum.

The electric field produced by a single laser beam is the light-like
static, spatially uniform electromagnetic field, and field
invariants $S$ and $P$ (\ref{lightlike}) vanish
\cite{1972PhRvD...6.2299T}
\begin{equation}
S= 0,\qquad P=0, \label{lightlike0}
\end{equation}
leading to $\varepsilon=\beta=0$ and no pair production
\cite{1951PhRv...82..664S, 1954PhRv...93..615S, 1954PhRv...94.1362S}, this can
be seen from Eqs.~(\ref{lightlike}), (\ref{wkbehfermion2}) and
(\ref{probabilityeh}). It is then required that two or more coherent laser
beams form a standing wave at their intersection spot, where a strong electric
field can hypothetically be produced without magnetic field.

We assume that each X-ray laser pulse is split into two equal parts and
recombined to form a standing wave with a frequency $\omega$, whose
electromagnetic fields are then given by
\begin{equation}
{\bf  E}(t)=[0,0,E_{\rm peak}\cos (\omega t)], \hskip0.3cm {\bf B}(t) =(0,0,0),
\label{oscillationfield}
\end{equation}
where the peak electric field is\cite{2001PhLB..510..107R, 2001hep.ph...12254R,
2003hep.ph....4139R},
\begin{equation}
E_{\rm peak}=\sqrt{\frac{P_{\rm laser}}{ \pi\sigma_{\rm laser}^2}}\simeq 1.1\cdot 10^{17}\left(\frac{P_{\rm laser}}{ 1 TW}\right)^{1/2}\left(\frac{0.1{\rm nm}}{ \sigma_{\rm laser}}\right)
\frac{V}{ {\rm m}},
\label{xfieldvalue}
\end{equation}
as expressed in terms of the laser power $P_{\rm laser}$ (1 TW=$10^{12}$W),
with the focus spot radius $\sigma_{\rm laser}$. Eq.~(\ref{xfieldvalue}) shows
that the peak electric energy density $E^2_{\rm peak}/2$ is created in a spot
of area $\pi\sigma_{\rm laser}^2$ by an X-ray laser of power $P_{\rm laser}$.
The laser beam intensity on the focused spot is then given by
\begin{equation}
I_{\rm laser}=\frac{P_{\rm laser}}{\pi\sigma_{\rm laser}^2} \simeq \frac{c}{4\pi}E_{\rm peak}^2.
\nonumber
\end{equation}
For a laser pulse with wavelength $\lambda$ about $1\mu m$ and the theoretical
diffraction limit $\sigma_{\rm laser}\simeq \lambda$ being reached, the
critical intensity laser beam can be defined as,
\begin{equation}
I^c_{\rm laser}=\frac{c}{ 4\pi} E_c^2 \simeq 4.6\cdot 10^{29}W/{\rm cm}^2,
\label{criticaldensity}
\end{equation}
which corresponds to the peak electric field approximately equal to
the critical value $E_c$ in (\ref{critical1}).

\subsubsection{Phenomenology of pair production in alternating fields}\label{alternatingfield}

To compute pair production rate in an alternating electric field
(\ref{oscillationfield}) of laser wave in a semi-classical manner, one assumes
the conditions that the peak electric field $E_{\rm peak}$ is much smaller than
the critical field $E_c$ (\ref{critical1}) and the energy $\hbar\omega$ of the
laser photons is much smaller than the rest energy of electron $m_ec^2$,
\begin{equation}
E_{\rm peak}\ll E_c,\quad \hbar\omega \ll m_ec^2. \label{condition}
\end{equation}
These conditions are well satisfied at realistic optical as well as X-ray
lasers \cite{2001PhLB..510..107R, 2001hep.ph...12254R, 2003hep.ph....4139R}.

The phenomenon of electron--positron pair production in alternating electric
fields was studied in Refs.~\cite{1970PhRvD...2.1191B, 1971ZhPmR..13..261P,
1972JETP...34..709P, 2001JETPL..74..133P, 1972JETP...35..659P,
1973JETPL..17..368M, Marinov1977, 1974JETP...38..427N, 1973ZhPmR..18..435P,
Mostepanenko:1974im, Popov1974, 1998PhRvD..58j5022D}. By using generalized JWKB method
\cite{1970PhRvD...2.1191B} and imaginary time method \cite{1971ZhPmR..13..261P,
1972JETP...34..709P, 2001JETPL..74..133P, 1973ZhPmR..18..435P, Popov1974} the
rate of pair production was computed. In Ref.~\cite{1970PhRvD...2.1191B}, the
rate of pair production was estimated to be (see Section~\ref{alternating}),
\begin{equation}
\tilde {\mathcal P}=\frac{c}{ 4\pi^3\lambda_C^4}\left(\frac{E_{\rm peak}}{ E_c}\right)^2
\frac{\pi}{ g(\eta)+\frac{1}{2\eta}g'(\eta)}
\exp\left[-\pi \frac{E_{\rm peak}}{ E_c}g(\eta)\right],
\label{orate}
\end{equation}
where the complex function $g(\eta)$ is given in
Refs.~\cite{1970PhRvD...2.1191B, 1971ZhPmR..13..261P, 1972JETP...34..709P,
2001JETPL..74..133P} (see Eq.~(\ref{xlasereta})),
\begin{equation}
g(\eta) = \frac{4}{\pi}\int_0^1du \Big[\frac{1-u^2}{ 1 + \eta^{-2} u^2}\Big]^{1/2}=\left\{\begin{array}{ll}
1-\frac{1}{ 8\eta^3 }+O(\eta^{-4}),
&\eta\gg 1\\
\frac{4\eta}{\pi}\ln\big(\frac{4}{ e\eta}\big) +O(\eta^3),& \eta\ll 1
\end{array}\right.
\label{xlasereta2}
\end{equation}
and the parameter $\eta$ is defined as the work done by the electric force $e
E_{\rm peak}$ in the Compton wavelength $\lambda_C$ in unit of laser photon
energy $\hbar\omega$,
\begin{equation}
\eta= \frac{eE_{\rm peak}\lambda_C}{ \hbar\omega}=
\frac{m_ec^2}{ \hbar\omega}\frac{E_{\rm peak}}{ E_c}.
\label{lasergamma}
\end{equation}
which is the same as $\eta$ in (\ref{bscattpro4}) in Section~\ref{alternating}
and agrees with its time-average (\ref{eta}) and (\ref{eta1}), over
one period of laser wave. The exponential factor in Eq.~(\ref{orate}) has been
confirmed by later works \cite{1971ZhPmR..13..261P, 1972JETP...34..709P,
2001JETPL..74..133P, 1973ZhPmR..18..435P, Popov1974}, which determine more
accurately the pre-exponential factor by taking into account also interference
effects. The parameter $\eta$ is related to the adiabaticity parameter
$\gamma=1/\eta$.

In the strong field and low frequency limit ($\eta\gg 1$), formula
(\ref{orate}) agrees to the Schwinger non-perturbative result
(\ref{probability1}) for a static and spatially uniform field, apart
from an ``inessential'' (see Ref.~\cite{1970PhRvD...2.1191B})
pre-exponential factor of $\pi$. This is similar to the adiabatic
approximation of a slowly varying electric field that we discuss in
Section~\ref{ad}. On the other hand, for $\eta\ll 1$, i.e. in high
frequency and weak field limit, Eq.~(\ref{orate}) yields
\cite{1970PhRvD...2.1191B},
\begin{equation}
\tilde {\mathcal P}\simeq \frac{c}{ 4\pi^3\lambda_C^4}\left(\frac{\hbar\omega}{
m_ec^2}\right)^2\frac{\pi\eta}{2\ln(4/\eta)}\left(\frac{e\eta}{ 4}\right)^{2\frac{2m_ec^2}{\hbar\omega}}
\left[1+O(\eta^2)\right],
\label{orateweak}
\end{equation}
corresponding to the $n$th order perturbative result, where $n$ is the minimum number
of quanta of laser field required to create an $e^+e^-$ pair:
\begin{equation}
n\gtrsim \frac{2m_ec^2}{ \hbar\omega }\gg 1.
\label{number}
\end{equation}
The pair production rate (\ref{orate}) interpolates analytically
between the adiabatic, non-perturbative tunneling mechanism
(\ref{probability1}) $(\eta \gg 1, \gamma\ll 1)$ and the
anti-adiabatic, perturbative multi-photon production mechanism
(\ref{orateweak}) $(\eta \ll 1, \gamma \gg 1)$.

In Refs.~\cite{1973ZhPmR..18..435P,Popov1974}, it was found that the
pair production rate, under the condition (\ref{condition}), can be
expressed as a sum of probabilities $w_n$ of many photon processes,
\begin{equation}
\tilde {\mathcal P}_p=\sum_{n>n_0}w_n, \hskip0.3cm {\rm with } \hskip0.3cm n_0=\frac{m_ec^2}{ \hbar\omega}\Delta_m,
\label{manyphoton}
\end{equation}
where $\Delta_m$ indicates an effective energy gap $m_ec^2\Delta_m$,
due to the transverse oscillations of the electron propagating in a
laser wave (see Section~\ref{lighttheoryquantum} and
Eq.~(\ref{ggprobability})). In the limiting cases of small and large
$\eta$, the result is given by \cite{Popov1974},
\begin{equation}
\tilde {\mathcal P}_p \simeq \frac{c}{4\pi^3\lambda_C^4}\left\{\begin{array}{ll}
\frac{\sqrt{2}}{\pi}\left(\frac{E_{\rm peak}}{ E_c}\right)^{5/2}\exp\Big[-\pi\left(\frac{E_c}{ E_{\rm peak}}\right)
\big(1-\frac{1}{8\eta^2} +O(\eta^{-4}\big)\Big],
&\eta\gg 1\\
\frac{\sqrt{2}}{2}\left(\frac{\hbar\omega}{ m_ec^2}\right)^{5/2}\sum_n
\left(\frac{e\eta}{4}\right)^{2n}
e^{-\phi} {\rm Erfi}\left(\phi^{1/2}\right),& \eta\ll 1,
\end{array}\right.
\label{wp}
\end{equation}
where $n>(2m_ec^2/\hbar\omega)$, $\phi=2(n-\frac{2m_ec^2}{\hbar\omega})$ and
Erfi$(x)$ is the imaginary error function \cite{1955htf..book.....B}. The range
of validity of results (\ref{orate}), (\ref{orateweak}), (\ref{manyphoton}) and
(\ref{wp}) is indicated by the conditions (\ref{condition}).

\subsubsection{Pair production in X-ray free electron lasers}\label{Xrayfree}

According to Eq.~(\ref{xfieldvalue}) for the electric field $E$ of an X-ray
laser, in order to obtain an observable effect of pair production we need to
have a large power $P$, a small laser focusing spot radius $\sigma_{\rm laser}$
and a long duration time $\Delta t$ of the coherent laser pulse. The power of
an X-ray free electron laser is limited by the current laser technology. The
focusing spot radii $\sigma_{\rm laser}$ are limited by the diffraction to the
order of the X-ray laser beam wavelength. In Ref.~\cite{2001PhLB..510..107R,
2001hep.ph...12254R, 2003hep.ph....4139R}, it was estimated that to produce at
least one pair of electron and positron, we need the minimum power of the X-ray
laser to be $\sim 2.5-4.5$TW corresponding to an electric field of $\sim
1.7-2.3\cdot 10^{15}V/{\rm cm}\sim 0.1 E_c$, provided the laser wavelength is
$\lambda \sim 0.1$nm and the theoretical diffraction limit $\sigma_{\rm
laser}\simeq \lambda$ is actually reached and the laser coherent duration time
$\Delta t\sim 10^{-(13\sim 16)}$ second. Based on these estimations, Ringwald
concluded \cite{2001PhLB..510..107R, 2001hep.ph...12254R, 2003hep.ph....4139R}
that with present available techniques, the power density and electric fields
of X-ray laser are far too small to produce a sizable
Sauter-Euler-Heisenberg-Schwinger effect. If the techniques for X-ray free
electron laser are considerably improved, so that the XFEL power can reach the
terawatt regime and the focusing spot radii can reach the theoretical
diffraction limit, we will still have the possibility of investigating the
Sauter-Euler-Heisenberg-Schwinger phenomenon by a future XFEL.

\subsubsection{Pair production by a circularly polarized laser beam}\label{circpolar}

Instead of a time varying electric field (\ref{oscillationfield}) that is
created by an intersection of more than two coherent laser beams, it was
suggested \cite{2004PhRvE..69c6408B, 2004PhLA..330....1N} to use a focused
circularly polarized laser beam having nonvanishing field invariants $S,P$
(\ref{lightlike}) and strong electromagnetic fields ${\bf E,B}$ for pair
production. It is well known that the electromagnetic field of a focused light
beam is not transverse, however, one can always represent the field of a
focused beam as a superposition of fields with transverse either electric
($e$-polarized) field or magnetic ($h$-polarized) field only, see e.g.,
\cite{1980poet.book.....B}.

In Ref.~\cite{2004PhLA..330....1N},  the $e$-polarized electric and
magnetic fields (${\bf E}^e, {\bf B}^e $) propagating in the $\hat
{\bf z}$-direction is described by the following exact solution of
Maxwell equations \cite{2000JETP...90..753N},
\begin{eqnarray}
{\bf E}^e &=&iE_{\rm peak}e^{-i\psi}\left[F_1({\bf e}_x\pm {\bf e}_y)-F_2e^{\pm 2i\phi}
({\bf e}_x\mp {\bf e}_y)\right];
\label{ecircular}\\
{\bf B}^e &=& \pm E_{\rm peak}e^{-i\psi}\Big\{\left( 1-i\delta^2\frac{\partial}{\partial\chi}\right)\nonumber\\
&&\left[F_1({\bf e}_x\pm {\bf e}_y)+F_2e^{\pm 2i\phi}
({\bf e}_x\mp {\bf e}_y)\right]
+2i\Delta e^{\pm 2i\phi}\frac{\partial F_1}{\partial\xi}{\bf e}_z \Big\},
\label{bcircular}
\end{eqnarray}
where $\psi=\omega (t -z/c)$, $e^{\pm 2i\phi}=(x+iy)/\rho$,
$\chi=z\Delta/R$, $\xi=\rho/R$ and $\rho=\sqrt{x^2+y^2}$. The
focusing parameter $\Delta =\lambda/(2\pi R)$ is expressed in terms
of laser's wavelength $\lambda$ and the focal spot radius $R$. The
functions $F_{1,2}(\xi,\chi,\Delta)$ obey differential equations
\cite{2000JETP...90..753N}, go to zero sufficiently fast when $\xi,
|\chi|\rightarrow \infty$ and conditions $F_1(0,0,\Delta)=1;
F_2(0,0,\Delta)=0$ are satisfied for $\Delta\rightarrow 0$
\cite{2004PhLA..330....1N}. The $h$-polarized electric and magnetic
fields ${\bf E}^h=\pm i{\bf B}^e$ and ${\bf B}^h=\mp i{\bf E}^e$
\cite{2000JETP...90..753N}.

Corresponding to electromagnetic fields (\ref{ecircular}),
(\ref{bcircular}), field invariants $S^e,P^e$ are given by
Eq.~(\ref{lightlike}) and $\varepsilon,\beta$ by
Eq.~(\ref{fieldinvariant}) in Section~\ref{semi}. The total number
of electron and positron pairs is given by Eq.~(\ref{probabilityeh})
for $n=1$ (see also Eq.~(\ref{nprobabilitym}),
\begin{equation}
N_{e^+e^-}\simeq \frac{\alpha}{ \pi}\int_V dV\int_0^\tau dt  \varepsilon \beta
\coth \frac{\pi \beta}{ \varepsilon}\exp\left(-\frac{\pi E_c}{ \varepsilon}\right),
\label{circularpair}
\end{equation}
where the integral is taken over the volume $V$ and duration $\tau$
of the laser pulse. The qualitative estimations and numerical
calculations of total number $N_{e^+e^-}$ of electron and positron
pairs in terms of laser intensity $I_{\rm laser}$ and focusing
parameter $\Delta$ are presented in Ref.~\cite{2004PhLA..330....1N}.
Two examples for $e$-polarized electromagnetic fields are as
follows:
\begin{enumerate}

\item for $E_{\rm peak}=E_c$, $\lambda =1 \mu m$, $\tau=10$fs and
$\Delta=0.1$ (the theoretical diffraction limit), the laser beam
intensity $I_{\rm laser}\simeq 1.5\cdot 10^{29}W/{\rm cm}^2\simeq
0.31 I^c_{\rm laser}$ the critical intensity
(\ref{criticaldensity}). The total number of pairs created
$N_{e^+e^-}\simeq 5\cdot 10^{20} $ according to the Schwinger
formula Eq.~(\ref{circularpair}) for pair production rate;\label{ec}

\item
with the same values of laser parameters $\Delta$, $\lambda$ and $\tau$, while
the laser pulse intensity
$I_{\rm laser}\simeq 5\cdot 10^{27}W/{\rm cm}^2 \sim 10^{-2}I^c_{\rm laser}$
corresponding to $E_{\rm peak}\sim 0.18 E_c$, Eq.~(\ref{circularpair}) gives
$N_{e^+e^-}\sim 20 $.

\end{enumerate}
Because the volume $V$ and duration $\tau$ of the laser pulse is much larger
than the Compton volume and time occupied by one pair, the average number of
pairs $N_{e^+e^-}\simeq 20 $ is large and possibly observable even if the peak
value of electric field is only $18 \%$ of the critical value. In addition,
pair production is much more effective by the $e$-polarized electric and
magnetic fields ${\bf E}^e,{\bf B}^e$ than by the $h$-polarized fields ${\bf
E}^h,{\bf B}^h$. The detailed analysis of the dependence of the number of pairs
$N_{e^+e^-}$ on the laser intensity $I_{\rm laser}$ and focusing parameter
$\Delta$ is given in \cite{2006JETP..102....9B}, and results are presented in
Fig.~(\ref{popovfigure}). In particular, it is shown that for the case of two
counter-propagating focused laser pulses with circular polarizations, pair
production becomes experimentally observable when the laser intensity $I_{\rm
laser}\sim 10^{26}W/{\rm cm}^2$ for each beam, which is about $1\sim 2$ orders
of magnitude lower than for a single focused laser pulse, and more than $3$
orders of magnitude lower than the critical intensity (\ref{criticaldensity}).
In these calculations the ``imaginary time'' method is useful \cite{Popov1974,
1973JETP...36..840P, 2006JETP..102....9B}, which gives a clear description of
tunneling of a quantum particles through a potential barrier. Recently the
process of electron--positron pair creation in the superposition of a nuclear
Coulomb and a strong laser field was studied in \cite{2006PhRvA..73f2106M}.

\begin{figure}[!ptb]
\begin{center}
\includegraphics[height=12cm,width=12cm]{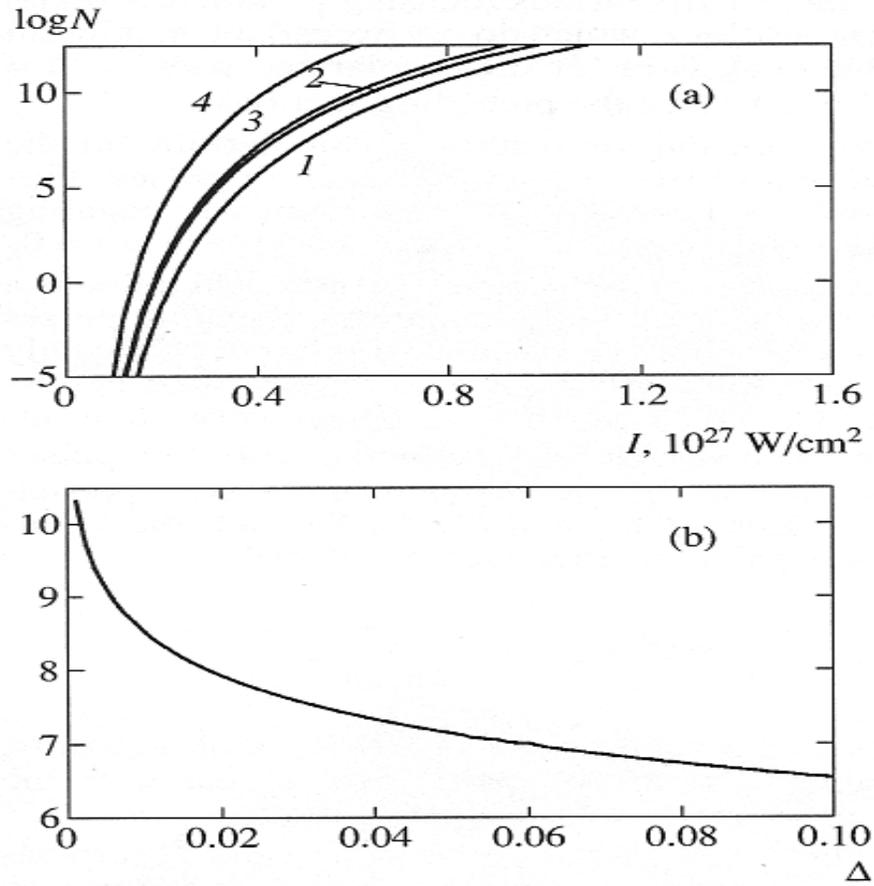}
\end{center}
\caption{Logarithm of the number of pairs $N_{e^+e^-}$ produced by the field of two counter-propagating laser-pulses
(circular polarization) is shown as functions of: (a) the beam intensity $I_{\rm laser}$ for the focusing parameters $\Delta=0.1,0.075,0.05$
and $0.01$ (the curves $1,2,3$ and $4$ correspondingly); (b) the focusing parameter $\Delta$ for the beam intensity
$I_{\rm laser}=4\cdot 10^{26}W/{\rm cm}^2$ and laser-pulse duration $\tau=10^{-14}$ sec.
This figure is reproduced from Fig.~6 in Ref.~\cite{2006JETP..102....9B}.}%
\label{popovfigure}%
\end{figure}

It was pointed \cite{2004PhLA..330....1N, 2006JETP..102....9B} that the
exploited method becomes inconsistent and one should take into account back
reaction of the pair production effect on the process of laser pulse focusing
at such high laser intensity and $E_{\rm peak}\sim E_c$. It has already been
argued in Refs.~\cite{1973JETPL..17..368M, Marinov1977, 1973ZhPmR..18..435P,
Popov1974} that for the superstrong field regime $E\gtrsim 0.1E_c$, such back
reaction of the produced electron--positron pairs on the external field and the
mutual interactions between these particles have to be considered. These back
reaction effects on pair production by laser beams leading to the formation of
plasma oscillation have been studied in Refs.~\cite{2002PhRvL..89o3901R,
2001PhRvL..87s3902A, 2005PhRvE..71a6404B}. Our studies
\cite{2003PhLB..559...12R, 2007PhLA..371..399R} show that the plasma
oscillation and electron--positron-photon collision are important for electric
fields $E\gtrsim 0.1E_c$, see Section~\ref{PO}.

\subsubsection{Availability of laser technology for pair production}\label{lasertech}

There are several ways to increase the electromagnetic fields of a laser beam.
One way is to increase the frequency of the laser radiation and then focus it
onto a tiny region. X-ray lasers can be used \cite{2001PhLB..510..107R,
2001hep.ph...12254R, 2003hep.ph....4139R, 2001PhRvL..87s3902A,
2003PlPhR..29..207T}. Another way is, clearly, to increase the intensities of
laser beams. The recent development of laser technology and the invention of
the chirped pulse amplification (CPA) method have led to a stunning increase of
the light intensity ($10^{22}$W/cm$^2$) in a laser focal spot
\cite{2002PhRvS...5c1301T, 1998PhT....51a..22M}. To achieve intensities of the
order $10^{24-25}$W/cm$^2$, a scheme was suggested in
Ref.~\cite{2002PhRvL..89A5004S}, where a quasi-soliton wave between two foils
is pumped by the external laser field up to an ultrahigh magnitude. Using the
method based on the simultaneous laser frequency upshifting and pulse
compression, another scheme for reaching critical intensities has also been
suggested in Ref.~\cite{Bulanov2003,2004PhRvL..92o9901B}, where the interaction
of the laser pulse with electron density modulations in a plasma produced by a
counter-propagating breaking wake plasma wave, results in the frequency upshift
and pulse focusing. In addition, it has been suggested
\cite{2002PhRvS...5c1301T} a path to reach the extremely high intensity level
of $10^{26-28}$W/cm$^2$ already in the coming decade. Such field intensities
are very close to the value of critical intensity $I^c_{\rm laser}$
(\ref{criticaldensity}). For a recent review, see
Ref.~\cite{2006RvMP...78..309M}. This technological situation has attracted the
attention of the theorists who involved in physics in strong electromagnetic
fields.

Currently available technologies allow for  intensities of the order
of $10^{20}$ W/cm$^2$ leading to abundant positron production
essentially through the Bethe-Heitler process (\ref{gi2p}) with
number densities of the order of $10^{16}$ cm$^-3$
\cite{2009PhRvL.102j5001C}.

\subsection{Phenomenology of pair production in electron beam-laser collisions}\label{light}

\subsubsection{Experiment of electron beam-laser collisions}\label{electronlaserexp}

After the availability of high dense and powerful laser beams, the
Breit--Wheeler process (\ref{2gammaee}) has been reconsidered in
Refs.~\cite{1962JMP.....3...59R, 1971PhRvL..26.1072R, Nikishov1964,
Nikishov1964a, Nikishov1965, 1967JETP...25.1135N, Nikishov1979, Nikishov1965a}
for high-energy multiple photon collisions. The phenomenon of $e^+e^-$ pair
production in multi-photon light-by-light scattering has been reported in
\cite{1996PhRvL..76.3116B, 1997PhRvL..79.1626B, Melissinos1999,
1999PhRvD..60i2004B} on the experiment SLAC-E-144 \cite{SLACe144,
1996NIMPA.383..309K}.

As described in Ref.~\cite{1997PhRvL..79.1626B}, such a large center of mass
energy ($2m_ec^2= 1.02$ MeV) can be possibly achieved in the collision of a
laser beam against another high-energy photon beam. With a laser beam of energy
$2.35$ eV, a high-energy photon beam of energy $111$ GeV is required for the
Breit--Wheeler reaction (\ref{2gammaee}) to be feasible. Such a high-energy
photon beam can be created for instance by backscattering the laser beam off a
high-energy electron beam, i.e., by inverse Compton scattering. With a laser
beam of energy $2.35$ eV (wavelength 527nm) backscattering off a high-energy
electron beam of energy $46.6$ GeV, as available at SLAC
\cite{1996NIMPA.383..309K}, the maximum energy acquired by
Compton-backscattered photon beam is only $29.2$GeV. This is still not enough
for the Breit--Wheeler reaction (\ref{2gammaee}) to occur, since such photon
energy is four times smaller than the needed energetic threshold.

Nevertheless in strong electromagnetic fields and a long coherent time-duration
$\Delta t=2\pi/\omega$ of the laser pulse, the number $n$ of laser photons
interacting with scattered electron becomes large, when the intensity parameter
of laser fields $\eta$ (\ref{eta}) approaches or even exceeds unity. Once this
number $n$ is larger than the critical number $n_0$ defined after
Eq.~(\ref{ggprobability}) in Section~\ref{light}, pair production by the
nonlinear Breit--Wheeler reaction (\ref{process2}) for high-energy multiple
photon collisions becomes feasible.

The probability of pair production by the processes (\ref{process1})
and (\ref{process2}) is given by Eqs.~(\ref{epprobability}) and
(\ref{ggprobability}) for any values of $\eta$ in
Section~\ref{lighttheoryquantum}. In high frequency and weak field
limit $\eta\ll 1$, the probability ${\mathcal P}_{e\gamma}$
(\ref{epprobability}) and ${\mathcal P}_{\gamma\gamma}$
(\ref{ggprobability}) for fairly small $n$ are proportional to
$\eta^{2n}$, i.e.
\begin{equation}
{\mathcal P}_{e\gamma}\propto \eta^{2n},\quad {\mathcal P}_{\gamma\gamma}\propto \eta^{2n}
\label{comparewithexp}
\end{equation}
(see Eqs.~(\ref{orateweak}), (\ref{wp})). This corresponds to the
anti-adiabatic, perturbative multi-photon production mechanism
(\ref{orateweak}), (\ref{wp}) for $(\eta \ll 1)$. In low frequency
and strong field limit $\eta\gg 1$, it essentially refers to process
in a constant and uniform field where $\bf E$ and $\bf B$ are
orthogonal and equal in magnitude. This corresponds to the adiabatic
limit of a slowly varying electromagnetic field discussed in
Section~\ref{alternatingfield}.

For $n\ge 5$ laser photons of energy $2.35$eV colliding with a photon of energy
$29$ GeV, the process of nonlinear Breit--Wheeler pair production becomes
energetically accessible. In Refs.~\cite{1996PhRvL..76.3116B,
1997PhRvL..79.1626B, 1999PhRvD..60i2004B}, it is reported that nonlinear
Compton scattering (\ref{process1}) and nonlinear Breit--Wheeler
electron--positron pair production (\ref{process2}) have been observed in the
collision of 46.6 GeV and 49.1 GeV electrons of the Final Focus Test Beam at
SLAC with terawatt pulse of 1053 nm (1.18 eV) and 527 nm (2.35 eV) wavelengths
from a Nd:glass laser. The rate of pair production, i.e., $R_{e^+}$ of
positrons/(laser shot) is measured in terms of the parameter $\eta$ ($\eta\ll
1$), as shown in Fig.~\ref{slacexp1}, where line represents a power law fit to
the data which gives \cite{1997PhRvL..79.1626B},
\begin{equation}
R_{e^+}\propto \eta^{2n},\hskip0.3cm {\rm with}
\hskip0.3cm n=5.1\pm 0.2({\rm stat})^{+0.5}_{-0.8}({\rm syst}).
\label{expresult}
\end{equation}
These experimental results are found to be in agreement with theoretical
predictions (\ref{comparewithexp}), i.e., (\ref{epprobability}),
(\ref{ggprobability}) for small $\eta$; as well as with (\ref{wp}) and
(\ref{orateweak}) for $\omega\rightarrow \gamma\omega$ in the frame of
reference where the electron beam is at rest. This shows that the pair
production of Breit--Wheeler type by the anti-adiabatic, perturbative
multi-photon production mechanism, described by Eqs.~(\ref{epprobability}),
(\ref{ggprobability}) or (\ref{orateweak}), (\ref{wp}) for small $\eta \ll 1$,
has been experimentally confirmed. However, one has not yet experimentally
observed the pair production by the adiabatic, non-perturbative tunneling
mechanism, described by Eqs.~(\ref{epprobability}), (\ref{ggprobability}) or
(\ref{orateweak}), (\ref{wp}) for large $\eta\gg 1$, i.e. for static and
constant electromagnetic fields. Nevertheless, pair production probabilities
Eqs.~(\ref{orate}), (\ref{wp}) and Eqs.~(\ref{epprobability}),
(\ref{ggprobability}) interpolates between both $\eta \ll 1$ and $\eta \gg 1$
regimes. Based on such analyticity of these probability functions in terms of
the laser intensity parameter $\eta$, we expect the pair production to be
observed in $\eta \gg 1$ regime.

\begin{figure}[!ptb]
\begin{center}
\includegraphics[height=8.4cm,width=12cm]{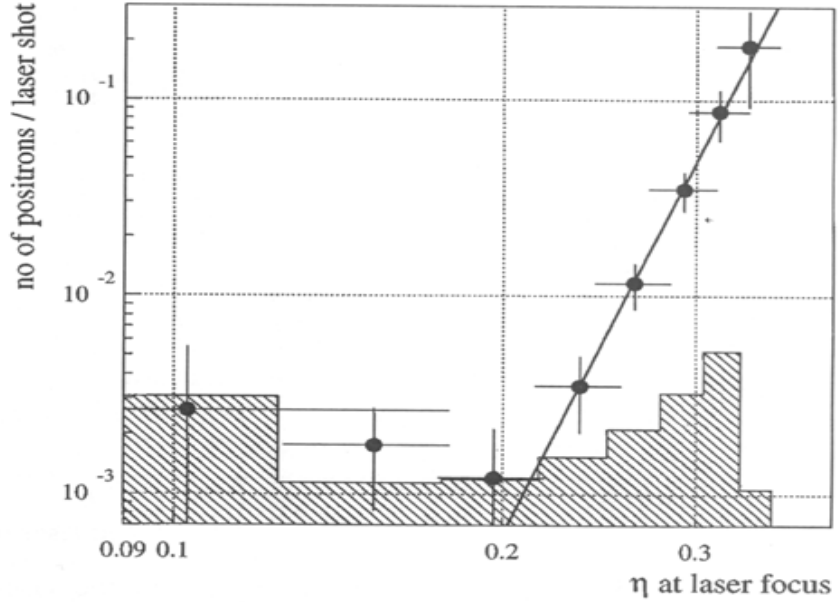}
\end{center}
\caption{Dependence of the positron rate per laser shot on the laser field-strength parameter $\eta$.
The line shows a power law fit to the data. The shaded distribution is the $95\%$ confidence limit on the
residual background from showers of lost beam particles after subtracting the laser-off positron rate. This figure
is reproduced from Fig.~4 in Ref.~\cite{1997PhRvL..79.1626B}}%
\label{slacexp1}%
\end{figure}

\subsubsection{Pair production viewed in the rest frame of electron beam}

In the reference frame where the electron beam is at rest, one can
discuss \cite{Melissinos1999} pair production in the processes
(\ref{process1}) and (\ref{process2}) by using pair production rate
Eqs.~(\ref{orate}), (\ref{orateweak}), (\ref{wp}) in
Section~\ref{Xray}. In the experiment of colliding $46.6$GeV
electron beam with $2.35$eV ($527$nm) laser wave, the field strength
in the laboratory is $E_{\rm lab}\simeq 6\cdot 10^{10}$V/cm  and
intensity $I\simeq 10^{19}W/{\rm cm}^2$ for $\eta=1$
\cite{1997PhRvL..79.1626B}. The Lorentz gamma factor of the electron
beam $\gamma={\mathcal E}_e/m_ec^2\simeq 9.32\cdot 10^4$ for
${\mathcal E}_e\simeq 46.6$GeV. In the rest frame of the electron
beam, the electric field is given by
\begin{equation}
{\bf E}_{\rm rest}=\gamma ({\bf E}_{\rm lab}+{\bf v}\times {\bf B}_{\rm lab})=
\gamma {\bf E}_{\rm lab}(1+|{\bf v}|)\simeq 2\gamma {\bf E}_{\rm lab}
\label{laserE}
\end{equation}
where laser's electromagnetic fields ${\bf E}\cdot {\bf B}=0$,
$|{\bf E}|= |{\bf B}|$, ${\bf B}\times \hat{\bf k}={\bf E}$, and
laser's wave vector $\hat{\bf k}=-\hat{\bf v}$, thus one has
\begin{equation}
E_{\rm rest}\simeq 2\gamma E_{\rm lab}\simeq 2\cdot 10^5 E_{\rm lab}\sim 0.86 E_c.
\label{Erest}
\end{equation}
The field of $2.35$eV laser wave is well defined coherent wave field
with wavelength $\lambda_{\rm lab}=5.27\cdot 10^{-5}$cm and
frequency $\omega_{\rm lab}=3.57\cdot 10^{15}$/sec (the period
$\Delta t_{\rm lab}=2\pi/\omega=1.76\cdot 10^{-15}$sec). In the rest
frame of electron beam, $\lambda_{\rm rest}= \gamma\lambda_{\rm
lab}=4.91$cm and $\Delta t_{\rm rest}=\Delta t_{\rm lab}/\gamma =
1.9\cdot 10^{-20}$sec. Comparing these wavelength and frequency of
laser wave field with the spatial length $\hbar/m_ec=3.86\cdot
10^{-11}$cm and timescale $\hbar/m_ec^2=1.29\cdot 10^{-21}$sec of
spontaneous pair production in vacuum, we are allowed to apply the
homogeneous and adiabatic approximation discussed in
Section~\ref{ad}, and use the rate of pair production (\ref{orate}),
(\ref{orateweak}), (\ref{wp}) in Section~\ref{Xray}.

\subsection{The Breit--Wheeler cutoff in high-energy $\gamma$-rays}\label{BWastro}

Having determined the theoretical basis as well as attempts to verify
experimentally the Breit--Wheeler formula we turn to a most important
application of the Breit--Wheeler process in the framework of cosmology. As
pointed out by Nikishov \cite{Nikishov1961} the existence of background photons
in cosmology puts a stringent cutoff on the maximum trajectory of the high
energy photons.

The Breit--Wheeler process for the photon-photon pair production is one of most
relevant elementary processes in high-energy astrophysics. In addition to the
importance of this process in dense radiation fields of compact objects
\cite{1971MNRAS.152...21B}, the essential role of this process in the context
of intergalactic absorption of high-energy $\gamma$-rays was first pointed out
by Nikishov \cite{Nikishov1961,1967PhRv..155.1408G}. The spectra of TeV
radiation observed from distant ($d> 100$ Mpc) extragalactic objects suffer
essential deformation during the passage through the intergalactic medium,
caused by energy-dependent absorption of primary $\gamma$-rays interacting with
the diffuse extragalactic background radiation, for the optical depth
$\tau_{\gamma\gamma}$ most likely significantly exceeding one
\cite{1967PhRv..155.1408G, 1992ApJ...390L..49S, 2000APh....12..217V,
1999APh....11...35C}. A relevant broad-band information about the cosmic
background radiation (CBR) is important for the interpretation of the observed
high-energy $\gamma$ spectra \cite{2000NewA....5..377A, 2002A&A...386....1K,
2005ApJ...618..657D, 2006Natur.440.1018A}. For details, readers are referred to
Refs.~\cite{2001ARA&A..39..249H, Aharonian2003}. In this section, we are
particularly interested in such absorption effect of high-energy $\gamma$-ray,
originated from cosmological sources, interacting with the Cosmic Microwave
Background (CMB) photons. Fazio and Stecker \cite{1970Natur.226..135F,
1977ApJ...214L..51S} were the first who calculated the cutoff energy versus
redshift for cosmological $\gamma$-rays. This calculation was applied to
further study of the optical depth of the Universe to high-energy $\gamma$-rays
\cite{1996SSRv...75..413M, 2004A&A...413..807K, 2006ApJ...648..774S}. With the
Fermi telescope, such study turns out to be important to understand the
spectrum of high-energy $\gamma$-ray originated from sources at cosmological
distance, we therefore offer the details of theoretical analysis as follow
\cite{Ruffini2008}.

We study the Breit--Wheeler process (\ref{2gammaee}) in the case that
high-energy photons $\omega_1$, originated from sources at cosmological
distance $z$, on their way to us, collide with CMB photons $\omega_2$ in the
rest frame of CMB photons, leading to electron--positron pair production. We
calculate the opacity and mean free path of these high-energy photons, find the
energy range of absorption as a function of the cosmological redshift $z$.

In general, a high-energy photon with a given energy $\omega_1$, collides with
background photons having all possible energies $\omega_2$. We assume that
$i$-type background photons have the spectrum distribution $f_i(\omega_2)$, the
opacity is then given by
\begin{equation}\label{opacity00}
\tau^i_{\gamma\gamma}(\omega_1,z)=\int^z_0\frac{dz'}{H(z')} \int_{m_e^2c^4/\omega_1}^\infty \frac{\omega_2^2d\omega_2}{\pi^2}f_i(\omega_2)
\sigma_{\gamma\gamma}\left(\frac{\omega_1\omega_2}{m_e^2c^4}\right),
\end{equation}
where $m_e^2c^4/\omega_1$ is the energy threshold (\ref{BW threshold}) above
which the Breit--Wheeler process (\ref{2gammaee}) can occur and the
cross-section $\sigma_{\gamma\gamma}$ is given by Eqs.~(\ref{BW section0});
$H(z)$ is the Hubble function, obeyed the Friedmann equation
\begin{equation}\label{free}
H(z)=H_0[\Omega_M(z+1)^3+\Omega_\Lambda]^{1/2}.
\end{equation}
We will assume $\Omega_M\simeq 0.3$ and $\Omega_M\simeq 0.7$ and $H_0=75$Km/s/Mpc.
The total opacity is then given by
\begin{equation}\label{topa}
\tau^{\rm total}_{\gamma\gamma}(\omega_1,z)=\sum_i\tau^i_{\gamma\gamma}(\omega_1,z),
\end{equation}
which the sum is over all types of photon backgrounds in the Universe.

In the case of CMB photons the ir %aga
distribution is black-body one
$f_{\rm CMB}(\omega_2/T)=1/(e^{\omega_2/T}-1)$ with the CMB
temperature $T$, the opacity is given by
\begin{equation}\label{opacity0}
\tau_{\gamma\gamma}(\omega_1,z)=\int^z_0\frac{dz'}{H(z')} \int_{m_e^2c^4/\omega_1}^\infty \frac{d\omega_2}{\pi^2}\frac{\omega_2^2}{e^{\omega_2/T}-1}
\sigma_{\gamma\gamma}(\frac{\omega_1\omega_2}{m_e^2c^4}),
\end{equation}
where the Boltzmann constant $k_B=1$. To simply
Eq.~(\ref{opacity0}), we set $x=\frac{\omega_1\omega_2}{m_e^2c^4}$.
In terms of CMB temperature and high-energy photons energy at the
present time,
\begin{equation}\label{rede}
T=(z+1)T^0;\quad \omega_{1,2}=(z+1)\omega^0_{1,2},
\end{equation}
we obtain,
\begin{equation}\label{opa1}
\tau_{\gamma\gamma}(\omega^0_1,z)=\frac{1}{R_0}\int^z_0\frac{dz'}{H(z')} \frac{}{(z+1)^3}\left(\frac{m_e^2c^4}{\omega^0_1}\right)^3\int_{1}^\infty
\frac{dx}{\pi^2}\frac{x^2}{\exp (x/\theta)-1}
\sigma_{\gamma\gamma}(x),
\end{equation}
where
\begin{equation}\label{dete}
\theta=x_0(z+1)^2; \quad x_0=\frac{\omega^0_1T^0}{m_e^2c^4},
\end{equation}
and $x_0$ is the energy $\omega^0_1$ in unit of $m_ec^2(m_ec^2/T^0)=1.11\cdot 10^{15}$eV.

The $\tau_{\gamma\gamma}(\omega^0_1,z)=1$ gives the relationship $\omega^0_1=\omega^0_1(z)$
that separates the optically thick $\tau_{\gamma\gamma}(\omega^0_1,z)>1$ and optically thin $\tau_{\gamma\gamma}(\omega^0_1,z)<1$ regimes in the $\omega^0_1-z$ plane.

\begin{figure}[!ptb]
%\begin{center}
%\begin{picture}(100,330)
%\put(-200,50){
\includegraphics[width=12cm]{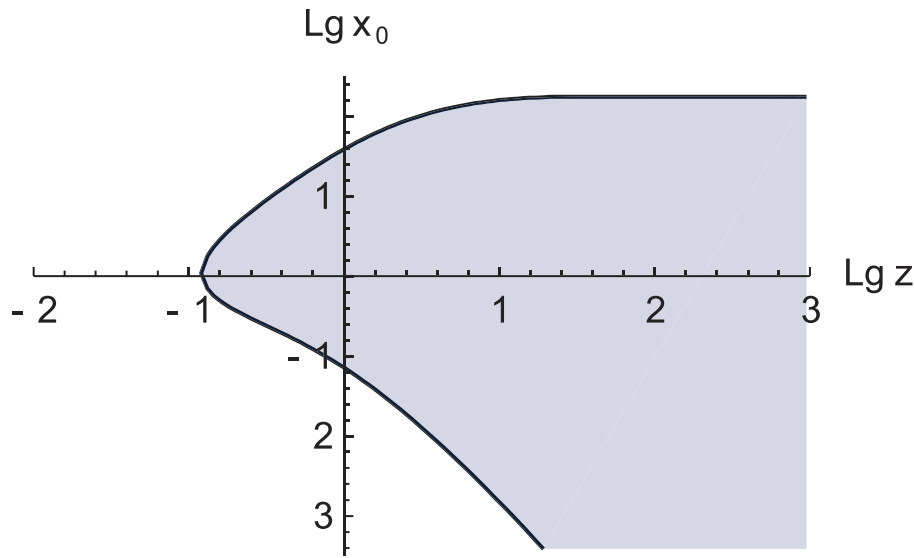}
%\put(50,230){$\tau_{\gamma\gamma}>1$}
%\put(-130,230){$\tau_{\gamma\gamma}<1$}
%\end{picture}
%\end{center}
\caption{This is a Log-Log plot for high energy photon energy $x_0$ in units of $1.11\cdot 10^{15}$ eV versus redshift $z$. The grey region represents optically thick case, while the white one is for optically thin case. The boundary between the two is the two-branch solution of Eq.~\ref{opa1} for $\tau_{\gamma\gamma}=1$. There is a critical redshift $z_c\simeq 0.1$ for a photon with arbitrary energy, which can reach the observer. The value of the photon energy corresponding to this critical redshift is $\omega^0_1\simeq 1.11\cdot 10^{15}$eV.}%
\label{e-z}%
\end{figure}
The integral (\ref{opa1}) is evaluated numerically and the result is presented in Fig.~\ref{e-z}. It clearly shows the following properties:
\begin{enumerate}
\item
for the redshift $z$ smaller than a critical value $z_c\simeq 0.1$
($z<z_c$), the CMB is transparent to photons with arbitrary energy,
this indicates a minimal mean free path of high-energy photons;

\item  for the redshift $z$ larger than the value ($z>z_c$), there
are two branches of solutions for
$\tau_{\gamma\gamma}(\omega^0_1,z)=1$, respectively corresponding to
the different energy dependence of the cross-section (\ref{BW
section0}): the cross-section $\sigma_{\gamma\gamma}(x)$ increases
with the center of mass energy $x={\omega_1\omega_2}^2_\gamma/%aga
(m_ec^2)^2$ from the energy threshold $x=1$ to $x\simeq 1.97$, and
decreases (\ref{bwsection3}) from $x\simeq 1.97$ to $x\rightarrow
\infty$. The energy of the CMB photon corresponding to the critical
redshift $z\simeq 0.1,\omega^0_1$ is $\simeq 1.15\cdot 10^{15}$eV
which separates two branches of the solution. The position of this
point in Fig. \ref{e-z} is determined by the maximal cross-section
at $x\simeq 1.97$. Due to existence of these two solutions for a
given redshift $z$, photons having energies in the grey region of
Fig. \ref{e-z} cannot reach the observer, while photons from the
white region of Fig. \ref{e-z} are observable.

\item above the critical redshift $z_c$ low-energy photons can reach us
    since their energies are smaller than the energetic threshold for the
    Breit--Wheeler process (\ref{2gammaee}). In addition, high-energy
    photons are also observable due to the fact that the cross-section of
    Breit--Wheeler process (\ref{2gammaee}) decreases with increasing
    energy of photons. For large redshifts $z\sim 10^3$, the Universe is
    opaque and we disregard this regime.
\end{enumerate}

In Section \ref{doub} we considered another relevant process which is double
pair production (\ref{gamma4ee}). This process contributes to the opacity at
very high energies and its effect has been computed in
\cite{1997ApJ...487L...9C}. We also computed the effect of this process on our
diagram in Fig. \ref{e-z}. This process becomes relevant at very high redshift
$z\sim10^3$.

Due to the fact that there are other radiation backgrounds contributing into
(\ref{topa}), the background of CMB photons gives the lower limit for opacity
for high-energy photons with respect to the Breit--Wheeler process
(\ref{2gammaee}). Finally, we point out that the small-energy solution for
large redshift in Fig.~(\ref{e-z}) agrees with the one found by Fazio and
Stecker \cite{1970Natur.226..135F,1977ApJ...214L..51S}.

\subsection{Theory of pair production in Coulomb potential}\label{Z137}

By far the major attention to build a critical electric field has
occurred in the physics of heavy nuclei and in heavy-ion collisions.
We recall in the following some of the basic ideas, calculations, as
well as experimental attempts to obtain the pair creation process in
nuclear physics.

\subsubsection{The $Z=137$ catastrophe}

Soon after the Dirac equation for a relativistic electron was
discovered \cite{1928RSPSA.117..610D,1947pqm..book.....D}, Gordon
\cite{1928ZPhy...48...11G,1928ZPhy...48..180G} (for all $Z< 137$)
and Darwin \cite{1928RSPSA.118..654D} (for $Z=1$) found its solution
in the point-like Coulomb potential $V(r)=-Z\alpha/r, \quad
0<r<\infty$. Solving the differential equations for the Dirac wave
function, they obtained the well-known Sommerfeld's formula
\cite{1916AnP...356....1S} for the energy spectrum,
\begin{equation}
{\mathcal E}(n,j)=m_ec^2\left[1+\left(\frac{Z\alpha}{ n-|K|+(K^2-Z^2\alpha^2)^{1/2}}\right)^2\right]^{-1/2}.
\label{dirac}
\end{equation}
Here the principle quantum number $n=1,2,3,\cdot\cdot\cdot$ and
\begin{equation}
K=\left\{\begin{array}{ll} -(j+1/2)= -(l+1), & {\rm if}\quad j=l+\frac{1}{2}, \quad l\ge 0 \\
 (j+1/2)= l, & {\rm if}\quad j=l-\frac{1}{2}, \quad l\ge 1
 \end{array}\right.
\label{dirac-k}
\end{equation}
where $l=0,1,2,\ldots$ is the orbital angular momentum corresponding
to the upper component of Dirac bi-spinor, $j$ is the total angular
momentum, and the states with $K=\mp 1,\mp 2,\mp 3,\cdot\cdot\cdot,
\mp (n-1)$ are doubly degenerate\footnote{This degeneracy is removed
by radiative corrections \cite{Schwinger1958,1982els..book.....B}.
The shift of the level $2S_{\frac{1}{2}}$ up, compared to the level
$2P_{\frac{1}{2}}$ (the famous Lamb shift) was discovered out of the
study of fine structure of the hydrogen spectrum.}, while the state
$K=-n$ is a singlet
\cite{1928ZPhy...48...11G,1928ZPhy...48..180G,1928RSPSA.118..654D}.
The integer values $n$ and $K$ label bound states whose energies are
${\mathcal E}(n,j)\in (0,m_ec^2)$. For the example, in the case of
the lowest-energy states, one has
\begin{eqnarray}
{\mathcal E}(1S_{\frac{1}{2}})&=& m_e c^2\sqrt {1-(Z\alpha)^2},\label{dirac-k1}\\
{\mathcal E}(2S_{\frac{1}{2}})&=&{\mathcal E}(2P_{\frac{1}{2}})
= m_e c^2\sqrt{\frac{1+\sqrt{1-(Z\alpha)^2}}{2}},\label{dirac-k2}\\
{\mathcal E}(2P_{\frac{3}{2}})&=& m_e c^2\sqrt {1-\frac{1}{4}(Z\alpha)^2}.
\label{dirac-k3}
\end{eqnarray}
For all states of the discrete spectrum, the binding energy $m_e
c^2-{\mathcal E}(n,j)$ increases as the nuclear charge $Z$
increases, as shown in Fig.~\ref{zspectrum}.  When $Z=137$,
${\mathcal E}(1S_{1/2})=0$, ${\mathcal E}(2S_{1/2})={\mathcal
E}(2P_{1/2})=(m_e c^2)/\sqrt{2}$ and ${\mathcal E}(2S_{3/2})=m_e c^2
\sqrt{3}/2$. Gordon noticed in his pioneer paper
\cite{1928ZPhy...48...11G,1928ZPhy...48..180G} that no regular
solutions with $n=1,j=1/2,l=0,$ and $K=-1$ (the $1S_{1/2}$ ground
state) are found beyond $Z=137$. This phenomenon is the so-called
``$Z=137$ catastrophe'' and it is associated with the assumption
that the nucleus is point-like in calculating the electronic energy
spectrum.

\begin{figure}[!ht]
\centering
\includegraphics[width=\hsize,clip]{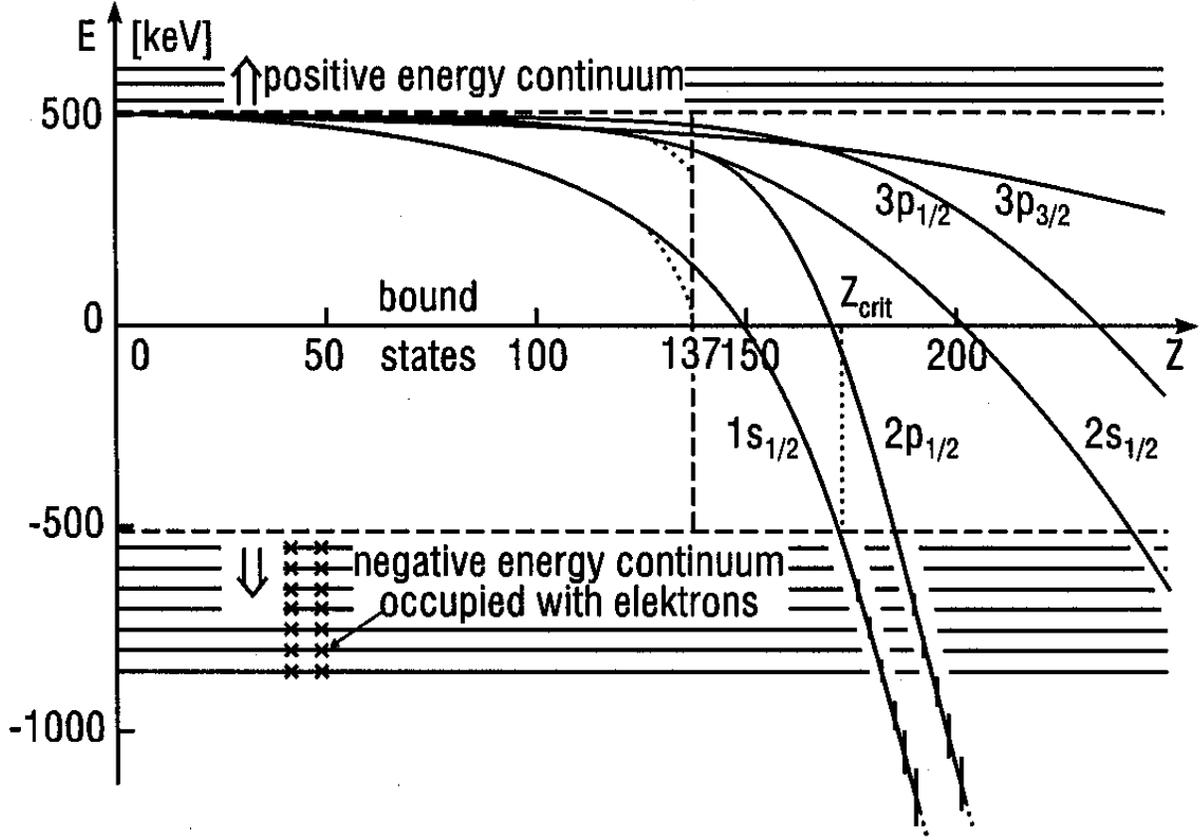}
\caption{Atomic binding energies as function of nuclear charge Z. This figure is reproduced from Fig.~1 in Ref.~\cite{Greiner1999}.}%
\label{zspectrum}%
\end{figure}
In fact, it was shown since the pioneering work of Pomeranchuk
\cite{Pomeranchuk1945} that in nature there cannot be a point-like
charged object with effective coupling constant $Z\alpha >1$ since
the entire electron shell will collapse to the center $r=0$.

\subsubsection{Semi-Classical description}\label{eff}

In order to have further understanding of this phenomenon, we study it in the
semi-classical scenario. For simplicity we treat relativistic electron as a
scalar particle fulfilling the Klein--Gordon equation, but still obeying
Fermi-Dirac statistics. Setting the origin of spherical coordinates
$(r,\theta,\phi)$ at the point-like charge, we introduce the vector potential
$A_\mu=({\bf A}, A_0)$, where ${\bf A}=0$ and $A_0$ is the Coulomb potential.
The motion of a relativistic ``electron'' with mass $m$ and charge $e$ is
described by its radial momentum $p_r$, angular momenta $p_\phi$ and the
Hamiltonian
\begin{eqnarray}
H_\pm &=& \pm
m_e c^2\sqrt{1+\left(\frac{p_r}{m_ec}\right)^2+\left(\frac{p_\phi}{m_e cr}\right)^2}+V(r),
\label{toth}
\end{eqnarray}
where the potential energy $V(r)=eA_0$, and $\pm$ corresponds for
positive and negative solutions. The states corresponding to
negative energy solutions are fully occupied. The angular momentum
$p_\phi$ is conserved, when the Hamiltonian is spherically
symmetric. For a given angular momentum $p_\phi$, the Hamiltonian
(\ref{toth}) describes electron's radial motion in the following
effective potential
\begin{eqnarray}
E_\pm &=& \pm m_e c^2\sqrt{1+\left(\frac{p_\phi}{m_e cr}\right)^2}+V(r).
\label{tote}
\end{eqnarray}
The Coulomb potential energy $V(r)$ is given by
\begin{equation}
V(r)=-\frac{Ze^2}{r}.
\label{potenoutside}
\end{equation}

In the classical scenario, given different values of angular momenta
$p_\phi$, the stable circulating orbits (states)
are determined by the minimum of the effective potential $E_+(r)$
(\ref{tote}).
Using $dE_+(r)/dr =0$, we obtain the stable
orbit location at the radius $R_L$ in the unit of the Compton length
$\lambda_C$,
\begin{equation}
R_L(p_\phi)=Z\alpha\lambda_C\sqrt{1-\left(\frac{Z\alpha}{p_\phi/\hbar}\right)^2},
\label{stable}
\end{equation}
where $\alpha=e^2/\hbar c$ and $p_\phi> Z\alpha$.
Substituting Eq.~(\ref{stable}) into Eq.~(\ref{tote}), we find the
energy of the electron at each stable orbit,
\begin{equation}
{\mathcal E}(p_\phi)\equiv {\rm min}(E_+) =
m_e c^2\sqrt{1-\left(\frac{Z\alpha}{p_\phi/\hbar}\right)^2}.
\label{mtote}
\end{equation}
The last stable orbits (minimal energy) are given by
\begin{equation}
p_\phi\rightarrow Z\alpha \hbar  + 0^+,
\quad R_L(p_\phi) \rightarrow 0^+,\quad {\mathcal E}(p_\phi) \rightarrow 0^+.
\label{mtote1}
\end{equation}
For stable orbits with $p_\phi/\hbar\gg 1$, the radii
$R_L/\lambda_C\gg 1$ and energies ${\mathcal E}\rightarrow m_e
c^2+0^-$; electrons in these orbits are critically bound since their
binding energy goes to zero. As the energy spectrum (\ref{dirac}),
see Eqs.~(\ref{dirac-k1},\ref{dirac-k2},\ref{dirac-k3}),
Eq.~(\ref{mtote}) shows, only positive or null energy solutions
(states) exist in the presence of a point-like nucleus.

In the semi-classical scenario, the discrete values of angular momentum
$p_\phi$ are selected by the Bohr-Sommerfeld quantization rule
\begin{equation}
\int p_\phi d\phi \simeq h (l+\frac{1}{2}),\quad \Rightarrow\quad
p_\phi(l) \simeq \hbar(l+\frac{1}{2}),\quad l=0,1,2,3,\ldots
\label{angq}
\end{equation}
describing the semi-classical states of radius and energy
\begin{eqnarray}
R_L(l) &\simeq &
(Z\alpha)^{-1}\lambda_C\sqrt{1-\left(\frac{2Z\alpha}{2l+1}\right)^2},\label{rlq}\\
{\mathcal E}(l) &\simeq &
m_e c^2\sqrt{1-\left(\frac{2Z\alpha}{2l+1}\right)^2}\label{elq}.
\end{eqnarray}
Other values of angular momentum $p_\phi$, radius $R_L$ and energy
${\mathcal E}$ given by Eqs.~(\ref{stable},\ref{mtote}) in the
classical scenario are not allowed. When these semi-classical states
are not occupied as required by the Pauli Principle, the transition
from one state to another with different discrete values ($l_1,l_2$
and $\Delta l=l_2-l_1=\pm 1$) is possible by emission or absorption
of a spin-1 ($\hbar$) photon. Following the energy and
angular-momentum conservations, photons emitted or absorbed in the
transition have angular momentum $p_\phi(l_2)-p_\phi(l_1)=\hbar
(l_2-l_1)=\pm\hbar$ and energy ${\mathcal E}(l_2)-{\mathcal
E}(l_1)$. As required by the Heisenberg uncertainty principle
$\Delta\phi\Delta p_\phi \simeq 4\pi p_\phi(l) \gtrsim h$, the
absolute ground state for minimal energy and angular momentum is
given by the $l=0$ state, $p_\phi\sim \hbar/2$, $R_L\sim
Z\alpha\lambda_C\sqrt{1-(2Z\alpha)^2}>0$ and ${\mathcal E} \sim  m_e
c^2\sqrt{1-(2Z\alpha)^2}>0$ for $Z\alpha \le 1/2$. Thus the
stability of all semi-classical states $l>0$ is guaranteed by the
Pauli principle. In contrast for $Z\alpha > 1/2$, there is not an
absolute ground state in the semi-classical scenario.

We see now how the lowest-energy states are selected by the
quantization rule in the semi-classical scenario out of the last
stable orbits (\ref{mtote1}) in the classical scenario. For the case
of $Z\alpha\le 1/2$, equating (\ref{mtote1}) to (\ref{angq}), we
find the selected state $l=0$ is only possible solution so that the
ground state $l=0$ in the semi-classical scenario corresponds to the
last stable orbits (\ref{mtote1}) in the classical scenario. On the
other hand for the case $Z\alpha > 1/2$, equating (\ref{mtote1}) to
(\ref{angq}), we find the selected state $l=\tilde l\equiv
(Z\alpha-1)/2>0$ in the semi-classical scenario corresponds to the
last stable orbits (\ref{mtote1}) in the classical scenario. This
state $l=\tilde l>0$ is not protected by the Heisenberg uncertainty
principle from quantum mechanically decaying in $\hbar$-steps to the
states with lower angular momentum and energy (correspondingly
smaller radius $R_L$ (\ref{rlq})) via photon emissions. This clearly
shows that the ``$Z=137$-catastrophe'' corresponds to
$R_L\rightarrow 0$, falling to the center of the Coulomb potential
and all semi-classical states ($l$) are unstable.

\subsubsection{The critical value of the nuclear charge $Z_{cr}=173$}\label{finitez173}

A very different situation is encountered when considering the fact that the
nucleus is not point-like and has an extended charge distribution
\cite{Pomeranchuk1945, 1950PhRv...80..797C, 1958PhRv..109..126W, Voronkov1961,
Popov1970, 1970JETPL..11..162P, 1971JETP...32..526P, 1971JETP...33..665P,
1971SvPhU..14..673Z}. In that case the $Z=137$ catastrophe disappears and the
energy levels ${\mathcal E}(n,j)$ of the bound states $1S$, $2P$ and $2S$,
$\cdot\cdot\cdot$ smoothly continue to drop toward the negative energy
continuum as $Z$ increases to values larger than $137$, as shown in
Fig.~\ref{zspectrum}. The reason is that the finite size $R$ of the nucleus
charge distribution provides a cutoff for the boundary condition at the origin
$r\rightarrow 0$ and the energy levels ${\mathcal E}(n,j)$ of the Dirac
equation are shifted due to this cutoff. In order to determine the critical
value $Z_{cr}$ when the negative energy continuum (${\mathcal E}<- m_ec^2$) is
encountered (see Fig.~\ref{zspectrum}), Zeldovich and Popov\cite{Popov1970,
1970JETPL..11..162P, 1971JETP...32..526P, 1971JETP...33..665P,
1971SvPhU..14..673Z} solved the Dirac equation corresponding to a nucleus of
finite extended charge distribution, i.e., the Coulomb potential is modified as
\begin{equation}
V(r)=\left\{\begin{array}{ll} -\frac{Ze^2}{ r}, &  r>R, \\
 -\frac{Ze^2}{ R}f\left(\frac{r}{ R}\right), &  r<R,
 \end{array}\right.
\label{extpotential}
\end{equation}
where $R\sim 10^{-12}$cm is the size of the nucleus. The form of the cutoff
function $f(x)$ depends on the distribution of the electric charge over the
volume of the nucleus $(x=r/R, 0<x<1$, with $f(1)=1)$. Thus, $f(x)=(3-x^2)/2$
corresponds to a constant volume density of charge.

Solving the Dirac equation with the modified Coulomb potential
(\ref{extpotential}) and calculating the corresponding perturbative
shift $\Delta {\mathcal E}_R$ of the lowest-energy level
(\ref{dirac-k1}) Popov obtains\cite{Popov1970,1971SvPhU..14..673Z}
\begin{equation}
\Delta {\mathcal E}_R= m_ec^2\frac{(\xi)^2(2\xi e^{-\Lambda})^{2\gamma_z}}{\gamma_z (1+2\gamma_z)}
\left[1-2\gamma_z\int_0^1f(x)x^{2\gamma_z} dx\right],
\label{zshift}
\end{equation}
where $\xi=Z\alpha$, $\gamma_z=\sqrt{1-\xi^2}$ and
$\Lambda=\ln(\hbar / m_ecR)\gg 1$ is a logarithmic parameter in the
problem under consideration. The asymptotic expressions for the
$1S_{1/2}$ energy that were obtained
are\cite{1971JETP...33..665P,1971SvPhU..14..673Z}
\begin{equation}
{\mathcal E}(1S_{1/2})=m_ec^2\left\{\begin{array}{ll} \sqrt {1-\xi^2}\coth(\Lambda\sqrt {1-\xi^2}), &  0<\xi<1, \\
 \Lambda^{-1}, &  \xi=1,\\
 \sqrt {\xi^2-1}\cot(\Lambda\sqrt {\xi^2-1}), &  \xi>1.
 \end{array}\right.
\label{easymptotic}
\end{equation}
As a result, the ``$Z=137$ catastrophe'' in Eq.~(\ref{dirac}) disappears and
${\mathcal E}(1S_{1/2})=0$ gives
\begin{equation}
\xi_0=1+\frac{\pi^2}{8\Lambda}+{\mathcal O}(\Lambda^{-4});
\label{0xi}
\end{equation}
the state $1S_{1/2}$ energy continuously goes down to the negative
energy continuum since $Z\alpha >1$, and ${\mathcal E}(1S_{1/2})=-1$
gives
\begin{equation}
\xi_{cr}=1+\frac{\pi^2}{2\Lambda(\Lambda +2)}+{\mathcal O}(\Lambda^{-4})
\label{0xicr}
\end{equation}
as shown in Fig.~\ref{zspectrum}. In Ref.~\cite{Popov1970, 1971SvPhU..14..673Z}
Popov and Zeldovich found that the critical value $\xi_c^{(n)}= Z_c\alpha$ for
the energy levels $nS_{1/2}$ and $nP_{1/2}$ reaching the negative energy
continuum is equal to
\begin{equation}
\xi_c^{(n)}= 1+ \frac{n^2\pi^2}{ 2\Lambda^2}+{\mathcal O}(\Lambda^{-3}).
\label{criticalxi}
\end{equation}
The critical value increases rapidly with increasing $n$. As a result, it is
found that $Z_{cr}\simeq 173$ is a critical value at which the lowest-energy
level of the bound state $1S_{1/2}$ encounters the negative energy continuum,
while other bound states encounter the negative energy continuum at
$Z_{cr}>173$ (see also Ref.~\cite{1958PhRv..109..126W} for a numerical
estimation of the same spectrum). The change in the vacuum polarization near a
high-Z nucleus arising from the finite extent of the nuclear charge density was
computed in \cite{1975PhRvD..12..581B, 1975PhRvD..12..596B,
1975PhRvD..12..609B} with all calculations done analytically, and to all orders
in $Z\alpha$. Note that for two nuclei with charges $Z_1$ and $Z_2$
respectively, if $Z_1>Z_2$ and $K$-shell of the $Z_1$-nucleus is empty, then
$Z_2$ may be a neutral atom. In this case two nuclei make a quasimolecular
state for which the ground term $(1s\sigma)$ is unoccupied by electrons: so
spontaneous production of positrons is also possible \cite{Gershtein1973,
Popov1973}. We refer the readers to Ref.~\cite{Popov1970, 1970JETPL..11..162P,
1971JETP...32..526P, 1971JETP...33..665P, 1971SvPhU..14..673Z,
2001PAN....64..367P} for mathematical and numerical details.

When $Z>Z_{cr}=173$, the  lowest-energy level of the bound state
$1S_{1/2}$ enters the negative energy continuum. Its energy level
can be estimated as follows
\begin{equation}
{\mathcal E}(1S_{1/2})=m_ec^2 - \frac{Z \alpha}{\bar r}<-m_ec^2,
\label{1S}
\end{equation}
where $\bar r $ is the average radius of the $1S_{1/2}$ state's orbit, and the
binding energy of this state satisfies $Z\alpha/\bar r > 2 m_ec^2$. If this
bound state is unoccupied, the bare nucleus gains a binding energy
$Z\alpha/\bar r$ larger than $2m_ec^2$, and becomes unstable against the
production of an electron--positron pair. Assuming this pair production occurs
around the radius $\bar r$, we have energies for the electron ($\epsilon_-$)
and positron ($\epsilon_+$) given by
\begin{equation}
\epsilon_-=\sqrt{|c{\bf p}_-|^2+m_e^2c^4}-\frac{Z \alpha}{\bar r};\
\quad \epsilon_+=\sqrt{|c{\bf p}_+|^2+m_e^2c^4}+\frac{Z \alpha}{\bar r},
\label{eofep}
\end{equation}
where ${\bf p}_\pm$ are electron and positron momenta, and ${\bf p}_-=-{\bf p}_+$.
The total energy required for the production of a pair is
\begin{equation}
\epsilon_{-+}=\epsilon_-+\epsilon_+=2\sqrt{|c{\bf p}_-|^2+m_e^2c^4},
\label{totaleofep}
\end{equation}
which is independent of the potential $V(\bar r)$. The potential energies $\pm
eV(\bar r)$ of the electron and positron cancel out each other and do not
contribute to the total energy (\ref{totaleofep}) required for pair production.
This energy (\ref{totaleofep}) is acquired from the binding energy
($Z\alpha/\bar r > 2 m_ec^2$) by the electron filling into the bound state
$1S_{1/2}$. A part of the binding energy becomes the kinetic energy of positron
that goes out. This is analogous to the familiar case when a proton ($Z=1$)
catches an electron into the ground state $1S_{1/2}$, and a photon is emitted
with the energy not less than 13.6 eV. In the same way, more electron--positron
pairs are produced, when $Z\gg Z_{cr}=173$ and the energy levels of the next
bound states $2P_{1/2},2S_{3/2},\ldots$ enter the negative energy continuum,
provided these bound states of the bare nucleus are unoccupied.

\subsubsection{Positron production}

Gershtein and Zeldovich \cite{1969JETP...30..358G, Gershtein1969} proposed that
when $Z>Z_{cr}$ the bare nucleus produces spontaneously pairs of electrons and
positrons: the two positrons\footnote{Hyperfine structure of $1S_{1/2}$ state:
single and triplet.} run off to infinity and the effective charge of the bare
nucleus decreases by two electrons, which corresponds exactly to filling the
K-shell\footnote{An assumption was made in Ref.~\cite{1969JETP...30..358G,
Gershtein1969} that the electron density of $1S_{1/2}$ state, as well as the
vacuum polarization density, is delocalized at $Z\rightarrow Z_{cr}$. Later it
was proved to be incorrect \cite{1970JETPL..11..162P, 1971JETP...32..526P,
1971SvPhU..14..673Z}.} A more detailed investigation was made for the solution
of the Dirac equation at $Z\sim Z_{cr}$, when the lowest electron level
$1S_{1/2}$ merges with the negative energy continuum, in Refs.~\cite{Popov1970,
1970JETPL..11..162P, 1971JETP...32..526P, 1971JETP...33..665P, Popov1971}. It
was there further clarified the situation, showing that at $Z\gtrsim Z_{cr}$,
an imaginary resonance energy of Dirac equation appears,
\begin{equation}
\epsilon = \epsilon_0 - i\frac{\Gamma_{\rm nucl}}{2},
\label{zimaginary}
\end{equation}
where
\begin{eqnarray}
\epsilon_0 &=&-m_e- a(Z-Z_{cr}),
\label{epsilon0}\\
\Gamma_{\rm nucl} &\sim& \theta(Z-Z_{cr})\exp \left(-b\sqrt\frac{Z_{cr}}{ Z-Z_{cr}}\right),
\label{zprobability}
\end{eqnarray}
and $a,b$ are constants, depending on the cutoff $\Lambda$ (for example,
$b=1.73$ for $Z=Z_{cr}=173$ \cite{1970JETPL..11..162P, 1971JETP...32..526P,
1971SvPhU..14..673Z}). The energy and momentum of emitted positrons are
$|\epsilon_0|$ and $|{\bf p}|=\sqrt{|\epsilon_0|-m_ec^2}$.

The kinetic energy of the two positrons at infinity is given by
\begin{equation}
\varepsilon_p = |\epsilon_0| - m_ec^2 = a(Z-Z_{cr})+\cdot\cdot\cdot,
\label{zkinetic}
\end{equation}
which is proportional to $Z-Z_{cr}$ (as long as $(Z-Z_{cr})\ll Z_{cr}$) and
tends to zero as $Z\rightarrow Z_{cr}$. The pair production resonance at the
energy (\ref{zimaginary}) is extremely narrow and practically all positrons are
emitted with almost same kinetic energy for $Z\sim Z_{cr}$, i.e. nearly
mono-energetic spectra (sharp line structure). Apart from a pre-exponential
factor, $\Gamma_{\rm nucl}$ in Eq.~(\ref{zprobability}) coincides with the
probability of positron production, i.e., the penetrability of the Coulomb
barrier (see Section~\ref{semi}). The related problems of vacuum charge density
due to electrons filling into the K-shell and charge renormalization due to the
change of wave function of electron states are discussed in
Refs.~\cite{Zeldovich1960, 1971TMP.....8..729S, Baz'1966, Migdal1971,
Perelomov1971}. An extensive and detailed review on this theoretical issue can
be found in Refs.~\cite{1971SvPhU..14..673Z, 2001PAN....64..367P,
2003spr..book.....G, Greiner1999}.

On the other hand, some theoretical work has been done studying the possibility
that pair production, due to bound states encountering the negative energy
continuum, is prevented from occurring by higher order processes of quantum
field theory, such as charge renormalization, electron self-energy and
nonlinearities in electrodynamics and even Dirac field itself
\cite{1978PhR....38..227R, 1976ARNPS..26..351M, 1977RPPh...40..219R,
1978scia.conf....3B, 1975NuPhA.244..497G, 1975PhRvA..12..748R,
1982PhRvL..48.1465S}. However, these studies show that various effects modify
$Z_{cr}$ by a few percent, but have no way to prevent the binding energy from
increasing to $2m_ec^2$ as $Z$ increases, without simultaneously contradicting
the existing precise experimental data on stable atoms
\cite{1982PhT....35h..24G}. Contrary claim \cite{1993NuPhA.560..973D} according
to which bound states are repelled by the lower continuum through some kind of
self-screening appear to be unfounded \cite{Greiner1999}.

It is worth noting that an overcritical nucleus ($Z\ge Z_{cr}$) can be formed
for example in the collision of two heavy nuclei \cite{Voronkov1961, Popov1971,
1969JETP...30..358G, Gershtein1969, 1969ZPhy..218..327P, 1972PhRvL..28.1235M,
1972ZPhy..257...62M, 1972ZPhy..257..183M}. To observe the emission of positrons
originated from pair production occurring near to an overcritical nucleus
temporally formed by two nuclei, the following necessary conditions have to be
full filled: (i) the atomic number of an overcritical nucleus is larger than
$Z_{cr}=173$; (ii) the lifetime of the overcritical nucleus must be much longer
than the characteristic time $(\hbar/m_ec^2)$ of pair production; (iii) the
inner shells (K-shell) of the overcritical nucleus should be unoccupied.

The collision of two Uranium nuclei with $Z=92$ was considered by Zeldovich,
Popov and Gershtein \cite{1971SvPhU..14..673Z, Gershtein1973}. The conservation
of energy in the collision reads
\begin{equation}
M_nv_0^2=(Ze)^2/R_{min},
\label{zmotion}
\end{equation}
where $v_0$ is the relative velocity of the nuclei at infinity, $R_{min}$ is
the smallest distance, and $M_n$ is the Uranium atomic mass. In order to have
$R_{min}\simeq 30$fm a fine tuning of the initial velocity narrowly peaked
around $v_0\simeq 0.034c$ is needed. The characteristic collision time would be
then $\Delta t_c= R_{min}/v_0\simeq 10^{-20}$s. The interesting possibility
then occurs, that the typical velocity of an electron in the inner shell ($r
\sim 115.8$fm) is $v \sim c$ and therefore its characteristic time
$\Delta\tau_0\sim r/v\sim 4 \cdot 10^{-22}$s. This means that the
characteristic collision time $\Delta t_c$ in which the two colliding nuclei
are brought into contact and separated again can be in principle much larger
than the timescale $\Delta\tau_0$ of electron evolution. This would give
justification for an adiabatic description of the collision in terms of {\it
quasimolecules}. The formation of ``quasimolecules'' could also be verified by
the characteristic molecular-orbital X-rays radiation due to the electron
transitions between ``quasimolecules'' orbits \cite{1982PhT....35h..24G,
Vincent1983, 1974PhLB...49..219M, 1978ZPhy..288..257A, 1974PhRvL..33..473G,
1974PhRvL..33..476K, 1975PhRvA..12.2641M, 1978PhRvA..18.1878V}. However, this
requires the above mentioned fine tuning in the bombarding energies
($M_nv_0^2/2$) close to the nuclear Coulomb barrier.

However, we notice that the above mentioned condition (ii) has never been
fulfilled in heavy-ion collisions. There has been up to now various
unsuccessful attempts to broaden this time of encounter by `sticking'
phenomena. Similarly, the condition (iii) is not sufficient for pair
production, since electrons that occupied outer shells of high energies must
undergo a rapid transition to occupy inner shells of lower energies, which is
supposed to be vacant and encountering the negative energy continuum. If such
transition and occupation take place faster than pair production, the pair
production process is blocked. As a consequence, it needs a larger value of
$Z>Z_{cr}=173$ to have stronger electric field for vacant out shells
encountering the negative energy continuum (see Eq.~(\ref{criticalxi})) so that
electrons produced can occupy outer shells. This makes pair production even
less probable to be observed, unless the overcritical charged nucleus is bare,
i.e. all shells are vacant.

\subsubsection{Homogeneous and adiabatic approximation}\label{ad}

There is a certain analogy between positron production by a nucleus with
$Z>Z_{cr}$ and pair production in a homogeneous electrostatic field. We note
that in a Coulomb potential of a nucleus with $Z=Z_{cr}$ the corresponding
electric field $E_{cr}=Z_{cr}|e|/r^2$ is comparable with the critical electric
field $E_c$, (\ref{critical1}), when $r\sim \lambda_C$. However, the condition
$E>E_c$ is certainly the necessary condition in order to have the pair creation
but not a sufficient one: the spatial extent of the region where $E>E_c$ occurs
must be larger than the de~Broglie wavelength of the created
electron---positron pair. If a pair production takes place, electrons should be
bound into the K-shell nucleus and positrons should run off to infinity. This
intuitive reasoning builds the connection between the phenomena of pair
production in the Coulomb potential at charge $Z>Z_{cr}$ and the one in an
external constant electric field which was treated in Section \ref{qedpair}.
The exact formula for pair production probability $W$ in an overcritical
Coulomb potential has not yet been obtained in the framework of QED. We cannot
expect a literal coincidence of formulas for the probability $W$ of pair
production in these different cases, since the Schwinger formula
(\ref{probability1}) is exactly derived for a homogeneous field, while the
Coulomb potential is strongly inhomogeneous at small distances. Some progress
in the treatment of this problem is presented in Section
\ref{inhomogeneousfield}.

All the discussions dealing with pair production in an external
homogeneous electric field or a Coulomb potential assume that the
electric field be static. Without the feedback of the particles
created on the field this will clearly lead to a divergence of the
number of pairs created. In the real description of the phenomenon
at $t\rightarrow -\infty$ we have an initial empty vacuum state. We
then have the turned on of an overcritical electric field and
ongoing process of pair creation with their feedback on a time
$\mathcal \tau$ on the electric field and a final state at remote
future $t\rightarrow +\infty$ with the electron and positron created
and the remaining subcritical electric field. To describe this very
different regimes a simplified ``adiabatic approximation'' can be
adopted by assuming the existence of a homogeneous field only during
a finite time interval $[-{\mathcal T},{\mathcal T}]$. That time
$\mathcal T$ should be of course shorter than the feedback time
$\mathcal \tau$. During that time interval the Schwinger formula
(\ref{probability1}) is assumed to be applicable and it is
appropriate to remark that the overcritical electric fields are
related to very high energy densities: $E_c^2/2=9.53\cdot
10^{26}$ergs/cm$^3$. In the adiabatic approximation an effective
spatial limitation to the electric field is also imposed. Therefore
the constant overcritical electric field and the application of the
Schwinger formula is limited both in space and time. Progress in
this direction has been presented in \cite{2007PhLA..371..399R}, see
Section \ref{electrofluidodynamics}. A significant amount of pairs
is only produced if the finite lifetime of the overcritical electric
field is larger than the characteristic time of pair production
($\hbar/m_ec^2$) and the spatial extent of the electric field is
larger than the tunneling length $a$ (\ref{tunnelinglength}).

We have already discussed in Secs.~\ref{Xray} and \ref{light} the experimental
status of electron--positron pair creation in X-ray free electron laser and an
electron-beam--laser collision, respectively. We now turn in
Section~\ref{heavy} to the multiyear attempts in creating electron--positron
pairs in heavy-ion collisions.

\subsection{Pair production in heavy-ion collisions}\label{heavy}

\subsubsection{A transient super heavy ``quasimolecules''}

There has been a multiyear effort to observe positrons from pair
production associated with the overcritical field of two colliding
nuclei, in heavy-ion collisions
\cite{Popov1971,1982PhT....35h..24G,1969JETP...30..358G,Gershtein1969,1972ZPhy..257...62M,1972ZPhy..257..183M,1974PhRvL..33..921G}.
The hope was to use heavy-ion collisions to form transient
superheavy ``quasimolecules'': a long-lived metastable nuclear
complex with $Z>Z_{cr}$. It was expected that the two heavy ions of
charges respectively $Z_1$ and $Z_2$ with $Z_1+Z_2>Z_{cr}$ would
reach small inter-nuclear distances well within the electron's
orbiting radii. The electrons would not distinguish between the two
nuclear centers and they would evolve as if they were bounded by
nuclear ``quasimolecules'' with nuclear charge $Z_1+Z_2$. Therefore,
it was expected that electrons would evolve quasi-statically through
a series of well defined nuclear ``quasimolecules'' states in the
two-center field of the nuclei as the inter-nuclear separation
decreases and then increases again.

When heavy-ion collision occurs the two nuclei come into contact and some deep
inelastic reaction occurs determining the duration $\Delta t_s$ of this
contact. Such ``sticking time'' is expected to depend on the nuclei involved in
the reaction and on the beam energy. Theoretical attempts have been proposed to
study the nuclear aspects of heavy-ion collisions at energies very close to the
Coulomb barrier and search for conditions, which would serve as a trigger for
prolonged nuclear reaction times, to enhance the amplitude of pair production.
The sticking time $\Delta t_s$ should be larger than $1\sim 2\cdot 10^{-21}$
sec \cite{Greiner1999} in order to have significant pair production, see
Fig.~\ref{example}. Up to now no success has been achieved in justifying
theoretically such a long sticking time. In reality the characteristic sticking
time has been found of the order of $\Delta t\sim 10^{-23}$ sec, hundred times
shorter than the one needed to activate the pair creation process. Moreover, it
is recognized that several other dynamical processes can make the existence of
a sharp line corresponding to an electron--positron annihilation very unlikely
\cite{Greiner1999, 1982PhT....35h..24G, 1978ZPhy..285...49R,
1981ZPhy..303..173R, 1988PhRvL..61.2831G}.

\begin{figure}[!ht]
\centering
\includegraphics[width=\hsize,clip]{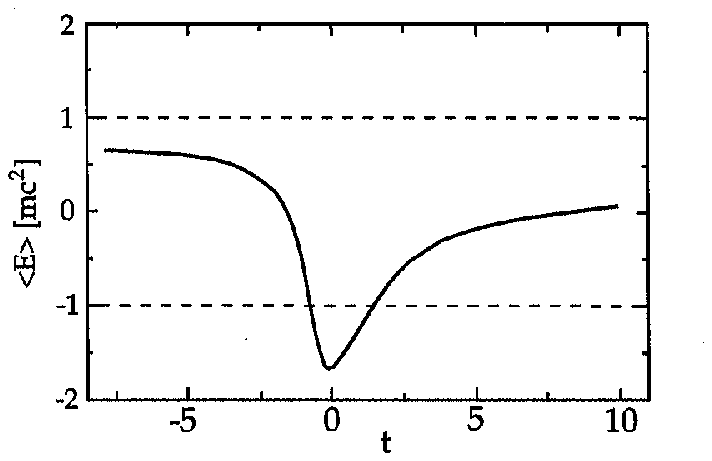}
\caption{Energy expectation values of the $1s\sigma$ state in a U+U collision
at 10 GeV/nucleon. The unit of time is $\hbar/m_ec^2$. This figure is
reproduced from Fig.~4 in Ref.~\cite{Greiner1999}.}%
\label{example}%
\end{figure}

It is worth noting that several other dynamical processes contribute to the
production of positrons in undercritical as well as in overcritical collision
systems \cite{1976ARNPS..26..351M, 1977RPPh...40..219R, 1978PhR....38..227R,
1978scia.conf....3B}. Due to the time-energy uncertainty relation (collision
broadening), the energy spectrum of such positrons has a rather broad and
oscillating structure, considerably different from a sharp line structure that
we would expect from pair production positron emission alone.

\subsubsection{Experiments}\label{heavyexp}

As remarked above, if the sticking time $\Delta t_s$ could be prolonged, the
probability of pair production in vacuum around the superheavy nucleus would be
enhanced. As a consequence, the spectrum of emitted positrons is expected to
develop a sharp line structure, indicating the spontaneous vacuum decay caused
by the overcritical electric field of a forming superheavy nuclear system with
$Z\ge Z_{cr}$. If the sticking time $\Delta t_s$ is not long enough and the
sharp line of pair production positrons has not yet well developed, in the
observed positron spectrum it is difficult to distinguish the pair production
positrons from positrons created through other different mechanisms. Prolonging
the ``sticking time'' and identifying pair production positrons among all other
particles \cite{1982peac.conf.....K,Vincent1983} created in the collision
process  has been an object of a very large experimental campaign
\cite{Kienle1983, Kienle1981, Greenberg1980, 1978PhRvL..40.1443B,
1979PhRvL..42..376K, Muller1983, Bokemeyer1983, Backe1983}.

For nearly 20 years the study of atomic excitation processes and in particular
of positron creation in heavy-ion collisions has been a major research topic at
GSI (Darmstadt) \cite{PhysRevLett.51.2261, 1996PhLB..389....4G,
1997PhLB..394...16L, 1998EPJA....1...27H}. The Orange and Epos groups at GSI
(Darmstadt) discovered narrow line structures (see Fig.~\ref{linestructure}) of
unexplained origin, first in the single positron energy spectra and later in
coincident electron--positron pair emission. Studying more collision systems
with a wider range of the combined nuclear charge $Z=Z_1+Z_2$ they found that
narrow line structures were essentially independent of $Z$. This has ruled out
the explanation as a pair production positron, since the line was expected to
be at the position of the $1s\sigma$ resonance, i.e., at a kinetic energy given
by Eq.~(\ref{zkinetic}), which is strongly $Z$ dependent. Attempts to link this
positron line to spontaneous pair production have failed. Other attempts to
explain this positron line in term of atomic physics and new particle scenario
were not successful as well \cite{Greiner1999}.

\begin{figure}[!ht]
\centering
\includegraphics[width=\hsize,clip]{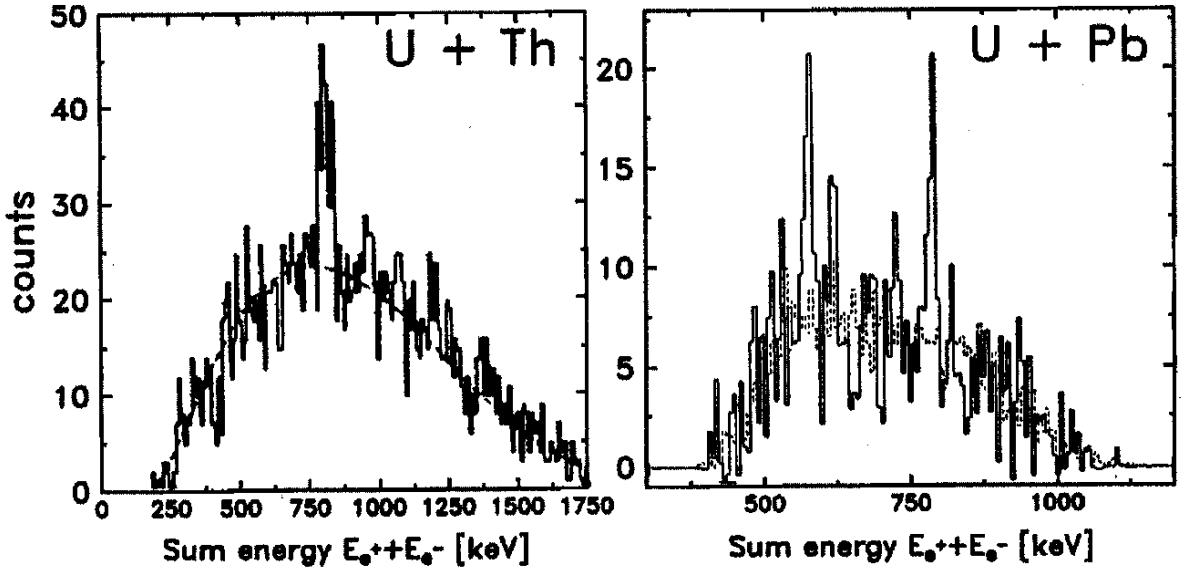}
\caption{Two typical example of coincident electron--positron spectra measured
by the
Epose group in the system U+Th (left) and by the Orange group in U+Pb collisions (right). When plotted as a function of the total energy of the electron and positron, very narrow line structures were observed. This figure is reproduced from Fig.~7 in Ref.~\cite{Greiner1999}.}%
\label{linestructure}%
\end{figure}

The anomalous positron line problem has perplexed experimentalists and
theorists alike for more than a decade. Moreover, later results obtained by the
Apex collaboration at Argonne National Laboratory showed no statistically
significant positron line structures \cite{1995PhRvL..75.2658A,
1997PhRvL..78..618A}. This is in strong contradiction with the former results
obtained by the Orange and Epos groups. However, the analysis of Apex data was
challenged in the comment by Ref.~\cite{PhysRevLett.77.2838,
1996PhRvL..77.2839A} pointing out that the Apex measurement would have been
less sensitive to extremely narrow positron lines. A new generation of
experiments (Apex at Argonne and the new Epos and Orange setups at GSI) with
much improved counting statistics has failed to reproduce the earlier results
\cite{Greiner1999}.

To overcome the problem posed by the short timescale of pair
production ($10^{-21}$ sec), hopes rest on the idea to select
collision systems in which a nuclear reaction with sufficient
sticking time occurs. Whether such a situation can be realized still
is an open question \cite{Greiner1999}. In addition, the anomalous
positron line problem and its experimental contradiction overshadow
the field of the pair production in heavy-ion collisions.

In summary, clear experimental signals for electron--positron pair production
in heavy-ion collisions are still missing \cite{Greiner1999} at the present
time. For more recent information on the pair production in the heavy-ion
collisions see
\cite{2005PhRvL..95g0403K,2006PhRvC..73c1602Z,2007JPhG...34....1Z} and for
complete references the resource letter \cite{2008AmJPh..76..509G}.

Having reviewed the situation of electron--positron pair creation by vacuum
polarization in Earth-bound experiments we turn now to the corresponding
problems in the realm of astrophysics. The obvious case is the one of black
holes where the existence of critical field is clearly predicted by the
analytic solutions of the Einstein-Maxwell field equations.

\section{The extraction of blackholic energy from a black hole by
vacuum polarization processes}
\label{blackhole}

It is becoming more and more clear that the theoretical description of the
gravitational collapse process to a Kerr--Newman black hole, with all the
aspects of nuclear physics and electrodynamics involved, is likely the most
complex problem in physics and astrophysics. Specific to this report is the
opportunity given by the process of gravitational collapse to study for the
first time the above mentioned three quantum processes simultaneously at work
under ultrarelativistic special and general relativistic regimes. The process
of gravitational collapse is characterized by the gravitational timescale
$\Delta t_{g}=GM/c^3\simeq 5\cdot 10^{-6}(M/M_\odot)$ sec, where $G$ is the
gravitational constant, $M$ is the mass of a collapsing object, and the energy
involved is of the order of $\Delta E=10^{54}M/M_\odot$ ergs. This is one of
the most energetic and most transient phenomena in physics and astrophysics and
needs for its correct description the identification of the basic constitutive
processes occurring in a highly time varying regime. Our approach in this
Section is to proceed with an idealized model which can give us estimates of
the basic energetics and some leading features of the real phenomenon. We shell
describe: (1) the basic energetic process of an already formed black hole; (2)
the vacuum polarization process a la Schwinger of an already formed
Kerr--Newman black hole; (3) the basic formula of the dynamics of the
gravitational collapse. We shall in particular recover the
Tolman-Oppenheimer-Snyder solutions in a more explicit form and give exact
analytic solution for the description of the gravitational collapse of charged
and uncharged shells. This will allow, among others, to recall the mass formula
of the black hole, to clarify the special role of the irreducible mass in that
formula, and to have a general derivation of the maximum extractable energy in
the process of gravitational collapse. We will as well address some conceptual
issues between general relativity and thermodynamics which have been of
interest to theoretical physicists in the last forty years. Of course in this
brief Section we will be only recalling some of these essential themes and
refer to the literature where in-depth analysis can be found. Since we are
interested in the gravitational collapse we are going to examine only processes
involving masses larger than the critical mass of neutron stars, for
convenience established in 3.2 $M_\odot$ \cite{1974PhRvL..32..324R}. We are
consequently not addressing the research on mini-black-holes
\cite{2008arXiv0806.3440B} which involves energetics $10^{41}$ times smaller
than the ones involved in gravitational collapse and discussed in this report.
This problematics implies a yet unknown physics of applying quantum mechanics
in conditions where the curvature of space-time is comparable to the wavelength
of the particle.

We recall here the basic steps leading to the study of the electrodynamics of a
Kerr--Newman black hole, indicating the relevant references. In this Section we
use the system of units $c=G=\hbar=1$.

\subsection{Test particles in Kerr--Newman geometries}\label{testKN}%aga

According to the uniqueness theorem for stationary, regular black holes (see
Ref.~\cite{1999magr.meet..136C}), the process of gravitational collapse of a
core whose mass is larger than the neutron star critical mass
\cite{1974PhRvL..32..324R} will generally lead to a black hole characterized by
\textit{all} the three fundamental parameters: the mass-energy $M$, the angular
momentum $L$, and the charge $Q$ (see \cite{1971PhT....24a..30R}). The creation
of critical electric fields and consequent process of pair creation by vacuum
polarization are expected to occur in the late phases of gravitational collapse
when the gravitational energy of the collapsing core is transformed into an
electromagnetic energy and eventually in electron--positron pairs. As of today
no process of the gravitational collapse either to a neutron star or to a black
hole has reached a satisfactory theoretical understanding. It is a fact that
even the theory of a gravitational collapse to a neutron star via a supernova
is not able to explain even the ejection of a supernova remnant
\cite{Mezzacappa2006}. In order to estimate the fundamental energetics of these
transient phenomena we recall first the metric of a Kerr--Newman black hole,
the role of the reversible and irreversible transformations in reaching the
mass formula as well as the role of the positive and negative energy states in
a quantum analog. We will then estimate the energy emission due to vacuum
polarization process. As we will see, such a process occurs on characteristic
quantum timescale of $t\sim \hbar/(m_e c^2)\sim 10^{-21}$ sec, which is many
orders of magnitude shorter than the characteristic gravitational collapse
timescale. Of course the astrophysical progenitor of the black hole will be a
neutral one, as all the astrophysical systems. Only during the process of the
gravitational collapse and for the above mentioned characteristic gravitational
timescale a process of charge separation will occur. The positively charged
core would give rise to the electrodynamical process approaching asymptotically
in time the horizon of a Kerr--Newman black hole.

A generally charged and rotating, black hole has been considered whose curved
space-time is described by the Kerr--Newman geometry
\cite{1965JMP.....6..918N}. In Kerr--Newman coordinates $(u,r,\theta,\phi)$ the
line element takes the form,
\begin{align}
ds^{2}=\  &  \Sigma d\theta^{2}-2a\sin^{2}\theta drd\phi+2drdu-2a\Sigma
^{-1}(2Mr-Q^{2})\sin^{2}\theta d\phi du\nonumber\\
&  +\Sigma^{-1}[(r^{2}+a^{2})^{2}-\Delta a^{2}\sin^{2}\theta]\sin^{2}\theta
d\phi^{2}-[(1-\Sigma^{-1}(2Mr-Q^{2})]du^{2} \label{kerrnewmann}%
\end{align}
where $\Delta=r^{2}-2Mr+a^{2}+Q^{2}$ and $\Sigma=r^{2}+a^{2}\cos^{2}\theta$,
$a=L/M$ being the angular momentum per unit mass of the black hole. The
Reissner--Nordstr\"{o}m and Kerr geometries are particular cases for a
nonrotating, $a=0$, and uncharged, $Q=0$, black holes respectively. The
Kerr--Newman space--time has a horizon at
\begin{equation}
r=r_{+}=M+(M^{2}-Q^{2}-a^{2})^{1/2}%
\end{equation}
where $\Delta=0$.

The electromagnetic vector potential around the Kerr--Newman black hole is
given by \cite{1965JMP.....6..918N}
\begin{equation}
\mathbf{A}=-\Sigma^{-1}Qr(du-a\sin^{2}\theta d\phi),
\end{equation}
the electromagnetic field tensor is then
\begin{align}
\mathbf{F}=  &  \ d\mathbf{A}=2Q\Sigma^{-2}[(r^{2}-a^{2}\cos^{2}%
\theta)dr\wedge du-2a^{2}r\cos\theta\sin\theta d\theta\wedge du\nonumber\\
&  -a\sin^{2}\theta(r^{2}-a^{2}\cos^{2}\theta)dr\wedge d\phi+2ar(r^{2}%
+a^{2})\cos\theta\sin\theta d\theta\wedge d\phi].
\end{align}
The equation of motion of a test particle of mass $m$ and charge $e$ in the
Kerr--Newman geometry reads
\begin{equation}
u^{\mu}\nabla_{\mu}u^{\nu}=(e/m)u^{\mu}{F_{\mu}}^{\nu}, \label{motione}%
\end{equation}
where $u^{\mu}$ is the 4--velocity of the particle. These equations may be
derived from the Lagrangian
\begin{equation}
L=\tfrac{1}{2}mg_{\mu\nu}u^{\mu}u^{\nu}+eA_{\mu}u^{\mu},
\end{equation}
or, equivalently, from the Hamiltonian
\begin{equation}
H=\tfrac{1}{2}g^{\mu\nu}(p_{\mu}-eA_{\mu})(p_{\nu}-eA_{\nu}),
\label{hamiltonian}%
\end{equation}
where we have introduced the 4--momentum of the particle
\begin{equation}
p_{\mu}=mu_{\mu}+eA_{\mu}.
\end{equation}
Note that Hamiltonian (\ref{hamiltonian}) is subject to the constraint
\begin{equation}
H=-\tfrac{1}{2}m^{2}. \label{H}%
\end{equation}
Carter \cite{Carter1968} firstly recognized that the corresponding
Hamilton--Jacobi equations
\begin{equation}
g^{\alpha\beta}\left(  {\frac{\partial S}{\partial x^{\alpha}}}+eA_{\alpha
}\right)  \left(  {\frac{\partial S}{\partial x^{\beta}}}+eA_{\beta}\right)
+m^{2}=0, \label{hamiltonj}%
\end{equation}
are separable. Correspondingly four integrals of the equation of motion
(\ref{motione}) can be found. Indeed, in addition to the constant of motion
(\ref{H}) which corresponds to conservation of the rest mass we have the two
first integrals%
\begin{align}
p_{u}  &  =-{\mathcal E}\label{E}\\
p_{\phi}  &  =\Phi\label{F}%
\end{align}
associated with the stationarity and the axial symmetry of Kerr--Newman
space-time respectively. ${\mathcal E}$ and $\Phi$ are naturally interpreted as
the energy and the angular momentum about the symmetry axis of the test
particle. It follows from the separability of Eq.~(\ref{hamiltonj}) that the
quantities
\begin{align}
p_{\theta}^{2}+(a{\mathcal E}\sin\theta-\Phi\sin^{-1}\theta)^{2}+a^{2}m^{2}\cos^{2}\theta
&  =K\label{K}\\
\Delta p_{r}^{2}-2[(r^{2}+a^{2}){\mathcal E}+eQr-a\Phi]p_{r}+m^{2}r^{2}  &  =-K
\label{-K}%
\end{align}
are conserved as well. Together with ${\mathcal E}$ and $\Phi$ they form a
complete set of first integrals of the motion and allow one to integrate
Eq.~(\ref{motione}). As an example consider the proper time derivative
$\dot{r}$ of the radial coordinate of the test particle. It follows from
Eqs.~(\ref{E}), (\ref{F}), (\ref{K}) and (\ref{-K}), that
\begin{equation}
\Sigma^{2}\dot{r}^{2}=({\mathcal E}(r^{2}+a^{2})+eQr-\Phi a)^{2}-\Delta(m^{2}r^{2}+K)
\label{rdot}%
\end{equation}
which can be numerically integrated using the effective potential technique
\cite{1973blho.conf..451R}.

\subsection{Reversible and irreversible transformations of a black hole:
the Christodoulou-Ruffini mass formula}\label{extractbh}

In 1969 Roger Penrose \cite{1969NCimR...1..252P} pointed out for the first time
the possibility to extract rotational energy from a Kerr black hole. The first
example of such an energy extraction was obtained by Ruffini and Wheeler who
also introduced the concept of the \emph{ergosphere} \cite{1974bhgw.book.....R,
Ruffini2009, 1971PhT....24a..30R}, the region between the horizon of the black
hole and the surface of infinite redshift. These works has been generalized by
Denardo and Ruffini in 1973 \cite{1973PhLB...45..259D} and Denardo, Hively and
Ruffini in 1974 \cite{1974PhLB...50..270D} to the case of a Kerr--Newman black
hole. The process described by Denardo, Hively and Ruffini can be described as
follows. A neutral particle $P_{0}$ approaches the black hole with positive
energy ${\mathcal E}_{0}$ and decays into two oppositely charged particles
$P_{1}$ and $P_{2}$ whose energies are ${\mathcal E}_{1}<0$ and ${\mathcal
E}_{2}>{\mathcal E}_{0}$ respectively. $P_{1}$ falls into the black hole while
$P_{2}$ is accelerated towards spatial infinity. Correspondingly, a positive
energy
\begin{equation}
\delta {\mathcal E}={\mathcal E}_{2}-{\mathcal E}_{0}%
\end{equation}
has been extracted from the black hole and deposited on $P_{2}$. The region
around the black hole where the energy extraction processes can occur is named
\emph{effective ergosphere} in Refs.~\cite{1973PhLB...45..259D},
\cite{1974PhLB...50..270D}. Note that, as the particle $P_{1}$ is swallowed,
the black hole undergoes a transformation since its energy, angular momentum
and charge change accordingly. When is the extracted energy maximal? In order
to answer this question note that the energy ${\mathcal E}$ of a particle of
angular momentum $\Phi$, charge $e$ and rest mass $m$ moving around a
Kerr--Newman black hole and having a turning point at $r$ is given by (see
Eq.~(\ref{rdot})) the quadratic equation
\begin{equation}
(r^{2}+a^{2})^{2}{\mathcal E}^{2}+2(eQr-\Phi a)(r^{2}+a^{2}){\mathcal E}+(eQr-\Phi a)^{2}%
-\Delta(m^{2}r^{2}+K)=0. \label{Energy}%
\end{equation}
As recalled in \cite[p. 352]{1975ctf..book.....L}, in the case of
$Q=0$ which corresponds to a pure Kerr solution, the explicit
integration of this equation was performed by Ruffini and Wheeler
\cite{1971ESRSP..52...45R}. They introduced the effective potential
energy defined by
\begin{eqnarray}
(r^{2}+a^{2})^{2}{\mathcal E}^{2}-2(\Phi a)(r^{2}+a^{2}){\mathcal E}+(\Phi a)^{2}%
-\Delta(m^{2}r^{2}+K)=0. \label{Energy1}%
\end{eqnarray}
The radii of stable orbits are determined by minimum of function
${\mathcal E}(r)$, i.e. by simultaneous solution of equations
${\mathcal E}(r)={\mathcal E}_0$, ${\mathcal E}'(r)=0$ for
${\mathcal E}''(r)>0$. The orbit closest to the center corresponds
to ${\mathcal E}''(r)_{\rm min}=0$; for $r<r_{\rm min}$, the
function ${\mathcal E}(r)$ has no minima. As a result
\begin{itemize}
\item When $\Phi<0$ (motion opposite to the direction of rotation of the collapsing object)
\begin{eqnarray}
\frac{r_{\rm min}}{2M}=\frac{9}{2},\quad \frac{{\mathcal E}_0}{m}=\frac{5}{3\sqrt{3}},\quad \frac{\Phi}{2mM}=\frac{11}{3\sqrt{3}}.
\end{eqnarray}
\item For $\Phi>0$ (motion in the direction of rotation of the collapsing object)
as $a\rightarrow M$ the radius $r_{\rm min}$ tends towards the
radius of the horizon. Setting $a=M(1+\delta)$, we find in the limit
$\delta\rightarrow 0$:
\begin{eqnarray}
\frac{r_{\rm hor}}{2M}=\frac{1}{2}(1+\sqrt{2\delta}),\quad \frac{r_{\rm min}}{2m}=\frac{1}{2}[1+(4\delta)^{1/3}].
\end{eqnarray}
Then
\begin{eqnarray}
\frac{{\mathcal E}_0}{m}=\frac{\Phi}{2mM}=\frac{1}{\sqrt{3}}[1+(4\delta)^{1/3}].
\end{eqnarray}
\end{itemize}
We call attention to the fact that $r_{\rm min}/r_{\rm hor}$ remainsgreater
than one throughout, i.e. the orbit does not go inside the horizon. This is as
it should be: the horizon is a null hypersurface, and no time-like world lines
of moving particles can lie on it. Although no general formula exists in the
case of the Kerr--Newman geometry the energy and the angular velocity of a test
particle in a circular orbit with radius $R$ in the Reissner--Nordstr\"om
geometry has been given by Ruffini and Zerilli \cite{1973blho.conf..451R}
\begin{eqnarray}
\dot\phi^2 &\!\!\!\!\!\!=\!\!\!\!\!\!& \frac{M}{R^3}-\frac{Q^2}{R^4}-\frac{e}{m}\,\frac{Q}{R^3}\left[\frac{e}{m}\,\frac{Q}{2R}+\left(1-\frac{3M}{R}+\frac{2Q^2}{R^2}+\frac{e}{m}\,\frac{Q^2}{4R^2}\right)^{1/2}\right], \\
\frac{{\mathcal E}}{m} &\!\!\!\!\!\!=\!\!\!\!\!\!& \left(1-\frac{2M}{R}+\frac{Q^2}{R^2}\right)/\left[\frac{e}{m}\,\frac{Q}{2R}+\left(1-\frac{3M}{R}+\frac{2Q^2}{R^2}+\frac{e}{m}\,\frac{Q^2}{4R^2}\right)^{1/2}\right]\\
&\!\!\!\!\!\!+\!\!\!\!\!\!&\frac{q}{m}\,\frac{Q}{R}\nonumber,
\end{eqnarray}
and the limiting cases are there treated.

Eq.~(\ref{Energy}) is not only relevant for understanding the fully
relativistic stable circular orbit but it also defines the \textquotedblleft
positive root states\textquotedblright\ and the \textquotedblleft negative root
states\textquotedblright\ for the particle \cite{1971PhRvD...4.3552C}. Such
states were first interpreted as limits of states of a quantum field by
Deruelle and Ruffini \cite{1974PhLB...52..437D}. Such an interpretation will be
discussed in the next section. Note that in the case $eQr-\Phi a <0$ there can
exist negative energy states of positive root solutions and, as a direct
consequence, energy can be extracted from a Kerr--Newman black hole via the
Denardo-Ruffini process. Such a process is most efficient when the reduction of
mass is greatest for a given reduction in angular momentum. To meet this
requirement the energy ${\mathcal E}_{1}$ must be as negative as possible. This
happens when $r=r_{+}$, that is the particle has a turning point at the horizon
of the black hole. When $r=r_{+}$, $\Delta=0$ and the separation between
negative and positive root states vanishes. This implies that capture processes
from such an orbit are reversible since they can be inverted bringing the black
hole to its original state. Correspondingly the energy of the incoming particle
is
\begin{equation}
{\mathcal E}_{1}={\frac{a\Phi+eQr_{+}}{a^{2}+r_{+}^{2}}}. \label{cartenergy}%
\end{equation}

If we apply the conservation of energy, angular momentum and charge to the
capture of the particle $P_{1}$ by the black hole, we find that $M$, $L$ and
$Q$ change as for the quantities
\begin{equation}
dM={\mathcal E}_{1},\quad dL=\Phi,\quad dQ=e.
\end{equation}
Thus Eq.~(\ref{cartenergy}) reads
\begin{equation}
dM={\frac{adL+r_{+}QdQ}{a^{2}+r_{+}^{2}}.}\label{diffener}%
\end{equation}
Integration of Eq.~(\ref{diffener}) gives
\begin{equation}
M^{2}c^{4}=\left(  M_{\mathrm{ir}}c^{2}+{\frac{c^{2}Q^{2}}{4GM_{\mathrm{ir}}}%
}\right)  ^{2}+{\frac{L^{2}c^{8}}{4G^{2}M_{\mathrm{ir}}^{2}}},\label{MassForm}%
\end{equation}
provided the condition is satisfied
\begin{equation}
\left(  {\frac{c^{2}}{16G^{2}M_{\mathrm{ir}}^{4}}}\right)  \left(
Q^{4}+4L^{2}c^{4}\right)  \leq1, \label{s1}
\end{equation}
where $M_{\mathrm{ir}}$ is an integration constant and we restored
the physical constants $c$ and $G$. Eq.~(\ref{MassForm}) is the
Christodoulou--Ruffini mass formula \cite{1971PhRvD...4.3552C} and
it expresses the contributions to the total energy of the black
hole. Extreme black holes satisfy equality (\ref{s1}). The
irreducible mass $M_{\mathrm{ir}}$ satisfies the equation
\cite{1971PhRvD...4.3552C}
\begin{equation}
S_a=\tfrac{16\pi G^{2}M_{\mathrm{ir}}^{2}}{c^{4}}\label{Area}%
\end{equation}
where $S_a$ is the surface area of the horizon of the black hole, and cannot be
decreased by classical processes. Any transformation of the black hole which
leaves fixed the irreducible mass (for instance, as we have seen, the capture
of a particle having a turning point at the horizon of the black hole) is
called reversible \cite{1971PhRvD...4.3552C}. Any transformation of the black
hole which increases its irreducible mass, for instance, the capture of a
particle with nonzero radial momentum at the horizon, is called irreversible.
In irreversible transformations there is always some kinetic energy that is
irretrievably lost behind the horizon. Note that energy can be extracted
approaching arbitrarily close to reversible transformations which are the most
efficient ones. Namely, from Eq.~(\ref{MassForm}) it follows that up to 29$\%$
of the mass-energy of an extreme Kerr black hole ($M^{2}=a^{2}$) can be stored
in its rotational energy term ${\frac{Lc^{4}}{2GM_{\mathrm{ir}}}}$ and can in
principle be extracted. Gedankenexperiments have been conceived to extract such
energy \cite{1969NCimR...1..252P, 1975PhRvD..12.2959R, 1977MNRAS.179..433B,
1978PhRvD..17.1518D, 1986bhmp.book....1P}. The first specific example of a
process of energy extraction from a black hole can be found in R.~Ruffini and
J.~A.~Wheeler, as quoted in \cite{1970PhRvL..25.1596C}, see also
\cite{1971Natur.229..177P}. Other processes of rotational energy extraction of
astrophysical interest based on magnetohydrodynamic mechanism occurring around
a rotating Black Hole have also been advanced \cite{1975PhRvD..12.2959R,
1977MNRAS.179..433B, 1978PhRvD..17.1518D, 1986bhmp.book....1P} though their
reversibility as defined in Ref.~\cite{1971PhRvD...4.3552C}, and consequently
their efficiency of energy extraction, has not been assessed. From the same
mass formula (\ref{MassForm}) follows that up to 50$\%$ of the mass energy of
an extreme black hole with $(Q=M)$ can be stored in the electromagnetic term
${\frac{c^{2}Q^{2}}{4GM_{\mathrm{ir}}}}$ and can be in principle extracted.
These extractable energies either rotational or electromagnetic will be
indicated in the following as blackholic energy and they can be the source of
some of the most energetic phenomena in the Universe like jets from active
galactic nuclei and GRBs.

\subsection{Positive and negative root states as limits of quantum field states}\label{druffini}
In 1974 Deruelle and Ruffini \cite{1974PhLB...52..437D} pointed out that
negative root solutions of Eq.~(\ref{Energy}) can be interpreted in the
framework of a fully relativistic quantum field theory as classical limits of
antimatter solutions. In this section we briefly review their analysis. The
equation of motion of a test particle in a Kerr--Newman geometry can be derived
by the Hamilton--Jacobi Eq.~(\ref{hamiltonj}). The first quantization of the
corresponding theory can be obtained by substituting the Hamilton--Jacobi
equation with the generalized Klein--Gordon equation
\begin{equation}
g^{\alpha\beta}\left(  {\nabla}_{\alpha}+ieA_{\alpha}\right)  \left(  {\nabla
}_{\beta}+ieA_{\beta}\right)  \Phi+m^{2}\Phi=0\label{GKG}%
\end{equation}
for the wave function $\Phi$. For simplicity we restrict to the Kerr case:
$Q=0$, when Eq.~(\ref{GKG}) reduces to
\begin{equation}
g^{\alpha\beta}{\nabla}_{\alpha}{\nabla}_{\beta}\Phi+m^{2}\Phi=0.\label{KG}%
\end{equation}
In order to solve Eq.~(\ref{KG}) we can separate the variables as follows:
\begin{equation}
\Phi=e^{-imEt}e^{ik\phi}S_{kl}(\theta)R(r)
\end{equation}
where $S_{kl}(\theta)$ are spheroidal harmonics. We thus obtain the radial
equation
\begin{align*}
\tfrac{d^{2}u}{dr^{\ast2}}   =&\ \left\{  -E^{2}m^{2}\left(  1+\tfrac{a^{2}%
}{r^{2}}+\tfrac{2Ma^{2}}{r^{3}}\right)  +\tfrac{4MakEm}{r^{3}}+m^{2}%
(1-\tfrac{2M}{r}+\tfrac{a^{2}}{r^{2}})-mk\tfrac{2M}{r^{3}}\right. \\ &-\tfrac{1}{r^{2}}-\tfrac{a^{2}}{r^{4}} \left.  -\tfrac{k^{2}a^{2}}{r^{4}}+\tfrac{2}{r^{6}}\left[  Mr^{3}%
-r^{2}(a^{2}+2M^{2})+3Ma^{2}r-a^{4}\right]  \right\}  u,
\end{align*}
where $u=R(r)r$ and $dr/dr{\ast}=\Delta/r^{2}$. It is natural to look for
\textquotedblleft resonances\textquotedblright\, states of the Klein--Gordon
equation corresponding to classical bound states (circular or elliptic orbits).
Then, impose as boundary conditions (a) an exponential decay of the wave
function for $r\rightarrow\infty$ and (b) a purely ingoing wave at the horizon
$r\rightarrow r_{+}$. The solutions of the corresponding problem can be found
numerically \cite{1974PhLB...52..437D}. The main conclusions of the integration
can be summarized as follows:

\begin{enumerate}
\item The continuum spectrum of the classical stable bound states is replaced
by a discrete spectrum of resonances with tunneling through the potential
barrier giving the finite probability of the particle to be captured by the
horizon.

\item In the classical limit $(GM/c^{2})/(\hbar/m_e c)\rightarrow\infty$ the
separation of the energy levels of the resonances tends to zero. The leakage
toward the horizon also decreases and the width of the resonance tends to zero.

\item The negative root solutions of Eq.~(\ref{Energy}) correspond to the
classical limit $(GM/c^{2})/(\hbar/m_e c)\rightarrow\infty$ of the negative
energy solutions of the Klein--Gordon Eq.~(\ref{KG}) and consequently they can
be thought of as antimatter solutions with an appropriate interchange of the sign of charge, the direction of time and the angular momentum.

\item  We can have positive root states of negative energy in the ergosphere, see e.g.
\cite{1974bhgw.book.....R}. In particular we can have crossing of positive and negative energy
root states. This corresponds, at the second quantized theory level to the
possibility of particle pair creation $\mathit{{\text{\`{a} l\`{a}}}}$ Klein,
Sauter, Heisenberg, Euler and Schwinger \cite{1929ZPhy...53..157K,1931ZPhy...73..547S,1931ZPhy...69..742S,1936ZPhy...98..714H,1951PhRv...82..664S,1954PhRv...93..615S,1954PhRv...94.1362S}.
\end{enumerate}

Similar considerations can be made in the Kerr--Newman case, $Q\neq0$, when the
generalized Klein Gordon Eq.~(\ref{GKG}) has to be integrated. The resonance
states can be obtained imposing the same boundary conditions as above. Once
again we can have level crossing inside the effective ergosphere
\cite{1973PhLB...45..259D,1974PhLB...50..270D} and therefore possible pair
creation.

\subsection{Vacuum polarization in Kerr--Newman geometries}\label{ruffini}

We discussed in the previous Sections the phenomenon of electron--positron pair
production in a strong electric field in a flat space-time. Nere we study the
same phenomenon occurring around a black hole endowed with mass $M$, charge $Q$
and the angular momentum $a$.

The space-time of a Kerr--Newman geometry is described by a metric which in
Boyer--Lindquist coordinates $(t,r,\theta,\phi)$ acquires the form
\begin{equation}
ds^{2}={\frac{\Sigma}{\Delta}}dr^{2}+\Sigma d\theta^{2}-{\frac{\Delta}{\Sigma
}}(dt-a\sin^{2}\theta d\phi)^{2}+{\frac{\sin^{2}\theta}{\Sigma}}\left[
(r^{2}+a^{2})d\phi-adt\right]  ^{2},\label{kerrnewmannBL}%
\end{equation}
where $\Delta$ and $\Sigma$ are defined following (\ref{kerrnewmann}). We
recall that the Reissner--Nordstr{\o}m geometry is the particular case $a=0$ of
a nonrotating black hole.

The electromagnetic vector potential around the Kerr--Newman black hole is
given in Boyer--Lindquist coordinates by
\begin{equation}
\mathbf{A}=-Q\Sigma^{-1}r(dt-a\sin^{2}\theta d\phi).
\label{potentialcurve}
\end{equation}
The electromagnetic field tensor is then
\begin{align}
\mathbf{F}= &\  d\mathbf{A}=   2Q\Sigma^{-2}[(r^{2}-a^{2}\cos^{2}\theta)dr\wedge
dt-2a^{2}r\cos\theta\sin\theta d\theta\wedge dt\nonumber\\
& -a\sin^{2}\theta(r^{2}-a^{2}\cos^{2}\theta)dr\wedge d\phi+2ar(r^{2}%
+a^{2})\cos\theta\sin\theta d\theta\wedge d\phi].
\end{align}

After some preliminary work in Refs.~\cite{1972JETP...35.1085Z,
1973JETP...37...28S, 1976PhRvD..14..870U}, the occurrence of pair production in
a Kerr--Newman geometry was addressed by Deruelle \cite{1977mgm..conf..483D}.
In a Reissner--Nordstr\"om geometry, QED pair production has been studied by
Zaumen \cite{1974Natur.247..530Z} and Gibbons \cite{1975CMaPh..44..245G}. The
corresponding problem of QED pair production in the Kerr--Newman geometry was
addressed by Damour and Ruffini \cite{1975PhRvL..35..463D}, who obtained the
rate of pair production with particular emphasis on:
\begin{itemize}
\item the limitations imposed by pair production on the strength of the
    electromagnetic field of a black hole \cite{1973blho.conf..451R};
\item the efficiency of extracting rotational and Coulomb energy (the ``blackholic''
    energy) from a black hole by pair production;
\item the possibility of having observational consequences of astrophysical interest.
\end{itemize}
In the following, we recall the main results of the work by Damour and Ruffini.

In order to study the pair production in the Kerr--Newman geometry, they
introduced at each event $(t,r,\theta,\phi)$ a local Lorentz frame associated
with a stationary observer ${\mathcal{O}}$ at the event $(t,r,\theta,\phi)$. A
convenient frame is defined by the following orthogonal tetrad
\cite{Carter1968}
\begin{align}
\boldsymbol{\omega}^{(0)} &  =(\Delta/\Sigma)^{1/2}(dt-a\sin^{2}\theta
d\phi),\label{tetrad1}\\
\boldsymbol{\omega}^{(1)} &  =(\Sigma/\Delta)^{1/2}dr,\label{tetrad2}\\
\boldsymbol{\omega}^{(2)} &  =\Sigma^{1/2}d\theta,\label{tetrad3}\\
\boldsymbol{\omega}^{(3)} &  =\sin\theta\Sigma^{-1/2}((r^{2}+a^{2}%
)d\phi-adt).\label{tetrad4}%
\end{align}
In this Lorentz frame, the electric potential $A_{0}$, the electric
field ${\bf E}$ and the magnetic field ${\bf B}$ are given by the following
formulas (c.e.g. Ref.~\cite{1973grav.book.....M}),
\begin{align*}
A_{0} &  =\boldsymbol{\omega}_{a}^{(0)}A^{a},\\
{\bf E}^{\alpha} &  =\boldsymbol{\omega}_{\beta}^{(0)}F^{\alpha\beta},\\
{\bf B}^{\beta} &  ={\frac{1}{2}}\boldsymbol{\omega}_{\gamma}^{(0)}%
\epsilon^{\alpha\gamma\delta\beta}F_{\gamma\delta}.
\end{align*}
We then obtain
\begin{equation}
A_{0}=-Qr(\Sigma\Delta)^{-1/2},\label{gaugepotential}%
\end{equation}
while the electromagnetic fields ${\bf E}$ and ${\bf B}$ are parallel to the
direction of $\boldsymbol{\omega}^{(1)}$ and have strengths given by
\begin{align}
E_{(1)} &  =Q\Sigma^{-2}(r^{2}-a^{2}\cos^{2}\theta),\label{e1}\\
B_{(1)} &  =Q\Sigma^{-2}2ar\cos\theta,\label{b1}%
\end{align}
respectively. The maximal strength $E_{\mathrm{\max}}$ of the electric field
is obtained in the case $a=0$ at the horizon of the black hole: $r=r_{+}$. We have
\begin{equation}
E_{\max}=Q/r_{+}^{2}\label{emax2}.%
\end{equation}
In the original paper a limit on the black hole mass $M_{\mathrm{max}} \simeq
7.2\cdot10^{6}M_{\odot}$ was established by requiring that the pair production
process would last less than the age of the Universe. For masses much smaller
than this absolute maximum mass the pair production process can drastically
modify the electromagnetic structure of black hole.

Both the gravitational and the electromagnetic background fields of the
Kerr--Newman black hole are stationary when considering the quantum field of
the electron. Since $m_eM\simeq 10^{14}\gg1$ the gravitational field of the
background black hole is practically constant over the Compton wavelength of
the electron characterizing the quantum field. As far as purely QED phenomena
such as pair production are concerned, it is possible to consider the electric
and magnetic fields defined by Eqs.~(\ref{e1},\ref{b1}) as constants in the
neighborhood of a few wavelengths around any events $(r,\theta,\phi,t)$. Thus,
the analysis and discussion on the Sauter-Euler-Heisenberg-Schwinger process
over a flat space-time can be locally applied to the case of the curved
Kerr--Newman geometry, based on the equivalence principle.

The rate of pair production around a Kerr--Newman black hole can be obtained
from the Schwinger formula (\ref{probabilityeh}) for parallel electromagnetic
fields $\varepsilon =E_{(1)}$ and $\beta= B_{(1)}$ as:
\begin{equation}
\frac{\tilde\Gamma}{V}={\frac{\alpha E_{(1)}B_{(1)}}{4\pi^{2}}}%
\sum_{n=1}^{\infty}{\frac{1}{n}}%
\coth\left(  {\frac{n\pi B_{(1)}}{E_{(1)}}}\right)  \exp\left(
-{\frac{n\pi E_{\mathrm{c}}}{E_{(1)}}}\right)  .\label{drw}%
\end{equation}
The total number of pairs produced in a region $D$ of the space-time is
\begin{equation}
N=\int_{D}d^{4}x\sqrt{-g}\frac{\tilde\Gamma}{V},\label{drn}%
\end{equation}
where $\sqrt{-g}=\Sigma\sin\theta$. In Ref.~\cite{1975PhRvL..35..463D}, it was
assumed that for each created pair the particle (or antiparticle) with the same
sign of charge as the black hole was expelled to infinity with charge $e$,
energy $\omega$ and angular momentum $l_{\phi}$ while the antiparticle was
absorbed by the black hole. This implies the decrease of charge, mass and
angular momentum of the black hole and a corresponding extraction of all three
quantities. These considerations, however, were profoundly modified later by
the introduction of the concept of dyadosphere which is presented in the next
section. The rates of change of the charge, mass and angular momentum were
estimated by
\begin{align}
\dot{Q} &  =-Re,\nonumber\\
\dot{M} &  =-R\langle\omega\rangle\label{damour1},\\
\dot{L} &  =-R\langle l_{\phi}\rangle,\nonumber
\end{align}
where $R=\dot{N}$ is the rate of pair production and $\langle\omega\rangle$ and
$\langle l_{\phi}\rangle$ represent some suitable mean values for the energy
and angular momentum carried by the pairs.

Supposing the maximal variation of black hole charge to be $\Delta Q=-Q$, one
can estimate the maximal number of pairs created and the maximal mass-energy
variation. It was concluded in Ref.~\cite{1975PhRvL..35..463D} that
the maximal mass-energy variation in the pair production process is larger than
$10^{41}$erg and up to $10^{58}$erg, depending on the black hole mass, see Table 1 in \cite{1975PhRvL..35..463D}.
They concluded at the time ``this work naturally leads to a most simple model for the explanation of the recently discovered $\gamma$-ray bursts''.

\subsection{The ``Dyadosphere'' in Reissner--Nordstr\"{o}m geometry}
\label{dyadosphere}

After the discovery in 1997 of the afterglow of GRBs \cite{1997Natur.387..783C}
and the determination of the cosmological distance of their sources, at once of
the order of $10^3$ theories explaining them were wiped out on energetic
grounds. On the contrary, it was noticed \cite{1998bhhe.conf..167R,
Ruffini2001} the coincidence between their observed energetics and the one
theoretically predicted by Damour and Ruffini \cite{1975PhRvL..35..463D} of
$10^{54}$ ergs per burst for $M=M_\odot$. Ruffini and collaborators therefore,
indirectly motivated by GRBs, returned to these theoretical results with
renewed interest developing some additional basic theoretical concepts
\cite{1998bhhe.conf..167R, 2003JKPSP, 1998A&A...338L..87P, 1999A&A...350..334R,
2000A&A...359..855R, 2008AIPC.1065..289R, 2008AIPC.1059...72R} such as the
dyadosphere and, more recently, the dyadotorus. In this Section we restore
constants $G$, $c$ and $\hbar$ for clarity.

As a first simplifying assumption the case of absence of rotation was
considered. The space-time is then described by the Reissner--Nordstr\"{o}m
geometry, see (\ref{kerrnewmannBL}) whose spherically symmetric metric is given
by
\begin{equation}
d^2s=g_{tt}(r)d^2t+g_{rr}(r)d^2r+r^2d^2\theta +r^2\sin^2\theta
d^2\phi ~,
\label{s}
\end{equation}
where $g_{tt}(r)= - \left[1-\frac{2GM}{c^2r}+\frac{Q^2G}{c^4r^2}\right] \equiv
- \alpha^2(r)$ and $g_{rr}(r)= \alpha^{-2}(r)$.

The first result obtained is that the pair creation process does not occur at
the horizon of the black hole: it extends over the entire region outside the
horizon in which the electric field exceeds the value $E^\star$ of the order of
magnitude of the critical value given by Eq.~(\ref{critical1}). We recall the
pair creation process is a quantum tunneling between the positive and negative
energy states, which needs a level crossing, can occur for $E^\star < E_c$ as
well, if the field extent to spatial dimension $D^*$ such that
$D^*E^*=2m_ec^2/e$. The probability of such pair creation process will be
exponentially damped by $exp(-\pi D^*/\lambda_c)$. Clearly, very intense
process of pair creation will occur for $E^*>E_c$. In order to give a scale of
the phenomenon, and for definiteness, in Ref.~\cite{1998A&A...338L..87P} it was
considered the case of $E^\star \equiv E_c$, although later in order to take
into due account the tunneling effects we have considered dyadosphere for
electric field in the range
\begin{equation}
E^\star=\kappa E_c,
\label{kappa}
\end{equation}
with $\kappa$ in the range 0.1-10. Since the electric field in the
Reissner--Nordstr\"{o}m geometry has only a radial component given by
\cite{1978pans.proc..287R}
\begin{equation}
E\left(r\right)=\frac{Q}{r^2}\, ,
\label{edir}
\end{equation}
this region extends from the horizon radius, for $\kappa=1$
\begin{eqnarray}
r_{+}&=&1.47 \cdot 10^5\mu (1+\sqrt{1-\xi^2})\hskip0.1cm {\rm cm}
\label{r+}
\end{eqnarray}
out to an outer radius \cite{1998bhhe.conf..167R}
\begin{align}
r^\star=\left(\frac{\hbar}{m_ec}\right)^{1/2}\left(\frac{GM}{
c^2}\right)^{1/2} \left(\frac{m_{\rm p}}{m_e}\right)^{1/2}\left(\frac{e}
{q_{\rm p}}\right)^{1/2}\left(\frac{Q}{\sqrt{G}M}\right)^{1/2}= \nonumber \\
=1.12\cdot 10^8\sqrt{\mu\xi} \hskip0.1cm {\rm cm},
\label{rc}
\end{align}
where we have introduced the dimensionless mass and charge parameters $\mu=\frac{M} {M_{\odot}}$, $\xi=\frac{Q}{(M\sqrt{G})}\le 1$, see Fig.~\ref{dyaon}.

The second result gave the local number density of electron and positron pairs created in this region as a function of radius
\begin{equation}
n_{e^+e^-}(r) = \frac{Q}{4\pi r^2\left(\frac{\hbar}{
m_e c}\right)e}\left[1-\left(\frac{r}{r^\star}\right)^2\right] ~,
\label{nd}
\end{equation}
and consequently the total number of electron and positron pairs in this region is
\begin{equation}
N^\circ_{e^+e^-}\simeq \frac{Q-Q_c}{e}\left[1+\frac{
(r^\star-r_+)}{\frac{\hbar}{m_e c}}\right],
\label{tn}
\end{equation}
where $Q_c = E_c r_+^2$.

\begin{figure}[!ht]
\includegraphics[width=10cm,clip]{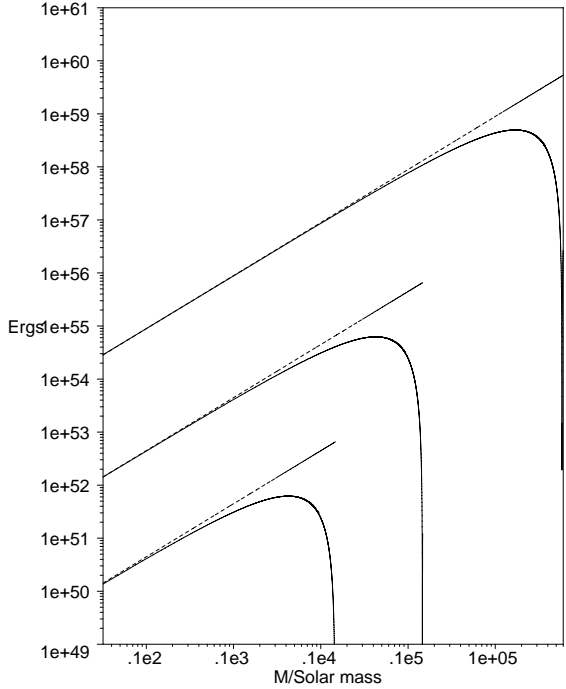}
\caption{The energy extracted by the process of vacuum polarization is plotted (solid lines) as a function of the mass $M$ in solar mass units for selected values of the charge parameter $\xi=1,0.1,0.01$ (from top to bottom) for a Reisner-Nordstr\"om black hole, the case $\xi=1$ reachable only as a limiting process. For comparison we have also plotted the maximum energy extractable from a black hole (dotted lines) given by Eq.~(\ref{MassForm}). Details in Ref.~\cite{2003JKPSP}.}
\label{prep}
\end{figure}

\begin{figure}[!ht]
\includegraphics[width=10cm,clip]{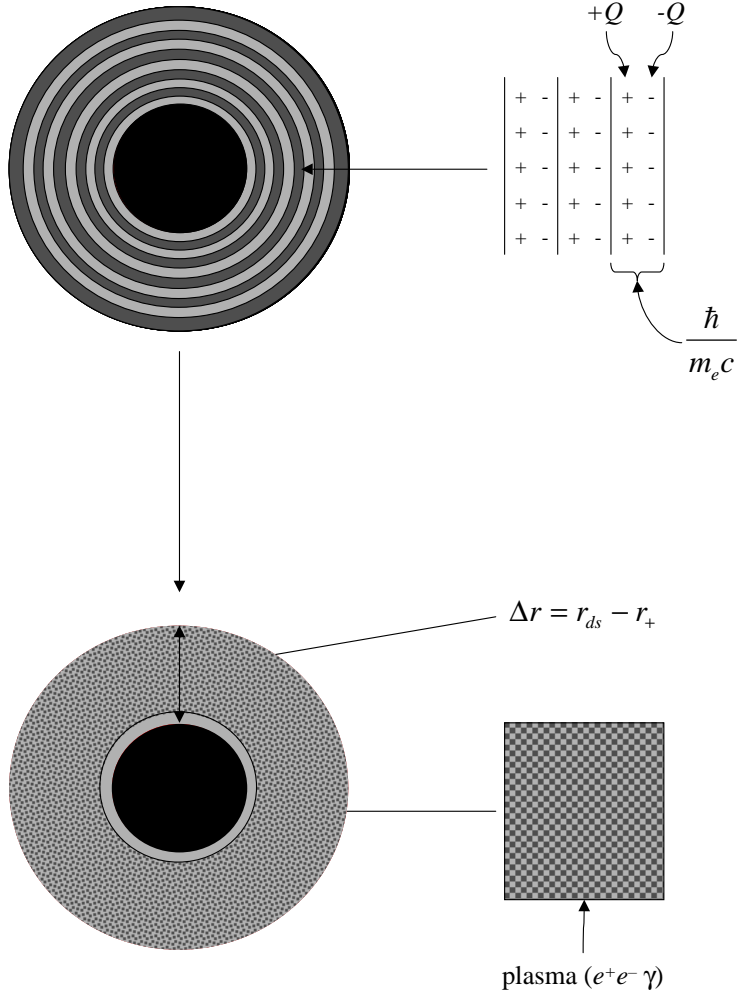}
\caption{The dyadosphere of a Reissner--Nordstr\"{o}m black hole can be
represented as equivalent to a concentric set of capacitor shells, each one of
thickness $\hbar/m_ec$ and producing a number of $e^+e^-$ pairs of the order of
$\sim Q/e$ on a time scale of $10^{-21}$ s, where $Q$ is the black hole charge.
The shells extend in a region of thickness $\Delta r$, from the horizon $r_{+}$
out to the Dyadosphere outer radius $r_{\rm ds}$ (see text). The system evolves
to a thermalized plasma configuration.} \label{dyaon}
\end{figure}

The total number of pairs is larger by an enormous factor
$r^{\star}/\left(\hbar/m_ec\right) > 10^{18}$ than the value $Q/e$ which a
naive estimate of the discharge of the black hole would have predicted. Due to
this enormous amplification factor in the number of pairs created, the region
between the horizon and $r^{\star}$ is dominated by an essentially high density
neutral plasma of electron--positron pairs. This region was defined
\cite{1998bhhe.conf..167R} as the dyadosphere of the black hole from the Greek
duas, duados for pairs. Consequently we have called $r^\star$ the dyadosphere
radius $r^\star \equiv r_{\rm ds}$ \cite{1998bhhe.conf..167R, 2003JKPSP,
1998A&A...338L..87P}. The vacuum polarization process occurs as if the entire
dyadosphere is subdivided into a concentric set of shells of capacitors each of
thickness $\hbar/m_ec$ and each producing a number of $e^+e^-$ pairs on the
order of $\sim Q/e$ (see Fig.~\ref{dyaon}). The energy density of the
electron--positron pairs is there given by
\begin{equation}
\epsilon(r) = \frac{Q^2}{8 \pi r^4} \biggl(1 - \biggl(\frac{r}{r_{\rm
ds}}\biggr)^4\biggr) ~, \label{jayet}
\end{equation}
(see Figs.~2--3 of Ref.~\cite{2003JKPSP}). The total energy of pairs
converted from the static electric energy and deposited within the
dyadosphere is then
\begin{equation}
{\mathcal E}_{\rm dya}=\frac{1}{2}\frac{Q^2}{r_+} \left(1-\frac{r_+}{r_{\rm
ds}}\right)\left[1-\left(\frac{r_+}{r_{\rm ds}}\right)^4\right] ~. \label{tee}
\end{equation}

In the limit $\frac{r_+}{r_{\rm ds}}\rightarrow 0$, Eq.~(\ref{tee})
leads to ${\mathcal E}_{\rm dya}\rightarrow
\frac{1}{2}\frac{Q^2}{r_+}$, which coincides with the energy
extractable from black holes by reversible processes ($M_{\rm
ir}={\rm const.}$), namely ${\mathcal E}_{BH}-M_{\rm
ir}=\frac{1}{2}\frac{Q^2}{r_+}$\cite{1971PhRvD...4.3552C}, see
Fig.~\ref{prep}. Due to the very large pair density given by
Eq.~(\ref{nd}) and to the sizes of the cross-sections for the
process $e^+e^-\leftrightarrow \gamma+\gamma$, the system has been
assumed to thermalize to a plasma configuration for which
\begin{equation}
n_{e^+}=n_{e^-} \sim n_{\gamma} \sim n^\circ_{e^+e^-},
\label{plasma}
\end{equation}
where $n^\circ_{e^+e^-}$ is the total number density of
$e^+e^-$-pairs created in the dyadosphere
\cite{2003JKPSP,1998A&A...338L..87P}.
\begin{figure}[!ht]
   \resizebox{\hsize}{8cm}{\includegraphics{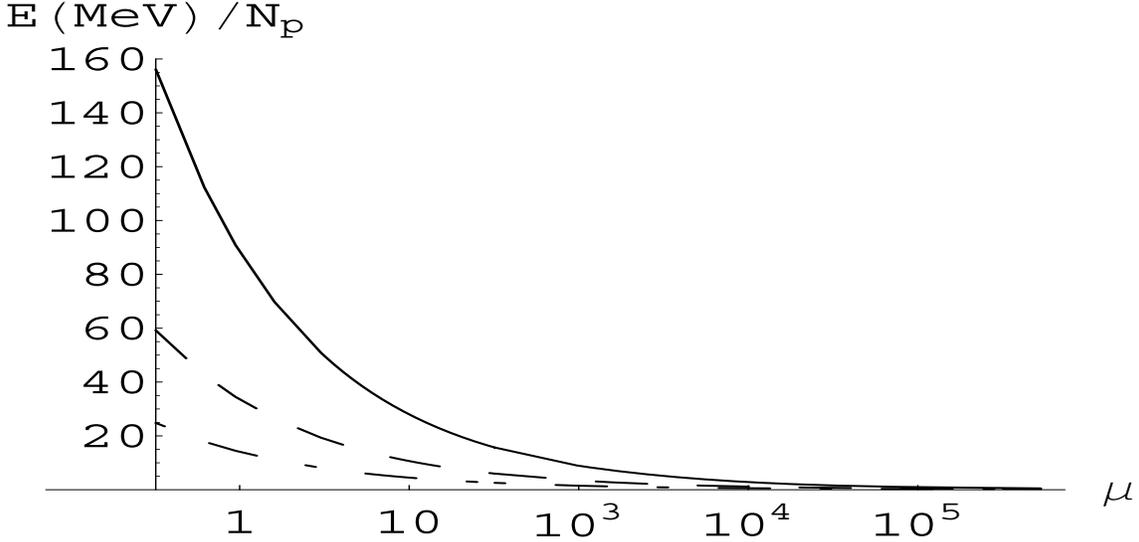}}
      \caption[]{The average energy per pair is shown here as a function of
the black hole mass in solar mass units
for $\xi=1$ (solid line), $\xi=0.5$ (dashed line) and
$\xi=0.1$ (dashed and dotted line).}
         \label{fig.4}
   \end{figure}
In Fig. \ref{fig.4} we show the average energy per pair as a
function of the black hole mass in solar mass units
\cite{1998A&A...338L..87P}. This assumption has been in the meantime
rigorously proven by Aksenov, Ruffini and Vereshchagin
\cite{2007PhRvL..99l5003A}, see Section \ref{aksenov}.

The third result, again introduced for simplicity, is that for a
given ${\mathcal E}_{\rm dya}$ it was assumed either a constant
average energy density over the entire dyadosphere volume, or a more
compact configuration with energy density equal to its peak value.
These are the two possible initial conditions for the evolution of
the dyadosphere (see Fig.~\ref{dens}).

\begin{figure}[!ht]
\centering
\includegraphics[width=0.49\hsize,clip]{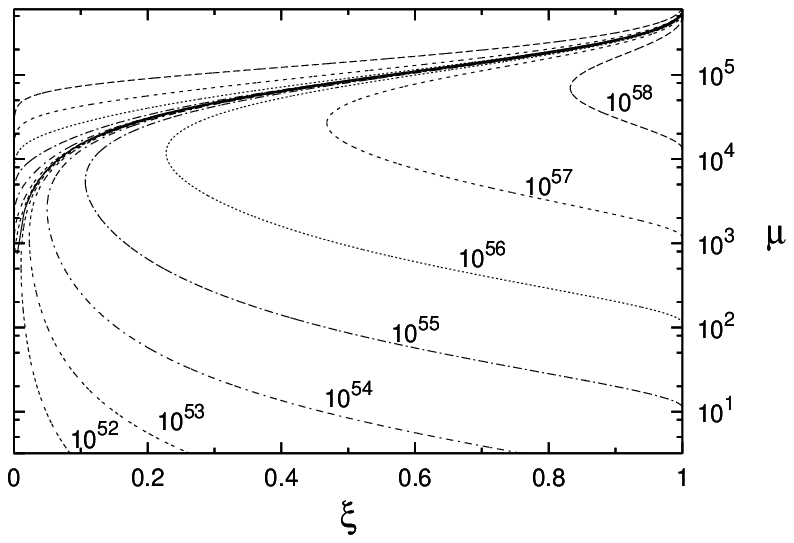}
\includegraphics[width=0.49\hsize,clip]{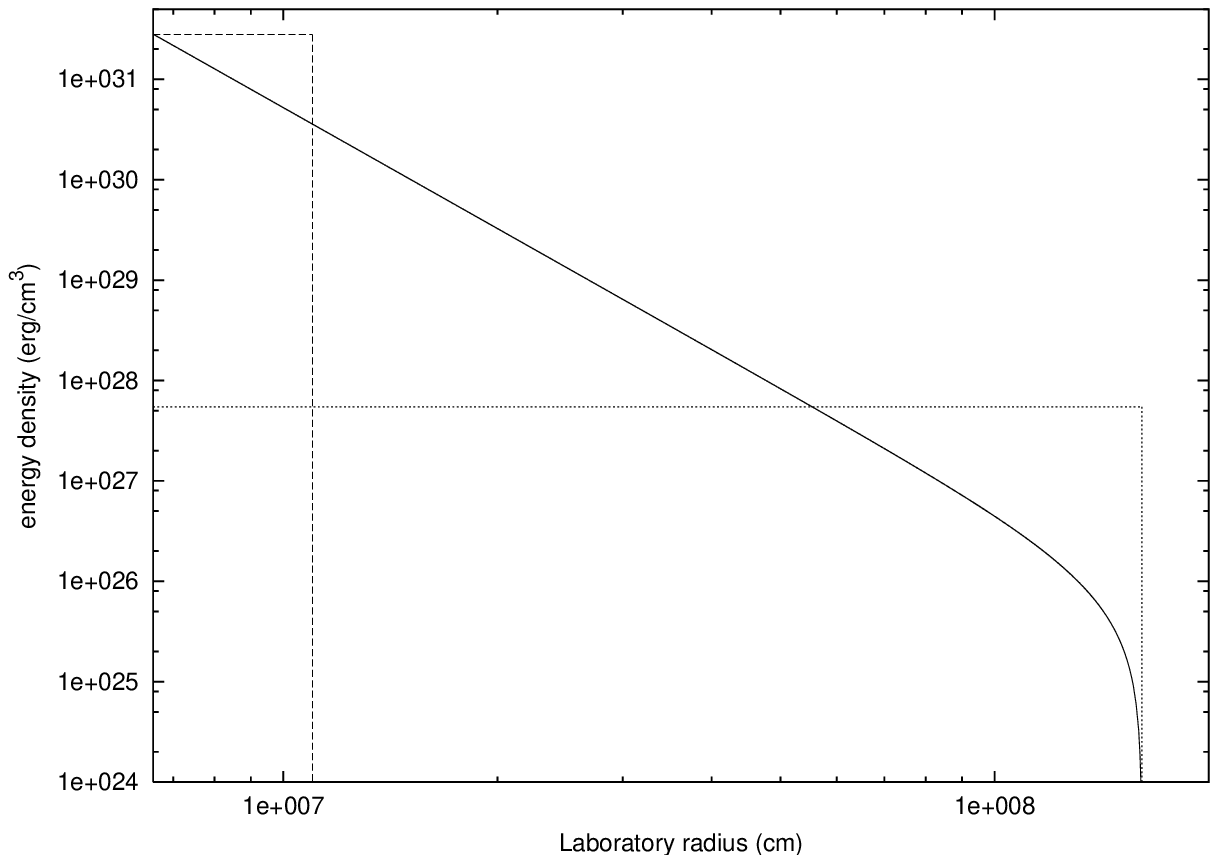}
\caption{{\bf Left)} Selected lines corresponding to fixed values of the ${\mathcal E}_{\rm dya}$ are given as a function of the two parameters $\mu$ $\xi$, only the solutions below the continuous heavy line are physically relevant. The configurations above the continuous heavy lines correspond to unphysical solutions with $r_{\rm ds} < r_+$. {\bf Right)} There are two different approximations for the energy density profile inside the Dyadosphere. The first one (dashed line) fixes the energy density equal to its peak value, and computes an ``effective'' Dyadosphere radius accordingly. The second one (dotted line) fixes the Dyadosphere radius to its correct value, and assumes a uniform energy density over the Dyadosphere volume. The total energy in the Dyadosphere is of course the same in both cases. The solid curve represents the real energy density profile. Details in \cite{Ruffini2003}.}
\label{muxi}
\label{dens}
\end{figure}

The above theoretical results permit a good estimate of the general
energetics processes originating in the dyadosphere, assuming an
already formed black hole and offer a theoretical framework to
estimate the general relativistic effect and characteristic time
scales of the approach to the black hole horizon
\cite{2002PhLB..545..226C,2002PhLB..545..233R,2003IJMPD..12..121R,2003PhLB..573...33R,2005IJMPD..14..131R,2006NCimB.121.1477F}.

\subsection{The ``dyadotorus''}\label{dyadotorus}

We turn now to examine how the presence of rotation modifies the geometry of
the surface containing the region where electron--positron pairs are created as
well as the conditions forthe existence of such a surface. Due to the axial
symmetry of the problem, this region was called the ``dyadotorus''
\cite{2008AIPC..966..123C,2009PhRvD..79l4002C}.

We shall follow the treatment of \cite{2008AIPC..966..123C,
2009PhRvD..79l4002C}. As in Damour \cite{1982mgm..conf..587D,
1978PhRvD..17.1518D} we introduce at each point of the space-time the
orthogonal Carter tetrad (\ref{tetrad1}-\ref{tetrad4}).

\begin{figure}[htp]
    \centering
        \includegraphics{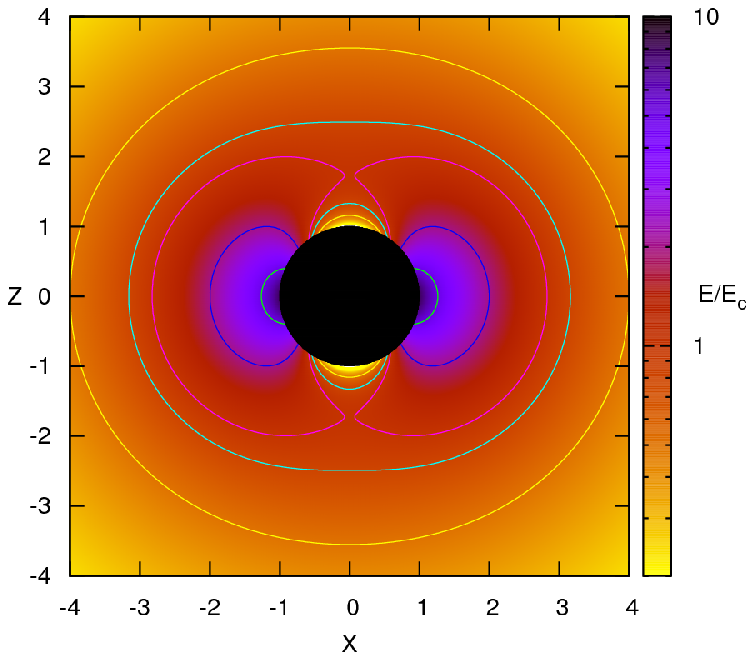}
        \includegraphics{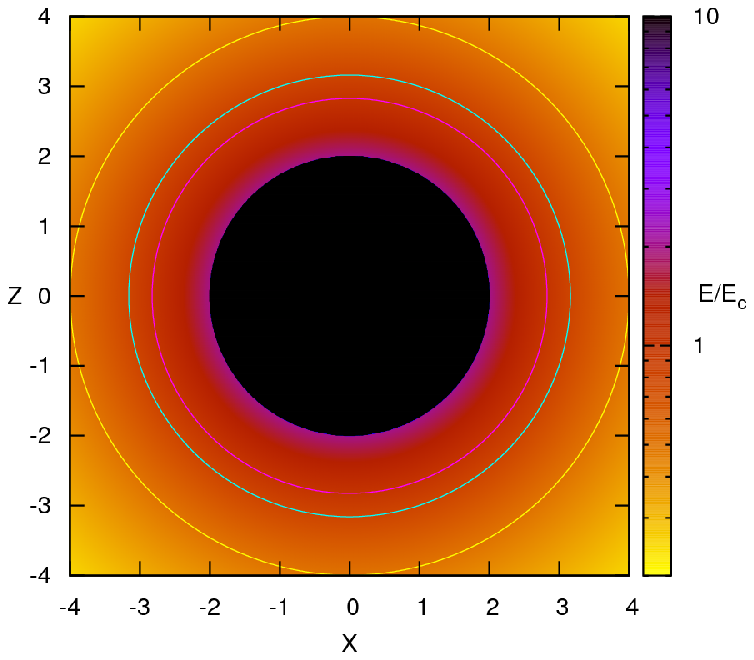}
    \caption{The projections of the dyadotorus on the X-Z plane corresponding to different values of the ratio $E/E_c=\kappa$ are shown (upper panel) for $\mu=10$ and $\lambda=1.49\times10^4$. The corresponding plot for the dyadosphere with the same mass energy and charge to mass ratio is shown in the lower panel for comparison. Reproduced from \cite{2009PhRvD..79l4002C}.}
    \label{figdyadosphere}
\end{figure}
From Eq. (\ref{e1}) we define the dyadotorus by the condition $|E_{(1)}|=\kappa E_c$, where
$10^{-1} \leq \kappa \leq 10$, see Fig. \ref{figdyadosphere}.  Solving for $r$ and introducing the
dimensionless quantities $\xi=Q/(\sqrt{G}M)$, $\mu=M/M_\odot$, $\tilde\alpha=ac^2/(GM)$, $\tilde {\mathcal E}=\kappa\,E_c\,M_\odot c^4/G^{3/2}$ and $\tilde{r}=rc^2/(GM)$ we get
\begin{equation}\label{dyadosurf}
\left(\frac{r^d_\pm c^2}{GM}\right)^2=\frac{\xi}{2\mu \tilde {\mathcal
E}}-\tilde\alpha^2\cos^2\theta\pm \sqrt{\frac{\xi^2}{4\mu^2 \tilde {\mathcal E}^2}
-\frac{2\xi}{\mu \tilde {\mathcal E}}\tilde\alpha^2\cos^2\theta}\, ,
\end{equation}
where the $\pm$ signs correspond to the two different parts of the surface.

The two parts of the surface join at the particular values $\theta^*$ and
$\pi-\theta^*$ of the polar angle where

\[
\theta^*=\arccos\left(\frac{1}{2\sqrt{2}\tilde\alpha}\sqrt{\frac{\xi}{\mu{\mathcal
E}}}\right).
\]

The requirement that $\cos\theta^*\leq1$ can be solved for instance for the
charge parameter $\xi$, giving a range of values of $\xi$ for which the
dyadotorus takes one of the shapes (see fig.\ref{figure1})
\begin{equation}
\textrm{surface}=
\begin{cases}
\textrm{ellipsoid--like} & \textrm{if}\,\, \xi \geq \xi_{*}\\
\textrm{thorus--like} & \textrm{if}\,\, \xi < \xi_{*}
\end{cases}
\end{equation}
where $\xi_{*}=8 \mu \tilde {\mathcal E} \tilde\alpha^2$.

In Fig.~\ref{figure1} we show some examples of the dyadotorus geometry for
different sets of parameters for an extreme Kerr--Newman black hole
($a^2c^8/G^2+Q^2/G=M^2$), we can see the transition from a toroidal geometry to
an ellipsoidal one depending on the value of the black hole charge.

\begin{figure}[!htbp]
\centering
$\begin{array}{cc}
\includegraphics[width=0.45\hsize]{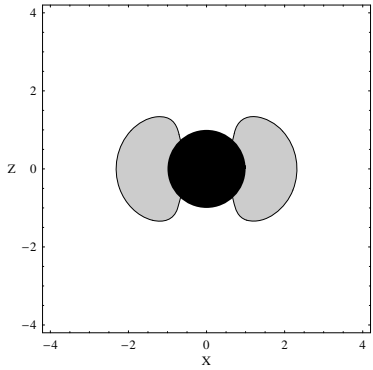}&\quad
\includegraphics[width=0.45\hsize]{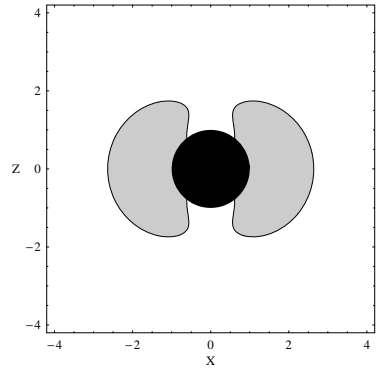}\\[0.4cm]
\mbox{(a)} & \mbox{(b)}\\[0.6cm]
\includegraphics[width=0.45\hsize]{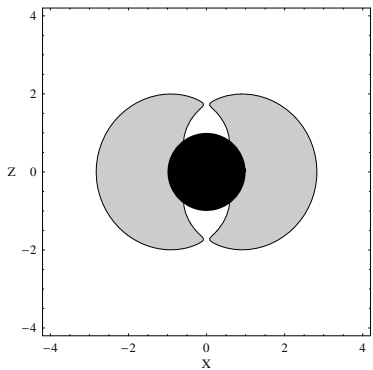}&\quad
\includegraphics[width=0.45\hsize]{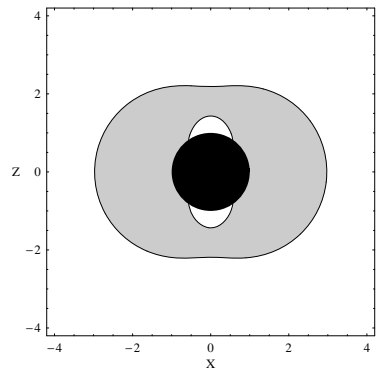}\\[0.4cm]
\mbox{(c)} & \mbox{(d)}\\
\end{array}$
\caption{The projection of the dyadotorus on the $X-Z$ plane ($X=r\sin \theta$,
$Z=r\cos \theta$ are Cartesian-like coordinates built up simply using the
Boyer-Lindquist radial and angular coordinates) is shown for an extreme
Kerr--Newman black hole with $\mu=10$ and different values of the charge
parameter $\xi=[1,1.3,1.49,1.65]\times10^{-4}$ (from (a) to (d) respectively).
 The black circle represents the black hole horizon. Details in \cite{2008AIPC..966..123C,2009PhRvD..79l4002C}.}\label{figure1}
\end{figure}

Fig.~\ref{figure2} shows the projections of the surfaces
corresponding to different values of the ratio $|E_{(1)}|/E_c\equiv \kappa$ for
the same choice of parameters as in Fig.~\ref{figure1} (b), as an example. We
see that the region enclosed by such surfaces shrinks for increasing values of
$\kappa$.

\begin{figure}[!ht]
\centering
\includegraphics[scale=0.45]{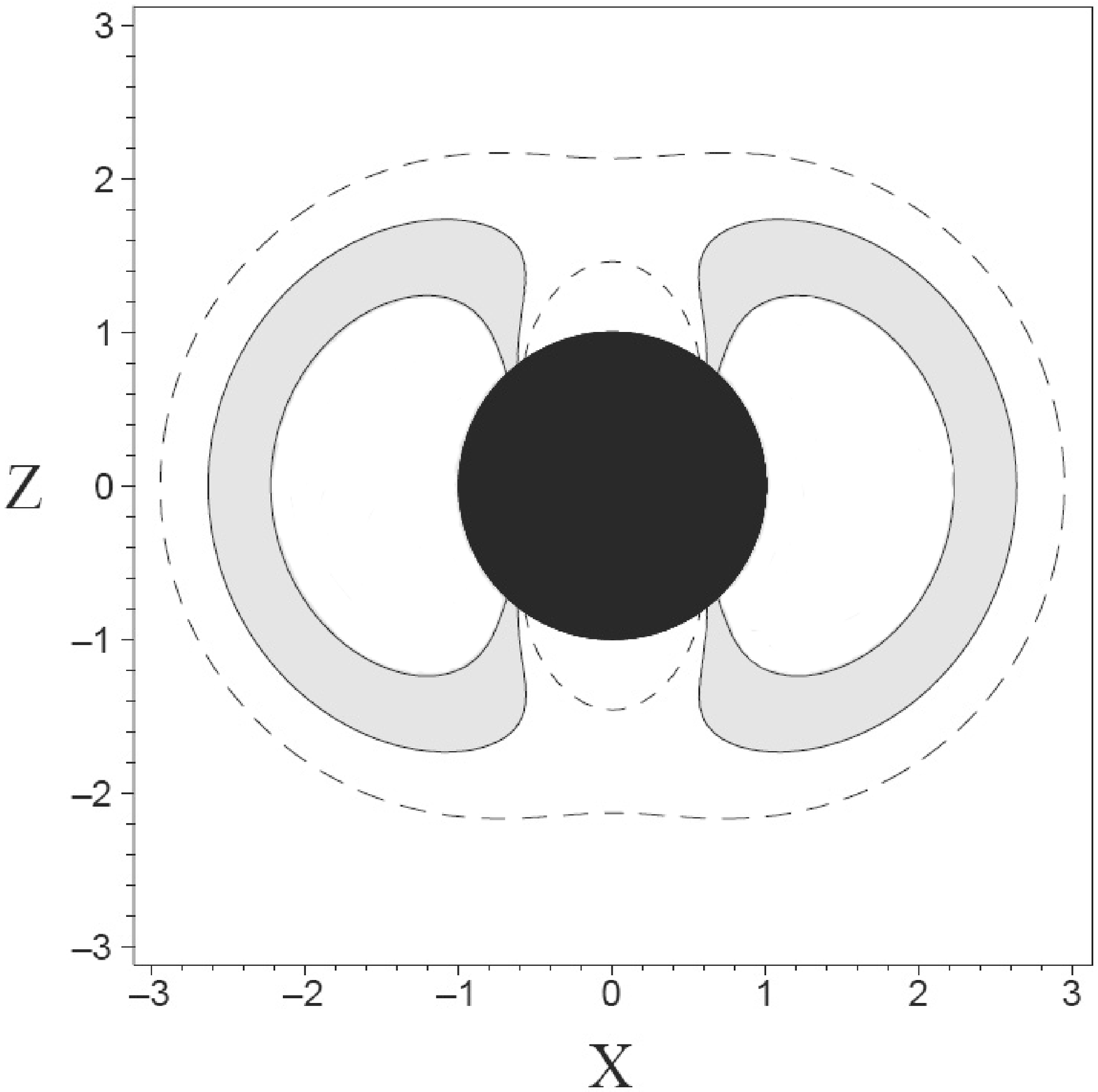}
\caption{The projections of the surfaces corresponding to different values of
the ratio $|E_{(1)}|/E_c\equiv \kappa$ are shown for the same choice of parameters
as in Fig.~\ref{figure1} (b), as an example.  The gray shaded region is part of
the ``dyadotorus'' corresponding to the case $\kappa=1$ as plotted in Fig.
\ref{figure1} (b).  The region delimited by dashed curves corresponds to
$\kappa=0.8$, i.e., to a value of the strength of the electric field smaller
than the critical one, and contains the dyadotorus; the latter in turn contains
the white region corresponding to $\kappa=1.4$, i.e., to a value of the strength
of the electric field greater than the critical one. Details in \cite{2008AIPC..966..123C}.} \label{figure2}
\end{figure}

Equating (\ref{e1}) and (\ref{kappa}) for $\theta=\pi/2$ and $\tilde\alpha=1$ we get
\begin{equation}
\mu=\frac{\xi}{\kappa}\times 5\times 10^5.
\end{equation}

\subsection{Geometry of gravitationally collapsing cores}\label{gravcoll}

In the previous Sections we have focused on the theoretically well
defined problem of pair creation in the electric field of an already
formed black hole. In this section we shall follow the treatment of
Cherubini et al. \cite{2002PhLB..545..226C} addressing some specific
issues on the dynamical formation of the black hole, recalling first
the Oppenheimer-Snyder solution and then considering its
generalization to the charged case using the classical work of W.
Israel and V. de la Cruz
\cite{1966NCimB..44....1I,1967NCimA..51..744C}.

\subsubsection{The Tolman-Oppenheimer-Snyder solution}\label{TOSsolution}

Oppenheimer and Snyder first found a solution of the Einstein equations
describing the gravitational collapse of spherically symmetric star of mass
greater than $\sim0.7M_{\odot}$. In this section we briefly review their
pioneering work as presented in Ref. \cite{1939PhRv...56..455O}.

In a spherically symmetric space-time such coordinates can be found
$(t,r,\theta,\phi)$ that the line element takes the form%aga
\begin{equation}
ds^{2}=e^{\nu}dt^{2}-e^{\lambda}dr^{2}-r^{2}d\Omega^{2},\label{E0}%
\end{equation}
$d\Omega^{2}=d\theta^{2}+\sin^{2}\theta d\phi^{2}$, $\nu=\nu(t,r)$,
$\lambda=\lambda(t,r)$. However the gravitational collapse problem is better
solved in a system of coordinates $(\tau,R,\theta,\phi)$ which are comoving
with the matter inside the star. In comoving coordinates the line element
takes the form
\[
ds^{2}=d\tau^{2}-e^{\sigma}dR^{2}-e^{\omega}d\Omega^{2},
\]
$\overline{\omega}=\overline{\omega}(\tau,R)$, $\omega=\omega(\tau,R)$.
Einstein equations read
\begin{align}
8\pi T_{1}^{1}  & =e^{-\omega}-e^{-\sigma}\tfrac{\omega^{\prime2}}{4}%
+\ddot{\omega}+\tfrac{3}{4}\dot{\omega}^{2}\label{E1}\\
8\pi T_{2}^{2}  & =8\pi T_{3}^{3}=-\tfrac{e^{-\sigma}}{4}\left(
2\omega^{\prime\prime}+\omega^{\prime2}-\sigma^{\prime}\omega^{\prime}\right)
\nonumber\\
& +\tfrac{1}{4}(2\ddot{\sigma}+\dot{\sigma}^{2}+2\ddot{\omega}+\dot{\omega
}^{2}+\dot{\sigma}\dot{\omega})\label{E2}\\
8\pi T_{4}^{4}  & =e^{-\omega}-e^{-\sigma}\left(  \omega^{\prime\prime}%
+\tfrac{3}{4}\omega^{\prime2}-\tfrac{\sigma^{\prime}\omega^{\prime}}%
{2}\right)  +\tfrac{\dot{\omega}^{2}}{4}+\tfrac{\dot{\sigma}\dot{\omega}}%
{2}\label{E3}\\
8\pi e^{\sigma}T_{4}^{1}  & =-8\pi T_{1}^{4}=\tfrac{1}{2}\omega^{\prime}%
(\dot{\omega}-\dot{\sigma})+\dot{\omega}^{\prime}.\label{E4}%
\end{align}
Where $T_{\mu\nu}$ is the energy--momentum tensor of the stellar
matter, a dot denotes a derivative with respect to $\tau$ and a
prime denotes a derivative with respect to $R$. Oppenheimer and
Snyder were only able to integrate Eqs. (\ref{E1})--(\ref{E4}) in
the case when the pressure $p$ of the stellar matter vanishes and no
energy is radiated outwards. In the following we thus put $p=0$.
Under this hypothesis
\[
T_{1}^{1}=T_{2}^{2}=T_{3}^{3}=T_{4}^{1}=T_{1}^{4}=0,\quad T_{4}^{4}=\rho
\]
where $\rho$ is the comoving density of the star. Eq.~(\ref{E4}) was first
integrated by Tolman in Ref. \cite{1934PNAS...20..169T}. The solution is
\begin{equation}
e^{\sigma}=e^{\omega}\omega^{\prime2}/4f^{2}(R),\label{E5}%
\end{equation}
where $f=f(R)$ is an arbitrary function. In Ref. \cite{1939PhRv...56..455O} the
case of $f(R)=1$ was studied. In Section~\ref{collapse} below the hypothesis
$f(R)=1$ is relaxed in the case of a shell of dust. Substitution of
Eq.~(\ref{E5}) into Eq.~(\ref{E1}) with $f(R)=1$ gives
\begin{equation}
\ddot{\omega}+\tfrac{3}{4}\dot{\omega}^{2}=0,\label{E6}
\end{equation}
which can be integrated to give
\begin{equation}
e^{\omega}=(F\tau+\tilde G)^{4/3},\label{E7}%
\end{equation}
where $F=F(R)$ and $\tilde G=\tilde G(R)$ are arbitrary functions.
Substitution of Eq.~(\ref{E5}) into Eq.~(\ref{E2}) gives
Eq.~(\ref{E6}) again. From Eqs.~(\ref{E3}), (\ref{E5}) and
(\ref{E7}) the density $\rho$ can be found as
\begin{equation}
8\pi\rho=\tfrac{4}{3}\left(  \tau+\tfrac{\tilde G}{F}\right)  ^{-1}\left(
\tau+\tfrac{\tilde G^{\prime}}{F^{\prime}}\right)  ^{-1}.\label{E8}%
\end{equation}
There is still the gauge freedom of choosing $R$ so to have
\[
\tilde G=R^{3/2}.
\]
Moreover, arbitrary initial density profile can be chosen, i.e., for
the density at the initial time $\tau=0$, $\rho_{0}=\rho_{0}(R)$.
Eq.~(\ref{E8}) then becomes
\[
FF^{\prime}=9\pi R^{2}\rho_{0}(R)
\]
whose solution contains only one arbitrary integration constant. It
is thus seen that the choice of Oppenheimer and Snyder of $f(R)=1$
allows one to assign only a 1-parameter family of functions for the
initial values $\dot{\rho}_{0}=\dot{\rho}_{0}(R)$ of $\dot{\rho}$.
However in general one should be able to assign the initial values
of $\dot{\rho}$ arbitrarily. This will be done in
Section~(\ref{collapse}) in the case of a shell of dust.

Choosing, for instance,
\[
\rho_{0}=\left\{
\begin{array}
[c]{cc}%
\mathrm{const}>0 & \text{if }R<R_{b}\\
0 & \text{if }R\geq R_{b}%
\end{array}
\right.  ,
\]
$R_{b}$ being the comoving radius of the boundary of the star, gives
\[
F=\left\{
\begin{array}
[c]{cc}%
-\tfrac{3}{2}r_{+}^{1/2}\left(  \tfrac{R}{R_{b}}\right)  ^{3/2} & \text{if
}R<R_{b}\\
-\tfrac{3}{2}r_{+}^{1/2} & \text{if }R\geq R_{b}%
\end{array}
\right.
\]
where $r_{+}=2M$ is the Schwarzschild radius of the star.

We are finally in the position of performing a coordinate
transformation from the comoving coordinates $(\tau,R,\theta,\phi)$
to new coordinates $(t,r,\theta,\phi)$ in which the line elements
looks like (\ref{E0}). The requirement that the line element be the
Schwarzschild one outside the star fixes the form of such a
coordinate transformation to be
\begin{align*}
r  & =(F\tau+G)^{2/3}\\
t  & =\left\{
\begin{array}
[c]{cc}%
\tfrac{2}{3}r_{+}^{-1/2}(R_{b}^{3/2}-r_{+}^{3/2}y^{3/2})-2r_{+}y^{1/2}%
+r_{+}\log\tfrac{y^{1/2}+1}{y^{1/2}-1} & \text{if }R<R_{b}\\
\tfrac{2}{3r_{+}^{1/2}}(R^{3/2}-r^{3/2})-2(rr_{+})^{1/2}+r_{+}\log
\tfrac{r^{1/2}+r_{+}^{1/2}}{r^{1/2}-r_{+}^{1/2}} & \text{if }R\geq R_{b}%
\end{array}
\right.  ,
\end{align*}
where
\[
y=\tfrac{1}{2}\left[  \left(  \tfrac{R}{R_{b}}\right)  ^{2}-1\right]
+\tfrac{R_{b}r}{r_{+}R}.
\]

\subsubsection{Gravitational collapse of charged and uncharged shells}
\label{collapse}

It is well known that the role of exact solutions has been fundamental in the
development of general relativity. In this section, we present these exact
solutions for a charged shell of matter collapsing into a black hole. Such
solutions were found in Ref.~\cite{2002PhLB..545..226C} and are new with
respect to the Tolman--Oppenheimer--Snyder class. For simplicity we consider
the case of zero angular momentum and spherical symmetry. This problem is
relevant on its own account as an addition to the existing family of
interesting exact solutions and also represents some progress in understanding
the role of the formation of the horizon and of the irreducible mass as will be
discussed in Section \ref{irrmassRV}, see e.g. \cite{2002PhLB..545..233R}. It
is also essential in improving the treatment of the vacuum polarization
processes occurring during the formation of a black hole discussed in
\cite{2003PhLB..573...33R, 2001ApJ...555L.107R, 2001ApJ...555L.113R,
2001ApJ...555L.117R, 2001NCimB.116...99R} and references therein.

W. Israel and V. de La Cruz \cite{1966NCimB..44....1I, 1967NCimA..51..744C}
showed that the problem of a collapsing charged shell can be reduced to a set
of ordinary differential equations. We reconsider here the following
relativistic system: a spherical shell of electrically charged dust which is
moving radially in the Reissner--Nordstr\"{o}m background of an already formed
nonrotating black hole of mass $M_{1}$ and charge $Q_{1}$, with $Q_{1}\leq
M_{1}$. The Einstein--Maxwell equations with a charged spherical dust as source
are
\begin{equation}
G_{\mu\nu}=8\pi\left[  T_{\mu\nu}^{\left(  \mathrm{d}\right)  }+T_{\mu\nu
}^{\left(  \mathrm{em}\right)  }\right]  ,\quad\nabla_{\mu}F^{\nu\mu}=4\pi
j^{\nu},\quad\nabla_{\lbrack\mu}F_{\nu\rho]}=0,
\end{equation}
where
\begin{equation}
T_{\mu\nu}^{\left(  \mathrm{d}\right)  }=\varepsilon u_{\mu}u_{\nu},\quad
T_{\mu\nu}^{\left(  \mathrm{em}\right)  }=\tfrac{1}{4\pi}\left(  F_{\mu}%
{}^{\rho}F_{\rho\nu}-\tfrac{1}{4}g_{\mu\nu}F^{\rho\sigma}F_{\rho\sigma
}\right)  ,\quad j^{\mu}=\sigma u^{\mu}.
\end{equation}
Here $T_{\mu\nu}^{\left(  \mathrm{d}\right)  }$, $T_{\mu\nu}^{\left(
\mathrm{em}\right)  }$ and $j^{\mu}$ are respectively the
energy-momentum tensor of the dust, the energy-momentum tensor of
the electromagnetic field $F_{\mu\nu}$ and the charge $4-$current.
The mass and charge density in the comoving frame are given by
$\varepsilon$, $\sigma$ and $u^{a}$ is the $4$-velocity of the dust.
In spherical--polar coordinates the line element is
\begin{equation}
ds^{2}\equiv g_{\mu\nu}dx^{\mu}dx^{\nu}=-e^{\nu\left(  r,t\right)  }%
dt^{2}+e^{\lambda\left(  r,t\right)  }dr^{2}+r^{2}d\Omega^{2},
\end{equation}
where $d\Omega^{2}=d\theta^{2}+\sin^{2}\theta d\phi^{2}$.

We describe the shell by using the four-dimensional Dirac
distribution $\delta^{\left(  4\right)  }$ normalized as
\begin{equation}
\int\delta^{\left(  4\right)  }\left(  x,x^{\prime}\right)  \sqrt{-g}d^{4}x=1
\end{equation}
where $g=\det\left\Vert g_{\mu\nu}\right\Vert $. We then have
\begin{align}
\varepsilon\left(  x\right)   &  =M_{0}\int\delta^{\left(  4\right)  }\left(
x,x_{0}\right)  r^{2}d\tau d\Omega,\label{eq1a}\\
\sigma\left(  x\right)   &  =Q_{0}\int\delta^{\left(  4\right)  }\left(
x,x_{0}\right)  r^{2}d\tau d\Omega. \label{eq2a}%
\end{align}
$M_{0}$ and $Q_{0}$ respectively are the rest mass and the charge of the shell
and $\tau$ is the proper time along the world surface $S:$ $x_{0}=x_{0}\left(
\tau,\Omega\right)  $ of the shell. $S$ divides the space-time into two
regions: an internal one $\mathcal{M}_{-}$ and an external one $\mathcal{M}%
_{+}$. As we will see in the next section for the description of the collapse
we can choose either $\mathcal{M}_{-}$ or $\mathcal{M}_{+}$. The two
descriptions, clearly equivalent, will be relevant for the physical
interpretation of the solutions.

Introducing the orthonormal tetrad
\begin{equation}
{\boldsymbol{\omega}}_{\pm}^{\left(  0\right)  }=f_{\pm}^{1/2}dt,\quad
{\boldsymbol{\omega}}_{\pm}^{\left(  1\right)  }=f_{\pm}^{-1/2}dr,\quad
{\boldsymbol{\omega}}^{\left(  2\right)  }=rd\theta,\quad{\boldsymbol{\omega}%
}^{\left(  3\right)  }=r\sin\theta d\phi;\quad
\end{equation}
we obtain the tetrad components of the electric field%
\begin{equation}
{\boldsymbol{E}}=E{\boldsymbol{\omega}}^{\left(
1\right)  }=\left\{
\begin{array}
[c]{l}%
\frac{Q}{r^{2}}\ {\boldsymbol{\omega}}_{+}^{\left(  1\right)  }\qquad
\text{outside the shell}\\
\frac{Q_{1}}{r^{2}}\ {\boldsymbol{\omega}}_{-}^{\left(  1\right)  }%
\qquad\text{inside the shell}%
\end{array}
\right.  , \label{E3a}%
\end{equation}
where $Q=Q_{0}+Q_{1}$ is the total charge of the system. From the $G_{tt}$
Einstein equation we get
\begin{equation}
ds^{2}=\left\{
\begin{array}
[c]{l}%
-f_{+}dt_{+}^{2}+f_{+}^{-1}dr^{2}+r^{2}d\Omega^{2}\qquad\text{outside the
shell}\\
-f_{-}dt_{-}^{2}+f_{-}^{-1}dr^{2}+r^{2}d\Omega^{2}\qquad\text{inside the
shell}%
\end{array}
\right.  , \label{E0a}%
\end{equation}
where $f_{+}=1-\tfrac{2M}{r}+\tfrac{Q^{2}}{r^{2}}$, $f_{-}=1-\tfrac{2M_{1}}%
{r}+\tfrac{Q_{1}^{2}}{r^{2}}$ and $t_{-}$ and $t_{+}$ are the
Schwarzschild-like time coordinates in $\mathcal{M}_{-}$ and $\mathcal{M}_{+}$
respectively. Here $M$ is the total mass-energy of the system formed by the
shell and the black hole, measured by an observer at rest at infinity.

Indicating by $r_{0}$ the Schwarzschild-like radial coordinate of the shell
and by $t_{0\pm}$ its time coordinate, from the $G_{tr}$ Einstein equation we
have
\begin{equation}
\tfrac{M_{0}}{2}\left[  f_{+}\left(  r_{0}\right)  \tfrac{dt_{0+}}{d\tau
}+f_{-}\left(  r_{0}\right)  \tfrac{dt_{0-}}{d\tau}\right]  =M-M_{1}%
-\tfrac{Q_{0}^{2}}{2r_{0}}-\tfrac{Q_{1}Q_{0}}{r_{0}}. \label{eq3a}%
\end{equation}
The remaining Einstein equations are identically satisfied. From (\ref{eq3a})
and the normalization condition $u_{\mu}u^{\mu}=-1$ we find
\begin{align}
\left(  \tfrac{dr_{0}}{d\tau}\right)  ^{2}  &  =\tfrac{1}{M_{0}^{2}}\left(
M-M_{1}+\tfrac{M_{0}^{2}}{2r_{0}}-\tfrac{Q_{0}^{2}}{2r_{0}}-\tfrac{Q_{1}Q_{0}%
}{r_{0}}\right)  ^{2}-f_{-}\left(  r_{0}\right) \nonumber\\
&  =\tfrac{1}{M_{0}^{2}}\left(  M-M_{1}-\tfrac{M_{0}^{2}}{2r_{0}}-\tfrac
{Q_{0}^{2}}{2r_{0}}-\tfrac{Q_{1}Q_{0}}{r_{0}}\right)  ^{2}-f_{+}\left(
r_{0}\right)  ,\label{EQUY}\\
\tfrac{dt_{0\pm}}{d\tau}  &  =\tfrac{1}{M_{0}f_{\pm}\left(  r_{0}\right)
}\left(  M-M_{1}\mp\tfrac{M_{0}^{2}}{2r_{0}}-\tfrac{Q_{0}^{2}}{2r_{0}}%
-\tfrac{Q_{1}Q_{0}}{r_{0}}\right)  . \label{EQUYa}%
\end{align}

We now define, as usual, $r_{\pm}\equiv M\pm\sqrt{M^{2}-Q^{2}}$: when $Q<M$,
$r_{\pm}$ are real and they correspond to the horizons of the new black hole
formed by the gravitational collapse of the shell. We similarly introduce the
horizons $r_{\pm}^{1}=M_{1}\pm\sqrt{M_{1}^{2}-Q_{1}^{2}}$ of the already
formed black hole. From (\ref{eq3a}) we have that the inequality
\begin{equation}
M-M_{1}-\tfrac{Q_{0}^{2}}{2r_{0}}-\tfrac{Q_{1}Q_{0}}{r_{0}}>0
\label{Constraint}%
\end{equation}
holds for $r_{0}>r_{+}$ if $Q<M$ and for $r_{0}>r_{+}^{1}$ if $Q>M$
since in these cases the left-hand side of (\ref{eq3a}) is clearly
positive. Eqs.~(\ref{EQUY}) and (\ref{EQUYa}) (together with
(\ref{E0a}), (\ref{E3a})) completely describe a 5-parameter ($M$,
$Q$, $M_{1}$, $Q_{1}$, $M_{0}$) family of solutions of the
Einstein-Maxwell equations.

For astrophysical applications \cite{2003PhLB..573...33R} the trajectory of the shell
$r_{0}=r_{0}\left(  t_{0+}\right)  $ is obtained as a function of the time
coordinate $t_{0+}$ relative to the space-time region $\mathcal{M}_{+}$. In
the following we drop the $+$ index from $t_{0+}$. From (\ref{EQUY}) and
(\ref{EQUYa}) we have
\begin{equation}
\tfrac{dr_{0}}{dt_{0}}=\tfrac{dr_{0}}{d\tau}\tfrac{d\tau}{dt_{0}}=\pm\tfrac
{F}{\Omega}\sqrt{\Omega^{2}-F}, \label{EQUAISRDLC}%
\end{equation}
where
\begin{equation}
F\equiv f_{+}\left(  r_{0}\right)  =1-\tfrac{2M}{r_{0}}+\tfrac{Q^{2}}%
{r_{0}^{2}},\quad\Omega\equiv\Gamma-\tfrac{M_{0}^{2}+Q^{2}-Q_{1}^{2}}%
{2M_{0}r_{0}},\quad\Gamma\equiv\tfrac{M-M_{1}}{M_{0}}.
\end{equation}
Since we are interested in an imploding shell, only the minus sign case in
(\ref{EQUAISRDLC}) will be studied. We can give the following physical
interpretation of $\Gamma$. If $M-M_{1}\geq M_{0}$, $\Gamma$ coincides with
the Lorentz $\gamma$ factor of the imploding shell at infinity; from
(\ref{EQUAISRDLC}) it satisfies
\begin{equation}
\Gamma=\tfrac{1}{\sqrt{1-\left(  \frac{dr_{0}}{dt_{0}}\right)  _{r_{0}=\infty
}^{2}}}\geq1.
\end{equation}
When $M-M_{1}<M_{0}$ then there is a \emph{turning point} $r_{0}^{\ast}$,
defined by $\left.  \tfrac{dr_{0}}{dt_{0}}\right\vert _{r_{0}=r_{0}^{\ast}}%
=0$. In this case $\Gamma$ coincides with the \textquotedblleft effective
potential\textquotedblright\ at $r_{0}^{\ast}$ :
\begin{equation}
\Gamma=\sqrt{f_{-}\left(  r_{0}^{\ast}\right)  }+M_{0}^{-1}\left(
-\tfrac{M_{0}^{2}}{2r_{0}^{\ast}}+\tfrac{Q_{0}^{2}}{2r_{0}^{\ast}}%
+\tfrac{Q_{1}Q_{0}}{r_{0}^{\ast}}\right)  \leq1.
\end{equation}

The solution of the differential equation (\ref{EQUAISRDLC}) is given by:
\begin{equation}
\int dt_{0}=-\int\tfrac{\Omega}{F\sqrt{\Omega^{2}-F}}dr_{0}. \label{GRYD}%
\end{equation}
The functional form of the integral (\ref{GRYD}) crucially depends on the
degree of the polynomial $P\left(  r_{0}\right)  =r_{0}^{2}\left(  \Omega
^{2}-F\right)  $, which is generically two, but in special cases has lower
values. We therefore distinguish the following cases:

\begin{enumerate}
\item {\boldmath$M=M_{0}+M_{1}$}; {\boldmath$Q_{1}=M_{1}$}; {\boldmath$Q=M$}:
$P\left(  r_{0}\right)  $ is equal to $0$, we simply have
\begin{equation}
r_{0}(t_{0})=\mathrm{{const}.}%
\end{equation}

\item {\boldmath$M=M_{0}+M_{1}$}; {\boldmath$M^{2}-Q^{2}=M_{1}^{2}-Q_{1}^{2}$%
}; {\boldmath$Q\neq M$}: $P\left(  r_{0}\right)  $ is a constant, we have
\begin{equation}
t_{0}=\mathrm{const}+\tfrac{1}{2\sqrt{M^{2}-Q^{2}}}\left[  \left(
r_{0}+2\right)  r_{0}+r_{+}^{2}\log\left(  \tfrac{r_{0}-r_{+}}{M}\right)
+r_{-}^{2}\log\left(  \tfrac{r_{0}-r_{-}}{M}\right)  \right]  . \label{CASO1}%
\end{equation}

\item {\boldmath$M=M_{0}+M_{1}$}; {\boldmath$M^{2}-Q^{2}\neq M_{1}^{2}%
-Q_{1}^{2}$}: $P\left(  r_{0}\right)  $ is a first order polynomial and
\begin{align}
t_{0}  &  =\mathrm{const}+2r_{0}\sqrt{\Omega^{2}-F}\left[  \tfrac{M_{0}r_{0}%
}{3\left(  M^{2}-Q^{2}-M_{1}^{2}+Q_{1}^{2}\right)  }\right. \nonumber\\
&  \left.  +\tfrac{\left(  M_{0}^{2}+Q^{2}-Q_{1}^{2}\right)  ^{2}%
-9MM_{0}\left(  M_{0}^{2}+Q^{2}-Q_{1}^{2}\right)  +12M^{2}M_{0}^{2}%
+2Q^{2}M_{0}^{2}}{3\left(  M^{2}-Q^{2}-M_{1}^{2}+Q_{1}^{2}\right)  ^{2}%
}\right] \nonumber\\
&  -\tfrac{1}{\sqrt{M^{2}-Q^{2}}}\left[  r_{+}^{2}\mathrm{arctanh}\left(
\tfrac{r_{0}}{r_{+}}\tfrac{\sqrt{\Omega^{2}-F}}{\Omega_{+}}\right)  -r_{-}%
^{2}\mathrm{arctanh}\left(  \tfrac{r_{0}}{r_{-}}\tfrac{\sqrt{\Omega^{2}-F}%
}{\Omega_{-}}\right)  \right]  , \label{CASO2}%
\end{align}

where $\Omega_{\pm}\equiv\Omega\left(  r_{\pm}\right)  $.

\item {\boldmath$M\neq M_{0}+M_{1}$}: $P\left(  r_{0}\right)  $ is a second
order polynomial and
\begin{align}
t_{0}  &  =\mathrm{const}-\tfrac{1}{2\sqrt{M^{2}-Q^{2}}}\left\{
\tfrac{2\Gamma\sqrt{M^{2}-Q^{2}}}{\Gamma^{2}-1}r_{0}\sqrt{\Omega^{2}-F}\right.
\nonumber\\
&  +r_{+}^{2}\log\left[  \tfrac{r_{0}\sqrt{\Omega^{2}-F}}{r_{0}-r_{+}}%
+\tfrac{r_{0}^{2}\left(  \Omega^{2}-F\right)  +r_{+}^{2}\Omega_{+}^{2}-\left(
\Gamma^{2}-1\right)  \left(  r_{0}-r_{+}\right)  ^{2}}{2\left(  r_{0}%
-r_{+}\right)  r_{0}\sqrt{\Omega^{2}-F}}\right] \nonumber\\
&  -r_{-}^{2}\log\left[  \tfrac{r_{0}\sqrt{\Omega^{2}-F}}{r_{0}-r_{-}}%
+\tfrac{r_{0}^{2}\left(  \Omega^{2}-F\right)  +r_{-}^{2}\Omega_{-}^{2}-\left(
\Gamma^{2}-1\right)  \left(  r_{0}-r_{-}\right)  ^{2}}{2\left(  r_{0}%
-r_{-}\right)  r_{0}\sqrt{\Omega^{2}-F}}\right] \nonumber\\
&  -\tfrac{\left[  2MM_{0}\left(  2\Gamma^{3}-3\Gamma\right)  +M_{0}^{2}%
+Q^{2}-Q_{1}^{2}\right]  \sqrt{M^{2}-Q^{2}}}{M_{0}\left(  \Gamma^{2}-1\right)
^{3/2}}\log\left[  \tfrac{r_{0}}{M}\sqrt{\Omega^{2}-F}\right. \nonumber\\
&  \left.  \left.  +\tfrac{2M_{0}\left(  \Gamma^{2}-1\right)  r_{0}-\left(
M_{0}^{2}+Q^{2}-Q_{1}^{2}\right)  \Gamma+2M_{0}M}{2M_{0}M\sqrt{\Gamma^{2}-1}%
}\right]  \right\}  . \label{CASO3}%
\end{align}

\end{enumerate}

In the case of a shell falling in a flat background ($M_{1}=Q_{1}=0$) it is of
particular interest to study the \emph{turning points} $r_{0}^{\ast}$ of the
shell trajectory. In this case equation (\ref{EQUY}) reduces to
\begin{equation}
\left(  \tfrac{dr_{0}}{d\tau}\right)  ^{2}=\tfrac{1}{M_{0}^{2}}\left(
M+\tfrac{M_{0}^{2}}{2r_{0}}-\tfrac{Q^{2}}{2r_{0}}\right)  ^{2}-1.
\label{EQUY2}%
\end{equation}
Case $(2)$ has no counterpart in this new regime and Eq.~(\ref{Constraint})
constrains the possible solutions to only the following cases:

\begin{enumerate}
\item {\boldmath$M=M_{0}$}; {\boldmath$Q=M_{0}$. }$r_{0}=r_{0}\left(
0\right)  $ constantly.

\item {\boldmath$M=M_{0}$}; {\boldmath$Q<M_{0}$. }There are no turning points,
the shell starts at rest at infinity and collapses until a
Reissner--Nordstr\"{o}m black hole is formed with horizons at $r_{0}=r_{\pm
}\equiv M\pm\sqrt{M^{2}-Q^{2}}$ and the singularity in $r_{0}=0$.

\item {\boldmath$M\neq M_{0}$. }There is one turning point $r_{0}^{\ast}$.

\begin{enumerate}
\item {\boldmath$M<M_{0}$}, then necessarily is {\boldmath$Q<M_{0}$}.
Positivity of the right-hand side of (\ref{EQUY2}) requires
$r_{0}\leq r_{0}^{\ast}$, where
$r_{0}^{\ast}=\frac{1}{2}\frac{Q^{2}-M_{0}^{2}}{M-M_{0}}$ is the
unique turning point. Then the shell starts from $r_{0}^{\ast}$ and
collapses until the singularity at $r_{0}=0$ is reached.

\item {\boldmath$M>M_{0}$}. The shell has finite radial velocity at infinity.

\begin{enumerate}
\item {\boldmath$Q\leq M_{0}$}. The dynamics are qualitatively analogous to
case (2).

\item {\boldmath$Q>M_{0}$}. Positivity of the right-hand side of (\ref{EQUY2}) and
(\ref{Constraint}) requires that $r_{0}\geq$ $r_{0}^{\ast}$, where
$r_{0}^{\ast}=\frac{1}{2}\frac{Q^{2}-M_{0}^{2}}{M-M_{0}}$. The shell
starts from infinity and bounces at $r_{0}=r_{0}^{\ast}$, reversing
its motion.
\end{enumerate}
\end{enumerate}
\end{enumerate}

In this regime the analytic forms of the solutions are given by
Eqs.~(\ref{CASO2}) and (\ref{CASO3}), simply setting $M_{1}=Q_{1}=0$.

Of course, it is of particular interest for the issue of vacuum polarization
the time varying electric field $E_{r_{0}}=\tfrac{Q}{r_{0}^{2}}$ on
the external surface of the shell. In order to study the variability of
$E_{r_{0}}$ with time it is useful to consider in the tridimensional
space of parameters $(r_{0},t_{0},E_{r_{0}})$ the parametric curve
$\mathcal{C}:\left(  r_{0}=\lambda,\quad t_{0}=t_{0}(\lambda),\quad
E_{r_{0}}=\tfrac{Q}{\lambda^{2}}\right)  $. In astrophysical
applications \cite{2003PhLB..573...33R} we are specially interested in the family of
solutions such that $\frac{dr_{0}}{dt_{0}}$ is 0 when $r_{0}=\infty$ which
implies that $\Gamma=1$. In Fig.~\ref{fig1a} we plot the collapse curves in the
plane $(t_{0},r_{0})$ for different values of the parameter $\xi\equiv\frac
{Q}{M}$, $0<\xi<1$. The initial data are chosen so that the integration
constant in Eq.~(\ref{CASO2}) is equal to 0. In all the cases we can
follow the details of the approach to the horizon which is reached in an
infinite Schwarzschild time coordinate.

\begin{figure}[!htp]
\begin{center}
\includegraphics[width=11cm]{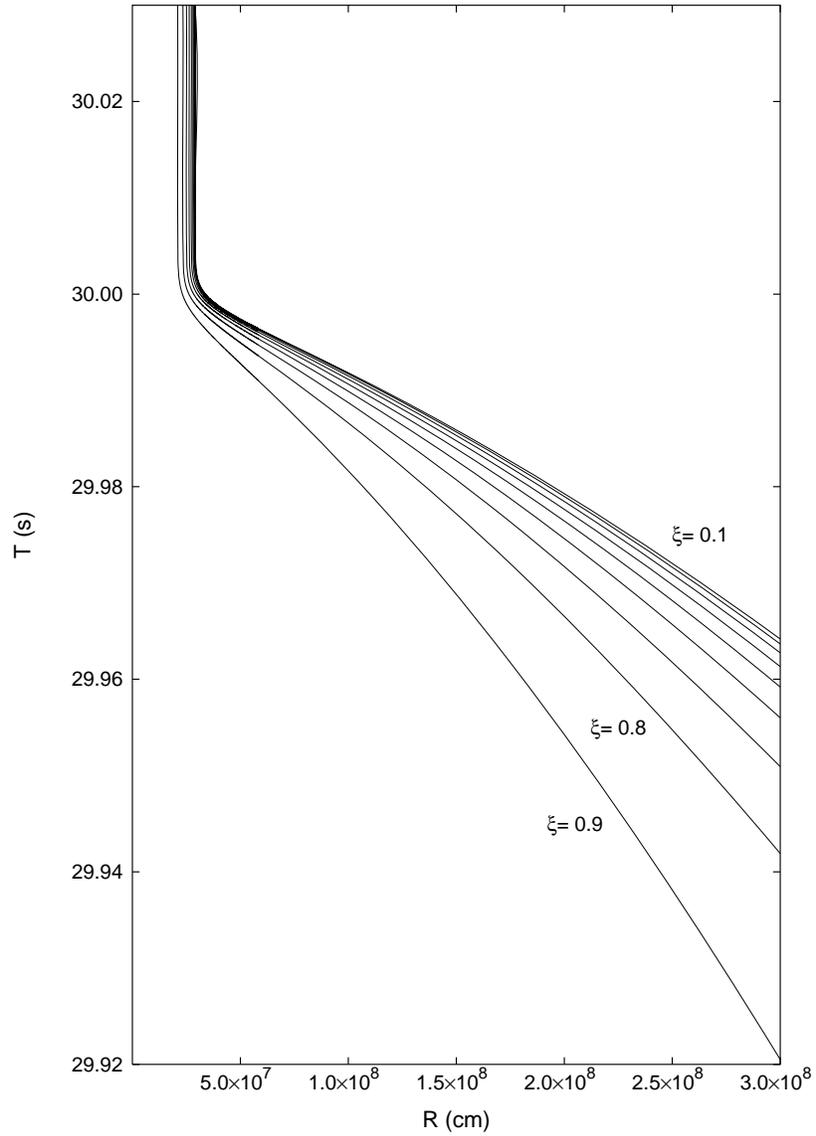}
\end{center}
\caption{Collapse curves in the plane $(T,R)$ for $M=20M_{\odot}$ and for
different values of the parameter $\xi$. The asymptotic behavior is the clear
manifestation of general relativistic effects as the horizon of the black hole is
approached. Details in \cite{2002PhLB..545..226C}.}%
\label{fig1a}%
\end{figure}

In Fig.~\ref{fig2a} we plot the parametric curves $\mathcal{C}$ in the space
$(r_{0},t_{0},E_{r_{0}})$ for different values of $\xi$. Again we
can follow the exact asymptotic behavior of the curves $\mathcal{C}$,
$E_{r_{0}}$ reaching the asymptotic value $\frac{Q}{r_{+}^{2}}$. The
detailed knowledge of this asymptotic behavior is of relevance for the
observational properties of the black hole formation, see e.g. \cite{2002PhLB..545..233R},
\cite{2003PhLB..573...33R}.

\begin{figure}[!htp]
\begin{center}
\includegraphics[width=11cm]{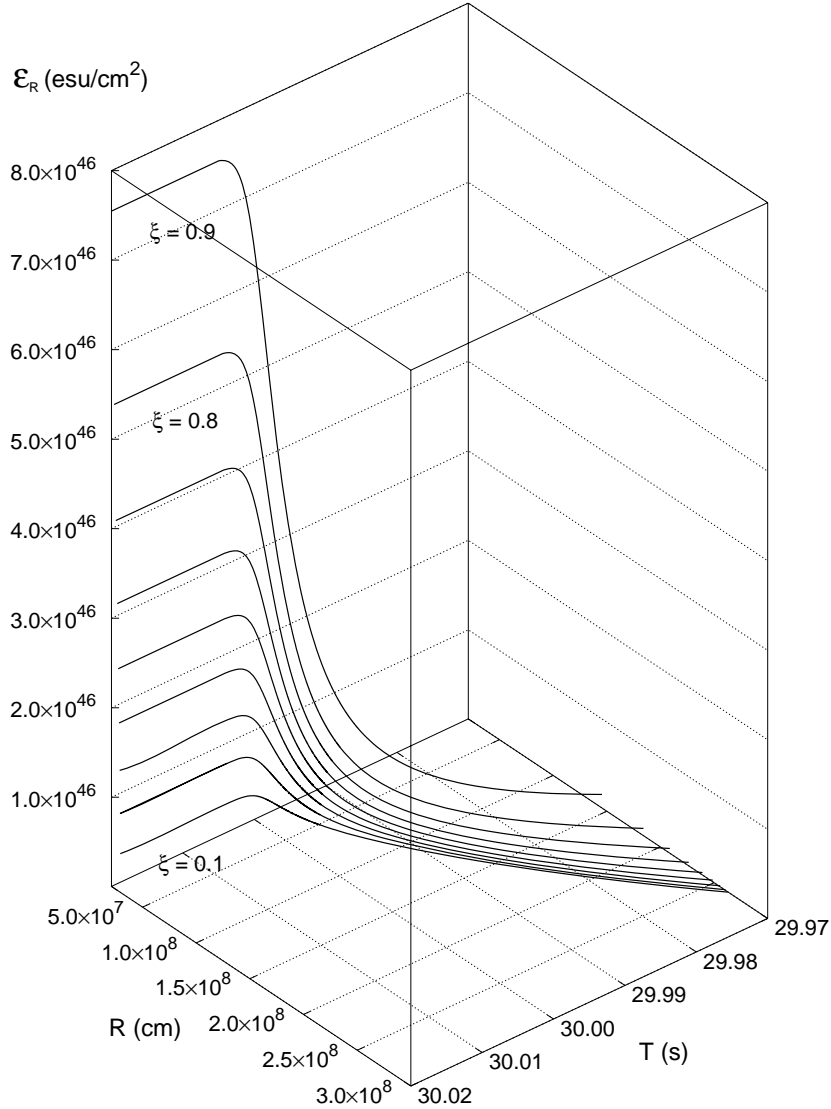}
\end{center}
\caption{Electric field behaviour at the surface of the shell for
$M=20M_{\odot}$ and for different values of the parameter $\xi$. The
asymptotic behavior is the clear manifestation of general relativistic effects
as the horizon of the black hole is approached. Details in \cite{2002PhLB..545..226C}.}%
\label{fig2a}%
\end{figure}

\subsection{The maximum energy extractable from a black hole}\label{energyextr}

The theoretical analysis of the collapsing shell considered in the previous
section allows to reach a deeper understanding of the mass formula of black
holes at least in the case of a Reissner--Nordstr\"om black hole. This allows
as well to give an expression of the irreducible mass of the black hole only in
terms of its kinetic energy of the initial rest mass undergoing gravitational
collapse and its gravitational energy and kinetic energy at the crossing of the
black hole horizon. It also allows to create a scenario for acceleration of the
ultrahigh energy cosmic rays with energy typically $10^{21}$ eV from black
holes, as opposed to the process of vacuum polarization producing pairs with
energies in the MeV region. We shall follow in this Section the treatment by
Ruffini and Vitagliano \cite{2003IJMPD..12..121R}.

\subsubsection{The formula of the irreducible mass of a black hole}\label{irrmassRV}

The main objective of this section is to clarify the interpretation
of the mass-energy formula \cite{1971PhRvD...4.3552C} for a black
hole. For simplicity we study the case of a nonrotating black hole
using the results presented in the previous section. As we saw
there, the collapse of a nonrotating charged shell can be described
by exact analytic solutions of the Einstein--Maxwell equations.
Consider to two complementary regions in which the world surface of
the shell divides the space-time: $\mathcal{M}_{-}$ and
$\mathcal{M}_{+}$. They are static space-times; we denote their
time-like Killing vectors by $\xi_{-}^{\mu}$ and $\xi_{+}^{\mu}$
respectively. $\mathcal{M}_{+}$ is foliated by the family $\left\{
\Sigma_{t}^{+}:t_{+}=t\right\}  $ of space-like hypersurfaces of
constant $t_{+}$.

The splitting of the space-time into the regions $\mathcal{M}_{-}$ and
$\mathcal{M}_{+}$ allows two physically equivalent descriptions of the
collapse and the use of one or the other depends on the question one is
studying. The use of $\mathcal{M}_{-}$ proves helpful for the identification
of the physical constituents of the irreducible mass while $\mathcal{M}_{+}$
is needed to describe the energy extraction process from black hole. The equation of
motion for the shell, Eq.~\ref{EQUY}, reduces in this case to
\begin{equation}
\left(  M_{0}\tfrac{dr_{0}}{d\tau}\right)  ^{2}=\left(  M+\tfrac{M_{0}^{2}%
}{2r_{0}}-\tfrac{Q^{2}}{2r_{0}}\right)  ^{2}-M_{0}^{2} \label{EQA}%
\end{equation}
in $\mathcal{M}_{-}$ and
\begin{equation}
\left(  M_{0}\tfrac{dr_{0}}{d\tau}\right)  ^{2}=\left(  M-\tfrac{M_{0}^{2}%
}{2r_{0}}-\tfrac{Q^{2}}{2r_{0}}\right)  ^{2}-M_{0}^{2}f_{+} \label{EQAb}%
\end{equation}
in $\mathcal{M}_{+}$. The constraint (\ref{Constraint}) becomes
\begin{equation}
M-\tfrac{Q^{2}}{2r_{0}}>0. \label{EQO}%
\end{equation}
Since $\mathcal{M}_{-}$ is a flat space-time we can interpret $-\tfrac
{M_{0}^{2}}{2r_{0}}$ in (\ref{EQA}) as the gravitational binding energy of the
system. $\tfrac{Q^{2}}{2r_{0}}$ is its electromagnetic energy. Then
Eqs.~(\ref{EQA}), (\ref{EQAb}) differ by the gravitational and electromagnetic
self-energy terms from the corresponding equations of motion of a test
particle.

Introducing the total radial momentum $P^r\equiv M_{0}u^{r}= M_{0}
\tfrac{dr_{0}}{d\tau}$ of the shell, we can express the kinetic energy of the
shell as measured by static observers in $\mathcal{M}_{-}$ as
$T\equiv-M_{0}u_{\mu} \xi_{-}^{\mu}-M_{0}=\sqrt{(P^r)^{2}+M_{0}^{2}}-M_{0}$.
Then from Eq.~(\ref{EQA}) we have
\begin{equation}
M=-\tfrac{M_{0}^{2}}{2r_{0}}+\tfrac{Q^{2}}{2r_{0}}+\sqrt{(P^r)^{2}+M_{0}^{2}%
}=M_{0}+T-\tfrac{M_{0}^{2}}{2r_{0}}+\tfrac{Q^{2}}{2r_{0}}. \label{EQC}%
\end{equation}
where we choose the positive root solution due to the constraint (\ref{EQO}).
Eq.~(\ref{EQC}) is the \emph{mass formula} of the shell, which depends on the
time-dependent radial coordinate $r_{0}$ and kinetic energy $T$. If $M\geq Q$,
a black hole is formed and we have
\begin{equation}
M=M_{0}+T_{+}-\tfrac{M_{0}^{2}}{2r_{+}}+\tfrac{Q^{2}}{2r_{+}}\,, \label{EQL}%
\end{equation}
where $T_{+}\equiv T\left(  r_{+}\right)  $ and $r_{+}=M+\sqrt{M^{2}-Q^{2}}$
is the radius of the external horizon of that
\begin{equation}
M=M_{\mathrm{ir}}+\tfrac{Q^{2}}{2r_{+}}, \label{irrmass}%
\end{equation}
so it follows that {}%
\begin{equation}
M_{\mathrm{ir}}=M_{0}-\tfrac{M_{0}^{2}}{2r_{+}}+T_{+}, \label{EQM}%
\end{equation}
namely that $M_{\mathrm{ir}}$ is the sum of only three
contributions: the rest mass $M_{0}$, the gravitational potential
energy and the kinetic energy of the rest mass evaluated at the
horizon. $M_{\mathrm{ir}}$ is independent of the electromagnetic
energy, a fact noticed by Bekenstein \cite{1971PhRvD...4.2185B}. We
have taken one further step here by identifying the independent
physical contributions to $M_{\mathrm{ir}}$. This has important
consequences for the energetics of black hole formation (see
\cite{2002PhLB..545..233R}).

Next we consider the physical interpretation of the electromagnetic term
$\tfrac{Q^{2}}{2r_{0}}$, which can be obtained by evaluating the Killing
integral
\begin{equation}
\int_{\Sigma_{t}^{+}}\xi_{+}^{\mu}T_{\mu\nu}^{\mathrm{(em)}}d\Sigma^{\nu}%
=\int_{r_{0}}^{\infty}r^{2}dr\int_{0}^{1}d\cos\theta\int_{0}^{2\pi}%
d\phi\ T^{\mathrm{(em)}}{}{}_{0}{}^{0}=\tfrac{Q^{2}}{2r_{0}}\,, \label{EQR}%
\end{equation}
where $\Sigma_{t}^{+}$ is the space-like hypersurface in $\mathcal{M}_{+}$
described by the equation $t_{+}=t=\mathrm{const}$, with $d\Sigma^{\nu}$ as
its surface element vector. The quantity in Eq.~(\ref{EQR}) differs from the
purely electromagnetic energy
\begin{equation}
\int_{\Sigma_{t}^{+}}n_{+}^{\mu}T_{\mu\nu}^{\mathrm{(em)}}d\Sigma^{\nu}%
=\tfrac{1}{2}\int_{r_{0}}^{\infty}dr\sqrt{g_{rr}}\tfrac{Q^{2}}{r^{2}},
\end{equation}
where $n_{+}^{\mu}=f_{+}^{-1/2}\xi_{+}^{\mu}$ is the unit normal to the
integration hypersurface and $g_{rr}=f_{+}$. This is similar to the analogous
situation for the total energy of a static spherical star of energy density
$\epsilon$ within a radius $r_{0}$, $m\left(  r_{0}\right)  =4\pi\int
_{0}^{r_{0}}dr\ r^{2}\epsilon$, which differs from the pure matter energy
\[
m_{\mathrm{p}}\left(  r_{0}\right)  =4\pi\int_{0}^{r_{0}}dr \sqrt{g_{rr}}%
r^{2}\epsilon
\]
 by the gravitational energy (see \cite{1973grav.book.....M}). Therefore the
term $\tfrac{Q^{2}}{2r_{0}}$ in the mass formula (\ref{EQC}) is the
\emph{total} energy of the electromagnetic field and includes its own
gravitational binding energy. This energy is stored throughout the region
$\mathcal{M}_{+}$, extending from $r_{0}$ to infinity.

\subsubsection{Extracting electromagnetic energy from a subcritical and
overcritical black hole}\label{subandoverbh}

We now turn to the problem of extracting the electromagnetic energy
from a black hole (see \cite{1971PhRvD...4.3552C}). We can
distinguish between two conceptually physically different processes,
depending on whether the electric field strength $E=\frac{Q}{r^{2}}$
is smaller or greater than the critical value
$E_{\mathrm{c}}$. The maximum value $E_{+}=\tfrac{Q}%
{r_{+}^{2}}$ of the electric field around a black hole is reached at
the horizon. In what follows we restore $G$, $\hbar$ and $c$.

For $E_{+}<E_{\mathrm{c}}$ the leading energy
extraction mechanism consists of a sequence of discrete elementary decay
processes of a particle into two oppositely charged particles.
The condition $E_{+}<E_{\mathrm{c}}$ implies
\begin{equation}
\xi\equiv\tfrac{Q}{\sqrt{G}M}\lesssim\left\{
\begin{array}
[c]{r}%
\tfrac{GM/c^{2}}{\lambda_{\mathrm{C}}}\left(  \tfrac{e}{\sqrt{G}m_{e}}\right)
^{-1}\sim10^{-6}\tfrac{M}{M_{\odot}}\quad\text{if }\tfrac{M}{M_{\odot}}%
\leq10^{6}\\
1\quad\quad\quad\quad\quad\quad\quad\quad\text{if }\tfrac{M}{M_{\odot}}>10^{6}%
\end{array}
\right.  , \label{critical3}%
\end{equation}
where $\lambda_{\mathrm{C}}$ is the Compton wavelength of the electron.
Denardo and Ruffini \cite{1973PhLB...45..259D} and Denardo, Hively and Ruffini \cite{1974PhLB...50..270D}
have defined as the \emph{effective ergosphere} the region around a black hole
where the energy extraction processes occur. This region extends from the
horizon $r_{+}$ up to a radius
\begin{equation}
r_{\mathrm{Eerg}}=\tfrac{GM}{c^{2}}\left[  1+\sqrt{1-\xi^{2}\left(
1-\tfrac{e^{2}}{G{m_{e}^{2}}}\right)  }\right]  \simeq\tfrac{e}{m_{e}}%
\tfrac{Q}{c^{2}}\,. \label{EffErg}%
\end{equation}
The energy extraction occurs in a finite number $N_{\mathrm{PD}}$ of such
discrete elementary processes, each one corresponding to a decrease of the black hole
charge. We have
\begin{equation}
N_{\mathrm{PD}}\simeq\tfrac{Q}{e}\,.
\end{equation}
Since the total extracted energy is (see Eq.~(\ref{irrmass})) ${\mathcal
E}^{\mathrm{tot} }=\tfrac{Q^{2}}{2r_{+}}$, we obtain for the mean energy per
accelerated particle $\left\langle {\mathcal E}\right\rangle
_{\mathrm{PD}}=\tfrac{{\mathcal E}^{\mathrm{tot}} }{N_{\mathrm{PD}}}$
\begin{equation}
\left\langle {\mathcal E}\right\rangle
_{\mathrm{PD}}=\tfrac{Qe}{2r_{+}}=
\tfrac{1}{2}\tfrac{\xi}{1+\sqrt{1-\xi^{2}}}\tfrac{e}{\sqrt{G}m_{e}}\
m_{e}c^{2} \simeq\tfrac{1}{2}\xi\tfrac{e}{\sqrt{G}m_{e}}\ m_{e}c^{2},
\end{equation}
which gives
\begin{equation}
\left\langle {\mathcal E}\right\rangle _{\mathrm{PD}}\lesssim\left\{
\begin{array}
[c]{r}%
\tfrac{M}{M_{\odot}}10^{21}eV\quad\text{if }\tfrac{M}{M_{\odot}}\leq10^{6}\\
10^{27}eV\quad\quad\text{if }\tfrac{M}{M_{\odot}}>10^{6}%
\end{array}
\right.  . \label{UHECR}%
\end{equation}

One of the crucial aspects of the energy extraction process from a black hole is
its back reaction on the irreducible mass expressed in \cite{1971PhRvD...4.3552C}. Although
the energy extraction processes can occur in the entire effective ergosphere
defined by Eq.~(\ref{EffErg}), only the limiting processes occurring on the
horizon with zero kinetic energy can reach the maximum efficiency while
approaching the condition of total reversibility (see Fig.~2 in \cite{1971PhRvD...4.3552C}
for details). The farther from the horizon that a decay occurs, the more it
increases the irreducible mass and loses efficiency. Only in the complete
reversibility limit \cite{1971PhRvD...4.3552C} can the energy extraction process from an
extreme black hole reach the upper value of $50\%$ of the total black hole energy.

For $E_{+}\geq E_{\mathrm{c}}$ the leading extraction process is the
\emph{collective} process based on the generation of the optically thick
electron--positron plasma by the vacuum polarization.
The condition $E_{+}\geq E%
_{\mathrm{c}}$ implies
\begin{equation}
\tfrac{GM/c^{2}}{\lambda_{\mathrm{C}}}\left(  \tfrac{e}{\sqrt{G}m_{e}}\right)
^{-1}\simeq2\cdot10^{-6}\tfrac{M}{M_{\odot}}\leq\xi\leq1\,.
\end{equation}
This vacuum polarization process can occur only for a black hole with mass
smaller than $5\cdot10^{5}M_{\odot}$. The electron--positron pairs are now
produced in the dyadosphere of the black hole. We have
\begin{equation}
r_{\mathrm{dya}}\ll r_{\mathrm{Eerg}}. \label{dya1}%
\end{equation}
The number of particles created \cite{1998A&A...338L..87P} is then
\begin{equation}
N_{\mathrm{dya}}=\tfrac{1}{3}\left(  \tfrac{r_{\mathrm{dya}}}{\lambda
_{\mathrm{C}}}\right)  \left(  1-\tfrac{r_{+}}{r_{\mathrm{dya}}}\right)
\left[  4+\tfrac{r_{+}}{r_{\mathrm{dya}}}+\left(  \tfrac{r_{+}}%
{r_{\mathrm{dya}}}\right)  ^{2}\right]  \tfrac{Q}{e}\simeq\tfrac{4}{3}\left(
\tfrac{r_{\mathrm{dya}}}{\lambda_{\mathrm{C}}}\right)  \tfrac{Q}{e}\,.
\label{numdya}%
\end{equation}
The total energy stored in the dyadosphere is \cite{1998A&A...338L..87P}
\begin{equation}
{\mathcal E}_{\mathrm{dya}}^{\mathrm{tot}}=\left(  1-\tfrac{r_{+}}{r_{\mathrm{dya}}%
}\right)  \left[  1-\left(  \tfrac{r_{+}}{r_{\mathrm{dya}}}\right)
^{4}\right]  \tfrac{Q^{2}}{2r_{+}}\simeq\tfrac{Q^{2}}{2r_{+}}\,.
\label{enedya}%
\end{equation}
The mean energy per particle produced in the dyadosphere
$\left\langle
{\mathcal E}\right\rangle _{\mathrm{dya}}=\tfrac{{\mathcal E}_{\mathrm{dya}}^{\mathrm{tot}}%
}{N_{\mathrm{dya}}}$ is then
\begin{equation}
\left\langle {\mathcal E}\right\rangle _{\mathrm{dya}}=\tfrac{3}{2}\tfrac{1-\left(
\tfrac{r_{+}}{r_{\mathrm{dya}}}\right)  ^{4}}{4+\tfrac{r_{+}}{r_{\mathrm{dya}%
}}+\left(  \tfrac{r_{+}}{r_{\mathrm{dya}}}\right)  ^{2}}\left(  \tfrac
{\lambda_{\mathrm{C}}}{r_{\mathrm{dya}}}\right)  \tfrac{Qe}{r_{+}}\simeq
\tfrac{3}{8}\left(  \tfrac{\lambda_{\mathrm{C}}}{r_{\mathrm{dya}}}\right)
\tfrac{Qe}{r_{+}}\,, \label{meanenedya}%
\end{equation}
which can be also rewritten as
\begin{equation}
\left\langle {\mathcal E}\right\rangle _{\mathrm{dya}}\simeq\tfrac{3}{8}\left(
\tfrac{r_{\mathrm{dya}}}{r_{+}}\right)  \ m_{e}c^{2}\sim\sqrt{\tfrac{\xi
}{M/M_{\odot}}}10^{5}keV\,. \label{GRB}%
\end{equation}
We stress again that the vacuum polarization around a black hole has been observed
to reach theoretically the maximum efficiency limit of $50\%$ of the total mass-energy of an
extreme black hole (see e.g. \cite{1998A&A...338L..87P}).

Let us now compare and contrast these two processes. We have
\begin{equation}
r_{\mathrm{Eerg}}\simeq\left(  \tfrac{r_{\mathrm{dya}}}{\lambda_{\mathrm{C}}%
}\right)  r_{\mathrm{dya}},\quad N_{\mathrm{dya}}\simeq\left(  \tfrac
{r_{\mathrm{dya}}}{\lambda_{\mathrm{C}}}\right)  N_{\mathrm{PD}}%
,\quad\left\langle {\mathcal E}\right\rangle _{\mathrm{dya}}\simeq\left(  \tfrac
{\lambda_{\mathrm{C}}}{r_{\mathrm{dya}}}\right)  \left\langle {\mathcal E}\right\rangle
_{\mathrm{PD}}.
\end{equation}
Moreover we see (Eqs.~(\ref{UHECR}), (\ref{GRB})) that $\left\langle
{\mathcal E}\right\rangle _{\mathrm{PD}}$ is in the range of
energies of UHECR (see \cite{2000RvMP...72..689N} and references
therein), while for $\xi\sim0.1$ and $M\sim 10M_{\odot}$,
$\left\langle {\mathcal E}\right\rangle _{\mathrm{dya}}$ is in the
$\gamma$-ray range. In other words, the discrete particle decay
process involves a small number of particles with ultrahigh energies
($\sim10^{21}eV$), while vacuum polarization involves a much larger
number of particles with lower mean energies ($\sim10MeV$).

\subsection{A theorem on a possible disagreement between black holes and thermodynamics}\label{bhthermodyn}

This analysis of vacuum polarization process around black holes is so general that it allows as well to look back to traditional results on black hole physics with an alternative point of view. We quote in particular a result which allows to overcome a claimed inconsistency between general relativity and thermodynamics in the field of black holes.

It is well known that if a spherically symmetric mass distribution without any
electromagnetic structure undergoes free gravitational collapse, its total
mass-energy $M$ is conserved according to the Birkhoff theorem: the increase
in the kinetic energy of implosion is balanced by the increase in the
gravitational energy of the system. If one considers the possibility that part
of the kinetic energy of implosion is extracted then the situation is very
different: configurations of smaller mass-energy and greater density can be
attained without violating Birkhoff theorem in view of the radiation process.

From a theoretical physics point of view it is still an open
question how far such a sequence can go: using causality
nonviolating interactions, can one find a sequence of braking and
energy extraction processes by which the density and the
gravitational binding energy can increase indefinitely and the
mass-energy of the collapsed object be reduced at will? This
question can also be formulated in the mass formula language
\cite{1971PhRvD...4.3552C} (see also
Ref.~\cite{2002PhLB..545..233R}): given a collapsing core of
nucleons with a given rest mass-energy $M_{0}$, what is the minimum
irreducible mass of the black hole which is formed?

Following the previous two sections, consider a spherical shell of rest mass
$M_{0}$ collapsing in a flat space-time. In the neutral case the irreducible
mass of the final black hole satisfies Eq.~\ref{EQM}. The minimum irreducible
mass $M_{\mathrm{irr}}^{\left(  {\mathrm{min}}\right)  }$ is obtained when the
kinetic energy at the horizon $T_{+}$ is $0$, that is when the entire kinetic
energy $T_{+}$ has been extracted. We then obtain, form Eq.~\ref{EQM}, the
simple result
\begin{equation}
M_{\mathrm{irr}}^{\left(  \mathrm{min}\right)  }=\tfrac{M_{0}}{2}.
\label{Mirrmin}%
\end{equation}
We conclude that in the gravitational collapse of a spherical shell of rest
mass $M_{0}$ at rest at infinity (initial energy $M_{\mathrm{i}}=M_{0}$), an
energy up to $50\%$ of $M_{0}c^{2}$ can in principle be extracted, by braking
processes of the kinetic energy. In this limiting case the shell crosses the
horizon with $T_{+}=0$. The limit $\tfrac{M_{0}}{2}$ in the extractable
kinetic energy can further increase if the collapsing shell is endowed with
kinetic energy at infinity, since all that kinetic energy is in principle extractable.

We have represented in Fig.~\ref{CC} the world lines of spherical
shells of the same rest mass $M_{0}$, starting their gravitational
collapse at rest at selected radii $r_{0}^{\ast}$. These initial
conditions can be implemented by performing suitable braking of the
collapsing shell and concurrent kinetic energy extraction processes
at progressively smaller radii (see also Fig.~\ref{fig3a}). The
reason for the existence of the minimum (\ref{Mirrmin}) in the black
hole mass is the \textquotedblleft self-closure\textquotedblright\
occurring by the formation of a horizon in the initial configuration
(thick line in Fig.~\ref{CC}).

\begin{figure}[!ptb]
\begin{center}
\includegraphics[height=7cm,width=11cm]{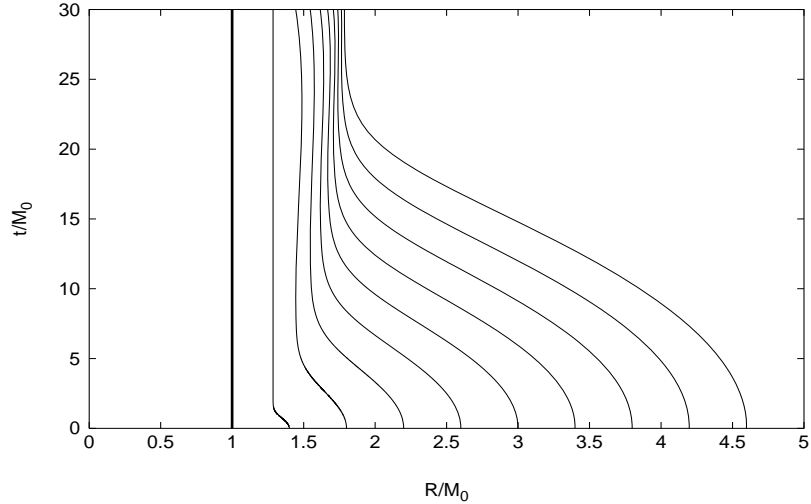}
\end{center}
\caption{Collapse curves for neutral shells with rest mass $M_{0}$ starting at
rest at selected radii $R^{\ast}$ computed by using the exact solutions given
in Ref.~\cite{2002PhLB..545..226C}. A different value of
$M_{\mathrm{irr}}$ (and therefore of $r_{+}$) corresponds to each curve. The
time parameter is the Schwarzschild time coordinate $t$ and the asymptotic
behaviour at the respective horizons is evident. The limiting configuration
$M_{\mathrm{irr}}=\tfrac{M_{0}}{2}$ (solid line) corresponds to the case in
which the shell is trapped, at the very beginning of its motion, by the
formation of the horizon.}%
\label{CC}%
\end{figure}

Is the limit $M_{\mathrm{irr}}\rightarrow\tfrac{M_{0}}{2}$ actually attainable
without violating causality? Let us consider a collapsing shell with charge
$Q$. If $M\geq Q$ a black hole is formed. As pointed out in the previous section
the irreducible mass of the final black hole does not depend on the charge $Q$.
Therefore Eqs.~(\ref{EQM}) and (\ref{Mirrmin}) still hold in the charged case.
In Fig.~\ref{fig3a} we consider the special case in which the shell is
initially at rest at infinity, i.e. has initial energy $M_{\mathrm{i}}=M_{0}$,
for three different values of the charge $Q$. We plot the initial energy
$M_{i}$, the energy of the system when all the kinetic energy of implosion has
been extracted as well as the sum of the rest mass energy and the
gravitational binding energy $-\tfrac{M_{0}^{2}}{2r_{0}}$ of the system (here
$r_{0}$ is the radius of the shell). In the extreme case $Q=M_{0}$, the shell
is in equilibrium at all radii (see Ref.~\cite{2002PhLB..545..226C}) and the kinetic energy
is identically zero. In all three cases, the sum of the extractable kinetic
energy $T$ and the electromagnetic energy $\tfrac{Q^{2}}{2r_{0}}$ reaches
$50\%$ of the rest mass energy at the horizon, according to Eq.~(\ref{Mirrmin}).

\begin{figure}[!ptb]
\begin{center}
\includegraphics[height=11cm]{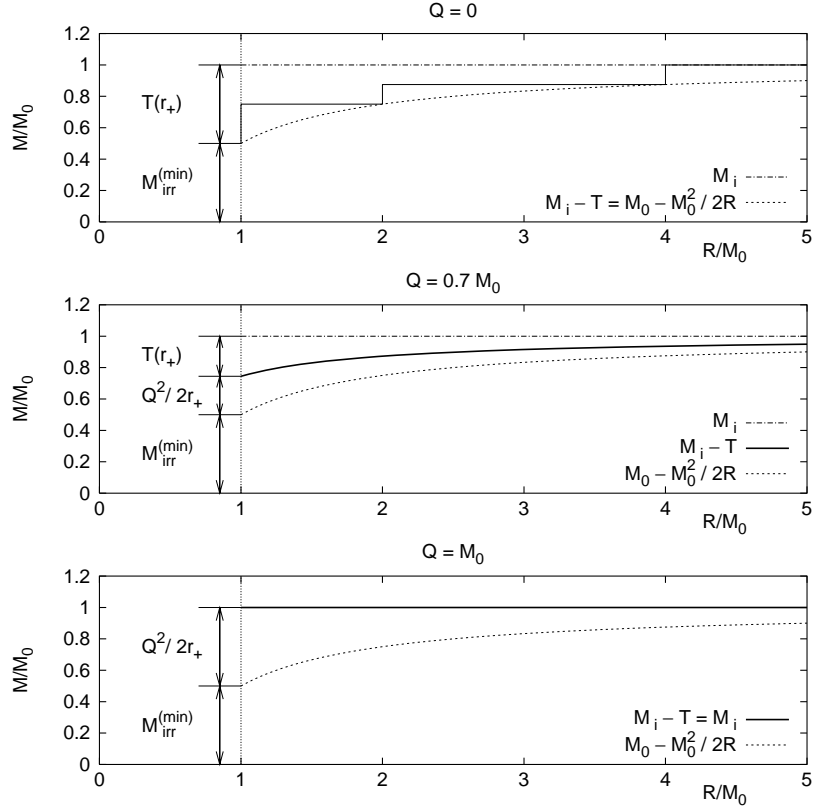}
\end{center}
\caption{Energetics of a shell such that $M_{\mathrm{i}}=M_{0} $, for selected
values of the charge. In the first diagram $Q=0$; the dashed line represents
the total energy for a gravitational collapse without any braking process as a
function of the radius $R$ of the shell; the solid, stepwise line represents a
collapse with suitable braking of the kinetic energy of implosion at selected
radii; the dotted line represents the rest mass energy plus the gravitational
binding energy. In the second and third diagram $Q/M_{0}=0.7$, $Q/M_{0}=1$
respectively; the dashed and the dotted lines have the same meaning as above;
the solid lines represent the total energy minus the kinetic energy. The
region between the solid line and the dotted line corresponds to the stored
electromagnetic energy. The region between the dashed line and the solid line
corresponds to the kinetic energy of collapse. In all the cases the sum of the
kinetic energy and the electromagnetic energy at the horizon is 50\% of
$M_{0}$. Both the electromagnetic and the kinetic energy are extractable. It
is most remarkable that the same underlying process occurs in the three cases:
the role of the electromagnetic interaction is twofold: a) to reduce the
kinetic energy of implosion by the Coulomb repulsion of the shell; b) to store
such an energy in the region around the black hole. The stored electromagnetic
energy is extractable as shown in Ref.~\cite{2002PhLB..545..233R}}%
\label{fig3a}%
\end{figure}

What is the role of the electromagnetic field here? If we consider the case of
a charged shell with $Q\simeq M_{0}$, the electromagnetic repulsion implements
the braking process and the extractable energy is entirely stored in the
electromagnetic field surrounding the black hole (see Ref.~\cite{2002PhLB..545..233R}). We emphasize here that the extraction of $50\%$ of the
mass-energy of a black hole is not specifically linked to the electromagnetic field
but depends on three factors: a) the increase of the gravitational energy
during the collapse, b) the formation of a horizon, c) the reduction of the
kinetic energy of implosion. Such conditions are naturally met during the
formation of an extreme black hole with $Q=M$, but as we have seen, they are more general and can indeed occur in a variety of different situations, e.g. during the formation of a Schwarzschild
black hole by a suitable extraction of the kinetic energy of implosion (see
Fig.~\ref{CC} and Fig.~\ref{fig3a}).

Before closing let us consider a test particle of mass $m$ in the gravitational field of an
already formed Schwarzschild black hole of mass $M$ and go through such a
sequence of braking and energy extraction processes. Kaplan \cite{1949ZhETF..19..951K} found
for the energy ${\mathcal E}$ of the particle as a function of the radius $r$
\begin{equation}
{\mathcal E}=m\sqrt{1-\tfrac{2M}{r}}. \label{pointtest}%
\end{equation}
It would appear from this formula that the entire energy of a particle could
be extracted in the limit $r\rightarrow2M$. Such $100\%$ efficiency of energy
extraction has often been quoted as evidence for incompatibility between
General Relativity and the second principle of Thermodynamics (see Ref.~\cite{1973PhRvD...7.2333B} and references therein). J. Bekenstein and S. Hawking have gone as
far as to consider General Relativity not to be a complete theory and to
conclude that in order to avoid inconsistencies with thermodynamics, the
theory should be implemented through a quantum description \cite{1973PhRvD...7.2333B,1974Natur.248...30H,1975CMaPh..43..199H,1977PhRvD..15.2752G}.
Einstein himself often expressed the opposite point of view (see, e.g., Ref.~\cite{Dyson2002} and references therein).

The analytic treatment presented in Section~\ref{collapse} can
clarify this fundamental issue. It allows to express the energy
increase ${\mathcal E}$ of a black hole of mass $M_{1}$ through the
accretion of a shell of mass $M_{0}$ starting its motion at rest at
a radius $r_{0}$ in the following formula which generalizes
Eq.~(\ref{pointtest}):
\begin{equation}
{\mathcal E}\equiv M-M_{1}=-\tfrac{M_{0}^{2}}{2r_{0}}+M_{0}\sqrt{1-\tfrac{2M_{1}}{r_{0}}%
},
\end{equation}
where $M=M_{1}+{\mathcal E}$ is clearly the mass-energy of the final
black hole. This formula differs from the Kaplan formula
(\ref{pointtest}) in three respects: (a) it takes into account the
increase of the horizon area due to the accretion of the shell; (b)
it shows the role of the gravitational self-energy of the imploding
shell; (c) it expresses the combined effects of (a) and (b) in an
exact closed formula.

The minimum value ${\mathcal E}_{\mathrm{\min}}$ of ${\mathcal E}$ is attained for the minimum value
of the radius $r_{0}=2M$: the horizon of the final black hole. This
corresponds to the maximum efficiency of the energy extraction. We have
\begin{equation}
{\mathcal E}_{\min}=-\tfrac{M_{0}^{2}}{4M}+M_{0}\sqrt{1-\tfrac{M_{1}}{M}}=-\tfrac
{M_{0}^{2}}{4(M_{1}+{\mathcal E}_{\min})}+M_{0}\sqrt{1-\tfrac{M_{1}}{M_{1}+{\mathcal E}_{\min}}},
\end{equation}
or solving the quadratic equation and choosing the positive solution for
physical reasons
\begin{equation}
{\mathcal E}_{\min}=\tfrac{1}{2}\left(  \sqrt{M_{1}^{2}+M_{0}^{2}}-M_{1}\right)  .
\end{equation}
The corresponding efficiency of energy extraction is
\begin{equation}
\eta_{\max}=\tfrac{M_{0}-{\mathcal E}_{\min}}{M_{0}}=1-\tfrac{1}{2}\tfrac{M_{1}}{M_{0}%
}\left(  \sqrt{1+\tfrac{M_{0}^{2}}{M_{1}^{2}}}-1\right)  , \label{efficiency}%
\end{equation}
which is strictly \emph{smaller than} 100\% for \emph{any} given $M_{0}\neq0$.
It is interesting that this analytic formula, in the limit $M_{1}\ll M_{0}$,
properly reproduces the result of equation (\ref{Mirrmin}), corresponding to
an efficiency of $50\%$. In the opposite limit $M_{1}\gg M_{0}$ we have
\begin{equation}
\eta_{\max}\simeq1-\tfrac{1}{4}\tfrac{M_{0}}{M_{1}}.
\end{equation}
Only for $M_{0}\rightarrow0$, Eq.~(\ref{efficiency}) corresponds to an
efficiency of 100\% and correctly represents the limiting reversible
transformations. It seems that the difficulties of reconciling General
Relativity and Thermodynamics are ascribable not to an incompleteness of
General Relativity but to the use of the Kaplan formula in a regime in which
it is not valid.

\subsection{Astrophysical gravitational collapse and black holes}

The time evolution of the gravitational collapse (occurring on characteristic
gravitational timescale $\tau=GM/c^3\simeq 5\times 10^{-5}M/M_\odot$ s) and the
associated electrodynamical process are too complex for direct description. We
addressed here a more confined problem: the vacuum polarization process around
an already formed Kerr--Newman black hole. This is a well defined problem which
deserves attention. It is theoretically expected to represent a physical state
asymptotically reached in the process of gravitational collapse. Such an
asymptotic configuration will be reached when all multipoles departing from the
Kerr--Newman geometry have been radiated away either by process of vacuum
polarization or electromagnetic and gravitational waves. What is most important
is that by performing this theoretical analysis we can have a direct evaluation
of the energetics and of the spectra and dynamics of the $e^+e^-$ plasma
created on the extremely short timescale due to the quantum phenomena of
$\Delta t=\hbar/(m_e c^2)\simeq10^{-21}$ s. This entire transient phenomena,
starting from an initial neutral condition of the core in the progenitor star,
undergoes the formation of the Kerr--Newman black hole by the collective
effects of gravitation, strong, weak, electromagnetic interactions during a
fraction of the above mentioned gravitational characteristic timescale of
collapse.

After the process of vacuum polarization all the electromagnetic energy of
incipient Kerr--Newman black hole will be radiated away and almost neutral Kerr
solution will be left and reached asymptotically in time. In a realistic
gravitational collapse the theoretical picture described above will be further
amplified by the presence of high-energy processes including neutrino emission
and gravitational waves emission with their electromagnetic coupling: the
gravitationally induced electromagnetic radiation and electromagnetically
induced gravitational radiation \cite{1973PhRvL..31.1317J,1974PhLB...49..185J}.

Similarly, in the next Section we proceed to a deeper understanding
of other collective plasma phenomena also studied in idealized
theoretically well defined cases. They will play an essential role
in the astrophysical description of the dynamical phase of
gravitational collapse.

\section{Plasma oscillations in electric fields}\label{time-independent}

We have seen in the previous Sections the application of the
Sauter-Heisenberg-Euler-Schwinger process for electron--positron pair
production in the heavy nuclei, in the laser and in the last Section in the
field of black holes. The case of black holes is drastically different from all
the previous ones. The number of electron--positron pairs created is of the
order of $10^{60}$, the plasma expected is optically thick and is very
different from the nuclear collisions and laser case where pairs are very few
and therefore optically thin. The following dynamical aspects need to be
addressed.

\begin{enumerate}
\item the back reaction of pair production on the external electric field;
\item the screening effect of pairs on the external electric field strengths;
\item the motion of pairs and their interactions.
\end{enumerate}

When these dynamical effects are considered, the pair production in an external
electric field is no longer only a process of quantum tunneling in a constant
static electric field. In fact, it turns out to be a much more complex process
during which all the three above mentioned effects play an important role. More
precisely, a phenomenon of electron--positron oscillation, {\it plasma
oscillation},  takes place. We are going to discuss such plasma oscillation
phenomenon in this Section. As we will see in this Section these phenomena can
become also relevant for heavy-ion collisions. After giving the basic equations
for description of plasma oscillations we give first some applications in the
field of heavy ions. In this Section in all formulas we use $c=\hbar=1$.

\subsection{Semiclassical theory of plasma oscillations in electric
fields}\label{semi-classical}

In the semi-classical QED \cite{1987PhRvD..36.3114C,1989PhRvD..40..456C}, one
quantizes only the Dirac field $\psi(x)$, while an external electromagnetic
field $A^\mu(x)$ is treated classically as a mean field.  This is the
self-consistent mean field or Gaussian approximation that can be formally
derived as the leading term in the large-$N$ limit of QED, where $N$ is the
number of charged matter fields \cite{1989PhRvD..40..456C, 1993PhRvD..47.2343B,
1994PhRvD..49.2769B, 1995PhRvD..51.4419B, 1996PhRvD..54.4013B}. The motion of
these electrons can be described by a Dirac equation in an external classical
electromagnetic potential $A^\mu(x)$
\begin{equation}
[\gamma_\mu(i\partial^\mu-eA^\mu)-m]\psi(x)=0
\label{equationofmotion}
\end{equation}
and the semi-classical Maxwell equation
\begin{equation}
\partial_\mu F^{\mu\nu}= \langle j^\nu(x)\rangle, \quad j^\nu(x)=i\frac{e}{2}[\bar\psi(x),\gamma^\nu\psi(x)],
\label{semimaxwell}
\end{equation}
where $j^\nu(x)$ is the electron and positron current and the expectation value
is with respect to the quantum states of the electron field. The dynamics that
these equations describe is not only the motion of electron and positron pairs,
but also their back reaction on the external electromagnetic field. The
resultant phenomenon is the so-called plasma oscillation that we will discuss
based on both a simplified model of semi-classical scalar QED and kinetic
Boltzmann-Vlasov equation as presented in Refs.~\cite{1987PhRvD..36.3114C,
1989PhRvD..40..456C, 1991PhRvL..67.2427K, 1993PhRvD..47.4639B,
1998PhRvD..58l5015K}.

A scheme for solving the back reaction problem in scalar QED was offered in
Refs.~\cite{1987PhRvD..36.3114C, 1989PhRvD..40..456C}. Based on this scheme, a
numerical analysis was made in (1+1)-dimensional case
\cite{1991PhRvL..67.2427K}. Eqs.~(\ref{equationofmotion}), (\ref{semimaxwell})
are replaced by the scalar QED coupled equations for a charged scalar field
$\Phi(x)$
\begin{equation}
[(i\partial^\mu-eA^\mu)^2-m_e^2]\Phi(x)=0.
\label{scalarqed}
\end{equation}
The current $j^\nu(x)$ of the charged scalar field in the
semi-classical Maxwell equations (\ref{semimaxwell}) is
\begin{equation}
j^\nu(x)=i\frac{e}{2}[\Phi^*(x)\partial^\nu\Phi(x)-\Phi(x)\partial^\nu\Phi^*(x)].
\label{scalarcurrent}
\end{equation}
Now, consider a spatially homogeneous electric field ${\bf
E}=E_z(t)\hat {\bf z}$ in the $\hat {\bf z}$-direction. A
corresponding gauge potential is ${\bf A}=A_z(t){\hat \bf z},
A_0=0$. Defining $E\equiv E_z$, $A\equiv A_z$ and $j\equiv j_z$, the
Maxwell equations (\ref{semimaxwell}) reduce to the single equation
\begin{equation}
\frac{d^2A}{dt^2}= \langle j(x)\rangle,
\label{11semimaxwell}
\end{equation}
for the potential and $E=-dA/dt$.

The quantized scalar field $\Phi(x)$ in Eq.~(\ref{scalarqed}) can be
expanded in terms of plane waves with operator-valued amplitudes
$f_{\bf k}(t)a_{\bf k}$ and $f_{-\bf k}^*(t)b^\dagger_{\bf k}$
\begin{equation}
\Phi(x)=\frac{1}{V^{1/2}}\sum_{\bf k}[f_{\bf k}(t)a_{\bf k}
+f_{-\bf k}^*(t)b^\dagger_{\bf k}]e^{-i{\bf kx}},
\label{11phi}
\end{equation}
where $V$ is the volume of the system and the time-independent
creation and annihilation operators obey the commutation relations
\begin{equation}
[a_{\bf k},a^\dagger_{\bf k'}]=[b_{\bf k},b^\dagger_{\bf k'}]=\delta_{\bf k,k'},
\label{abasis}
\end{equation}
and each ${\bf k}$-mode function $f_{\bf k}$ obeys the Wronskian condition,
\begin{equation}
f_{\bf k}\dot f^*_{\bf k}-\dot f_{\bf k}f_{\bf k}^*=i.%\hbar
\label{abasis1}
\end{equation}
The time dependency in this basis $(a_{\bf k},b^\dagger_{\bf k})$
(\ref{11phi},\ref{abasis}) is carried by the complex mode functions $f_{\bf
k}(t)$ that satisfy the following equation of motion, as demanded from the QED
coupled Eq.~(\ref{scalarqed}) of Klein--Gordon type
\begin{equation}
\left(\frac{d^2}{dt^2} + \omega_{\bf k}^2(t)\right)f_{\bf k}(t)=0,
\label{11equation}
\end{equation}
where the time-dependent frequency $\omega^2_{\bf k}(t)$ is given by
\begin{equation}
\omega^2_{\bf k}(t)\equiv [{\bf k}-e{\bf A}]^2+m_e^2=[k-eA(t)]^2+{\bf k}^2_\perp +m_e^2.
\label{11frequency}
\end{equation}
Here $k$ is the {\it constant canonical} momentum in the ${\bf \hat
z}$-direction which should be distinguished from the gauge
invariant, but {\it time-dependent kinetic } momentum
\begin{equation}
p(t)=k-eA(t),\quad \frac{dp}{dt}=eE,
\label{11p}
\end{equation}
which reflects the acceleration of the charged particles due to the electric field,
while in the directions transverse to the electric field the kinetic and canonical momenta are the same
${\bf k}_\perp={\bf p}_\perp$.

The mean value of electromagnetic current (\ref{scalarcurrent}) in
the ${\bf \hat z}$-direction is then
\begin{equation}
\langle j(t)\rangle = 2e\int \frac{d^3{\bf k}}{(2\pi)^3}[k-eA(t)]|f_{\bf k}(t)|^2[1+N_+({\bf k})+N_-(-{\bf k})],
\label{qcurrent}
\end{equation}
where $N_+({\bf k})=\langle a^\dagger_{\bf k}a_{\bf k}\rangle$ and
$N_-({\bf k})=\langle b^\dagger_{\bf k}b_{\bf k}\rangle$ are the
mean numbers of particles and antiparticles in the time-independent
basis (\ref{11phi},\ref{abasis}). The mean charge density must
vanish
\begin{equation}
\langle j^0(t)\rangle =e\int \frac{d^3{\bf k}}{(2\pi)^3}[N_+({\bf k})-N_-(-{\bf k})]=0,\quad\int \frac{d^3{\bf k}}{(2\pi)^3}
 \equiv \frac{1}{V}\sum_{\bf k}
\nonumber
\end{equation}
by the Gauss law for a spatially homogeneous electric field (i.e.,
${\bf \nabla\cdot E}=0$). As a result, $N_+({\bf k})=N_-(-{\bf
k})\equiv N_{\bf k}$. For the vacuum state, $N_{\bf k}=0$. The
Maxwell equation (\ref{11semimaxwell}) for the evolution of electric
field becomes
\begin{equation}
\frac{d^2A}{dt^2}= 2e\int \frac{d^3{\bf k}}{(2\pi)^3}[k-eA(t)]
|f_{\bf k}(t)|^2\sigma_{\bf k},\quad \sigma_{\bf k}=(1+2N_{\bf k}).
\label{semimaxwellnum}
\end{equation}
These two scalar QED coupled Eqs.~(\ref{11equation}) and
(\ref{semimaxwellnum}) in (1+1)-dimensional case were numerically
integrated in Ref. \cite{1991PhRvL..67.2427K}. The results are shown
in Fig.~\ref{11oscilla}, where the time evolutions of the scaled
electric field $\tilde E\equiv E/E_c$ and current $\tilde j\equiv
j\hbar/(E_cm_ec^2)$ are shown as functions of time $\tau\equiv
(m_ec^2/\hbar)t$ in unit of Compton time $(\hbar/m_ec^2)$. Starting
with a strong electric field, one clearly finds the phenomenon of
oscillating electric field $E(t)$ and current $j(t)$, i.e., plasma
oscillation.

\begin{figure}[!ht]
\begin{center}
\includegraphics[height=11cm,width=12cm]{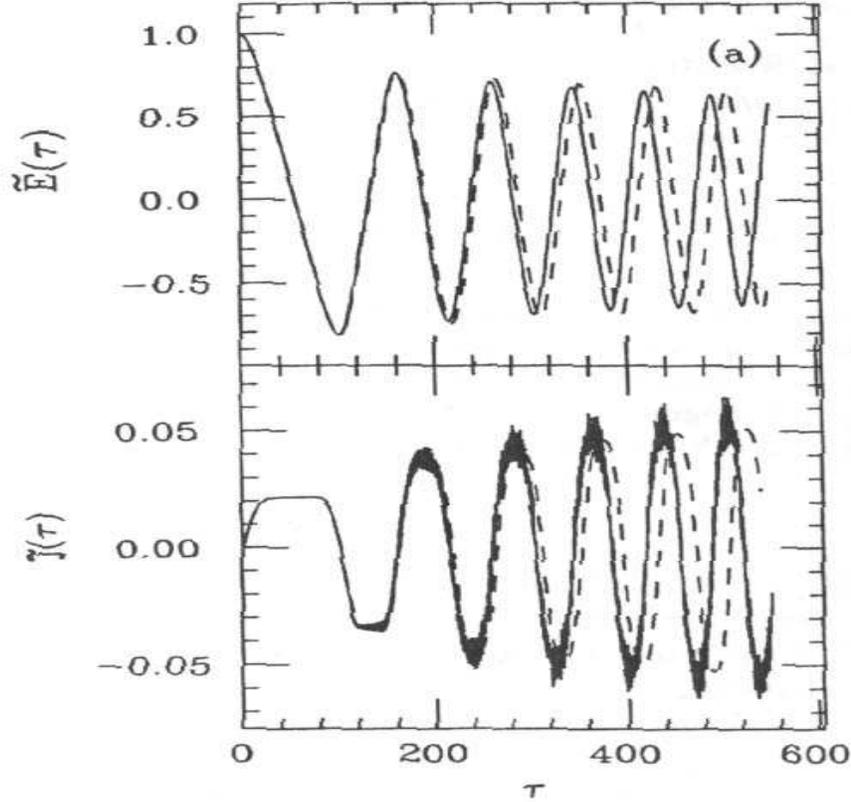}
\end{center}
\caption{Time evolution of scaled electric field $\tilde E$ and current
$\tilde j$, with initial value $\tilde E=1.0$ and coupling $e^2/m_e^2=0.1$.
The solid line is semi classical scalar QED, and the dashed line is the Boltzmann-Vlasov model.
This figure is reproduced from Fig.~1 (a) in Ref.~\cite{1991PhRvL..67.2427K}.}%
\label{11oscilla}%
\end{figure}

This phenomenon of plasma oscillation is shown in
Fig.~\ref{11oscilla} and is easy to understand as follows. In a
classical kinetic picture, we have the electric current
$j=2en\langle v\rangle $ where $n$ is the density of electrons (or
positrons) and $\langle v\rangle$ is their mean velocity. Driven by
the external electric field, the velocity $\langle v\rangle$ of
electrons (or positrons) continuously increases, until the electric
field of electron and positron pairs screens the external electric
field down to zero, and the kinetic energy of electrons (or
positrons) reaches its maximum. The electric current $j$ saturates
as the velocity $\langle v\rangle$ is close to the speed of light.
Afterward, these electrons and positrons continuously move apart
from each other further, their electric field, whose direction is
opposite to the direction of the external electric field, increases
and decelerates electrons and positrons themselves. Thus the
velocity $\langle v\rangle$ of electrons and positrons decreases,
until the electric field reaches negative maximum and the velocity
vanishes. Then the velocity $\langle v\rangle$ of electrons and
positrons starts to increase in backward direction and the electric
field starts to decrease for another oscillation cycle.

\subsection{Kinetic theory of plasma oscillations in electric fields}\label{kineticplasma}

In describing the same system as in the previous section it can be
used, alternatively to the semi-classical theory, a phenomenological
model based on the following relativistic Boltzmann-Vlasov equation
\cite{1991PhRvL..67.2427K}
\begin{equation}
\frac{d {\mathcal F}}{d t}\equiv\frac{\partial {\mathcal F}}{\partial t}+eE\frac{\partial {\mathcal F}}{\partial p}=\frac{dN}{dtdVd^3{\bf p}},
\label{11bv}
\end{equation}
where the ${\bf x}$-independent function ${\mathcal F}({\bf p},t)$
is a classical distribution function of particle and antiparticle
pairs in phase space. The source term in the right-hand side of
Eq.~(\ref{11bv}) is related to the Schwinger rate ${\mathcal P}_{\rm
boson}$ (\ref{bosonrate}) for the pair production of spin-$0$
particle and antiparticle
\begin{equation}
\frac{dN}{dtdVd^3{\bf p}}=[1+2{\mathcal F}({\bf p},t)]{\mathcal P}_{\rm boson}\delta^3({\bf p}),
\label{11bosonrate}
\end{equation}
where the $\delta$-function $\delta^3({\bf p})$ expresses the fact
that particles are produced at rest and the factor $[1+2{\mathcal
F}({\bf p},t)]$ accounts for stimulated pair production (Bose
enhancement). The Boltzmann-Vlasov equation (\ref{11bv}) for the
distribution function ${\mathcal F}({\bf p},t)$ is in fact
attributed to the conservation of particle number in phase space.

In the field equation (\ref{11semimaxwell}) for the classical gauge
potential $A$, the electric current $\langle j(x)\rangle$ is
contributed from the conduction current
\begin{equation}
j_{\rm cond}=2e\int \frac{d^3{\bf p}}{(2\pi)^3}\frac{[p-eA(t)]}{\omega_{\bf p}} {\mathcal F}({\bf p},t),
\label{11cond}
\end{equation}
and the polarization current \cite{1987PhRvD..36..114G}
\begin{equation}
j_{\rm pol}=\frac{2}{E}\int \frac{d^3{\bf p}}{(2\pi)^3} \omega_{\bf p} \frac{dN}{dtdVd^3{\bf p}}.
\label{11pol}
\end{equation}
The relativistic Boltzmann-Vlasov equation (\ref{11bv}) and field
equation (\ref{11semimaxwell}) with the conduction current
(\ref{11cond}) and the polarization current (\ref{11pol}) were
numerically integrated \cite{1991PhRvL..67.2427K} in
(1+1)-dimensional case. The numerical integration shows that the
system undergoes plasma oscillations. In Fig.~\ref{11oscilla} the
results of the semi-classical analysis and the numerical integration
of the Boltzmann equation are compared. We see that they are in good
quantitative agreement. The discrepancies are because in addition to
spontaneous pair production, the quantum theory takes into account
pair production via bremsstrahlung (``induced'' pair production),
which are neglected in Eq. (\ref{11bv})).

In Refs.~\cite{1992PhRvD..45.4659K,1993IJMPE...2..333K}, the study
of plasma oscillation was extended to the fermionic case. On the
basis of semi-classical theory of spinor QED, expressing the
solution of the Dirac equation (\ref{equationofmotion}) as
\begin{equation}
\psi(x)=[\gamma_\mu(i\partial^\mu-eA^\mu)+m_e]\phi(x),
\nonumber
\end{equation}
where $\phi(x)$ is a four-component spinor, one finds that $\phi(x)$
satisfies the quadratic Dirac equation,
\begin{equation}
\left[(i\partial^\mu-eA^\mu)^2-\frac{e}{2}\sigma^{\mu\nu}F_{\mu\nu}-m_e^2\right]\phi(x)=0.
\label{quadraticdirac}
\end{equation}
The electric current of spinor field $\phi(x)$ couples to the
external electric field that obeys the field equation
(\ref{11semimaxwell}).

The source term in the right-hand side of the kinetic
Boltzmann-Vlasov equation (\ref{11bv}) is changed to the Schwinger
rate ${\mathcal P}_{\rm fermion}$ (\ref{probability1}) for the pair
production of electrons and positrons,
\begin{equation}
\frac{dN}{dtdVd^3{\bf p}}=[1-2{\mathcal F}({\bf p},t)]{\mathcal P}_{\rm fermion}\delta^3({\bf p}),
\label{11fermionrate}
\end{equation}
where the Pauli blocking is taken into account by the factor
$[1-2{\mathcal F}({\bf p},t)]$. Analogously to the scalar QED case,
both semi-classical theory of spinor QED and kinetic
Boltzmann-Vlasov equation have been analyzed and numerical
integration was made in the (1+1)-dimensional case
\cite{1992PhRvD..45.4659K,1993IJMPE...2..333K}. The numerical
results show that plasma oscillations of electric field, electron
and positron currents are similar to that plotted in
Fig.~\ref{11oscilla}.

\subsection{Plasma oscillations in the color electric field of heavy ions}

The Relativistic Heavy-Ion Collider (RIHC) at Brookhaven National Laboratory
and Large Hadron Collider (LHC) at CERN are designed with the goal of producing
a phase of deconfined hadronic matter: the quark-gluon plasma. A popular
theoretical model for studying high-energy heavy-ion collisions begins with the
creation of a flux tube containing a strong color electric field
\cite{Biro1984}. The field energy is converted into particles such as quark and
antiquark pairs and gluons that are created by the
Sauter--Euler--Heisenberg--Schwinger quantum tunneling mechanism. A
relativistic Boltzmann--Vlasov equation coupling to such particle creation
source is phenomenologically adopted in a kinetic theory model for the
hydrodynamics of quark and gluon plasma \cite{1987PhRvD..36..114G,
1984PhRvD..30.2371B, 1985PhRvD..31..198B, 1985ZPhyC..28..255B,
1986NuPhB.267..242B, 1986PhRvL..56..219K, 1988NuPhB.296..611B}.

In the collision of heavy-ion beams, one is clearly dealing with a situation
that is not spatially homogeneous. However, particle production in the central
rapidity region can be modeled as hydrodynamical system with longitudinal boost
invariance \cite{Landau1953, 1975PhRvD..11..192C, 1975PhRvD..12..163L,
1975PhRvL..34.1286N, 1983PhR....97...31A}. To express the longitudinal boost
invariance of the hydrodynamical system, one introduces the comoving
coordinates: fluid proper time $\tau$ and rapidity $\eta$ by the relationships
\cite{1993PhRvD..48..190C}
\begin{equation}
z=\tau\sinh(\eta),\quad t=\tau\cosh\eta,
\label{plasmacomoving}
\end{equation}
in terms of the Minkowski time $t$ and the coordinate $z$ along the
beam direction ${\bf \hat z}$ in the ordinary laboratory frame. The
line element and metric tensor in these coordinates are given by
\begin{equation}
ds^2=d\tau^2-dx^2-dy^2-\tau^2d\eta^2, \quad g_{\mu\nu}={\rm diag}(1,-1,-1,-\tau^2),
\label{heavyionline}
\end{equation}
and
\begin{equation}
g_{\mu\nu}=V^a_\mu V^b_\nu \eta_{ab}, \quad V^a_\mu={\rm diag}(1,1,1,\tau).
\nonumber
\end{equation}
where the vierbein $V^a_\mu$ transforms the curvilinear coordinates to
Minkowski coordinates and ${\rm det}(V)=\sqrt{-g}=\tau$. The covariant
derivative on fermion field $\psi(x)$ is given by \cite{1972gcpa.book.....W}
\begin{equation}
\nabla_\mu\psi(x)\equiv [(i\partial_\mu-eA_\mu)+\Gamma_\mu]\psi(x)
\label{vierbein}
\end{equation}
and the spin connection $\Gamma_\mu$ is
\begin{equation}
\Gamma_\mu=\frac{1}{2}\Sigma^{ab}V_{a\nu}
(\partial_\mu V_b^\nu +\Gamma^\nu_{\mu\lambda}V_b^\lambda),
\quad \Sigma^{ab}=\frac{1}{4}[\gamma^a,\gamma^b],
\nonumber
\end{equation}
with $\Gamma^\nu_{\mu\lambda}$ the usual Christoffel symbols and $\gamma^a$ the
usual coordinate-independent Dirac gamma matrices. The coordinate-dependent
gamma matrices $\tilde \gamma^\mu$ are obtained via
\begin{equation}
\tilde \gamma^\mu=\gamma^a V_a^\mu.
\nonumber
\end{equation}
In the curved space time (\ref{heavyionline}), the Dirac equation
(\ref{equationofmotion}) and semi-classical Maxwell equation
(\ref{semimaxwell}) are modified as
\begin{equation}
[i\tilde\gamma_\mu\nabla^\mu-m_e]\psi(x)=0
\label{curvequationofmotion}
\end{equation}
and
\begin{equation}
\frac{1}{\sqrt{-g}}\partial_\mu \sqrt{-g} F^{\mu\nu}= \langle j^\nu(x)\rangle, \quad j^\nu(x)=\frac{e}{2}[\bar\psi(x),\tilde\gamma^\nu\psi(x)].
\label{curvsemimaxwell}
\end{equation}

The phenomenological Boltzmann--Vlasov equation in (3+1)-dimensions can be also
written covariantly as
\begin{equation}
\frac{D {\mathcal F}}{D \tau}\equiv p^\mu\frac{\partial {\mathcal F}}{\partial q^\mu}
-ep^\mu F_{\mu\nu}\frac{\partial {\mathcal F}}{\partial p_ \nu}=\frac{dN}{\sqrt{-g}dq^0d^3{\bf q}d{\bf p}},
\label{curv11bv}
\end{equation}
where $D/D\tau$ is the total proper time derivative. This kinetic transport
equation is written in the comoving coordinates and their conjugate momenta:
\begin{equation}
q^\mu=(\tau,x,y,\eta),\quad p_\mu=(p_\tau,p_x,p_y,p_\eta).
\nonumber
\end{equation}
Due to the longitudinal boost invariance, energy density and color electric
field are spatially homogeneous, i.e., they are functions of the proper time
$\tau$ only \cite{1993PhRvD..48..190C}. Consequently, the approach for
spatially homogeneous electric field presented in
Refs.~\cite{1989PhRvD..40..456C, 1991PhRvL..67.2427K, 1992PhRvD..45.4659K,
1993IJMPE...2..333K} and discussed in the previous Section~\ref{kineticplasma}
is applicable to the phenomenon of plasma oscillations in ultrarelativistic
heavy-ion collisions \cite{1993PhRvD..48..190C} using
Eq.~\ref{curvequationofmotion} (resp.~Eq.~\ref{curv11bv}) in the place of
Eq.~\ref{equationofmotion} (resp.~Eq.~\ref{11bv}).

\subsection{Quantum Vlasov equation}\label{quanvlasov}

To understand the connection between the two frameworks of semi-classical field
theory and classical kinetic theory, both of which describe the plasma
oscillations, one can try to study a quantum transport equation in the
semi-classical theory \cite{1998PhRvD..58l5015K, 1997hep.ph...12377S,
1998IJMPE...7..709S, 2009PhRvL.102o0404H}. For this purpose, a Bogoliubov
transformation from the time-independent number basis ($a_{\bf
k},b^\dagger_{\bf k}$) (\ref{11phi}, \ref{abasis}) to a time-dependent number
basis [$\tilde a_{\bf k}(t),\tilde b^\dagger_{\bf k}(t)$] is introduced
\cite{1998PhRvD..58l5015K, 1997hep.ph...12377S, 1998IJMPE...7..709S},
\begin{eqnarray}
f_{\bf k}(t)&=&\alpha_{\bf k}(t)\tilde f_{\bf k}(t)+
\beta_{\bf k}(t)\tilde f^*_{\bf k}(t);\label{bogo0}\\
\dot f_{\bf k}(t)&=&-i\omega_{\bf k}\alpha_{\bf k}(t)\tilde f_{\bf k}(t)
+i\omega_{\bf k}\beta_{\bf k}(t)\tilde f^*_{\bf k}(t),
\label{bogo1}
\end{eqnarray}
and
\begin{eqnarray}
a_{\bf k}(t)&=&\alpha^*_{\bf k}(t)\tilde a_{\bf k}(t)-
\beta^*_{\bf k}(t)\tilde b^\dagger_{-\bf k}(t);\label{bogo1.5}\\
b^\dagger_{-\bf k}(t)&=&\alpha_{\bf k}(t)\tilde b^\dagger_{-\bf k}(t)
-\beta_{\bf k}(t)\tilde a_{\bf k}(t),
\label{bogo2}
\end{eqnarray}
where $\alpha_{\bf k}(t)$ and $\beta_{\bf k}(t)$ are the Bogoliubov
coefficients. They obey
\begin{equation}
|\alpha_{\bf k}(t)|^2-|\beta_{\bf k}(t)|^2=1,
\nonumber
\end{equation}
for each mode {\bf k}. In the limit of very slowly varying $\omega_{\bf k}(t)$
as a function of time $t$ (\ref{11frequency}), i.e., $\dot\omega_{\bf k}\ll
\omega^2_{\bf k}$ and $\ddot\omega_{\bf k}\ll \omega^3_{\bf k}$, the adiabatic
number basis [$\tilde a_{\bf k}(t),\tilde b^\dagger_{\bf k}(t)$] is defined by
first constructing the adiabatic mode functions,
\begin{equation}
\tilde f_{\bf k}(t)=\left(\frac{\hbar}{2\omega_{\bf k}}\right)^{1/2}
\exp\left[-i\Theta_{\bf k}(t)\right],\quad
\Theta_{\bf k}(t)=\int^t\omega_{\bf k}(t')dt'.
\label{adia}
\end{equation}
The particle number ${\mathcal N}_{\bf k}(t)$ in the time-dependent adiabatic
number basis is given by
\begin{eqnarray}
{\mathcal N}_{\bf k}(t) &= &\langle \tilde a^\dagger_{\bf k}(t)\tilde a_{\bf k}(t)\rangle
=\langle \tilde b^\dagger_{-\bf k}(t)\tilde b_{-\bf k}(t)\rangle\nonumber\\
&=&|\alpha_{\bf k}(t)|^2N_{\bf k}+|\beta_{\bf k}(t)|^2[1+N_{\bf k}],
\label{timeN}
\end{eqnarray}
which though time-dependent, is an adiabatic invariant of the
motion. Consequently, it is a natural candidate for a particle
number density distribution function ${\mathcal F}({\bf p},t)$ in
the phase space, that is needed in a kinetic description.

By differentiating ${\mathcal N}_{\bf k}(t)$ (\ref{timeN}) and using the basic
relationships (\ref{11frequency},\ref{bogo1},\ref{adia}), one obtains,
\begin{equation}
\dot{\mathcal N}_{\bf k}(t)=
\frac{\dot\omega_{\bf k}}{\omega_{\bf k}}
{\rm Re}\left\{{\mathcal C}_{\bf k}(t)
\exp[-2i\Theta_{\bf k}(t)]\right\},\quad
{\mathcal C}_{\bf k}(t)=(1+2N_{\bf k})\alpha_{\bf k}(t)\beta^*_{\bf k}(t),
\label{vtimeN}
\end{equation}
and
\begin{equation}
\dot{\mathcal C}_{\bf k}(t)= \frac{\dot\omega_{\bf k}}{2\omega_{\bf k}}
[1+2{\mathcal N}_{\bf k}(t)] \exp[2i\Theta_{\bf k}(t)]. \label{derivC}
\end{equation}
These two equations give rise to the quantum Vlasov equations
\cite{1998PhRvD..58l5015K, 1997hep.ph...12377S, 1998IJMPE...7..709S},
\begin{eqnarray}
\dot{\mathcal N}_{\bf k}(t)&=&{\mathcal S}_{\bf k}(t),\label{quanbv0}\\
{\mathcal S}_{\bf k}(t)&=&\frac{\dot\omega_{\bf k}}{2\omega_{\bf k}}\int^t_{-\infty}dt'
\frac{\dot\omega_{\bf k}}{\omega_{\bf k}}(t')[1\pm 2{\mathcal N}_{\bf k}(t')]
\cos[2\Theta_{\bf k}(t)-2\Theta_{\bf k}(t')],
\label{quanbv}
\end{eqnarray}
describing the time evolution of the adiabatic particle number ${\mathcal
N}_{\bf k}(t)$ of the mean field theory. ${\mathcal S}_{\bf k}(t)$ describes
the quantum creation rate of particle number in an arbitrary slowly varying
mean field. The Bose enhancement and Pauli blocking factors $[1\pm 2{\mathcal
N}_{\bf k}(t')]$ appear in Eq.~(\ref{quanbv}) so that both spontaneous and
induced particle creation are included automatically in the quantum treatment.
The most important feature of Eq.~(\ref{quanbv}) is that the source term
${\mathcal S}_{\bf k}(t)$ is nonlocal in time, indicating the particle creation
rate depending on the entire history of the system. This means that the time
evolution of the particle number ${\mathcal N}_{\bf k}(t)$ governed by the
quantum Vlasov equation is a non-Markovian process.

The mean electric current $\langle j(t)\rangle$ (\ref{qcurrent}) in the basis
of adiabatic number $[\tilde a_{\bf k}(t),\tilde b^\dagger_{\bf k}(t)]$
(\ref{bogo1},\ref{bogo2}) can be rewritten as,
\begin{eqnarray}
\langle j(t)\rangle &=& j_{\rm cond}+j_{\rm pol},\label{qcurrent2}\\
j_{\rm cond}&=&2e\int \frac{d^3{\bf k}}{(2\pi)^3}
\frac{[k-eA(t)]}{\omega_{\bf k}}{\mathcal N}_{\bf k}(t),\label{qcurrent21}\\
j_{\rm pol}&=&\frac{2}{E}\int \frac{d^3{\bf k}}{(2\pi)^3}
\omega_{\bf k}\dot{\mathcal N}_{\bf k}(t),\label{qcurrent22}
\end{eqnarray}
by using Eqs.~(\ref{timeN}), (\ref{vtimeN}). This means electric current
$\langle j(t)\rangle$ enters into the right-hand side of the field equation
(\ref{11semimaxwell}).

For the comparison between the quantum Vlasov equation (\ref{quanbv}) and the
classical Boltzmann--Vlasov equation (\ref{11bv}), the adiabatic particle
number ${\mathcal N}_{\bf k}(t)$ has to be understood as the counterpart of the
classical distribution function ${\mathcal F}({\bf p},t)$ of particle number in
the phase space. The source term, that is composed by the Schwinger rate of
pair production and the factor $[1\pm 2{\mathcal N}_{\bf k}(t')]$ for either
the Bose enhancement or Pauli blocking, is phenomenologically added into in the
Boltzmann--Vlasov equation (\ref{11bv}). Such source term  is local in time,
indicating that the time evolution of the classical distribution function
${\mathcal F}({\bf p},t)$ is a Markovian process. In the limit of a very slowly
varying uniform electric field $E$ and at very large time $t$ the source term
${\mathcal S}_{\bf k}(t)$ (\ref{quanbv}) integrated over momenta ${\bf k}$
reduces to the source term in the Boltzmann--Vlasov equation (\ref{11bv})
\cite{1998PhRvD..58l5015K, 1997hep.ph...12377S, 1998IJMPE...7..709S}. As a
result, the conduction current $j_{\rm cond}$ and the polarization current
$j_{\rm pol}$ in Eq.~(\ref{qcurrent2}) are reduced to their counterparts
Eqs.~(\ref{11cond}), (\ref{11pol}) in the phenomenological model of kinetic
theory. In Ref.~\cite{1998PhRvD..58l5015K, 1999PhRvD..60k6011B,
1999PhRvD..59i4005S, 2001EPJC...22..341V, 2002PhRvL..89o3901R}, the quantum
Vlasov equation has been numerically studied to show the plasma oscillations
and the non-Markovian effects. They are also compared with the
Boltzmann--Vlasov equation (\ref{11bv}) that corresponds to the Markovian
limit.

\subsection{Quantum decoherence in plasma oscillations}\label{quantumdecoherence}

As showed in Fig.~\ref{11oscilla}, the collective oscillations of
electric field $E(t)$ and associate electric current $\langle
j(t)\rangle$ are damped in their amplitude. Moreover, as time
increases, a decoherence in their oscillating frequency occurs
\cite{1991PhRvL..67.2427K,1996PhRvL..76.4660H}. This indicates that
plasma oscillations decay in time. This effective energy dissipation
or time irreversibility is the phenomenon of quantum decoherence
\cite{1991PhT....44j..36Z} in the process of creation and
oscillation of particles, in the sense that energy flows from
collective motion of the classical electromagnetic field to the
quantum fluctuations of charged matter fields without returning back
over times of physical interest
\cite{1996PhRvL..76.4660H,1997PhRvD..55.6471C}. This means that the
characteristic frequency $\omega_{\bf k}$ of the quantum fluctuation
mode ``${\bf k}$'' is much larger than the frequency $\omega_{pl}$
of the classical electric field: $\omega_{\bf k}\gg \omega_{pl}$ and
$\omega_{pl}^2\sim 2e^2\hbar n_p/(m_ec^2)$
\cite{1998PhRvD..58l5015K,1996PhRvL..76.4660H}, where $n_p$ is the
number density of particles and antiparticles. The study of quantum
decoherence and energy dissipation associated with particle
production to understand the plasma oscillation frequency and
damping can be found in
Refs.~\cite{1998PhRvD..58l5015K,1996PhRvL..76.4660H,1997PhRvD..55.6471C}.

To understand the energy dissipation from the collective oscillation
of classical mean fields to rapid fluctuations of quantum fields, it
is necessary to use the Hamiltonian formalism of semi-classical
theory. One defines the quantum fluctuation $\xi_{\bf k}(t)$ upon
classical mean field $\langle\Phi_{\bf k}(t)\rangle$ in the
semi-classical scalar theory
\cite{1998PhRvD..58l5015K,1996PhRvL..76.4660H,1997PhRvD..55.6471C},
\begin{equation}
\xi^2_{\bf k}(t)=\langle[\Phi_{\bf k}(t)-\langle\Phi_{\bf k}(t)\rangle]^2\rangle
=\langle\Phi_{\bf k}(t)\Phi^*_{\bf k}(t)\rangle-[\langle\Phi_{\bf k}(t)\rangle]^2,
\label{fluctuation}
\end{equation}
where $\Phi_{\bf k}(t)$ is the Fourier ${\bf k}-$component of the
quantized scalar field $\Phi(x)$ (\ref{11phi}). In the
time-independent basis (\ref{abasis},\ref{abasis1}), one has (see
Section~\ref{time-independent}),
\begin{equation}
\xi^2_{\bf k}(t)=\sigma_{\bf k}|f_{\bf k}(t)|^2,
\label{fluctuationfield}
\end{equation}
and the mode equation (\ref{11equation}) for $f_{\bf k}(t)$ can be rewritten as
\begin{equation}
\dot\eta_{\bf k}(t)=\ddot\xi_{\bf k}(t)=-\omega^2_{\bf k}(t)\xi_{\bf k}(t)
+\frac{\hbar^2\sigma^2_{\bf k}}{4\xi_{\bf k}^3(t)},
\label{equationH}
\end{equation}
where $\eta_{\bf k}(t)=\dot\xi_{\bf k}(t)$ is the momentum canonically
conjugate to $\xi_{\bf k}(t)$. Moreover, the semi-classical Maxwell equation
(\ref{semimaxwellnum}) is rewritten as
\begin{equation}
\frac{d^2A}{dt^2}= 2e\int \frac{d^3{\bf k}}{(2\pi)^3}[k-eA(t)]\xi^2_{\bf k}(t).
\label{semimaxwellnumH}
\end{equation}
Eqs.~(\ref{equationH}) and (\ref{semimaxwellnumH}) actually
are Hamilton equations of motion,
\begin{equation}
\dot\eta_{\bf k}(t)=-\frac{\delta {\mathcal H}_{\rm eff}}{\delta\xi_{\bf k}(t)},
\quad \dot P_A(t)=-\frac{\delta {\mathcal H}_{\rm eff}}{\delta A(t)},
\label{11equationH}
\end{equation}
for a closed system with Hamiltonian,
\begin{equation}
{\mathcal H}_{\rm eff}(A,P_A,\xi,\eta,\sigma)=V\frac{E^2}{2}+
V\int\frac{d^3{\bf k}}{(2\pi)^3}\left(\eta_{\bf k}^2+\omega^2_{\bf k}(A)\xi_{\bf k}^2
+\frac{\hbar^2\sigma^2_{\bf k}}{4\xi_{\bf k}^2}\right),
\label{dissH}
\end{equation}
where $P_A=\dot A=-E$ is the momentum canonically conjugate to $A$,
$\omega_{\bf k}(A)$ is the field-dependent frequency of quantum
fluctuations given by Eq.~(\ref{11frequency}) and the value of mean
field (\ref{fluctuation}) vanishes, $\langle\Phi_{\bf k}(t)\rangle
=0$, for each ${\bf k}$-mode. In Eq.~(\ref{dissH}), the first term
is the electric energy and the second term is the energy of quantum
fluctuations of charged matter field, interacting with electric
field.

Quantum decoherence can be studied within this Hamiltonian
framework. If one considers only the time evolution of classical
electric field $A(t)$, that is influenced by the quantum fluctuating
modes $f_{\bf k}(t)$, the latter can be treated as a heat bath
``environment''. Quantitative information about the quantum
decoherence is contained in the so-called influence functional,
which is a functional of two time evolution trajectories $A_1(t)$
and $A_2(t)$ \cite{1996PhRvL..76.4660H}
\begin{equation}
F_{12}(t)=\exp[i\Gamma_{12}(t)]={\rm Tr}(|A_1(t)\rangle\langle A_2(t)|),
\label{f12}
\end{equation}
where $|A_{1,2}(t)\rangle$ are different time evolution states determined by
Eq.~(\ref{11equationH}), starting with the same initial state $|A(0)\rangle$
and initial vacuum condition $N_{\bf k}=0, \sigma_{\bf k}=1$
(\ref{semimaxwellnum}). One finds \cite{1996PhRvL..76.4660H}
\begin{equation}
\Gamma_{12}(t)=-\frac{i}{2}\ln\left[\frac{i\hbar}{|f_1f_2|}
\left(\frac{f_1f_2^*}{f_1\dot
f_2^*-\dot f_1f_2^*}\right)\right],
\label{gamma12}
\end{equation}
in terms of the two sets of mode functions $\{f_1(t)\}$ and
$\{f_2(t)\}$ (the subscript ${\bf k}$ is omitted). This
$\Gamma_{12}$ (\ref{gamma12}) is precisely the closed time path
(CTP) effective action functional which generates the connected real
$n$-points vertices in the quantum theory
\cite{1989PhRvD..40..656C,1994PhRvD..49.6636C,1994PhRvD..50.2848C,
1961JMP.....2..407S,1963JMP.....4....1B,1963JMP.....4...12B,
Keldysh1964,1985PhR...118....1C}.
The absolute value of $F_{12}$ (\ref{f12}) measures the influence of
quantum fluctuations $f_{\bf k}(t)$ on the time evolution of the
classical electric field $A(t)$, i.e., the effect of quantum
decoherence. If there is no influence of quantum fluctuations on
$A(t)$, then $|A_1(t)\rangle=|A_2(t)\rangle$ and $|F_{12}|=1$,
otherwise $A_2(t)$ deviates from $A_1(t)$,
$|A_1(t)\rangle\not=|A_2(t)\rangle$ and $|F_{12}|<1$. Numerical
results about the damping and the decoherence of the electric field
are presented in
Refs.~\cite{1991PhRvL..67.2427K,1996PhRvL..76.4660H} (see
Figs.~\ref{11oscilla} and \ref{dephasing}).

\begin{figure}[!ht]
\begin{center}
\includegraphics[height=11cm,width=12cm]{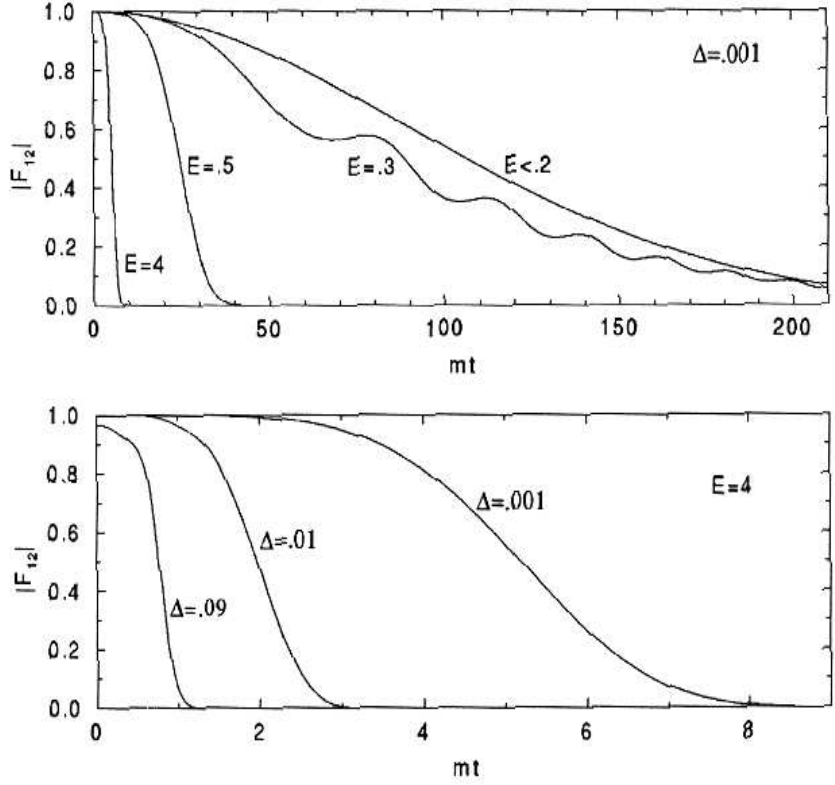}
\end{center}
\caption{Absolute values of the decoherence functional $|F_{12}|$ as a function of time. The two field values
are $E$ and $E-\Delta$. The top figure shows (for fixed $\Delta$) the sharp dependence of decoherence on
particle production when $|E|\ge 0.2E_c$. The second illustrates the relatively milder dependence on $\Delta$.
These figures are reproduced from Fig.~2 in Ref.~\cite{1996PhRvL..76.4660H}.}%
\label{dephasing}%
\end{figure}

The effective damping discussed above is certainly collisionless,
since the charged particle modes $f_{\bf k}(t)$ interact only with
the electric field but not directly with each other. The damping of
plasma oscillation attributed to the collisions between charged
particle modes $f_{\bf k}(t)$ will be discussed in the next section.

\subsection{Collision decoherence in plasma oscillations}\label{collisiondecoherence}

If there are interactions between different modes ${\bf k}$ and species of
particles, the time evolutions of electric field and the distribution function
${\mathcal F}({\bf p},t)$ of particle number in the phase space are certainly
changed. This can be phenomenologically studied in the relativistic
Boltzmann-Vlasov equation (\ref{11bv}) by adding collision terms ${\mathcal
C}({\bf p},t)$,
\begin{equation}
\frac{d {\mathcal F}}{d t}\equiv\frac{\partial {\mathcal F}}{\partial t}+eE\frac{\partial {\mathcal F}}{\partial p}
={\mathcal S}_{\bf p}(t) + {\mathcal C}({\bf p},t).
\label{11bvc}
\end{equation}
These collision terms ${\mathcal C}({\bf p},t)$ describe not only the
interactions of different modes ${\bf k}$ of particles, but also interactions
of different species of particles, for example, electron and positron
annihilation to two photons and {\it vice versa}. In
Refs.~\cite{1999PhRvD..60k6011B, 2001EPJC...22..341V}, the following
equilibrating collision terms were considered,
\begin{equation}
{\mathcal C}({\bf p},t)=\frac{1}{\tau_{\rm r}}[{\mathcal F}^{\rm eq}({\bf p},t,T)-{\mathcal F}({\bf p},t)],
\label{thermalcoll}
\end{equation}
where ${\mathcal F}^{\rm eq}({\bf p},t,T)$ is the thermal (Fermi or
Bose) distribution function of particle number (fermions or bosons)
at temperature $T$, the relaxation time $\tau_{\rm r}$ is determined
by the mean free path $\lambda(t)$ and mean velocity $\bar v(t)$ of
particles, through
\begin{equation}
\tau_{\rm r}(t)=\tau_c\frac{\lambda(t)}{\bar v(t)},
\label{relatime}
\end{equation}
and $\tau_c$ is a dimensionless parameter. $\lambda(t)$ is computed from the number density $n(t)$ of particles,
\begin{equation}
\lambda(t)= \frac{1}{[n(t)]^{1/3}},\quad n(t)=\int \frac{d^3{\bf k}}{(2\pi)^3}{\mathcal F}({\bf p},t).
\label{clambda}
\end{equation}
Whereas, the mean velocity of particles is given by $\bar v(t)=\bar
p(t)/\bar\epsilon(t)$, expressed in terms of mean kinetic momentum
$\bar p(t)$ and energy $\bar\epsilon(t)$ of particles. The mean
values of momentum $\bar p(t)$ and energy $\bar\epsilon(t)$ are
computed \cite{2001EPJC...22..341V} by using distribution function
${\mathcal F}({\bf p},t)$ regularized via the procedure described in
Ref.~\cite{1999PhRvD..60k6011B} that yields the renormalized
electric current (\ref{qcurrent2}). The temperature $T(t)$ in
Eq.~(\ref{thermalcoll}) is the ``instantaneous temperature'', which
is determined by requiring that at each time $t$ the mean particle
energy $\bar\epsilon(t)$ is identical to that in an equilibrium
distribution ${\mathcal F}^{\rm eq}({\bf k},t,T)$ at the temperature
$T(t)$,
\begin{equation}
\bar\epsilon(t)=\int \frac{d^3{\bf k}}{(2\pi)^3}
\epsilon({\bf k}){\mathcal F}^{\rm eq}[{\bf k},t,T(t)].
\label{cenergy}
\end{equation}
This system of two coupled equations: (i) the field equation
(\ref{11semimaxwell}) with renormalized electric currents (\ref{qcurrent2});
(ii) the relativistic Boltzmann-Vlasov equation (\ref{11bvc}) with the source
term ${\mathcal S}_{\bf p}(t)$ (\ref{quanbv}) and equilibrating collision term
${\mathcal C}({\bf p},t)$ (\ref{thermalcoll}), are numerically integrated in
Refs.~\cite{1999PhRvD..60k6011B, 2001EPJC...22..341V}. One of these numerical
results is presented in Fig.~\ref{11oscilladamping}. It shows
\cite{1999PhRvD..60k6011B} that when the collision timescale $\tau_{\rm r}$
(\ref{relatime}) is much larger than the plasma oscillation timescale
$\tau_{pl}$, $\tau_{\rm r}\gg \tau_{pl}$, the plasma oscillations are
unaffected. On the other hand, when $\tau_{\rm r}\sim \tau_{pl}$ the collision
term has a significant impact on both the amplitude and the frequency of the
oscillations that result damped. There is a value of $\tau_{\rm r}$ below which
no oscillations arise and the system evolves quickly and directly to thermal
equilibrium. It is worthwhile to contrast this collision damping of plasma
oscillation with the collisionless damping effect due to rapid quantum
fluctuations described in Section~\ref{quantumdecoherence}.

\begin{figure}[!ht]
\begin{center}
\includegraphics[height=10cm,width=12cm]{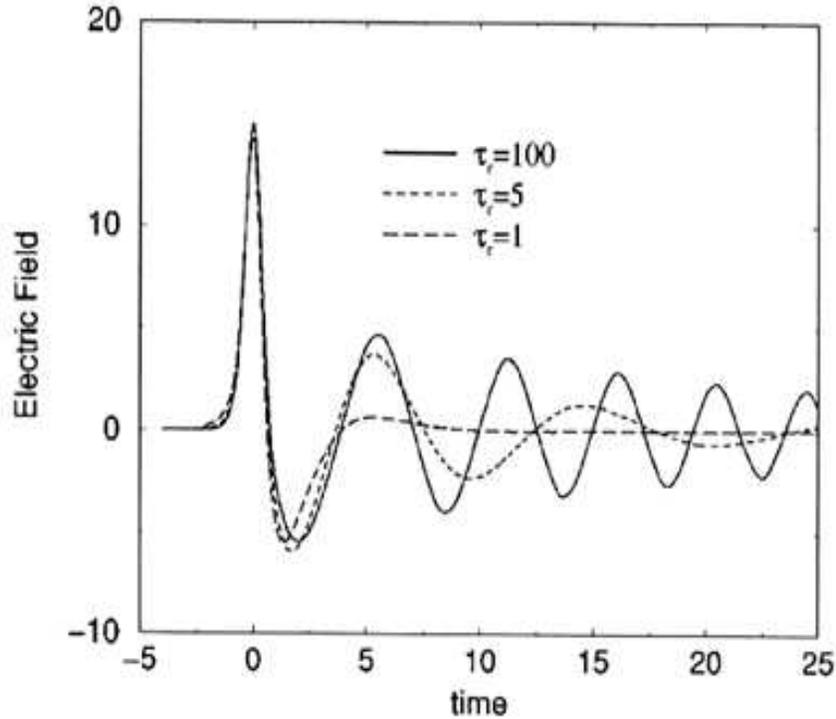}
\end{center}
\caption{Time evolution for electric field obtained using different relaxation times $\tau_{\rm r}$
in the collision term of Eq.~(\ref{thermalcoll}) and with the impulse external field
$E_{ex}(t)=-A_0[b\cosh^2(t/b)]^{-1}$, where $A_0=0.7$ and $b=0.5$. All dimensioned
quantities are given in units of the mass-scale $m$.
This Figure is reproduced from Fig.~6 in Ref.~\cite{1999PhRvD..60k6011B}.}%
\label{11oscilladamping}%
\end{figure}

\subsection{$e^+e^-\gamma$ interactions in plasma oscillations in electric fields} \label{PO}

In this section, a detailed report of the studies \cite{2003PhLB..559...12R} of
the relativistic Boltzmann-Vlasov equations for electrons, positrons and
photons with collision terms originated from annihilation of electrons and
positrons pair into two photons and {\it vice versa} is presented. These
collision terms lead to the damping of plasma oscillation and possibly to
energy equipartition between different types of particles.%aga
We focus on the evolution of a system of $e^{+}e^{-}$ pairs created in a
strongly overcritical electric field ($E\sim 10 E_{\mathrm{c}}$), explicitly
taking into account the process $e^{+}e^{-}\rightleftarrows\gamma\gamma$. Since
it is far from equilibrium the collisions cannot be modeled by an effective
relaxation time term in the transport equations, as discussed in the previous
section. Rather the actual, time varying, collision integrals have to be used.

Furthermore we are mainly interested in a system in which the
electric field varies on macroscopic length scale and therefore one
can approximate an electric field as a homogeneous one. Also,
transport equations are used for electrons, positrons and photons,
with collision terms, coupled to Maxwell equations, as introduced in
Section~\ref{semi-classical} and \ref{kineticplasma}. There is no
free parameter here: the collision terms can be exactly computed,
since the QED cross-sections are known. Starting from a regime which
is far from thermal equilibrium, one finds that collisions do not
prevent plasma oscillations in the initial phase of the evolution
and analyze the issue of the timescale of the approach to an
$e^{+}e^{-}\gamma$ plasma equilibrium configuration.

As discussed in Section~\ref{semi-classical} and \ref{kineticplasma}
one can describe positrons (electrons) created by vacuum
polarization in a strong homogeneous electric field $\mathbf{E}$
through the distribution function $f_{e^{+}}$ ($f_{e^{-}}$) in the
phase space of positrons (electrons). Because of homogeneity
$f_{e^{+}}$ ($f_{e^{-}}$) only depend on the time $t$ and the
positron (electron) 3--momentum $\mathbf{p}$:
\begin{equation}
f_{e^{+,-}}=f_{e^{+,-}}\left(  t,\mathbf{p}\right)  .
\end{equation}
Moreover, because of particle--antiparticle symmetry, one also has
\begin{equation}
f_{e^{+}}\left(  t,\mathbf{p}\right)  =f_{e^{-}}\left(  t,-\mathbf{p}\right)
\equiv f_{e}\left(  t,\mathbf{p}\right)  .
\end{equation}
Analogously photons created by pair annihilation are described through the
distribution function $f_{\gamma}$ in the phase space of photons.
$\mathbf{k}$ is the photon 3--momentum , then
\begin{equation}
f_{\gamma}=f_{\gamma}\left(  t,\mathbf{k}\right)  .
\end{equation}
$f_{e}$ and $f_{\gamma}$ are normalized so that
\begin{align}
\int\tfrac{d^{3}\mathbf{p}}{\left(  2\pi\right)  ^{3}}\ f_{e}\left(
t,\mathbf{p}\right)   &  =n_{e}\left(  t\right)  ,\\
\int\tfrac{d^{3}\mathbf{k}}{\left(  2\pi\right)  ^{3}}\ f_{\gamma}\left(
t,\mathbf{k}\right)   &  =n_{\gamma}\left(  t\right)  ,
\end{align}
where $n_{e}$ and $n_{\gamma}$ are number densities of positrons (electrons)
and photons, respectively. For any function of the momenta, one can denote by
\begin{equation}
\left\langle F\left(  \mathbf{p}_{1},...,\mathbf{p}_{n}\right)  \right\rangle
_{e}\equiv n_{e}^{-n}\int\tfrac{d^{3}\mathbf{p}_{1}}{\left(  2\pi\right)
^{3}}...\tfrac{d^{3}\mathbf{p}_{n}}{\left(  2\pi\right)  ^{3}}\ F\left(
\mathbf{p}_{1},...,\mathbf{p}_{n}\right)  \cdot f_{e}\left(  \mathbf{p}%
_{1}\right)  \cdot...\cdot f_{e}\left(  \mathbf{p}_{n}\right)  ,
\end{equation}
or
\begin{equation}
\left\langle G\left(  \mathbf{k}_{1},...,\mathbf{k}_{l}\right)  \right\rangle
_{\gamma}\equiv n_{\gamma}^{-l}\int\tfrac{d^{3}\mathbf{k}_{1}}{\left(
2\pi\right)  ^{3}}...\tfrac{d^{3}\mathbf{k}_{l}}{\left(  2\pi\right)  ^{3}%
}\ G\left(  \mathbf{k}_{1},...,\mathbf{k}_{l}\right)  \cdot f_{\gamma}\left(
\mathbf{k}_{1}\right)  \cdot...\cdot f_{\gamma}\left(  \mathbf{k}_{l}\right)
,
\end{equation}
its mean value in the phase space of positrons (electrons) or photons, respectively.

The motion of positrons (electrons) is the result of three
contributions: the pair creation, the electric acceleration and the
annihilation damping. The probability density rate
$\mathcal{S}\left(  \mathbf{E},\mathbf{p}\right)  $ for the creation
of a pair with 3--momentum $\mathbf{p}$ in the electric field
$\mathbf{E}$ is given by the Schwinger formula (see also
Refs.~\cite{1991PhRvL..67.2427K,1992PhRvD..45.4659K}):
\begin{align}
\mathcal{S}\left(  \mathbf{E},\mathbf{p}\right)   &  =\left(  2\pi\right)
^{3}\tfrac{dN}{dtd^{3}\mathbf{x}d^{3}\mathbf{p}}\nonumber\\
&  =-\left\vert e\mathbf{E}\right\vert \log\left[  1-\exp\left(  -\tfrac
{\pi(m_{e}^{2}+\mathbf{p}_{\perp}^{2})}{\left\vert e\mathbf{E}\right\vert
}\right)  \right]  \delta(p_{\parallel}), \label{S}%
\end{align}
where $p_{\parallel}$ and $\mathbf{p}_{\perp}$ are the components of the
3-momentum $\mathbf{p}$ parallel and orthogonal to $\mathbf{E}$, respectively.
Also the energy is introduced
\begin{equation}
\epsilon_{\mathbf{p}}=\left(  \mathbf{p}\cdot\mathbf{p}+m_{e}^{2}\right)
^{1/2}%
\end{equation}
of an electron of 3-momentum $\mathbf{p}$ and the energy
\begin{equation}
\epsilon_{\mathbf{k}}=\left(  \mathbf{k}\cdot\mathbf{k}\right)  ^{1/2}%
\end{equation}
of a photon of 3-momentum $\mathbf{k}$. Then, the probability density rate
$\mathcal{C}_{e}\left(  t,\mathbf{p}\right)  $ for the creation (destruction)
of a fermion with 3-momentum $\mathbf{p}$ is given by
\begin{align}
\mathcal{C}_{e}\left(  t,\mathbf{p}\right)   &  \simeq\tfrac{1}{\epsilon
_{\mathbf{p}}}\int\tfrac{d^{3}\mathbf{p}_{1}}{\left(  2\pi\right)
^{3}\epsilon_{\mathbf{p}_{1}}}\tfrac{d^{3}\mathbf{k}_{1}}{\left(  2\pi\right)
^{3}\epsilon_{\mathbf{k}_{1}}}\tfrac{d^{3}\mathbf{k}_{2}}{\left(  2\pi\right)
^{3}\epsilon_{\mathbf{k}_{2}}}\left(  2\pi\right)  ^{4}\delta^{\left(
4\right)  }\left(  p+p_{1}-k_{1}-k_{2}\right) \nonumber\\
&  \times\left\vert \mathcal{M}_{e^{+}\left(  \mathbf{p}\right)  e^{-}\left(
\mathbf{p}_{1}\right)  \rightarrow\gamma\left(  \mathbf{k}_{1}\right)
\gamma\left(  \mathbf{k}_{2}\right)  }\right\vert ^{2}\left[  f_{e}\left(
\mathbf{p}\right)  f_{e}\left(  \mathbf{p}_{1}\right)  -f_{\gamma}\left(
\mathbf{k}_{1}\right)  f_{\gamma}\left(  \mathbf{k}_{2}\right)  \right]  ,
\label{Ce}%
\end{align}
where
\begin{equation}
\mathcal{M}_{e^{+}\left(  \mathbf{p}_{1}\right)  e^{-}\left(  \mathbf{p}%
_{2}\right)  \rightarrow\gamma\left(  \mathbf{k}_{1}\right)  \gamma\left(
\mathbf{k}_{2}\right)  }=\mathcal{M}_{e^{+}\left(  \mathbf{p}_{1}\right)
e^{-}\left(  \mathbf{p}_{2}\right)  \leftarrow\gamma\left(  \mathbf{k}%
_{1}\right)  \gamma\left(  \mathbf{k}_{2}\right)  }\equiv\mathcal{M}%
\end{equation}
is the matrix element for the process
\begin{equation}
e^{+}\left(  \mathbf{p}_{1}\right)  e^{-}\left(  \mathbf{p}_{2}\right)
\rightarrow\gamma\left(  \mathbf{k}_{1}\right)  \gamma\left(  \mathbf{k}%
_{2}\right)
\end{equation}
and as a first approximation, Pauli blocking and Bose enhancement (see, for
instance, Ref.~\cite{1992PhRvD..45.4659K}) are neglected. Analogously the
probability density rate $\mathcal{C}_{\gamma}\left( t,\mathbf{p}\right)  $ for
the creation (annihilation) of a photon with 3-momentum $\mathbf{k}$ is given
by
\begin{align}
\mathcal{C}_{\gamma}\left(  t,\mathbf{k}\right)   &  \simeq\tfrac{1}%
{\epsilon_{\mathbf{k}}}\int\tfrac{d^{3}\mathbf{p}_{1}}{\left(  2\pi\right)
^{3}\epsilon_{\mathbf{p}_{1}}}\tfrac{d^{3}\mathbf{p}_{2}}{\left(  2\pi\right)
^{3}\epsilon_{\mathbf{p}_{2}}}\tfrac{d^{3}\mathbf{k}_{1}}{\left(  2\pi\right)
^{3}\epsilon_{\mathbf{k}_{1}}}\left(  2\pi\right)  ^{4}\delta^{\left(
4\right)  }\left(  p_{1}+p_{2}-k-k_{1}\right) \nonumber\\
\times &  \left\vert \mathcal{M}_{e^{+}\left(  \mathbf{p}_{1}\right)
e^{-}\left(  \mathbf{p}_{2}\right)  \rightarrow\gamma\left(  \mathbf{k}%
\right)  \gamma\left(  \mathbf{k}_{1}\right)  }\right\vert ^{2}\left[
f_{e}\left(  \mathbf{p}_{1}\right)  f_{e}\left(  \mathbf{p}_{2}\right)
-f_{\gamma}\left(  \mathbf{k}\right)  f_{\gamma}\left(  \mathbf{k}_{1}\right)
\right]  , \label{Cf}%
\end{align}
Finally the evolution of the pairs is governed by the transport
Boltzmann--Vlasov equations
\begin{align}
\partial_{t}f_{e}+e\mathbf{E\cdot\nabla}_{\mathbf{p}}f_{e}  &  =\mathcal{S}%
\left(  \mathbf{E},\mathbf{p}\right)  -\mathcal{C}_{e}\left(  t,\mathbf{p}%
\right)  ,\label{pairs}\\
\partial_{t}f_{\gamma}  &  =2\mathcal{C}_{\gamma}\left(  t,\mathbf{k}\right)
, \label{photons}%
\end{align}
Note that the collisional terms (\ref{Ce}) and (\ref{Cf}) are negligible, when
created pairs do not produce a dense plasma.

Because pair creation back reacts on the electric field, as seen in
Section~\ref{semi-classical} and \ref{kineticplasma}, Vlasov
equations are coupled with the homogeneous Maxwell equations, which
read
\begin{equation}
\partial_{t}\mathbf{E}=-\mathbf{j}_{p}\left(  \mathbf{E}\right)
-\mathbf{j}_{c}\left(  t\right)  , \label{Maxwell}%
\end{equation}
where
\begin{equation}
\mathbf{j}_{p}\left(  \mathbf{E}\right)  =2\tfrac{\mathbf{E}}{\mathbf{E}^{2}%
}\int\tfrac{d^{3}\mathbf{p}}{\left(  2\pi\right)  ^{3}}\epsilon_{\mathbf{p}%
}\mathcal{S}\left(  \mathbf{E},\mathbf{p}\right)
\end{equation}
is the polarization current and
\begin{equation}
\mathbf{j}_{c}\left(  t\right)  =2en_{e}\int\tfrac{d^{3}\mathbf{p}}{\left(
2\pi\right)  ^{3}}\tfrac{\mathbf{p}}{\epsilon_{\mathbf{p}}}f_{e}\left(
\mathbf{p}\right)
\end{equation}
is the conduction current (see Ref.~\cite{1987PhRvD..36..114G}).

Eqs.~(\ref{pairs}), (\ref{photons}) and (\ref{Maxwell}) describe the dynamical
evolution of the electron--positron pairs, the photons and the strong
homogeneous electric field due to the Schwinger process of pair creation, the
pair annihilation into photons and the two photons annihilation into pairs. It
is hard to (even numerically) solve this system of integral and partial
differential equations. It is therefore useful to introduce a simplification
procedure of such a system through an approximation scheme. First of all note
that Eqs.~(\ref{pairs}) and (\ref{photons}) can be suitably integrated over the
phase spaces of positrons (electrons) and photons to get differential equations
for mean values. The following exact equations for mean values are obtained:
\begin{align}
\tfrac{d}{dt}n_{e}  &  =S\left(  \mathbf{E}\right)  -n_{e}^{2}\left\langle
\sigma_{1}v^{\prime}\right\rangle _{e}+n_{\gamma}^{2}\left\langle \sigma
_{2}v^{\prime\prime}\right\rangle _{\gamma},\nonumber\\
\tfrac{d}{dt}n_{\gamma}  &  =2n_{e}^{2}\left\langle \sigma_{1}v^{\prime
}\right\rangle _{e}-2n_{\gamma}^{2}\left\langle \sigma_{2}v^{\prime\prime
}\right\rangle _{\gamma},\nonumber\\
\tfrac{d}{dt}n_{e}\left\langle \epsilon_{\mathbf{p}}\right\rangle _{e}  &
=en_{e}\mathbf{E}\cdot\left\langle \mathbf{v}\right\rangle _{e}+\tfrac{1}%
{2}\mathbf{E\cdot j}_{p}-n_{e}^{2}\left\langle \epsilon_{\mathbf{p}}\sigma
_{1}v^{\prime\prime}\right\rangle _{e}+n_{\gamma}^{2}\left\langle
\epsilon_{\mathbf{k}}\sigma_{2}v^{\prime\prime}\right\rangle _{\gamma
},\nonumber\\
\tfrac{d}{dt}n_{\gamma}\left\langle \epsilon_{\mathbf{k}}\right\rangle
_{\gamma}  &  =2n_{e}^{2}\left\langle \epsilon_{\mathbf{p}}\sigma_{1}%
v^{\prime}\right\rangle _{e}-2n_{\gamma}^{2}\left\langle \epsilon_{\mathbf{k}%
}\sigma_{2}v^{\prime\prime}\right\rangle _{\gamma},\nonumber\\
\tfrac{d}{dt}n_{e}\left\langle \mathbf{p}\right\rangle _{e}  &  =en_{e}%
\mathbf{E}-n_{e}^{2}\left\langle \mathbf{p}\sigma_{1}v^{\prime}\right\rangle
_{e},\nonumber\\
\tfrac{d}{dt}\mathbf{E}  &  =-2en_{e}\left\langle \mathbf{v}\right\rangle
_{e}-\mathbf{j}_{p}\left(  \mathbf{E}\right)  , \label{System1}%
\end{align}
where
\begin{align}
S\left(  \mathbf{E}\right)   &  =\int\tfrac{d^{3}\mathbf{p}}{\left(
2\pi\right)  ^{3}}\mathcal{S}\left(  \mathbf{E},\mathbf{p}\right) \\
&  \equiv\tfrac{dN}{dtd^{3}\mathbf{x}},
\end{align}
is the total probability rate for Schwinger pair production. In
Eqs.~(\ref{System1}), $v^{\prime\prime}=c$ the velocity of light and
$v^{\prime}=2|{\mathbf p}/\epsilon_{\mathbf{p}}^{\mathrm{CoM}}|$ is
the relative velocity between electrons and positrons in the
reference frame of the center of mass, where
$\mathbf{p}=|\mathbf{p}_{e^{\pm}}|$, $\mathbf{p}
_{e^{-}}=-\mathbf{p}_{e^{+}}$ are 3--momenta of electron and
positron and
$\epsilon_{\mathbf{p}_{e^{\mp}}}=\epsilon_{\mathbf{p}}^{\mathrm{CoM}}$
are their energies. $\sigma_{1}=\sigma_{1}\left(
\epsilon_{\mathbf{p}}^{\mathrm{CoM}}\right)$ is the total
cross-section for the process $e^{+}e^{-}\rightarrow\gamma\gamma$,
and $\sigma_{2}=\sigma_{2}\left(
\epsilon_{\mathbf{k}}^{\mathrm{CoM}}\right)$ is the total
cross-section for the process $\gamma\gamma\rightarrow e^{+}e^{-}$,
here $\epsilon^{\mathrm{CoM}}$ is the energy of a particle in the
reference frame of the center of mass.

In order to evaluate the mean values in system (\ref{System1}) some
further hypotheses on the distribution functions are needed. One defines $\bar
{p}_{\parallel}$, $\bar{\epsilon}_{\mathbf{p}}$ and $\mathbf{\bar{p}}_{\perp
}^{2}$ such that
\begin{equation}
\left\langle p_{\parallel}\right\rangle _{e}\equiv\bar{p}_{\parallel},
\end{equation}%
\begin{equation}
\left\langle \epsilon_{\mathbf{p}}\right\rangle _{e}    \equiv\bar{\epsilon
}_{\mathbf{p}}
 \equiv(\bar{p}_{\parallel}^{2}+\mathbf{\bar{p}}_{\perp}^{2}+\ m_{e}^{2})^{1/2}.
\end{equation}
%\begin{align}
%\left\langle \epsilon_{\mathbf{p}}\right\rangle _{e}  &  \equiv\bar{\epsilon
%}_{\mathbf{p}}\nonumber\\
%&  \equiv(\bar{p}_{\parallel}^{2}+\mathbf{\bar{p}}_{\perp}^{2}+\ m_{e}%
%^{2})^{1/2}.
%\end{align}
It is assumed
\begin{equation}
f_{e}\left(  t,\mathbf{p}\right)  \propto n_{e}\left(  t\right)  \delta\left(
p_{\parallel}-\bar{p}_{\parallel}\right)  \delta\left(  \mathbf{p}_{\perp}%
^{2}-\mathbf{\bar{p}}_{\perp}^{2}\right)  . \label{fe}%
\end{equation}
Since in the scattering $e^{+}e^{-}\rightarrow\gamma\gamma$ the coincidence of
the scattering direction with the incidence direction is statistically
favored, it is also assumed
\begin{equation}
f_{\gamma}\left(  t,\mathbf{k}\right)  \propto n_{\gamma}\left(  t\right)
\delta\left(  \mathbf{k}_{\perp}^{2}-\mathbf{\bar{k}}_{\perp}^{2}\right)
\left[  \delta\left(  k_{\parallel}-\bar{k}_{\parallel}\right)  +\delta\left(
k_{\parallel}+\bar{k}_{\parallel}\right)  \right]  , \label{fgamma}%
\end{equation}
where $k_{\parallel}$ and $\mathbf{k}_{\perp}$ have analogous meaning as
$p_{\parallel}$ and $\mathbf{p}_{\perp}$ and the terms $\delta\left(
k_{\parallel}-\bar{k}_{\parallel}\right)  $ and $\delta\left(  k_{\parallel
}+\bar{k}_{\parallel}\right)  $ account for the probability of producing,
respectively, forwardly scattered and backwardly scattered photons. Since the
Schwinger source term (\ref{S}) implies that the positrons (electrons) have
initially fixed $p_{\parallel}$, namely $p_{\parallel}=0$, assumption
(\ref{fe}) ((\ref{fgamma})) means that the distribution of $p_{\parallel}$
($k_{\parallel}$) does not spread too much with time and, analogously, that the
distribution of energies is sufficiently peaked to be describable by a
$\delta$--function. As long as this condition is fulfilled, approximations
(\ref{fe}) and (\ref{fgamma}) are applicable. The actual dependence on the
momentum of the distribution functions has been discussed in
Ref.~\cite{1992PhRvD..45.4659K, 1998PhRvD..58l5015K}. If Eqs.~(\ref{fe}) and
(\ref{fgamma}) are substituted into the system (\ref{System1}) one gets a new
system of ordinary differential equations. One can introduce the inertial
reference frame which on average coincides with the center of mass frame for
the processes $e^{+}e^{-}\rightleftarrows \gamma\gamma$, and has
$\epsilon^{\mathrm{CoM}}\simeq\bar{\epsilon}$ for each species, and therefore
substituting Eqs.~(\ref{fe}) and (\ref{fgamma}) into Eqs.~(\ref{System1}) one
finds
\begin{align}
\tfrac{d}{dt}n_{e}  &  =S\left(  E\right)  -2n_{e}^{2}\sigma_{1}%
\rho_{e}^{-1}\left|  {\pi}_{e\parallel}\right|  +2n_{\gamma}^{2}\sigma
_{2},\nonumber\\
\tfrac{d}{dt}n_{\gamma}  &  =4n_{e}^{2}\sigma_{1}\rho_{e}^{-1}\left|  {\pi
}_{e\parallel}\right|  -4n_{\gamma}^{2}\sigma_{2},\nonumber\\
\tfrac{d}{dt}\rho_{e}  &  =en_{e}E\rho_{e}^{-1}\left|  {\pi
}_{e\parallel}\right|  +\tfrac{1}{2}Ej_{p}-2n_{e}\rho_{e}\sigma
_{1}\rho_{e}^{-1}\left|  {\pi}_{e\parallel}\right|  +2n_{\gamma}\rho_{\gamma
}\sigma_{2},\nonumber\\
\tfrac{d}{dt}\rho_{\gamma}  &  =4n_{e}\rho_{e}\sigma_{1}\rho_{e}^{-1}\left|
{\pi}_{e\parallel}\right|  -4n_{\gamma}\rho_{\gamma}\sigma_{2},\nonumber\\
\tfrac{d}{dt}{\pi}_{e\parallel}  &  =en_{e}E-2n_{e}{\pi}%
_{e\parallel}\sigma_{1}\rho_{e}^{-1}\left|  {\pi}_{e\parallel}\right|
,\nonumber\\
\tfrac{d}{dt}E  &  =-2en_{e}\rho_{e}^{-1}\left|  {\pi}_{e\parallel
}\right|  -j_{p}\left(  E\right)  , \label{System2}%
\end{align}
where
\begin{align}
\rho_{e}  &  =n_{e}\bar{\epsilon}_{\mathbf{p}},\\
\rho_{\gamma}  &  =n_{\gamma}\bar{\epsilon}_{\mathbf{k}},\\
{\pi}_{e\parallel}  &  =n_{e}\bar{p}_{\parallel}%
\end{align}
are the energy density of positrons (electrons), the energy density of photons
and the density of ``parallel momentum'' of positrons (electrons),
$E$ is the electric field strength and $j_{p}$ the unique component
of $\mathbf{j}_{p}$ parallel to $\mathbf{E}$. $\sigma_{1}$ and $\sigma_{2}$
are evaluated at $\epsilon^{\mathrm{CoM}}=\bar{\epsilon}$ for each species.
Note that Eqs.~(\ref{System2}) are ``classical'' in the sense that the only
quantum information is encoded in the terms describing pair creation and
scattering probabilities. Finally Eqs.~(\ref{System2}) are duly consistent
with energy density conservation:
\begin{equation}
\tfrac{d}{dt}\left(  \rho_{e}+\rho_{\gamma}+\tfrac{1}{2}E%
^{2}\right)  =0.
\end{equation}

Eqs.~(\ref{System2}) have to be integrated with the following initial
conditions
\begin{align*}
n_{e}  &  =0,\\
n_{\gamma}  &  =0,\\
\rho_{e}  &  =0,\\
\rho_{\gamma}  &  =0,\\
\pi_{e\parallel}  &  =0,\\
E  &  =E_{0}.
\end{align*}
In Fig.~\ref{Oscillation} the results of the numerical integration for
$E_{0}=9E_{\mathrm{c}}$ is showed. The integration stops
at $t=150\ \tau_{\mathrm{C}}$ (where, as usual, $\tau_{\mathrm{C}}=\hbar
/m_{e}c^2$ is the Compton time of the electron). Each quantity is represented in
units of $m_{e}$ and $\lambda_{\mathrm{C}}=\hbar/m_{e}c$, the Compton length of
the electron.

\begin{figure}[!ptb]
\begin{center}
\includegraphics[width=8cm]{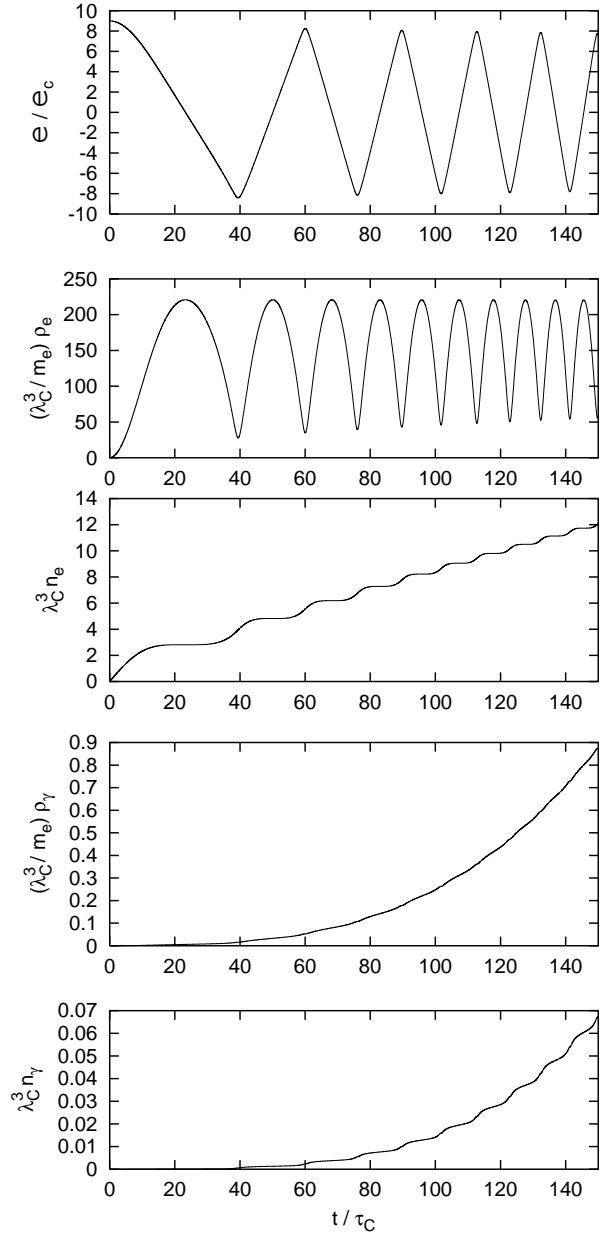}
\end{center}
\caption{Plasma oscillations in a strong homogeneous electric field: early
times behavior. Setting $E_{0}=9E_{\mathrm{c}}$,
$t<150\tau_{\mathrm{C}}$ and it is plotted, from the top to the bottom panel: a)
electromagnetic field strength; b) electrons energy density; c) electrons
number density; d) photons energy density; e) photons number density as
functions of time.}%
\label{Oscillation}%
\end{figure}

The numerical integration confirms \cite{1991PhRvL..67.2427K,
1992PhRvD..45.4659K} that the system undergoes plasma oscillations:

\begin{enumerate}
\item the electric field does not abruptly reach the equilibrium value but
rather oscillates with decreasing amplitude;

\item electrons and positrons oscillates in the electric field direction,
reaching ultrarelativistic velocities;

\item the role of the $e^{+}e^{-}\rightleftarrows$ $\gamma\gamma$ scatterings
is marginal in the early time of the evolution.
\end{enumerate}

This last point can be easily explained as follows: since the electrons are
too extremely relativistic, the annihilation probability is very low and
consequently the density of photons builds up very slowly (see details in
Fig.~\ref{Oscillation}).

At late times the system is expected to relax to an equilibrium configuration
and assumptions (\ref{fe}) and (\ref{fgamma}) have to be generalized to take
into account quantum spreading of the distribution functions. It is
nevertheless instructive to look at the solutions of Eqs.~(\ref{System2}) in
this regime. Moreover, such a solution should give information at least at the
order of magnitude level. In Fig.~\ref{OscillationLT} the numerical solution of
Eqs.~(\ref{System2}) is plotted, but the integration extends here all the way
up to $t=7000\ \tau _{\mathrm{C}}$ (the timescale of oscillations is not
resolved in these plots).

\begin{figure}[!ptb]
\begin{center}
\includegraphics[width=8cm]{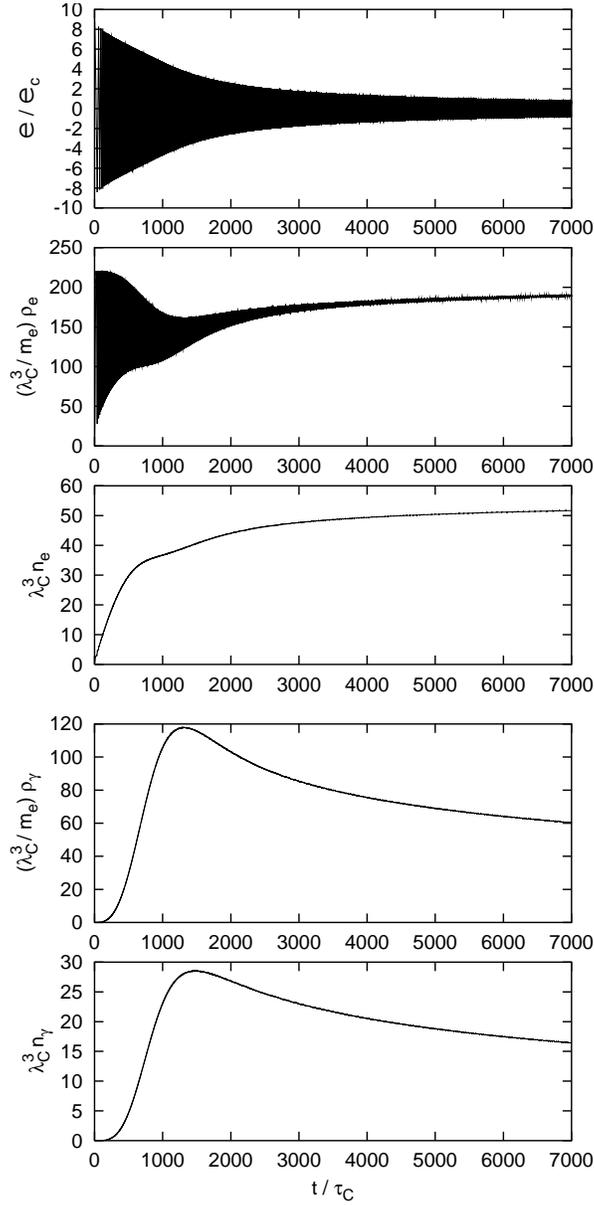}
\end{center}
\caption{Plasma oscillations in a strong homogeneous electric field:\ late
time expected behaviour. Setting $E_{0}=9E_{\mathrm{c}}$,
$t<7000\tau_{\mathrm{C}}$ and it is plotted, from the top to the bottom panel: a)
electromagnetic field strength; b) electrons energy density; c) electrons
number density; d) photons energy density; e) photons number density as
functions of time -- the oscillation period is not resolved in these plots.
The model should have a breakdown at a time much earlier than $7000\tau
_{\mathrm{C}}$ and therefore this plot contains no more than qualitative
informations.}%
\label{OscillationLT}%
\end{figure}

It is interesting that the leading term recovers the expected asymptotic behavior:

\begin{enumerate}
\item the electric field is screened to about the critical value:
$E\simeq E_{\mathrm{c}}$ for $t\sim10^{3}-10^{4}%
\tau_{\mathrm{C}}\gg\tau_{\mathrm{C}}$;

\item the initial electromagnetic energy density is distributed over
electron--positron pairs and photons, indicating energy equipartition;

\item photons and electron--positron pairs number densities are asymptotically
comparable, indicating number equipartition.
\end{enumerate}

At such late times a regime of thermalized electrons--positrons--photons
plasma is expected to begin (as qualitatively indicated by points 2 and 3
above) during which the system is describable by hydrodynamic equations
\cite{2003PhLB..573...33R,1999A&A...350..334R}.

Let us summarize the results in this section. A very simple
formalism is provided to describe simultaneously the creation of
electron--positron pairs by a strong electric field $E\gtrsim
E_{\mathrm{c}}$ and the pairs annihilation into photons. As
discussed in literature, one finds plasma oscillations. In
particular the collisions do not prevent such a feature. This is
because the momentum of electrons (positrons) is very high,
therefore the cross-section for the process
$e^{+}e^{-}\rightarrow\gamma \gamma$ is small and the annihilation
into photons is negligible in the very first phase of the evolution.
As a result, the system takes some time
($t\sim10^{3}-10^{4}\tau_{\mathrm{C}}$) to reach an equilibrium
$e^{+}e^{-}\gamma$ plasma configuration.

\subsection{Electro-fluidodynamics of the pair plasma}\label{electrofluidodynamics}

In the previous section, collisional terms in the Vlasov-Boltzmann equation are
introduced, describing interaction of pairs and photons via the reaction
$e^+e^-\leftrightarrow\gamma\gamma$. These results have been considered of
interest in the studies of pair production in free electron lasers
\cite{2001PhLB..510..107R, 2001hep.ph...12254R, 2003hep.ph....4139R,
2006JETP..102....9B, 2004PhLA..330....1N}, in optical lasers
\cite{2006PhRvL..96n0402B}, of millicharged fermions in extensions of the
standard model of particle physics \cite{2006EL.....76..794G}, electromagnetic
wave propagation in a plasma \cite{2005PhRvE..71a6404B}, as well in
astrophysics \cite{2003PhLB..573...33R}.

In this section, following \cite{2007PhLA..371..399R, 2008AIPC..966..207R}, the
case of undercritical electric field is explored. It is usually expected that
for $E<E_{c}$ back reaction of the created electrons and positrons on the
external electric field can be neglected and electrons and positrons would move
as test particles along lines of force of the electric field. Here it is shown
that this is not the case in a uniform unbounded field. This work is important
since the first observation of oscillations effects should be first detectable
in experiments for the regime $E<E_{c}$, in view of the rapid developments in
experimental techniques, see e.g. \cite{2002PhRvS...5c1301T, Bulanov2003,
2004PhRvL..92o9901B}.

An approach is introduced based on continuity, energy-momentum conservation and
Maxwell equations in order to account for the back reaction of the created
pairs. This approach is more simple than the one, presented in the previous
section. However, the final equations coincide with (\ref{System2}) when the
interaction with photons can be neglected. By this treatment one can analyze
the new case of undercritical field, $E<E_{c}$, and recover the old results for
overcritical field, $E>E_{c}$. In particular, the range $0.15E_{c}<E<10E_{c}$
is focused.

It is generally assumed that electrons and positrons are created at rest in
pairs, due to vacuum polarization in uniform electric field with strength $E$
\cite{1931ZPhy...69..742S, 1936ZPhy...98..714H, 1951PhRv...82..664S,
1954PhRv...93..615S, 1954PhRv...94.1362S, Narozhnyi:1970uv, Greiner1985,
1980ato..book.....G}, with the average rate per unit volume and per unit time
(\ref{nprobabilitym1})
\begin{equation}
S(E)\equiv\frac{dN}{dVdt}=\frac{m_e^{4}}{4\pi^{3}}\left( \frac{E}{E_{c}}\right)
^{2}\exp\left( -\pi\frac{E_{c}}{E}\right) .   \label{rate}
\end{equation}
This formula is derived for uniform constant in time electric field. However,
it still can be used for slowly time varying electric field provided the
inverse adiabaticity parameter \cite{Greiner1985, 1980ato..book.....G,
1970PhRvD...2.1191B, 1971ZhPmR..13..261P, 1972JETP...34..709P,
2001JETPL..74..133P, 1973ZhPmR..18..435P}, see Eq. (\ref{lasergamma}), is much
larger than one,
\begin{equation}
\eta=\frac{m_e}{\omega}\frac{E_{peak}}{E_{c}}=\tilde{T}\tilde{E}_{peak}\gg1,
\label{eta2}
\end{equation}
where $\omega$ is the frequency of oscillations, $\tilde{T}=m_e/\omega$ is
dimensionless period of oscillations. Equation (\ref{eta2}) implies that time
variation of the electric field is much slower than the rate of pair
production. In two specific cases considered in this section, $E=10E_{c}$ and $%
E=0.15E_{c}$ one finds for the first oscillation $\eta=334$ and $\eta
=3.1\times10^{6}$ respectively. This demonstrates applicability of the
formula (\ref{rate}) in this case.

From the continuity, energy-momentum conservation and Maxwell equations
written for electrons, positrons and electromagnetic field one can have
\begin{align}
\frac{\partial\left( \bar{n}U^{\mu}\right) }{\partial x^{\mu}} & =S,
\label{cont} \\
\frac{\partial T^{\mu\nu}}{\partial x^{\nu}} & =-F^{\mu\nu}J_{\nu },
\label{em} \\
\frac{\partial F^{\mu\nu}}{\partial x^{\nu}} & =-4\pi J^{\mu},   \label{me}
\end{align}
where $\bar{n}$ is the comoving number density of electrons, $T^{\mu\nu}$ is
energy-momentum tensor of electrons and positrons
\begin{equation}
T^{\mu\nu}=m_e\bar{n}\left(
U_{(+)}^{\mu}U_{(+)}^{\nu}+U_{(-)}^{\mu}U_{(-)}^{\nu}\right) ,
\label{emten}
\end{equation}
$F^{\mu\nu}$ is electromagnetic field tensor, $J^{\mu}$ is the total 4-current
density, $U^{\mu}$ is four velocity respectively of positrons
and electrons%
\begin{equation}
U_{(+)}^{\mu}=U^{\mu}=\gamma\left( 1,v,0,0\right) ,\qquad U_{(-)}^{\mu
}=\gamma\left( 1,-v,0,0\right) ,
\end{equation}
$v$ is the average velocity of electrons, $\gamma=\left(1-v^{2}\right)^{-1/2}
$ is relativistic Lorentz factor. Electrons and positrons move along the
electric field lines in opposite directions.

One can choose a coordinate frame where pairs are created at rest. Electric
field in this frame is directed along $x$-axis. In spatially homogeneous case
from (\ref{cont}) one has for coordinate number density $n=\bar{n}\gamma$%aga
\begin{equation}
\dot{n}=S.
\end{equation}
With definitions (\ref{emten}) from (\ref{em}) and equation of motion
for positrons and electrons%
\begin{equation}
m_e\frac{\partial U_{(\pm)}^{\mu}}{\partial x^{\nu}}=\mp eF_{\nu}^{\mu},
\end{equation}
one finds
\begin{equation}
\frac{\partial T^{\mu\nu}}{\partial x^{\nu}}=-e\bar{n}\left( U_{(+)}^{\nu
}-U_{(-)}^{\nu}\right) F_{\nu}^{\mu}+m_eS\left( U_{(+)}^{\mu}+U_{(-)}^{\mu
}\right) =-F_{\nu}^{\mu}J^{\nu},
\end{equation}
where the total 4-current density is the sum of conducting $J_{cond}^{\mu}$
and polarization $J_{pol}^{\mu}$ currents \cite{1987PhRvD..36..114G} densities%
\begin{align}
\qquad J^{\mu} & =J_{cond}^{\mu}+J_{pol}^{\mu}, \\
J_{cond}^{\mu} & =e\bar{n}\left( U_{(+)}^{\mu}-U_{(-)}^{\mu}\right) , \\
J_{pol}^{\mu} & =\frac{2m_eS}{E}\gamma\left( 0,1,0,0\right) .
\end{align}

Energy-momentum tensor in (\ref{em}) and electromagnetic field tensor in (%
\ref{me}) change for two reasons: 1)\ electrons and positrons acceleration
in the electric field, given by the term $J_{cond}^{\mu}$, 2)\ particle
creation, described by the term $J_{pol}^{\mu}$. Equation (\ref{cont}) is
satisfied separately for electrons and positrons.

Defining energy density of positrons
\begin{equation}
\rho=\frac{1}{2}T^{00}=m_en\gamma,
\end{equation}
one can find from (\ref{em})
\begin{equation}
\dot{\rho}=envE+ m_e\gamma S.
\end{equation}
Due to homogeneity of the electric field and plasma, electrons and positrons
have the same energy and absolute value of the momentum density $p$, but their
momenta have opposite directions. The definitions also imply for velocity and
momentum densities of electrons and positrons
\begin{equation}
v=\frac{p}{\rho},   \label{veleq}
\end{equation}
and%
\begin{equation}
\rho^{2}=p^{2}+m_e^{2}n^{2},   \label{rhopn}
\end{equation}
which is just relativistic relation between the energy, momentum and mass
densities of particles.

Gathering together the above equations one then has the following equations
\begin{align}
\dot{n}& =S,  \label{ndot} \\
\dot{\rho}& =E\left( env+\frac{m_e\gamma S}{E}\right) ,  \label{rhodot} \\
\dot{p}& =enE+m_ev\gamma S,  \label{pdot} \\
\dot{E}& =-8\pi \left( env+\frac{m_e\gamma S}{E}\right) .  \label{Edot}
\end{align}%
From (\ref{rhodot}) and (\ref{Edot}) one obtains the energy conservation
equation
\begin{equation}
\frac{E_{0}^{2}-E^{2}}{8\pi }+2\rho =0,  \label{energy}
\end{equation}%
where $E_{0}$ is the constant of integration, so the particle energy density
vanishes for initial value of the electric field, $E_{0}$.

These equations give also the maximum number of the pair density
asymptotically attainable consistently with the above rate equation and
energy conservation%
\begin{equation}
n_{0}=\frac{E_{0}^{2}}{8\pi m_e}.   \label{n0}
\end{equation}

For simplicity  dimensionless variables $n=m_e^{3}\tilde{n}$, $%
\rho=m_e^{4}\tilde{\rho}$, $p=m_e^{4}\tilde{p}$, $E=E_{c}\tilde{E}$, and $%
t=m_e^{-1}\tilde{t}$ are introduced. With these variables the system of equations (\ref{ndot}%
)-(\ref{Edot}) takes the form
\begin{align}
\frac{d\tilde{n}}{d\tilde{t}} & =\tilde{S},  \notag \\
\frac{d\tilde{\rho}}{d\tilde{t}} & =\tilde{n}\tilde{E}\tilde{v}+\tilde{\gamma%
}\tilde{S},  \label{numsys} \\
\frac{d\tilde{p}}{d\tilde{t}} & =\tilde{n}\tilde{E}+\tilde{\gamma}\tilde {v}%
\tilde{S},  \notag \\
\frac{d\tilde{E}}{d\tilde{t}} & =-8\pi\alpha\left( \tilde{n}\tilde{v}+\frac{%
\tilde{\gamma}\tilde{S}}{\tilde{E}}\right) ,  \notag
\end{align}
where $\tilde{S}=\frac{1}{4\pi^{3}}\tilde{E}^{2}\exp\left( -\frac{\pi}{%
\tilde{E}}\right) $, $\tilde{v}=\frac{\tilde{p}}{\tilde{\rho}}$ and $\tilde{%
\gamma}=\left( 1-\tilde{v}^{2}\right) ^{-1/2}$, $\alpha=e^{2}/(\hbar c)$ as
before.

The system of equations (\ref{numsys}) is solved numerically with the initial
conditions $n(0)=\rho(0)=v(0)=0$, and the electric field $E(0)=E_{0}$.

In fig. \ref{fig1}, electric field strength, number
density, velocity and Lorentz gamma factor of electrons as functions of
time, are presented for initial values of the electric field $E_{0}=10E_{c}$ (left column)
and $E_{0}=0.15E_{c}$ (right column). Slowly decaying plasma oscillations
develop in both cases. The half-life of oscillations to be $%
10^{3}t_{c}$ for $E_{0}=10E_{c}$ and $10^{5}t_{c}$ for $E_{0}=0.8E_{c}$
are estimated respectively. The period of the fist oscillation is $50t_{c}$ and $%
3\times10^{7}t_{c}$, the Lorentz factor of electrons and positrons in the
first oscillation equals $75$ and $3\times10^{5}$ respectively for $%
E_{0}=10E_{c}$ and $E_{0}=0.15E_{c}$. Therefore, in contrast to the case $%
E>E_{c}$, for $E<E_{c}$ plasma oscillations develop on a much longer
timescale, electrons and positrons reach extremely relativistic velocities.
\begin{figure}[!ptb]
\begin{center}
\includegraphics[width=10cm]{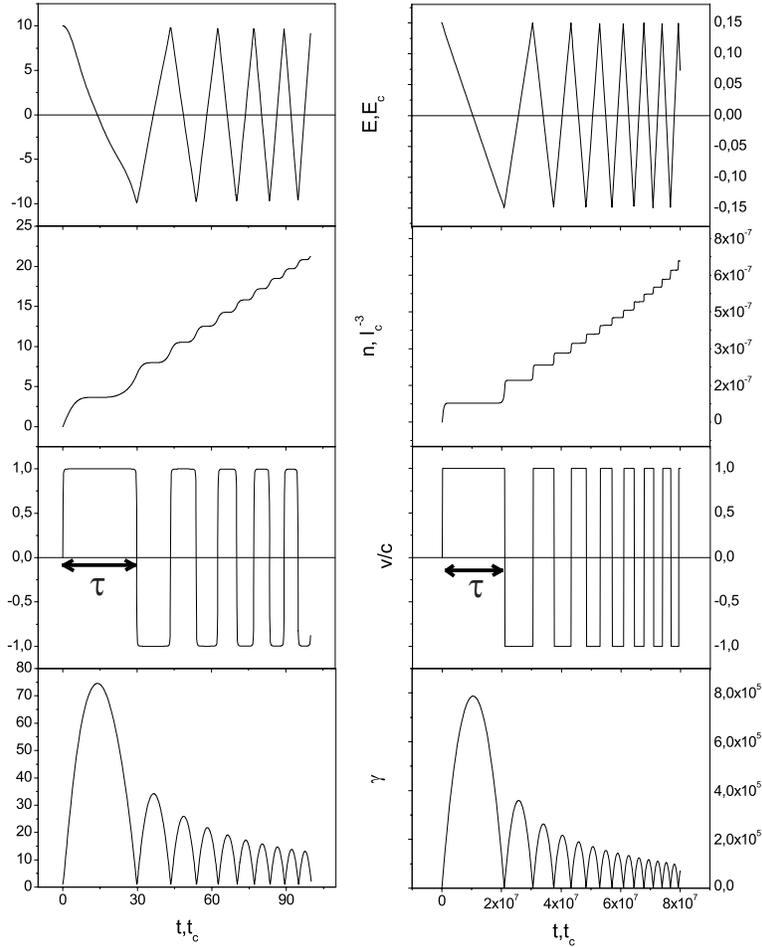}
\end{center}
\caption{Electric field strength, number density of electrons, their
velocity and Lorentz gamma factor depending on time with $E_{0}=10E_{c}$
(left column) and $E_{0}=0.15E_{c}$ (right column). Electric field, number
density and velocity of positron are measured respectively in terms of the
critical field $E_{c}$, Compton volume $\lambda_{C}^{3}=\left( \frac{\hbar}{m_e c}%
\right) ^{3}$, and the speed of light $c$. The length of
oscillation is defined as $D=c\protect\tau$, where $\protect\tau$ is the time needed
for the first half-oscillation, shown above.}
\label{fig1}
\end{figure}

In fig. \ref{fig2} the characteristic length of oscillations is shown
together with the distance between the pairs at the moment of their
creation. For constant electric field the formation length for the
electron--positron pairs, or the quantum tunneling length, is not simply $%
m_e/(eE)$, as expected from a semi-classical approximation, but
\cite{Nikishov1969,2000NCimB.115..761K}
\begin{equation}
D^{\ast}=\frac{m_e}{eE}\left( \frac{E_{c}}{E}\right) ^{1/2}.   \label{D}
\end{equation}

Thus, given initial electric field strength one can define two characteristic
distances: $D^{\ast}$, the distance between created pairs, above which pair
creation is possible, and the length of oscillations, $D=c\tau$, above which
plasma oscillations occur in a uniform electric field. The length of
oscillations is the maximal distance between two turning points in the
motion of electrons and positrons (see fig. \ref{fig2}). From fig. \ref{fig2}
it is clear that $D\gg D^{\ast}$. In the oscillation phenomena the larger
electric field is, the larger becomes the density of pairs and therefore the
back reaction, or the screening effect, is stronger. Thus the period of
oscillations becomes shorter. Note that the frequency of oscillation is not
equal to the plasma frequency, so it cannot be used as the measure of the
latter. Notice that for $E\ll E_{c}$ the length of oscillations becomes
macroscopically large.
\begin{figure}[!thf]
\begin{center}
\includegraphics[width=11cm]{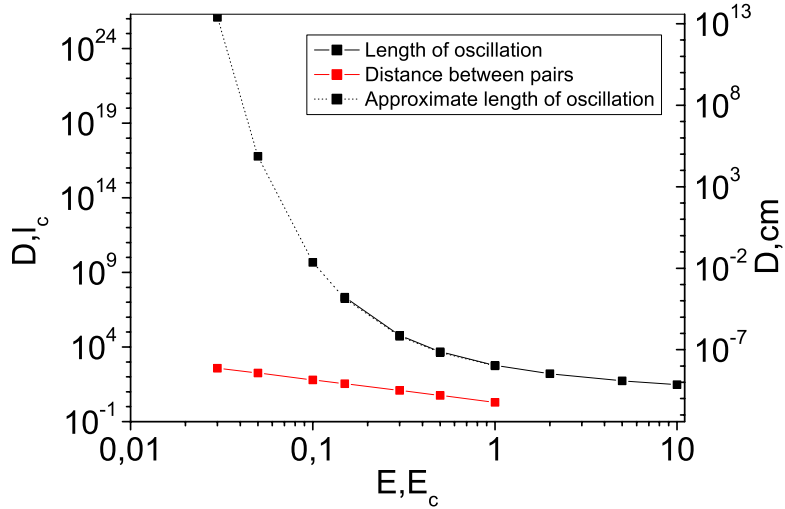}
\end{center}
\caption{Maximum length of oscillations (black curves) together with the
distance between electron and positron in a pair (red curve) computed from (%
\protect\ref{D}), depending on initial value of electric field strength. The
solid black curve is obtained from solutions of exact equations (\protect\ref%
{numsys}), while the dotted black curve corresponds to solutions of
approximate equation (\protect\ref{eeq}).}
\label{fig2}
\end{figure}
\begin{figure}[!bhf]
\begin{center}
\includegraphics[width=11cm]{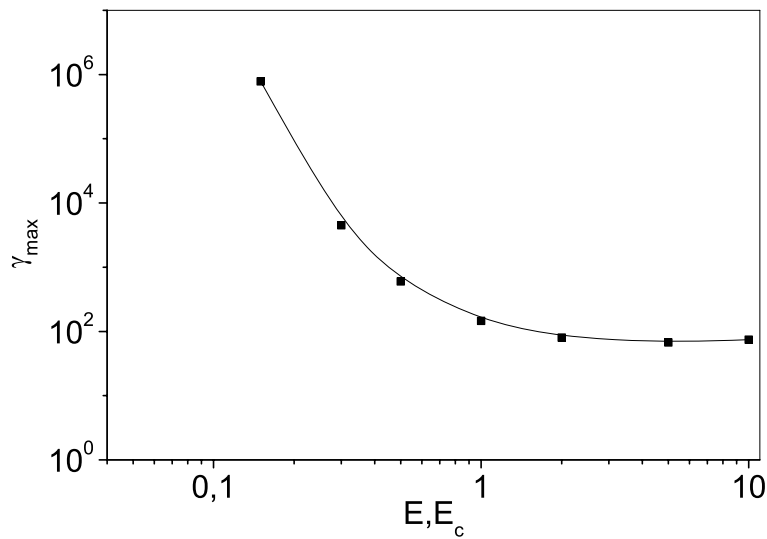}
\end{center}
\caption{Maximum Lorentz gamma factor $\protect\gamma$ reached at the first
oscillation depending on initial value of the electric field strength.}
\label{fig3}
\end{figure}

At fig. \ref{fig3} maximum Lorentz gamma factor in the first oscillation is
presented depending on initial value of the electric field. Since in the
successive oscillations the maximal value of the Lorentz factor is
monotonically decreasing (see fig. \ref{fig1}) It is concluded that for every
initial value of the electric field there exists a maximum Lorentz factor
attainable by the electrons and positrons in the plasma. It is interesting
to stress the dependence of the Lorentz factor on initial electric field
strength. The kinetic energy contribution becomes overwhelming in the $%
E<E_{c}$ case. On the contrary, in the case $E>E_{c}$ the electromagnetic
energy of the field goes mainly into the rest mass energy of the pairs.

This diagram clearly shows that never in this process the test particle
approximation for the electrons and positrons motion in uniform electric field
can be applied. Without considering back reaction on the initial field,
electrons and positrons moving in a uniform electric field would experience
constant acceleration reaching $v\sim c$ for $E=E_{c}$ on the timescale $t_{c}$
and keep that speed thereafter. Therefore, the back
reaction effects in a uniform field are essential both in the case of $%
E>E_{c}$ and $E<E_{c}$.

The average rate of pair creation is compared for two cases: when the
electric field value is constant in time (an external energy source keeps
the field unchanged) and when it is self-regulated by equations (\ref{numsys}%
). The result is represented in fig. \ref{fig4}. It is clear from fig. \ref%
{fig4} that when the back reaction effects are taken into account, the
effective rate of the pair production is smaller than the corresponding rate
(\ref{rate}) in a uniform field $E_{0}$. At the same time, discharge of the
field takes much longer time. In order to quantify this effect we need to
compute the efficiency of the pair production defined as $\epsilon
=n(t_{S})/n_{0}$ where $t_{S}$ is the time when pair creation with the constant
rate $S(E_{0}) $\ would stop, and $n_{0}$ is defined above, see (\ref{n0}). For
$E_{0}=E_{c}
$ one finds $\epsilon =14$\%, while for $E_{0}=0.3E_{c}$ one has $\epsilon =1$%
\%.
\begin{figure}[!tph]
\centering
\includegraphics[width=11cm]{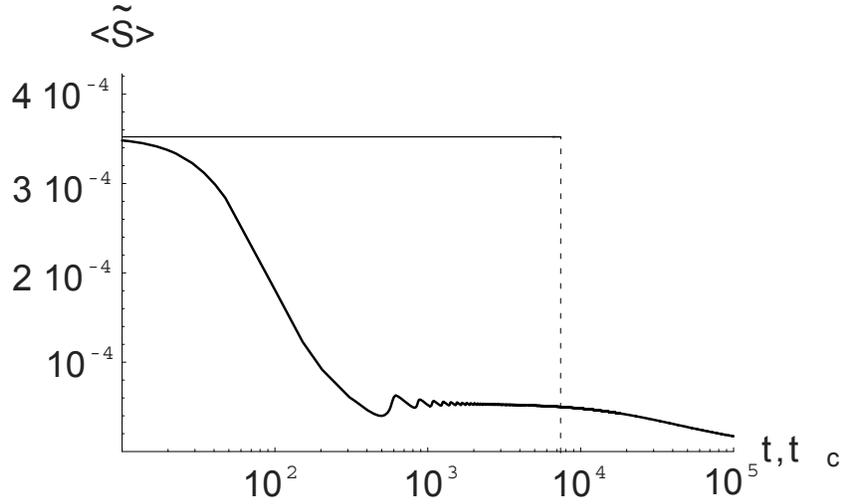}
\caption{The average rate of pair production $n/t$ is shown as function of
time (thick curve), comparing to its initial value $S(E_{0})$ (thin line)
for $E_{0}=E_{c}$. The dashed line marks the time when the energy of
electric field would have exhaused if the rate kept constant.}
\label{fig4}
\end{figure}

It is clear from the structure of the above equations that for $E<E_{c}$ the
number of pairs is small, electrons and positrons are accelerated in
electric field and the conducting current dominates. Assuming electric field
to be weak, the polarization current is neglected in energy conservation (\ref%
{rhodot}) and in Maxwell equation (\ref{Edot}). This means energy density
change due to acceleration is much larger than the one due to pair creation,%
\begin{equation}
Eenv\gg m_e\gamma S.   \label{weak}
\end{equation}
In this case oscillations equations (\ref{ndot})-(\ref{Edot}) simplify. From
(\ref{rhodot}) and (\ref{pdot}) one has $\dot{\rho}=v\dot{p}$, and using (%
\ref{veleq}) obtains $v=\pm1$. This is the limit when rest mass energy is
much smaller than the kinetic energy, $\gamma\gg1$.

One may therefore use only the first and the last equations from the above
set. Taking time derivative of the Maxwell equation one arrives to a single
second order differential equation
\begin{equation}
\ddot{E}+\frac{2em_e^{4}}{\pi^{2}}\left( \frac{E}{E_{c}}\right) \left\vert
\frac{E}{E_{c}}\right\vert \exp\left( -\pi\left\vert \frac{E_{c}}{E}%
\right\vert \right) =0.   \label{eeq}
\end{equation}
Equation (\ref{eeq}) is integrated numerically to find the length of
oscillations shown in fig. \ref{fig2} for $E<E_{c}$. Notice that condition (%
\ref{weak}) means ultrarelativistic approximation for electrons and
positrons, so that although according to (\ref{ndot}) there is creation of
pairs with rest mass $2m$ for each pair, the corresponding increase of
plasma energy is neglected, as can be seen from (\ref{weak}).

Now one can turn to qualitative properties of the system
(\ref{ndot})--(\ref{Edot}). These nonlinear ordinary differential equations
describe certain dynamical system which can be studied by using methods of
qualitative analysis of dynamical systems. The presence of the two integrals
(\ref{rhopn}) and (\ref{energy}) allows reduction of the system to two
dimensions. It is useful to work with the variables $v$ and $E$. In these
variables one has
\begin{align}
\frac{d\tilde{v}}{d\tilde{t}} & =\left( 1-\tilde{v}^{2}\right) ^{3/2}\tilde{E%
}, \\
\frac{d\tilde{E}}{d\tilde{t}} & =-\frac{1}{2}\tilde{v}\left( 1-\tilde {v}%
^{2}\right) ^{1/2}\left( \tilde{E}_{0}^{2}-\tilde{E}^{2}\right) -8\pi\alpha%
\frac{\tilde{S}}{\tilde{E}\left( 1-\tilde{v}^{2}\right) ^{1/2}}.
\end{align}
Introducing the new time variable $\tau$
\begin{equation}
\frac{d\tau}{d\tilde{t}}=\left( 1-\tilde{v}^{2}\right) ^{-1/2}
\end{equation}
one arrives at%
\begin{align}
\frac{d\tilde{v}}{d\tau} & =\left( 1-\tilde{v}^{2}\right) ^{2}\tilde {E},
\label{vdeq} \\
\frac{d\tilde{E}}{d\tau} & =-\frac{1}{2}\tilde{v}\left( 1-\tilde{v}%
^{2}\right) \left( \tilde{E}_{0}^{2}-\tilde{E}^{2}\right) -8\pi\alpha \frac{%
\tilde{S}}{\tilde{E}}.   \label{edeq}
\end{align}
Clearly the phase space is bounded by the two curves $\tilde{v}=\pm1$.
Moreover, physical requirement $\rho\geq0$ leads to existence of two other
bounds $\tilde{E}=\pm\tilde{E}_{0}$. This system has only one singular point
in the physical region, of the type focus at $\tilde{E}=0$ and $\tilde{v}=0$.

The phase portrait of the dynamical system (\ref{vdeq}),(\ref{edeq}) is
represented at fig. {\ref{fig5}}.
\begin{figure}[!ptb]
\centering
\includegraphics[width=11cm]{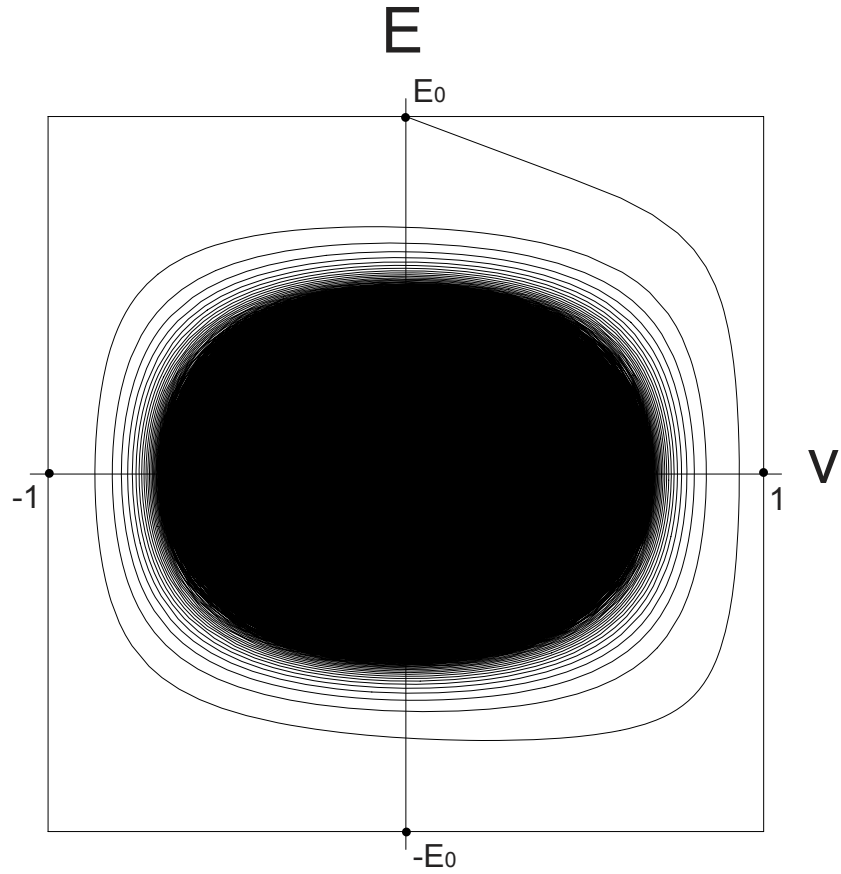}
\caption{Phase portrait of the two-dimensional dynamical system (\protect\ref%
{vdeq}),(\protect\ref{edeq}). Tildes are omitted. Notice that phase
trajectories are not closed curves and with each cycle they approach the
point with $\tilde{E}=0$ and $\tilde{v}=0$.}
\label{fig5}
\end{figure}
Thus, every phase trajectory tends asymptotically to the only singular point
at $\tilde{E}=0$ and $\tilde{v}=0$. This means oscillations stop only when
electric field vanishes. At that point clearly
\begin{equation}
\rho=m_en.   \label{rest}
\end{equation}
is valid. i.e. all the energy in the system transforms just to the rest mass
of the pairs.

In order to illustrate details of the phase trajectories shown at fig. \ref%
{fig5} only 1.5 cycles are plotted at fig. \ref{fig5a}. One can see that the
deviation from closed curves, representing undamped oscillations and shown by
dashed curves, is maximal when the field peaks, namely when the pair production
rate is maximal.
\begin{figure}[!ptb]
\centering
\includegraphics[width=11cm]{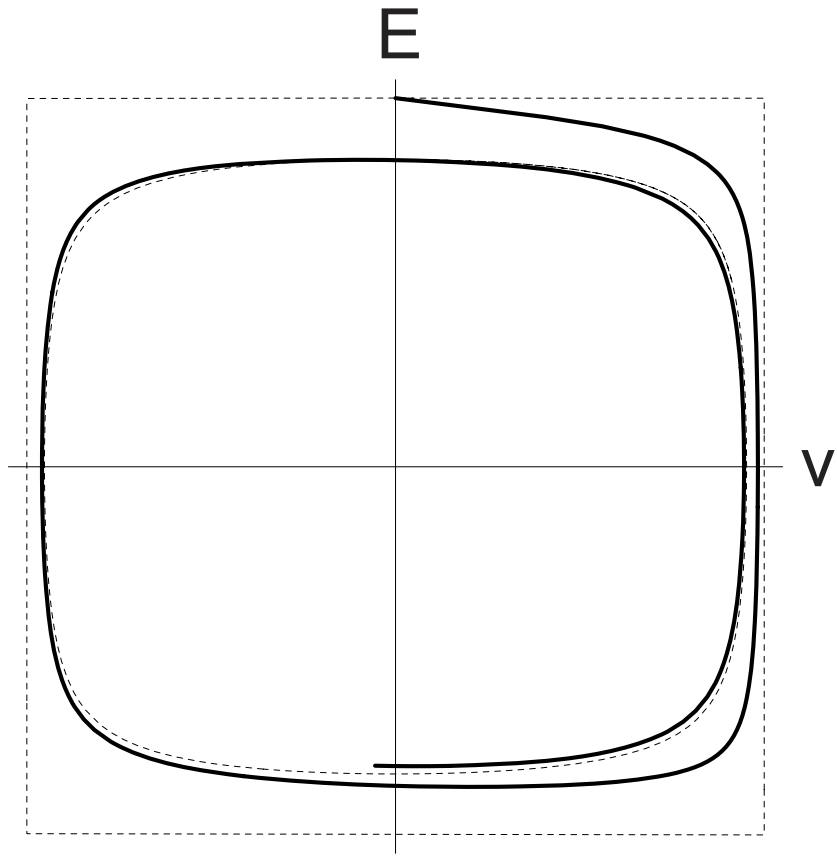}
\caption{Phase trajectory for 1.5 cycles (thick curve) compared with
solutions where the Schwinger pair production is switched off (dashed
curves).}
\label{fig5a}
\end{figure}

The above treatment has been done by considering uniquely back reaction of the
electron--positron pairs on the external uniform electric field. The only
source of damping of the oscillations is pair production, i.e. creation of
mass. As analysis shows the damping in this case is exponentially weak.
However, since electrons and positrons are strongly accelerated in electric
field the bremsstrahlung radiation may give significant contribution to the
damping of oscillations and further reduce the pair creation rate. Therefore,
the effective rate shown in fig. \ref{fig4} will represent an upper limit. In
order to estimate the effect of bremsstrahlung, the
classical formula for the radiation loss in electric field is recalled%
\begin{equation}
I=\frac{2}{3}\frac{e^{4}}{m_e^{2}}E^{2}=\frac{2}{3}\alpha m_e^{2}\left( \frac{E}{%
E_{c}}\right) ^{2}.
\end{equation}%
Thus the equations (\ref{rhodot}) and (\ref{pdot}), generalized for
bremsstrahlung, are%
\begin{align}
\dot{\rho}& =E\left( env+\frac{m_e\gamma S}{E}\right) -\frac{2}{3}e^{4}m_eE^{2},
\label{energy_br} \\
\dot{p}& =enE+m_ev\gamma S-\frac{2}{3}e^{4}m_eE^{2}v.  \label{momentum_br}
\end{align}%
while equations (\ref{ndot}) and (\ref{Edot}) remain unchanged. Assuming
that new terms are small, relations (\ref{rhopn}) and (\ref{energy}) are
still approximately satisfied.

Now damping of the oscillations is caused by two terms:%
\begin{equation}
\frac{\tilde{\gamma}}{4\pi ^{2}}\tilde{E}^{2}\exp \left( -\frac{\pi }{\tilde{%
E}}\right) \text{ \ \ \ \ \ and \ \ \ }\frac{2}{3}\alpha \tilde{E}^{2}.
\label{terms}
\end{equation}

\begin{figure}[!tbp]
\centering
\includegraphics[width=11cm]{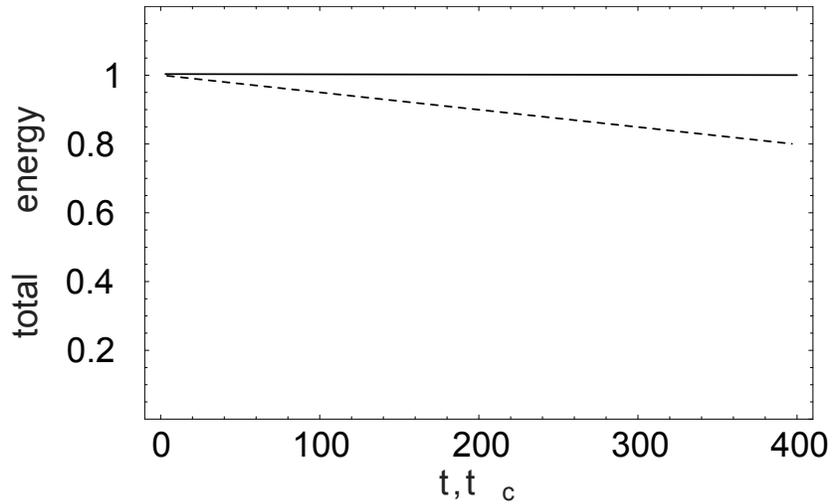}
\caption{Losses of the energy due to classical bremsstrahlung radiation. The
energy density of the system of electrons, positrons and the electric field
normalized to the initial energy density is shown without (solid line) and
with (dashed line) the effect of bremsstrahlung.}
\label{fig6}
\end{figure}

The modified system of equations is integrated, taking into account radiation
loss, starting with $E_{0}=10E_{c}$. The results are presented in
Fig.~\ref{fig6} where the sum of the energy of electric field and
electron--positron pairs normalized to the initial energy is shown as a
function of time. The energy loss reaches 20 percent for 400 Compton times.
Thus the effect of bremsstrahlung is as important as the effect of collisions
considered in \cite{2003PhLB..559...12R} for $E>E_{c}$, leading to comparable
energy loss for pairs on the same timescale. For $E<E_{c}$ one expects that the
damping due to bremsstrahlung dominates, but the correct description in this
case requires Vlasov--Boltzmann treatment \cite{2007PhRvL..99l5003A}.

The damping of the plasma oscillations due to electron--positron annihilation
into photons has been addressed in \cite{2003PhLB..559...12R}. There it was
found that the system evolves towards an electron--positron-photon plasma
reaching energy equipartition. Such a system undergoes self-acceleration
process following the work of \cite{1999A&A...350..334R}.

Therefore the following conclusions are reached:

\begin{itemize}
\item It is usually assumed that for $E<E_{c}$ electron--positron pairs,
created by the vacuum polarization process, move as charged particles in
external uniform electric field reaching arbitrary large Lorentz factors. The
existence of plasma oscillations of the electron--positron pairs also for
$E\lesssim E_{c}$ is demonstrated. The corresponding results for $E>E_c$ are
well known in the literature. For both cases the maximum Lorentz factors
$\gamma _{\max }$ reached by electrons and positrons are determined. The length
of oscillations is 10 $\hbar /(m_e c)$ for $E_{0}=10E_{c}$, and 10$^{7}$ $\hbar
/(m_e c)$ for $E_{0}=0.15E_{c}$. The asymptotic behavior in time,
$t\rightarrow\infty$, of the plasma oscillations by the phase portrait
technique is also studied.

\item For $E>E_{c}$ the vacuum polarization process transforms the
electromagnetic energy of the field mainly in the rest mass of pairs, with
moderate contribution to their kinetic energy:  for $E_{0}=10E_{c}$ one finds
$\gamma _{\max }=76$. For $E<E_{c}$ the kinetic energy contribution is
maximized with respect to the rest mass of pairs: $\gamma _{\max }=8\times
10^{5}$ for $E_{0}=0.15E_{c}$.

\item In the case of oscillations the effective rate of pair production is
smaller than the rate in uniform electric field constant in time, and
consequently, the discharge process lasts longer. The half-life of oscillations
is $10^{3}t_{c}$ for $E_{0}=10E_{c}$ and $10^{5}t_{c}$ for $E_{0}=0.8E_{c}$.
The efficiency of pair production is computed with respect to the one in a
uniform constant field. For $E=0.3E_{c}$ the efficiency is reduced to one
percent, decreasing further for smaller initial electric field.

\end{itemize}

All these considerations apply to a uniform electric field unbounded in space.
The presence of a boundary or a gradient in electric field would require the
use of partial differential equations, in contrast to the ordinary differential
equations used here. This topic needs further study. The effect of
bremsstrahlung for $E>E_{c}$ is also estimated, and it is found that it
represents comparable contribution to the damping of the plasma oscillations
caused by collisions \cite{2003PhLB..559...12R}. It is therefore clear, that
the effect of oscillations introduces a new and firm upper limit to the rate of
pair production which would be further reduced if one takes into account
bremsstrahlung, collisions and boundary effects.

\section{Thermalization of the mildly relativistic pair plasma}
\label{aksenov}

An electron--positron plasma is of interest in many fields of physics and
astrophysics. In the early Universe \cite{1972gcpa.book.....W,
1990eaun.book.....K} during the lepton era, ultrarelativistic
electron--positron pairs contributed to the matter contents of the Universe. In
GRBs electron--positron pairs play an essential role in the dynamics
of expansion \cite{1986ApJ...308L..47G, 1999PhR...314..575P}%
,\cite{1999A&A...350..334R}. Indications exist on the presence of the pair
plasma also in active galactic nuclei \cite{1998Natur.395..457W}, in the
center of our Galaxy \cite{2005MNRAS.357.1377C},\ around hypothetical quark
stars \cite{1998PhRvL..80..230U}. In the laboratory pair plasma is expected to
appear in the fields of ultraintense lasers \cite{2006PhRvL..96n0402B}.

In many stationary astrophysical sources the pair plasma is thought to be in
thermodynamic equilibrium. A detailed study of the relevant processes
\cite{1971SvA....15...17B, 1976PhRvA..13.1563W, 1982ApJ...253..842L,
1982ApJ...254..755G,1983MNRAS.204.1269S, 1990MNRAS.245..453C}, radiation
mechanisms \cite{1981ApJ...251..713L}, possible equilibrium configurations
\cite{1982ApJ...253..842L, 1982ApJ...258..335S} and spectra
\cite{1984ApJ...283..842Z} in an optically thin pair plasma has been carried
out. Particular attention has been given to collisional relaxation process
\cite{1981PhFl...24..102G, 1983MNRAS.202..467S}, pair production and
annihilation \cite{1982ApJ...258..321S}, relativistic bremsstrahlung
\cite{1980ApJ...238.1026G, 1985A&A...148..386H}, double Compton scattering
\cite{1981ApJ...244..392L, 1984ApJ...285..275G}.

An equilibrium occurs if the sum of all reaction rates vanishes. For instance,
electron--positron pairs are in equilibrium when the net pair production
(annihilation) rate is zero. This can be achieved by variety of ways and the
corresponding condition can be represented as a system of algebraic equations
\cite{1984MNRAS.209..175S}. However, the main assumption made in all the above
mentioned works is that the plasma is in thermodynamic equilibrium.

At the same time, in some cases considered above the pair plasma can be
optically thick. Although moderately thick plasmas have been considered in the
literature \cite{1985MNRAS.212..523G}, only qualitative description is
available for large optical depths. Assumption of thermal equilibrium is often
adopted for rapidly evolving systems without explicit
proof \cite{1986ApJ...308L..47G,1999PhR...314..575P}%
,\cite{1999A&A...350..334R,2004ApJ...601...78I}. Then hydrodynamic
approximation is usually applied both for leptons and photons. However,
particles may not be in equilibrium initially. Moreover, it is very likely
situation, especially in the early Universe or in transient events when the
energy is released on a very short timescale and there is not enough time for
the system to relax to thermal equilibrium configuration.

Ultrarelativistic expansion of GRB sources is unprecedented in astrophysics.
There are indications that relativistic jets in X-ray binaries have Lorentz
factors $\gamma\sim2-10$ while in active galactic nuclei $\gamma\sim10-20$
\cite{2006MNRAS.367.1432M}, but some bursts sources have $\gamma\sim400$ and
possibly larger \cite{2007Ap&SS.311..197V}. There is a consensus in the
literature that the acceleration required to reach ultrarelativistic velocity
in astrophysical flows comes from the radiation pressure, namely from photons
and electron--positron pairs. Therefore, the source does not move as a whole,
but expands from a compact region, almost reaching the speed of light. The bulk
of radiation is emitted far from the region of formation of the plasma, when it
becomes transparent for photons, trapped initially inside by the huge optical
depth. Thus the plasma is optically thick at the moment of its formation and
intense interactions between electrons, positrons and photons take place in it.
Even if initially the energy is released in the form of only photons, or only
pairs, the process of creation and annihilation of pairs soon redistribute the
energy between particles in such a way that the final state will be a mixture
of pairs and photons. The main question arises: \emph{what is the initial
state, prior to expansion, of the pair plasma?} Is it in a kind of equilibrium
and, if so, is it thermal equilibrium, as expected from the optically thick
plasma? Stationary sources in astrophysics have enough time for such an
equilibrium to be achieved. On the contrary, for transient sources with the
timescale of expansion of the order of milliseconds it is not at all clear that
equilibrium can be reached.

In the literature there is no consensus on this point. Some authors
considered thermal equilibrium as the initial state prior to
expansion \cite{1986ApJ...308L..47G,1999A&A...350..334R}, while
others did not \cite{1978MNRAS.183..359C}. In fact, the study of the
pair plasma equilibrium configurations in detail, performed in
\cite{1982ApJ...258..335S}, cannot answer this question, because
essentially nonequilibrium processes have to be considered.

Thus, observations provide motivation for theoretical analysis of physical
conditions taking place in the sources of GRBs, and more generally, in
nonequilibrium optically thick pair plasma. Notice that there is substantial
difference between the ion-electron plasma on the one hand and
electron--positron plasma on the other hand. Firstly, the former is
collisionless in the wide range of parameters, while collisions are always
essential in the latter. Secondly, when collisions are important relevant
interactions in the former case are Coulomb scattering of particles which are
usually described by the classical Rutherford cross-section. In contrast,
interactions in the pair plasma are described by quantum cross-sections even if
the plasma itself can be still considered as classical one.

The study reported in \cite{2007PhRvL..99l5003A,2008AIPC..966..191A}
in the case of pure pair plasma clarified the issue of initial state
of the pair plasma in GRB sources. The numerical calculations show
that the pair plasma quickly reach thermal equilibrium prior to
expansion, due to intense binary and triple collisions. In this
Section details about the computational scheme adopted in
\cite{2007PhRvL..99l5003A} are given. Generalization to the presence
of proton loading is given in \cite{2009PhRvD..79d3008A}.

\subsection{Qualitative description of the pair plasma}\label{qualitplasma}

First of all the domain of parameters characterizing the pair plasma
considered in this Section is specified. It is convenient to use
dimensionless parameters usually adopted for this purpose.

Mildly relativistic pair plasma is considered, thus the average energy per
particle $\epsilon$ brackets the electron rest mass energy%
\begin{equation}
0.1\;\mathrm{MeV}\lesssim\epsilon\lesssim10\;\mathrm{MeV.}\label{avenergy}%
\end{equation}
The lower boundary is required for significant concentrations of pairs, while
the upper boundary is set to avoid substantial production of other particles
such as muons.

The plasma parameter is $\mathfrak{g}=(n_{-}d^{3})^{-1}$, where
$d=\sqrt{\frac{k_{B}T_{-}}{4\pi
e^{2}n_{-}}}=\frac{c}{\omega}\sqrt{\theta_{-}} $ is the Debye
length, $k_{B}$ is Boltzmann's constant, $n_{-}$ and $T_{-} $ are
electron number density and temperature respectively,
$\theta_{-}=k_{B}T_{-}/(m_e c^{2})$ is dimensionless temperature,
$\omega=\sqrt{4\pi e^{2}n_{-}/m_e}$ is the plasma frequency. To
ensure applicability of kinetic approach it is necessary that the
plasma parameter is small, $\mathfrak{g}\ll1$. This condition means
that kinetic energy of particles dominates their potential energy
due to mutual interaction. For the pair plasma considered in this
Section this condition is satisfied.

Further, the classicality parameter, defined as $\varkappa=e^{2}/(\hbar
v_{r})=\alpha/\beta_{r}$, where $v_{r}=\beta_{r}c$ is mean
relative velocity of particles, see (\ref{gammarel}). The
condition $\varkappa\gg1$ means that particles collisions can be considered
classically, while for $\varkappa\ll1$ quantum description is required. Both for pairs and protons quantum cross-sections are used since
$\varkappa<1$.

The strength of screening of the Coulomb interactions is characterized by the
Coulomb logarithm $\Lambda=m_edv_{r}/\hbar$. Coulomb logarithm varies
with mean particle velocity, and it cannot be set a constant as in most of
studies of the pair plasma.

Finally, pair plasma is considered with linear dimensions $R$ exceeding the
mean free path of photons $l=\left(  n_{-}\sigma\right)  ^{-1}$, where $n_{-}$
is concentration of electrons and $\sigma$ is the corresponding total
cross-section. Thus the optical depth $\tau=n\sigma R\gg1$ is large, and
interactions between photons and other particles have to be taken in due
account. These interaction are reviewed in the next section.

Note that natural parameter for perturbative expansion in the problem under
consideration is the fine structure constant $\alpha$.

Pure pair plasma composed of electrons $e^{-}$, positrons $e^{+}$, and photons
$\gamma$ is considered. It is assumed that pairs or photons appear by some
physical process in the region with a size $R$ and on a timescale $t<R/c$. We
assume that distribution functions of particles depend on neither spatial
coordinates nor direction of momentum $f_{i}=f_{i}(\epsilon,t)$, i.e. isotropic
in momentum space and uniform plasma is considered.

To make sure that classical kinetic description is adequate the
dimensionless degeneracy temperature is estimated%
\begin{equation}
\theta_{F}=\left[  \left(  \frac{\hbar}{m_e c}\right)  ^{2}\left(  3\pi^{2}%
n_{-}\right)  ^{\frac{2}{3}}+1\right]  ^{1/2}-1,
\end{equation}
and compared with the estimated temperature in thermal equilibrium. With
initial conditions (\ref{avenergy}) the degeneracy temperature is always
smaller than the temperature in thermal equilibrium and therefore the classical
kinetic approach is applied. Besides, since ideal plasma is considered with the
plasma parameter $\mathfrak{g}\sim10^{-3}$ it is possible to use one-particle
distribution functions. These considerations justify the computational approach
based on classical relativistic Boltzmann equation. At the same time the
right-hand side of Boltzmann equations contains collisional integrals with
quantum and not classical matrix elements, as discussed above.

Relativistic Boltzmann equations \cite{1956DAN...107...807B}%
,\cite{1984oup..book.....M} in spherically symmetric case are%
\begin{equation}
\frac{1}{c}\frac{\partial f_{i}}{\partial t}+\beta_{i}\left(  \mu
\frac{\partial f_{i}}{\partial r}+\frac{1-\mu^{2}}{r}\frac{\partial f_{i}%
}{\partial\mu}\right)  -\mathbf{\nabla}U\frac{\partial f_{i}}{\partial
\mathbf{p}}=\sum_{q}\left(  \eta_{i}^{q}-\chi_{i}^{q}f_{i}\right)
,\label{bol}%
\end{equation}
where $\mu=\cos\vartheta$, $\vartheta$ is the angle between the radius vector
$\mathbf{r}$ from the origin and the particle momentum $\mathbf{p}$, $U$ is a
potential due to some external force, $\beta_{i}=v_{i}/c$ are particles
velocities, $f_{i}(\epsilon,t)$ are their distribution functions, the index $i$
denotes the type of particle, $\epsilon$ is their energy, and $\eta _{i}^{q}$
and $\chi_{i}^{q}$ are the emission and the absorption coefficients for the
production of a particle of type \textquotedblleft$i$" via the physical process
labeled by $q$. This is a coupled system of partial--integro-differential
equations. For homogeneous and isotropic distribution functions of electrons,
positrons and photons Eqs. (\ref{bol}) reduce to
\begin{equation}
\frac{1}{c}\frac{\partial f_{i}}{\partial t}=\sum_{q}\left(  \eta_{i}^{q}%
-\chi_{i}^{q}f_{i}\right)  ,\label{BE}%
\end{equation}
which is a coupled system of integro-differential equations. In (\ref{BE}) the
Vlasov term $\mathbf{\nabla}U\frac{\partial f_{i}}{\partial\mathbf{p}}$ is
explicitly neglected.

Therefore, the left-hand side of the Boltzmann equation is reduced to partial
derivative of the distribution function with respect to time. The right-hand
side contains collisional integrals, representing interactions between
electrons, positrons and photons.

Differential probability for all processes per unit time and unit volume
\cite{1982els..book.....B} is defined as
\begin{align}
dw =c(2\pi\hbar)^{4}\delta^{(4)}\left(  \mathfrak{p}_{f}-\mathfrak{p}%
_{i}\right)  \left\vert M_{fi}\right\vert ^{2}V\times
\left[  \prod\limits_{b}\frac{\hbar c}{2\epsilon_{b}V}\right]  \left[
\prod_{a}\frac{d\mathbf{p}_{a}^{\prime}}{(2\pi\hbar)^{3}}\frac{\hbar
c}{2\epsilon_{a}^{\prime}}\right]  ,
\label{dp}
\end{align}
where $\mathbf{p}_{a}^{\prime}$ and $\epsilon_{a}^{\prime}$ are respectively
momenta and energies of outgoing particles, $\epsilon_{b}$ are energies of
particles before interaction, $M_{fi}$ are the corresponding matrix elements,
$\delta^{(4)}$ stands for energy-momentum conservation, $V$ is the
normalization volume. The matrix elements are related to the scattering
amplitudes by%
\begin{equation}
M_{fi}=\left[  \prod\limits_{b}\frac{\hbar c}{2\epsilon_{b}V}\right]  \left[
\prod_{a}\frac{\hbar c}{2\epsilon_{a}^{\prime}V}\right]  T_{fi}.
\end{equation}

As example consider absorption coefficient for Compton scattering which is
given by
\begin{equation}
\chi^{\mathrm{\gamma e}^{\mathrm{\pm}}\mathrm{\rightarrow\gamma^{\prime}%
e^{\pm\prime}}}f_{\gamma}=\int d\mathbf{k}^{\prime}d\mathbf{p}d\mathbf{p}%
^{\prime}w_{\mathbf{k}^{\prime},\mathbf{p}^{\prime};\mathbf{k},\mathbf{p}%
}f_{\gamma}(\mathbf{k},t)f_{\pm}(\mathbf{p},t),\label{comptonemission}%
\end{equation}
where $\mathbf{p}$ and $\mathbf{k}$ are momenta of electron (positron) and
photon respectively, $d\mathbf{p}=d\epsilon_{\pm}do\epsilon_{\pm}^{2}%
\beta_{\pm}/c^{3}$, $d\mathbf{k}^{\prime}=d\epsilon_{\gamma}^{\prime}%
\epsilon_{\gamma}^{\prime2}do_{\gamma}^{\prime}/c^{3}$ and the differential
probability $w_{\mathbf{k}^{\prime},\mathbf{p}^{\prime};\mathbf{k},\mathbf{p}%
}$ is given by (\ref{df-abs}).

In (\ref{comptonemission}) one can perform one integration over $d\mathbf{p}%
^{\prime}$
\begin{equation}
\int d\mathbf{p}^{\prime}\delta(\mathbf{k}+\mathbf{p}-\mathbf{k}^{\prime
}-\mathbf{p}^{\prime})\rightarrow1,
\end{equation}
but it is necessary to take into account the momentum conservation in the next
integration over $d\mathbf{k}^{\prime}$, so
\begin{gather}
\int d\epsilon_{\gamma}^{\prime}\delta(\epsilon_{\gamma}+\epsilon_{\pm
}-\epsilon_{\gamma}^{\prime}-\epsilon_{\pm}^{\prime})=\\
=\int d(\epsilon_{\gamma}^{\prime}+\epsilon_{\pm}^{\prime})\frac{1}%
{|\partial(\epsilon_{\gamma}^{\prime}+\epsilon_{\pm}^{\prime})/\partial
\epsilon_{\gamma}^{\prime}|}\delta(\epsilon_{\gamma}+\epsilon_{\pm}%
-\epsilon_{\gamma}^{\prime}-\epsilon_{\pm}^{\prime})\rightarrow\frac
{1}{|\partial(\epsilon_{\gamma}^{\prime}+\epsilon_{\pm}^{\prime}%
)/\partial\epsilon_{\gamma}^{\prime}|}\equiv J_{\mathrm{cs}},\nonumber
\end{gather}
where the Jacobian of the transformation is%
\begin{equation}
J_{\mathrm{cs}}=\frac{1}{1-\beta_{\pm}^{\prime}\mathbf{b}_{\gamma}^{\prime
}\mathbf{\cdot b}_{\pm}^{\prime}},
\end{equation}
where $\mathbf{b}_{i}=\mathbf{p}_{i}/p$, $\mathbf{b}_{i}^{\prime}%
=\mathbf{p}_{i}^{\prime}/p^{\prime}$, $\mathbf{b}_{\pm}^{\prime}=(\beta_{\pm
}\epsilon_{\pm}\mathbf{b}_{\pm}+\epsilon_{\gamma}\mathbf{b}_{\gamma}%
-\epsilon_{\gamma}^{\prime}\mathbf{b}_{\gamma}^{\prime})/(\beta_{\pm}^{\prime
}\epsilon_{\pm}^{\prime})$.

Finally, for the absorption coefficient
\begin{equation}
\chi^{\mathrm{cs}}f_{\gamma}=-\int do_{\gamma}^{\prime}d\mathbf{p}%
\frac{\epsilon_{\gamma}^{\prime}|M_{fi}|^{2}\hbar^{2}c^{2}}{16\epsilon_{\pm
}\epsilon_{\gamma}\epsilon_{\pm}^{\prime}}J_{\mathrm{cs}}f_{\gamma}%
(\mathbf{k},t)f_{\pm}(\mathbf{p},t),\label{comptonabscoef}%
\end{equation}
where the matrix element here is dimensionless. This integral is evaluated
numerically.

For all binary interactions exact QED matrix elements are used which can be
found in the standard textbooks, e.g. in \cite{1982els..book.....B}%
,\cite{2003spr..book.....G,1981nau..book.....A}, and are given
below.

In order to account for charge screening in Coulomb scattering the minimal
scattering angles are introduced following \cite{1988A&A...191..181H}. This
allows to apply the same scheme for the computation of emission and absorption
coefficients even for Coulomb scattering, while many treatments in the
literature use Fokker-Planck approximation \cite{1997ApJ...486..903P}.

For such a dense plasma collisional integrals in (\ref{BE})\ should include
not only binary interactions, having order $\alpha^{2}$ in Feynman diagrams,
but also triple ones, having order $\alpha^{3}$ \cite{1982els..book.....B}.
Consider relativistic bremsstrahlung
\begin{equation}
e_{1}+e_{2}\leftrightarrow e_{1}^{\prime}+e_{2}^{\prime}+\gamma^{\prime
}.\label{brems}%
\end{equation}
For the time derivative, for instance, of the distribution function $f_{2}$ in
the direct and in the inverse reactions (\ref{brems}) one has
\begin{gather}
\dot{f}_{2}=\int d\mathbf{p}_{1}d\mathbf{p}_{1}^{\prime}d\mathbf{p}%
_{2}^{\prime}d\mathbf{k}^{\prime}\left[  W_{\mathbf{p}_{1}^{\prime}%
,\mathbf{p}_{2}^{\prime},\mathbf{k}^{\prime};\mathbf{p}_{1},\mathbf{p}_{2}%
}f_{1}^{\prime}f_{2}^{\prime}f_{k}^{\prime}-  W_{\mathbf{p}_{1},
\mathbf{p}_{2};\mathbf{p}_{1}^{\prime}%
,\mathbf{p}_{2}^{\prime},\mathbf{k}^{\prime}}f_{1}f_{2}\right]  =\\
\int d\mathbf{p}_{1}d\mathbf{p}_{1}^{\prime}d\mathbf{p}_{2}^{\prime}d\mathbf{k}%
^{\prime}\frac{c^{6}\hbar^{3}}{(2\pi)^{2}}\frac{\delta^{(4)}(P_{f}-P_{i})
|M_{fi}|^{2}}{2^{5}\epsilon_{1}%
\epsilon_{2}\epsilon_{1}^{\prime}\epsilon_{2}^{\prime}\epsilon_{\gamma
}^{\prime}}\left[  f_{1}^{\prime}f_{2}^{\prime}f_{k}^{\prime}-\frac{1}%
{(2\pi\hbar)^{3}}f_{1}f_{2}\right]  ,\nonumber
\end{gather}
where%
\begin{align*}
d\mathbf{p}_{1}d\mathbf{p}_{2}W_{\mathbf{p}_{1}^{\prime},\mathbf{p}%
_{2}^{\prime},\mathbf{k}^{\prime};\mathbf{p}_{1},\mathbf{p}_{2}}  & \equiv
V^{2}dw_{1},\\
d\mathbf{p}_{1}^{\prime}d\mathbf{p}_{2}^{\prime}d\mathbf{k}^{\prime
}W_{\mathbf{p}_{1},\mathbf{p}_{2};\mathbf{p}_{1}^{\prime},\mathbf{p}%
_{2}^{\prime},\mathbf{k}^{\prime}}  & \equiv Vdw_{2},
\end{align*}
and $dw_{1}$ and $dw_{2}$\ are differential probabilities given by (\ref{dp}).
The matrix element here has dimensions of the length squared.

In the case of the distribution functions (\ref{dk}), see below, there are
multipliers proportional to $\exp\frac{\varphi}{k_{B}T}$ in front of the
integrals, where $\varphi$ are chemical potentials. The calculation of emission
and absorption coefficients is then reduced to the well known thermal
equilibrium case \cite{1984MNRAS.209..175S}. In fact, since reaction rates of
triple interactions are $\alpha$ times smaller than binary reaction rates, it
is expected that binary reactions come to detailed balance first. Only when
binary reactions are all balanced, triple interactions become important. In
addition, when binary reactions come into balance, distribution functions
already acquire the form (\ref{dk}). Although there is no principle difficulty
in computations using exact matrix elements for triple reactions as well, the
simplified scheme allows for much faster numerical computation.

All possible binary and triple interactions between electrons,
positrons and photons are considered as summarized in Tab.~\ref{tab1}.%

%TCIMACRO{\TeXButton{B}{\begin{table}[tbp] \centering}}%
%BeginExpansion
\begin{table}[tbp] \centering
%EndExpansion%
\begin{tabular}
[c]{|c|c|}\hline
Binary interactions & Radiative and pair producing variants\\\hline\hline
\multicolumn{2}{|c|}{Reactions with pairs}\\\hline\hline
{M{\o }ller and Bhabha scattering} & {Bremsstrahlung}\\
{$e_{1}^{\pm}{e_{2}^{\pm}\longrightarrow e_{1}^{\pm}}^{\prime}$}${e_{2}^{\pm}%
}^{\prime}$ & {$e_{1}^{\pm}e_{2}^{\pm}{\leftrightarrow}e_{1}^{\pm\prime}%
e_{2}^{\pm\prime}\gamma$}\\
{$e^{\pm}{e^{\mp}\longrightarrow e^{\pm\prime}}$}${e^{\mp\prime}}$ & {$e^{\pm
}e^{\mp}{\leftrightarrow}e^{\pm\prime}e{^{\mp\prime}}\gamma$}\\\hline
Single {Compton scattering} & {Double Compton scattering}\\
{\ $e^{\pm}\gamma{\longrightarrow}e^{\pm}\gamma^{\prime}$} & {$e^{\pm}%
\gamma{\leftrightarrow}e^{\pm\prime}\gamma^{\prime}\gamma^{\prime\prime}$%
}\\\hline
{Pair production} & Radiative pair production\\
and annihilation & and three photon annihilation\\
{$\gamma\gamma^{\prime}{\leftrightarrow}e^{\pm}e^{\mp}$} & $\gamma
\gamma^{\prime}${${\leftrightarrow}e^{\pm}e^{\mp}$}$\gamma^{\prime\prime}$\\
& {$e^{\pm}e^{\mp}{\leftrightarrow}\gamma\gamma^{\prime}$}$\gamma
^{\prime\prime}$\\\hline
& $e^{\pm}\gamma${${\leftrightarrow}e^{\pm\prime}{e^{\mp}}e^{\pm\prime\prime}
$}\\\hline
\end{tabular}
\caption{Microphysical processes in the pair plasma.}\label{tab1}%
%TCIMACRO{\TeXButton{E}{\end{table}} }%
%BeginExpansion
\end{table}
%EndExpansion
Each of the above mentioned reactions is characterized by the corresponding
timescale and optical depth. For Compton scattering of a photon, for instance%
\begin{equation}
t_{\mathrm{cs}}=\frac{1}{\sigma_{T}n_{\pm}c},\qquad\tau_{\mathrm{cs}}%
=\sigma_{T}n_{\pm}R,\label{ttau}%
\end{equation}
where $\sigma_{T}=\frac{8\pi}{3}\alpha^{2}(\frac{\hbar}{m_e c})^{2}$ is the
Thomson cross-section. There are two timescales in the problem that
characterize the condition of detailed balance between direct and inverse
reactions, $t_{\mathrm{cs}}$ for binary and $\alpha^{-1}t_{\mathrm{cs}}$ for
triple interactions respectively.

Notice, that electron--positron pair can annihilate into neutrino channel with
the main contribution from the reaction {$e^{\pm}{e^{\mp}\longrightarrow}$%
}${{\nu}{\bar{\nu}}}$. By this process the energy could leak out from the
plasma if it is transparent for neutrinos. The optical depth and energy loss
for this process can be estimated following \cite{1967ApJ...150..979B}\ by
using Fermi theory, see also \cite{1972PhRvD...6..941D,2006PhRvD..74d3006M} for
calculations within electroweak theory.

The optical depth is given by (\ref{ttau}) with the cross-section%
\begin{equation}
\sigma_{{{\nu}{\bar{\nu}}}}\sim\frac{g^{2}}{\pi}\left(  \frac{\hbar}%
{m_e c}\right)  ^{2},
\end{equation}
where $g\simeq10^{-12}$ is the weak interaction coupling constant and it is
assumed that typical energies of electron and positron are $\sim m_e c^{2}$ and
their relative velocities $v\sim c$. Numerically $\sigma_{{{\nu}{\bar{\nu}}}%
}/\sigma_{T}=\frac{3}{8\pi^{2}}\left(  g/\alpha\right)  ^{2}\simeq7\;10^{-22}%
$. For astrophysical sources the plasma may be both transparent and opaque
to neutrino production. The energy loss when pairs are relativistic and
nondegenerate is%
\begin{equation}
\frac{d\rho}{dt}=\frac{128g^{2}}{\pi^{5}}\zeta(5)\zeta(4)\theta^{9}m_e c^{2}\left(
\frac{m_e c}{\hbar}\right)  ^{3}\left(  \frac{m_e c^{2}}{\hbar}\right)  .
\end{equation}

The ratio between the energy lost due to neutrinos and the energy of photons
in thermal equilibrium is then%
\begin{equation}
\frac{1}{\rho_{\gamma}}\frac{d\rho}{dt}\Delta t=\frac{128g^{2}}{\pi^{3}}%
\zeta(5)\zeta(4)\theta^{5}\left(  \frac{m_e c^{2}}{\hbar}\right)  \Delta
t\simeq3.6\;10^{-3}\theta^{5}\frac{\Delta t}{1\sec}.
\end{equation}
For astrophysical sources with the dynamical time $\Delta t\sim10^{-3}$
sec, the energy loss due to neutrinos becomes relevant
\cite{2005MNRAS.364..934K} for high temperatures $\theta>10$. However, on the
timescale of relaxation to thermal equilibrium $\Delta t\sim10^{-12}$ sec the
energy loss is negligible.

Starting from arbitrary distribution functions a common development is found:
at the time $t_{\mathrm{cs}}$ the distribution functions always have evolved
in a functional form on the entire energy range, and depend only on two
parameters. In fact it is found for the distribution functions
\begin{equation}
f_{i}(\varepsilon)=\frac{2}{(2\pi\hbar)^{3}}\exp\left(  -\frac{\varepsilon
-\nu_{i}}{\theta_{i}}\right)  ,\label{dk}%
\end{equation}
with chemical potential $\nu_{i}\equiv\frac{\varphi_{i}}{m_e c^{2}}$
and temperature $\theta_{i}\equiv\frac{k_{B}T_{i}}{m_{e}c^{2}}$,
where $\varepsilon\equiv\frac{\epsilon}{m_{e}c^{2}}$ is the energy
of the particle. Such a configuration corresponds to a kinetic
equilibrium
\cite{1990eaun.book.....K,1997ApJ...486..903P,1973rela.conf....1E}
in which all particles acquire a common temperature and nonzero
chemical potentials. Triple interactions become essential for
$t>t_{\mathrm{cs}}$, after the establishment of kinetic equilibrium.
In strict mathematical sense the sufficient condition for reaching
thermal equilibrium is when all direct reactions are exactly
balanced with their inverse. Therefore, in principle, not only
triple, but also four-particle, five-particle and so on reaction
have to be accounted for in equation (\ref{BE}). The timescale for
reaching thermal equilibrium will be then determined by the slowest
reaction which is not balanced with its inverse. The necessary
condition here is the detailed balance at least in triple
interactions, since binary reactions do not change chemical
potentials at all.

Notice that similar method to ours was applied in \cite{1997ApJ...486..903P}
in order to compute spectra of particles in kinetic equilibrium. However, it
was never shown how particles evolve down to thermal equilibrium.

In the case of pure pair plasma chemical potentials in (\ref{dk}) represent
deviations from the thermal equilibrium through the relation%
\begin{equation}
\nu=\theta\ln(n/n_{th}),
\end{equation}
where $n_{th}$ are concentrations of particles in thermal equilibrium.

\subsection{The discretization procedure and the computational scheme}\label{computsch}

\label{discretization}

In order to solve equations (\ref{BE}) a finite difference method is used by
introducing a computational grid in the phase space to represent the
distribution functions and to compute collisional integrals following
\cite{2004ApJ...609..363A}. The goal is to construct the scheme implementing
energy, baryon number and electric charge conservation laws. For
this reason instead of distribution functions
$f_{i}$, spectral energy densities are used
\begin{equation}
E_{i}(\epsilon_{i})=\frac{4\pi\epsilon_{i}^{3}\beta_{i}f_{i}}{c^{3}%
},\label{Efuncs}%
\end{equation}
where $\beta_{i}=\sqrt{1-(m_{i}c^{2}/\epsilon_{i})^{2}}$, in the phase space
$\epsilon_{i}$. Then
\begin{equation}
\epsilon_{i}f_{i}(\mathbf{p},t)d\mathbf{r}d\mathbf{p}=\frac{4\pi\epsilon
^{3}\beta_{i}f_{i}}{c^{3}}\mathbf{r}d\epsilon_{i}=E_{i}d\mathbf{r}%
d\epsilon_{i}%
\end{equation}
is the energy in the volume of the phase space $d\mathbf{r}d\mathbf{p}$. The
particle density is
\begin{equation}
n_{i}=\int f_{i}d\mathbf{p}=\int\frac{E_{i}}{\epsilon_{i}}d\epsilon_{i},\qquad
dn_{i}=f_{i}d\mathbf{p},
\end{equation}
while the corresponding energy density is%
\[
\rho_{i}=\int\epsilon_{i}f_{i}d\mathbf{p}=\int E_{i}d\epsilon_{i}.
\]
Boltzmann equations (\ref{BE}) can be rewritten in the form
\begin{equation}
\frac{1}{c}\frac{\partial E_{i}}{\partial t}=\sum_{q}(\tilde{\eta}_{i}%
^{q}-\chi_{i}^{q}E_{i}),\label{EBoltzmannEq}%
\end{equation}
where $\tilde{\eta}_{i}^{q}=(4\pi\epsilon_{i}^{3}\beta_{i}/c^{3})\eta_{i}^{q}
$.

The computational grid for phase space is $\{\epsilon_{i},\mu ,\phi\}$, where
$\mu=\cos\vartheta$, $\vartheta$ and $\phi$ are angles between radius vector
$\mathbf{r}$ and the particle momentum $\mathbf{p}$. The zone boundaries are
$\epsilon_{i,\omega\mp1/2}$, $\mu_{k\mp1/2}$, $\phi_{l\mp1/2}$ for
$1\leq\omega\leq\omega_{\mathrm{max}}$, $1\leq k\leq k_{\mathrm{max}}$, $1\leq
l\leq l_{\mathrm{max}}$. The length of the $i$th interval is
$\Delta\epsilon_{i,\omega}\equiv\epsilon_{i,\omega+1/2}-\epsilon
_{i,\omega-1/2}$.\ On the finite grid the functions (\ref{Efuncs}) become
\begin{equation}
E_{i,\omega}\equiv\frac{1}{\Delta\epsilon_{i,\omega}}\int_{\Delta
\epsilon_{i,\omega}}d\epsilon E_{i}(\epsilon).
\end{equation}

Now the collisional integrals in (\ref{EBoltzmannEq}) are replaced by the
corresponding sums.

After this procedure the set of ordinary differential equations (ODE's) is
obtained, instead of the system of partial differential equations for the
quantities $E_{i,\omega}$ to be solved. There are several characteristic times
for different processes in the problem, and therefore the system of
differential equations is stiff. (Eigenvalues of Jacobi matrix differs
significantly, and the real parts of eigenvalues are negative.) Gear's method
\cite{1976oup..book.....H} is used to integrate ODE's numerically. This high
order implicit method was developed for the solution of stiff ODE's.

In this method exact energy conservation law is satisfied. For binary
interactions the particles number conservation law is satisfied as
interpolation of grid functions $E_{i,\omega}$ inside the energy intervals is
adopted.

\subsection{Conservation laws}\label{conslaws}

Conservation laws consist of charge and energy conservations.
In addition, in binary reactions particle number is conserved.

Energy conservation law can be rewritten for the spectral density%
\begin{equation}
\frac{d}{dt}\sum_{i}\rho_{i}=0,\quad\mathrm{or}\quad\frac{d}{dt}\sum
_{i,\omega}Y_{i,\omega}=0,\label{energycons}%
\end{equation}
where%
\begin{equation}
Y_{i,\omega}=\int_{\epsilon_{i,\omega}-\Delta\epsilon_{i,\omega}/2}%
^{\epsilon_{i,\omega}+\Delta\epsilon_{i,\omega}/2}E_{i}d\epsilon.
\end{equation}
Particle's conservation law in binary reactions reduces to%
\begin{equation}
\frac{d}{dt}\sum_{i}n_{i}=0,\quad\mathrm{or}\quad\frac{d}{dt}\sum_{i,\omega
}\frac{Y_{i,\omega}}{\epsilon_{i,\omega}}=0.\label{numbercons}%
\end{equation}
For electrically neutral plasma considered in this Section charge
conservation implies
\begin{equation}
n_{-}=n_{+}.\label{chargecons}%
\end{equation}

\subsection{Determination of temperature and chemical potentials in kinetic
equilibrium}

Consider distribution functions for photons and pairs in the most general form
(\ref{dk}). If one supposes that reaction rate for the Bhabha scattering
vanishes, i.e. there is equilibrium with respect to reaction%
\begin{equation}
e^{+}+e^{-}\leftrightarrow+e^{+}{^{\prime}}+e^{-\prime},
\end{equation}
and the corresponding condition can be written in the following way
\begin{equation}
f_{+}(1-f_{+}{^{\prime}})f_{-}(1-f_{-}{^{\prime}})=f_{+}{^{\prime}}%
(1-f_{+})f_{-}{^{\prime}}(1+f_{-}),
\end{equation}
where Bose-Einstein enhancement along with Pauli blocking factors are taken
into account, it can be shown that electrons and positrons have the same
temperature%
\begin{equation}
\theta_{+}=\theta_{-}\equiv\theta_{\pm},\label{kineticTe}%
\end{equation}
and they have arbitrary chemical potentials.

With (\ref{kineticTe}) analogous consideration for the Compton scattering
\begin{equation}
e^{\pm}+\gamma\leftrightarrow+e^{\pm}{^{\prime}}+\gamma^{\prime},
\end{equation}
gives%
\begin{equation}
f_{\pm}(1-f_{\pm}{^{\prime}})f_{\gamma}(1+f_{\gamma}{^{\prime}})=f_{\pm
}{^{\prime}}(1-f_{\pm})f_{\gamma}{^{\prime}}(1+f_{\gamma}),
\end{equation}
and leads to the same temperature of pairs and photons
\begin{equation}
\theta_{\pm}=\theta_{\gamma}\equiv\theta_{k},\label{kineticT}%
\end{equation}
with arbitrary chemical potentials. If, in addition, reaction rate in the
pair creation and annihilation process%
\begin{equation}
e^{\pm}+e^{\mp}\leftrightarrow\gamma+\gamma^{\prime}%
\end{equation}
vanishes too, i.e. there is equilibrium with respect to pair production and
annihilation, with the corresponding condition,
\begin{equation}
f_{+}f_{-}(1+f_{\gamma})(1+f_{\gamma}{^{\prime}})=f_{\gamma}f_{\gamma
}{^{\prime}}(1-f_{+})(1-f_{-}),
\end{equation}
it turns out that also chemical potentials for pairs and photons satisfy the
following condition for the chemical potentials%
\begin{equation}
\nu_{+}+\nu_{-}=2\nu_{\gamma}.\label{chempotcond}%
\end{equation}
However, since in general $\nu_{\gamma}\neq0$ the condition (\ref{chempotcond}%
) does not imply $\nu_{+}=\nu_{-}$.

In general, the detailed balance conditions for different reactions lead to
relations between temperatures and chemical potentials summarized in table
\ref{tab2}.%

%TCIMACRO{\TeXButton{B}{\begin{table}[tbp] \centering}}%
%BeginExpansion
\begin{table}[tbp] \centering
%EndExpansion%
\begin{tabular}
[c]{|c|c|c|}\hline\hline
& Interaction & Parameters of DFs\\\hline\hline
I & $e^{+}e^{-}$ scattering & $\theta_{+}=\theta_{-}$, $\forall\nu_{+}$%
,$\nu_{-}$\\\hline
II & $e^{\pm}\gamma$ scattering & $\theta_{\gamma}=\theta_{\pm}$, $\forall
\nu_{\gamma}$,$\nu_{\pm}$\\\hline
III & pair production & $\nu_{+}+\nu_{-}=2\nu_{\gamma}$, if $\theta_{\gamma
}=\theta_{\pm}$\\\hline
IV & Tripe interactions & $\nu_{\gamma}$, $\nu_{\pm}=0$, if $\theta_{\gamma
}=\theta_{\pm}$\\\hline\hline
\end{tabular}
\caption{Relations between parameters of equilibrium DFs fulfilling detailed
balance conditions for the reactions shown in Tab.~\ref{tab1}.}\label{tab2}
%TCIMACRO{\TeXButton{E}{\end{table}}}%
%BeginExpansion
\end{table}%
%EndExpansion

Kinetic equilibrium\ is first established simultaneously for electrons,
positrons and photons. Thus they reach the same temperature, but with chemical
potentials different from zero. Later on, protons reach the same temperature.

In order to find temperatures and chemical potentials the
following constraints are implemented:\ energy conservation (\ref{energycons}), particle
number conservation (\ref{numbercons}), charge conservation (\ref{chargecons}%
), condition for the chemical potentials (\ref{chempotcond}).

Given (\ref{dk}) it is found for photons%
\begin{equation}
\frac{\rho_{\gamma}}{n_{\gamma}m_e c^{2}}=3\theta_{\gamma},\quad n_{\gamma}%
=\frac{1}{V_{0}}\exp\left(  \frac{\nu_{\gamma}}{\theta_{\gamma}}\right)
2\theta_{\gamma}^{3},\label{kineqphotons}%
\end{equation}
and for pairs%
\begin{equation}
\frac{\rho_{\pm}}{n_{\pm}m_e c^{2}}=j_{2}(\theta_{\pm}),\quad n_{\pm}=\frac
{1}{V_{0}}\exp\left(  \frac{\nu_{\pm}}{\theta_{\pm}}\right)  j_{1}(\theta
_{\pm}),\label{kineqpairs}%
\end{equation}
where the Compton volume is%
\begin{equation}
V_{0}=\frac{1}{8\pi}\left(  \frac{2\pi\hbar}{m_e c}\right)  ^{3}%
\end{equation}
and functions $j_{1}$ and $j_{2}$\ are defined as%
\begin{align}
j_{1}(\theta)  & =\theta K_{2}(\theta^{-1})\rightarrow\left\{
\begin{array}
[c]{cc}%
\sqrt{\frac{\pi}{2}}e^{-\frac{1}{\theta}}\theta^{3/2}, & \theta\rightarrow0\\
2\theta^{3}, & \theta\rightarrow\infty
\end{array}
\right.  ,\\
j_{2}(\theta)  & =\frac{3K_{3}(\theta^{-1})+K_{1}(\theta^{-1})}{4K_{2}%
(\theta^{-1})}\rightarrow\left\{
\begin{array}
[c]{cc}%
1+\frac{3\theta}{2}, & \theta\rightarrow0\\
3\theta, & \theta\rightarrow\infty
\end{array}
\right.  .
\end{align}

For pure electron--positron-photon plasma in kinetic equilibrium, summing up
energy densities in (\ref{kineqphotons}),(\ref{kineqpairs}) and using
(\ref{kineticTe}),(\ref{kineticT}) and (\ref{chempotcond}) it is found%
\begin{equation}
\sum_{e^{+},e^{-},\gamma}\rho_{i}=\frac{2m_ec^{2}}{V_{0}}\exp\left(  \frac
{\nu_{\mathrm{k}}}{\theta_{\mathrm{k}}}\right)  \left[  3\theta^{4}%
+j_{1}(\theta_{k})j_{2}(\theta_{\mathrm{k}})\right]  ,
\end{equation}
and analogously for number densities%
\begin{equation}
\sum_{e^{+},e^{-},\gamma}n_{i}=\frac{2}{V_{0}}\exp\left(  \frac{\nu
_{\mathrm{k}}}{\theta_{\mathrm{k}}}\right)  \left[  \theta_{\mathrm{k}}%
^{3}+j_{1}(\theta_{\mathrm{k}})\right]  .
\end{equation}
Therefore, two unknowns, $\nu_{k}$ and
$\theta_{\mathrm{k}}$ can be found.

In thermal equilibrium $\nu_{\gamma}$ vanishes and one has%
\begin{equation}
\nu_{+}=\nu_{-}=0.%
\label{thermalnu}%
\end{equation}

\subsection{Binary interactions}\label{binaryint}

In this section the expressions for emission and absorption coefficients in
Compton scattering, pair creation and annihilation with two photons, M{\o}ller
and Bhabha scattering are obtained.

\subsubsection{Compton scattering}

\label{compton}

The time evolution of the distribution functions of photons and pair particles
due to Compton scattering may be described by \cite{1981els..book.....L}%
,\cite{1979MAt..book.....O}%
\begin{gather}
\left(  \frac{\partial f_{\gamma}(\mathbf{k},t)}{\partial t}\right)  _{\gamma
e^{\pm}\rightarrow\gamma^{\prime}e^{\pm\prime}}=\int d\mathbf{k}^{\prime
}d\mathbf{p}d\mathbf{p}^{\prime}Vw_{\mathbf{k}^{\prime},\mathbf{p}^{\prime
};\mathbf{k},\mathbf{p}}\times\nonumber\\ \times[f_{\gamma}({\mathbf{k}^{\prime},t)}f_{\pm
}({\mathbf{p}^{\prime}},t)-f_{\gamma}(\mathbf{k},t)f_{\pm}(\mathbf{p}%
,t)],\label{Comptongamma}%
\end{gather}%
\begin{gather}
\left(  \frac{\partial f_{\pm}(\mathbf{p},t)}{\partial t}\right)  _{\gamma
e^{\pm}\rightarrow\gamma^{\prime}e^{\pm\prime}}=\int d\mathbf{k}%
d\mathbf{k}^{\prime}d\mathbf{p}^{\prime}Vw_{\mathbf{k}^{\prime},\mathbf{p}%
^{\prime};\mathbf{k},\mathbf{p}}\times\nonumber\\ \times[f_{\gamma}({\mathbf{k}^{\prime}},t)f_{\pm
}({\mathbf{p}^{\prime}},t)-f_{\gamma}(\mathbf{k},t)f_{\pm}(\mathbf{p}%
,t)],\label{Comptone}%
\end{gather}
where
\begin{equation}
w_{\mathbf{k}^{\prime},\mathbf{p}^{\prime};\mathbf{k},\mathbf{p}}=\frac
{\hbar^{2}c^{6}}{(2\pi)^{2}V}\delta(\epsilon_{\gamma}-\epsilon_{\pm}%
-\epsilon_{\gamma}^{\prime}-\epsilon_{\pm}^{\prime})\delta(\mathbf{k}%
+\mathbf{p}-\mathbf{k}^{\prime}-\mathbf{p}^{\prime})\frac{|M_{fi}|^{2}%
}{16\epsilon_{\gamma}\epsilon_{\pm}\epsilon_{\gamma}^{\prime}\epsilon_{\pm
}^{\prime}},\label{df-abs}%
\end{equation}
is the probability of the process,
\begin{align}
|M_{fi}|^{2}  & =2^{6}\pi^{2}\alpha^{2}\left[  \frac{m_e^{2}c^{2}}{s-m_e^{2}c^{2}%
}+\frac{m_e^{2}c^{2}}{u-m_e^{2}c^{2}}+\left(  \frac{m_e^{2}c^{2}}{s-m_e^{2}c^{2}%
}+\frac{m_e^{2}c^{2}}{u-m_e^{2}c^{2}}\right)  ^{2}\right. \nonumber\\
& \left.  -\frac{1}{4}\left(  \frac{s-m_e^{2}c^{2}}{u-m_e^{2}c^{2}}+\frac
{u-m_e^{2}c^{2}}{s-m_e^{2}c^{2}}\right)  \right]  ,\label{M_fi_gamma1}%
\end{align}
is the square of the matrix element, $s=(\mathfrak{p}+\mathfrak{k})^{2}$ and
$u=(\mathfrak{p}-\mathfrak{k}^{\prime})^{2}$ are invariants, $\mathfrak{k}%
=(\epsilon_{\gamma}/c)(1,\mathbf{e}_{\gamma})$ and $\mathfrak{p}%
=(\epsilon_{\pm}/c)(1,\beta_{\pm}\mathbf{e}_{\pm})$ are energy-momentum four
vectors of photons and electrons, respectively, $d\mathbf{p}=d\epsilon_{\pm
}do\epsilon_{\pm}^{2}\beta_{\pm}/c^{3}$, $d\mathbf{k}^{\prime}=d\epsilon
_{\gamma}^{\prime}\epsilon_{\gamma}^{\prime2}do_{\gamma}^{\prime}/c^{3}$ and
$do=d\mu d\phi$.

The energies of photon and positron (electron) after the scattering are%
\begin{equation}
\epsilon_{\gamma}^{\prime}=\frac{\epsilon_{\pm}\epsilon_{\gamma}(1-\beta_{\pm
}\mathbf{b}_{\pm}\mathbf{\cdot}\mathbf{b}_{\gamma})}{\epsilon_{\pm}%
(1-\beta_{\pm}\mathbf{b}_{\pm}\mathbf{\cdot}\mathbf{b}_{\gamma}^{\prime
})+\epsilon_{\gamma}(1-\mathbf{b}_{\gamma}\mathbf{\cdot}\mathbf{b}_{\gamma
}^{\prime})}\,,\qquad\,\epsilon_{\pm}^{\prime}=\epsilon_{\pm}+\epsilon
_{\gamma}-\epsilon_{\gamma}^{\prime}\,,
\end{equation}
$\mathbf{b}_{i}=\mathbf{p}_{i}/p$, $\mathbf{b}_{i}^{\prime}=\mathbf{p}%
_{i}^{\prime}/p^{\prime}$, $\mathbf{b}_{\pm}^{\prime}=(\beta_{\pm}%
\epsilon_{\pm}\mathbf{b}_{\pm}+\epsilon_{\gamma}\mathbf{b}_{\gamma}%
-\epsilon_{\gamma}^{\prime}\mathbf{b}_{\gamma}^{\prime})/(\beta_{\pm}^{\prime
}\epsilon_{\pm}^{\prime})$.

For photons, the absorption coefficient (\ref{comptonabscoef}) in the
Boltzmann equations (\ref{BE}) is
\begin{equation}
\chi_{\gamma}^{\gamma e^{\pm}\rightarrow\gamma^{\prime}e^{\pm\prime}}%
f_{\gamma}=-\frac{1}{c}\left(  \frac{\partial f_{\gamma}}{\partial t}\right)
_{\gamma e^{\pm}\rightarrow\gamma^{\prime}e^{\pm\prime}}^{\mathrm{abs}}=\int
dn_{\pm}do_{\gamma}^{\prime}J_{\mathrm{cs}}\frac{\epsilon_{\gamma}^{\prime
}|M_{fi}|^{2}\hbar^{2}c^{2}}{16\epsilon_{\pm}\epsilon_{\gamma}\epsilon_{\pm
}^{\prime}}f_{\gamma},\label{chi-fgamma}%
\end{equation}
where $dn_{i}=d\epsilon_{i}do_{i}\epsilon_{i}^{2}\beta_{i}f_{i}/c^{3}%
=d\epsilon_{i}do_{i}E_{i}/(2\pi\epsilon_{i})$.

From equations (\ref{Comptongamma}) and (\ref{chi-fgamma}), the
absorption coefficient for photon energy density $E_{\gamma}$ averaged over
the $\epsilon,\mu$-grid with zone numbers $\omega$ and $k$ is
\begin{align}
(\chi E)_{\gamma,\omega}^{\gamma e^{\pm}\rightarrow\gamma^{\prime}e^{\pm
\prime}}  & \equiv\frac{1}{\Delta\epsilon_{\gamma,\omega}}\int_{\epsilon
_{\gamma}\in\Delta\epsilon_{\gamma,\omega}}d\epsilon_{\gamma}d\mu_{\gamma
}(\chi E)_{\gamma}^{\gamma e^{\pm}\rightarrow\gamma^{\prime}e^{\pm\prime}%
}=\nonumber\\
& =\frac{1}{\Delta\epsilon_{\gamma,\omega}}\int_{\epsilon_{\gamma}\in
\Delta\epsilon_{\gamma,\omega}}dn_{\gamma}dn_{\pm}do_{\gamma}^{\prime
}J_{\mathrm{cs}}\frac{\epsilon_{\gamma}^{\prime}|M_{fi}|^{2}\hbar^{2}c^{2}%
}{16\epsilon_{\pm}\epsilon_{\pm}^{\prime}},\label{chiE1}%
\end{align}
where the Jacobian of the transformation is%
\begin{equation}
J_{\mathrm{cs}}=\frac{\epsilon_{\gamma}^{\prime}\epsilon_{\pm}^{\prime}%
}{\epsilon_{\gamma}\epsilon_{\pm}\left(  1-\beta_{\pm}\mathbf{b}_{\gamma
}\mathbf{\cdot b}_{\pm}\right)  }.
\end{equation}
Similar integrations can be performed for the other terms of equations
(\ref{Comptongamma}), (\ref{Comptone}), and%
\begin{align}
\eta_{\gamma,\omega}^{\gamma e^{\pm}\rightarrow\gamma^{\prime}e^{\pm\prime}}
& =\frac{1}{\Delta\epsilon_{\gamma,\omega}}\int_{\epsilon_{\gamma}^{\prime}%
\in\Delta\epsilon_{\gamma,\omega}}dn_{\gamma}dn_{\pm}do_{\gamma}^{\prime
}J_{\mathrm{cs}}\frac{\epsilon_{\gamma}^{\prime2}|M_{fi}|^{2}\hbar^{2}c^{2}%
}{16\epsilon_{\pm}\epsilon_{\gamma}\epsilon_{\pm}^{\prime}},\label{chiE2}\\
\eta_{\pm,\omega}^{\gamma e^{\pm}\rightarrow\gamma^{\prime}e^{\pm\prime}}  &
=\frac{1}{\Delta\epsilon_{\pm,\omega}}\int_{\epsilon_{\pm}^{\prime}\in
\Delta\epsilon_{\pm,\omega}}dn_{\gamma}dn_{\pm}do_{\gamma}^{\prime
}J_{\mathrm{cs}}\frac{\epsilon_{\gamma}^{\prime}|M_{fi}|^{2}\hbar^{2}c^{2}%
}{16\epsilon_{\pm}\epsilon_{\gamma}},\label{chiE3}\\
(\chi E)_{\pm,\omega}^{\gamma e^{\pm}\rightarrow\gamma^{\prime}e^{\pm\prime}}
& =\frac{1}{\Delta\epsilon_{\pm,\omega}}\int_{\epsilon_{\pm}\in\Delta
\epsilon_{\pm,\omega}}dn_{\gamma}dn_{\pm}do_{\gamma}^{\prime}J_{\mathrm{cs}%
}\frac{\epsilon_{\gamma}^{\prime}|M_{fi}|^{2}\hbar^{2}c^{2}}{16\epsilon
_{\gamma}\epsilon_{\pm}^{\prime}}.\label{chiE4}%
\end{align}

In order to perform integrals (\ref{chiE1})-(\ref{chiE4}) numerically over
$\phi$ ($0\leq\phi\leq2\pi$) a uniform grid $\phi_{l\mp1/2}$ is introduced with
$1\leq l\leq l_{\mathrm{max}}$ and $\Delta\phi_{l}=(\phi_{l+1/2}-\phi
_{l-1/2})/2=2\pi/l_{\mathrm{max}}$. It is assumed that any function of $\phi$
in equations (\ref{chiE1})-(\ref{chiE4}) in the interval $\Delta\phi_{j}$ is
equal to its value at $\phi=\phi_{j}=(\phi_{l-1/2}+\phi_{l+1/2})/2$. It is
necessary to integrate over $\phi$ only once at the beginning of calculations.
The number of intervals of the $\phi$-grid depends on the average energy of
particles and is typically taken as $l_{\mathrm{max}}=2k_{\mathrm{max}}=64$.

\subsubsection{Pair creation and annihilation}
\label{pair}

The rates of change of the distribution function due to pair creation and
annihilation are
\begin{equation}
\left(  \frac{\partial f_{\gamma_{j}}(\mathbf{k}_{i},t)}{\partial t}\right)
_{\gamma_{1}\gamma_{2}\rightarrow e^{-}e^{+}}=-\int d\mathbf{k}_{j}%
d\mathbf{p}_{-}d\mathbf{p}_{+}Vw_{\mathbf{p}_{-},\mathbf{p}_{+};\mathbf{k}%
_{1},\mathbf{k}_{2}}f_{\gamma_{1}}(\mathbf{k}_{1},t)f_{\gamma_{2}}%
(\mathbf{k}_{2},t)\,,\label{fgamma1}
\end{equation}
\begin{equation}
\left(  \frac{\partial f_{\gamma_{i}}(\mathbf{k}_{i},t)}{\partial t}\right)
_{e^{-}e^{+}\rightarrow\gamma_{1}\gamma_{2}}=\int d\mathbf{k}_{j}%
d\mathbf{p}_{-}d\mathbf{p}_{+}Vw_{\mathbf{k}_{1},\mathbf{k}_{2};\mathbf{p}%
_{-},\mathbf{p}_{+}}f_{-}(\mathbf{p}_{-},t)f_{+}(\mathbf{p}_{+},t)\,,
\end{equation}
for $i=1,~j=2$, and for $j=1,~i=2$.
\begin{equation}
\left(  \frac{\partial f_{\pm}(\mathbf{p}_{\pm},t)}{\partial t}\right)
_{\gamma_{1}\gamma_{2}\rightarrow e^{-}e^{+}}=\int d\mathbf{p}_{\mp
}d\mathbf{k}_{1}d\mathbf{k}_{2}Vw_{\mathbf{p}_{-},\mathbf{p}_{+}%
;\mathbf{k}_{1},\mathbf{k}_{2}}f_{\gamma}(\mathbf{k}_{1},t)f_{\gamma
}(\mathbf{k}_{2},t)\,,
\end{equation}
\begin{equation}
\left(  \frac{\partial f_{\pm}(\mathbf{p}_{\pm},t)}{\partial t}\right)
_{e^{-}e^{+}\rightarrow\gamma_{1}\gamma_{2}}=-\int d\mathbf{p}_{\mp
}d\mathbf{k}_{1}d\mathbf{k}_{2}Vw_{\mathbf{k}_{1},\mathbf{k}_{2}%
;\mathbf{p}_{-},\mathbf{p}_{+}}f_{-}(\mathbf{p}_{-},t)f_{+}(\mathbf{p}%
_{+},t)\,,\label{fe+}
\end{equation}
where
\begin{equation}
w_{\mathbf{p}_{-},\mathbf{p}_{+};\mathbf{k}_{1},\mathbf{k}_{2}}=\frac
{\hbar^{2}c^{6}}{(2\pi)^{2}V}\delta(\epsilon_{-}+\epsilon_{+}-\epsilon
_{1}-\epsilon_{2})\delta(\mathbf{p}_{-}+\mathbf{p}_{+}-\mathbf{k}%
_{1}-\mathbf{k}_{2})\frac{|M_{fi}|^{2}}{16\epsilon_{-}\epsilon_{+}\epsilon
_{1}\epsilon_{2}}.
\end{equation}
Here, the matrix element $|M_{fi}|^{2}$ is given by equation
(\ref{M_fi_gamma1}) with the new invariants $s=(\mathfrak{p}_{-}%
-\mathfrak{k}_{1})^{2}$ and $u=(\mathfrak{p}_{-}-\mathfrak{k}_{2})^{2}$, see
\cite{1982els..book.....B}.

The energies of photons created via annihilation of a $e^{\pm}$ pair are
\begin{equation}
\epsilon_{1}(\mathbf{b}_{1})=\frac{m^{2}c^{4}+\epsilon_{-}\epsilon_{+}%
(1-\beta_{-}\beta_{+}\mathbf{b}_{-}\mathbf{\cdot}\mathbf{b}_{+})}{\epsilon
_{-}(1-\beta_{-}\mathbf{b}_{-}\mathbf{\cdot}\mathbf{b}_{1})+\epsilon
_{+}(1-\beta_{+}\mathbf{b}_{+}\mathbf{\cdot}\mathbf{b}_{1})}\,,\qquad
\epsilon_{2}(\mathbf{b}_{1})=\epsilon_{-}+\epsilon_{+}-\epsilon_{1}\,,
\end{equation}
while the energies of pair particles created by two photons are found from%
\begin{equation}
\epsilon_{-}(\mathbf{b}_{-})=\frac{B\mp\sqrt{B^{2}-AC}}{A}\,,\qquad
\epsilon_{+}(\mathbf{b}_{-})=\epsilon_{1}+\epsilon_{2}-\epsilon_{-}%
\,,\label{A27}%
\end{equation}
where $A=(\epsilon_{1}+\epsilon_{2})^{2}-[(\epsilon_{1}\mathbf{b}_{1}%
+\epsilon_{2}\mathbf{b}_{2})\mathbf{\cdot}\mathbf{b}_{-}]^{2}$, $B=(\epsilon
_{1}+\epsilon_{2})\epsilon_{1}\epsilon_{2}(1-\mathbf{b}_{1}\mathbf{\cdot
}\mathbf{b}_{2})$, $C=m_{e}^{2}c^{4}[(\epsilon_{1}\mathbf{b}_{1}+\epsilon
_{2}\mathbf{b}_{2})\mathbf{\cdot}\mathbf{b}_{-}]^{2}+\epsilon_{1}^{2}%
\epsilon_{2}^{2}(1-\mathbf{b}_{1}\mathbf{\cdot}\mathbf{b}_{2})^{2}$. Only one
root in equation (\ref{A27}) has to be chosen. From energy-momentum
conservation
\begin{equation}
\mathfrak{k}_{1}+\mathfrak{k}_{2}-\mathfrak{p}_{-}=\mathfrak{p}_{+},
\end{equation}
taking square from the energy part leads to%
\begin{equation}
\epsilon_{1}^{2}+\epsilon_{2}^{2}+\epsilon_{-}^{2}+2\epsilon_{1}\epsilon
_{2}-2\epsilon_{1}\epsilon_{-}-2\epsilon_{2}\epsilon_{-}=\epsilon_{+}^{2},
\end{equation}
and taking square from the momentum part%
\begin{equation}
\epsilon_{1}^{2}+\epsilon_{2}^{2}+\epsilon_{-}^{2}\beta_{-}^{2}+2\epsilon
_{1}\epsilon_{2}\mathbf{b}_{1}\mathbf{\cdot b}_{2}-2\epsilon_{1}\epsilon
_{-}\beta_{-}\mathbf{b}_{1}\mathbf{\cdot b}_{-}-2\epsilon_{2}\epsilon_{-}%
\beta_{-}\mathbf{b}_{2}\mathbf{\cdot b}_{-}=(\epsilon_{+}\beta_{+})^{2}.
\end{equation}
There are no additional roots because of the arbitrary $\mathbf{e}_{+}$%
\begin{gather}
\epsilon_{1}\epsilon_{2}(1-\mathbf{b}_{1}\mathbf{\cdot b}_{2})-\epsilon
_{1}\epsilon_{-}(1-\beta_{-}\mathbf{b}_{1}\mathbf{\cdot b}_{-})-\epsilon
_{2}\epsilon_{-}(1-\beta\mathbf{b}_{2}\mathbf{\cdot b}_{-})=0,\\
\epsilon_{-}\beta_{-}(\epsilon_{1}\mathbf{b}_{1}+\epsilon_{2}\mathbf{b}%
_{2})\mathbf{\cdot b}_{-}=\epsilon_{-}(\epsilon_{1}+\epsilon_{2})-\epsilon
_{1}\epsilon_{2}(1-\mathbf{b}_{1}\mathbf{\cdot b}_{2}).\nonumber
\end{gather}
Eliminating $\beta$ it is obtained%
\begin{align}
&  \epsilon_{1}^{2}\epsilon_{2}^{2}(1-\mathbf{b}_{1}\mathbf{\cdot b}_{2}%
)^{2}-2\epsilon_{1}\epsilon_{2}(1-\mathbf{b}_{1}\mathbf{\cdot b}_{2}%
)(\epsilon_{1}+\epsilon_{2})\epsilon_{-}+\nonumber\\& +\left\{  (\epsilon_{1}+\epsilon
_{2})^{2}-\left[  (\epsilon_{1}\mathbf{b}_{1}+\epsilon_{2}\mathbf{b}%
_{2})\mathbf{\cdot b}_{-}\right]  ^{2}\right\}  \epsilon_{-}^{2}=\nonumber\\
&  =\left[  (\epsilon_{1}\mathbf{b}_{1}+\epsilon_{2}\mathbf{b}_{2}%
)\mathbf{\cdot b}_{-}\right]  (-m^{2}),
\end{align}
Therefore, the condition to be checked reads%
\begin{align}
\epsilon_{-}\beta_{-}\left[  (\epsilon_{1}\mathbf{b}_{1}+\epsilon
_{2}\mathbf{b}_{2})\mathbf{\cdot b}_{-}\right]  ^{2}&=\left[  \epsilon
_{-}(\epsilon_{1}+\epsilon_{2})-(\epsilon_{1}\epsilon_{2})(1-\mathbf{b}%
_{1}\mathbf{\cdot b}_{2})\right] \times\nonumber\\ \times \left[  (\epsilon_{1}\mathbf{b}_{1}%
+\epsilon_{2}\mathbf{b}_{2})\mathbf{\cdot b}_{-}\right]  \geq0.
\end{align}

Finally, integration of equations (\ref{fgamma1})-(\ref{fe+}) yields%
\begin{align}
\eta_{\gamma,\omega}^{e^{-}e^{+}\rightarrow\gamma_{1}\gamma_{2}}  & =\frac
{1}{\Delta\epsilon_{\gamma,\omega}}\left(  \int_{\epsilon_{1}\in\Delta
\epsilon_{\gamma,\omega}}d^{2}n_{\pm}J_{\mathrm{ca}}\frac{\epsilon_{1}%
^{2}|M_{fi}|^{2}\hbar^{2}c^{2}}{16\epsilon_{-}\epsilon_{+}\epsilon_{2}}\right)+\nonumber\\
&+\frac
{1}{\Delta\epsilon_{\gamma,\omega}}\left(\int_{\epsilon_{2}\in\Delta\epsilon_{\gamma,\omega}}d^{2}n_{\pm
}J_{\mathrm{ca}}\frac{\epsilon_{1}|M_{fi}|^{2}\hbar^{2}c^{2}}{16\epsilon
_{-}\epsilon_{+}}\right)  ,\label{A28}\\
(\chi E)_{e,\omega}^{e^{-}e^{+}\rightarrow\gamma_{1}\gamma_{2}}  & =\frac
{1}{\Delta\epsilon_{e,\omega}}\left(  \int_{\epsilon_{-}\in\Delta
\epsilon_{e,\omega}}d^{2}n_{\pm}J_{\mathrm{ca}}\frac{\epsilon_{1}|M_{fi}%
|^{2}\hbar^{2}c^{2}}{16\epsilon_{+}\epsilon_{2}}\right)+\nonumber\\&+\frac
{1}{\Delta\epsilon_{e,\omega}}\left(\int_{\epsilon_{+}\in
\Delta\epsilon_{e,\omega}}d^{2}n_{\pm}J_{\mathrm{ca}}\frac{\epsilon_{1}%
|M_{fi}|^{2}\hbar^{2}c^{2}}{16\epsilon_{-}\epsilon_{2}}\right)  ,\label{A29}\\
(\chi E)_{\gamma,\omega}^{\gamma_{1}\gamma_{2}\rightarrow e^{-}e^{+}}  &
=\frac{1}{\Delta\epsilon_{\gamma,\omega}}\left(  \int_{\epsilon_{1}\in
\Delta\epsilon_{\gamma,\omega}}d^{2}n_{\gamma}J_{\mathrm{ca}}\frac
{\epsilon_{-}\beta_{-}|M_{fi}|^{2}\hbar^{2}c^{2}}{16\epsilon_{2}\epsilon_{+}%
}\right)+\nonumber\\&+\frac{1}{\Delta\epsilon_{\gamma,\omega}}\left(\int_{\epsilon_{2}\in\Delta\epsilon_{\gamma,\omega}}d^{2}n_{\gamma
}J_{\mathrm{ca}}\frac{\epsilon_{-}\beta_{-}|M_{fi}|^{2}\hbar^{2}c^{2}%
}{16\epsilon_{1}\epsilon_{+}}\right)  ,\label{A30}\\
\eta_{e,\omega}^{\gamma_{1}\gamma_{2}\rightarrow e^{-}e^{+}}  & =\frac
{1}{\Delta\epsilon_{e,\omega}}\left(  \int_{\epsilon_{-}\in\Delta
\epsilon_{e,\omega}}d^{2}n_{\gamma}J_{\mathrm{ca}}\frac{\epsilon_{-}^{2}%
\beta_{-}|M_{fi}|^{2}\hbar^{2}c^{2}}{16\epsilon_{1}\epsilon_{2}\epsilon_{+}%
}\right)+\nonumber\\&+\frac
{1}{\Delta\epsilon_{e,\omega}}\left(\int_{\epsilon_{+}\in\Delta\epsilon_{e,\omega}}d^{2}n_{\gamma}%
J_{\mathrm{ca}}\frac{\epsilon_{-}\beta_{-}|M_{fi}|^{2}\hbar^{2}c^{2}%
}{16\epsilon_{1}\epsilon_{2}}\right)  ,\label{A31}%
\end{align}
where $d^{2}n_{\pm}=dn_{-}dn_{+}do_{1},d^{2}n_{\gamma}=dn_{\gamma_{1}%
}dn_{\gamma_{2}}do_{-},$ $dn_{\pm}=d\epsilon_{\pm}do_{\pm}\epsilon_{\pm}%
^{2}\beta_{\pm}f_{\pm}$, $dn_{\gamma_{1,2}}=d\epsilon_{1,2}do_{1,2}%
\epsilon_{1,2}^{2}f_{\gamma_{1,2}}$ and the Jacobian is%
\begin{equation}
J_{\mathrm{ca}}=\frac{\epsilon_{+}\beta_{-}}{\left(  \epsilon_{+}+\epsilon
_{-}\right)  \beta_{-}-\left(  \epsilon_{1}\mathbf{b}_{1}+\epsilon
_{2}\mathbf{b}_{2}\right)  \mathbf{\cdot b}_{-}}.
\end{equation}

\subsubsection{M{\o }ller scattering of electrons and positrons}

\label{moller}

The time evolution of the distribution functions of electrons (or positrons)
is described by%
\begin{gather}
\left(  \frac{\partial f_{i}(\mathbf{p}_{i},t)}{\partial t}\right)
_{e_{1}e_{2}\rightarrow e_{1}^{\prime}e_{2}^{\prime}}=\int d\mathbf{p}%
_{j}d\mathbf{p}_{1}^{\prime}d\mathbf{p}_{2}^{\prime}Vw_{\mathbf{p}_{1}%
^{\prime},\mathbf{p}_{2}^{\prime};\mathbf{p}_{1},\mathbf{p}_{2}}%
\times\nonumber\\ \times[f_{1}(\mathbf{p}_{1}^{\prime},t)f_{2}(\mathbf{p}_{2}^{\prime},t)-f_{1}%
(\mathbf{p}_{1},t)f_{2}(\mathbf{p}_{2},t)]\,,\label{A32}%
\end{gather}
with $i=1,~j=2$, and with $j=1,~i=2$, and where%
\begin{align}
w_{\mathbf{p}_{1}^{\prime},\mathbf{p}_{2}^{\prime};\mathbf{p}_{1}%
,\mathbf{p}_{2}} &  =\frac{\hbar^{2}c^{6}}{(2\pi)^{2}V}\delta(\epsilon
_{1}+\epsilon_{2}-\epsilon_{1}^{\prime}-\epsilon_{2}^{\prime})\delta
(\mathbf{p}_{1}+\mathbf{p}_{2}-\mathbf{p}_{1}^{\prime}-\mathbf{p}_{2}^{\prime
})\frac{|M_{fi}|^{2}}{16\epsilon_{1}\epsilon_{2}\epsilon_{1}^{\prime}%
\epsilon_{2}^{\prime}},\\
|M_{fi}|^{2} &  =2^{6}\pi^{2}\alpha^{2}\left\{  \frac{1}{t^{2}}\left[
\frac{s^{2}+u^{2}}{2}+4m_e^{2}c^{2}(t - m_e^{2}c^{2})\right]  +\right. \\
&  +\frac{1}{u^{2}}\left[  \frac{s^{2}+t^{2}}{2}+4m_e^{2}c^{2}%
(u-m_e^{2}c^{2})\right]  + \\ & + \left.\frac{4}{tu}\left(  \frac{s}{2}-m_e^{2}c^{2}\right)
\left(  \frac{s}{2}-3m_e^{2}c^{2}\right)  \right\}  ,\label{M_fi_e}%
\end{align}
with $s=(\mathfrak{p}_{1}+\mathfrak{p}_{2})^{2}=2(m_e^{2}c^{2}+\mathfrak{p}%
_{1}\mathfrak{p}_{2})$, $t=(\mathfrak{p}_{1}-\mathfrak{p}_{1}^{\prime}%
)^{2}=2(m_e^{2}c^{2}-\mathfrak{p}_{1}\mathfrak{p}_{1}^{\prime})$, and
$u=(\mathfrak{p}_{1}-\mathfrak{p}_{2}^{\prime})^{2}=2(m_e^{2}c^{2}%
-\mathfrak{p}_{1}\mathfrak{p}_{2}^{\prime})$ \cite{1982els..book.....B}.

The energies of final state particles are given by (\ref{A27}) with new
coefficients $\tilde{A}=(\epsilon_{1}+\epsilon_{2})^{2}-(\epsilon_{1}\beta
_{1}\mathbf{b}_{1}\mathbf{\cdot b}_{1}^{\prime}+\epsilon_{2}\beta
_{2}\mathbf{b}_{2}\mathbf{\cdot b}_{1}^{\prime})^{2}$, $\tilde{B}%
=(\epsilon_{1}+\epsilon_{2})[m_e^{2}c^{4}+\epsilon_{1}\epsilon_{2}(1-\beta
_{1}\beta_{2}\mathbf{b}_{1}\mathbf{b}_{2})]$, and $\tilde{C}=m_e^{2}%
c^{4}(\epsilon_{1}\beta_{1}\mathbf{b}_{1}\mathbf{\cdot b}_{1}^{\prime
}+\epsilon_{2}\beta_{2}\mathbf{b}_{2}\mathbf{\cdot b}_{1}^{\prime})^{2}%
+[m_e^{2}c^{4}+\epsilon_{1}\epsilon_{2}(1-\beta_{1}\beta_{2}\mathbf{b}%
_{1}\mathbf{\cdot b}_{2})]^{2}$. The condition to be checked is%
\begin{equation}
\left[  \epsilon_{1}^{\prime}(\epsilon_{1}+\epsilon_{2})-m_e^{2}c^{4}%
-(\epsilon_{1}\epsilon_{2})(1-\beta_{1}\beta_{2}\mathbf{b}_{1}\mathbf{\cdot
b}_{2})\right]  \left[  (\epsilon_{1}\beta_{1}\mathbf{b}_{1}+\epsilon_{2}%
\beta_{2}\mathbf{b}_{2})\mathbf{\cdot b}_{1}^{\prime}\right]  \geq
0.\label{A34}%
\end{equation}

Integration of equations (\ref{A32}), similar to the case of Compton
scattering in Section \ref{compton} yields
\begin{align}
\eta_{e,\omega}^{e_{1}e_{2}\rightarrow e_{1}^{\prime}e_{2}^{\prime}}  &
=\frac{1}{\Delta\epsilon_{e,\omega}}\left(  \int_{\epsilon_{1}^{\prime}%
\in\Delta\epsilon_{e,\omega}}d^{2}nJ_{\mathrm{ms}}\frac{\epsilon_{1}^{\prime
2}\beta_{1}^{\prime}|M_{fi}|^{2}\hbar^{2}c^{2}}{16\epsilon_{1}\epsilon
_{2}\epsilon_{2}^{\prime}}\right)+\nonumber\\ &+ \frac{1}{\Delta\epsilon_{e,\omega}}\left(\int_{\epsilon_{2}^{\prime}\in\Delta\epsilon
_{e,\omega}}d^{2}nJ_{\mathrm{ms}}\frac{\epsilon_{1}^{\prime}\beta_{1}^{\prime
}|M_{fi}|^{2}\hbar^{2}c^{2}}{16\epsilon_{1}\epsilon_{2}}\right)  ,\\
(\chi E)_{e,\omega}^{e_{1}e_{2}\rightarrow e_{1}^{\prime}e_{2}^{\prime}}  &
=\frac{1}{\Delta\epsilon_{e,\omega}}\left(  \int_{\epsilon_{1}\in
\Delta\epsilon_{e,\omega}}d^{2}nJ_{\mathrm{ms}}\frac{\epsilon_{1}^{\prime
}\beta_{1}^{\prime}|M_{fi}|^{2}\hbar^{2}c^{2}}{16\epsilon_{2}\epsilon
_{2}^{\prime}}\right)+\nonumber\\ & +\frac{1}{\Delta\epsilon_{e,\omega}}\left(\int_{\epsilon_{2}\in\Delta\epsilon_{e,\omega}}d^{2}%
nJ_{\mathrm{ms}}\frac{\epsilon_{1}^{\prime}\beta_{1}^{\prime}|M_{fi}|^{2}%
\hbar^{2}c^{2}}{16\epsilon_{1}\epsilon_{2}^{\prime}}\right)  ,
\end{align}
where $d^{2}n=dn_{1}dn_{2}do_{1}^{\prime}$, $dn_{1,2}=d\epsilon_{1,2}%
do_{1,2}\epsilon_{1,2}^{2}\beta_{1,2}f_{_{1,2}}$, and the Jacobian is%
\begin{equation}
J_{\mathrm{ms}}=\frac{\epsilon_{2}^{\prime}\beta_{2}^{\prime}}{(\epsilon
_{1}^{\prime}+\epsilon_{2}^{\prime})\beta_{1}^{\prime}-(\epsilon_{1}\beta
_{1}\mathbf{b}_{1}+\epsilon_{2}\beta_{2}\mathbf{b}_{2})\mathbf{\cdot b}%
_{1}^{\prime}}.
\end{equation}

\subsubsection{Bhaba scattering of electrons on positrons}

The time evolution of the distribution functions of electrons and positrons
due to Bhaba scattering is described by%

\begin{align}
\left(  \frac{\partial f_{\pm}(\mathbf{p}_{\pm},t)}{\partial t}\right)
_{e^{-}e^{+}\rightarrow e^{-\prime}e^{+\prime}} = \int d\mathbf{p}_{\mp
}d\mathbf{p}_{-}^{\prime}d\mathbf{p}_{+}^{\prime}Vw_{\mathbf{p}_{-}^{\prime
},\mathbf{p}_{+}^{\prime};\mathbf{p}_{-},\mathbf{p}_{+}}\times\nonumber \\ \times [f_{-}(\mathbf{p}%
_{-}^{\prime},t)f_{+}(\mathbf{p}_{+}^{\prime},t)-f_{-}(\mathbf{p}_{-}%
,t)f_{+}(\mathbf{p}_{+},t)],\label{A38}%
\end{align}
where
\begin{equation}
w_{\mathbf{p}_{-}^{\prime},\mathbf{p}_{+}^{\prime};\mathbf{p}_{-}%
,\mathbf{p}_{+}}=\frac{\hbar^{2}c^{6}}{(2\pi)^{2}V}\delta(\epsilon
_{-}+\epsilon_{+}-\epsilon_{-}^{\prime}-\epsilon_{+}^{\prime})\delta
(\mathbf{p}_{-}+\mathbf{p}_{+}-\mathbf{p}_{-}^{\prime}-\mathbf{p}_{+}^{\prime
})\frac{|M_{fi}|^{2}}{16\epsilon_{-}\epsilon_{+}\epsilon_{-}^{\prime}%
\epsilon_{+}^{\prime}},
\end{equation}
and $|M_{fi}|$ is given by the equation (\ref{M_fi_e}), but the invariants are
$s=(\mathfrak{p}_{-}-\mathfrak{p}_{+}^{\prime})^{2}$, $t=(\mathfrak{p}%
_{+}-\mathfrak{p}_{+}^{\prime})^{2}$ and $u=(\mathfrak{p}_{-}+\mathfrak{p}%
_{+})^{2}$. The final energies $\epsilon_{-}^{\prime}$, $\epsilon_{+}^{\prime
}$ are functions of the outgoing particle directions in a way similar to that
in Section \ref{moller}, see also \cite{1982els..book.....B}.

Integration of equations (\ref{A38}) yields
\begin{align}
\eta_{\pm,\omega}^{e^{-}e^{+}\rightarrow e^{-\prime}e^{+\prime}}  & =\frac
{1}{\Delta\epsilon_{\pm,\omega}}\left(  \int_{\epsilon_{-}^{\prime}\in
\Delta\epsilon_{e,\omega}}d^{2}n_{\pm}^{\prime}J_{\mathrm{bs}}\frac
{\epsilon_{-}^{\prime2}\beta_{-}^{\prime}|M_{fi}|^{2}\hbar^{2}c^{2}%
}{16\epsilon_{-}\epsilon_{+}\epsilon_{+}^{\prime}}\right)+\nonumber\\&+\frac
{1}{\Delta\epsilon_{\pm,\omega}}\left(\int_{\epsilon_{+}^{\prime
}\in\Delta\epsilon_{e,\omega}}d^{2}n_{\pm}^{\prime}J_{\mathrm{bs}}%
\frac{\epsilon_{-}^{\prime}\beta_{-}^{\prime}|M_{fi}|^{2}\hbar^{2}c^{2}%
}{16\epsilon_{-}\epsilon_{+}}\right)  ,\\
(\chi E)_{\pm,\omega}^{e^{-}e^{+}\rightarrow e^{-\prime}e^{+\prime}}  &
=\frac{1}{\Delta\epsilon_{\pm,\omega}}\left(  \int_{\epsilon_{-}\in
\Delta\epsilon_{e,\omega}}d^{2}n_{\pm}^{\prime}J_{\mathrm{bs}}\frac
{\epsilon_{-}^{\prime}\beta_{-}^{\prime}|M_{fi}|^{2}\hbar^{2}c^{2}}%
{16\epsilon_{+}\epsilon_{+}^{\prime}}\right)+\nonumber\\ & +\frac{1}{\Delta\epsilon_{\pm,\omega}}\left(\int_{\epsilon_{+}\in\Delta
\epsilon_{e,\omega}}d^{2}n_{\pm}^{\prime}J_{\mathrm{bs}}\frac{\epsilon
_{-}^{\prime}\beta_{-}^{\prime}|M_{fi}|^{2}\hbar^{2}c^{2}}{16\epsilon
_{-}\epsilon_{+}^{\prime}}\right)  ,
\end{align}
where $d^{2}n_{\pm}^{\prime}=dn_{-}dn_{+}do_{-}^{\prime}$, $dn_{\pm}%
=d\epsilon_{\pm}do_{\pm}\epsilon_{\pm}^{2}\beta_{\pm}f_{\pm}$, and the
Jacobian is%
\begin{equation}
J_{\mathrm{bs}}=\frac{\epsilon_{+}^{\prime}\beta_{+}^{\prime}}{(\epsilon
_{-}^{\prime}+\epsilon_{+}^{\prime})\beta_{-}^{\prime}-(\epsilon_{-}\beta
_{-}\mathbf{b}_{-}+\epsilon_{+}\beta_{+}\mathbf{b}_{+})\mathbf{\cdot b}%
_{-}^{\prime}}.
\end{equation}

Analogously to the case of pair creation and annihilation in Section
(\ref{pair}) the energies of final state particles are given by (\ref{A27})
with the coefficients $\breve{A}=(\epsilon_{-}+\epsilon_{+})^{2}-(\epsilon
_{-}\beta_{-}\mathbf{b}_{-}\mathbf{\cdot b}_{-}^{\prime}+\epsilon_{+}\beta
_{+}\mathbf{b}_{+}\mathbf{\cdot b}_{-}^{\prime})^{2}$, $\breve{B}%
=(\epsilon_{-}+\epsilon_{+})\left[  m_e^{2}c^4+\epsilon_{-}\epsilon_{+}(1-\beta
_{-}\beta_{+}\mathbf{b}_{-}\mathbf{\cdot b}_{+})\right]  $, $\breve{C}=\left[
m_e^{2}c^4+\epsilon_{-}\epsilon_{+}(1-\beta_{-}\beta_{+}\mathbf{b}_{-}\mathbf{\cdot
b}_{+})\right]  ^{2}+m_e^{2}c^4\left[  \epsilon_{-}\beta_{-}\mathbf{b}%
_{-}\mathbf{\cdot b}_{-}^{\prime}+\epsilon_{+}\beta_{+}\mathbf{b}%
_{+}\mathbf{\cdot b}_{-}^{\prime}\right]  ^{2}$. In order to select the
correct root one has to check the condition (\ref{A34}) changing the
subscripts $1\rightarrow-$, $2\rightarrow+$.

\subsection{Three-body processes}\label{tripleint}

As we discussed above, for the collisional integrals for three-body interactions we assume that particles already reached kinetic equilibrium. In that case one can use the corresponding expressions, obtained in the literature for the thermal equilibrium case, and multiply the collisional integrals by the exponents, containing the chemical potentials of particles.

Emission coefficients for triple interactions in thermal equilibrium may be
computed by averaging of the differential cross-sections given in Section
\ref{chap-pair-theory} of the corresponding processes over the thermal
distributions of particles. Analytic results as a rule exist only for
nonrelativistic and/or ultrarelativistic cases. The only way to get approximate
analytical expressions is then to find the fitting formulas, connecting the two
limiting cases with reasonable accuracy. This work has been done by Svensson
\cite{1984MNRAS.209..175S}, also for reactions with protons, and in what
follows we adopt the emission and absorption coefficient for triple
interactions given in that paper.

Bremsstrahlung%
\begin{align}
\eta_{\gamma}^{e^{\mp}e^{\mp}\rightarrow e^{\mp}e^{\mp}\gamma}=(n_{+}%
^{2}+n_{-}^{2})\frac{16}{3}\frac{\alpha c}{\varepsilon}\left(  \frac{e^{2}%
}{m_e c^{2}}\right)  ^{2}\times\nonumber\\\times\ln\left[  4\xi(11.2+10.4\theta^{2})\frac{\theta
}{\varepsilon}\right]  \frac{\frac{3}{5}\sqrt{2}\theta+2\theta^{2}}%
{\exp(1/\theta)K_{2}(1/\theta)},
\end{align}%
\begin{align}
\eta_{\gamma}^{e^{-}e^{+}\rightarrow e^{-}e^{+}\gamma}=n_{+}n_{-}\frac{16}%
{3}\frac{2\alpha c}{\varepsilon}\left(  \frac{e^{2}}{m_e c^{2}}\right)  ^{2}%
\times\nonumber\\\times\ln\left[  4\xi(1+10.4\theta^{2})\frac{\theta}{\varepsilon}\right]
\frac{\sqrt{2}+2\theta+2\theta^{2}}{\exp(1/\theta)K_{2}(1/\theta)},
\end{align}%
where $\xi=e^{-0.5772}$, and $K_{2}(1/\theta)$ is the modified Bessel function
of the second kind of order 2.

Double Compton scattering%
\begin{align}
\eta_{\gamma}^{e^{\pm}\gamma\rightarrow e^{\pm\prime}\gamma^{\prime}%
\gamma^{\prime\prime}}=(n_{+}+n_{-})n_{\gamma}\frac{128}{3}\frac{\alpha
c}{\varepsilon}\times\nonumber\\\times\left(  \frac{e^{2}}{m_e c^{2}}\right)  ^{2}\frac{\theta^{2}%
}{1+13.91\theta+11.05\theta^{2}+19.92\theta^{3}},
\end{align}

Three photon annihilation%

\begin{equation}
\eta_{\gamma}^{e^{\pm}e^{\mp}\rightarrow\gamma\gamma^{\prime}\gamma
^{\prime\prime}}=n_{+}n_{-}\alpha c\left(  \frac{e^{2}}{m_e c^{2}}\right)
^{2}\frac{1}{\varepsilon}\frac{\frac{4}{\theta}\left(  2\ln^{2}2\xi
\theta+\frac{\pi^{2}}{6}-\frac{1}{2}\right)  }{4\theta+\frac{1}{\theta^{2}%
}\left(  2\ln^{2}2\xi\theta+\frac{\pi^{2}}{6}-\frac{1}{2}\right)  },
\end{equation}
where two limiting approximations \cite{1984MNRAS.209..175S} are joined together.

Radiative pair production%

\begin{equation}
\eta_{e}^{\gamma\gamma^{\prime}\rightarrow\gamma^{\prime\prime}e^{\pm}e^{\mp}%
}=\eta_{\gamma}^{e^{\pm}e^{\mp}\rightarrow\gamma\gamma^{\prime}\gamma
^{\prime\prime}}\frac{n_{\gamma}^{2}}{n_{+}n_{-}}\left[  \frac{K_{2}%
(1/\theta)}{2\theta^{2}}\right]  ^{2}.
\end{equation}

Electron-photon pair production%
\begin{equation}
\eta_{\gamma}^{e_{1}^{\pm}\gamma\rightarrow e_{1}^{\pm\prime}e^{\pm}e^{\mp}%
}=\left\{
\begin{array}
[c]{cc}%
(n_{+}+n_{-})n_{\gamma}\alpha c\left(  \frac{e^{2}}{m_e c^{2}}\right)  ^{2}%
\exp\left(  -\frac{2}{\theta}\right)  16.1\theta^{0.541}, & \theta\leq2,\\
(n_{+}+n_{-})n_{\gamma}\alpha c\left(  \frac{e^{2}}{m_e c^{2}}\right)
^{2}\left(  \frac{56}{9}\ln2\xi\theta-\frac{8}{27}\right)  \frac
{1}{1+0.5/\theta}, & \theta>2.
\end{array}
\right.
\end{equation}

The absorption coefficient for three-body processes is written as
\begin{equation}
\chi_{\gamma}^{\mathrm{3p}}=\eta_{\gamma}^{\mathrm{3p}}/E_{\gamma
}^{\mathrm{eq}}\,,
\end{equation}
where $\eta_{\gamma}^{\mathrm{3p}}$ is the sum of the emission coefficients of
photons in the three-particle processes, $E_{\gamma}^{\mathrm{eq}}%
=2\pi\epsilon^{3}f_{\gamma}^{\mathrm{eq}}/c^{3}$, where $f_{\gamma
}^{\mathrm{eq}}$ is given by (\ref{dk}).

From equation (\ref{EBoltzmannEq}), the law of energy conservation in the
three-body processes is
\begin{equation}
\int{\sum_{i}(\eta_{i}^{\mathrm{3p}}-\chi_{i}^{\mathrm{3p}}E_{i})d\mu
d\epsilon}=0\,.
\end{equation}
For exact conservation of energy in these processes the following
coefficients of emission and absorption for electrons are introduced:
\begin{equation}
\chi_{e}^{\mathrm{3p}}=\frac{\int(\eta_{\gamma}^{\mathrm{3p}}-\chi_{\gamma
}^{\mathrm{3p}}E_{\gamma})d\epsilon d\mu}{\int E_{e}d\epsilon d\mu},\qquad
\eta_{e}^{\mathrm{3p}}=0,\qquad\int(\eta_{\gamma}^{\mathrm{3p}}-\chi_{\gamma
}^{\mathrm{3p}}E_{\gamma})d\epsilon d\mu>0\,,
\end{equation}
or
\begin{equation}
\frac{\eta_{e}^{\mathrm{3p}}}{E_{e}}=-\frac{\int(\eta_{\gamma}^{\mathrm{3p}%
}-\chi_{\gamma}^{\mathrm{3p}}E_{\gamma})d\epsilon d\mu}{\int E_{e}d\epsilon
d\mu},\qquad\chi_{e}^{\mathrm{3p}}=0,\qquad\int(\eta_{\gamma}^{\mathrm{3p}%
}-\chi_{\gamma}^{\mathrm{3p}}E_{\gamma})d\epsilon d\mu<0\,.
\end{equation}

\subsection{Cutoff in the Coulomb scattering}\label{coulcut}

\label{cutoff}

Denote quantities in the center of mass (CM) frame with index $0$, and with
prime after interaction. Suppose there are two particles with masses $m_{1}$ and
$m_{2}$. The change of the angle of the first particle in CM system is
\begin{equation}
\theta_{10}=\arccos(\mathbf{b}_{10}\mathbf{\cdot b}_{10}^{\prime
}),\label{angle}%
\end{equation}
the numerical grid size is $\Delta\theta_{\mathrm{g}}$, the minimal angle at
the scattering is $\theta_{\mathrm{min}}$.

By definition in the in CM frame%
\begin{equation}
\mathbf{p}_{10}+\mathbf{p}_{20}=0,
\end{equation}
where%
\begin{equation}
\mathbf{p}_{i0}=\mathbf{p}_{i}+\left[  (\Gamma-1)(\mathbf{N}\mathbf{p}%
_{i})-\Gamma\frac{V}{c}\frac{\epsilon_{i}}{c}\right]  \mathbf{N},\quad i=1,2,
\end{equation}
and%
\begin{equation}
\epsilon_{i}=\Gamma(\epsilon_{i0}+\mathbf{V}\mathbf{p}_{i0}).
\end{equation}
Then for the velocity of the CM\ frame
\begin{equation}
\frac{\mathbf{V}}{c}=c\frac{\mathbf{p}_{1}+\mathbf{p}_{2}}{\epsilon
_{1}+\epsilon_{2}},\quad\mathbf{N}=\frac{\mathbf{V}}{V},\quad\Gamma=\frac
{1}{\sqrt{1-\left(  \frac{V}{c}\right)  ^{2}}}.
\end{equation}
By definition
\begin{equation}
\mathbf{b}_{10}=\mathbf{b}_{20},\quad\mathbf{b}_{10}^{\prime}=\mathbf{b}%
_{20}^{\prime},
\end{equation}
and then%
\begin{gather}
\left\vert \mathbf{p}_{10}\right\vert =\left\vert \mathbf{p}_{20}\right\vert
=p_{0}\equiv\nonumber\\
\equiv\frac{1}{c}\sqrt{\epsilon_{10}^{2}-m_{1}^{2}c^{4}}=\frac{1}{c}%
\sqrt{\epsilon_{20}^{2}-m_{2}^{2}c^{4}},
\end{gather}
where%
\begin{align}
\epsilon_{10}  & =\frac{(\epsilon_{1}+\epsilon_{2})^{2}-\Gamma^{2}(m_{2}%
^{2}-m_{1}^{2})c^{4}}{2(\epsilon_{1}+\epsilon_{2})\Gamma},\\
\epsilon_{20}  & =\frac{(\epsilon_{1}+\epsilon_{2})^{2}+\Gamma^{2}(m_{2}%
^{2}-m_{1}^{2})c^{4}}{2(\epsilon_{1}+\epsilon_{2})\Gamma}.
\end{align}
Haug \cite{1988A&A...191..181H} gives the minimal scattering angle in the
center of mass system%
\begin{equation}
\theta_{\mathrm{min}}=\frac{2\hbar}{m_ecD}\frac{\gamma_{r}}%
{(\gamma_{r}+1)\sqrt{2(\gamma_{r}-1)}},
\end{equation}
where the maximum impact parameter (neglecting the effect of protons) is
\begin{equation}
D=\frac{c^{2}}{\omega}\frac{p_{0}}{\epsilon_{10}},
\end{equation}
and the invariant Lorentz factor of relative motion (e.g.
\cite{1988A&A...191..181H}) is%
\begin{equation}
\gamma_{r}=\frac{1}{\sqrt{1-\left(  \frac{v_{r}}{c}\right)  ^{2}}}%
=\frac{\epsilon_{1}\epsilon_{2}-\mathbf{p}_{1}\cdot\mathbf{p}_{2}c^{2}}{m_{1}%
m_{2}c^{4}}.\label{gammarel}%
\end{equation}

Finally, in the CM frame%
\[
t_{\mathrm{min}}=2\left[  \left(  m_e c\right)  ^{2}-\left(  \frac{\epsilon_{10}%
}{c}\right)  ^{2}\left(  1-\beta_{10}^{2}\cos\theta_{\mathrm{min}}\right)
\right]
\]

Since it is invariant, $t$ in the denominator of $|M_{fi}%
|^{2}$ in (\ref{M_fi_e}) is replaced by the value
$t\sqrt{1+t_{\mathrm{min}}^{2}/t^{2}}$ to implement the cutoff scheme. Also at
the scattering of equivalent particles $u$ in the denominator of $|M_{fi}|^{2}$
in (\ref{M_fi_e}) is changed to the value
$u\sqrt{1+t_{\mathrm{min}}^{2}/u^{2}}$.%aga

\subsection{Numerical results}\label{numresp}

The results of numerical simulations are reported below. Two limiting initial
conditions with flat spectra are chosen: (i) electron--positron pairs with a
$10^{-5}$ energy fraction of photons and (ii) the reverse case, i.e., photons
with a $10^{-5}$ energy fraction of pairs. The grid consists of 60 energy
intervals and $16\times32$ intervals for two angles characterizing the
direction of the particle momenta. In both cases the total energy density is
$\rho=10^{24}\mbox{ erg}\cdot\mbox{cm}^{-3}$. In the first case initial
concentration of pairs is $3.1\cdot10^{29}\mbox{ cm}^{-3}$, in the second case
the concentration of photons is $7.2\cdot10^{29}\mbox{ cm}^{-3}$.
\begin{figure}[!ht]
\begin{center}
\includegraphics[width=15cm]{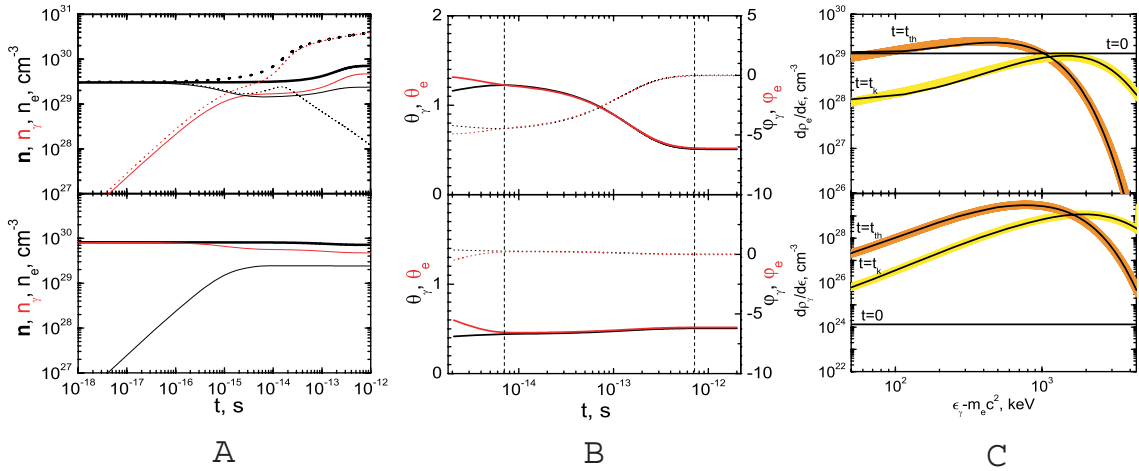}
\caption{{\bf A:} Dependence on time of concentrations of pairs (black), photons (red)
and both (thick) when all interactions take place (solid). Upper (lower)
figure corresponds to the case when initially there are mainly pairs
(photons). Dotted curves on the upper figure show concentrations when inverse
triple interactions are neglected. In this case an enhancement of the pairs
occurs with the corresponding increase in photon number and thermal
equilibrium is never reached. {\bf B:} Time dependence of temperatures, measured on the left axis (solid), and chemical potentials, measured on the right axis (dotted), of electrons
(black) and photons (red). The dashed lines correspond to the reaching of the
kinetic ($\sim10^{-14}$sec) and the thermal ($\sim10^{-12}$sec) equilibria.
Upper (lower) figure corresponds to the case when initially there are mainly
pairs (photons). {\bf C:} Spectra of pairs (upper figure) and photons (lower figure) when
initially only pairs are present. The black curve represents the results of
numerical calculations obtained successively at $t=0$, $t=t_{k}$ and
$t=t_{th}$ (see the text). Both spectra of photons and pairs are initially
taken to be flat. The yellow curves indicate the spectra obtained form
(\ref{dk}) at $t=t_{k}$. The perfect fit of the two curves is most evident in
the entire energy range leading to the first determination of the temperature
and chemical potential both for pairs and photons. The orange curves indicate
the final spectra as thermal equilibrium is reached.}
\label{ntot}
\end{center}
\end{figure}

In Fig.~\ref{ntot}, panel A concentrations of photons and pairs as well as
their sum for both initial conditions are shown. After calculations begin,
concentrations and energy density of photons (pairs) increase rapidly with
time, due to annihilation (creation) of pairs. Then, in the kinetic
equilibrium phase, concentrations of each component stay almost constant, and
the sum of concentrations of photons and pairs remains unchanged. Finally,
both individual components and their sum reach stationary values. If one
compares and contrasts both cases as reproduced in Fig.~\ref{ntot} A one can see
that, although the initial conditions are drastically different, in both cases
the same asymptotic values of the concentration are reached.

One can see in Fig.~\ref{ntot}, panel C that the spectral density of photons
and pairs can be fitted already at
$t_{k}\approx20t_{\mathrm{Cs}}\approx7\cdot10^{-15}$ sec by distribution
functions (\ref{dk}) with definite values of temperature
$\theta_{k}(t_{\mathrm{Cs}})\approx1.2$ and chemical potential $\phi_{k}%
(t_{k})\approx-4.5$, common for pairs and photons. As expected, after $t_{k}$
the distribution functions preserve their form (\ref{dk}) with the values of
temperature and chemical potential changing in time, as shown in
Fig.~\ref{ntot}, panel B. As one can see from this figure, the chemical
potential evolves with time and reaches zero at the moment $t_{\mathrm{th}%
}\approx\alpha^{-1}t_{k}\approx7\cdot10^{-13}$ sec , corresponding to the
final stationary solution. Condition (\ref{kineticT}) is satisfied in kinetic equilibrium.

The necessary condition for thermal equilibrium in the pair plasma is the
detailed balance between direct and inverse triple interactions. This point is
usually neglected in the literature where there are claims that the thermal
equilibrium may be established with only binary interactions
\cite{1983MNRAS.202..467S}. In order to demonstrate it explicitly
in Fig.~\ref{ntot}, panel A the dependence of concentrations of pairs and photons when
inverse triple interactions are artificially switched off is also shown. In this case (see
dotted curves in the upper part of Fig.~\ref{ntot}, panel A), after kinetic equilibrium
is reached concentrations of pairs decrease monotonically with time, and
thermal equilibrium is never reached.%

The existence of a non-null chemical potential for photons indicates the
departure of the distribution function from the one corresponding to thermal
equilibrium. Negative (positive) value of the chemical potential generates an
increase (decrease) of the number of particles in order to approach the one
corresponding to the thermal equilibrium state. Then, since the total number
of particles increases (decreases), the energy is shared between more (less)
particles and the temperature decreases (increases). Clearly, as thermal
equilibrium is approached, the chemical potential of photons is zero.

In this example with the energy density $10^{24}\mbox{ erg}\cdot\mbox{cm}^{-3}$
the thermal equilibrium is reached at $\sim 7\cdot10^{-13}$ sec with the final
temperature $T_{\mathrm{th}}=0.26$ MeV. For a larger energy density the
duration of the kinetic equilibrium phase, as well as of the thermalization
timescale, is smaller. Recall, that in entire temperature range the plasma is
nondegenerate.

The results, obtained for the case of an uniform plasma, can only be adopted
for a description of a physical system with dimensions $R_{0}\gg\frac
{1}{n\sigma_{T}}=4.3\;10^{-5}$cm.

The assumption of the constancy of the energy density is only valid if the
dynamical timescale $t_{dyn}=\left(  \frac{1}{R}\frac{dR}{dt}\right)  ^{-1}$
of the plasma is much larger than the above timescale $t_{th}$ which is indeed
true in all the cases of astrophysical interest.

Since thermal equilibrium is obtained already on the timescale $t_{th}%
\lesssim10^{-12}$sec, and such a state is independent of the initial
distribution functions for electrons, positrons and photons, the sufficient
condition to obtain an isothermal distribution on a causally disconnected
spatial scale $R>ct_{th}=10^{-2}$cm is the request of constancy of the energy
density on such a scale as well as, of course, the invariance of the physical laws.

To summarize, the evolution of an initially nonequilibrium optically thick
electron--positron-photon plasma is considered up to reaching thermal
equilibrium. Starting from arbitrary initial conditions kinetic equilibrium is
obtained from first principles, directly solving the relativistic Boltzmann
equations with collisional integrals computed from QED matrix elements. The
essential role of direct and inverse triple interactions in reaching thermal
equilibrium is demonstrated. These results can be applied in the theories of
the early Universe and of astrophysical sources, where thermal equilibrium is
postulated at the very early stages. These results can in principle be tested
in laboratory experiments in the generation of electron--positron pairs.

%%%%%%%%%%%%%%%%%%%%%%%%%%%%%%%%%%%%%%%%%%%%%%%%%%%%%%%%%%%%%%%%%%%%%%%%%%%%%%%%%%%%%%%%%%%

\section{Concluding remarks}\label{remarks}

We have reviewed three fundamental quantum processes which have highlighted
some of the greatest effort in theoretical and experimental physics in last
seventy years. They all deal with creation and annihilation of
electron--positron pairs. We have followed the original path starting from the
classical works of Dirac, on the process $e^+e^-\rightarrow 2\gamma$, and the
inverse process, $2\gamma\rightarrow e^+e^-$, considered by Breit--Wheeler. We
have then reviewed the $e^+e^-$ pair creation in a critical electric field
$E_c=m_e^2c^3/(\hbar e)$ and the Sauter-Heisenberg-Euler-Schwinger description
of this process both in Quantum Mechanics and Quantum Electro-Dynamics. We have
also taken this occasion to reconstruct the exciting conceptual developments,
initiated by the Sauter work, enlarged by the Born-Infeld nonlinear
electrodynamical approach, finally leading to the Euler and Euler-Heisenberg
results. We were guided in this reconstruction by the memories of many
discussions of one of us (RR) with Werner Heisenberg. We have then reviewed the
latest theoretical developments deriving the general formula for pair
production rate in electric fields varying in space and in time,  compared with
one in a constant electric field approximation originally studied by Schwinger
within QED.  We also reviewed recent studies of pair production rates in
selected electric fields varying both in space and in time, obtained in the
literature using instanton and  JWKB methods. Special attention has been given
to review the pair production rate in electric fields alternating periodically
in time, early derived by  Brezin, Itzykson and Popov, and the nonlinear
Compton effect in the processes of electrons and photons colliding with laser
beams, studied by Nikishov and Narozhny. These theoretical results play an
essential role in Laboratory experiments to observe the pair production
phenomenon using laser technologies.

We then reviewed the different level of verification of these three processes
in experiments carried all over the world. We stressed the success of
experimental verification of the Dirac process, by far one of the most prolific
and best tested process in the field of physics. We also recalled the study of
the hadronic branch in addition to the pure electrodynamical branch originally
studied by Dirac, made possible by the introduction of $e^+e^-$ storage rings
technology. We then turned to the very exciting current situation which sees
possibly the Breit--Wheeler formula reaching its first experimental
verification. This result is made possible thanks to the current great
developments of laser physics. We reviewed as well the somewhat traumatic
situation in the last forty years of the heavy-ion collisions in Darmstadt and
Brookhaven, yet unsuccessfully attempting to observe the creation of
electron--positron pairs. We also reviewed how this vast experimental program
was rooted in the theoretical ideas of Zeldovich, Popov, Greiner and their
schools.

We have also recalled the novelty in the field of relativistic astrophysics
where we are daily observing the phenomenon of Gamma Ray Bursts
\cite{1999PhR...314..575P, 2005RvMP...76.1143P, 2006RPPh...69.2259M,
Ruffini2003, 2007AIPC..910...55R}. These bursts of photons occur in energy
range keV to MeV, last about one second and come from astrophysical sources
located at a cosmological distance \cite{1997Natur.387..783C,1997Natur.386..686V, 1998Natur.393...35K, 1998Natur.393...41H,
1998Natur.393...43R}. The energy released is up to $\sim 10^{55}$ ergs,
equivalent to all the energy emitted by all the stars of all the galaxies of
the entire visible Universe during that second. It is generally agreed that the
energetics of these GRB sources is dominated by a dense plasma of electrons,
positrons and photons created during the process of gravitational collapse
leading to a Black Hole, see e.g.
\cite{1975PhRvL..35..463D,1999A&A...350..334R,2000A&A...359..855R,1998A&A...338L..87P}
and references therein. The Sauter-Heisenberg-Euler-Schwinger vacuum
polarization process, we have considered in the first part of the report, is a
classic theoretical model to study the creation of an electron--positron
optically thick plasma. Similarly the Breit--Wheeler and the Dirac processes we
have discussed, are essential in describing the further evolution of such an
optically thick electron--positron plasma. The GRBs present an unique
opportunity to test new unexplored regime of ultrahigh energy physics with
Lorentz factor $\gamma\sim 100-1000$ and relativistic field theories in the
strongest general relativistic domain.

The aim in this report, in addition to describe the above mentioned three basic
quantum processes, has been to identify and review three basic relativistic
regimes dealing with an optically thin and optically thick electron--positron
plasma.

The first topic contains the basic results of the physics of black holes, of
their energetics and of the associated process of vacuum polarization. We
reviewed the procedures to generalize in a Kerr--Newman geometry the QED
treatment of Schwinger and the creation of enormous number of $10^{60}$
electron--positron pairs in such a process.

The second topic is the back reaction of a newly created electron--positron
plasma on an overcritical electric field. Again we reviewed the Breit--Wheeler
and Dirac processes applied in the wider context of the
Vlasov--Boltzmann--Maxwell equations. To discuss the back reaction of
electron--positron pair on external electric fields, we reviewed semi-classical
and kinetic theories describing the plasma oscillations using respectively the
Dirac-Maxwell equations and the Boltzmann--Vlasov equations. We also reviewed
the discussions of plasma oscillations damping due to quantum decoherence and
collisions, described by respectively the quantum Boltzmann--Vlasov equation
and Boltzmann--Vlasov equation with particle collisions terms. We particularly
addressed the study of the influence of the collision processes
$e^{+}e^{-}\rightleftarrows \gamma\gamma$ on the plasma oscillations in
supercritical electric field $E > E_c$.  After $10^{3-4}$ Compton times, the
oscillating electric field is damped to its critical value with a large number
of photons created. An equipartition of number and energy between
electron--positron pairs and photons is reached. For the plasma oscillation
with undercritical electric field $E \lesssim E_c$,  we recalled that
electron--positron pairs, created by the vacuum polarization process, move as
charged particles in external electric field reaching a maximum Lorentz factor
at finite length of oscillations, instead of arbitrary large Lorentz factors,
as traditionally assumed. Finally we point out some recent results which
differentiate the case $E>E_{c}$  from the one $E<E_{c}$ with respect to the
creation of the rest mass of the pair versus its kinetic energy. For $E>E_{c}$
the vacuum polarization process transforms the electromagnetic energy of the
field mainly in the rest mass of pairs, with moderate contribution to their
kinetic energy. Such phenomena, certainly fundamental on astrophysical scales,
may become soon directly testable in the verification of the Breit--Wheeler
process tested in laser experiments in the laboratory.

As the third topic we have reviewed the recent progress in the understanding of
thermalization process of an optically thick electron--positron-photon plasma.
Numerical integration of relativistic Boltzmann equation with collisional
integrals for binary and triple interactions is used to follow the time
evolution of such a plasma, in the range of energies per particle between 0.1
and 10 MeV, starting from arbitrary nonequilibrium configuration. It is
recalled that there exist two types of equilibria in such a plasma: kinetic
equilibrium, when all particles are at the same temperature, but have different
nonzero chemical potential of photons, and thermal equilibrium, when chemical
potentials vanish. The crucial role of direct and inverse binary and triple
interactions in reaching thermal equilibrium is emphasized.

In a forthcoming report we will address how the above mentioned three
relativistic processes can be applied to a variety of astrophysical systems
including neutron stars formation and gravitational collapse, supernovae
explosions and GRBs.

This report is dedicated to the progress of theoretical physics in extreme
regimes of relativistic field theories which are on the verge of finding their
experimental and observational verification in physics and astrophysics. It is
then possible from our review and the many references we have given to gain a
basic understanding of this new field of research. The three topics which we
have reviewed are closely linked to the three quantum processes currently being
tested in precision measurement in the laboratories. The experiments in the
laboratories and the astrophysical observations cover complementary aspects
which may facilitate a deeper and wider understanding of the nuclear and laser
physics processes, of heavy-ion collisions as well as neutron stars formation
and gravitational collapse, supernovae and GRBs phenomena. We shall return on
such an astrophysical and observational topics in a dedicated forthcoming
report.

\vskip1cm
\begin{center}
*\hskip3cm *\hskip3cm *
\end{center}

\vskip1cm

We are witnessing in these times some enormous experimental and observational
successes which are going to be the natural ground to test some of the
theoretical works which we have reviewed in this report. Among the many
experimental progresses being done in particle accelerators worldwide we like
to give special mention to two outstanding experimental facilities which are
expected to give results in the forthcoming years. We refer here to the
National Ignition Facility at the Lawrence Livermore National Laboratory to be
soon becoming operational, see e.g. \cite{2009Sci...324..326C} as well as the
corresponding facility in France, the Mega Joule project
\cite{2006JPhy4.133..631G}.

In astrophysics these results will be tested in galactic and extragalactic
black holes observed in binary X-ray sources, active galactic nuclei,
microquasars and in the process of gravitational collapse to a neutron star and
also of two neutron stars to a black hole in GRBs. The progress there is
equally remarkable. In the last few days after the completion of this report
thanks to the tremendous progress in observational technology for the first
time a massive hypergiant star has been identified as the progenitor of the
supernova SN 2005gl \cite{Gal-Yam2009}. In parallel the joint success of
observations of the flotilla of X-ray observatories and ground-based large
telescopes \cite{Ruffini2009} have allowed to identify the first object ever
observed at $z\approx8$ the GRB090423 \cite{2009arXiv0905.0001P}.

To follow the progress of this field we are planning a new report which will be
directed to the astrophysical nature of the progenitors and the initial
physical conditions leading to the process of the gravitational collapse. There
the electrodynamical structure of neutron stars, the phenomenon of the
supernova explosion as well as theories of Gamma-Ray Bursts (GRBs) will be
discussed. Both in the case of neutron stars and the case of black holes there
are fundamental issues still to be understood about the process of
gravitational collapse especially with the electrodynamical conditions at the
onset of the process. The major difficulties appear to be connected with the
fact that all fundamental interactions, the gravitational, the electromagnetic,
the strong, the weak interactions appear to participate in essential way to
this process which appear to be therefore one of the most interesting
fundamental process of theoretical physics. Current progress is presented in
the following works \cite{2007IJMPD..16....1R, 2008AIPC..966..147R,
2008arXiv0804.3197R, 2008AIPC.1053..243R, 2008APS..APR8HE093R,
2008APS..APR8HE095R, 2009arXiv0903.3727P, 2009arXiv0903.4095R,
2009AIPC..........P, 2009AIPC..........R, 2009PhRvD........R,
2009PhRvDa.......R}. What is important to recall at this stage is only that
both the supernovae and GRBs processes are among the most energetic and
transient phenomena ever observed in the Universe: a supernova can reach energy
of $\sim 10^{52}$ ergs (hypernovae) on a time scale of a few months and GRBs
can have emission of up to $\sim 10^{55}$ ergs \cite{2009Sci...323.1688A} in a
time scale as short as of a few seconds. The central role in their description
of neutron stars, for supernova as well as of black holes and the
electron--positron plasma discussed in this report, for GRBs, are widely
recognized. The reason which makes this last research so important can be seen
in historical prospective: the Sun has been the arena to understand the
thermonuclear evolution of stars \cite{1968QB464.B46......}, Cyg X-1 has
evidenced the gravitational energy role in explaining an astrophysical system
\cite{2003RvMP...75..995G}, the GRBs are promising to prove the existence for
the first time of the ``blackholic energy''. These three quantum processes
described in our report reveal the basic phenomena in the process of
gravitational collapse predicted by Einstein theory of General Relativity
\cite{Ruffini2009}.

\vspace{0.3in}

\section*{Notes added in proof}

\begin{itemize}

\item
In Ref. \cite{1998PhRvD..58j5022D} two terms related to the initial and final frequencies are missing in the effective action (real part and imaginary part). These have been corrected in \cite{2008PhRvD..78j5013K,2009arXiv0910.3363K}.

\item
Some new results were obtained for QED in quasi constant in time electric field, in particular, the distributions of particles created is discussed in \cite{1996PhRvD..53.7162G,2008NuPhB.795..645G}; consistency restrictions on maximal electric field strength in QFT are discussed in \cite{2008PhRvL.101m0403G}. One-loop energy-momentum tensor in QED is obtained in \cite{2008PhRvD..78d5017G}. The exact rate of pair production by a smooth potential proportional to $tanh(kz)$ in three dimensions is obtained in \cite{2009PhRvD..80f5010C}.

\item
A recent review on the muon anomalous magnetic moment (muon g-2), offers the possibility, by making most precise measurement of muon g-2 in low-energies, to infer virtual hadronic vacuum polarization and light-by-light scattering effects due to virtual quark-pairs production in high-energies. In recent BNL E821 experiment\footnote{http://www.g-2.bnl.gov}, the muon anomalous magnetic moment can be rather accurately measured. In the theory of Standard Model for elementary particles, the muon anomalous magnetic moment $a_\mu$ receive leptonic QED-contributions $(e,\mu,\tau)$ to $a_\mu^{\rm QED}$, which has been calculated, see for a review \cite{2009PhR...477....1J}, up to 5-loop contributions ${\mathcal O}(\alpha^5)$
\begin{equation}
a_\mu^{\rm QED}\sim663(20)(4.6)\left(\frac{\alpha}{\pi}\right)^5.\nonumber
\end{equation}
While hadronic ($u,d,sc,\cdot\cdot\cdot$) contributions to the muon anomalous magnetic moment $a_\mu^{\rm had}$ contain, due to the strong interaction, both perturbative and non-perturbative parts, the ${\mathcal O}(\alpha^2)$ contribution to $a_\mu^{\rm had}$,
\begin{equation}
a_\mu^{(4)}({\rm vap,had})=\left(\frac{\alpha m_\mu}{3\pi}\right)^2\left(\int_{m_{\pi^0}^2}^{E_{\rm cut}^2}ds\frac{R_{\rm had}^{\rm data}(s){\hat K}(s)}{s^2}+\int_{E_{\rm cut}^2}^\infty ds\frac{R_{\rm had}^{\rm pQCD}(s){\hat K}(s)}{s^2}\right),\nonumber
\end{equation}
where $R(s)$ is given by
\begin{equation}
R_{\rm had}(s)=\sigma(e^+e^-\rightarrow {\rm hadrons})/\frac{4\pi\alpha(s)^2}{3s},\nonumber
\end{equation}
and $K(s)$ is the vacuum polarization contribution,
\begin{equation}
K(s)=\int_0^1 dx\frac{x^2(1-x)}{x^2+(s/m_\mu^2)(1-x)},\nonumber
\end{equation}
and a cut $E_{\rm cut}$ in the energy, separating the non-perturvative part to be evaluated from data and the perturbative high energy tail to be calculated using perturvative QCD (pQCD), analogously to QED-calculations. The pQCD calculations may only be trusted above 2 GeV and away from threshold and resonances. In the report \cite{2009PhR...477....1J}, authors resort to a semiphenomenological approach using dispersion relations together with the optical theorem and experimental data. The ``measurements of $R_{\rm had}(s)$'' get more difficult as increasing energies more and more channels open for meson-resonances. In addition, the most problematic set of hadronic corrections is that related to hadronic light-by-light scattering, which for the first time show up at order ${\mathcal O}(\alpha^3)$ via the diagrams with insertion of a box with four photon lines. As a contribution to the anomalous magnetic moment three of four photons are virtual and to be integrated over all four-momentum space, such that a direct experimental input for the non-perturbative dressed four-photon correlator is not available. In this case one has to resort the low energy effective description of QCD like {\it chiral perturbation theory} (CHPT) extended to include vector-mesons, which is reviewed in detail \cite{2009PhR...477....1J}.
Furthermore, the Electroweak corrections of weak virtual process including intermediate bosons $W^\pm$ and $Z$ to $g-2$ are important and now almost three standard deviations, and without it the deviation between theory and experiment would be the $6\sigma$ level. The test of the weak contribution is actually one of the milestones achieved by Brookhaven experiment E821 (see the footnote above). These studies and experiments are crucial to include the contributions from all known particles and interactions such that from a possible deviation between theory and experiment we may get a hint of the yet unknown physics.

\item
Some recent results on $e^+e^-$ annihilation cross sections in the GeV region are obtained in the following experiments: KLOE \cite{2005PhLB..606...12K,2009PhLB..670..285K,2008NuPhS.181..106K}, CMD-2 \cite{2004PhLB..578..285A,2009NuPhS.189..239L} and SND \cite{2009NuPhS.189..239L}.

\item
We would like to point out that in our report, one-loop vacuum polarization and light-by-light scattering effects (effective Euler-Heisenberg Lagrangian), as well as their high-order corrections in low-energies are considered in Section \ref{qedpair}; non-linear Compton effect is discussed in both Section \ref{lighttheoryquantum} (theory) and Sections \ref{Xray},\ref{light} (experiments); the Breit-Wheeler cutoff in high-energy $\gamma$-rays for astrophysics in Section \ref{BWastro}. All these discussions are limited in the leptonic sector for low-energies. The reason of recording here these hadronic contributions in the high-energy region is motivated by our expectation that these effects will be possibly soon detected by direct measurements of gamma ray emission from high-energy astrophysical processes. We shall return on this topic in our forthcoming report already mentioned above.

\end{itemize}

\vspace{0.3in}

\textbf{Acknowledgment.}

We would like to thank A.G. Aksenov, G. Altarelli, D. Arnett, A. Ashtekar, G. Barbiellini, V.A. Belinski, C.L. Bianco, D. Bini, P. Chardonnet, C. Cherubini, F. Fraschetti, R. Giacconi, W. Greiner, F. Guerra, R. Jantzen, G. 't Hooft, I.B. Khriplovich, H. Kleinert, P. Menotti, S. Mercuri, N.B. Narozhny, A. Pelster, V.S. Popov, G. Preparata, B. Punsly, J. Rafelski, J. Reihardt, A. Ringwald, M. Rotondo, J. Rueda, M. Testa, L. Titarchuk, L. Vitagliano, J.R. Wilson for many important discussions.

\end{document}